%% file: Paper2014.tex
\documentclass[11pt]{article}
\usepackage{etex}
\usepackage[T2A,T1]{fontenc}
\usepackage[ansinew]{inputenc}
\usepackage[russian,ngerman,english]{babel}
\usepackage{lmodern}
\usepackage[toc,page]{appendix}
\usepackage{tocbasic}
\usepackage[german]{authblk}
\usepackage{a4wide}
\usepackage{setspace}
\usepackage{array}
\usepackage{enumitem}
\usepackage{booktabs}
\usepackage{fancyhdr}
\usepackage{framed}
\usepackage{lettrine}
\usepackage{longtable}
\usepackage{marvosym}
\usepackage{microtype}
\usepackage{multicol}
\usepackage{multirow}
\usepackage{pbox}
\usepackage{rotating}
\usepackage{arydshln}
\usepackage{amsmath}
\usepackage{amssymb}
\usepackage{amsthm}
\usepackage{amsfonts}
\usepackage{bm}
\usepackage{braket}
\usepackage{dsfont}
\usepackage{latexsym}
\usepackage{mathrsfs}
\usepackage[sumlimits,intlimits]{mathtools}
\usepackage{stmaryrd}
\usepackage{slashed}
\usepackage{tensor}
\usepackage{units}
\usepackage{yfonts}
\usepackage{aurical}
\usepackage{suetterl}
\usepackage{pgothic}
\usepackage{upgreek}
\usepackage{empheq}
\usepackage{graphicx}
\usepackage[arrow, matrix, curve]{xy}
\usepackage{color}
\usepackage{float}
\usepackage[lflt]{floatflt}
\usepackage{picture}
\usepackage{wrapfig}
\usepackage[hang]{subfigure}
\usepackage{tikz}
\usetikzlibrary{3d}
\usetikzlibrary{calc}
\usetikzlibrary{external}
\usetikzlibrary{intersections}
\usetikzlibrary{matrix}
\usetikzlibrary{trees}
\usepackage{epstopdf}
\usepackage{pdfpages}
\usepackage[
square,comma,numbers]{natbib}
\usepackage{url}
\usepackage{feynmp}
\usepackage{lipsum}
\usepackage{hyperref}
\hypersetup{
    bookmarks=true,         
    unicode=true,          
    pdftoolbar=true,        
    pdfmenubar=true,        
    pdffitwindow=false,     
    pdfstartview={FitH},    
    pdftitle={My title},    
    pdfauthor={Author},     
    pdfsubject={Subject},   
    pdfcreator={Creator},   
    pdfproducer={Producer}, 
    pdfkeywords={keywords}, 
    pdfnewwindow=true,      
    colorlinks=true,       
    linkcolor=red,          
    citecolor=blue,        
    filecolor=magenta,      
    urlcolor=cyan           
}
\usepackage[{
		a4paper, 
		top=25mm, 
		left=25mm, 
		right=25mm, 
		bottom=25mm, 
		headsep=12mm, 
		footskip=12mm,
	}]{geometry}
\DeclareFontFamily{OT1}{pzc}{}
\DeclareFontShape{OT1}{pzc}{m}{it}{<-> s * [1.10] pzcmi7t}{}
\DeclareMathAlphabet{\mathpzc}{OT1}{pzc}{m}{it}

\newcommand{\LC}{\textnormal{\scriptsize \textsc{lc}}}

\newcommand{\intd}{\mathrm{d}}

\DeclareMathOperator{\tr}{tr}

\numberwithin{equation}{section}

\begin{document}

\pagestyle{empty}
\title{
\begin{flushright}
\normalsize{MITP/15-067}
\vspace{1cm}
\end{flushright}
\bf The metric on field space, functional renormalization, and metric-torsion quantum gravity}
\author[]{Martin Reuter\thanks{reuter@thep.physik.uni-mainz.de}}
\author[]{Gregor M. Schollmeyer\thanks{schollmeyer@thep.physik.uni-mainz.de} }
\date{}
\affil[]{\small\textit{PRISMA Cluster of Excellence \& Institute of Physics,} \\ 
\textit{Johannes Gutenberg University, D-55099 Mainz, Germany}}
\maketitle

\begin{abstract}
Searching for new non-perturbatively renormalizable quantum gravity theories, functional renormalization group (RG) flows are studied on a theory space of action functionals depending on the metric and the torsion tensor, the latter parameterized by three irreducible component fields. A detailed comparison with Quantum Einstein-Cartan Gravity (QECG), Quantum Einstein Gravity (QEG), and ``tetrad-only'' gravity, all based on different theory spaces, is performed. It is demonstrated that, over a generic theory space, the construction of a functional RG equation (FRGE) for the effective average action requires the specification of a metric on the infinite-dimensional field manifold as an additional input. A modified FRGE is obtained if this metric is scale-dependent, as it happens in the metric-torsion system considered.
\end{abstract}
\clearpage

\tableofcontents
\clearpage

\pagestyle{plain}

\section{Introduction}\label{ch:intro}
\input{Intro}

\section{The FRGE and the metric in field space}\label{ch:fieldmetric}
\input{FRGEandMetric}

\section{The Metric-torsion theory space and its truncation}\label{ch:holst}
\input{HolstTruncation}

\section{Einstein-Hilbert Truncation with Torsion}\label{ch:EHTor}
\input{EHTruncation}

\section{Analysis of the RG flow}\label{sec:analysis}
\input{Results}

\section{Summary and Conclusions}\label{sec:summary}
\input{Summary}

\appendix
\input{AppendixComp}

\addcontentsline{toc}{section}{References}

\input{Bibliography}
\end{document}

%% file: Intro.tex
One of the most remarkable features of the Asymptotic Safety scenario is that, in the Effective Average Action (EAA) approach, it is completely independent of any prejudice about an underlying classical theory, in particular the nature of its degrees of freedom and their dynamical laws. Rather than a tool for quantizing a given classical system, the Asymptotic Safety program can be seen as a systematic way to search for microscopic theories (``bare actions'') which are non-perturbatively renormalizable at a non-Gaussian renormalization group (RG) fixed point. The very same device which is used to spot such fixed points, the functional RG equation, can also be used to find the RG trajectories originating there, each of which constitutes a consistent fundamental quantum field theory. For the time being this ``Background Independence of the second kind'', meaning that a priori no special action functional plays a distinguished role, is still somewhat obscured by the practical necessity to restrict ourselves to truncated calculations. Nevertheless, conceptually speaking the only piece of information that needs to be supplied to the EAA-based algorithm is the \textit{theory space}, i.e. the manifold of action functionals on which the coarse-graining flow generated by the functional RG equation (FRGE) lives. Once such a space of functionals is selected, consisting of actions with prescribed symmetry properties and depending on a given set of fields, at least in principle one can ``turn the crank'' and test whether this particular theory space can support asymptotically safe theories.

The basic idea behind the functional renormalization group is as follows: instead of by one single (and therefore complicated) integration, the quantum fluctuations to be integrated over in the pertinent path integral are taken into account in a piecewise, step-by-step manner. Applying this strategy directly at the level of the path integral, dividing it into several ``momentum shells'', Wilson pioneered this approach and the related theory of non-perturbative renormalization in the seventies \cite{Wilson:1974mb, Wilson:1973jj}. Instead of using a sharp momentum cutoff we might also use smooth regulator functions, and we can implement the underlying idea at the level of the effective action $\Gamma[\Phi]$ rather than bare action, $S[\Phi]$. Adding to $S[\Phi]$ a scale-dependent mode suppression term $\Delta S_{k}[\Phi]$ leads to the scale-dependent generalization of the effective action, the effective average action $\Gamma_{k}$. The mass scale $k$ acts as an effective IR cutoff: the contributions with momentum $p \lesssim k$ are suppressed, while all others are integrated out. The effective average action (EAA) satisfies an exact FRGE from which it can be computed, when an initial action is specified.

The existence of \textit{fundamental}, as opposed to effective quantum field theories is closely linked to the existence of \textit{complete} RG trajectories, i.e. solutions to the FRGE with both a well defined UV limit $(k \rightarrow \infty)$ and IR limit $(k \rightarrow 0)$. The theories which are non-perturbatively renormalizable by the Asymptotic Safety construction have their UV limit taken at a suitable non-Gaussian fix point (NGFP) in theory space such that they are free from divergences and remain predictive at arbitrarily short distances. 

Within the Asymptotic Safety approach we do not follow the traditional path of ``quantization'' which starts off from a classical hamiltonian system which we want to turn into a quantum mechanical one. At most we draw inspiration from classical systems as for their field content $\Phi$ and their symmetry or gauge group, $\textbf{G}$. These data constitute the key ingredients for fixing a specific ``theory space'':
\begin{equation}
	\mathcal{T} = \bigl\{ S:\Phi \mapsto S[\Phi] \ | \ S\ \text{invariant under} \ \textbf{G} \bigr\} \ .
\end{equation}
It contains all action functionals $S$ that depend on a given set of fields $\Phi$ and are invariant under the action of the transformation group $\textbf{G}$. The theory space $\mathcal{T}$ represents the arena on which the RG dynamics takes place. Geometrically speaking, the FRGE should be seen as defining a vector field on $\mathcal{T}$ whose integral curves are the ``RG trajectories'' $k \mapsto \Gamma_k$.

In our search for a suitable theory space of gravity the logical first step is to take advantage of the remarkably rich variety of variational principles giving rise to Einstein's equation, or equivalent equations expressed in terms of different field variables. The best known example is the Einstein-Hilbert action, either expressed in terms of the metric, $S_{\mathrm{EH}}[g_{µ \nu}]$, or the tetrad, $S_{\mathrm{EH}}[\tensor{e}{^{a}_{µ}}]$, the two variants being connected by $g_{µ \nu} = \eta_{a b} \tensor{e}{^{a}_{µ}} \tensor{e}{^{a}_{\nu}}$. In absence of spinning matter the first-order Hilbert-Palatini action $S_{\mathrm{HP}}[\tensor{e}{^{a}_{µ}},\tensor{\omega}{^{a b}_{µ}}]$, additionally depending on the spin connection $\tensor{\omega}{^{a b}_{µ}}$, offers another classically equivalent formulation. These settings, as well as various variations and generalizations thereof, are the basis for most modern investigations in quantum gravity. They constitute the main motivation for a detailed exploration of the Asymptotic Safety conjecture on the following, potentially relevant theory spaces:\\
\\
\textbf{(A) Einstein gravity.} In the standard metric description of gravity the dynamical field variable is the metric tensor $\Phi = g_{µ \nu}$, and the gauge group consists of the diffeomorphisms on the (Euclidean) spacetime manifold, $\textbf{G} = \mathrm{Diff}(\mathcal{M})$. This setting is referred to as Einstein Gravity; it is characterized by the theory space
\begin{equation}
	\mathcal{T}_{\mathrm{E}} = \bigl\{ S[g_{µ \nu}] \ | \ S \ \text{inv. under} \ \textbf{G} = \mathrm{Diff}(\mathcal{M}) \bigr\}.
\end{equation}
The original work on the EAA approach to asymptotically safe quantum gravity \cite{Reuter:1996cp}, as well as most further investigations of the Asymptotic Safety Scenario have been carried out on this theory space. For a general overview see the review articles \cite{Niedermaier:2006wt,Reuter:2012id,Ashtekar:2014kba}. Systematic extensions of this space include the addition of matter fields \cite{Dou:1997fg,Percacci:2002ie,Percacci:2003jz,Vacca:2010mj,Dona:2013qba} and the incorporation of surface terms \cite{Becker:2012js}. Besides these ``single metric'' calculations also bimetric generalizations were analyzed \cite{Manrique:2009uh,Manrique:2010mq,Manrique:2010am,Becker:2014qya}. In all of these studies a NGFP has indeed been found, allowing for a consistent Asymptotic Safety construction.\\
\\
\textbf{(B) Tetrad gravity.} Switching from the metric to the tetrads, due to the relation between the metric and the vielbeins not being unique, there exists an entire $\mathrm{O}(d)$ manifold of tetrads that correspond to the same metric. Therefore the pertinent action functionals admit an additional $\mathrm{O}(d)_{\mathrm{loc}}$ gauge freedom. Thus the ``tetrad-only'' theory space is:
\begin{equation}
	\mathcal{T}_{\mathrm{tet}} = \bigl\{ S[\tensor{e}{^{a}_{µ}}] \ | \ S \ \text{inv. under} \ \textbf{G} = \mathrm{Diff}(\mathcal{M}) \ltimes \mathrm{O}(d)_{\mathrm{loc}} \bigr\}.
\end{equation}
First RG explorations of this theory space have been performed in \cite{Harst:2012ni,Harst:2012phd}. \\
\\
\textbf{(C) Einstein-Cartan gravity.} The classical Einstein-Cartan theory suggests to generalize the tetrad-only description of gravity by treating the spin connection $\tensor{\omega}{^{a b}_{µ}}$ as an independent field variable, while leaving the symmetry group unchanged. Hence we are motivated to analyze the RG flow on the theory space
\begin{equation}
	\mathcal{T}_{\mathrm{EC}} = \bigl\{ S[\tensor{e}{^{a}_{µ}}, \tensor{\omega}{^{a b}_{µ}}] \ | \ S \ \text{inv. under} \ \textbf{G} = \mathrm{Diff}(\mathcal{M}) \ltimes \mathrm{O}(d)_{\mathrm{loc}} \bigr\}.
\end{equation}
Using two different functional RG equations, non-perturbative RG investigations on this theory space have been conducted in \cite{Daum:2010phd,Daum:2011bs,Daum:2010qt,Daum:2013fu} and \cite{Harst:2012phd,Harst:2014vca}, respectively\footnote{In \cite{Harst:2012phd,Harst:TBA} a variant restricting the spin connection to be (anti-)selfdual has also been carried out.}. Both RG analyses on $\mathcal{T}_{\mathrm{EC}}$ shared the same truncation, which was inspired by the classical Holst action \cite{Holst:1995pc}. It is defined in 4 dimensions only:
\begin{equation}
	S_{\mathrm{Ho}} = - \frac{1}{16 \pi G} \int \mathrm{d}^{4}x \ e \left[ \tensor{e}{_{a}^{µ}} \tensor{e}{_{b}^{\nu}} \left( \tensor{F}{^{ab}_{µ \nu}} - \frac{1}{2 \gamma} \tensor{\epsilon}{^{ab}_{cd}} \tensor{F}{^{cd}_{µ \nu}} \right) - 2 \Lambda \right] .
\end{equation} 
It generalizes the Hilbert-Palatini action of classical Einstein-Cartan gravity by the addition of a third field monomial which introduces the Immirzi parameter $\gamma$ as a new free constant. This term vanishes on spacetimes without torsion and hence has no counterpart in metric or tetrad-only gravity. 

The Holst action is the starting point for several approaches to quantum gravity, most importantly the canonical quantization in Ashtekar's variables \cite{Ashtekar:2004eh,Rovelli:2004tv}, Loop Quantum Gravity (LQG) \cite{Thiemann:2008} and spin foam models \cite{Perez:2003vx}. Particularly in LQG the dimensionless Immirzi parameter $\gamma$ plays a distinguished role, as it enters the spectrum of area and volume operators and thus affects the entropy of black holes \cite{Rovelli:2004tv}. If we include fermionic matter in the action the fermions act as a source of torsion in the classical field equations, and the presence of the Immirzi term then gives rise to a CP violating four-fermion interaction. The strength of this interaction depends on the value of the Immirzi parameter, thus leading to an (at least in principle) observable effect caused by the Immirzi term \cite{Percacci:2002ie}. 

A fully non-perturbative investigation of a Holst-type truncation in the Asymptotic Safety context is of great interest therefore and may help to shed some light on the role played by the Immirzi parameter\footnote{Effects of the Holst action in perturbative quantum gravity have been explored in \cite{Benedetti:2011nd,Benedetti:2011yb} and \cite{Shapiro:2014kma} to 1-loop order.}. In the EAA approach the Immirzi parameter is treated as a scale dependent coupling that generically will display a non-trivial RG running.\\
\\
\textbf{(D) Torsion gravity.} Switching back to the metric formalism, a different extension is to regard the torsion tensor $\tensor{T}{^{\lambda}_{µ \nu}}$ as an additional independent field variable. This is analogous to the Einstein-Cartan theory space, treating however the torsion part of the connection as the additional field variable\footnote{To make the analogy even closer we could alternatively take the contorsion tensor $\tensor{K}{^{\lambda}_{µ \nu}}$ rather than $\tensor{T}{^{\lambda}_{µ \nu}}$ as the new field variable.}. As we do not introduce a new redundancy here, such as the $\mathrm{O}(d)_{\mathrm{loc}}$ invariance in the case of tetrad gravity, the gauge group is the same as in Einstein gravity. This leads to the following theory space:
\begin{equation}
	\mathcal{T}_{\mathrm{tor}} = \bigl\{ S[g_{µ \nu}, \tensor{T}{^{\lambda}_{µ \nu}}] \ | \ S \ \text{inv. under} \ \textbf{G} = \mathrm{Diff}(\mathcal{M})\bigr\}.
\end{equation} 
For physical clarity and to simplify concrete calculations it is actually advantageous to further decompose the torsion field.\footnote{For the additional inclusion of non-metricity and the corresponding invariants see \cite{Pagani:2015ema}.} \\
\\
\textbf{(E) Decomposed torsion gravity.} Starting from the full torsion tensor, we can make use of the familiar decomposition \cite{Shapiro:2001rz}
\begin{equation}
	\tensor{T}{^{\lambda}_{µ \nu}} = \frac{1}{3} \left( \tensor{\delta}{^{\lambda}_{\nu}} T_{µ} - \tensor{\delta}{^{\lambda}_{µ}} T_{\nu} \right) + \frac{1}{6} \frac{\tensor{\epsilon}{^{\lambda}_{µ \nu \alpha}}}{\sqrt{g}}  S^{\alpha} + \tensor{q}{^{\lambda}_{µ \nu}}
	\label{eq:Tdecompfirsttime}
\end{equation}
and use the irreducible pieces $\left( S_µ , T_µ , q_{\lambda µ \nu} \right)$, rather than $\tensor{T}{^\lambda_{µ \nu}}$, as the independent field variables which accompany the metric $g_{µ \nu}$.
We denote the corresponding theory space by
\begin{equation}
	\mathcal{T}_{\mathrm{dtor}} = \bigl\{ S[g_{µ \nu}, S_{µ}, T_{µ}, \tensor{q}{_{\lambda µ \nu}}] \ | \ S \ \text{inv. under} \ \textbf{G} = \mathrm{Diff}(\mathcal{M}) \bigr\}.
\end{equation}
It is this theory space, $\mathcal{T}_{\mathrm{dtor}}$, which we are going to explore in the present paper, carrying out a first EAA analysis in the $(S,T,q)$ field basis. 

The purpose of the present paper is twofold. It has both a concrete physical motivation and a much more general, technical one:

\noindent \textbf{(i)} By considering a Holst-type truncation in metric variables including torsion, $\Phi \equiv \left\{ g_{µ \nu}, S_µ , T_µ , q_{\lambda µ \nu} \right\}$, we hope not only to further illuminate the impact of the Immirzi term on the Newton and cosmological coupling, but also to extend the existing picture of potentially relevant theory spaces and thus gain a better understanding of the field parametrization dependence. This work is in the same vein as the studies on $\mathcal{T}_{\mathrm{tet}}$ carried out in \cite{Harst:2012phd,Harst:2012ni}, where the question whether differences between results obtained from dissimilar theory spaces are mainly due to the use of the different truncations, different field variables, or even both, was likewise considered.

\noindent \textbf{(ii)} In order to write down functional flow equations on the various theory spaces it is necessary to generalize them by allowing the corresponding actions to depend on non-dynamical background fields also, as well as on the Faddeev-Popov ghosts needed to deal with gauge symmetries. As a result, the actions in $\mathcal{T}_{\mathrm{dtor}}$, for instance, are actually functionals not only of the ``dynamical'' fields $\bigl\{g_{µ \nu}, S_{µ}, T_{µ}, \tensor{q}{_{\lambda µ \nu}}\bigr\}$, but also the corresponding background fields $\bigl\{\bar{g}_{µ \nu}, \bar{S}_{µ}, \bar{T}_{µ}, \tensor{\bar{q}}{_{\lambda µ \nu}}\bigr\}$, and the diffeomorphism ghosts.

As we shall see, the four basis fields $g_{µ \nu}$, $S_µ$, $T_µ$, and $q_{\lambda µ \nu}$, respectively, possess different canonical mass dimensions which prevents us from straightforwardly applying the usual FRGE of the EEA. In Section \ref{ch:fieldmetric} we demonstrate that a conceptually clean way of uniformizing the dimensions by means of a mass parameter, $\bar{µ}$, requires introducing a nontrivial metric on the infinite dimensional manifold of all fields. This discussion is completely general and applies to arbitrary theory spaces. In particular we derive the correct form of the FRGE for metrics on field space that are scale dependent. 

Then, in Section \ref{ch:holst}, the classical prerequisites of the gravity-torsion theory space are introduced and in Section \ref{ch:EHTor} the FRGE is evaluated for a truncation inspired by the Holst action, and the corresponding $\beta$-functions are obtained. In Section \ref{sec:analysis} we describe the properties of the resulting RG flow and perform a detailed comparison with earlier work on Quantum Einstein-Cartan Gravity (QECG), ``tetrad-only'' gravity, and Quantum Einstein Gravity (QEG), all of which are based on different theory space, employing different sets of fundamental fields. Finally, Section \ref{sec:summary} contains a brief summary and the conclusions. 

Parts of the material are relegated to various appendices; in particular the phase portraits for the various parameter choices discussed in the main part of the paper are collected systematically in Appendix \ref{sec:PP}.

%% file: FRGEandMetric.tex
The main tool in our search for an asymptotically safe quantum field theory of gravity is the functional renormalization group equation for the Effective Average Action. For a generic theory space, involving a multiplet of dynamical fields $\Phi$ and corresponding background fields $\bar{\Phi}$, it is usually written as
\begin{equation}
	\partial_t \Gamma_{k} [\Phi,\bar{\Phi}] = \frac{1}{2} \operatorname{STr} \left[ \left(\Gamma_{k}^{(2)}[\Phi,\bar{\Phi}] + \mathcal{R}_k [\bar{\Phi}] \right)^{-1} \partial_t \mathcal{R}_k [\bar{\Phi}] \right] \ .
\end{equation}
Here the operatorial supertrace ``$\operatorname{STr}$'' includes a sum over all field components, and in the case of Grassmann-odd fields, takes care of the additional minus sign. A central role is played by the Hessian operator $\Gamma_{k}^{(2)}$, which is related to, but not identical with the second functional derivative of $\Gamma_k$ with respect to $\Phi$. Indeed, as we shall discuss in detail, special attention has to be paid to its construction. While it is straightforward (albeit often quite tedious) to differentiate $\Gamma_k$ twice with respect to the dynamical fields, the hidden difficulty lies in defining the actual Hessian \textit{operator}. 

In this work we will highlight the conceptual aspects of its construction which, in simpler theories, are usually overlooked or brushed over. In particular, as an illustrative model calculation, we determine the $\beta$-functions for the RG flow on $\mathcal{T}_{\mathrm{dtor}}$. As it will turn out, they are explicitly dependent on how precisely the Hessian operator is related to the \textit{quadratic form} constituted by the second functional derivatives of $\Gamma_k$. 

As a necessary preparation we briefly discuss, in subsection \ref{subsec:fieldspace}, metrics in the space of fields $\Phi$, before we describe the relationship between the Hessian operator and the underlying quadratic form in subsection \ref{subsec:QFvsOp}. As a new result, we obtain a generalized form of the FRGE which is valid also \textit{when the metric on field space is $k$-dependent}. This will indeed be the relevant case later on.

\subsection[The field space \texorpdfstring{$\mathcal{F}$}{F}]{The field space \texorpdfstring{$\bm{\mathcal{F}}$}{F}}\label{subsec:fieldspace}
Because our focus in this section is on formal properties that apply to any quantum or statistical field theory, it is convenient to adopt a condensed index notation following DeWitt \cite{DeWitt:1965jb}. In this convention the field variables are denoted $\varphi^{i}$ where the index $i$ stands not only for all discrete labels of the field, but also the spacetime argument. Consider for example a gauge field $\tensor{A}{^{a}_{µ}}(x)$ whose labels are a group index $a$ and a spacetime index $µ$ in conventional notation. Then, in condensed form, $\varphi^{i} \leftrightarrow \tensor{A}{^{a}_{µ}}(x)$ where the index $i$ stands for the tuple $i \leftrightarrow (a,µ,x)$. The standard summation convention is extended to include an integration over repeated continuous labels. For example, an expression such as $M_{ij}\varphi^{i}\varphi^{j}$ is shorthand for
\begin{equation}
	M_{ij}\varphi^{i}\varphi^{j} = \int \intd^d x \int \intd^d x' \tensor{M}{_{ab}^{µ \nu}}(x,x') \tensor{A}{^{a}_{µ}}(x)\tensor{A}{^{b}_{\nu}}(x') \ .
\end{equation}

At the purely formal level it is straightforward to generalize the notion of manifolds to the case of infinite dimensionality. In this sense we regard fields $\varphi^{i}$ as local coordinates of points in a field manifold $\mathcal{F}$, analogous to $x^µ$ being local coordinates of points in a spacetime manifold $\mathcal{M}$. Thus we are led to investigate the geometry of field space and, for example, to introduce the notion of a metric on $\mathcal{F}$, as well as the ensuing geometrical structures, such as curvature, for instance. A metric on the space of fields, $\mathcal{F}$, amounts to a line element
\begin{equation}
	\intd \text{\Large{\Fontlukas s}}^2 = \text{\Fontlukas G}_{ij} \, \intd \varphi^{i} \, \intd \varphi^{j} 
\end{equation} 
when written in the ``coordinate'' basis. The metric defines an inner product
\begin{equation}
	\left( v_1 , v_2 \right) \equiv \text{\Fontlukas G}_{ij} \, v_1^i \, v_2^j \ ,
	\label{eq:innerproduct2}	
\end{equation}
which in particular applies to the fibers of the cotangent bundle over $\mathcal{F}$,
\begin{equation}
	\left( \intd \varphi , \intd \psi \right) \equiv \text{\Fontlukas G}_{ij} \, \intd \varphi^i \, \intd \psi^j \ .
\label{eq:innerproduct}
\end{equation}
The consequences and applications of a field space  metric have been studied by DeWitt \cite{DeWitt:1967yk,DeWitt:1967ub,DeWitt:1967uc} and Vilkovisky \cite{Vilkoviskii:1984un,Vilkovisky:1984st} in the context of canonical gravity and the ``universal off-shell effective action'' of general gauge theories; a comprehensive review can be found in \cite{Parker:2009uva}.

If, for instance, $\mathcal{F}$ is the space of all metrics on a given spacetime manifold $\mathcal{M}$, we identify $\varphi^{i} \leftrightarrow g_{µ \nu}(x)$, and the ensuing line element reads
\begin{equation}
	\intd \text{\Large{\Fontlukas s}}^2 = \text{\Fontlukas G}_{ij} \, \intd \varphi^{i} \intd \varphi^{j} = \int_{\mathcal{M}} \intd^d x \int_{\mathcal{M}} \intd^d x' \, G^{µ \nu \rho \sigma}(x,x') \, \intd g_{µ \nu}(x) \, \intd g_{\rho \sigma}(x') \ .
\end{equation}
An important special case are the ultralocal metrics on $\mathcal{F}$. According to DeWitt \cite{DeWitt:1967yk,DeWitt:1967ub} they are given by the one-parameter family 
\begin{equation}
	G^{µ \nu \rho \sigma}(x,x') = \sqrt{g(x)} \left[ g^{µ(\rho}(x)g^{\sigma)\nu}(x) + \frac{c}{2} g^{µ \nu}(x)g^{\rho \sigma}(x) \right] \delta (x-x')
	\label{eq:vilkoviskydewittmetric}
\end{equation}
where $c$ is an arbitrary dimensionless constant. Note that the metric on $\mathcal{F}$ is ``position dependent'', i.e. $\text{\Fontlukas G}_{ij} \equiv \text{\Fontlukas G}_{ij}[g_{µ \nu}]$ exhibits a dependence on the field itself.

\subsection{Quadratic forms vs. operators in the FRGE}\label{subsec:QFvsOp}

In this subsection we discuss how the metric on field space can appear in the various building blocks that make up the functional RG equation.\\
\\
\textbf{(A)} We consider a generic theory with a partition function $Z[\bar{\Phi}] = \int \mathcal{D}\widehat{\Phi}\, e^{-S[\hat{\Phi},\bar{\Phi}]}$ for an arbitrary set of fields $\widehat{\Phi} \equiv \{\widehat{\Phi}^i\}$\footnote{To keep the notation simple we leave the book keeping of the sign factors due to fields with odd Grassmann parity implicit here. They are easy to retrieve when needed. For a systematic treatment of anticommuting fields using a superspace formalism, see \cite{DeWitt:1985bc,Rogers:1979vp}.}. Besides these dynamical fields, the bare action $S$ may also depend on an analogous set of background fields $\bar{\Phi} \equiv \{\bar{\Phi}^i\}$. In the case of gauge theories, $S$ contains gauge-fixing and ghost terms and $\widehat{\Phi}$ is assumed to include the corresponding Faddeev-Popov ghosts. Performing a linear split $\widehat{\Phi}^i = \bar{\Phi}^i + \widehat{\varphi}^i$ we interpret the fluctuation fields $\widehat{\varphi}^i$ as the actual integration variables: $Z[\bar{\Phi}] = \int \mathcal{D}\widehat{\varphi}\, e^{-S[\hat{\varphi},\bar{\Phi}]}$. Expectation values will be denoted $\Phi^i \equiv \Braket{\widehat{\Phi}^i}$ and $\varphi^i \equiv \Braket{\widehat{\varphi}^i}$, respectively, whence $\Phi^i = \bar{\Phi}^i + \varphi^i$. Furthermore, we augment the functional integral by a mode-suppression factor $e^{-\Delta S_k}$ and a coupling to external sources, yielding the generating functional 
\begin{equation}
	Z_k [J;\bar{\Phi}] \equiv e^{W_k [J;\bar{\Phi}]} = \int \mathcal{D}\widehat{\varphi}\, \operatorname{exp} \left( - S[\widehat{\varphi};\bar{\Phi}] - \Delta S_k [\widehat{\varphi};\bar{\Phi}] + J_i \widehat{\varphi}^i  \right) \ .
	\label{eq:scaledepgenfunc}
\end{equation}
We also redefine the summation convention to include the integration over continous indices with the invariant background volume element $\intd \bar{v}_x \equiv \intd^d x \sqrt{\bar{g}(x)}$, so the source term reads
\begin{equation}
	J_i \widehat{\varphi}^i \equiv \int \intd^d x \sqrt{\bar{g}(x)} \, J_A (x) \widehat{\varphi}^A (x) \ ,
\end{equation}
where $'A'$ denotes the sub-set of discrete indices contained in the index $i$. The cutoff action $\Delta S_k$ is a $\bar{\Phi}$-dependent quadratic form with respect to the fluctuations:
\begin{equation}
	\Delta S_k [\widehat{\varphi};\bar{\Phi}] = \frac{1}{2} \left( \mathcal{R}_k [\bar{\Phi}] \right)_{ij} \widehat{\varphi}^i \widehat{\varphi}^j \ .
	\label{eq:cutoffaction}
\end{equation}
The mode suppression kernel $( \mathcal{R}_k [\bar{\Phi}] )_{ij}$  has to be adapted such that $\Delta S_k$ provides a $k$-dependent mass-term for the IR fluctuation modes while vanishing for the corresponding UV modes that get integrated out unsuppressed. Given $W_k$, we can express the expectation value of the fluctuations as derivatives $\varphi^i \equiv \frac{\delta W_k}{\delta J_i}$, and likewise we obtain for the connected two-point function:
\begin{equation}
	G^{ij} \equiv \Braket{\widehat{\varphi}^i \widehat{\varphi}^j} - \varphi^i \varphi^j = \frac{\delta^2 W_k}{\delta J_i \delta J_j} \ .
	\label{eq:2pointfunc}
\end{equation}
In our conventions all derivatives come with inverse powers of the background volume element, so the above definitions when written explicitly read
\begin{equation}
	\varphi^i \equiv \varphi^{A}(x) = \frac{1}{\sqrt{\bar{g}(x)}} \frac{\delta W_k}{\delta J_A (x)} \quad , \quad G^{ij} \equiv G^{AB}(x,y) = \frac{1}{\sqrt{\bar{g}(x)}} \frac{1}{\sqrt{\bar{g}(y)}} \frac{\delta^2 W_k}{\delta J_A (x) \delta J_B (y)} .
\end{equation}
The labels $A,B$ again denote the sub-set of discrete indices. In this language, the unit operator $\mathds{1}$ has the matrix elements
\begin{equation}
	\tensor{\mathds{1}}{^i_j} \equiv \tensor{\delta}{^A_B} \delta^x_y \quad \text{with} \quad 	\delta^x_y \equiv \Braket{x|y} \equiv \frac{\delta(x-y)}{\sqrt{\bar{g}(y)}}.
\end{equation}
In fact, $\sum_i \tensor{\mathds{1}}{^i_j} \equiv \sum_A \tensor{\delta}{^A_B} \int \intd^d x \delta (x-y) = 1 $, as it should be, since the factors of $\sqrt{\bar{g}}$ cancel. 

Let $\tilde{\Gamma}_k[\varphi;\bar{\Phi}]$ denote the Legendre transform of $W_k[J;\bar{\Phi}]$ with respect to $J_i$ at fixed $\bar{\Phi}^i$. By definition, one obtains the effective average action $\Gamma_k$ from $\tilde{\Gamma}_k$ by subtracting the mode suppression term $\Delta S_{k}$ with the classical fields replacing the quantum ones:
\begin{equation}
	\Gamma_{k}[\varphi; \bar{\Phi}] \equiv \tilde{\Gamma}_{k}[\varphi; \bar{\Phi}] - \Delta S_{k}[\varphi; \bar{\Phi}] \ .
\end{equation}

\noindent \textbf{(B)} With the ultimate goal of deriving a flow equation for the effective action let us now introduce the following Hessian matrix built from the  second functional derivatives of  $\tilde{\Gamma}_k$:
\begin{equation}
	\bigl(\tilde{\Gamma}_{k}^{(2)}\bigr)_{ij} \equiv \frac{\delta^2 \tilde{\Gamma}_{k} }{\delta \varphi^{i} \delta \varphi^{j}} \ .
	\label{eq:hessian1}
\end{equation}
Since the identity \cite{Wipf:2013vp}
\begin{equation}
	\frac{\delta^2 W_k}{\delta J_i \delta J_l} \, \frac{\delta^2 \tilde{\Gamma}_{k} }{\delta \varphi^{l} \delta \varphi^{j}} = \tensor{\delta}{^i_j} \ ,
\end{equation}
is well known to be valid for every pair of functionals related by a Legendre transformation, we see that the connected two-point function and the Hessian satisfy
\begin{equation}
\boxed{
	G^{il} \, \bigl(\tilde{\Gamma}_{k}^{(2)}\bigr)_{lj} = \tensor{\delta}{^i_j} 
	} \ .
	\label{eq:combination}
\end{equation}
The equation (\ref{eq:combination}) is an important relation which is needed in the derivation of the flow equation for the EAA. Usually it is interpreted by saying that $G^{ij}$ and $\bigl(\tilde{\Gamma}_{k}^{(2)}\bigr)_{ij}$ are mutually inverse operators, and one rewrites (\ref{eq:combination}) as a statement about operators:
\begin{equation}
	\bm{G} \, \bm{\tilde{\Gamma}_{k}^{(2)}} = \bm{\mathds{1}} \quad \text{or} \quad \bm{G} = \bigl(\bm{\tilde{\Gamma}_k^{(2)}}\bigr)^{-1} \ .
	\label{eq:operators}
\end{equation}
In the following we shall see that this is actually not quite correct, and that the latter representation actually requires extra input. \\
\\
\textbf{(C)} A priori the components $\bigl\{ \bigl(\tilde{\Gamma}_{k}^{(2)}\bigr)_{ij} \bigr\}$ which form the Hessian matrix merely define a \textit{quadratic form} $v \mapsto \bigl(\tilde{\Gamma}_{k}^{(2)}\bigr)_{ij} v^i v^j$ mapping ``vectors'' $\{v^i \} \equiv v$ onto real numbers, but they do not, in a natural way, give rise to an \textit{operator}, i.e. a linear map of vectors onto vectors. 

Using boldface letters to denote operators, they have the structure $\bm{W} : v \mapsto \bm{W}(v) \equiv \bm{W} v$, $(\bm{W} v)^i = \tensor{\bm{W}}{^i_j} v^j$, where the coefficient matrix $\{ \tensor{\bm{W}}{^i_j}\}$ characterizes the abstract operator $\bm{W}$ with respect to a chosen basis. The composition of two such maps, say $\bm{V}$ and $\bm{W}$, is described by the corresponding matrix product $\tensor{(\bm{V}\bm{W})}{^i_j} = \tensor{\bm{V}}{^i_k}\tensor{\bm{W}}{^k_j}$.

On the other hand, quadratic forms $Q: v \mapsto Q(v) = Q_{ij} v^i v^j$, assuming values in the real numbers, cannot be composed or iterated. Therefore, even though their coordinate representation, too, involves a 2-index object, the ``matrix'' $Q_{ij}$, it has no intrinsic meaning to multiply two such objects according to the rules of matrix multiplication. This is in sharp contradistinction to the set of coefficients $\tensor{\bm{W}}{^i_j}$ which form a matrix-representation of $\bm{W}$. 

It is clear therefore that there is no intrinsic (coordinate independent) way of interpreting the quadratic form provided by $\tilde{\Gamma}_k$, i.e. $Q_{ij} = \bigl(\tilde{\Gamma}_{k}^{(2)}\bigr)_{ij}$, as an operator. An analogous remark applies to $G^{ij}$ which defines a quadratic form for ``dual vectors'': $\alpha \equiv (\alpha_i) \mapsto G^{ij} \alpha_i \alpha_j$. Already the different tensorial properties with respect to $\mathrm{Diff}(\mathcal{M})$ make it impossible to regard $Q_{ij}$ as the component representation of an operator.

If we insist on turning $Q_{ij}$ into an operator, \textit{additional information must be supplied}, namely an isomorphism which relates vectors $v^i$ to dual vectors $v_i$, i.e. a prescription for pulling indices up and down, loosely speaking. 

In principle there are many ways of establishing an isomorphism of this kind. Here and in the following we employ the field space metric $\text{\Fontlukas G}_{ij}$ and its inverse, $\text{\Fontlukas G}^{ij}$, for this purpose. We define the operator $\bm{Q}$ which is associated to a given quadratic form $\{ Q_{ij} \}$ in terms of its matrix elements
\begin{equation}
\boxed{
 \tensor{\bm{Q}}{^i_j} \equiv \text{\Fontlukas G}^{il} Q_{lj} 
} \ . 
\end{equation}

Using the above inner product, $\left( v_1 , v_2 \right) \equiv \text{\Fontlukas G}_{ij} \, v_1^i v_2^j$, we can then rewrite the quadratic form $Q(v) \equiv Q_{ij} v^i v^j$ in terms of the operator $\bm{Q}$, according to 
\begin{equation}
	Q(v) = (v,\bm{Q}v) \ . 
\end{equation}
In this latter representation the $\text{\Fontlukas G}_{ij}$-dependencies implicit in both $\bm{Q}$ and the inner product cancel, and $Q(v)$ becomes independent of the metric in field space, as it should be. 

Returning now to equation (\ref{eq:combination}), with the metric on field space as the extra geometric input, we define the operators $\bm{\tilde{\Gamma}_{k}^{(2)}}$ and $\bm{G}$ by
\begin{equation}
	\tensor{\bigl( \bm{\tilde{\Gamma}_{k}^{(2)}} \bigr)}{^i_j} \equiv \overline{\text{\Fontlukas G}}^{il} \, \bigl( \tilde{\Gamma}_{k}^{(2)} \bigr)_{lj} \quad \text{and} \quad \tensor{\bigl(\bm{G}\bigr)}{^i_j} \equiv G^{il} \ \overline{\text{\Fontlukas G}}_{lj} \ .
	\label{eq:hessianoperator}
\end{equation}
Recalling the field split $\Phi^i = \bar{\Phi}^i + \varphi^i$, the metric on field space is always understood to be evaluated at the background fields: $\overline{\text{\Fontlukas G}}_{ij} \equiv (\text{\Fontlukas G}[\bar{\Phi}])_{ij}$. 

With $\bm{G}$ and $\bm{\tilde{\Gamma}_{k}^{(2)}}$ defined in this way, the Hessian relationship (\ref{eq:combination}) indeed becomes equivalent to the operator equation (\ref{eq:operators}) now. In terms of their matrix elements it reads
\begin{equation}
	\tensor{\bigl(\bm{G}\bigr)}{^i_l} \, \tensor{\bigl( \bm{\tilde{\Gamma}_{k}^{(2)}} \bigr)}{^l_j} = \tensor{\bigl( \bm{\mathds{1}} \bigr)}{^i_j} \ ,
\end{equation}
or in full detail:
\begin{equation}
	\int \mathrm{d}^{d}z\ \sqrt{\bar{g}(z)}\ \tensor{G[J;\bar{\Phi}]}{_A^C}(x,z) \ \tensor{\tilde{\Gamma}_{k}^{(2)}[\varphi;\bar{\Phi}]}{_C^B}(z,y) = \tensor{\delta}{_A^B} \frac{\delta(x-y)}{\sqrt{\bar{g}(y)}}
\end{equation}
\textbf{(D)} Having defined and understood the status of the operator statement $\bm{G} = \bigl(\bm{\tilde{\Gamma}_k^{(2)}}\bigr)^{-1}$, let us now derive the flow equation for the EAA. Differentiating (\ref{eq:scaledepgenfunc}) with respect to the RG-time $t \equiv \ln k$ we obtain
\begin{align}
	- \partial_t W_k [J;\bar{\Phi}] &= \frac{1}{2} \bigl( \partial_t \mathcal{R}_k [\bar{\Phi}] \bigr)_{ij} \Braket{\widehat{\varphi}^i \widehat{\varphi}^j} \\
	&= \frac{1}{2} \bigl( \partial_t \mathcal{R}_k [\bar{\Phi}] \bigr)_{ij} \ G^{ij} + \frac{1}{2} \bigl( \partial_t \mathcal{R}_k [\bar{\Phi}] \bigr)_{ij} \varphi^i \varphi^j \\
	&= \frac{1}{2} \bigl( \partial_t \mathcal{R}_k [\bar{\Phi}] \bigr)_{ij} \ G^{ij} + \partial_t \Delta S_k [\varphi;\bar{\Phi}] \ .
\end{align}
Here we have employed (\ref{eq:2pointfunc}) in the second, and (\ref{eq:cutoffaction}) in the third line. Using $\Gamma_k = \tilde{\Gamma}_k - \Delta S_k$ and exploiting the standard properties of the Legendre transformation leads to 
\begin{equation}
	\partial_t \Gamma_k [\varphi;\bar{\Phi}] = \frac{1}{2} \bigl( \partial_t \mathcal{R}_k [\bar{\Phi}] \bigr)_{ij} \ G^{ij} \ .
	\label{eq:before}
\end{equation}

This brings us to the crucial step which usually receives little attention, namely the promotion of the quadratic forms appearing in (\ref{eq:before}), $\bigl( \partial_t \mathcal{R}_k [\bar{\Phi}] \bigr)_{ij}$ and $G^{ij}$, respectively, to operators. Assuming that we are given a (possibly $k$-dependent) metric on field space, we perform this transition by inserting a unit with matrix elements $\delta^i_j = \overline{\text{\Fontlukas G}}^{il}\overline{\text{\Fontlukas G}}_{lj}$ in between the two forms in (\ref{eq:before}). With (\ref{eq:hessianoperator}), and the following definition of the operator $\bm{\bigl( \partial_t \mathcal{R}_k [\bar{\Phi}] \bigr)}$,
\begin{equation}
	\tensor{\bm{\bigl( \partial_t \mathcal{R}_k [\bar{\Phi}] \bigr)}}{^i_j} \equiv \overline{\text{\Fontlukas G}}^{il} \, \bigl( \partial_t \mathcal{R}_k [\bar{\Phi}] \bigr)_{lj}
	\label{eq:dtRKoperator}
\end{equation}
this leads us to the desired structure of an operator trace:
\begin{equation}
	\partial_t \Gamma_k [\varphi;\bar{\Phi}] = \frac{1}{2} \tensor{\bigl( \partial_t \mathcal{R}_k [\bar{\Phi}] \bigr)}{_i^j} \tensor{G}{_j^i} = \frac{1}{2} \operatorname{STr} \Bigl[ \bm{ \bigl( \partial_t \mathcal{R}_k \bigr)} \ \bm{G} \Bigr] \ .
	\label{eq:after}
\end{equation}
Now finally we are in the position to employ $\bm{G} \bm{\tilde{\Gamma}_{k}^{(2)}} = \bm{\mathds{1}}$ and replace $\bm{G}$ with the inverse of the operator associated to the Hessian:
\begin{equation}
	\partial_t \Gamma_k [\varphi;\bar{\Phi}] =  \frac{1}{2} \operatorname{STr} \Bigl[ \bm{\bigl( \partial_t \mathcal{R}_k \bigr)} \ (\bm{\tilde{\Gamma}_{k}^{(2)}})^{-1} \Bigr] \ .
	\label{eq:dontknowwhattolabelthis}
\end{equation}

The mode suppression kernel $\bigl( \partial_t \mathcal{R}_k [\bar{\Phi}] \bigr)_{ij} = \frac{\delta^2 \Delta S_k [\varphi;\bar{\Phi}]}{\delta \varphi^i \delta \varphi^j}$ can be seen as a Hessian as well, having an associated operator $\bm{\mathcal{R}_k} \equiv \bm{\mathcal{R}_k [\bar{\Phi}]}$ with matrix elements $\tensor{\bm{\bigl(\mathcal{R}_k\bigr)}}{^i_j} = \overline{\text{\Fontlukas G}}^{il} \bigl( \mathcal{R}_k [\bar{\Phi}] \bigr)_{lj}$. Hence the definition of the EAA implies $\bigl(\Gamma_{k}^{(2)} \bigr)_{ij} = \bigl(\tilde{\Gamma}_{k}^{(2)} \bigr)_{ij} - \bigl( \mathcal{R}_k \bigr)_{ij}$ and therefore $\bm{\tilde{\Gamma}_{k}^{(2)}} = \bm{\Gamma_{k}^{(2)}} + \bm{\mathcal{R}_k}$. Plugging this relation into (\ref{eq:dontknowwhattolabelthis}) we arrive at the following functional flow equation:
\begin{equation}
\boxed{
 \partial_t \Gamma_k [\varphi;\bar{\Phi}] =  \frac{1}{2} \operatorname{STr} \Bigl[ \bm{\bigl( \partial_t \mathcal{R}_k \bigr)} \ (\bm{\Gamma_{k}^{(2)}} + \bm{ \mathcal{R}_k} )^{-1} \Bigr] 
} \ .
\label{eq:operatorflowequation}
\end{equation}

\noindent \textbf{(E)} While this equation has the familiar appearance of the FRGE as it is written in the literature, the above derivation highlights the precise meaning of its building blocks, namely the operators $\bm{\Gamma_{k}^{(2)}}$, $\bm{ \mathcal{R}_k}$, and $\bm{\bigl( \partial_t \mathcal{R}_k \bigr)}$. All of them depend on the metric in field space which we are free to choose in whatever way we like. Usually, in particular in pure matter theories on a classical flat spacetime, this enormous freedom is not exploited and not even recognized as such, since typically there exists a unique natural, or ``canonical'' choice of $\overline{\text{\Fontlukas G}}_{ij}$ that is fixed by the symmetries at hand and other basic requirements such as ultra-locality. 

Under the condition that the initial quadratic form $\bigl( \mathcal{R}_k \bigr)_{ij}$ has no dependence on $\overline{\text{\Fontlukas G}}_{ij}$, it is clear from our derivation that \textit{even though $\bm{\Gamma_{k}^{(2)}}$, $\bm{ \mathcal{R}_k}$, and $\bm{\bigl( \partial_t \mathcal{R}_k \bigr)}$ separately do depend on $\overline{\text{\Fontlukas G}}_{ij}$, the RHS of the FRGE, and therefore $\partial_t \Gamma_k$, is independent of it.} In fact, in going from (\ref{eq:before}) to (\ref{eq:after}) by inserting a factor of unity no overall $\overline{\text{\Fontlukas G}}_{ij}$-dependence was introduced, since the metric and its inverse cancel between $\bm{\bigl( \partial_t \mathcal{R}_k \bigr)}$ and $\bm{G}$. 

If, for reasons unrelated to the transition from quadratic forms to operators, already the $\bigl( \partial_t \mathcal{R}_k \bigr)_{ij}$ or $G^{ij}$ on RHS of (\ref{eq:before}) have a certain $\overline{\text{\Fontlukas G}}_{ij}$-dependence, then the operator trace in (\ref{eq:after}) has exactly the same.\\
\\
\textbf{(F)} The new FRGE of (\ref{eq:operatorflowequation}) is valid even for \textit{scale-dependent} metrics, $\overline{\text{\Fontlukas G}}_{ij} \equiv \overline{\text{\Fontlukas G}}_{ij}(k)$. In this case it is actually \textit{not} equivalent to the standard one, the reason being that \textit{the operator $\bm{\bigl( \partial_t \mathcal{R}_k \bigr)}$ does not coincide with $\partial_t$ applied to the operator $\bm{\mathcal{R}_k}$, that is, $\bm{\bigl( \partial_t \mathcal{R}_k \bigr)} \neq \partial_t \bigl( \bm{\mathcal{R}_k} \bigr)$.} In fact, applying $\partial_t$ to $\bm{\mathcal{R}_k}$ yields an operator, denoted $\partial_t \bm{\mathcal{R}_k}$, which has the matrix elements
\begin{equation}
	\tensor{\bigl( \partial_t \bm{\mathcal{R}_k} \bigr)}{^i_j} \equiv \partial_t \tensor{\bigl( \bm{\mathcal{R}_k} \bigr)}{^i_j} = \partial_t \left[ \overline{\text{\Fontlukas G}}^{il}\bigl( \mathcal{R}_k \bigr)_{lj} \right] \ .
\end{equation}
Here $\partial_t$ acts on both $\overline{\text{\Fontlukas G}}^{il}$ and $\bigl( \mathcal{R}_k \bigr)_{lj}$, whereas in the definition of $\bm{\bigl( \partial_t \mathcal{R}_k \bigr)}$, eq. (\ref{eq:dtRKoperator}), only the second factor is differentiated: $\tensor{\bm{\bigl( \partial_t \mathcal{R}_k [\bar{\Phi}] \bigr)}}{^i_j} \equiv \overline{\text{\Fontlukas G}}^{il} \bigl( \partial_t \mathcal{R}_k [\bar{\Phi}] \bigr)_{lj}$. Thus the two operators $\partial_t \bm{\mathcal{R}_k}$ and $\bm{\bigl( \partial_t \mathcal{R}_k \bigr)}$ differ by a term which involves the scale derivative of the metric in field space. Explicitly,
\begin{equation}
	\bm{\bigl( \partial_t \mathcal{R}_k \bigr)} = \partial_t \bm{\mathcal{R}_k} + \overline{\text{\Fontlukas G}}^{-1} \bigl( \partial_t \, \overline{\text{\Fontlukas G}} \bigr) \, \bm{\mathcal{R}_k}
	\label{eq:dtRkexplicit}
\end{equation}
where $\overline{\text{\Fontlukas G}}^{-1}\bigl(\partial_t \, \overline{\text{\Fontlukas G}}\bigr)$ is the matrix with elements $\tensor{\left(\overline{\text{\Fontlukas G}}^{-1}\bigl(\partial_t \, \overline{\text{\Fontlukas G}}\bigr) \right)}{^i_j} = \overline{\text{\Fontlukas G}}^{il} \partial_t \, \overline{\text{\Fontlukas G}}_{lj}$.

The object we are used to deal with in setting up flow equations is the ordinary derivative of $\mathcal{R}_k$, i.e. $\partial_t \bm{\mathcal{R}_k}$, and not $\bm{\bigl( \partial_t \mathcal{R}_k \bigr)}$, the operator associated to the differentiated quadratic form. However, it is the latter operator in terms of which the FRGE has its standard form (\ref{eq:operatorflowequation}). Thus, employing $\partial_t \bm{\mathcal{R}_k}$ with its familiar meaning of a differentiated operator, the flow equation reads
\begin{equation}
\boxed{
	\partial_t \Gamma_k [\varphi;\bar{\Phi}] = \frac{1}{2} \operatorname{STr} \left[ \left( \partial_t \bm{\mathcal{R}_k} + \overline{\text{\Fontlukas G}}^{-1} \bigl( \partial_t \, \overline{\text{\Fontlukas G}} \bigr) \bm{\mathcal{R}_k} \right) \bigl(\bm{\Gamma_{k}^{(2)}} + \bm{ \mathcal{R}_k} \bigr)^{-1} \right]
	} \ .
	\label{eq:newflowequation}
\end{equation}
This is the final form of the generalized FRGE for the Effective Average Action. It applies also to the case of $k$-dependent metrics on field space.

It is a matter of taste whether one wants to use the flow equation in the form of (\ref{eq:operatorflowequation}) or (\ref{eq:newflowequation}). If one prefers to work with the conventional-looking FRGE a practical rule is as follows: If the primary, $\overline{\text{\Fontlukas G}}$-independent object is $\bigl( \mathcal{R}_k \bigr)_{ij}$ and we use $\overline{\text{\Fontlukas G}}^{ij}$ to find $\bm{\mathcal{R}_k}$, the decisive ingredient $\bm{\bigl( \partial_t \mathcal{R}_k \bigr)}$ can be calculated by applying a scale derivative to $\mathcal{R}_k$ and simply deleting ``by hand'' all terms in which $\partial_t$ hits a component of the metric.\\
\\
\textbf{(G)} One might wonder whether $\beta$-functions computed on the basis of (\ref{eq:operatorflowequation}) or (\ref{eq:newflowequation}) really can depend on $\overline{\text{\Fontlukas G}}$. After all, in \textbf{(E)} above we observed that this metric cancels among the various building blocks of the FRGE, $\bm{\Gamma_{k}^{(2)}}$, $\bm{\mathcal{R}_k}$, and $\bm{\bigl( \partial_t \mathcal{R}_k \bigr)}$, respectively. 

We emphasize that this cancellation can take place only under the condition that the quadratic form $\bigl( \mathcal{R}_k \bigr)_{ij}$, or equivalently the second functional derivatives of $\Delta S_k$, are the primary objects and hence, by assumption, are independent of $\overline{\text{\Fontlukas G}}$; the operator $\tensor{\bigl( \bm{\mathcal{R}_k} \bigr)}{^i_j}$ is a derived object then, and so it does depend on $\overline{\text{\Fontlukas G}}$. 

If instead the operator $\tensor{\bigl( \bm{\mathcal{R}_k} \bigr)}{^i_j}$ is specified as the primary, i.e. $\overline{\text{\Fontlukas G}}$-independent object, while $\bigl( \mathcal{R}_k \bigr)_{ij}$ is derived from it by ``pulling down'' an index with $\overline{\text{\Fontlukas G}}$, the situation is different. Then in setting up the FRGE, only $\Gamma_{k}^{(2)}$ remains to be converted to an operator by means of an explicit factor of $\overline{\text{\Fontlukas G}}{}^{-1}$, which no longer can cancel against a corresponding factor from the cutoff. Thus the conclusion is that if the cutoff \textit{operator} is chosen freely, without recourse to a metric on $\mathcal{F}$, as this is usually done in practical FRGE computations, then $\overline{\text{\Fontlukas G}}$ affects only the map $\bigl( \Gamma^{(2)}_{k} \bigr)_{ij} \mapsto \tensor{\bigl( \bm{\Gamma_{k}^{(2)}} \bigr)}{^i_j}$, and as a consequence \textit{the $\overline{\text{\Fontlukas G}}$-dependence can make its way into the $\beta$-functions}.

%% file: HolstTruncation.tex
In this section we introduce the truncation of $\mathcal{T}_{\mathrm{dtor}}$ whose RG flow we shall analyze later on. As we study special, geometrically motivated field variables and the inclusion of torsion we will begin, in Section \ref{subsec:RiemmannGeometry}, with a brief overview of the relevant geometric concepts and related definitions. The ensuing subsections are then devoted to, respectively, the full, un-truncated theory space $\mathcal{T}_{\mathrm{dtor}}$, its ``Holst truncation'', and the parity-even sector of the latter on which we shall focus thereafter.

\subsection{The classical setting}\label{subsec:RiemmannGeometry}

\textbf{(A) Connections.} We consider a $4$-dimensional spacetime manifold $\mathcal{M}$ equipped with an arbitrary Euclidean metric $g_{µ \nu}$ and an affine connection $D$, which we demand to be metric compatible, $D_\kappa g_{µ \nu} = 0$, but not necessarily torsion-free. Hence, locally, $\mathcal{M}$ carries two independent fields, namely $\{ g_{µ \nu} , \tensor{\Gamma}{^\lambda_{µ \nu}}\}$, or equivalently $\{ g_{µ \nu} , \tensor{T}{^\lambda_{µ \nu}}\}$. Here $\tensor{\Gamma}{^\lambda_{µ \nu}}$ and $\tensor{T}{^\lambda_{µ \nu}}$ denote the connection coefficients and the torsion tensor, respectively, the latter being defined by
\begin{equation}
	\tensor{T}{^{\lambda}_{µ \nu}} = \tensor{\Gamma}{^{\lambda}_{µ \nu}} - \tensor{\Gamma}{^{\lambda}_{\nu µ}} \ .
\label{eq:torsiontensor}
\end{equation}
The connection can be expressed as\footnote{We use the following conventions for (anti-)symmetrization: $T_{[µ_{1} \ldots µ_{n}]} = \frac{1}{n!} \sum_{\pi} \delta_{\pi} T_{µ_{\pi(1)} \ldots µ_{\pi(n)} }$ and $T_{(µ_{1} \ldots µ_{n})} = \frac{1}{n!} \sum_{\pi} T_{µ_{\pi(1)} \ldots µ_{\pi(n)} }$. Furthermore, all geometrical objects defined with the Levi-Civita connection will be denoted by the subscript $_{\mathrm{LC}}$.}
\begin{align}
	\begin{split}
		\tensor{\Gamma}{^{\lambda}_{µ \nu}} &= \tensor{\Gamma}{^{\lambda}_{(µ \nu)}} + \tensor{\Gamma}{^{\lambda}_{[µ \nu]}} \\
		&= \frac{1}{2} g^{\lambda \kappa} \left( \partial_{µ} g_{\nu \kappa} + \partial_{\nu} g_{µ \kappa} - \partial_{\kappa} g_{µ \nu} \right) + \frac{1}{2} \left( \tensor{T}{_{\nu}^{\lambda}_{µ}} + \tensor{T}{_{µ}^{\lambda}_{\nu}} \right) + \frac{1}{2} \tensor{T}{^{\lambda}_{µ \nu}} \\
		&\equiv \tensor{(\Gamma_{\LC})}{^{\lambda}_{µ \nu}} + \tensor{K}{^{\lambda}_{µ \nu}}.
	\end{split}
\end{align}
Here we introduced the Levi-Civita connection given by the Christoffel symbols,
\begin{equation}
	\tensor{(\Gamma_{\LC})}{^{\lambda}_{µ \nu}} = \frac{1}{2} g^{\lambda \kappa} \left( \partial_{µ} g_{\nu \kappa} + \partial_{\nu} g_{µ \kappa} - \partial_{\kappa} g_{µ \nu} \right) \ ,
\end{equation}
as well as the contorsion tensor
\begin{equation}
	\tensor{K}{^{\lambda}_{µ \nu}} = \frac{1}{2} \left( \tensor{T}{^{\lambda}_{µ \nu}} + \tensor{T}{_{\nu}^{\lambda}_{µ}} + \tensor{T}{_{µ}^{\lambda}_{\nu}} \right) \ .
	\label{eq:contorsion}
\end{equation}
The torsion and contorsion tensor, respectively, exhibit the following symmetry properties:
\begin{equation}
	\tensor{T}{^{\lambda}_{µ \nu}} = - \tensor{T}{^{\lambda}_{\nu µ}} \ , \quad
	K_{\lambda µ \nu} = - K_{\nu µ \lambda}.
	\label{eq:contrel}
\end{equation}

\noindent\textbf{(B) Curvature and torsion.} The components of the Riemann curvature tensor are
\begin{equation}
\tensor{R}{^{\kappa}_{\lambda µ \nu}} = \partial_µ \tensor{\Gamma}{^{\kappa}_{\nu \lambda}} - \partial_{\nu} \tensor{\Gamma}{^{\kappa}_{µ \lambda}} + \tensor{\Gamma}{^{\rho}_{\nu \lambda}} \tensor{\Gamma}{^{\kappa}_{µ \rho}} - \tensor{\Gamma}{^{\rho}_{µ \lambda}} \tensor{\Gamma}{^{\kappa}_{\nu \rho}}
\end{equation}
and satisfy 
\begin{equation}
	\tensor{R}{^{\kappa}_{\lambda µ \nu}} = - \tensor{R}{^{\kappa}_{\lambda \nu µ}}.
\end{equation}
By contraction we obtain the Ricci tensor, $\mathrm{Ric}_{µ \nu} = \tensor{R}{^{\lambda}_{µ \lambda \nu}}$, and the scalar curvature, $R = g^{µ \nu} \mathrm{Ric}_{µ \nu}$. (Our conventions for the Riemann and Ricci tensor are the same as in \cite{Reuter:1996cp}.)

The Riemann curvature tensor for the Levi-Civita connection,
\begin{equation}
	\tensor{(R_{\LC})}{^{\kappa}_{\lambda µ \nu}} = \partial_µ \tensor{(\Gamma_{\LC})}{^{\kappa}_{\nu \lambda}} - \partial_{\nu} \tensor{(\Gamma_{\LC})}{^{\kappa}_{µ \lambda}} + \tensor{(\Gamma_{\LC})}{^{\rho}_{\nu \lambda}} \tensor{(\Gamma_{\LC})}{^{\kappa}_{µ \rho}} - \tensor{(\Gamma_{\LC})}{^{\rho}_{µ \lambda}} \tensor{(\Gamma_{\LC})}{^{\kappa}_{\nu \rho}}
\end{equation}
admits the following additional symmetries:
\begin{equation}
	(R_{\LC})_{\kappa \lambda µ \nu} = - (R_{\LC})_{\kappa \lambda \nu µ} \ , \ (R_{\LC})_{\kappa \lambda µ \nu} = - (R_{\LC})_{\lambda \kappa µ \nu} \ , \ (R_{\LC})_{\kappa \lambda µ \nu} = (R_{\LC})_{µ \nu \kappa \lambda} \ .
\end{equation}
As a results of these additional symmetries the corresponding Ricci tensor is symmetric:  $\tensor{(\mathrm{Ric}_{\LC})}{_{µ \nu}} = \tensor{(\mathrm{Ric}_{\LC})}{_{\nu µ}}$. The curvature tensor of the Levi-Civita connection also satisfies the two Bianchi identities:
\begin{align}
	\tensor{(R_{\LC})}{^{\kappa}_{\lambda µ \nu}} + \tensor{(R_{\LC})}{^{\kappa}_{µ \nu \lambda}} + \tensor{(R_{\LC})}{^{\kappa}_{\nu \lambda µ}} &= 0 \label {eq:fbianchi} \\
	(D_{\LC})_{\kappa} \tensor{(R_{\LC})}{^{\alpha}_{\lambda µ \nu}} + (D_{\LC})_{µ} \tensor{(R_{\LC})}{^{\alpha}_{\lambda \nu \kappa}} + (D_{\LC})_{\nu} \tensor{(R_{\LC})}{^{\alpha}_{\lambda \kappa µ}} &= 0 \ .
\end{align}

The Riemann curvature tensor, Ricci tensor and curvature scalar pertaining to a generic connection can be decomposed into their respective Levi-Civita counterparts plus torsion terms:
\begin{align}
	\tensor{R}{^{\kappa}_{\lambda µ \nu}} &= \tensor{(R_{\LC})}{^{\kappa}_{\lambda µ \nu}} + 2 \tensor{(D_{\LC})}{_{ [ µ }} \tensor{K}{^{\kappa}_{\nu ] \lambda}} + 2 \tensor{K}{^{\kappa}_{ [ µ | \rho |}} \tensor{K}{^{\rho}_{\nu ] \lambda}} \label{eq:Riemdecomp} \\
	\tensor{\mathrm{Ric}}{_{µ \nu}} &= \tensor{(\mathrm{Ric}_{\LC})}{_{µ \nu}} + 2 \tensor{(D_{\LC})}{_{ [ \kappa }} \tensor{K}{^{\kappa}_{\nu ] µ}} + 2 \tensor{K}{^{\kappa}_{ [ \kappa | \lambda |}} \tensor{K}{^{\lambda}_{\nu ] µ}} \label{eq:Ricdecomp} \\
	R &= R_{\LC} + 2 \tensor{(D_{\LC})}{_{ \kappa }} \tensor{K}{^{\kappa}_{\lambda}^{\lambda}} + 2 \tensor{K}{^{\kappa}_{ [ \kappa | \rho |}} \tensor{K}{^{\rho}_{\lambda ]}^{\lambda}} \label{eq:Rdecomb}
\end{align}
This decomposition will prove useful later on in the construction of the truncation ansatz.\\
\\
\textbf{(C) Decomposition of the torsion tensor.} The torsion tensor $\tensor{T}{^{\lambda}_{µ \nu}}$ can be decomposed into three tensors which are irreducible with respect to the Lorentz group \cite{Baekler:2011jt,Capozziello:2001mq,Shapiro:2001rz}:
\begin{equation}
	\tensor{T}{^{\lambda}_{µ \nu}} = \frac{1}{3} \left( \tensor{\delta}{^{\lambda}_{\nu}} T_{µ} - \tensor{\delta}{^{\lambda}_{µ}} T_{\nu} \right) + \frac{1}{6} \frac{\tensor{\epsilon}{^{\lambda}_{µ \nu \alpha}}}{\sqrt{g}}  S^{\alpha} + \tensor{q}{^{\lambda}_{µ \nu}} \ .
	\label{eq:Tdecomp}
\end{equation}
The trace part $T_µ \equiv \tensor{T}{^{\kappa}_{µ \kappa}}$ and the pseudo-trace $S^\alpha \equiv \frac{\epsilon^{\rho \sigma \beta \alpha}}{\sqrt{g}} T_{\rho \sigma \beta}$ transform as a vector and a pseudo-vector, respectively. The ``remainder'' $\tensor{q}{^{\lambda}_{µ \nu}}$ is a tensor with vanishing trace, $\tensor{q}{^{\lambda}_{µ \lambda}} = 0$, and totally antisymmetric part, $\epsilon^{µ \nu \rho \sigma} q_{\nu \rho \sigma} = 0$. It is antisymmetric in its last two indices $\tensor{q}{^{\lambda}_{µ \nu}} = - \tensor{q}{^{\lambda}_{\nu µ}}$ which implies that also the second possible contraction is trivial, $\tensor{q}{^\lambda_µ^µ} = 0$. In consequence of these properties all contractions of two indices vanish and the following identity holds:
\begin{equation}
	q^{[µ \nu] \rho} = - q^{\rho µ \nu}
	\label{eq:qrel}
\end{equation}  

In $d=4$ dimensions the torsion tensor $\tensor{T}{^{\lambda}_{µ \nu}}$ has 24 independent components. These components divide among the three irreducible tensors as follows: $T_µ$ and $S_µ$ have 4 components each, and $\tensor{q}{^{\lambda}_{µ \nu}}$ contains the remaining 16 components. In consequence, the newly introduced tensors $(S_µ , T_µ , \tensor{q}{^{\lambda}_{µ \nu}})$ are indeed in one-to-one correspondence with the torsion tensor.\\
\\
\textbf{(D) Invariants built from $\bm{(S,T,q)}$.} It is straightforward to identify all independent invariants which are quadratic in the irreducible torsion components $(S,T,q)$.

There exist only three \textit{parity-even} invariant combinations, as $S^µ$ is a pseudo-vector and $T^µ$ as well as $q^{µ \nu \rho}$ are (true) tensors that can only couple to themselves to form a scalar. Contracting $T^µ$ with $q^{µ \nu \rho}$ the other two indices of $q$ have to be contracted as well and thus the combination vanishes. If we combine $S$ and $q$ using an $\epsilon$-density the resulting combinations vanish as $q$ has no totally antisymmetric part. Hence, we are left with the three parity-even invariants
\begin{equation}
	I_1 = T_{µ}T^{µ}, \quad I_2 = S_{µ}S^{µ}, \quad I_3 = q_{µ \nu \rho} q^{µ \nu \rho}.
	\label{eq:invariant1}
\end{equation}
Any additional $q^2$-contractions are not independent from $I_3$. Starting with 6 contractions corresponding to the 6 permutations of the indices of the second $q$-factor, the terms with odd permutations are related to the remaining cyclic permutations by the antisymmetry of $q$ in its last two indices. Using (\ref{eq:qrel}) we obtain for the cyclic permutations
\begin{equation}
	q_{µ \nu \rho} q^{\nu \rho µ} = - \frac{1}{2} q_{µ \nu \rho} q^{µ \nu \rho}, \quad q_{µ \nu \rho} q^{\rho µ \nu} = - \frac{1}{2} q_{µ \nu \rho} q^{µ \nu \rho}
\end{equation}
such that only $I_3$ remains independent.

For the \textit{parity-odd} combinations only the two invariants
\begin{equation}
	I_4 = S_{µ}T^{µ}, \quad I_5 = \frac{\tensor{\epsilon}{_{µ \nu}^{\rho \sigma}}}{\sqrt{g}} q^{µ \nu \lambda} q_{\rho \sigma \lambda}
	\label{eq:invariant2}
\end{equation}
are independent. Any other $\epsilon q^2$ combinations are related to $I_5$ according to
\begin{equation}
	\frac{\tensor{\epsilon}{_{µ \nu}^{\rho \sigma}}}{\sqrt{g}} q^{\lambda µ \nu} q_{\lambda \rho \sigma} = 4 I_5 , \quad \frac{\tensor{\epsilon}{_{µ \nu}^{\rho \sigma}}}{\sqrt{g}} q^{\lambda µ \nu} q_{\rho \sigma \lambda} = - 2 I_5
\end{equation}
and therefore do not yield additional independent invariants.\\
\\
\textbf{(E) The Holst action in metric variables.} The Holst action as given in \cite{Holst:1995pc} and used as basis for the truncation ansatz in \cite{Daum:2010phd,Daum:2013fu} and \cite{Harst:2012phd,Harst:2014vca}, respectively, was formulated in terms of the vielbein $\tensor{e}{^{a}_{µ}}$ and the spin connection $\tensor{\omega}{^{a b}_{µ}}$ where it takes the form
\begin{equation}
\label{eq:holstactionoriginal}
	S_{\mathrm{Ho}} = - \frac{1}{16 \pi G} \int \mathrm{d}^{4}x \ e \left[ \tensor{e}{_{a}^{µ}} \tensor{e}{_{b}^{\nu}} \left( \tensor{F(e,\omega)}{^{ab}_{µ \nu}} - \frac{1}{2 \gamma} \tensor{\epsilon}{^{ab}_{cd}} \tensor{F(e,\omega)}{^{cd}_{µ \nu}} \right) - 2 \lambda \right].
\end{equation}
Because here we are interested in a metric formulation we express the field strength $F(e,\omega)$ in terms of the Riemann curvature tensor $R \equiv R(\Gamma)$, with the connection coefficients $\Gamma$ including torsion. Performing this change of variables results in 
\begin{equation}
	S_{\mathrm{Ho}} = - \frac{1}{16 \pi G} \int \mathrm{d}^{4}x\ \sqrt{g} \left[ R - \frac{1}{2 \gamma \sqrt{g}} \tensor{\epsilon}{^{µ \nu}_{\rho}^{\sigma}} \tensor{R}{_{µ \nu}^{\rho}_{\sigma}} - 2 \lambda \right].
\label{eq:holstmetric}
\end{equation}
In Appendix \ref{ch:holstaction} we decompose the Riemann tensor and the curvature scalar in $S_{\mathrm{Ho}}$ into their Levi-Civita counterparts and explicit torsion terms, using (\ref{eq:Riemdecomp}) and (\ref{eq:Rdecomb}), and then further decompose the torsion tensor into its irreducible parts according to (\ref{eq:Tdecompfirsttime}). This yields
\begin{multline}
 S_{\mathrm{Ho}} = - \frac{1}{16 \pi G} \int \mathrm{d}^{4}x\ \sqrt{g} \left[ \vphantom{\frac{\tensor{\epsilon}{^{µ \nu}_{\rho}^{\sigma}}}{\sqrt{g}}} R_{\LC} + 2 \tensor{(D_{\LC})}{_{ µ }} T^{µ} - \frac{2}{3} T_µ T^µ - \frac{1}{24} S_µ S^µ + \frac{1}{2} q_{µ \nu \rho} q^{µ \nu \rho} \right.
 \\ \left. - \frac{1}{\gamma} \left( \frac{1}{2} \tensor{(D_{\LC})}{_{ µ }} S^µ - \frac{1}{3} T_µ S^µ + \frac{\tensor{\epsilon}{^{µ \nu \rho \sigma}}}{\sqrt{g}} \tensor{q}{_{µ \nu}^{\tau}} q_{\rho \sigma \tau} \right) - 2 \lambda \right].
 \label{eq:holstaction}
\end{multline}
This variant of the Holst action constitutes the classical ``inspiration'' for the truncated action functional we are going to discuss in the present paper. Note that in (\ref{eq:holstaction}) the terms proportional to $\frac{1}{\gamma}$ are odd under parity, while all others are even. Note also that the contributions $\propto (D_{\LC})_µ T^µ$ and $(D_{\LC})_µ S^µ$ give rise to a parity-even and -odd boundary term, respectively.

\subsection[The untruncated theory space \texorpdfstring{$\mathcal{T}_{\mathrm{dtor}}$}{Tdtor}]{The untruncated theory space \texorpdfstring{$\bm{\mathcal{T}_{\mathrm{dtor}}}$}{Tdtor}}\label{subsec:untruncspace}

The field content and gauge invariance of the classical action in (\ref{eq:holstaction}) motivates us to study RG flows on a theory space which, to be fully explicit, has the following structures:
\begin{multline}
	\mathcal{T}_{\mathrm{dtor}} = \left\{ (\Phi, \bar{\Phi}) \mapsto S[\Phi,\bar{\Phi}] \right. \\
	\text{with} \ \Phi \equiv \left\{ g_{µ \nu},S_µ , T_µ , q_{\lambda µ \nu}, \xi^µ , \bar{\xi}_µ \right\} \ \text{and} \ \bar{\Phi} \equiv \left\{ \bar{g}_{µ \nu},\bar{S}_µ , \bar{T}_µ , \bar{q}_{\lambda µ \nu}, \overline{\xi^µ} , \overline{\bar{\xi}_µ} \right\} \\
	\left. | \ S[\Phi,\bar{\Phi}] \ \text{inv. under} \ \textbf{G} = \mathrm{Diff}(\mathcal{M}) \right\}
\end{multline}
Several remarks are in order here.

\noindent \textbf{(1)} The actions $S$ ``living'' in $\mathcal{T}_{\mathrm{dtor}}$ depend on the dynamical fields $g_{µ \nu},S_µ , T_µ , q_{\lambda µ \nu}$ all of which are assumed to carry \textit{lower indices only}. Referring back to the functional integral formalism it is clear that using $\widehat{S}_µ$ as the fundamental field, or rather $\widehat{S}^µ$, where the index has been raised with an inverse metric, yields different measures and potentially inequivalent quantum theories. In the FRGE framework, the RG flow is potentially different for actions depending on $S_µ$ or $S^µ$. So, to be very careful, one should examine the different cases separately. An obvious source of differences to which we shall come back below is the fact that $S_µ$ and $S^µ$ have different canonical dimensions if the metric components are considered dimensionful. 

\noindent \textbf{(2)} A generic action $S \in \mathcal{T}_{\mathrm{dtor}}$ can depend on a classical background for any of the dynamical fields. We denote those fields $\bar{g}_{µ \nu},\bar{S}_µ , \bar{T}_µ , \bar{q}_{\lambda µ \nu}$, and for consistency we assume that all background tensor fields, too, are written with lower indices only. 

\noindent \textbf{(3)} Actions in $\mathcal{T}_{\mathrm{dtor}}$ will also depend on Faddeev-Popov ghosts $\xi^µ$ and $\bar{\xi}_µ$, with backgrounds $\overline{\xi^µ}$ and $\overline{\bar{\xi}_µ}$, since the FRGE coarse-grains \textit{gauge fixed} actions. In the case at hand, the pertinent background gauge transformations under which $S$ is required to be invariant are spacetime diffeomorphisms which act via the standard tensor transformation laws on all arguments of $S$ simultaneously.

\subsection[The ``Holst truncation'' of \texorpdfstring{$\mathcal{T}_{\mathrm{dtor}}$}{Tdtor}]{The ``Holst truncation'' of \texorpdfstring{$\bm{\mathcal{T}_{\mathrm{dtor}}}$}{Tdtor}}
We observe that the classical Holst action (\ref{eq:holstaction}) written in terms of the new variables $(g,S,T,q)$ contains all three parity-even invariants $I_1$, $I_2$, $I_3$ of (\ref{eq:invariant1}) and the parity-odd ones $I_4$, $I_5$ from (\ref{eq:invariant2}). The relative weights of $I_1$, $I_2$, and $I_3$, for instance, are precisely fixed by the prefactors $+ \frac{2}{3}$, $+\frac{1}{24}$, $-\frac{1}{2}$. They follow from the fact that (\ref{eq:holstaction}) descends from the ``geometrically natural'' precursor (\ref{eq:holstactionoriginal}) in which the component fields appear combined into $\tensor{F}{^{ab}_{µ \nu}}$. However, generically a FRGE which is formulated in terms of the irreducible components and cuts off their fluctuations individually will \textit{not} respect these specific ratios. The prefactors of $S^2$ and $T^2$, say, will RG-evolve independently if no further symmetries (beyond the diffeomorphisms) are assumed. 

This motivates us to consider the following truncation ansatz:
\begin{equation}
	\Gamma_{k} \bigl[g,\bar{g},S,T,q,\xi,\bar{\xi} \bigr] = \Gamma_{k}^{\mathrm{Ho}}\bigl[ g,\bar{g},S,T,q \bigr] + \Gamma_{k}^{\mathrm{gf}}\bigl[g,\bar{g}\bigr] + S_{\mathrm{gh}}\bigl[g,\bar{g},\xi,\bar{\xi}\bigr].
	\label{eq:trunctype}
\end{equation}
This functional depends on the set of dynamical fields $\Phi \equiv \left\{ g_{µ \nu},S_{µ},T_{µ},q_{\lambda µ \nu},\xi^{µ},\bar{\xi}_µ \right\}$ and background fields $\bar{\Phi} \equiv \left\{ \bar{g}_{µ \nu} \right\}$. Here the background value of all fields except the metric is chosen to vanish. On the RHS of (\ref{eq:trunctype}), the term  $\Gamma_{k}^{\mathrm{Ho}}$ is a $k$-dependent variant of the Holst action\footnote{Note that we have shifted the minus sign, previously in front of $S_{\mathrm{Ho}}$, into the integral.}:
\begin{multline}
\label{eq:holsttruncation}
	\Gamma_{k}^{\mathrm{Ho}} \left[ g_{µ \nu},S_µ,T_µ,q_{µ \nu \rho} \right] = \frac{1}{16 \pi G_k} \int \mathrm{d}^{4}x\, \sqrt{g} \left[ \vphantom{\frac{\tensor{\epsilon}{^{µ \nu}_{\rho}^{\sigma}}}{\sqrt{g}}} - R_{\LC} + 2 \bar{\lambda}_k \right. \\
  + \frac{1}{24} f_{k}^{S} \Bigl( g^{µ \nu} S_µ S_{\nu} \Bigr) + \frac{2}{3} f_{k}^{T} \Bigl( g^{µ \nu} T_µ T_{\nu} \Bigr) - \frac{1}{2} f_{k}^{q} \Bigl( g^{µ \nu} g^{\rho \sigma} g^{\alpha \beta} q_{µ \rho \alpha} q_{\nu \sigma \beta} \Bigr) \\ 
  - \left.\frac{1}{\gamma_k} \left\{ \frac{1}{3} F_{k}^{ST} \Bigl( g^{µ \nu} T_µ S_{\nu} \Bigr) - F_{k}^{q} \left( \frac{\tensor{\epsilon}{^{µ \nu \rho \sigma}}}{\sqrt{g}} g^{\alpha \beta} q_{µ \nu \alpha} q_{\rho \sigma \beta} \right) \right\} \right] .
\end{multline}
We replaced the Newton and the cosmological constant, as well as the Immirzi parameter by $k$-dependent functions $\{G_k , \bar{\lambda}_k , \gamma_k \}$, and omitted the surface terms present in (\ref{eq:holstaction}). Assuming spacetime manifolds without boundary from now on, they are of no importance for our analysis. Moreover we have allowed for scale dependent prefactors $f_{k}$ for the parity-even, and $F_{k}$ for the parity-odd torsion invariants. They are normalized such that the ``Holst point'' in theory space has the coordinates $f^S = f^T = f^q = 1$ and $F^{ST} = F^q = 1$.

As for the gauge fixing term in (\ref{eq:trunctype}), $\Gamma_{k}^{\mathrm{gf}} \propto \int \intd^d x (\mathcal{F}_µ)^2$, the condition $\mathcal{F}_µ = 0$ must gauge fix the diffeomorphisms only, exactly as in metric gravity without torsion. This is one of the main differences with respect to the Einstein-Cartan theory where the local frame rotations must be gauge fixed in addition. A priori $\mathcal{F}_µ \equiv \mathcal{F}_µ [g,\bar{g},S,T,q,\ldots]$ may depend arbitrarily on all dynamical and background fields as long as it transforms tensorically under background gauge transformations \cite{Reuter:1996cp}. In the case at hand we opt for a torsion-independent gauge fixing, namely
\begin{equation}
	\mathcal{F}_µ = \frac{1}{\sqrt{16 \pi \bar{G}}} \left[ \delta_{µ}^{\beta} \bar{g}^{\alpha \gamma} (\bar{D}_{\LC})_{\gamma} - \frac{1}{2} \bar{g}^{\alpha \beta} (\bar{D}_{\LC})_µ \right] h_{\alpha \beta} \ .
	\label{eq:gaugecondition}
\end{equation}
Note that in (\ref{eq:gaugecondition}) we decided to employ the (background) Levi-Civita connection $\bar{D}_{\LC}$ rather than the equally admissible full connection $\bar{D}$. As a consequence, it represents exactly the same harmonic coordinate condition that is frequently used in Einstein gravity \cite{Reuter:1996cp}. As (\ref{eq:gaugecondition}) is indeed covariant under background gauge transformations, the resulting effective average action is invariant under simultaneous gauge transformations of all field arguments including the Faddeev-Popov ghosts $\xi^µ$ and $\bar{\xi}_µ$. The choice (\ref{eq:gaugecondition}) leads to the following ghost action:
\begin{multline}
	S_{\mathrm{gh}}\bigl[g,\bar{g},\xi,\bar{\xi}\bigr] = - \sqrt{2} \int \mathrm{d}^{4}x\, \sqrt{\bar{g}}\ \bar{\xi}_µ \left[ \vphantom{\frac{1}{2}} \bar{g}^{µ \rho} \bar{g}^{\sigma \lambda} (\bar{D}_{\LC})_{\lambda} \Bigl( g_{\rho \nu} (D_{\LC})_{\sigma} + g_{\sigma \nu} (D_{\LC})_{\rho} \Bigr) \right. \\ \left. \vphantom{\Bigl( g_{\rho \nu} (D_{\mathrm{LC}})_{\sigma} + g_{\sigma \nu} (D_{\mathrm{LC}})_{\rho} \Bigr)} - \bar{g}^{\rho \sigma} \bar{g}^{µ \lambda} (\bar{D}_{\LC})_{\lambda} g_{\sigma \nu} (D_{\LC})_{\rho} \vphantom{\frac{1}{2}} \right] \xi^{\nu}.
	\label{eq:ghostaction2}
\end{multline}
Thus, the gauge fixing and ghost sector of the truncation (\ref{eq:holsttruncation}) which we are going to use in the following sections is exactly the same as in \cite{Reuter:1996cp} where the Einstein-Hilbert truncation of purely metric gravity was studied.

\subsection{The parity-even sector}\label{subsec:parevensec}

To illustrate the general points we would like to make in this paper it is sufficient to reduce the above Holst truncation to a particularly transparent subsystem. We are going to consider only the limit of infinite Immirzi parameter, $\gamma \rightarrow \infty$, so that the remaining subsystem is solely comprised of the parity even part of the Holst truncation:
\begin{multline}
	\Gamma_{k}^{\mathrm{Tor}} \left[ g,S,T,q \right] = \frac{1}{16 \pi G_k} \int \mathrm{d}^{4}x\, \sqrt{g} \left[ - R_{\LC} + 2 \bar{\lambda}_k + \frac{1}{24} f_{k}^{S} \Bigl( g^{µ \nu} S_µ S_{\nu} \Bigr)  \right. \\ \left. + \frac{2}{3} f_{k}^{T} \Bigl( g^{µ \nu} T_µ T_{\nu} \Bigr) - \frac{1}{2} f_{k}^{q} \Bigl( g^{µ \nu} g^{\rho \sigma} g^{\alpha \beta} q_{µ \rho \alpha} q_{\nu \sigma \beta} \Bigr) \right]
	\label{eq:ehtrunctor}
\end{multline}
Restricting our attention to the parity even part of the Holst truncation, henceforth denoted $\Gamma_{k}^{\mathrm{Tor}}$, we of course loose the ability to investigate the RG flow of the Immirzi parameter itself, but the truncation is still general enough to study the influence of the torsion squared terms on the running of the Newton and cosmological coupling. And indeed, it is this influence which will be most illuminating to compare to the earlier calculations on $\mathcal{T}_{\mathrm{EC}}$. 

The limit $\gamma \rightarrow \infty$ is also of special importance in Quantum Einstein-Cartan Gravity (QECG) as it corresponds to the \textit{$(\lambda,g)$-subsystem}\footnote{Obviously the $(\lambda,g)$-subsystem does not only describe the case $\gamma \rightarrow \infty$, but is comprised of the general case $\gamma = \mathrm{const}$. However, the limit of infinite Immirzi parameter has generated some of the most interesting results, and it has proven to be especially robust and, therefore, reliable, see \cite{Daum:2010phd,Daum:2013fu} and \cite{Harst:2012phd,Harst:2014vca}.} which has turned out most suitable for a direct comparison with the metric case, Quantum Einstein Gravity (QEG). Therefore, in both existing analyses of $\mathcal{T}_{\mathrm{EC}}$, \cite{Daum:2010phd,Daum:2013fu} and \cite{Harst:2012phd,Harst:2014vca} respectively, the planes of infinite Immirzi parameter were investigated particularly thoroughly.

In the present paper, we try to build a bridge between QEG on one side, and QECG in the plane of infinite Immirzi parameter on the other. In \cite{Harst:2012phd,Harst:2012ni} ``tetrad only'' gravity was investigated as an intermediary between these theory spaces, $\mathcal{T}_{\mathrm{E}}$ and $\mathcal{T}_{\mathrm{EC}}$, and here we aim to supplement it in this role with the analysis of another closely related space, $\mathcal{T}_{\mathrm{dtor}}$.

%% file: EHTruncation.tex
With the restriction to the parity even part of the Holst truncation, $\Gamma_k$ is now of the form of (\ref{eq:trunctype}) with $\Gamma_{k}^{\mathrm{Ho}}$ replaced by $\Gamma_{k}^{\mathrm{Tor}}$:
\begin{align}
	\begin{split}
		\Gamma_{k} \left[g,\bar{g}, S ,T , q, \xi , \bar{\xi} \right] &= \Gamma_{k}^{\mathrm{Tor}} \bigl[ g,S,T,q \bigr] + \Gamma_{k}^{\mathrm{gf}}\bigl[g,\bar{g}\bigr] + S_{\mathrm{gh}}\bigl[g,\bar{g},\xi,\bar{\xi}\bigr] \\
		&\equiv \breve{\Gamma}_{k} \bigl[ g,\bar{g},S,T,q \bigr] + S_{\mathrm{gh}}\bigl[g,\bar{g},\xi,\bar{\xi}\bigr]
	\end{split}
	\label{eq:thetruncation}
\end{align}
Here we abbreviated $\breve{\Gamma}_{k} = \Gamma_{k}^{\mathrm{Tor}} + \Gamma_{k}^{\mathrm{gf}}$ whereby $\Gamma_{k}^{\mathrm{gf}}$ and $S_{\mathrm{gh}}$ are as specified above. The FRGE (\ref{eq:operatorflowequation}) for a truncation ansatz of this form where the ghost sector is treated $k$-independent decomposes into two parts which stem from the Grassmann-even fields and the ghosts, respectively:
\begin{equation}
	\partial_{t} \Gamma_{k} = \frac{1}{2} \operatorname{Tr} \left[ \frac{\bm{\left( \partial_t \breve{\mathcal{R}}_k \bigl( \Delta \bigr) \right)}}{\bm{\left( \breve{\Gamma}_{k}^{(2)} + \breve{\mathcal{R}}_k \bigl( \Delta \bigr) \right)}} \right] - \operatorname{Tr} \left[ \frac{\bm{\left( \partial_t \mathcal{R}_{k}^{\mathrm{gh}} \bigl( \Delta \bigr) \right)}}{\bm{\left( S_{\mathrm{gh}}^{(2)} + \mathcal{R}_{k}^{\mathrm{gh}} \bigl( \Delta \bigr) \right)}} \right].
	\label{eq:flowbrevegamma}
\end{equation}
We choose the respective cutoff operators $\bm{\breve{\mathcal{R}}_k}$ and $\bm{\mathcal{R}_{k}^{\mathrm{gh}}}$ to depend on the (negative) Laplacian $\Delta \equiv - \bar{g}^{µ \nu} (\bar{D}_{\LC})_µ (\bar{D}_{\LC})_{\nu}$ constructed from the background metric.

In this section we discuss in detail the various ingredients that enter into the flow equation (\ref{eq:flowbrevegamma}), with a particular emphasis on the field space metric, and we explicitly compute the $\beta$-functions for Newton's constant and the cosmological constant. They will serve as our main testing ground for studying the impact of the metric on $\mathcal{F}$. In Section \ref{subsec:metricdtor} we shall introduce our choice of $\text{\Fontlukas G}_{ij}$, before we then embark on the construction of the Hessian operator in Sections \ref{subsec:hessian1} and \ref{subsec:hessian2}, and the calculation of the traces in Section \ref{sec:H0}.

 To evaluate (\ref{eq:flowbrevegamma}), we only have to calculate the ``bosonic'' part $\breve{\Gamma}_{k}^{(2)}$ of the full Hessian $\Gamma_{k}^{(2)}$; the Hessian in the ghost sector, $S_{\mathrm{gh}}^{(2)}$, can be read off directly from the ghost action (\ref{eq:ghostaction2}) after setting $h=0$ , as it is already quadratic in the ghost fields.

For the rest of the Section we only deal with objects constructed from the Levi-Civita connection and therefore drop the subscript $_{\mathrm{LC}}$ unless we have to explicitly distinguish between torsion-full and torsion-free tensors. 

\subsection{A field space metric for decomposed torsion gravity}\label{subsec:metricdtor}

We want to use the theory space $\mathcal{T}_{\mathrm{dtor}}$ as an example to study the influence of the field space metric on the RG flow. In this section we specify the corresponding metric on $\mathcal{F}$.\\
\\
\textbf{(A)} Let $\mathcal{F}$ be the space of all fields $\Phi^i \equiv \left\{ g_{µ\nu}(x),S_{µ}(x),T_{µ}(x),q_{\lambda µ \nu}(x) \right\}$ on a given spacetime manifold $\mathcal{M}$. There are infinitely many covariant tensor densities which can be constructed out of the $\Phi^i$'s and which might serve as metrics on $\mathcal{F}$. Here we select the \textit{minimal choice} for such a metric retaining all fields.
Let $\mathcal{F}$ be defined as the product field space manifold $\mathcal{F} \equiv \mathcal{F}_{g} \times \mathcal{F}_S \times \mathcal{F}_T \times \mathcal{F}_q $ and we assume a line element of the general form
\begin{multline}
\label{eq:metricthefirst}
	\mathrm{d}\text{\Large{\Fontlukas s}}^2 = \int \mathrm{d}^{4}x \, \sqrt{g(x)} \int \mathrm{d}^{4}x' \, \sqrt{g(x')} \Bigl( G^{µ \nu \rho \sigma}_{gg}(x,x') \, \intd g_{µ \nu}(x) \, \intd g_{\rho \sigma}(x') \Bigr. \\
	+ G^{µ \nu}_{SS}(x,x') \, \intd S_µ (x) \, \intd S_{\nu} (x') + G^{µ \nu}_{TT}(x,x') \, \intd T_µ (x) \, \intd T_{\nu} (x') \\
	\Bigl. + G^{µ \nu \rho \alpha \beta \sigma}_{qq}(x,x') \, \intd q_{µ \nu \rho}(x) \, \intd q_{\alpha \beta \sigma}(x') \Bigr) \ .
\end{multline}
The coefficient functions $G^{µ \nu \rho \sigma}_{gg}, \cdots$ will be given in a moment.\footnote{This type of field space metric is analogous to the class of warped product merics \cite{ONeill:1983} on product manifolds in (semi-)Riemannian geometry.}\\
\\
\textbf{(B) Mass dimensions.} A remark on the canonical mass dimensions is in order here. We employ dimensionless coordinates, $\bigl[ x^µ \bigr] = 0$, so that the metric coefficients carry the entire dimensionality of $\bigl[ \mathrm{d}s^2 \bigr] = -2$, implying $\bigl[ g_{µ \nu} \bigr]  = -2$ and $\bigl[ g^{µ \nu} \bigr]  = +2$. One easily verifies that then $\bigl[ R \bigr] = +2$, as it should be, so that the couplings in the Einstein-Hilbert action have the expected dimensions $\bigl[ G \bigr] = -2$ and $\bigl[ \bar{\lambda} \bigr] = +2$, respectively. As for the torsion couplings in $\Gamma_{k}^{\mathrm{Tor}}$, they must be assumed dimensionless, 
\begin{equation}
	\bigl[ f_{k}^{S} \bigr] = \bigl[ f_{k}^{T} \bigr] = \bigl[ f_{k}^{q} \bigr] = 0 \ ,
\end{equation}
since we associate the dimensionless coordinate values $f_{k}^{S}=f_{k}^{T}=f_{k}^{q}=1$ with the Holst point in theory space. Deducing now the dimensions of the irreducible torsion fields that make the $(S_µ)^2$, $(T_µ)^2$, and $(q_{µ \nu \rho})^2$ terms in (\ref{eq:ehtrunctor}) dimensionally consistent, we find
\begin{equation}
 \bigl[ S_µ \bigr] = 0 \ , \ \bigl[ T_µ \bigr] = 0 \ , \ \bigl[ q_{µ \nu \rho} \bigr] = -2.
\end{equation}
The origin of the different dimensions is easy to understand: to form the ``squares'' $(S_µ)^2 \equiv g^{µ \nu} S_µ S_\nu$, and likewise for $T_µ$, we need only one factor of $g^{µ \nu}$, while $(q_{µ \nu \rho})^2$ requires three of them, each of which contributes $+2$ to the counting of dimensions. It is also clear that after the background split the components $\bar{\Phi}^i$ and $\varphi^i$, respectively, cannot be assigned uniform dimensions either:
\begin{align}
	\begin{split}
		\bigl[ \bar{g}_{µ \nu} \bigr] = -2 \ , \ \bigl[ \bar{S}_µ \bigr] = \bigl[ \bar{T}_µ \bigr] = 0 \ , \ \bigl[ \bar{q}_{µ \nu \rho} \bigr] = -2 \ ; \\
		\bigl[ h_{µ \nu} \bigr] = -2 \ , \ \bigl[ \mathpzc{S}_µ \bigr] = \bigl[ \mathpzc{T}_µ \bigr] = 0 \ , \ \bigl[ \mathpzc{q}_{µ \nu \rho} \bigr] = -2 \ .
	\end{split}
	\label{eq:dimensionsagain}
\end{align}
An obvious consequence is that the various partial functional derivatives of the EAA with respect to $\varphi^i = \{ h,\mathpzc{S},\mathpzc{T},\mathpzc{q} \}$ unavoidably have nonuniform dimensions, the prime example being the elements of the Hessian $\bigl( \breve{\Gamma}_{k}^{(2)} \bigr)_{ij}$.\\
\\
\textbf{(C)} The line element (\ref{eq:metricthefirst}) is well defined only if all sectors of $\mathcal{F}$ supply a term of the same dimension to $\mathrm{d}\text{\Large{\Fontlukas s}}^2$. Therefore we have to adjust the mass dimensions of certain field metrics $G_{\Phi \Phi}$. To achieve this we introduce a constant $\bar{µ}$ which has the dimension of a mass. Concretely we make the ultralocal choice
\begin{multline}
	\mathrm{d}\text{\Large{\Fontlukas s}}^2 
	\equiv \int \mathrm{d}^{4}x \, \sqrt{g(x)} \int \mathrm{d}^{4}x' \, \sqrt{g(x')} \, \left\{ \left[ \vphantom{\frac{1}{3}} g^{µ (\rho}(x) g^{\sigma) \nu}(x) \right] \intd g_{µ \nu}(x) \intd g_{\rho \sigma}(x') \right. \\
		 + \left[ \frac{g^{µ \nu} (x)}{\bar{µ}^2} \right] \intd S_µ (x) \intd S_{\nu} (x') + \left[ \frac{g^{µ \nu}(x)}{\bar{µ}^2} \right] \intd T_µ (x) \intd T_{\nu} (x') \\
	\left. + \left[ \frac{g^{µ \alpha }(x) g^{\nu [\beta}(x) g^{\sigma ] \rho}(x)}{\bar{µ}^2} \right] \intd q_{µ \nu \rho}(x) \intd q_{\alpha \beta \sigma}(x') \right\} \, \frac{\delta (x-x')}{\sqrt{g(x')}}
\label{eq:metricfieldspace2}
\end{multline}
While the matrix elements $\text{\Fontlukas G}_{ij}$ can be read off from (\ref{eq:metricfieldspace2}), the relation $\text{\Fontlukas G}_{ij}\text{\Fontlukas G}^{jl} = \delta_i^l$ implies for the components of the inverse $\text{\Fontlukas G}^{ij}$:
\begin{subequations}
	\begin{align}
		G_{\rho \sigma \kappa \lambda}^{gg}(x,x') &= \Bigl[ g_{\rho (\kappa}(x) g_{\lambda) \sigma}(x) \Bigr] \frac{\delta (x-x')}{\sqrt{g(x')}} \\
		G_{\nu \lambda}^{SS}(x,x') &= \Bigl[ \bar{µ}^2 \, g_{\nu \lambda}(x) \Bigr] \frac{\delta (x-x')}{\sqrt{g(x')}} \\
		G_{\nu \lambda}^{TT}(x,x') &= \Bigl[ \bar{µ}^2 \, g_{\nu \lambda}(x) \Bigr] \frac{\delta (x-x')}{\sqrt{g(x')}} \\
		G_{\alpha \beta \sigma \kappa \lambda \tau}^{qq}(x,x') &= \Bigl[ \bar{µ}^2 \, g_{\alpha \kappa}(x) g_{\beta [ \lambda}(x) g_{\tau ] \sigma}(x) \Bigr] \frac{\delta (x-x')}{\sqrt{g(x')}}
	\end{align}
\end{subequations}
Using (\ref{eq:metricfieldspace2}) to pull down the field index of $\Phi^i$ we obtain the following rule:
\begin{equation}
	\Phi^i \equiv \Bigl( g_{µ \nu} , S_µ , T_µ , q_{\lambda µ \nu} \Bigr) \quad \Longleftrightarrow \quad	\Phi_i \equiv \text{\Fontlukas G}_{ij} \Phi^j \equiv \left( g^{µ \nu} , \frac{S^µ}{\bar{µ}^2} , \frac{T^µ}{\bar{µ}^2} , \frac{q^{\lambda µ \nu}}{\bar{µ}^2} \right) \ .
\end{equation} 

\noindent \textbf{(D)} As we use a linear split of the fields into background fields and fluctuations, we have to specify how the metric acts on each part. Leaving the ghosts aside now we use the shorthand notation
\begin{subequations}
\begin{align}
	\Phi^i = \bar{\Phi}^i + \varphi^i &= \left( \vphantom{\bar{S}} g_{µ \nu},S_µ,T_µ,q_{\lambda µ \nu} \right) \\
	\text{with} \quad \bar{\Phi}^i &= \left( \bar{g}_{µ \nu},\bar{S}_µ,\bar{T}_µ,\bar{q}_{\lambda µ \nu} \right) \\
	\text{and} \quad \varphi^i &= \left( \vphantom{\bar{S}} h_{µ \nu}, \mathpzc{S}_µ, \mathpzc{T}_µ, \mathpzc{q}_{\lambda µ \nu} \right) \ .
	\label{eq:shorthand}
\end{align}
\end{subequations}
We leave the $x$-dependence implicit, unless specifically needed. Note that in the present case the abstract notation $\varphi^i$, with an \textit{upper} superindex, corresponds to the fields $h_{µ \nu}(x),\mathpzc{S}_{µ}(x),\mathpzc{T}_{µ}(x),\mathpzc{q}_{\lambda µ \nu}(x)$ with \textit{lower} spacetime indices throughout. Explicitly, the fields are split according to $g_{µ \nu} = \bar{g}_{µ \nu} + h_{µ \nu}$, together with
\begin{equation}
	S_µ = \bar{S}_µ + \mathpzc{S}_µ \ , \ T_µ = \bar{T}_µ + \mathpzc{T}_µ \ , \ q_{\lambda µ \nu} = \bar{q}_{\lambda µ \nu} + \mathpzc{q}_{\lambda µ \nu} \ .
\end{equation}
Here we decide that \textit{all indices are raised and lowered with the background field metric}, $\overline{\text{\Fontlukas G}}_{ij}$, i.e. the field space metric (\ref{eq:metricfieldspace2}) evaluated on the background fields, $\mathrm{d}\overline{\text{\Large{\Fontlukas s}}}^2 
	\equiv \bigl(\text{\Fontlukas G}[\bar{\Phi}]\bigr)_{ij} \, \intd \Phi^i \intd \Phi^j \equiv \overline{\text{\Fontlukas G}}_{ij} \, \intd \Phi^i \intd \Phi^j$. It reads explicitly
\begin{multline}
	\mathrm{d}\overline{\text{\Large{\Fontlukas s}}}^2 
	\equiv \int \mathrm{d}^{4}x \, \sqrt{\bar{g}(x)} \int \mathrm{d}^{4}x' \, \sqrt{\bar{g}(x')} \, \left\{ \left[ \vphantom{\frac{1}{3}} \bar{g}^{µ (\rho}(x) \bar{g}^{\sigma) \nu}(x) \right] \intd g_{µ \nu}(x) \intd g_{\rho \sigma}(x') \right. \\
		 + \left[ \frac{\bar{g}^{µ \nu} (x)}{\bar{µ}^2} \right] \intd S_µ (x) \intd S_{\nu} (x') + \left[ \frac{\bar{g}^{µ \nu}(x)}{\bar{µ}^2} \right] \intd T_µ (x) \intd T_{\nu} (x') \\
	\left. + \left[ \frac{\bar{g}^{µ \alpha }(x) \bar{g}^{\nu [\beta}(x) \bar{g}^{\sigma ] \rho}(x)}{\bar{µ}^2} \right] \intd q_{µ \nu \rho}(x) \intd q_{\alpha \beta \sigma}(x') \right\} \, \frac{\delta(x-x')}{\sqrt{\bar{g}(x')}} \ ,
\label{eq:metricfieldspace}
\end{multline}
As a consequence,
\begin{subequations}
\begin{align}
	\bar{\Phi}^i &\equiv \Bigl( \bar{g}_{µ \nu} , \bar{S}_µ , \bar{T}_µ , \bar{q}_{\lambda µ \nu} \Bigr) \quad \Longleftrightarrow \quad	\bar{\Phi}_i \equiv \left( \bar{g}^{µ \nu} , \frac{\bar{S}^µ}{\bar{µ}^2} , \frac{\bar{T}^µ}{\bar{µ}^2} , \frac{\bar{q}^{\lambda µ \nu}}{\bar{µ}^2} \right) \ , \\
	\varphi^i &\equiv \Bigl( h_{µ \nu} , \mathpzc{S}_µ , \mathpzc{T}_µ , \mathpzc{q}_{\lambda µ \nu} \Bigr) \quad \Longleftrightarrow \quad	\varphi_i \equiv \left( h^{µ \nu} , \frac{\mathpzc{S}^µ}{\bar{µ}^2} , \frac{\mathpzc{T}^µ}{\bar{µ}^2} , \frac{\mathpzc{q}^{\lambda µ \nu}}{\bar{µ}^2} \right) \ .
\end{align} 
\end{subequations}
We shall come back to these rules for raising and lowering field indices in a moment. We stress that the above metric is not the only viable choice. 

\subsection[The Hessian \texorpdfstring{$\breve{\Gamma}_{k}^{(2)}$}{Gamma}: Promoting the quadratic form to an operator]{The Hessian \texorpdfstring{$\bm{\breve{\Gamma}_{k}^{(2)}}$}{Gamma}: Promoting the quadratic form to an operator}\label{subsec:hessian1}
Next we turn to the evaluation of the truncated flow equation. To obtain $\bigl(\breve{\Gamma}_{k}^{(2)}\bigr)_{ij}$ and the operator associated to it, we first differentiate $\breve{\Gamma}_{k} = \Gamma_{k}^{\mathrm{Tor}} + \Gamma_{k}^{\mathrm{gf}}$ with respect to the fluctuation fields $\varphi$, keeping the background field configuration $\bar{\Phi}$ fixed:
\begin{equation}
	\frac{\delta^2}{\delta \Phi^i \delta \Phi^j} \, \breve{\Gamma}_{k} [\Phi,\bar{\Phi}] \equiv \frac{\delta^2}{\delta \varphi^i \delta \varphi^j} \, \breve{\Gamma}_{k} [\bar{\Phi} + \varphi,\bar{\Phi}]
\end{equation} 
In practice, we shall expand $\breve{\Gamma}_{k}$ in a Taylor series about $\bar{\Phi}$ to second order in the fluctuation fields,
\begin{equation}
	\breve{\Gamma}_{k} [ \bar{\Phi} + \varphi, \bar{\Phi} ] = \breve{\Gamma}_{k} [ \bar{\Phi}, \bar{\Phi} ] + \mathcal{O} ( \varphi ) + \breve{\Gamma}^{\mathrm{quad}}_{k} [ \varphi; \bar{\Phi} ] + \mathcal{O} ( \varphi^3  ) \ .
\end{equation}
It is then easy to read off the Hessian, $\bigl(\breve{\Gamma}_{k}^{(2)}\bigr)_{ij}$, from the functional bilinear in $\varphi$, that is, the quadratic form $\breve{\Gamma}^{\mathrm{quad}}_{k}$:
\begin{equation}
	\breve{\Gamma}^{\mathrm{quad}}_{k} [ \varphi; \bar{\Phi} ] \equiv \frac{1}{2} \, \varphi^i \, \bigl(\breve{\Gamma}_{k}^{(2)}\bigr)_{ij} \, \varphi^j \quad \text{with} \quad \breve{\Gamma}_{k}^{(2)} \equiv \breve{\Gamma}_{k}^{(2)} [0;\bar{\Phi}] .
	\label{eq:gammaquadagainapparantly}
\end{equation}

In a second step, we shall use the metric in field space in order to construct the associated operator, $\bm{\breve{\Gamma}_{k}^{(2)}}$, such that
\begin{equation}
	\breve{\Gamma}_{k}^{\mathrm{quad}} \equiv \frac{1}{2} \, \varphi_i \, \tensor{\bm{\bigl( \breve{\Gamma}_{k}^{(2)} \bigr)}}{^i_j} \, \varphi^j \equiv \frac{1}{2} \left(\varphi , \bm{ \breve{\Gamma}_{k}^{(2)}} \, \varphi \right) \quad \text{with} \quad \varphi_i \equiv \text{\Fontlukas G}_{ij} \varphi^j .
\end{equation}
Acceptable choices of $\text{\Fontlukas G}_{ij}$ are constrained by the condition that \textit{all matrix elements $\tensor{\bm{\bigl( \breve{\Gamma}_{k}^{(2)} \bigr)}}{^i_j}$ should have the same canonical mass dimension.} As we shall see, our above choice of $\text{\Fontlukas G}_{ij}$ does have this property, albeit only after a mandatory rescaling of the $\varphi^i$'s.

\subsubsection[The bilinear functional \texorpdfstring{$\breve{\Gamma}_{k}^{\mathrm{quad}}$}{Gammaquad}]{The bilinear functional \texorpdfstring{$\bm{\breve{\Gamma}_{k}^{\mathrm{quad}}}$}{Gammaquad}}\label{subsubsec:bilinfunc}
Explicitly, the quadratic form $\breve{\Gamma}^{\mathrm{quad}}_{k}$ is given by the following sum of both diagonal and off-diagonal terms, reflecting the block structure of the Hessian in field-space:
\begin{multline}
\label{eq:gammaquadratic}	\breve{\Gamma}^{\mathrm{quad}}_{k}[h,\mathpzc{S},\mathpzc{T},\mathpzc{q};\bar{g},\bar{S},\bar{T},\bar{q}] = \frac{1}{32 \pi G_k} \int \intd^{4}x\, \sqrt{\bar{g}} \left\{ \vphantom{\frac{1}{2}} h_{µ \nu} \left[ \vphantom{\frac{1}{2}} \mathcal{U}^{µ \nu \rho \sigma} - \mathcal{K}^{µ \nu \rho \sigma} \bar{g}^{\kappa \lambda} \bar{D}_{\kappa} \bar{D}_{\lambda} \right] h_{\rho \sigma} \right. \\
	+ 2 h_{µ \nu} \left[ \vphantom{\frac{1}{2}} \mathcal{L}^{µ \nu \lambda} \right] \mathpzc{S}_{\lambda} + 2 h_{µ \nu} \left[ \vphantom{\frac{1}{2}} \mathcal{M}^{µ \nu \lambda} \right] \mathpzc{T}_{\lambda} + 2 h_{µ \nu} \left[ \vphantom{\frac{1}{2}} \mathcal{N}^{µ \nu \lambda \rho \sigma} \right] \mathpzc{q}_{\lambda \rho \sigma} \\
	+ \left. \mathpzc{S}_{µ} \left[ \vphantom{\frac{1}{2}} \frac{1}{12} f_{k}^{S} \bar{g}^{µ \nu} \right] \mathpzc{S}_{\nu} + \mathpzc{T}_{µ} \left[ \vphantom{\frac{1}{2}} \frac{4}{3} f_{k}^{T} \bar{g}^{µ \nu} \right] \mathpzc{T}_{\nu} + \mathpzc{q}_{µ \rho \alpha} \left[ \vphantom{\frac{1}{2}} - f_{k}^{q} \bar{g}^{µ \nu} \bar{g}^{\rho [ \sigma} \bar{g}^{\beta ] \alpha} \right] \mathpzc{q}_{\nu \sigma \beta} \vphantom{\frac{1}{2}} \right\} \ .
\end{multline}
The contributions to $\breve{\Gamma}^{\mathrm{quad}}_{k}$ are bilinear in the fluctuation fields and contain various tensors, designated as $\bigl[ \, \cdots \bigr]$, which are built from the backgrounds $(\bar{g},\bar{S},\bar{T},\bar{q})$. The $hh$-diagonal term, for example, involves the 4\textsuperscript{th} rank tensors $\mathcal{U}$ and $\mathcal{K}$ given by
{\allowdisplaybreaks 
\begin{subequations}
\label{eq:kernels1}
\begin{align}
	\begin{split}
		\mathcal{U}^{µ \nu \rho \sigma} = & \left[ \frac{1}{2} \bar{g}^{µ (\rho|} \bar{g}^{\nu | \sigma)} - \frac{1}{4} \bar{g}^{µ \nu} \bar{g}^{\rho \sigma} \right] \left( \Bigl( \bar{R} - 2 \bar{\lambda}_k \Bigr) - \frac{1}{24} f_{k}^{S} \Bigl( \bar{g}^{\kappa \lambda} \bar{S}_{\kappa} \bar{S}_{\lambda} \Bigr) \right. \\ 
		& \phantom{~~~~~} \left. - \frac{2}{3} f_{k}^{T} \Bigl( \bar{g}^{\kappa \lambda} \bar{T}_{\kappa} \bar{T}_{\lambda} \Bigr)  + \frac{1}{2} f_{k}^{q} \Bigl( \bar{g}^{\kappa \lambda} \bar{g}^{\gamma \delta} \bar{g}^{\alpha \beta} \bar{q}_{\kappa \gamma \alpha} \bar{q}_{\lambda \delta \beta} \Bigr) \right) \\
		& + \frac{1}{2} \left[ \vphantom{\frac{1}{2}} \bar{g}^{µ \nu} \bar{R}^{\rho \sigma} + \bar{g}^{\rho \sigma} \bar{R}^{µ \nu} \right] \\
		& - \frac{1}{48} f_{k}^{S} \left[ \vphantom{\frac{1}{2}} \bar{g}^{µ \nu} \bar{g}^{\kappa \rho} \bar{g}^{\lambda \sigma} + \bar{g}^{\rho \sigma} \bar{g}^{\kappa µ} \bar{g}^{\lambda \nu} \right] \Bigl( \bar{S}_{\kappa} \bar{S}_{\lambda} \Bigr) \\
		& - \frac{1}{3} f_{k}^{T} \left[ \vphantom{\frac{1}{2}} \bar{g}^{µ \nu} \bar{g}^{\kappa \rho} \bar{g}^{\lambda \sigma} + \bar{g}^{\rho \sigma} \bar{g}^{\kappa µ} \bar{g}^{\lambda \nu} \right] \Bigl( \bar{T}_{\kappa} \bar{T}_{\lambda} \Bigr) \\
		& + \frac{1}{4} f_{k}^{q} \left[ \vphantom{\frac{1}{2}} \bar{g}^{µ \nu} \bar{g}^{\kappa (\rho|} \bar{g}^{\lambda |\sigma)} \bar{g}^{\gamma \delta} \bar{g}^{\alpha \beta} + \bar{g}^{\rho \sigma} \bar{g}^{\kappa (µ|} \bar{g}^{\lambda |\nu)} \bar{g}^{\gamma \delta} \bar{g}^{\alpha \beta} \right] \Bigl( \bar{q}_{\kappa \gamma \alpha}\, \bar{q}_{\lambda \delta \beta} \Bigr) \\
		& + \frac{1}{2} f_{k}^{q} \left[ \vphantom{\frac{1}{2}} \bar{g}^{µ \nu} \bar{g}^{\kappa (\rho|} \bar{g}^{\lambda |\sigma)} \bar{g}^{\gamma \delta} \bar{g}^{\alpha \beta} + \bar{g}^{\rho \sigma} \bar{g}^{\kappa (µ|} \bar{g}^{\lambda |\nu)} \bar{g}^{\gamma \delta} \bar{g}^{\alpha \beta} \right] \Bigl( \bar{q}_{\gamma \kappa \alpha}\, \bar{q}_{\delta \lambda \beta} \Bigr) \\
		& - \frac{1}{2} \left[ \vphantom{\frac{1}{2}} \bar{R}^{µ (\rho| \nu |\sigma)} + \bar{R}^{\nu (\rho| µ |\sigma)} + \bar{R}^{µ (\rho|} \bar{g}^{\nu |\sigma)} + \bar{R}^{\nu (\rho|} \bar{g}^{µ |\sigma)} \right] \\
		& - f_{k}^{q} \left[ \vphantom{\frac{1}{2}} \bar{g}^{(µ|\kappa} \bar{g}^{|\nu)\lambda} \bar{g}^{(\rho|\gamma} \bar{g}^{|\sigma)\delta} \bar{g}^{\alpha \beta} + \bar{g}^{(\rho|\kappa} \bar{g}^{|\sigma)\lambda} \bar{g}^{(µ|\gamma} \bar{g}^{|\nu)\delta} \bar{g}^{\alpha \beta} \right] \Bigl( \bar{q}_{\kappa \gamma \alpha}\, \bar{q}_{\lambda \delta \beta} \Bigr) \\
		& - \frac{1}{2} f_{k}^{q} \left[ \vphantom{\frac{1}{2}} \bar{g}^{(µ| \kappa} \bar{g}^{|\nu) \lambda} \bar{g}^{(\rho| \gamma} \bar{g}^{|\sigma) \delta} \bar{g}^{\alpha \beta} + \bar{g}^{(\rho| \kappa} \bar{g}^{|\sigma) \lambda} \bar{g}^{(µ| \gamma} \bar{g}^{|\nu) \delta} \bar{g}^{\alpha \beta} \right] \Bigl( \bar{q}_{\alpha \kappa \gamma}\, \bar{q}_{\beta \lambda \delta} \Bigr)
	\end{split} \\
	\mathcal{K}^{µ \nu \rho \sigma} = & \, \left[ \frac{1}{2} \bar{g}^{µ (\rho|} \bar{g}^{\nu | \sigma)} - \frac{1}{4} \bar{g}^{µ \nu} \bar{g}^{\rho \sigma} \right] \quad .
\end{align}
\end{subequations}}
The off-diagonal terms coupling $h_{µ \nu}$ to the torsion fluctuations are governed by the following three tensors of rank 3 and 5, respectively:
{\allowdisplaybreaks 
\begin{subequations}
\label{eq:kernelsagain}
	\begin{align}
		\mathcal{L}^{µ \nu \lambda} = & \, f_{k}^{S} \left[ \frac{1}{24} \bar{g}^{µ \nu} \bar{g}^{\kappa \lambda} - \frac{1}{12} \bar{g}^{(µ| \kappa} \bar{g}^{|\nu) \lambda} \right] \bar{S}_{\kappa} \\
		\mathcal{M}^{µ \nu \lambda} = & \, f_{k}^{T} \left[ \frac{2}{3} \bar{g}^{µ \nu} \bar{g}^{\kappa \lambda} - \frac{4}{3} \bar{g}^{(µ| \kappa} \bar{g}^{|\nu) \lambda} \right] \bar{T}_{\kappa} \\
		\begin{split}
		\mathcal{N}^{µ \nu \lambda \rho \sigma} = & \, f_{k}^{q} \left[ - \frac{1}{2} \bar{g}^{µ \nu} \bar{g}^{\kappa \lambda} \bar{g}^{\alpha [ \rho} \bar{g}^{\sigma ] \beta} \right] q_{\kappa \alpha \beta} + f_{k}^{q} \left[ \vphantom{\frac{1}{2}}\bar{g}^{(µ| \kappa} \bar{g}^{|\nu) \lambda} \bar{g}^{\alpha [ \rho} \bar{g}^{\sigma ] \beta} \right] q_{\kappa \alpha \beta} \\
	& + f_{k}^{q} \left[\vphantom{\frac{1}{2}} 2 \bar{g}^{(µ| \alpha} \bar{g}^{|\nu) [ \rho|} \bar{g}^{\kappa \lambda} \bar{g}^{\beta |\sigma]} \right] q_{\kappa \alpha \beta} \quad .
		\end{split}
	\end{align}
\end{subequations}}
As it stands, (\ref{eq:gammaquadratic}) represents $\breve{\Gamma}^{\mathrm{quad}}$ in the style of (\ref{eq:gammaquadagainapparantly}), i.e. as 
\begin{equation}
	\breve{\Gamma}^{\mathrm{quad}}_{k} \equiv \frac{1}{2} \, \bigl( \breve{\Gamma}_{k}^{(2)} \bigr)_{ij} \, \varphi^i \varphi^j \ .
\end{equation}
After symmetrizing the off-diagonal terms it is easy to read off the components $\bigl(\breve{\Gamma}_{k}^{(2)}\bigr)_{ij} = \bigl(\breve{\Gamma}_{k}^{(2)}\bigr)_{ji}$.

\subsubsection[Uniform dimensions of \texorpdfstring{$\bigl( \breve{\Gamma}_{k}^{(2)} \bigr)^i{}_j$}{Gammaop} ?]{Uniform dimensions of \texorpdfstring{$\bm{\bigl( \breve{\Gamma}_{k}^{(2)} \bigr)^i{}_j}$}{Gammaop} ?}\label{subsubsec:uniformdimensions}

Our next task is to turn the quadratic form $\bigl( \breve{\Gamma}_{k}^{(2)} \bigr)_{ij}$ into an operator $\bm{\bigl( \breve{\Gamma}_{k}^{(2)} \bigr)}$ whose matrix elements $\tensor{\bm{\bigl( \breve{\Gamma}_{k}^{(2)} \bigr)}}{^i_j}$ have the same mass dimension for all $i$ and $j$. Setting $\tensor{\bm{\bigl( \breve{\Gamma}_{k}^{(2)} \bigr)}}{^i_j} = \text{\Fontlukas G}^{il} \bigl( \breve{\Gamma}_{k}^{(2)} \bigr)_{lj}$ this requirement clearly implies certain conditions on $\text{\Fontlukas G}^{ij}$. Let us see whether the metric chosen above satisfies them. 

Utilizing the metric in field space defined in (\ref{eq:metricfieldspace}), we rewrite $\breve{\Gamma}^{\mathrm{quad}}_{k}$ in the style of an ``expectation value'' $\breve{\Gamma}_{k}^{\mathrm{quad}} \equiv \frac{1}{2} \, \varphi_i \, \tensor{\bm{\bigl( \breve{\Gamma}_{k}^{(2)} \bigr)}}{^i_j} \, \varphi^j$ from which the operator matrix elements can be read off. We find
\begin{multline}
\label{eq:gammaoperator}	\breve{\Gamma}^{\mathrm{quad}}_{k}[h,\mathpzc{S},\mathpzc{T},\mathpzc{q};\bar{g},\bar{S},\bar{T},\bar{q}] = \frac{1}{32 \pi G_k} \int \intd^{4}x\, \sqrt{\bar{g}} \left\{ \vphantom{\frac{1}{2}} h^{µ \nu} \left[ \vphantom{\frac{1}{2}} \tensor{\mathcal{U}}{_{µ \nu}^{\rho \sigma}} - \tensor{\mathcal{K}}{_{µ \nu }^{\rho \sigma}} \bar{g}^{\kappa \lambda} \bar{D}_{\kappa} \bar{D}_{\lambda} \right] h_{\rho \sigma} \right. \\
	+ h^{µ \nu} \left[ \vphantom{\frac{1}{2}} \tensor{\mathcal{L}}{_{µ \nu}^{\lambda}} \right] \mathpzc{S}_{\lambda} + h^{µ \nu} \left[ \vphantom{\frac{1}{2}} \tensor{\mathcal{M}}{_{µ \nu}^{\lambda}} \right] \mathpzc{T}_{\lambda} + h^{µ \nu} \left[ \vphantom{\frac{1}{2}} \tensor{\mathcal{N}}{_{µ \nu}^{\lambda \rho \sigma}} \right] \mathpzc{q}_{\lambda \rho \sigma} \\
	+ \frac{\mathpzc{S}^{\lambda}}{\bar{µ}^2} \left[ \vphantom{\frac{1}{2}} \tensor{\mathcal{A}}{^{µ \nu}_{\lambda}} \right] h_{µ \nu} + \frac{\mathpzc{T}^{\lambda}}{\bar{µ}^2} \left[ \vphantom{\frac{1}{2}} \tensor{\mathcal{B}}{^{µ \nu}_{\lambda}} \right] h_{µ \nu} + \frac{\mathpzc{q}^{\lambda \rho \sigma}}{\bar{µ}^2} \left[ \vphantom{\frac{1}{2}} \tensor{\mathcal{C}}{^{µ \nu}_{\lambda \rho \sigma}} \right] h_{µ \nu} \\
	+ \left. \frac{\mathpzc{S}^{µ}}{\bar{µ}^2} \left[ \vphantom{\frac{1}{2}} \frac{\bar{µ}^2}{12} f_{k}^{S} \delta^{\nu}_{µ} \right] \mathpzc{S}_{\nu} + \frac{\mathpzc{T}^{µ}}{\bar{µ}^2} \left[ \vphantom{\frac{1}{2}} \frac{4 \bar{µ}^2}{3} f_{k}^{T} \delta^{\nu}_{µ} \right] \mathpzc{T}_{\nu} + \frac{\mathpzc{q}^{µ \rho \alpha}}{\bar{µ}^2} \left[- \bar{µ}^2 f_{k}^{q} \delta^{\nu}_{µ} \delta^{\![\sigma}_{\rho} \delta^{\beta]}_{\alpha}\vphantom{\frac{\bar{µ}^2}{2}} \right] \mathpzc{q}_{\nu \sigma \beta} \vphantom{\frac{1}{2}} \right\} \ .
\end{multline}
The corresponding kernels are given by:
{\allowdisplaybreaks
\begin{subequations}
\begin{align}
	\begin{split}
		\tensor{\mathcal{U}}{_{µ \nu}^{\rho \sigma}} = & \left[ \frac{1}{2} \delta^{\!(\rho}_{µ} \delta^{\sigma)}_{\nu} - \frac{1}{4} \bar{g}_{µ \nu} \bar{g}^{\rho \sigma} \right] \left( \Bigl( \bar{R} - 2 \bar{\lambda}_k \Bigr) - \frac{1}{24} f_{k}^{S} \Bigl( \bar{g}^{\kappa \lambda} \bar{S}_{\kappa} \bar{S}_{\lambda} \Bigr) \right. \\ 
		& \phantom{~~~~~} \left. - \frac{2}{3} f_{k}^{T} \Bigl( \bar{g}^{\kappa \lambda} \bar{T}_{\kappa} \bar{T}_{\lambda} \Bigr)  + \frac{1}{2} f_{k}^{q} \Bigl( \bar{g}^{\kappa \lambda} \bar{g}^{\gamma \delta} \bar{g}^{\alpha \beta} \bar{q}_{\kappa \gamma \alpha} \bar{q}_{\lambda \delta \beta} \Bigr) \right) \\
		& + \frac{1}{2} \left[ \vphantom{\frac{1}{2}} \bar{g}_{µ \nu} \bar{R}^{\rho \sigma} + \bar{g}^{\rho \sigma} \bar{R}_{µ \nu} \right] - \tensor{\bar{R}}{_{\!(µ}^{\!(\rho}_{\!\nu)}^{\!\sigma)}} - \tensor{R}{_{\!(µ}^{\!(\rho}} \delta_{\nu)}^{\sigma)} \\
		& - \frac{1}{48} f_{k}^{S} \left[ \vphantom{\frac{1}{2}} \bar{g}_{µ \nu} \bar{g}^{\kappa \rho} \bar{g}^{\lambda \sigma} \bar{S}_{\kappa} \bar{S}_{\lambda} + \bar{g}^{\rho \sigma} \bar{S}_{µ} \bar{S}_{\nu} \right]  \\
		& - \frac{1}{3} f_{k}^{T} \left[ \vphantom{\frac{1}{2}} \bar{g}_{µ \nu} \bar{g}^{\kappa \rho} \bar{g}^{\lambda \sigma} \bar{T}_{\kappa} \bar{T}_{\lambda} + \bar{g}^{\rho \sigma} \bar{T}_{µ} \bar{T}_{\nu} \right] \\
		& + \frac{1}{4} f_{k}^{q} \left[ \vphantom{\frac{1}{2}} \bar{g}_{µ \nu} \bar{g}^{\kappa (\rho|} \bar{g}^{\lambda |\sigma)} \bar{g}^{\gamma \delta} \bar{g}^{\alpha \beta} \bar{q}_{\kappa \gamma \alpha}\, \bar{q}_{\lambda \delta \beta} + \bar{g}^{\rho \sigma} \bar{g}^{\gamma \delta} \bar{g}^{\alpha \beta} \bar{q}_{(µ| \gamma \alpha}\, \bar{q}_{|\nu) \delta \beta} \right] \\
		& + \frac{1}{2} f_{k}^{q} \left[ \vphantom{\frac{1}{2}} \bar{g}_{µ \nu} \bar{g}^{\kappa (\rho|} \bar{g}^{\lambda |\sigma)} \bar{g}^{\gamma \delta} \bar{g}^{\alpha \beta} \bar{q}_{\gamma \kappa \alpha}\, \bar{q}_{\delta \lambda \beta} + \bar{g}^{\rho \sigma} \bar{g}^{\gamma \delta} \bar{g}^{\alpha \beta} \bar{q}_{\gamma (µ| \alpha}\, \bar{q}_{\delta |\nu) \beta} \right] \\
		& - f_{k}^{q} \left[ \vphantom{\frac{1}{2}} \bar{g}^{(\rho|\kappa} \bar{g}^{|\sigma) \lambda} \bar{g}^{\alpha \beta} \right] \Bigl( \bar{q}_{\kappa (µ| \alpha}\, \bar{q}_{\lambda |\nu) \beta} + \bar{q}_{(µ| \kappa \alpha}\, \bar{q}_{|\nu) \lambda \beta} \Bigr) \\
		& - \frac{1}{2} f_{k}^{q} \left[ \vphantom{\frac{1}{2}} \bar{g}^{(\rho| \kappa} \bar{g}^{|\sigma) \lambda} \bar{g}^{\alpha \beta} \right] \Bigl( \bar{q}_{\alpha \kappa (µ|}\, \bar{q}_{\beta \lambda |\nu)} + \bar{q}_{\alpha (µ| \kappa}\, \bar{q}_{\beta |\nu) \lambda} \Bigr)
	\end{split} \\
	\tensor{\mathcal{K}}{_{µ \nu}^{\rho \sigma}} = & \left[ \frac{1}{2} \delta^{\!(\rho}_{µ} \delta^{\sigma)}_{\nu} - \frac{1}{4} \bar{g}_{µ \nu} \bar{g}^{\rho \sigma} \right] \\
	\tensor{\mathcal{L}}{_{µ \nu}^{\lambda}} = & \, f_{k}^{S} \left[ \frac{1}{24} \bar{g}_{µ \nu} \bar{g}^{\kappa \lambda} - \frac{1}{12} \delta^{\kappa}_{\!(µ} \delta^{\lambda}_{\nu)} \right] \bar{S}_{\kappa} \\
	\tensor{\mathcal{M}}{_{µ \nu}^{\lambda}} = & \, f_{k}^{T} \left[ \frac{2}{3} \bar{g}_{µ \nu} \bar{g}^{\kappa \lambda} - \frac{4}{3} \delta^{\kappa}_{\!(µ} \delta^{\lambda}_{\nu)} \right] \bar{T}_{\kappa} \\
	\tensor{\mathcal{N}}{_{µ \nu}^{\lambda \rho \sigma}} = & \, f_{k}^{q} \left[ - \frac{1}{2} \bar{g}_{µ \nu} \bar{g}^{\kappa \lambda} \bar{g}^{\alpha [ \rho} \bar{g}^{\sigma ] \beta} + \delta_{\!(µ}^{\kappa} \delta_{\nu)}^{\lambda} \bar{g}^{\alpha [ \rho} \bar{g}^{\sigma ] \beta} + 2 \delta_{\!(µ}^{\alpha} \delta_{\nu)}^{\![\rho|} \bar{g}^{\kappa \lambda} \bar{g}^{\beta |\sigma]} \right] q_{\kappa \alpha \beta} \\
	\tensor{\mathcal{A}}{^{µ \nu}_{\lambda}} = & \, \bar{µ}^2 f_{k}^{S} \left[ \frac{1}{24} \bar{g}^{µ \nu} \bar{S}_{\lambda} - \frac{1}{12} \delta_{\lambda}^{\!(µ} \bar{g}^{|\nu) \kappa} \bar{S}_{\kappa} \right] \\
	\tensor{\mathcal{B}}{^{µ \nu}_{\lambda}} = & \, \bar{µ}^2 f_{k}^{T} \left[ \frac{2}{3} \bar{g}^{µ \nu} \bar{T}_{\lambda} - \frac{4}{3} \delta_{\lambda}^{\!(µ} \bar{g}^{|\nu) \kappa} \bar{T}_{\kappa} \right] \\
	\tensor{\mathcal{C}}{^{µ \nu}_{\lambda \rho \sigma}} = & \, \bar{µ}^2 f_{k}^{q} \left[ -\frac{1}{2} \bar{g}^{µ \nu} \bar{q}_{\lambda \rho \sigma} + \delta_{\lambda}^{\!(µ} \bar{g}^{\nu)\kappa} \bar{q}_{\kappa \rho \sigma} + 2 \delta_{\![\rho|}^{\!(µ} \bar{g}^{\nu) \kappa} \bar{q}_{\lambda \kappa |\sigma]} \right] \quad .
\end{align}
\end{subequations}}
It is easily checked that the two formulae for $\breve{\Gamma}^{\mathrm{quad}}_{k}$, (\ref{eq:gammaoperator}) and (\ref{eq:gammaquadratic}), are equivalent. 

Unfortunately, however, our goal of furnishing the matrix elements of the Hessian operator $\tensor{\bm{\bigl( \breve{\Gamma}_{k}^{(2)} \bigr)}}{^i_j}$, to be read off from the terms in the square brackets in (\ref{eq:gammaoperator}), with equal mass dimensions has not yet been accomplished. To make this explicit, we have collected the mass dimensions of all relevant expression in Table \ref{tab:massdimensions}. The reader can easily check that the relevant dimensions as given in the last column do not agree with each other, illustrating that our procedure has not yet fully solved the problem.

\begin{table}[H]
	\centering
		\begin{tabular}{c||c|c||c|c}
		couplings & cov. fields & contrav. fields & $\bigl( \breve{\Gamma}_{k}^{(2)} \bigr)_{ij}$ tensors & $\tensor{\bm{\bigl( \breve{\Gamma}_{k}^{(2)} \bigr)}}{^i_j}$ tensors \\
		\hline \hline
		$\begin{aligned}[t]
		[G_k] &= -2 \\
		[\bar{\lambda}_k] &= +2 \\
		[f_k^S] &= 0 \\
		[f_k^T] &= 0 \\
		[f_k^q] &= 0
		\end{aligned}$ & $\begin{aligned}[t]
		[g_{µ \nu}] &= -2 \\
		[S_{µ}] &= 0 \\
		[T_{µ}] &= 0 \\
		[q_{\lambda µ \nu}] &= -2 
		\end{aligned}$ & $\begin{aligned}[t]
		[g^{µ \nu}] &= +2 \\
		[S^{µ}] &= +2 \\
		[T^{µ}] &= +2 \\
		[q^{\lambda µ \nu}] &= +4 
		\end{aligned}$ & $\begin{aligned}[t]
		[\mathcal{U}^{µ \nu \rho \sigma}] &= +6 \\
		[\mathcal{K}^{µ \nu \rho \sigma} \bar{D}^2] &= +6 \\
		[\mathcal{L}^{µ \nu \lambda}] &= +4 \\
		[\mathcal{M}^{µ \nu \lambda}] &= +4 \\
		[\mathcal{N}^{µ \nu \lambda \rho \sigma}] &= +6 
		\end{aligned}$ & $\begin{aligned}[t]
		[\tensor{\mathcal{U}}{_{µ \nu }^{\rho \sigma}}] &= +2 \\
		[\tensor{\mathcal{K}}{_{µ \nu }^{\rho \sigma}} \bar{D}^2] &= +2 \\
		[\tensor{\mathcal{L}}{_{µ \nu }^{\lambda}}] &= 0 \\
		[\tensor{\mathcal{A}}{^{µ \nu }_{\lambda}}] &= +4 \\
		[\tensor{\mathcal{M}}{_{µ \nu }^{\lambda}}] &= 0 \\
		[\tensor{\mathcal{B}}{^{µ \nu }_{\lambda}}] &= +4 \\
		[\tensor{\mathcal{N}}{_{µ \nu }^{\lambda \rho \sigma}}] &= +2 \\
		[\tensor{\mathcal{C}}{^{µ \nu }_{\lambda \rho \sigma}}] &= +2
		\end{aligned}$
		\end{tabular}
	\caption{Mass dimensions of the various expressions comprising the quadratic form (\ref{eq:gammaquadratic}) and its operatorial analog (\ref{eq:gammaoperator}). Comparing them note also that $[\sqrt{g}] = -4 , [R^{µ \nu \rho \sigma}] = +6 , [R^{µ \nu}] = +4 ,[R] = +2 $.}
	\label{tab:massdimensions}
\end{table}

Comparing the various blocks in (\ref{eq:gammaoperator}) we find that the off-diagonal terms $\propto h-\mathpzc{S},\mathpzc{S}-h$ and $\propto h-\mathpzc{T},\mathpzc{T}-h$ still deviate from the desired mass dimension $+2$. On a positive note, however, the metric we chose managed to bring all diagonal terms to an uniform mass dimension. A closer look reveals that it is unfeasible to align all sectors of $\tensor{\bm{\bigl( \breve{\Gamma}_{k}^{(2)} \bigr)}}{^i_j}$, with a metric in field space alone, as long as the fields $\Phi^i$ have different mass dimensions.

To remedy this defeat, we have to take an extra step, which we describe next.

\subsubsection[The Hessian \texorpdfstring{$\breve{\Gamma}_{k}^{(2)}$}{Gamma} after rescaling the fields]{The Hessian \texorpdfstring{$\bm{\breve{\Gamma}_{k}^{(2)}}$}{Gamma} after rescaling the fields}\label{subsec:hessian2}

We saw that the dimensionally adjusted metric allows for a uniform mass dimension of all diagonal-terms in the operator $\tensor{\bm{\bigl( \breve{\Gamma}_{k}^{(2)} \bigr)}}{^i_j}$, \textit{as well as all terms stemming from fields with the same mass dimension.} The two blocks, $\tensor{\mathcal{N}}{_{µ \nu }^{\lambda \rho \sigma}}$ and $\tensor{\mathcal{C}}{_{µ \nu }^{\lambda \rho \sigma}}$, both agreed with the dimension of the diagonal ones, due to the underlying basis fields, namely $g_{µ \nu}$ and $q_{\lambda µ \nu}$, both being of mass dimension $-2$. Only if all fields $\Phi^i$ have the same mass dimension we can manage to redistribute the mass dimensions from the quadratic form, $\breve{\Gamma}^{\mathrm{quad}}_{k} \equiv \frac{1}{2} \, \bigl( \breve{\Gamma}_{k}^{(2)} \bigr)_{ij} \, \varphi^i \varphi^j$, to the operator form, $	\breve{\Gamma}_{k}^{\mathrm{quad}} \equiv \frac{1}{2} \, \varphi_i \, \tensor{\bm{\bigl( \breve{\Gamma}_{k}^{(2)} \bigr)}}{^i_j} \, \varphi^j$, in such a way that all elements $\tensor{\bm{\bigl( \breve{\Gamma}_{k}^{(2)} \bigr)}}{^i_j}$ have uniform mass dimension as well.

Thus we introduce rescaled torsion fields, $\widetilde{S}$ and $\widetilde{T}$, according to
\begin{equation}
	S_µ \equiv \bar{µ}^2 \, \widetilde{S}_µ \quad \text{and} \quad T_µ \equiv \bar{µ}^2 \, \widetilde{T}_µ \ ,
\label{eq:rescaledtorsion}
\end{equation} 
and analogously for the background fields, $\bar{S}_µ \equiv \bar{µ}^2 \, \widetilde{\bar{S}}_µ$, $\bar{T}_µ \equiv \bar{µ}^2 \, \widetilde{\bar{T}}_µ$, and the fluctuations, $\mathpzc{S}_µ \equiv \bar{µ}^2 \, \widetilde{\mathpzc{S}}_µ$, $\mathpzc{T}_µ \equiv \bar{µ}^2 \, \widetilde{\mathpzc{T}}_µ$. Leaving the metric and the rank-3 tensor untouched the new field multiplets are
\begin{subequations}
\begin{align}
	\widetilde{\Phi}^i &= \left\{ \vphantom{\bar{S}} g_{µ \nu},\widetilde{S}_µ,\widetilde{T}_µ,q_{\lambda µ \nu} \right\} \\
	\text{with} \quad \widetilde{\bar{\Phi}}{}^i &= \left\{ \bar{g}_{µ \nu},\widetilde{\bar{S}}_µ,\widetilde{\bar{T}}_µ,\bar{q}_{\lambda µ \nu} \right\} \\
	\text{and} \quad \widetilde{\varphi}^i &= \left\{ h_{µ \nu},\widetilde{\mathpzc{S}}_µ,\widetilde{\mathpzc{T}}_µ,\mathpzc{q}_{\lambda µ \nu} \right\} \ .
	\label{eq:shorthand2}
\end{align}
\end{subequations}
We emphasize that we retain the the usual form of the linear background split here\footnote{In an alternative setting, sometimes adopted in perturbation theory, one could give different dimensions to the background fields and the corresponding fluctuations, defining, say, $\Phi^i = \bar{\Phi}^i + c^i \widetilde{\varphi}^i$ where now $[c^i] \neq 0$. We shall not follows this route here.}: $\widetilde{\Phi}^i = \widetilde{\bar{\Phi}}{}^i + \widetilde{\varphi}^i$.

All components of the rescaled background, fluctuation, and full fields are of the same dimensionality now: $[\widetilde{\Phi}^i] = [\widetilde{\bar{\Phi}}{}^i] = [\widetilde{\varphi}^i] = - 2$, or concretely
\begin{equation}
\boxed{
	[g_{µ \nu}] = [\widetilde{S}_µ] = [\widetilde{T}_µ] = [q_{\lambda µ \nu}] = -2 
	} \ .
\end{equation}

We have to affirm that one might as well rescale the opposite blocks and we made our choice to have a closer correspondence with earlier investigations. By using a different selection of the dimensionality of the basis, e.g. having the coordinates carry the dimensionality, one also changes the required mass dimensions of the $G_{\Phi \Phi}$'s, implying yet another diverse metric in field space. The choices we made in defining (\ref{eq:metricfieldspace2}) and (\ref{eq:rescaledtorsion}) seem the most natural and compatible with our approach.

It is straightforward now to adapt the truncation ansatz (\ref{eq:ehtrunctor}) and the resulting quadratic form (\ref{eq:gammaquadratic}) - (\ref{eq:kernelsagain}) to the new field basis. The changes boil down to replacing in all formulae $S_µ \rightarrow \widetilde{S}_µ$ and $f_k^S \rightarrow \bar{µ}^4 f_k^S$ and analogously $T_µ \rightarrow \widetilde{T}_µ$, $f_k^T \rightarrow \bar{µ}^4 f_k^T$.

We also must adapt the metric in field space to the new coordinates on $\mathcal{F}$. Adjusting the $S$- and $T$-field metrics, $G^{µ \nu}_{SS} \rightarrow G^{µ \nu}_{\widetilde{S}\widetilde{S}}$, $G^{µ \nu}_{TT} \rightarrow G^{µ \nu}_{\widetilde{T}\widetilde{T}}$, the new metric in field space reads:
\begin{multline}
	\mathrm{d}\widetilde{\text{\Large{\Fontlukas s}}}^2 
	\equiv \int \mathrm{d}^{4}x \, \sqrt{g(x)} \int \mathrm{d}^{4}x' \, \sqrt{g(x')} \, \left\{ \left[ \vphantom{\frac{1}{3}} g^{µ (\rho}(x) g^{\sigma) \nu}(x) \right] \intd g_{µ \nu}(x) \intd g_{\rho \sigma}(x') \right. \\
		 + \left[ \vphantom{\frac{1}{2}} \bar{µ}^2 g^{µ \nu}(x) \right] \intd \widetilde{S}_µ (x) \intd \widetilde{S}_{\nu} (x') + \left[ \vphantom{\frac{1}{2}} \bar{µ}^2 g^{µ \nu}(x) \right] \intd \widetilde{T}_µ (x) \intd \widetilde{T}_{\nu} (x') \\
	\left. + \left[ \frac{g^{µ \alpha }(x) g^{\nu [\beta}(x) g^{\sigma ] \rho}(x)}{\bar{µ}^2} \right] \intd q_{µ \nu \rho}(x) \intd q_{\alpha \beta \sigma}(x') \right\} \, \frac{\delta (x-x')}{\sqrt{g(x')}} .
\label{eq:metricfieldspaceredefined}
\end{multline}
The inverse of the new blocks is found to be
\begin{subequations}
	\begin{align}
		G_{\nu \lambda}^{\widetilde{S}\widetilde{S}}(x,x') &= \left[ \frac{g_{\nu \lambda}(x)}{\bar{µ}^2} \right] \frac{\delta (x-x')}{\sqrt{g(x')}} \\
		G_{\nu \lambda}^{\widetilde{T}\widetilde{T}}(x,x') &= \left[ \frac{g_{\nu \lambda}(x)}{\bar{µ}^2} \right] \frac{\delta (x-x')}{\sqrt{g(x')}} \ .
	\end{align}
\end{subequations}
By changing the metric we also change the set of fields with a lowered index
\begin{equation}
	\widetilde{\Phi}_i \equiv \left( g^{µ \nu} , \bar{µ}^2 \widetilde{S}^µ , \bar{µ}^2 \widetilde{T}^µ , \frac{q^{\lambda µ \nu}}{\bar{µ}^2} \right) \ ,
\end{equation} 
and likewise for $\widetilde{\varphi}_i$.

Taking the above changes into account we recalculate the ``expectation value'' form of the quadratic action, 
\begin{equation}
	\breve{\Gamma}_{k}^{\mathrm{quad}} [\widetilde{\varphi};\widetilde{\bar{\Phi}}] \equiv \frac{1}{2} \, \widetilde{\varphi}_i \, \tensor{\bm{\bigl( \breve{\Gamma}_{k}^{(2)} \bigr)}}{^i_j} \, \widetilde{\varphi}^j \equiv \frac{1}{2} \, \widetilde{\varphi}^m \, \overline{\text{\Fontlukas G}}_{mi} \, \overline{\text{\Fontlukas G}}^{in} \, \bigl(\breve{\Gamma}_{k}^{(2)}\bigr)_{nj} \, \widetilde{\varphi}^j \ , 
\end{equation}
and are lead to the following final result
\begin{multline}
\label{eq:gammanewoperator}
\breve{\Gamma}^{\mathrm{quad}}_{k}[h,\widetilde{\mathpzc{S}},\widetilde{\mathpzc{T}},\mathpzc{q};\bar{g},\widetilde{\bar{S}},\widetilde{\bar{T}},\bar{q}]
	= \frac{1}{32 \pi G_k} \int \intd^{4}x\, \sqrt{\bar{g}} \left\{ \vphantom{\frac{1}{2}} 
	h^{µ \nu} \left[ \vphantom{\frac{1}{2}} \tensor{\widetilde{\mathcal{U}}}{_{µ \nu}^{\rho \sigma}} - \tensor{\mathcal{K}}{_{µ \nu }^{\rho \sigma}} \bar{g}^{\kappa \lambda} \bar{D}_{\kappa} \bar{D}_{\lambda} \right] h_{\rho \sigma} \right. \\
	+ h^{µ \nu} \left[ \vphantom{\frac{1}{2}} \tensor{\widetilde{\mathcal{L}}}{_{µ \nu}^{\lambda}} \right] \widetilde{\mathpzc{S}}_{\lambda} 
	+ h^{µ \nu} \left[ \vphantom{\frac{1}{2}} \tensor{\widetilde{\mathcal{M}}}{_{µ \nu}^{\lambda}} \right] \widetilde{\mathpzc{T}}_{\lambda} 
	+ h^{µ \nu} \left[ \vphantom{\frac{1}{2}} \tensor{\mathcal{N}}{_{µ \nu}^{\lambda \rho \sigma}} \right] \mathpzc{q}_{\lambda \rho \sigma} \\
	+ \bar{µ}^2 \widetilde{\mathpzc{S}}^{\lambda} \left[ \vphantom{\frac{1}{2}} \tensor{\widetilde{\mathcal{A}}}{^{µ \nu}_{\lambda}} \right] h_{µ \nu} 
	+ \bar{µ}^2 \widetilde{\mathpzc{T}}^{\lambda} \left[ \vphantom{\frac{1}{2}} \tensor{\widetilde{\mathcal{B}}}{^{µ \nu}_{\lambda}} \right] h_{µ \nu} 
	+ \frac{\mathpzc{q}^{\lambda \rho \sigma}}{\bar{µ}^2} \left[ \vphantom{\frac{1}{2}} \tensor{\mathcal{C}}{^{µ \nu}_{\lambda \rho \sigma}} \right] h_{µ \nu} \\
	+ \left. \bar{µ}^2 \widetilde{\mathpzc{S}}^µ \left[ \frac{\bar{µ}^2}{12} f_{k}^{S} \delta^{\nu}_{µ} \right] \widetilde{\mathpzc{S}}_{\nu} 
	+ \bar{µ}^2 \widetilde{\mathpzc{T}}^µ \left[ \frac{4 \bar{µ}^2}{3} f_{k}^{T} \delta^{\nu}_{µ} \right] \widetilde{\mathpzc{T}}_{\nu} 
	+ \frac{\mathpzc{q}^{µ \rho \alpha}}{\bar{µ}^2} \left[- \bar{µ}^2 f_{k}^{q} \delta^{\nu}_{µ} \delta^{\![\sigma}_{\rho} \delta^{\beta]}_{\alpha}\vphantom{\frac{\bar{µ}^2}{2}} \right] \mathpzc{q}_{\nu \sigma \beta} \vphantom{\frac{1}{2}} \right\} \ .
\end{multline}
The matrix elements of the new operator $\bm{\bigl( \breve{\Gamma}_{k}^{(2)} \bigr)}$ involve the following building blocks:
{\allowdisplaybreaks
\begin{subequations}
\begin{align}
	\begin{split}
		\tensor{\widetilde{\mathcal{U}}}{_{µ \nu}^{\rho \sigma}} = & \left[ \frac{1}{2} \delta^{\!(\rho}_{µ} \delta^{\sigma)}_{\nu} - \frac{1}{4} \bar{g}_{µ \nu} \bar{g}^{\rho \sigma} \right] \left( \Bigl( \bar{R} - 2 \bar{\lambda}_k \Bigr) - \frac{\bar{µ}^4}{24} f_{k}^{S} \Bigl( \bar{g}^{\kappa \lambda} \widetilde{\bar{S}}_{\kappa} \widetilde{\bar{S}}_{\lambda} \Bigr) \right. \\ 
		& \phantom{~~~~~} \left. - \frac{2 \bar{µ}^4}{3} f_{k}^{T} \Bigl( \bar{g}^{\kappa \lambda} \widetilde{\bar{T}}_{\kappa} \widetilde{\bar{T}}_{\lambda} \Bigr)  + \frac{1}{2} f_{k}^{q} \Bigl( \bar{g}^{\kappa \lambda} \bar{g}^{\gamma \delta} \bar{g}^{\alpha \beta} \bar{q}_{\kappa \gamma \alpha} \bar{q}_{\lambda \delta \beta} \Bigr) \right) \\
		& + \frac{1}{2} \left[ \vphantom{\frac{1}{2}} \bar{g}_{µ \nu} \bar{R}^{\rho \sigma} + \bar{g}^{\rho \sigma} \bar{R}_{µ \nu} \right] - \tensor{\bar{R}}{_{\!(µ}^{\!(\rho}_{\!\nu)}^{\!\sigma)}} - \tensor{R}{_{\!(µ}^{\!(\rho}} \delta_{\nu)}^{\sigma)} \\
		& - \frac{\bar{µ}^4}{48} f_{k}^{S} \left[ \vphantom{\frac{1}{2}} \bar{g}_{µ \nu} \bar{g}^{\kappa \rho} \bar{g}^{\lambda \sigma} \widetilde{\bar{S}}_{\kappa} \widetilde{\bar{S}}_{\lambda} + \bar{g}^{\rho \sigma} \widetilde{\bar{S}}_{µ} \widetilde{\bar{S}}_{\nu} \right]  \\
		& - \frac{\bar{µ}^4}{3} f_{k}^{T} \left[ \vphantom{\frac{1}{2}} \bar{g}_{µ \nu} \bar{g}^{\kappa \rho} \bar{g}^{\lambda \sigma} \widetilde{\bar{T}}_{\kappa} \widetilde{\bar{T}}_{\lambda} + \bar{g}^{\rho \sigma} \widetilde{\bar{T}}_{µ} \widetilde{\bar{T}}_{\nu} \right] \\
		& + \frac{1}{4} f_{k}^{q} \left[ \vphantom{\frac{1}{2}} \bar{g}_{µ \nu} \bar{g}^{\kappa (\rho|} \bar{g}^{\lambda |\sigma)} \bar{g}^{\gamma \delta} \bar{g}^{\alpha \beta} \bar{q}_{\kappa \gamma \alpha}\, \bar{q}_{\lambda \delta \beta} + \bar{g}^{\rho \sigma} \bar{g}^{\gamma \delta} \bar{g}^{\alpha \beta} \bar{q}_{(µ| \gamma \alpha}\, \bar{q}_{|\nu) \delta \beta} \right] \\
		& + \frac{1}{2} f_{k}^{q} \left[ \vphantom{\frac{1}{2}} \bar{g}_{µ \nu} \bar{g}^{\kappa (\rho|} \bar{g}^{\lambda |\sigma)} \bar{g}^{\gamma \delta} \bar{g}^{\alpha \beta} \bar{q}_{\gamma \kappa \alpha}\, \bar{q}_{\delta \lambda \beta} + \bar{g}^{\rho \sigma} \bar{g}^{\gamma \delta} \bar{g}^{\alpha \beta} \bar{q}_{\gamma (µ| \alpha}\, \bar{q}_{\delta |\nu) \beta} \right] \\
		& - f_{k}^{q} \left[ \vphantom{\frac{1}{2}} \bar{g}^{(\rho|\kappa} \bar{g}^{|\sigma) \lambda} \bar{g}^{\alpha \beta} \right] \Bigl( \bar{q}_{\kappa (µ| \alpha}\, \bar{q}_{\lambda |\nu) \beta} + \bar{q}_{(µ| \kappa \alpha}\, \bar{q}_{|\nu) \lambda \beta} \Bigr) \\
		& - \frac{1}{2} f_{k}^{q} \left[ \vphantom{\frac{1}{2}} \bar{g}^{(\rho| \kappa} \bar{g}^{|\sigma) \lambda} \bar{g}^{\alpha \beta} \right] \Bigl( \bar{q}_{\alpha \kappa (µ|}\, \bar{q}_{\beta \lambda |\nu)} + \bar{q}_{\alpha (µ| \kappa}\, \bar{q}_{\beta |\nu) \lambda} \Bigr)
	\end{split} \\
	\tensor{\mathcal{K}}{_{µ \nu}^{\rho \sigma}} = & \left[ \frac{1}{2} \delta^{\!(\rho}_{µ} \delta^{\sigma)}_{\nu} - \frac{1}{4} \bar{g}_{µ \nu} \bar{g}^{\rho \sigma} \right] \\
	\tensor{\widetilde{\mathcal{L}}}{_{µ \nu}^{\lambda}} = & \, \bar{µ}^4 f_{k}^{S} \left[ \frac{1}{24} \bar{g}_{µ \nu} \bar{g}^{\kappa \lambda} - \frac{1}{12} \delta^{\kappa}_{\!(µ} \delta^{\lambda}_{\nu)} \right] \widetilde{\bar{S}}_{\kappa} \\
	\tensor{\widetilde{\mathcal{M}}}{_{µ \nu}^{\lambda}} = & \, \bar{µ}^4 f_{k}^{T} \left[ \frac{2}{3} \bar{g}_{µ \nu} \bar{g}^{\kappa \lambda} - \frac{4}{3} \delta^{\kappa}_{\!(µ} \delta^{\lambda}_{\nu)} \right] \widetilde{\bar{T}}_{\kappa} \\
	\tensor{\mathcal{N}}{_{µ \nu}^{\lambda \rho \sigma}} = & \, f_{k}^{q} \left[ - \frac{1}{2} \bar{g}_{µ \nu} \bar{g}^{\kappa \lambda} \bar{g}^{\alpha [ \rho} \bar{g}^{\sigma ] \beta} + \delta_{\!(µ}^{\kappa} \delta_{\nu)}^{\lambda} \bar{g}^{\alpha [ \rho} \bar{g}^{\sigma ] \beta} + 2 \delta_{\!(µ}^{\alpha} \delta_{\nu)}^{\![\rho|} \bar{g}^{\kappa \lambda} \bar{g}^{\beta |\sigma]} \right] q_{\kappa \alpha \beta} \\
	\tensor{\widetilde{\mathcal{A}}}{^{µ \nu}_{\lambda}} = & \, \bar{µ}^2 f_{k}^{S} \left[ \frac{1}{24} \bar{g}^{µ \nu} \widetilde{\bar{S}}_{\lambda} - \frac{1}{12} \delta_{\lambda}^{\!(µ} \bar{g}^{|\nu) \kappa} \widetilde{\bar{S}}_{\kappa} \right] \\
	\tensor{\widetilde{\mathcal{B}}}{^{µ \nu}_{\lambda}} = & \, \bar{µ}^2 f_{k}^{T} \left[ \frac{2}{3} \bar{g}^{µ \nu} \widetilde{\bar{T}}_{\lambda} - \frac{4}{3} \delta_{\lambda}^{\!(µ} \bar{g}^{|\nu) \kappa} \widetilde{\bar{T}}_{\kappa} \right] \\
	\tensor{\mathcal{C}}{^{µ \nu}_{\lambda \rho \sigma}} = & \, \bar{µ}^2 f_{k}^{q} \left[ -\frac{1}{2} \bar{g}^{µ \nu} \bar{q}_{\lambda \rho \sigma} + \delta_{\lambda}^{\!(µ} \bar{g}^{\nu)\kappa} \bar{q}_{\kappa \rho \sigma} + 2 \delta_{\![\rho|}^{\!(µ} \bar{g}^{\nu) \kappa} \bar{q}_{\lambda \kappa |\sigma]} \right] \quad .
\end{align}
\end{subequations}}
In Table \ref{tab:massdimensions2} we collect the dimensions of these tensors and the other building blocks appearing in (\ref{eq:gammanewoperator}) for a direct comparison.

\begin{table}[H]
	\centering
		\begin{tabular}{c||c|c||c|c}
		couplings & cov. fields & contrav. fields & $\bigl( \breve{\Gamma}_{k}^{(2)} \bigr)_{ij}$ tensors & $\tensor{\bm{\bigl( \breve{\Gamma}_{k}^{(2)} \bigr)}}{^i_j}$ tensors \\
		\hline \hline
		$\begin{aligned}[t]
		[G_k] &= -2 \\
		[\bar{\lambda}_k] &= +2 \\
		[f_k^S] &= 0 \\
		[f_k^T] &= 0 \\
		[f_k^q] &= 0
		\end{aligned}$ & $\begin{aligned}[t]
		[g_{µ \nu}] &= -2 \\
		[\widetilde{S}_{µ}] &= -2 \\
		[\widetilde{T}_{µ}] &= -2 \\
		[q_{\lambda µ \nu}] &= -2 
		\end{aligned}$ & $\begin{aligned}[t]
		[g^{µ \nu}] &= +2 \\
		[\widetilde{S}^{µ}] &= 0 \\
		[\widetilde{T}^{µ}] &= 0 \\
		[q^{\lambda µ \nu}] &= +4 
		\end{aligned}$ &  $\begin{aligned}[t]
		[\mathcal{U}^{µ \nu \rho \sigma}] &= +6 \\
		[\mathcal{K}^{µ \nu \rho \sigma} \bar{D}^2] &= +6 \\
		[\mathcal{L}^{µ \nu \lambda}] &= +4 \\
		[\mathcal{M}^{µ \nu \lambda}] &= +4 \\
		[\mathcal{N}^{µ \nu \lambda \rho \sigma}] &= +6 
		\end{aligned}$ & $\begin{aligned}[t]
		[\tensor{\widetilde{\mathcal{U}}}{_{µ \nu }^{\rho \sigma}}] &= +2 \\
		[\tensor{\mathcal{K}}{_{µ \nu }^{\rho \sigma}} \bar{D}^2] &= +2 \\
		[\tensor{\widetilde{\mathcal{L}}}{_{µ \nu }^{\lambda}}] &= +2 \\
		[\tensor{\widetilde{\mathcal{A}}}{^{µ \nu }_{\lambda}}] &= +2 \\
		[\tensor{\widetilde{\mathcal{M}}}{_{µ \nu }^{\lambda}}] &= +2 \\
		[\tensor{\widetilde{\mathcal{B}}}{^{µ \nu }_{\lambda}}] &= +2 \\
		[\tensor{\mathcal{N}}{_{µ \nu }^{\lambda \rho \sigma}}] &= +2 \\
		[\tensor{\mathcal{C}}{^{µ \nu }_{\lambda \rho \sigma}}] &= +2
		\end{aligned}$
		\end{tabular}
	\caption{Mass dimensions of the various building blocks comprising the new operatorial representation (\ref{eq:gammanewoperator}). (The metric quantities retain their respective dimensions, $[\sqrt{g}] = -4 , [R^{µ \nu \rho \sigma}] = +6 , [R^{µ \nu}] = +4 ,[R] = +2 $.)}
	\label{tab:massdimensions2}
\end{table}

\noindent We observe that, finally, all matrix elements $\tensor{\bm{\bigl( \breve{\Gamma}_{k}^{(2)} \bigr)}}{^i_j}$ exhibit the same mass dimension, namely $+2$.

\textit{Only the combination of redefining the fields to give them a uniform mass dimension, and dimensionally adapting the metric in field space was able to achieve this goal.} Solely adjusting the metric in field space is not enough, as seen in the previous section, because it fails to capture all off-diagonal elements. Conversely, had we only redefined the torsion fields, but not modified the field space metric, the diagonal elements would not equalize. Hence we are forced to implement \textit{both} procedures to realize a uniform mass dimension of $\bm{\breve{\Gamma}_{k}^{(2)}}$.

The uniform mass dimension comes with the cost of introducing a multitude of additional mass parameters. We presented the required process for the special case that 1.) the background and fluctuation fields are rescaled equally and 2.) all inserted mass parameters are identified with each other. Generally speaking each field redefinition and each sector of the metric (\ref{eq:metricfieldspaceredefined}) is independent of each other and in principle comes with its own mass parameter $\bar{µ}_1 , \bar{µ}_2 , \cdots$. 

Having said this, it should actually be called into question, if these parameters truly have an impact on the physical contents of the RG flow. In the case of pure field redefinitions the mass parameters may well enter a generic $\beta$-function but should not affect observable (on shell) quantities.


\subsection[Derivation of the \texorpdfstring{$µ$}{mu}-dependent \texorpdfstring{$\beta$}{beta}-functions]{Derivation of the \texorpdfstring{$\bm{µ}$}{mu}-dependent \texorpdfstring{$\bm{\beta}$}{beta}-functions}\label{sec:H0}

In the sequel we explore the ramifications the externally introduced mass parameter(s) have at the level of the RG flow. As above, we set all such mass parameters equal, so only one $\bar{µ} \equiv \bar{µ}_1 = \bar{µ}_2 = \cdots$ appears in the following. 

We analyze the parity-even Holst truncation, and furthermore choose vanishing background fields in the torsion sector. As already mentioned in the definition of the truncation (\ref{eq:holsttruncation}), the background metric is the only non-vanishing background field. As a consequence we only need to retain the diagonal blocks in our Hessian. Thus, to keep the notation simple we switch back to the representation in (\ref{eq:gammaoperator}), as it sufficient for those terms. The rescaling of the fields, the second step above, plays no role here.

For the case considered, the projection of the evolution equation is onto the $2$-dimensional subspace of $\mathcal{T}_{\mathrm{dtor}}$ spanned by the operators $\sqrt{g}$ and $\sqrt{g}R$. What can we learn from a calculation after having cut down the truncation in this manner? While we loose the ability to investigate the RG flow of the torsion couplings $f_{k}$, our truncation, or strictly speaking our Hessian (\ref{eq:gammaoperator}), is still sufficiently general for a detailed analysis of how the mass parameter $\bar{µ}$ that entered via $\text{\Fontlukas G}_{ij}$ affects the $\beta$-functions of the cosmological constant and Newton's constant, and thus will lead to interesting conclusions about the properties of their non-Gaussian fixed point.

The parameter $\bar{µ}$, in various guises, has arisen before in several Asymptotic Safety studies. In the case of tetrad gravity $(\mathcal{T}_{\mathrm{tet}})$, as well as Einstein-Cartan gravity $(\mathcal{T}_{\mathrm{EC}})$, this parameter had to be introduced to guarantee a uniform mass dimension of the respective Hessians. In the previous works dedicated to these theory spaces, \cite{Harst:2012phd,Harst:2012ni} and \cite{Daum:2010phd,Daum:2013fu,Harst:2012phd,Harst:2014vca}, respectively, this mass parameter was found to change the resulting RG trajectories quite significantly. Our present calculation is still general enough to make contact with the kind of $\bar{µ}$-dependences found there, while at the same time it reduces the calculations to a manageable and illuminating level. In particular $\left. \breve{\Gamma}_{k}^{(2)} \right|_{\bar{S}=\bar{T}=\bar{q}=0}$ still features all terms quadratic in the torsion fluctuation, so we can test their individual effects on the RG trajectories. 

\subsubsection{Vanishing background torsion}\label{subsubsec:vanishingbtorsion}
Specializing for vanishing background torsion, $\bar{S}_µ = \bar{T}_µ = \bar{q}_{\lambda µ \nu} = 0$, we abbreviate $H_k \bigl[h;\bar{g} \bigr] \equiv \left. \breve{\Gamma}_{k}^{(2)} \right|_{\bar{S}=\bar{T}=\bar{q}=0}$ for the Hessian stemming from the action (\ref{eq:ehtrunctor}) and the gauge fixing term (\ref{eq:gaugecondition}). The FRGE (\ref{eq:flowbrevegamma}) therefore reduces to
\begin{equation}
	\left.\vphantom{\frac{1}{2}} \partial_{t} \Gamma_{k} \right|_{\bar{S}=\bar{T}=\bar{q}=0} = \frac{1}{2} \operatorname{Tr} \left[ \frac{\bm{\left( \partial_t \breve{\mathcal{R}}_k \bigl( -\bar{D}^2 \bigr) \right)}}{\bm{\left( H_k + \breve{\mathcal{R}}_k \bigl( -\bar{D}^2 \bigr) \right)}} \right] - \operatorname{Tr} \left[ \frac{\bm{\left( \partial_t \mathcal{R}_{k}^{\mathrm{gh}} \bigl( -\bar{D}^2 \bigr) \right)}}{\bm{\left( S_{\mathrm{gh}}^{(2)} + \mathcal{R}_{k}^{\mathrm{gh}} \bigl( -\bar{D}^2 \bigr) \right)}} \right] \ ,
	\label{eq:floweqh0}
\end{equation}
where the Hessian in the ghost sector remains untouched. After having calculated the inverse propagator $H_k +  \breve{\mathcal{R}}_k$, we may set $\bar{g}_{µ \nu} = g_{µ \nu}$, as we are not considering a bimetric truncation \cite{Manrique:2009uh,Becker:2014qya}. On the left hand side of the evolution equation we have then 
\begin{equation}
	\left.\vphantom{\frac{1}{2}} \partial_{t} \Gamma_{k} \bigl[g,g\bigr] \right|_{\bar{S}=\bar{T}=\bar{q}=0} = \frac{1}{16 \pi \bar{G}} \int \mathrm{d}^{4}x\, \sqrt{g} \Bigl\{- R\, \partial_{t}Z_{Nk} + 2\, \partial_{t}\bigl(Z_{Nk}\bar{\lambda}_k \bigr) \Bigr\}.
	\label{eq:lhs}
\end{equation}
In order to evaluate the right hand side of (\ref{eq:floweqh0}), we expand the traces up to terms of second order in the derivatives and thus retain only the terms proportional to $\int\sqrt{g}$ and $\int\sqrt{g}R$. Equating the result to (\ref{eq:lhs}) determines the $\beta$-functions as the prefactors of those invariants.

The quadratic form corresponding to $H_k$ is given by (\ref{eq:gammaquadratic}) with $\bar{S}=\bar{T}=\bar{q}=0$:
\begin{multline}
	H_{k}^{\mathrm{quad}} \equiv \frac{1}{2} \, \varphi^i \, \bigl(H_{k}\bigr)_{ij} \, \varphi^j = \frac{1}{32 \pi G_k} \int \mathrm{d}^{4}x\, \sqrt{\bar{g}} \left\{ h_{µ \nu} \left[ \vphantom{\frac{1}{2}} U^{µ \nu \rho \sigma} - K^{µ \nu \rho \sigma} \bar{g}^{\kappa \lambda} \bar{D}_\kappa \bar{D}_\lambda \right] h_{\rho \sigma} \right. \\ 
	\left. + \mathpzc{S}_{µ} \left[ \vphantom{\frac{1}{2}} \frac{1}{12} f_{k}^{S} \bar{g}^{µ \nu} \right] \mathpzc{S}_{\nu} + \mathpzc{T}_{µ} \left[ \vphantom{\frac{1}{2}} \frac{4}{3} f_{k}^{T} \bar{g}^{µ \nu} \right] \mathpzc{T}_{\nu} + \mathpzc{q}_{µ \rho \alpha} \left[- f_{k}^{q} \bar{g}^{µ \nu} \bar{g}^{\rho [\sigma} \bar{g}^{\beta] \alpha} \vphantom{\frac{1}{2}} \right] \mathpzc{q}_{\nu \sigma \beta} \right\} .
	\label{eq:Hkquad}
\end{multline} 
Here $U^{µ \nu \rho \sigma} \equiv \left. \mathcal{U}^{µ \nu \rho \sigma}\right|_{\bar{S}=\bar{T}=\bar{q}=0}$ and $K^{µ \nu \rho \sigma} \equiv \mathcal{K}^{µ \nu \rho \sigma}$. Likewise, the operator form is obtained from (\ref{eq:gammaoperator}):
\begin{multline}
	H_{k}^{\mathrm{quad}} \equiv \frac{1}{2} \, \varphi_i \, \tensor{\bm{\bigl(H_{k}\bigr)}}{^i_j} \, \varphi^j = \frac{1}{32 \pi G_k} \int \mathrm{d}^{4}x\, \sqrt{\bar{g}} \left\{ h^{µ \nu} \left[ \vphantom{\frac{\bar{µ}^2}{12}} \tensor{U}{_{µ \nu}^{\rho \sigma}} - \tensor{K}{_{µ \nu}^{\rho \sigma}} \bar{D}^{2} \right] h_{\rho \sigma} \right. \\ \left. + \mathpzc{S}^{µ} \left[ \vphantom{\frac{1}{2}} \frac{\bar{µ}^2}{12} f_{k}^{S} \delta^{\nu}_{µ} \right] \mathpzc{S}_{\nu} + \mathpzc{T}^{µ} \left[ \vphantom{\frac{1}{2}} \frac{4 \bar{µ}^2}{3} f_{k}^{T} \delta^{\nu}_{µ} \right] \mathpzc{T}_{\nu} + \mathpzc{q}^{µ \rho \alpha} \left[- \bar{µ}^2 f_{k}^{q} \delta^{\nu}_{µ} \delta^{\![\sigma}_{\rho} \delta^{\beta]}_{\alpha}\vphantom{\frac{\bar{µ}^2}{2}} \right] \mathpzc{q}_{\nu \sigma \beta} \right\} .
	\label{eq:h0full}
\end{multline}
The tensors that make up the operator matrix elements in (\ref{eq:h0full}) read
\begin{subequations}
	\begin{align}
	\begin{split}
		\tensor{U}{_{µ \nu}^{\rho \sigma}} = & \left[ \frac{1}{2} \delta^{\!(\rho}_{µ} \delta^{\sigma)}_{\nu} - \frac{1}{4} \bar{g}_{µ \nu} \bar{g}^{\rho \sigma} \right] \Bigl( \bar{R} - 2 \bar{\lambda}_k \Bigr) + \frac{1}{2} \left[ \vphantom{\frac{1}{2}} \bar{g}_{µ \nu} \bar{R}^{\rho \sigma} + \bar{g}^{\rho \sigma} \bar{R}_{µ \nu} \right] - \tensor{\bar{R}}{_{\!(µ}^{\!(\rho}_{\!\nu)}^{\!\sigma)}} - \tensor{\bar{R}}{_{\!(µ}^{\!(\rho}} \delta_{\nu)}^{\sigma)}
	\end{split} \\
	\tensor{K}{_{µ \nu}^{\rho \sigma}} = & \left[ \frac{1}{2} \delta^{\!(\rho}_{µ} \delta^{\sigma)}_{\nu} - \frac{1}{4} \bar{g}_{µ \nu} \bar{g}^{\rho \sigma} \right] \ .
	\end{align}
\end{subequations}
Note that aside from the index position these kernels are the same as in \cite{Reuter:1996cp}, where they are given in the form $\tensor{U}{^{µ \nu}_{\rho \sigma}}$ and $\tensor{K}{^{µ \nu}_{\rho \sigma}}$.

In order to partially diagonalize (\ref{eq:h0full}) we decompose the metric fluctuations $h_{µ \nu}$ into a symmetric traceless tensor $\mathpzc{h}_{µ \nu}$ and a trace part $\phi_{h}$, according to 
\begin{equation}
	h_{µ \nu} = \mathpzc{h}_{µ \nu} + \frac{1}{d}\, \bar{g}_{µ \nu} \phi_{h} \qquad \text{with} \quad \phi_{h} = \bar{g}^{µ \nu} h_{µ \nu} \quad \text{and} \quad \bar{g}^{µ \nu} \mathpzc{h}_{µ \nu} = 0 \ .
	\label{eq:hsplit}
\end{equation}
Also, we specify the background spacetime to be a maximally symmetric Einstein space:
\begin{equation}
		\bar{R}_{µ \nu \rho \sigma} = \frac{1}{d(d-1)} \bigl[ \bar{g}_{µ \rho} \bar{g}_{\nu \sigma} - \bar{g}_{µ \sigma} \bar{g}_{\nu \rho} \bigr] \bar{R} \quad , \quad	\bar{R}_{µ \nu} = \frac{1}{d}\, \bar{g}_{µ \nu} \bar{R} \ .
	\label{eq:symmetricspace}
\end{equation}
This spacetime suffices to identify the contributions to the invariants $\int\sqrt{\bar{g}}$ and $\int\sqrt{\bar{g}}\bar{R}$ and distinguish them from higher order terms in the derivative expansion unambiguously. From here on the curvature scalar $\bar{R}$ parametrizes the family of metrics inserted and thus should be regarded as a remnant of the $\bar{g}_{µ \nu}$-argument of $\Gamma_k$ rather than a functional of the metric. 

Using the relations (\ref{eq:hsplit}) and (\ref{eq:symmetricspace}) the Hessian boils down to
\begin{multline}
		H_{k}^{\mathrm{quad}} = \frac{1}{2}\, \frac{1}{32 \pi G_k} \int \mathrm{d}^{4}x\, \sqrt{\bar{g}} \left\{ \vphantom{\frac{\bar{µ}^2}{2}}\mathpzc{h}^{µ \nu} \right. \left[ \vphantom{\frac{1}{2}} \left(- \bar{D}^2 - 2 \bar{\lambda}_k + C_{t}\bar{R} \right) \delta^{\!(\rho}_{µ} \delta^{\sigma)}_{\nu} \right] \mathpzc{h}_{\rho \sigma} \\
		- \left( \frac{d-2}{2d} \right) \phi_{h} \left[ - \bar{D}^2 - 2 \bar{\lambda}_k + C_{s}\bar{R} \right] \phi_{h} \\
		+ \mathpzc{S}^{µ} \left[ \vphantom{\frac{1}{2}} \frac{\bar{µ}^2}{6} f_{k}^{S} \delta^{\nu}_{µ} \right] \mathpzc{S}_{\nu} + \mathpzc{T}^{µ} \left[ \vphantom{\frac{1}{2}} \frac{8 \bar{µ}^2}{3} f_{k}^{T} \delta^{\nu}_{µ} \right] \mathpzc{T}_{\nu} + \mathpzc{q}^{µ \rho \alpha} \left[-2 \bar{µ}^2 f_{k}^{q} \delta^{\nu}_{µ} \delta^{\![\sigma}_{\rho} \delta^{\beta]}_{\alpha}\vphantom{\frac{1}{2}} \right] \mathpzc{q}_{\nu \sigma \beta} \left. \vphantom{\left[ \vphantom{\frac{1}{2}} \frac{8 \bar{µ}^2}{3} f_{k}^{T} \delta_{\nu}^{µ} \right]}\right\} \ ,
	\label{eq:h0final}
\end{multline}
with the constants
\begin{equation}
	C_{t} \equiv \frac{d(d-3)+4}{d(d-1)}\, , \quad C_{s} \equiv \frac{d-4}{d} \ .
\end{equation}
Whereas the symmetric traceless tensor $\mathpzc{h}_{µ \nu}$ has a standard positive definite kinetic term, the trace part $\phi_{h}$ is well known to have a “wrong sign” kinetic term in $d>2$. 

\subsubsection[The ``tachyonic'' \texorpdfstring{$q_{\lambda µ \nu}$}{q} field]{The ``tachyonic'' \texorpdfstring{$\bm{q_{\lambda µ \nu}}$}{q} field}\label{subsec:tachyonfield}
Equation (\ref{eq:h0final}) shows that in generalizing the Einstein-Hilbert to the Holst action there arises \textit{a second field which is plagued by an instability}: the $\mathpzc{q}_{\lambda µ \nu}$-torsion component is a ``tachyon'', that is, the $\mathpzc{q}^2$-term has a negative prefactor. The influence of the crucial minus sign in (\ref{eq:h0final}) on the RG flow will be investigated in detail later on. 

A delicate issue related to this instability is the precise form of the cutoff operator $\breve{\mathcal{R}}_k$. In the various sectors of field space it has the generic form
\begin{equation}
	\breve{\mathcal{R}}_k = \breve{\mathcal{Z}}_{k} k^2 R^{(0)}\bigl( - \tfrac{\bar{D}^2}{k^2} \bigr) \ ,
\end{equation}
where $\breve{\mathcal{Z}}_k$ is a diagonal matrix in field space and $R^{(0)}(z)$ is a dimensionless shape function, interpolating smoothly between $R^{(0)}(0) = 1$ and $ \lim\limits_{z \rightarrow \infty} R^{(0)}(z) = 0$. 

The entries $\mathcal{Z}_k$ of $\breve{\mathcal{Z}}_k$ are usually fixed by the so-called $\mathcal{Z}_k = \zeta_k$-rule: $\mathcal{Z}_k$ should be chosen such that for any eigenmode of $\Gamma_{k}^{(2)}$, with eigenvalue $\zeta_k \, p^2$, the sum $\Gamma_{k}^{(2)} + \mathcal{R}_k$ has the eigenvalue $\zeta_k \bigl[ p^2 + k^2 R^{(0)}\bigl( \tfrac{p^2}{k^2} \bigr) \bigr]$. It has been advocated to apply this kind of adaptation even for $\zeta_k < 0$, as divergences that may arise due to the conformal factor problem can be circumvented by this choice; for a detailed discussion see \cite{Reuter:1996cp,Lauscher:2002sq}.

In the present case we follow \cite{Reuter:1996cp} and implement this rule for the modes of $\mathpzc{h}_{µ \nu}$ and $\phi_{h}$. Their the kinetic operators $\bigl[ - \bar{D}^2 + \cdots \bigr]$ get replaced by $\bigl[  - \bar{D}^2 + k^2 R^{(0)}\bigl( -\tfrac{\bar{D}^2}{k^2}\bigr) + \cdots \bigr]$ if we set
\begin{equation}
	\tensor{\left[\bigl( \breve{\mathcal{Z}}_k \bigr)_{\mathpzc{h}\mathpzc{h}}\right]}{_{µ}^{\rho}_{\nu}^{\sigma}} = \frac{1}{32 \pi G_k} \delta^{\!(\rho}_{µ} \delta^{\sigma)}_{\nu} \ , \quad \bigl( \breve{\mathcal{Z}}_k \bigr)_{\phi_{h}\phi_{h}} = - \frac{1}{32 \pi G_k} \left( \frac{d-2}{2d} \right) \ .
\end{equation}

The torsion fields $\mathpzc{S}_µ , \mathpzc{T}_µ , \mathpzc{q}_{\lambda µ \nu}$ are somewhat unusual in that they possess no true kinetic term, their action is entirely of the potential type. 

Especially ``pathological'' is the $\mathpzc{q}_{\lambda µ \nu}$-field with its negative $\mathpzc{q}^2$-term. A priori both a positively and negatively chosen cutoff could be plausible choices. For this reason, we shall explore both options. 

\noindent \textbf{(i)} To begin with, we equip all torsion fields with a \textit{positive} cutoff action $\Delta S_k > 0$ and fix $\breve{\mathcal{Z}}_k$ by the ``$\mathcal{Z}_k = | \zeta_k |$-rule''. That is, for the torsion modes we require that adding $\breve{\mathcal{R}}_k$ to the terms inside the square brackets of (\ref{eq:h0final}), symbolically of the form $\bigl[ \pm \bar{µ}^2 f^{\bullet}_k \bigr]$, results in $\bigl[+ k^2 R^{(0)}\bigl( -\tfrac{\bar{D}^2}{k^2}\bigr) \pm \bar{µ}^2 f^{\bullet}_k \bigr]$. 

\noindent \textbf{(ii)} Later on, in Section \ref{subsec:negacut}, we apply the ``$\mathcal{Z}_k = \zeta_k$-rule'', allowing now a cutoff action for the $\mathpzc{q}_{\lambda µ \nu}$-field which is \textit{negative}. We require that adding $\breve{\mathcal{R}}_k$ to $H_k$ replaces $\bigl[ \pm \bar{µ}^2 f^{\bullet}_k \bigr]$ by $\bigl[ \pm k^2 R^{(0)}\bigl( -\tfrac{\bar{D}^2}{k^2}\bigr) \pm \bar{µ}^2 f^{\bullet}_k \bigr]$.

We will see that while both choices are technically equally viable they lead to quite different flows, making one of the choices more preferred. 

Taking the overall prefactor into account we thus obtain the following matrix elements of $\breve{\mathcal{Z}}_k$: 
\begin{equation}
\label{eq:zdefine}
	\tensor{\left[ \bigl( \breve{\mathcal{Z}}_k \bigr)_{\mathpzc{S}\mathpzc{S}} \right]}{_{µ}^{\nu}} = \frac{1}{32 \pi G_k} \delta_{µ}^{\nu} , \ \tensor{\left[\bigl( \breve{\mathcal{Z}}_k \bigr)_{\mathpzc{T}\mathpzc{T}}\right]}{_{µ}^{\nu}} = \frac{1}{32 \pi G_k} \delta_{µ}^{\nu} , \ \tensor{\left[\bigl( \breve{\mathcal{Z}}_k \bigr)_{\mathpzc{q}\mathpzc{q}}\right]}{_{µ}^{\nu}_{\rho}^{\sigma}_{\alpha}^{\beta}} = \frac{\xi}{32 \pi G_k} \delta^{\nu}_{µ} \delta^{\![\sigma}_{\rho} \delta^{\beta]}_{\alpha}
\end{equation}
Here $\xi = \pm 1$ determines the sign of the $\mathpzc{q}_{\lambda µ \nu}$-cutoff action: $\xi = +1$ and $\xi = -1$ correspond to the options \textbf{(i)} and \textbf{(ii)}, respectively. With this choice, the operator form $\frac{1}{2} \, \varphi_i \, \tensor{\bm{\bigl(H_{k} + \breve{\mathcal{R}}_k \bigr)}}{^i_j} \, \varphi^j$ contains the following kernels in the various sectors of field space:
\begin{subequations}
\label{eq:sectors}
	\begin{align}
		\tensor{\left[ \bm{\left( H_k + \breve{\mathcal{R}}_k \right)}_{\mathpzc{h}\mathpzc{h}} \right]}{_{µ}^{\rho}_{\nu}^{\sigma}} &= \frac{1}{32 \pi G_k} \left[ - \bar{D}^2 + k^2 R^{(0)}\bigl( -\tfrac{\bar{D}^2}{k^2}\bigr) - 2 \bar{\lambda}_k + C_{t}\bar{R} \right] \delta^{\!(\rho}_{µ} \delta^{\sigma)}_{\nu} \\
		\bm{\left( H_k + \breve{\mathcal{R}}_k \right)}_{\phi_{h}\phi_{h}} &= - \frac{1}{32 \pi G_k}\left( \frac{d-2}{2d} \right) \left[ - \bar{D}^2 + k^2 R^{(0)}\bigl( -\tfrac{\bar{D}^2}{k^2}\bigr) - 2 \bar{\lambda}_k + C_{s}\bar{R} \right] \\
		\tensor{\left[ \bm{\left( H_k + \breve{\mathcal{R}}_k \right)}_{\mathpzc{S}\mathpzc{S}} \right]}{_{µ}^{\nu}} &= \frac{1}{32 \pi G_k} \left[ \frac{\bar{µ}^2}{6} f_{k}^{S} +  k^2 R^{(0)} \bigl(- \tfrac{\bar{D}^2}{k^2} \bigr) \right] \delta_{µ}^{\nu} \\
		\tensor{\left[ \bm{\left( H_k + \breve{\mathcal{R}}_k \right)}_{\mathpzc{T}\mathpzc{T}} \right]}{_{µ}^{\nu}} &= \frac{1}{32 \pi G_k} \left[ \frac{8\bar{µ}^2}{3} f_{k}^{T} +  k^2 R^{(0)} \bigl(- \tfrac{\bar{D}^2}{k^2} \bigr) \right] \delta_{µ}^{\nu} \\
		\tensor{\left[ \bm{\left( H_k + \breve{\mathcal{R}}_k \right)}_{\mathpzc{q}\mathpzc{q}} \right]}{_{µ}^{\nu}_{\rho}^{\sigma}_{\alpha}^{\beta}} &= \frac{1}{32 \pi G_k} \left[\vphantom{\frac{1}{2}}-2\bar{µ}^2 f_{k}^{q} + \xi k^2 R^{(0)} \bigl(- \tfrac{\bar{D}^2}{k^2} \bigr) \right] \delta_{µ}^{\nu} \delta_{\rho}^{\![\sigma} \delta_{\alpha}^{\beta]} \ .
	\end{align}
\end{subequations}
From now on we may set $\bar{g}_{µ \nu} = g_{µ \nu}$ and omit the bars for the rest of the calculation. 

The last piece missing for the evaluation of the FRGE (\ref{eq:floweqh0}) is the Hessian in the ghost sector. From the ghost action (\ref{eq:ghostaction2}) we obtain for the maximally symmetric background $\tensor{\left(	S_{\mathrm{gh}}^{(2)} \right)}{_µ^\nu} = \delta_{µ}^{\nu} \left[ - D^2 + C_{v}R \vphantom{S_{\mathrm{gh}}^{(2)}} \right]$ with the constant $C_{v} \equiv - \frac{1}{d}$. Since we do not take into account any renormalization effects in the ghost sector, we set there $\tensor{\left[\mathcal{Z}_{k}^{\mathrm{gh}}\right]}{_{µ}^{\nu}} = \delta_{µ}^{\nu}$, yielding
\begin{equation}
	\tensor{\left[ \bm{\left( S_{\mathrm{gh}}^{(2)} + \mathcal{R}_{k}^{\mathrm{gh}} \right)} \right]}{_{µ}^{\nu}} = \delta_{µ}^{\nu} \left[ - D^2 + k^2 R^{(0)}\bigl( -\tfrac{D^2}{k^2}\bigr) + C_{v}R \right] \ .
\end{equation}
As a result, the second trace on the right hand side of (\ref{eq:floweqh0}) coincides exactly with the ghost contribution to the $\beta$-functions in pure metric gravity without torsion. We may therefore take over the corresponding results from \cite{Reuter:1996cp}.

\subsubsection{Traces and threshold functions}\label{subsubsec:tracesandthreshold}
The right hand side of the FRGE (\ref{eq:floweqh0}) is now composed of four parts: 
\begin{equation}
\label{eq:yetanotherflowequation}
	\frac{1}{16 \pi \bar{G}} \int \mathrm{d}^{4}x\, \sqrt{g} \, \Bigl\{- R\, \partial_{t}Z_{Nk} + 2\, \partial_{t}\bigl(Z_{Nk}\bar{\lambda}_k \bigr) \Bigr\} = \Delta \mathcal{S}^{\mathrm{EH}} + \Delta \mathcal{B}_{SS}^{\mathrm{Tor}} + \Delta \mathcal{B}_{TT}^{\mathrm{Tor}} + \Delta \mathcal{B}_{qq}^{\mathrm{Tor}} \ .
\end{equation}
In this equation we have written
\begin{multline}
	\Delta \mathcal{S}^{\mathrm{EH}} = \frac{1}{2} \operatorname{Tr}_t \left[ \frac{ \bm{\left( \partial_t \breve{\mathcal{R}}_k \right)}_{\mathpzc{h}\mathpzc{h}} }{ \bm{\left( H_k + \breve{\mathcal{R}}_k \right)}_{\mathpzc{h}\mathpzc{h}} } \right] + \frac{1}{2} \operatorname{Tr}_s \left[ \frac{ \bm{\left( \partial_t \breve{\mathcal{R}}_k \right)}_{\phi_{h}\phi_{h}} }{ \bm{\left( H_k + \breve{\mathcal{R}}_k \right)}_{\phi_{h}\phi_{h}} } \right] - \operatorname{Tr}_v \left[ \frac{\bm{\left( \partial_t \mathcal{R}_{k}^{\mathrm{gh}} \right)}}{\bm{\left( S_{\mathrm{gh}}^{(2)} + \mathcal{R}_{k}^{\mathrm{gh}} \right)}} \right]
	\label{eq:deltaseh}
\end{multline}
for the contributions stemming from the metric and the ghosts, and 
\begin{subequations}
\label{eq:deltab}
	\begin{align}
		\Delta \mathcal{B}_{SS}^{\mathrm{Tor}} &= \frac{1}{2} \operatorname{Tr}_{\mathpzc{S}} \left[ \frac{ \bm{\left( \partial_t \breve{\mathcal{R}}_k \right)}_{\mathpzc{S}\mathpzc{S}} }{ \bm{\left( H_k + \breve{\mathcal{R}}_k \right)}_{\mathpzc{S}\mathpzc{S}} } \right] \\
		\Delta \mathcal{B}_{TT}^{\mathrm{Tor}} &= \frac{1}{2} \operatorname{Tr}_{\mathpzc{T}} \left[ \frac{ \bm{\left( \partial_t \breve{\mathcal{R}}_k \right)}_{\mathpzc{T}\mathpzc{T}} }{ \bm{\left( H_k + \breve{\mathcal{R}}_k \right)}_{\mathpzc{T}\mathpzc{T}} } \right] \\
		\Delta \mathcal{B}_{qq}^{\mathrm{Tor}} &= \frac{1}{2} \operatorname{Tr}_{\mathpzc{q}} \left[ \frac{ \bm{\left( \partial_t \breve{\mathcal{R}}_k \right)}_{\mathpzc{q}\mathpzc{q}} }{ \bm{\left( H_k + \breve{\mathcal{R}}_k \right)}_{\mathpzc{q}\mathpzc{q}} } \right]
	\end{align}
\end{subequations}
for the torsion parts. The subscripts of the traces in (\ref{eq:deltab}) indicate on which type of tensors the covariant Laplacian is acting. The remaining computation consists in projecting out the invariants $\int\sqrt{g}$ and $\int\sqrt{g}R$ from the functionals defined by the traces. In the metric sector the evaluation of $\Delta \mathcal{S}^{\mathrm{EH}}$ corresponds exactly to the calculations for the familiar Einstein-Hilbert truncation in Quantum Einstein Gravity \cite{Reuter:1996cp}. Specialising the general results obtained in \cite{Reuter:1996cp}, valid in any number of spacetime dimensions, to the case $d=4$, we have
\begin{multline}
	\Delta \mathcal{S}^{\mathrm{EH}} = \frac{1}{(4 \pi)^2} \left\{ k^4 \left( 10 \Phi_{2}^{1} (- \tfrac{2 \bar{\lambda}_k}{k^2}) - 5 \eta_N \tilde{\Phi}_{2}^{1} (- \tfrac{2 \bar{\lambda}_k}{k^2}) - 8 \Phi_{2}^{1}(0) \right) \int \mathrm{d}^4 x \sqrt{g} \right. \\
	+ k^{2} \left( \frac{5}{3} \Phi_{1}^{1} (- \tfrac{2 \bar{\lambda}_k}{k^2}) - \frac{5}{6} \eta_N \tilde{\Phi}_{1}^{1} (- \tfrac{2 \bar{\lambda}_k}{k^2}) - \frac{4}{3} \Phi_{1}^{1}(0) \right. \\
	\left. \left. - 6 \Phi_{2}^{2} (- \tfrac{2 \bar{\lambda}_k}{k^2}) + 3 \eta_N \tilde{\Phi}_{2}^{2} (- \tfrac{2 \bar{\lambda}_k}{k^2}) - 2 \Phi_{2}^{2}(0) \right) \int \mathrm{d}^4 x \sqrt{g} R \right\}.
	\label{eq:ehtraces}
\end{multline}
This result is expressed in terms of the anomalous dimension $\eta_{N}(k) \equiv - \partial_t \ln Z_{Nk}$ and the familiar threshold functions \cite{Reuter:1996cp}
\begin{subequations}
\begin{align}
	\Phi_{n}^{p}(w) &= \frac{1}{\Gamma(n)} \int_{0}^{\infty} \mathrm{d}^{d}x\ x^{n-1} \frac{R^{(0)}(x)-x R^{(0)}\vphantom{a}'(x)}{\left[x + R^{(0)}(x) + w \right]^p} \\
	\tilde{\Phi}_{n}^{p}(w) &= \frac{1}{\Gamma(n)} \int_{0}^{\infty} \mathrm{d}^{d}x\ x^{n-1} \frac{R^{(0)}(x)}{\left[x + R^{(0)}(x) + w \right]^p}
\end{align}
for $n>0$, and
\begin{equation}
	\Phi_{0}^{p}(w) = \tilde{\Phi}_{0}^{p}(w) = \frac{1}{\left( 1 + w \right)^p}
\end{equation}
\end{subequations}
for $n=0$.

The evaluation of the torsion traces requires threshold functions of a new type. Taking the explicit form of $\bm{\left( \partial_t \breve{\mathcal{R}}_k \right)}$ in the various sectors into account we are led to 
\allowdisplaybreaks{
\begin{subequations}
\label{eq:threetraces}
\begin{align}
	\Delta \mathcal{B}_{SS}^{\mathrm{Tor}} &= \operatorname{Tr}_{\mathpzc{S}} \left[ \frac{R^{(0)} \bigl(- \tfrac{D^2}{k^2} \bigr) + \frac{D^2}{k^2} R^{(0)}\vphantom{a}' \bigl(- \tfrac{D^2}{k^2} \bigr)}{\left[ \frac{\bar{µ}^2}{6 k^2} f_{k}^{S} + R^{(0)} \bigl(- \tfrac{D^2}{k^2} \bigr) \right]} \right] - \frac{\eta_N}{2} \operatorname{Tr}_{\mathpzc{S}} \left[ \frac{R^{(0)} \bigl(- \tfrac{D^2}{k^2} \bigr)}{\left[ \frac{\bar{µ}^2}{6 k^2} f_{k}^{S} + R^{(0)} \bigl(- \tfrac{D^2}{k^2} \bigr) \right]} \right] \\
	\Delta \mathcal{B}_{TT}^{\mathrm{Tor}} &= \operatorname{Tr}_{\mathpzc{T}} \left[ \frac{R^{(0)} \bigl(- \tfrac{D^2}{k^2} \bigr) + \frac{D^2}{k^2} R^{(0)}\vphantom{a}' \bigl(- \tfrac{D^2}{k^2} \bigr)}{\left[ \frac{8 \bar{µ}^2}{3 k^2} f_{k}^{T} + R^{(0)} \bigl(- \tfrac{D^2}{k^2} \bigr) \right]} \right] - \frac{\eta_N}{2} \operatorname{Tr}_{\mathpzc{T}} \left[ \frac{R^{(0)} \bigl(- \tfrac{D^2}{k^2} \bigr)}{\left[ \frac{8 \bar{µ}^2}{3 k^2} f_{k}^{T} + R^{(0)} \bigl(- \tfrac{D^2}{k^2} \bigr) \right]} \right] \\
	\Delta \mathcal{B}_{qq}^{\mathrm{Tor}} &= \operatorname{Tr}_{\mathpzc{q}} \left[ \frac{R^{(0)} \bigl(- \tfrac{D^2}{k^2} \bigr) + \frac{D^2}{k^2} R^{(0)}\vphantom{a}' \bigl(- \tfrac{D^2}{k^2} \bigr)}{\left[ - \frac{2 \bar{µ}^2}{\xi k^2} f_{k}^{q} + R^{(0)} \bigl(- \tfrac{D^2}{k^2} \bigr) \right]} \right] - \frac{\eta_N}{2} \operatorname{Tr}_{\mathpzc{q}} \left[ \frac{R^{(0)} \bigl(- \tfrac{D^2}{k^2} \bigr)}{\left[ - \frac{2 \bar{µ}^2}{\xi k^2} f_{k}^{q} + R^{(0)} \bigl(- \tfrac{D^2}{k^2} \bigr) \right]} \right].
\end{align}
\end{subequations}}
Note the different signs in the denominators appearing here. The traces in (\ref{eq:threetraces}) are all over operators which are functions of the Laplacian. They can be evaluated by means of the ``master formula'' derived in \cite{Reuter:1996cp},
\begin{equation}
	\operatorname{Tr}\left[ W(- D^2) \right] = (4 \pi)^{- \frac{d}{2}} \operatorname{tr}(\mathds{1}) \left\{ Q_{\frac{d}{2}}[W] \int \mathrm{d}^{d}x\, \sqrt{g} + \frac{1}{6} Q_{\frac{d}{2}-1}[W] \int \mathrm{d}^{d}x\, \sqrt{g} R + \mathcal{O}(R^2) \right\} .
	\label{eq:heatkernelexp}
\end{equation}
using the standard heat-kernel expansion \cite{Vassilevich:2003xt} for the covariant Laplacian:
\begin{equation}
	\operatorname{Tr}\left[ e^{-i s D^2 } \right] = \left( \frac{i}{4 \pi s} \right)^{\frac{d}{2}} \operatorname{tr}(\mathds{1}) \int \mathrm{d}^{d}x\, \sqrt{g} \left(1 - \frac{1}{6} i s R + \mathcal{O}(R^2) \right).
	\label{eq:heatkernelexpansiongeneral}
\end{equation}
The ``$Q$-functionals'' are given by
\begin{align}
\begin{split}
	Q_{0}[W] &= W(0) \\
	Q_{n}[W] &= \frac{1}{\Gamma(n)} \int_{0}^{\infty} \mathrm{d}z\ z^{n-1} W(z) \quad \text{for} \ n>0 \ .
	\label{eq:Qfunc2}
\end{split}
\end{align}
The type of tensor field under consideration enters in the present case only via the trace of the unit operator, $\tr \mathds{1}$, which equals, respectively, 4, 4, and 16 in the case of $S_µ$, $T_µ$, and $q_{\lambda µ \nu}$, respectively. 

Applying (\ref{eq:heatkernelexp}) to (\ref{eq:threetraces}) we are led to define the following new type of threshold functions
\begin{subequations}
\label{eq:thresholdcheck}
\begin{align}
	\check{\Phi}_{n}^{p}(w) &= \frac{1}{\Gamma(n)} \int_{0}^{\infty} \mathrm{d}^{d}x\ x^{n-1} \frac{R^{(0)}(x)-x R'^{(0)}(x)}{\left[R^{(0)}(x) + w \right]^p} \\
	\check{\tilde{\Phi}}_{n}^{p}(w) &= \frac{1}{\Gamma(n)} \int_{0}^{\infty} \mathrm{d}^{d}x\ x^{n-1} \frac{R^{(0)}(x)}{\left[R^{(0)}(x) + w \right]^p}
\end{align}
for $n>0$, and
\begin{equation}
	\check{\Phi}_{0}^{p}(w) = \check{\tilde{\Phi}}_{0}^{p}(w) = \frac{1}{\left( 1 + w \right)^p}
\end{equation}
\end{subequations}
for $n=0$. (In a different context they were first employed in \cite{Harst:2012phd}.) It is straightforward to express the $Q$-functionals appearing in the torsion traces by these new threshold functions, and to arrive at
\begin{subequations}
\label{eq:thresholdcheck4d}
\begin{multline}
	\Delta \mathcal{B}_{SS}^{\mathrm{Tor}} = \frac{1}{(4 \pi)^2} \left\{ k^4 \left( 4 \check{\Phi}_{2}^{1}(\tfrac{\bar{µ}^2}{6 k^2} f_{k}^{S}) - 2 \eta_N \check{\tilde{\Phi}}_{2}^{1}(\tfrac{\bar{µ}^2}{6 k^2} f_{k}^{S}) \right) \int \mathrm{d}^4 x \sqrt{g} \right.  \\
	\left. + k^2 \left( \frac{2}{3} \check{\Phi}_{1}^{1}(\tfrac{\bar{µ}^2}{6 k^2} f_{k}^{S}) - \frac{1}{3} \eta_N \check{\tilde{\Phi}}_{1}^{1}(\tfrac{\bar{µ}^2}{6 k^2} f_{k}^{S}) \right) \int \mathrm{d}^4 x \sqrt{g} R \right\}
\end{multline}
\begin{multline}
	\Delta \mathcal{B}_{TT}^{\mathrm{Tor}} = \frac{1}{(4 \pi)^2} \left\{ k^4 \left( 4 \check{\Phi}_{2}^{1}(\tfrac{8 \bar{µ}^2}{3 k^2} f_{k}^{T}) - 2 \eta_N \check{\tilde{\Phi}}_{2}^{1}(\tfrac{8 \bar{µ}^2}{3 k^2} f_{k}^{T}) \right) \int \mathrm{d}^4 x \sqrt{g} \right.  \\
	\left. + k^2 \left( \frac{2}{3} \check{\Phi}_{1}^{1}(\tfrac{8 \bar{µ}^2}{3 k^2} f_{k}^{T}) - \frac{1}{3} \eta_N \check{\tilde{\Phi}}_{1}^{1}(\tfrac{8 \bar{µ}^2}{3 k^2} f_{k}^{T}) \right) \int \mathrm{d}^4 x \sqrt{g} R \right\}
\end{multline}
\begin{multline}
	\Delta \mathcal{B}_{qq}^{\mathrm{Tor}} = \frac{1}{(4 \pi)^2} \left\{ k^4 \left( 16 \check{\Phi}_{2}^{1}(-\tfrac{2\bar{µ}^2}{\xi k^2} f_{k}^{q}) - 8 \eta_N \check{\tilde{\Phi}}_{2}^{1}(-\tfrac{2 \bar{µ}^2}{\xi k^2} f_{k}^{q}) \right) \int \mathrm{d}^4 x \sqrt{g} \right.  \\
	\left. + k^2 \left( \frac{8}{3} \check{\Phi}_{1}^{1}(-\tfrac{2\bar{µ}^2}{\xi k^2} f_{k}^{q}) - \frac{4}{3} \eta_N \check{\tilde{\Phi}}_{1}^{1}(-\tfrac{2\bar{µ}^2}{\xi k^2} f_{k}^{q}) \right) \int \mathrm{d}^4 x \sqrt{g} R \right\}.
\end{multline}
\end{subequations}

\subsubsection{The RG equations}\label{subsubsec:RGequations}
Inserting (\ref{eq:thresholdcheck4d}) together with the metric and ghost contributions (\ref{eq:ehtraces}) into the flow equation (\ref{eq:yetanotherflowequation}) we can derive the scale derivatives $\partial Z_{Nk}$ and $\partial_t \left( Z_{Nk} \bar{\lambda}_k \right)$. We reexpress them in terms of the dimensionless running couplings and mass parameter, respectively:
\begin{subequations}
\begin{align}
	g_{k} &\equiv k^2 G_k \equiv k^2 \frac{\bar{G}}{Z_{Nk}} \quad , \quad  \lambda_k \equiv k^{-2}  \bar{\lambda}_k \ , \\
	µ^2 &\equiv \frac{\bar{µ}^2}{k^2} \quad .
\end{align}
\end{subequations}
This leads us to the final system of coupled, dimensionless RG equations:
\begin{subequations}
	\begin{align}
		\partial_t g_k &= \beta_g \left(g_{k},\lambda_{k};µ^2\right) \\
		\partial_t \lambda_k &= \beta_\lambda \left(g_{k},\lambda_{k};µ^2\right)
	\end{align}
	\label{eq:system}
\end{subequations}
This system involves the $µ$-dependent $\beta$-functions
\begin{subequations}
\label{eq:set}
	\begin{equation}
	\boxed{
		\beta_{g}\left(g,\lambda;µ^2\right) = \Bigl[ 2 + \eta_{N} \Bigr] g
		}
		\label{eq:betag}
	\end{equation}
	\begin{equation}
	\boxed{
		\beta_{\lambda}\left(g,\lambda;µ^2\right) = - \Bigl( 2 - \eta_{N} \Bigr) \lambda + \frac{g}{2 \pi} \left[ \vphantom{\frac{1}{3}} \left( B_{3}(\lambda) + \check{B}_{3}(µ^2) \right) + \eta_{N} \left( B_{4}(\lambda) + \check{B}_{4}(µ^2)  \right) \right]
		}
	\label{eq:betal}
	\end{equation}
	The anomalous dimension $\eta_{N}\equiv\eta_{N}\left(g,\lambda;µ^2\right)$ appearing in these functions is given by the following ($µ$ dependent) scalar function on theory space:
	\begin{equation}
	\boxed{
		\eta_{N}\left(g,\lambda;µ^2\right) = \frac{g \left( B_{1}(\lambda) + \check{B}_{1}(µ^2) \right)}{1 - g \left( B_{2}(\lambda) + \check{B}_{2}(µ^2) \right)} 
		} \ .
		\label{eq:anomalousdimension}
	\end{equation}
\end{subequations}
This formula for the anomalous dimension contains the following functions of the dimensionless cosmological constant and $µ^2$, respectively:
\begin{subequations}
\label{eq:B12}
\begin{align}
	B_{1}(\lambda) &= \frac{1}{3 \pi} \left( 5 \Phi_{1}^{1} ( - 2 \lambda ) - 18 \Phi_{2}^{2} (- 2 \lambda) - 4 \Phi_{1}^{1}(0) - 6 \Phi_{2}^{2}(0) \right) \\
	\check{B}_{1}(µ^2) &= \frac{1}{3 \pi} \left( 2 \check{\Phi}_{1}^{1}(\tfrac{1}{6} µ^2) + 2 \check{\Phi}_{1}^{1}(\tfrac{8}{3 } µ^2) + 8 \check{\Phi}_{1}^{1}(- 2 \xi µ^2) \right) \\
	B_{2}(\lambda) &= - \frac{1}{6 \pi} \left( 5 \tilde{\Phi}_{1}^{1} (- 2 \lambda) - 18 \tilde{\Phi}_{2}^{2} (- 2 \lambda) \right) \\
	\check{B}_{2}(µ^2) &= - \frac{1}{3 \pi} \left( \check{\tilde{\Phi}}_{1}^{1}(\tfrac{1}{6} µ^2) + \check{\tilde{\Phi}}_{1}^{1}(\tfrac{8}{3} µ^2) + 4 \check{\tilde{\Phi}}_{1}^{1}(-2 \xi µ^2) \right).
\end{align}
\end{subequations}
The $\beta$-function for the cosmological constant contains two more functions of $\lambda$ itself, $B_3$ and $B_4$, and of the mass parameter $µ$, $\check{B}_{3}$ and $\check{B}_{4}$:
\begin{subequations}
\label{eq:B34}
\begin{align}
	B_{3}(\lambda) &= 10 \Phi_{2}^{1} ( - 2 \lambda )  - 8 \Phi_{2}^{1}(0) \\
	\check{B}_{3}(µ^2) &= 4 \check{\Phi}_{2}^{1}(\tfrac{1}{6} µ^2) + 4 \check{\Phi}_{2}^{1}(\tfrac{8}{3} µ^2) + 16 \check{\Phi}_{2}^{1}(-2 \xi µ^2) \\
	B_{4}(\lambda) &= - 5 \tilde{\Phi}_{2}^{1} ( - 2 \lambda ) \\
	\check{B}_{4}(µ^2) &=  - \left( 2 \check{\tilde{\Phi}}_{2}^{1}(\tfrac{1}{6} µ^2) + 2 \check{\tilde{\Phi}}_{2}^{1}(\tfrac{8}{3} µ^2) + 8 \check{\tilde{\Phi}}_{2}^{1}(-2 \xi µ^2) \right).
\end{align}
\end{subequations}
The system (\ref{eq:system}) is the set of coupled ordinary differential equations we wanted to derive for our investigations. Although they were derived from a comparatively restrictive truncation, the above evolution equations still encapsulate non-perturbative effects, including contributions from all orders of perturbation theory.

%% file: Results.tex
Now that we explicitly know the system of RG equations, (\ref{eq:system}), let us focus on its analysis. We start in Section \ref{subsec:general} with some comments about the general features caused by the new type of threshold functions. In Section \ref{subsec:anal} we perform the analysis of the $\beta$-functions with the optimized cutoff in detail, discussing the fixed points and critical exponents of the resulting RG flow. To test the cutoff scheme dependence of the RG flow we repeat the investigation of its features with the generalized exponential cutoff in Section \ref{subsec:shape}. In Section \ref{subsec:inditorinv} we switch to an examination of the influence the \textit{individual} torsion invariants exert on the RG phase portrait. In Section \ref{subsec:planck}, we discuss a special identification of the mass parameter by relating $µ$ to the running Planck scale. The analysis in these four sections, \ref{subsec:anal}-\ref{subsec:planck}, is performed with a completely positive cutoff action, i.e. $\xi=+1$. Finally, Section \ref{subsec:negacut} details the changes when using a negative cutoff action, $\xi = -1$, for the $\mathpzc{q}_{\lambda µ \nu}$-sector.

\subsection[General Features of the \texorpdfstring{$\beta$}{beta}-functions]{General Features of the \texorpdfstring{$\bm{\beta}$}{beta}-functions}\label{subsec:general}
\textbf{(A)} We will carry out most of the analysis of the $\beta$-functions for the \textit{optimized shape function} \cite{Litim:2000ci}, $R^{(0)}_{\mathrm{opt}}(z) = (1-z)\Theta(1-z)$, for which all necessary threshold functions $\left\{ \Phi, \tilde{\Phi}, \check{\Phi}, \check{\tilde{\Phi}} \right\}$ can be computed in closed form. The standard threshold functions needed in $d=4$, already encountered in absence of torsion, are defined for all $w \in \mathds{R}\setminus\{-1\}$ and take the form:
\begin{subequations}
	\begin{alignat}{3}
		&\Phi_{1}^{1}(w) = \frac{1}{1+w} \ , \quad  &\Phi_{2}^{1}(w) = \frac{1}{2(1+w)} \ , \quad &\Phi_{2}^{2}(w) = \frac{1}{2 (1+w)^2} \ , \\
		&\tilde{\Phi}_{1}^{1}(w) = \frac{1}{2(1+w)} \ , \quad  &\tilde{\Phi}_{2}^{1}(w) = \frac{1}{6 (1+w)} \ , \quad &\tilde{\Phi}_{2}^{2}(w) = \frac{1}{6 (1+w)^2} \ .
	\end{alignat}
\end{subequations}
These functions are well known to have poles of first and second order at $w = -1$ \cite{Reuter:2001ag}. 
Computing the new threshold functions with $R^{(0)}_{\mathrm{opt}}$ we obtain the following result:
\begin{subequations}
	\begin{align}
		&\check{\Phi}_{1}^{1}(w) = \ln\left( 1 + \frac{1}{w} \right) \ , \quad &{}&\check{\Phi}_{2}^{1}(w) = -1 +(1+w)\ln\left(1+ \frac{1}{w} \right) , \\
		&\check{\tilde{\Phi}}_{1}^{1}(w) = 1- w \ln\left(1+ \frac{1}{w} \right) \ , \quad &{}&\check{\tilde{\Phi}}_{2}^{1}(w) = \frac{1}{2}+w-\left(w+w^2\right)\ln\left(1+ \frac{1}{w}\right).
	\end{align}
	\label{eq:newthresholdfunctions}
\end{subequations}
Here a further complication arises. We observe that the functions in (\ref{eq:newthresholdfunctions}) are not even well defined at all points of the interval $(-1,+ \infty)$: they are not real for $-1 < w < 0$, in addition to the singularities at $w = -1$ and $w=0$. The argument of the logarithms becomes negative if $w \in (-1,0)$. As this would make the $\beta$-functions and in turn the couplings complex we are forced to exclude this region from parameter space. Implementing these restrictions we find that, if we fix $\xi = +1$ the factor $\ln\left(1- \tfrac{1}{2µ^2} \right)$ appears in all $\check{B}$-functions. As a consequence the parameter $µ$ can assume values only in the range $µ > \frac{1}{\sqrt{2}}$. For the opposite choice, $\xi = -1$, all factors in the $\check{B}$-functions are of the form $\ln\left(1+ \tfrac{\text{const.}}{µ^2} \right)$, thus leading to an unrestricted mass parameter.

For the full set of $B$- and $\check{B}$-functions evaluated using the optimized cutoff see Appendix \ref{ch:threshold}.\\
\\
\textbf{(B)} We shall also employ the \textit{generalized exponential cutoff}, $R^{(0)}_{\mathrm{exp}}(z;s) = \frac{s z}{e^{s z}-1}$. Whilst the integrals defining $\Phi_{n}^{p}$ can be carried out analytically for vanishing argument \cite{Lauscher:2002sq}, for non-vanishing arguments no such representation exists and we have to rely on numerical approaches. Therefore, we have to pay attention to numerical instabilities that may plague our computations. We refrain from writing down all considered threshold functions individually, and just give the full set of $B$-functions and $\check{B}$-functions in Appendix \ref{ch:threshold}.

\begin{figure}[htbp]
	\centering
	\subfigure[$µ^2 = 1$ and $R^{(0)}_{\mathrm{opt}}$.]{
		\centering
		\includegraphics[width=0.3\textwidth]{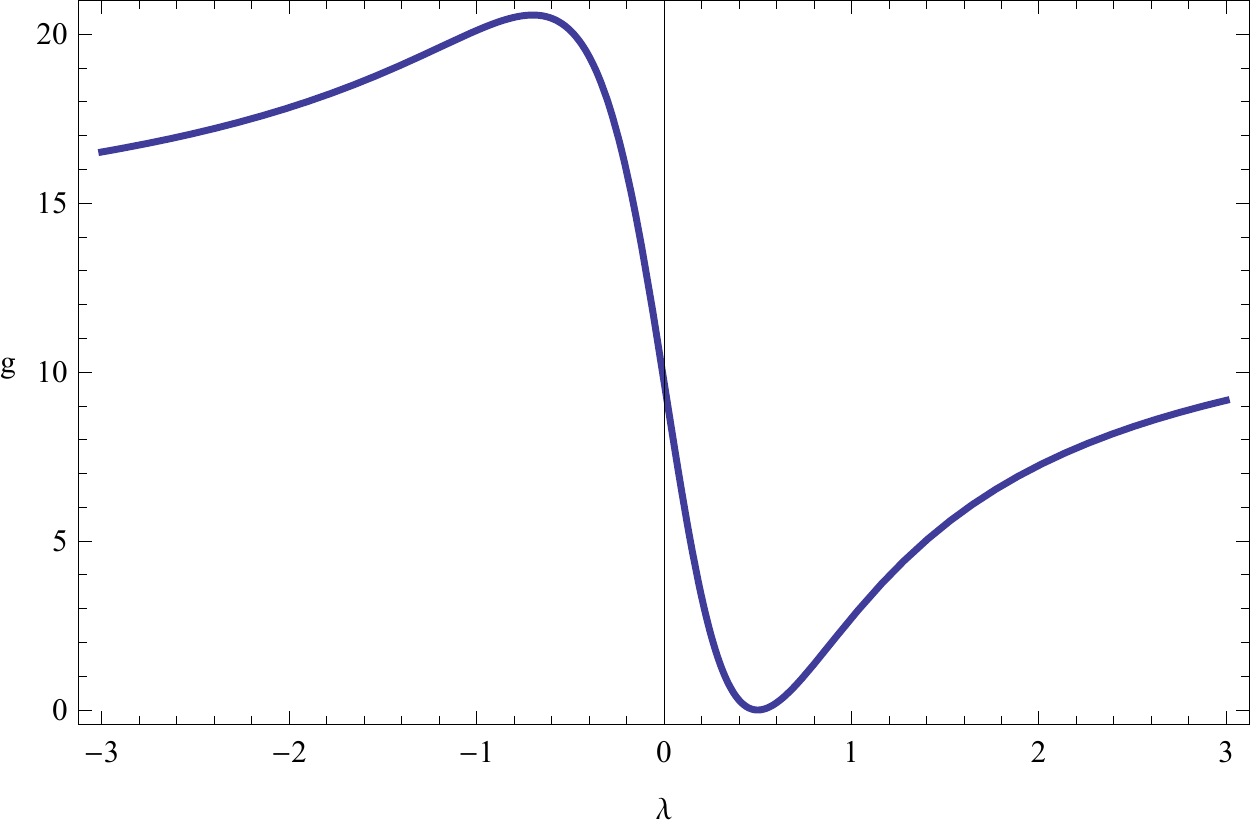}
		\label{fig:adim02}}
	\subfigure[$µ^2 = \frac{3}{10}$ and $R^{(0)}_{\mathrm{opt}}$; $S_µ$-contribution only.]{
		\centering
		\includegraphics[width=0.3\textwidth]{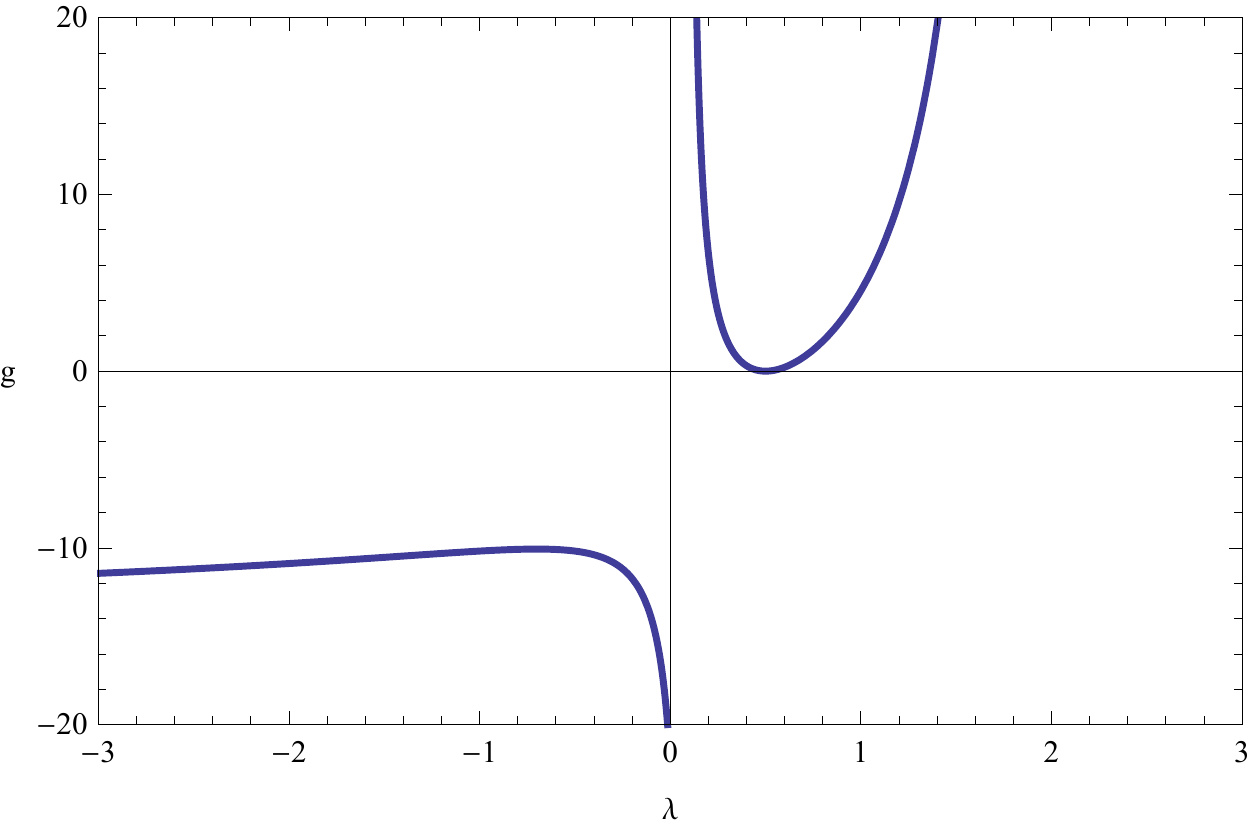}
		\label{fig:adim03v2}}
	\subfigure[$µ^2 = \frac{1}{1.9}$ and $R^{(0)}_{\mathrm{exp}}$.]{
		\centering
		\includegraphics[width=0.3\textwidth]{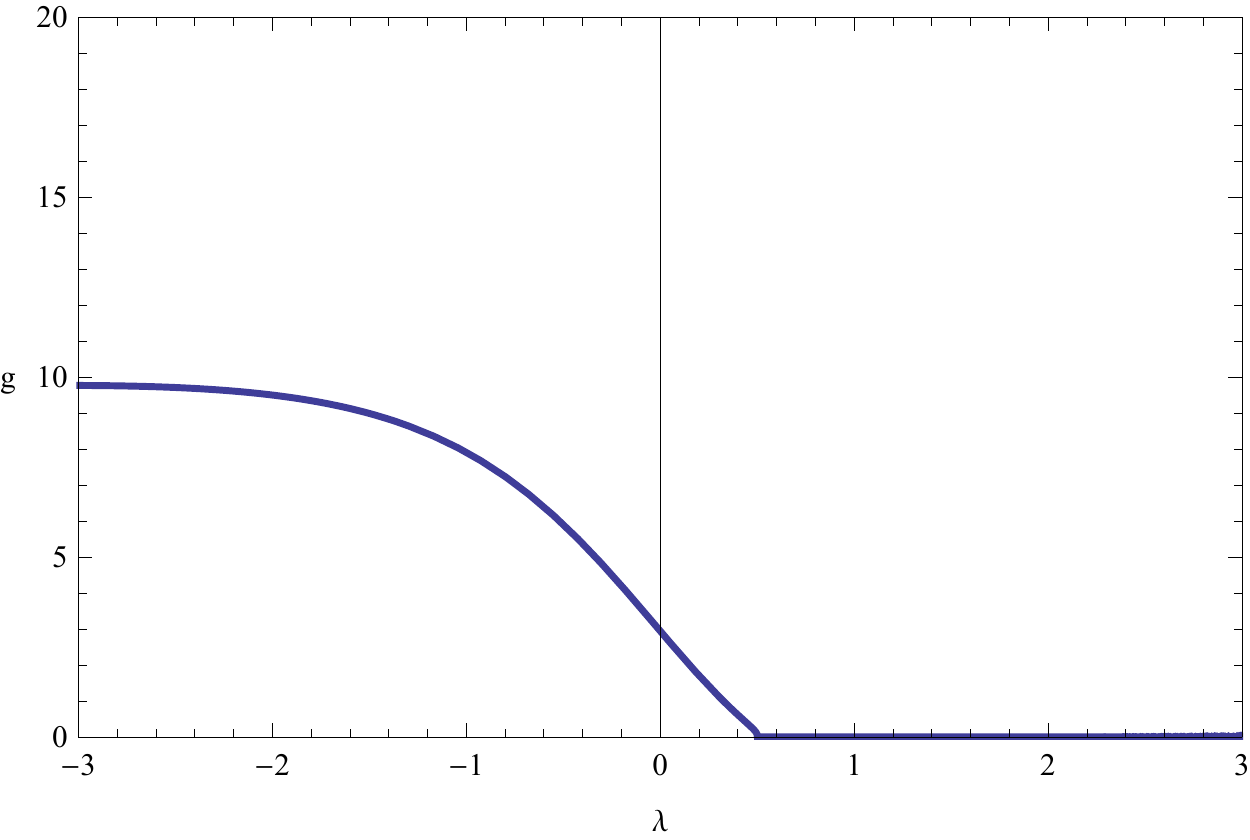}
		\label{fig:adim04}}
	\caption{Lines $g \equiv G(\lambda)$ with divergent anomalous dimension $\eta_{N} $ for various choices of $µ^2$ and the shape function $R^{(0)}$.}
	\label{fig:anomalousdimension1}
\end{figure}

\begin{figure}[htbp]
	\centering
	\subfigure[$µ^2 = 1$ and $R^{(0)}_{\mathrm{opt}}$.]{
		\centering
		\includegraphics[width=0.3\textwidth]{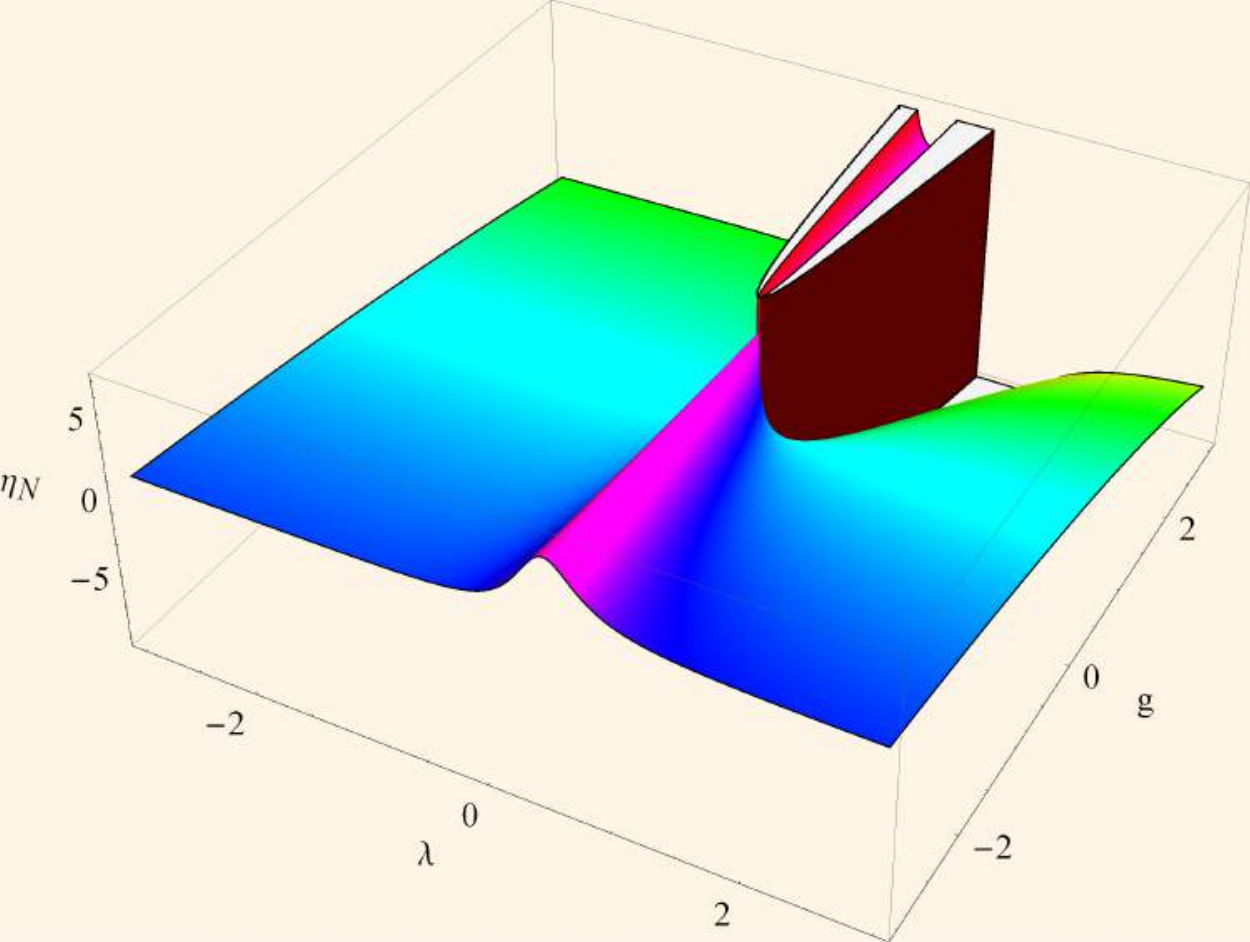}
		\label{fig:adim3d}}
	\subfigure[$µ^2 = \frac{3}{10}$ and $R^{(0)}_{\mathrm{opt}}$; $S_µ$-contribution only.]{
		\centering
		\includegraphics[width=0.3\textwidth]{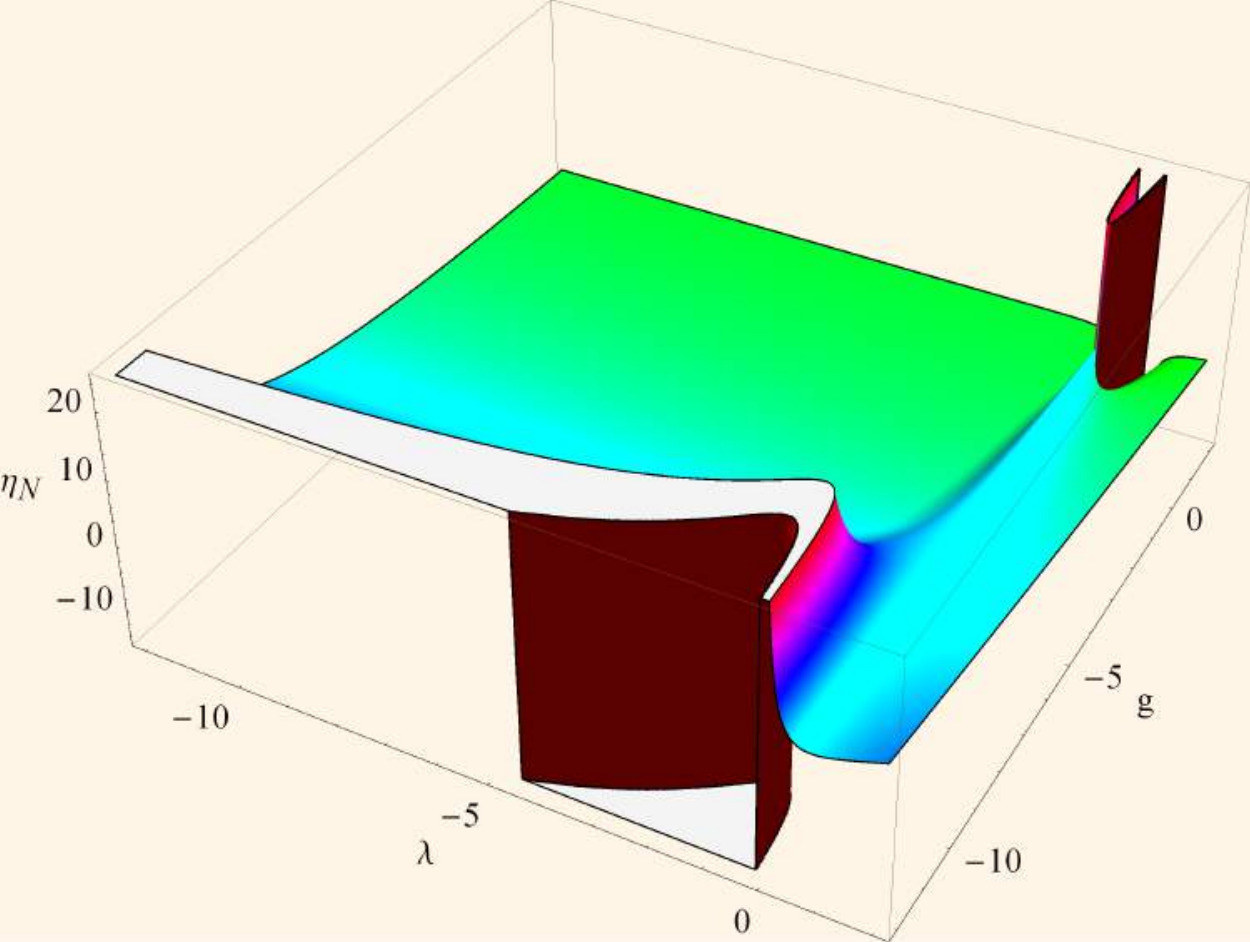}
		\label{fig:adim3d2}}
	\subfigure[$µ^2 = \frac{1}{1.9}$ and $R^{(0)}_{\mathrm{exp}}$.]{
		\centering
		\includegraphics[width=0.3\textwidth]{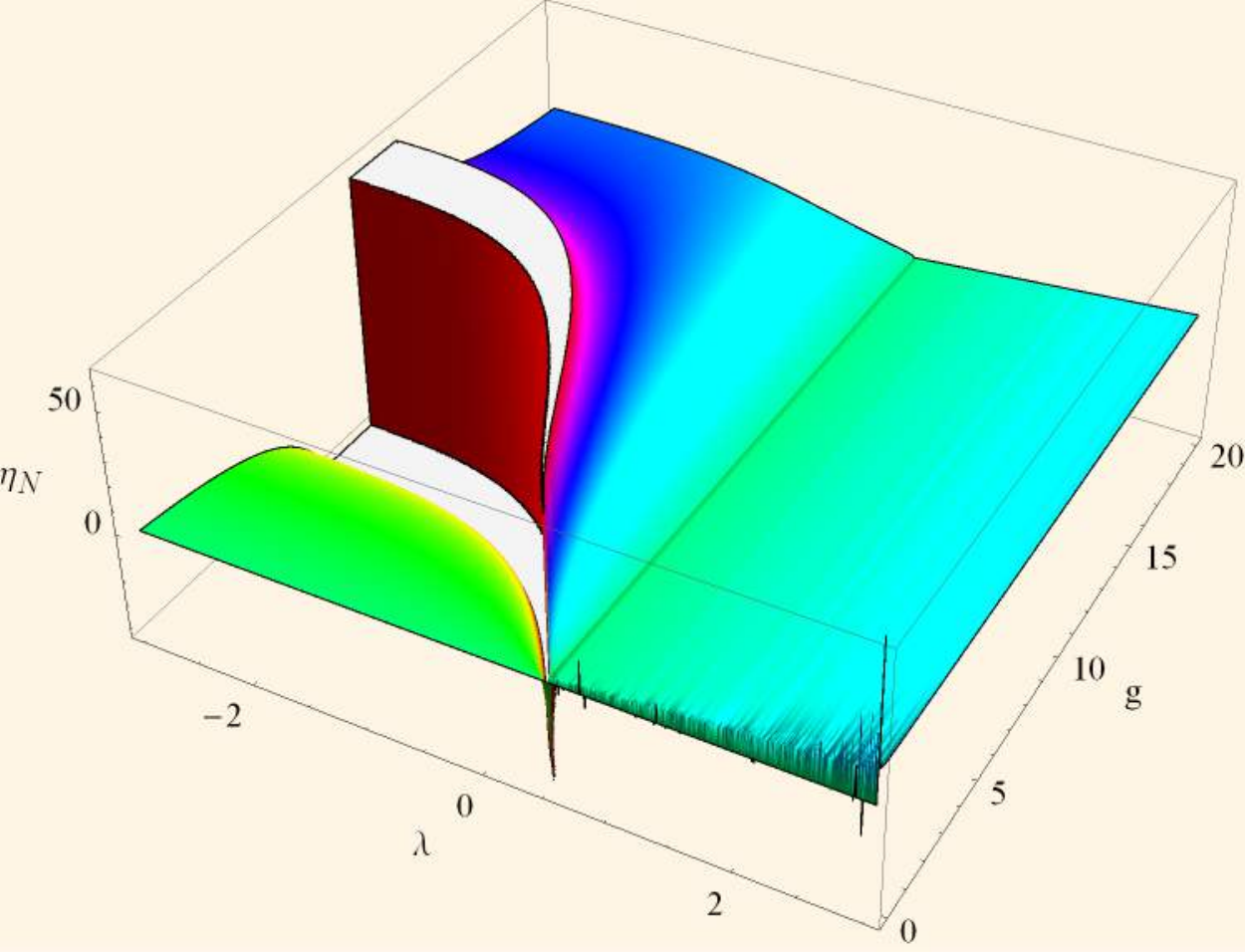}
		\label{fig:adim3d3}}
	\caption{Anomalous dimension $\eta_{N}$ as a function of $g$ and $\lambda$ for various choices of $µ^2$ and $R^{(0)}$.}
	\label{fig:anomalousdimension2}
\end{figure}

A numerical analysis shows that the problem of a well defined value of the new threshold functions $\check{\Phi}_{n}^{p}(w)$ in the range $w \in [-1,0]$ also manifests itself using the generalized exponential cutoff $R^{(0)}_{\mathrm{exp}}$. We believe that this is not a sign of an improperly chosen shape function $R^{(0)}$ but rather a general feature of the threshold functions $\check{\Phi}$. \\
\\
\textbf{(C)} Let us have a closer look at the \textit{anomalous dimension $\eta_{N}$}. For certain combinations of $g$ and $\lambda$ values the anomalous dimension diverges, $|\eta_{N}| \rightarrow \infty$. This generates a singular boundary of the ``physical'' theory space, namely a line which RG trajectories can not cross, thus separating different domains in the $g-\lambda$-plane from each other. This phenomenon is well known in principle \cite{Reuter:2001ag}. From the expression (\ref{eq:anomalousdimension}) we learn that the anomalous dimension $\eta_{N}$ diverges if $g$ and $\lambda$ are such that
\begin{equation}
	\frac{1}{B_{2}(\lambda) + \check{B}_{2}(µ^2)} = g \ .
\end{equation}
In Figure \ref{fig:anomalousdimension1} we plot the line of points $(g,\lambda)$ satisfying this relation in the form $g\equiv g(\lambda)$ for various fixed values of $µ^2$. We see that the singularity changes shape quite drastically when varying the mass parameter $µ^2$ or changing the shape function $R^{(0)}$. In the case of the generalized exponential cutoff, additional numerical instabilities, arise.
\\
In Figure \ref{fig:anomalousdimension2} we plot $\eta_{N}$ as a function of $\lambda$ and $g$ to better visualize the singularities of the anomalous dimension.

In the phase portraits we are going to present later on we shall indicate regions that are hidden ``behind'' the singularities of the anomalous dimension - because any RG trajectory emanating from within that region runs into a singularity - by a gray filling of the area.

\subsection{Analytical analysis of the RG flow (optimized cutoff)}\label{subsec:anal}

To start with, let us stick to the optimized shape function $R^{(0)}_{\mathrm{opt}}$ and analyze the $µ$-dependence of the RG flow for $\xi = +1$ in the allowed domain $µ^2 > \frac{1}{2}$. 

The fixed points of the flow correspond to common zeros ($\lambda^{*},g^{*}$) of the two $\beta$-functions (\ref{eq:betag}) and (\ref{eq:betal}). There is clearly a Gaussian fixed point, henceforth denoted \textbf{GFP}, at vanishing couplings, $\lambda^{*}=g^{*}=0$. It has one attractive, and one repulsive eigen-direction.

The most important finding is that additionally to the \textbf{GFP} there exists a non-Gaussian fixed point, denoted \textbf{NGFP}, for any allowed value of the dimensionless mass parameter $µ$. One might expect its critical exponents $\theta = \theta' + i \theta''$ to be approximately universal quantities, alongside the product of its coordinates, $\lambda^{*} g^{*}$. 

In Table \ref{tab:properties1} we give an overview of these quantities at the \textbf{NGFP}; it will be discussed in detail in a moment. 

\begin{figure}[H]
	\centering
	\subfigure[$\theta'$ (blue solid), $\theta''$ (red dashed) and $\lambda^{*}g^{*}$ (yellow dot-dashed) as a function of $µ^2$.]{
		\centering
		\includegraphics[width=0.47\textwidth]{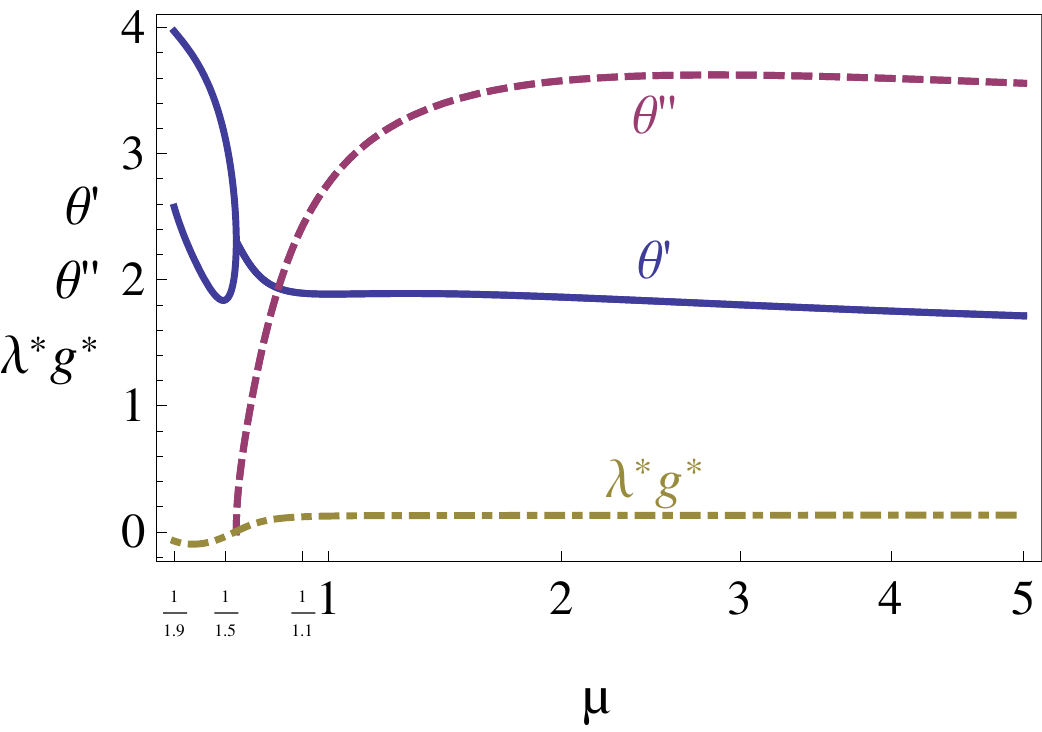}
		\label{fig:CriticalExponentsOpt}}
	\subfigure[$\theta'$]{
		\centering
		\includegraphics[width=0.45\textwidth]{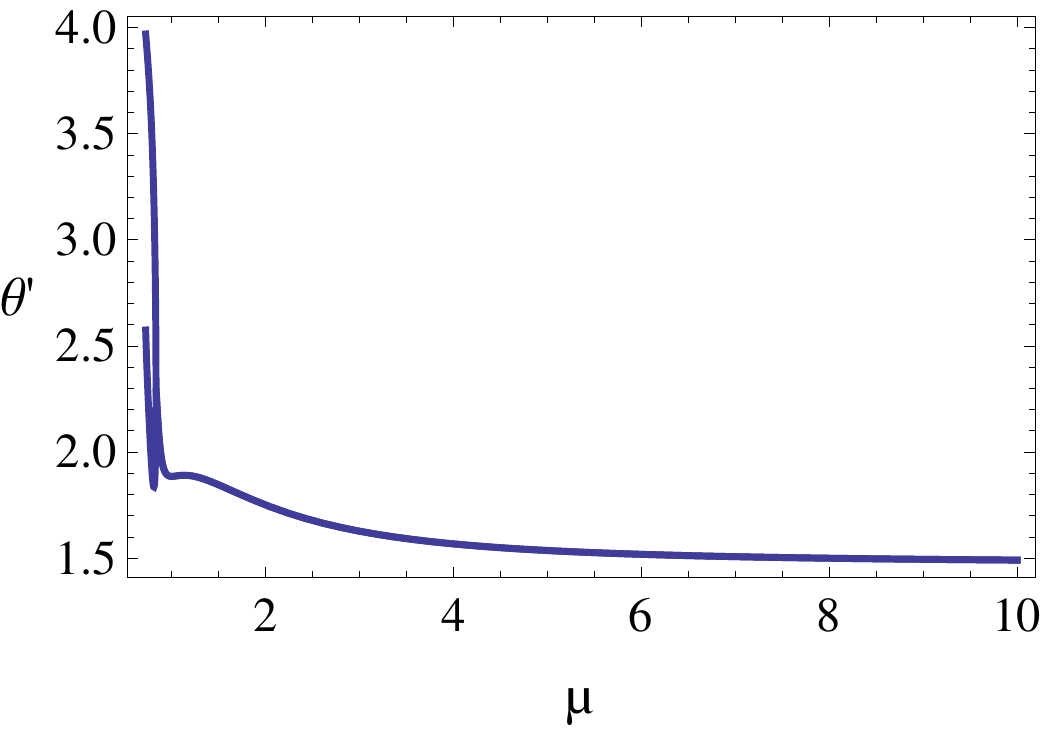}
		\label{fig:CriticalExponentsReOptv2}}
	\subfigure[$\theta''$]{
		\centering
		\includegraphics[width=0.45\textwidth]{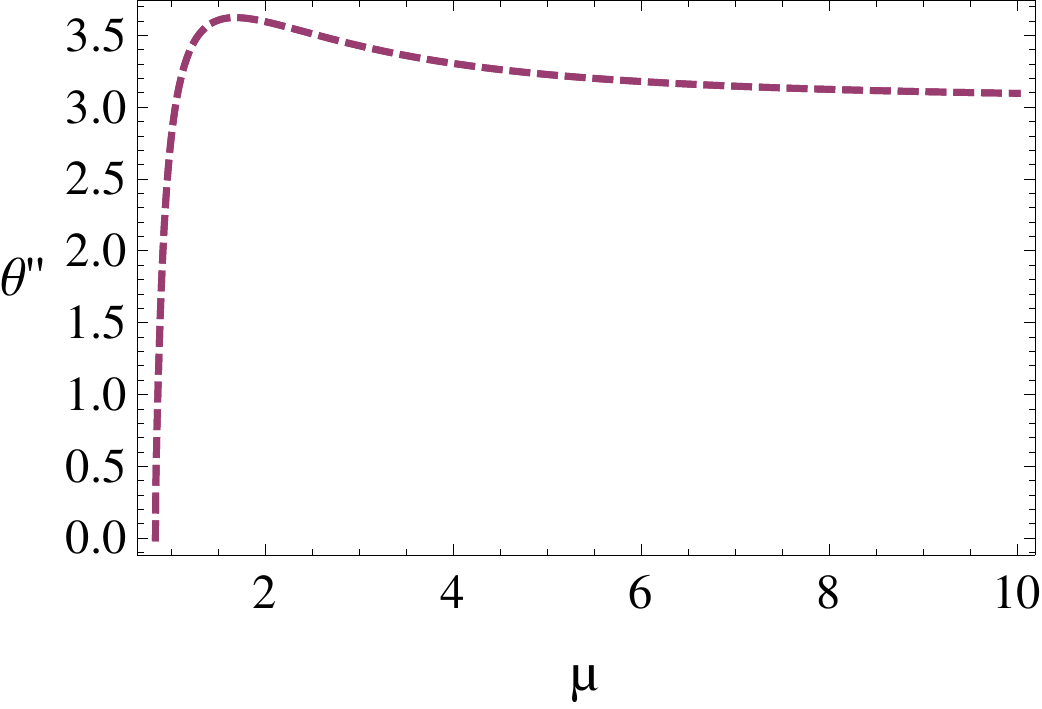}
		\label{fig:CriticalExponentsImOptv2}}
	\subfigure[$\lambda^{*}g^{*}$]{
		\centering
		\includegraphics[width=0.5\textwidth]{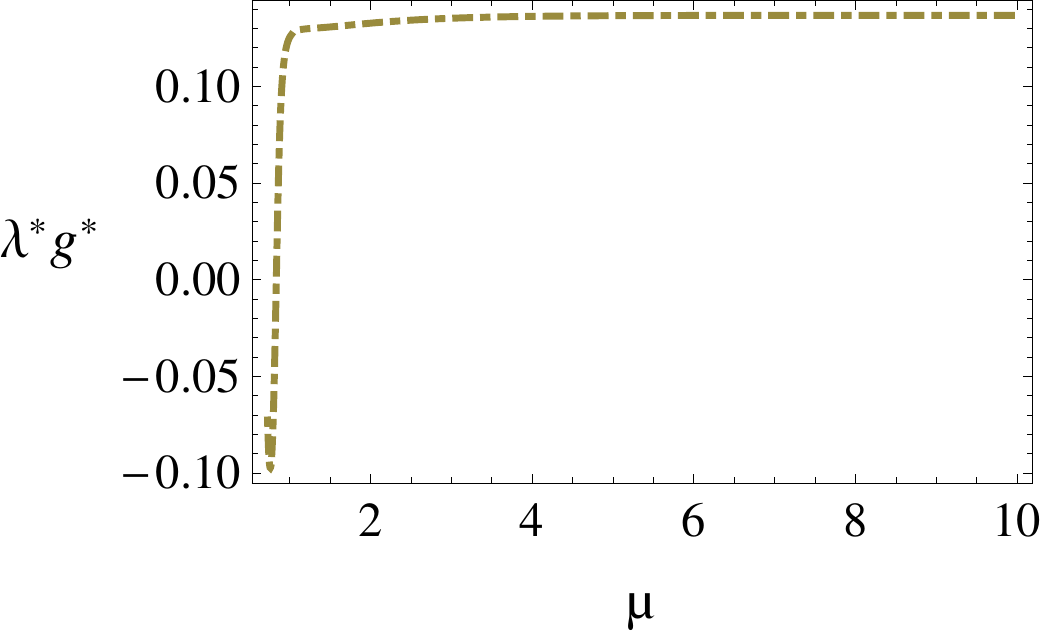}
		\label{fig:lgplotoptv2}}
	\caption{Critical exponents and $\lambda^{*}g^{*}$ of the \textbf{NGFP} as functions of $µ$. At $µ = µ_{\mathrm{crit}}$ the critical exponents $\theta = \theta' + i \theta''$ split into real and imaginary part (blue and red line, respectively), and the product $\lambda^{*}g^{*}$ (yellow line) changes its sign accordingly.}
	\label{fig:critexpopt}
\end{figure}

Figure \ref{fig:critexpopt} shows the $µ$-dependence of the three quantities we expect to be universal, $\theta'$, $\theta''$, and $\lambda^{*} g^{*}$. 

The full set of phase portraits of the RG flow can be found in Figures \ref{fig:NGFPfullTor} and \ref{fig:NGFPfullTor2} of Appendix \ref{sec:PP}. 
{\allowdisplaybreaks
\begin{table}[H]
	\centering
		\begin{tabular}{r|cc|ccc|l}
		$µ^2$ & $\lambda^{*}$ & $g^{*}$ & $\lambda^{*} g^{*}$ & $\theta'$ & $\theta''$ & phase portrait \\
		\hline \hline
		$\frac{1}{1.9}$ & -0.158758 & 0.456914 & -0.072539 & $\begin{array}{c} 3.96944 \\ 2.57512 \end{array}$  & ~ & Fig. \ref{fig:PDopt119} \\
		$\frac{1}{1.5}$ & -0.0401966 & 0.905493 & -0.0363977 & $\begin{array}{c} 3.11744 \\ 1.83953 \end{array}$ & ~ & Fig. \ref{fig:PDopt115} \\
		\hline
		$\approx 0.695309$ & 0 & 0.91875 & 0 & $\begin{array}{c} 2.55544 \\ 2.08823 \end{array}$ & ~ & Fig. \ref{fig:PDopttest} \\
		\hline
		$\approx 0.698591$ & 0.00461372 & 0.918447 & 0.00423746 & $\begin{array}{c} 2.30474 \\ 2.30474 \end{array}$ & ~ & Fig. \ref{fig:PDoptReal} \\
		\hline
		$\frac{1}{1.1}\vphantom{\frac{\check{h}}{\check{h}}}$ & 0.16717 & 0.715096 & 0.119543 & 1.893 & 2.42284 & Fig. \ref{fig:PDopt111v3} \\
		1 & 0.187746 & 0.668119 & 0.125437 & 1.88548 & 2.76685 & Fig. \ref{fig:PDopt1v3} \\
		2 & 0.228324 & 0.57135 & 0.130453 & 1.86164 & 3.57737 & Fig. \ref{fig:PDopt2v3} \\
		3 & 0.229112 & 0.574684 & 0.131667 & 1.80055 & 3.62287 & Fig. \ref{fig:PDopt3v3} \\
		4 & 0.226895 & 0.584904 & 0.132712 & 1.75107 & 3.59617 & Fig. \ref{fig:PDopt4v3} \\
		5 & 0.22437 & 0.595076 & 0.133517 & 1.71334 & 3.55794 & Fig. \ref{fig:PDopt5v3} \\
		6 & 0.222017 & 0.604143 & 0.13413 & 1.68418 & 3.52025 & Fig. \ref{fig:PDopt6v3} \\
		7 & 0.219922 & 0.612038 & 0.1346 & 1.66113 & 3.48592 & Fig. \ref{fig:PDopt7v3} \\
		8 & 0.218077 & 0.618896 & 0.134967 & 1.64252 & 3.45535 & Fig. \ref{fig:PDopt8v3} \\
		9 & 0.216453 & 0.624877 & 0.135256 & 1.62721 & 3.42828 & Fig. \ref{fig:PDopt9v3} \\
		10 & 0.215018 & 0.630124 & 0.135488 & 1.61441 & 3.40429 & Fig. \ref{fig:PDopt10v3} \\
		\hline
		25 & 0.204452 & 0.668012 & 0.136577 & 1.53616 & 3.22776 & Fig. \ref{fig:PDopt25v3} \\
		50 & 0.199415 & 0.685726 & 0.136744 & 1.50653 & 3.14452 & Fig. \ref{fig:PDopt50v3} \\
		100 & 0.196479 & 0.695964 & 0.136742 & 1.49109 & 3.09645 & Fig. \ref{fig:PDopt100v3}
		\end{tabular}
	\caption{Properties of the \textbf{NGFP} for various values of $µ^2$.}
	\label{tab:properties1}
\end{table}}

\noindent \textbf{(A) Fixed point properties.} The very existence of the \textbf{NGFP} seems indeed to be a universal feature of the flow. Its properties depend on the value of $µ$ though. For $µ < µ_{\mathrm{crit}}$ with the ``crititcal'' value $µ_{\mathrm{crit}}^2 \approx 0.69$ we find an \textit{UV attractive} fixed point with two \textit{real} critical exponents. They turn into a complex conjugated pair $\theta = \theta' \pm i \theta''$ at $µ > µ_{\mathrm{crit}}$. Although the critical exponents change from real to complex, the fixed point stays UV attractive throughout, for all $µ > µ_{\mathrm{crit}}$. 
Also at $µ = µ_{\mathrm{crit}}$, the product $\lambda^{*}g^{*}$ changes its sign from negative to positive. 

For $µ$-values well above $µ_{\mathrm{crit}}$ the dependence on the mass parameter weakens for all three quantities, and the expected approximate universality is realized.

As for giving a numerical value to $µ = \frac{\bar{µ}}{k}$, we have decide to choose the \textit{dimensionless} parameter, $µ$, to be $k$-independent, which is in the same spirit as expressing dimensionfull couplings in units of the cutoff scale. This, then, gives a linear $k$-dependence to the dimensionful one: $\bar{µ}= µ \cdot k$. Therefore, a $µ$-value of the order of 1 would be most natural, as any other choice would introduce an additional unphysical scale other than $k$. 

Encouragingly, the situation in the region near $µ=1$ is very similar to QEG, i.e. metric gravity without torsion: We find the \textbf{NGFP} to be UV attractive, with a positive product $\lambda^{*}g^{*}$ and a pair of complex conjugate critical exponents. In the case of QEG the critical exponents assume values in the range $1.4 \lesssim \theta'\lesssim 1.8$ and $2.3 \lesssim \theta''\lesssim 4$, respectively, and the coordinate product is about $\lambda^{*}g^{*} \approx 0.12$ \cite{Reuter:2012id}. Comparing these values with those in Tab. \ref{tab:properties1} we observe they are in remarkable agreement.\\
\\
\textbf{(B) The phase portrait.} Let us have a closer look at the global properties of the RG flow. We observe that the case $µ \approx 1$, represented in Figures \ref{fig:PDopt1v3} and \ref{fig:PDopt2v3}, is indeed very similar to the flow of QEG in the Einstein-Hilbert truncation \cite{Reuter:1996cp,Reuter:2012id}. The \textbf{NGFP} is located in the  first quadrant\footnote{Here we follow the usual naming convention for the four quadrants of a two-dimensional plane: going counter-clockwise, the quadrants are denoted I $(+,+)$, II $(-,+)$, III $(-,-)$ and IV $(+,-)$.} with two attractive directions. The RG trajectories spiral into the fixed point due to the nonzero imaginary part of the critical exponents. The \textbf{NGFP} is connected to the Gaussian fixed point via a ``separatrix'', the RG trajectory hitting the \textbf{GFP} at $k \rightarrow 0$, hence separating trajectories with positive or negative IR values for the cosmological constant $\lambda$. 

Additionally we note that the IR attractive eigen-direction of the \textbf{GFP} points into the positive $\lambda$-halfplane. As $µ$ is lowered, this direction turns from the first to second quadrant. For sufficiently small values of $µ$, it points into the negative $\lambda$-halfplane, as the \textbf{NGFP} itself now lies at $\lambda^{*} <0$, i.e. in the second quadrant. Also the critical exponents change from a complex conjugated pair to two positive real exponents. The fixed points stays UV attractive though, even for the $µ^2 \approx \frac{1}{2}$. Qualitatively the phase portrait for small mass parameter Figure \ref{fig:PDopt119} resembles the RG flow of Quantum Einstein-Cartan Gravity (QECG) in the planes of infinite or vanishing Immirzi parameter, as found in \cite{Daum:2010phd,Daum:2013fu}. The fixed point moves with increasing $µ$ in the direction of positive $\lambda$ and crosses the $g$-axis at $µ \approx 0.833851$ (or $µ^2 \approx 0.695309$), resulting in a vanishing cosmological constant $\lambda^{*}=0$, see Figure \ref{fig:PDopttest}.  

For $µ$ larger than about $µ^2 = 2$, from Figure \ref{fig:PDopt3v3} onward, the mass parameter dependence becomes very weak and the entire phase portrait only minimally deviates from the metric case. While the fixed point position $(g^{*},\lambda^{*})$ changes slightly, but with an almost constant product $\lambda^{*}g^{*}$, the qualitative features of the flow are remarkably similar for all choices of $µ$. The \textbf{NGFP} seems to converge and stabilize at a point close to the fixed point of the purely metric theory. \\
\\
\textbf{(C) Comparison with Tetrad Gravity.} The sequence of phase portraits in Figures \ref{fig:NGFPfullTor} and \ref{fig:NGFPfullTor2} which is labeled by $µ$ strongly resembles the RG flow of Tetrad Gravity (for the case of $\xi = 1$\footnote{Here the parameter $\xi$ is merely a mathematical tool that enables the study of a continuous class of field redefinitions at a time according to $g_{µ \nu} = \xi^{-1} \tensor{e}{^{a}_{µ}}\tensor{e}{^{b}_{\nu}} \eta_{ab}$.}) studied in \cite{Harst:2012ni,Harst:2012phd}. There one also found a non-Gaussian fixed point in the second quadrant for small $µ$, which moved towards positive $\lambda$-values when $µ$ was increased, and arrived in the first quadrant for $µ \approx 1$. However, two distinct differences occur. In Tetrad Gravity, the IR attractive eigen-direction of the GFP points into the negative $\lambda$-halfplane, even if the NGFP itself lies at $\lambda^{*}>0$. The ``separatrix'' thus crosses the $g$-axis and hits the Gaussian fixed point from the second quadrant. 

More importantly, for larger $µ$ the RG flow exhibits a significantly different behavior. While in the $\mathcal{T}_{\mathrm{dtor}}$-truncation at hand the RG flow does not substantially deviate from the metric case for $µ > 1$, in Tetrad Gravity one observes the formation of a limit cycle for $µ \gtrsim 1.35$. The NGFP of Tetrad Gravity, while still located in the first quadrant, is UV repulsive now, as the critical exponents form a complex conjugate pair with a negative real part. 

In \cite{Harst:2012ni,Harst:2012phd} both of these features were accredited to the $µ$-dependence of the flow, stemming there from the need to rescale the Faddeev-Popov ghosts taking care of the local Lorentz invariance.

One might wonder, why we do not observe the formation of an analogous (unphysical) limit cycle within our truncation. Taking a closer look at the $\beta$-functions of Tetrad Gravity obtained in \cite{Harst:2012ni,Harst:2012phd}, we observe that besides $µ$ the parameter $\xi$ also appears in the arguments of the threshold functions, an example being $\Phi^{1}_{\frac{d}{2}-1}\bigl(-2 \lambda \frac{d-2 + \xi}{d-2} \bigr)$. The global structure of the flow is determined by the interplay with the cosmological constant $\lambda$. Tests indicate that the deviating arguments (in the example, $\Phi_{1}^{1}(-3 \lambda)$ for Tetrad Gravity instead of $\Phi_{1}^{1}(-2 \lambda)$ for QEG in the case of $d=4, \xi =1$) holds the actual responsibility for the limit cycle. So its formation is only indirectly related to the magnitude of $µ$. We therefore conjecture that, while for small $µ$ the corresponding threshold functions $\check{\Phi}_{n}^{p}(w)$ can suppress the limit cycle, they vanish for $w \rightarrow \infty$ thus enabling the formation of the limit cycle. \\
\\
\textbf{(D) Summary.} The RG flow on $\mathcal{T}_{\mathrm{dtor}}$ displays a strong $µ$-dependence of the quantities $\theta'$, $\theta''$, $\lambda^{*}g^{*}$ for small $µ$ which subsides considerably for larger values of $µ$, leading to a stabilization of the entire flow. A NGFP is found to exist for any value of $µ$. Furthermore, we find features comparable with QECG, which was our original choice for a comparison. Finally, the $\mathcal{T}_{\mathrm{dtor}}$ flow for mass parameters $µ \gtrsim 1$ is in good agreement with the results of QEG. 

To confirm the picture, we still have to carefully survey the cutoff scheme dependence of these $R^{(0)}_{\mathrm{opt}}$-based results and test their robustness under changes of the shape function $R^{(0)}$.

\subsection{Numerical analysis of the RG flow (generalized exponential cutoff)}\label{subsec:shape}

In this section we explore the cutoff scheme dependence of the presumably universal quantities that were studied in the previous section with the optimized cutoff $R^{(0)}_{\mathrm{opt}}$. Employing now the generalized exponential cutoff, the one parameter family of shape functions $R^{(0)}_{\mathrm{exp}}(z;s) = \frac{s z}{e^{s z}-1}$, the resulting critical exponents and the product $\lambda^{*}g^{*}$ of the \textbf{NGFP} are determined for shape parameters $s$ ranging from $\frac{1}{2}$ to $20$. As already mentioned in Section \ref{subsec:general}, we hereby heavily rely on numerical methods, since the threshold functions can not be computed analytically.

\begin{figure}[H]
	\centering
	\subfigure[$\theta'$ (blue) and $\theta''$ (red) for $µ^2 = \frac{1}{1.9}$.]{
		\centering
		\includegraphics[width=0.45\textwidth]{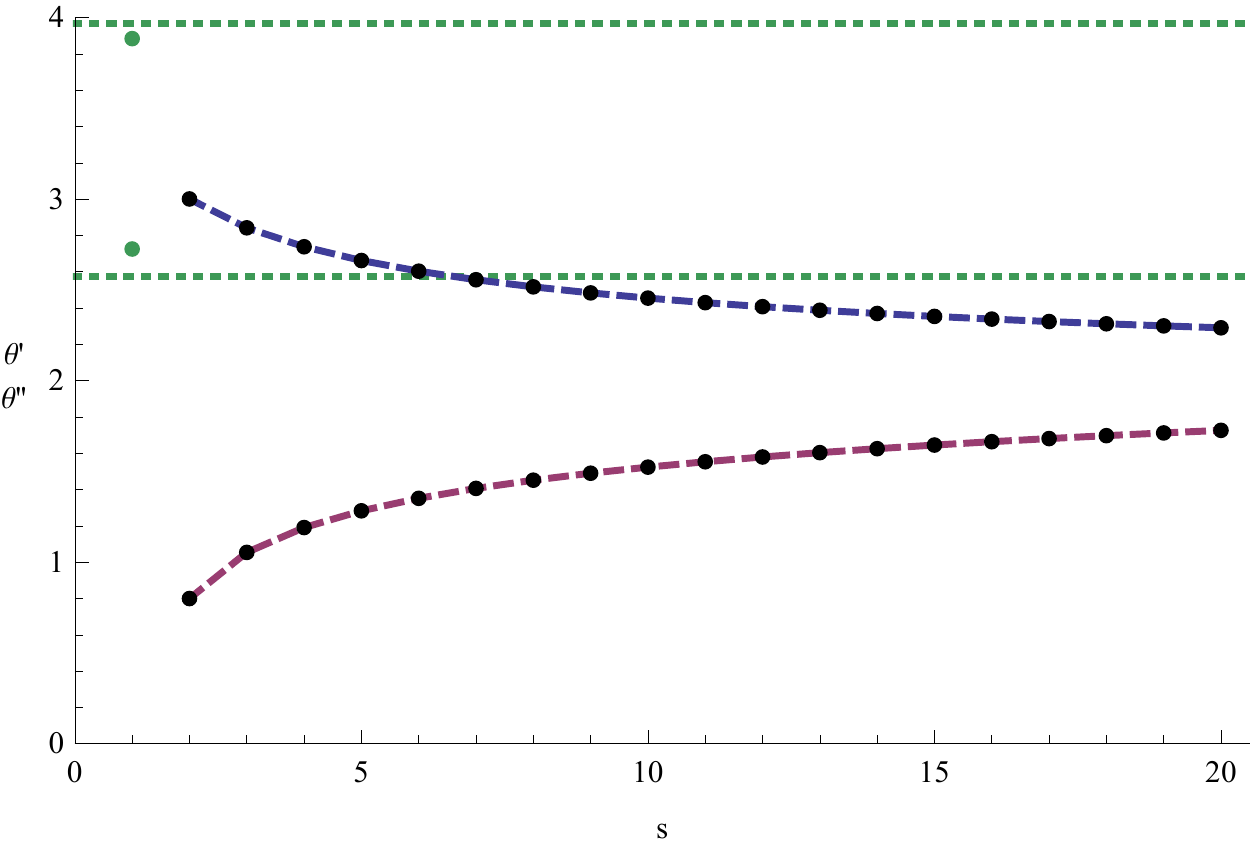}
		\label{fig:thetam119}}
	\subfigure[$\theta'$ (blue) and $\theta''$ (red) for $µ^2 = 1$.]{
		\centering
		\includegraphics[width=0.45\textwidth]{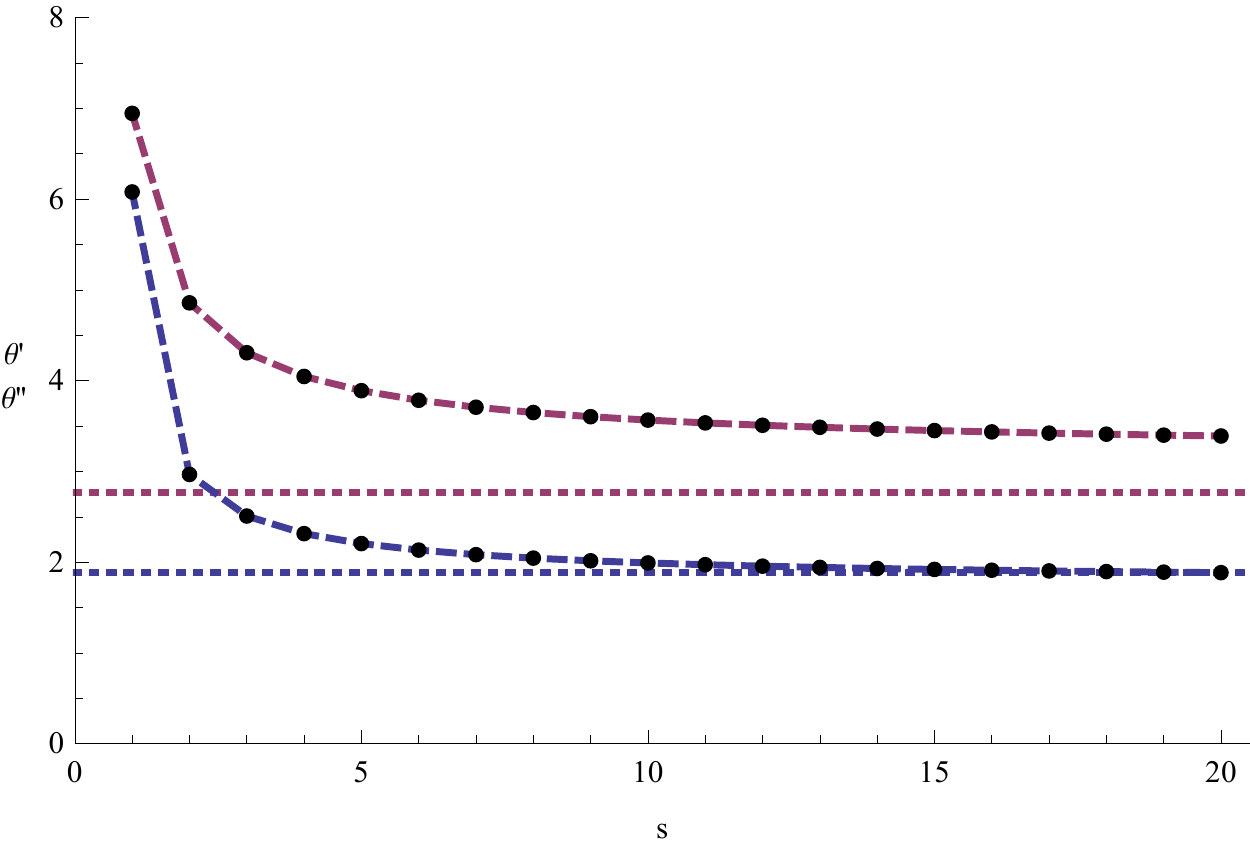}
		\label{fig:thetam1}}
	\subfigure[$\theta'$ (blue) and $\theta''$ (red) for $µ^2 = 2$.]{
		\centering
		\includegraphics[width=0.45\textwidth]{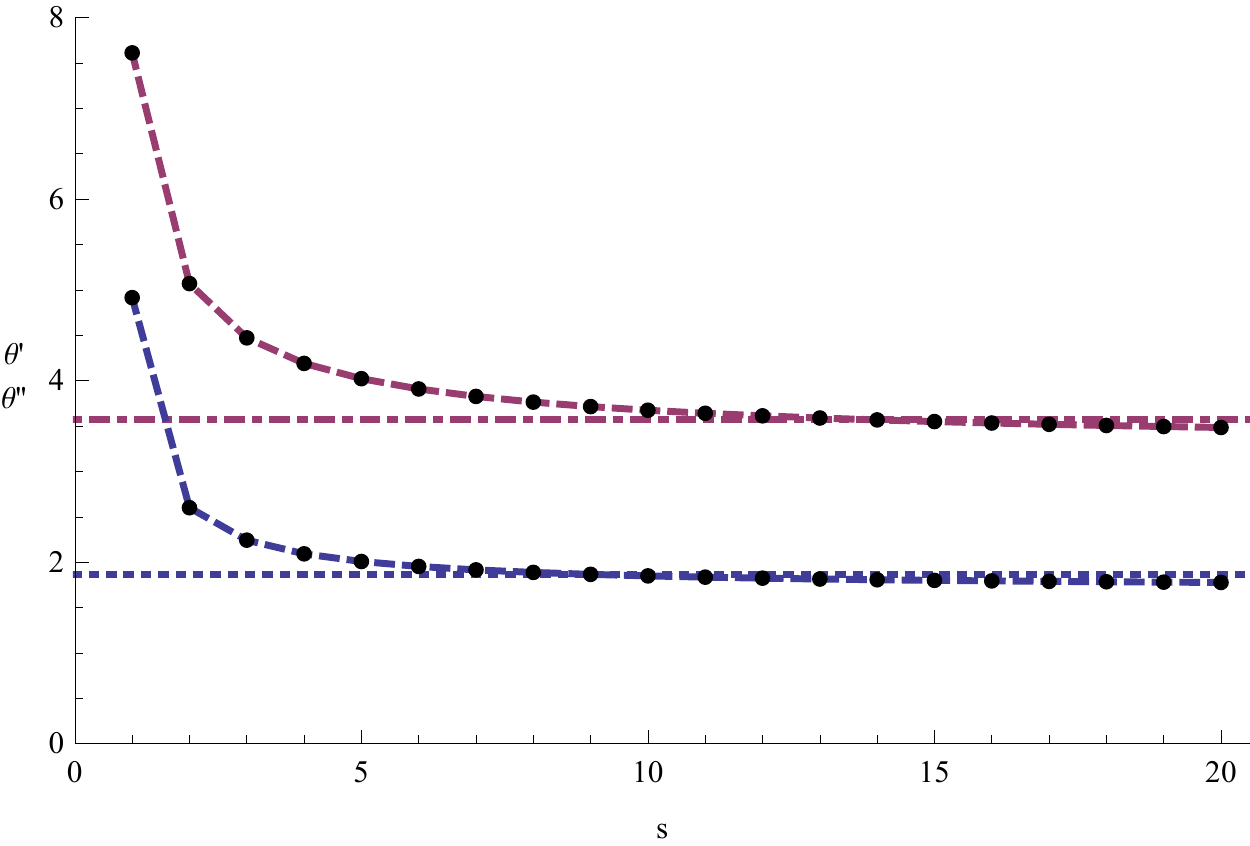}
		\label{fig:thetam2}}
	\subfigure[$\theta'$ (blue) and $\theta''$ (red) for $µ^2 = 5$.]{
		\centering
		\includegraphics[width=0.45\textwidth]{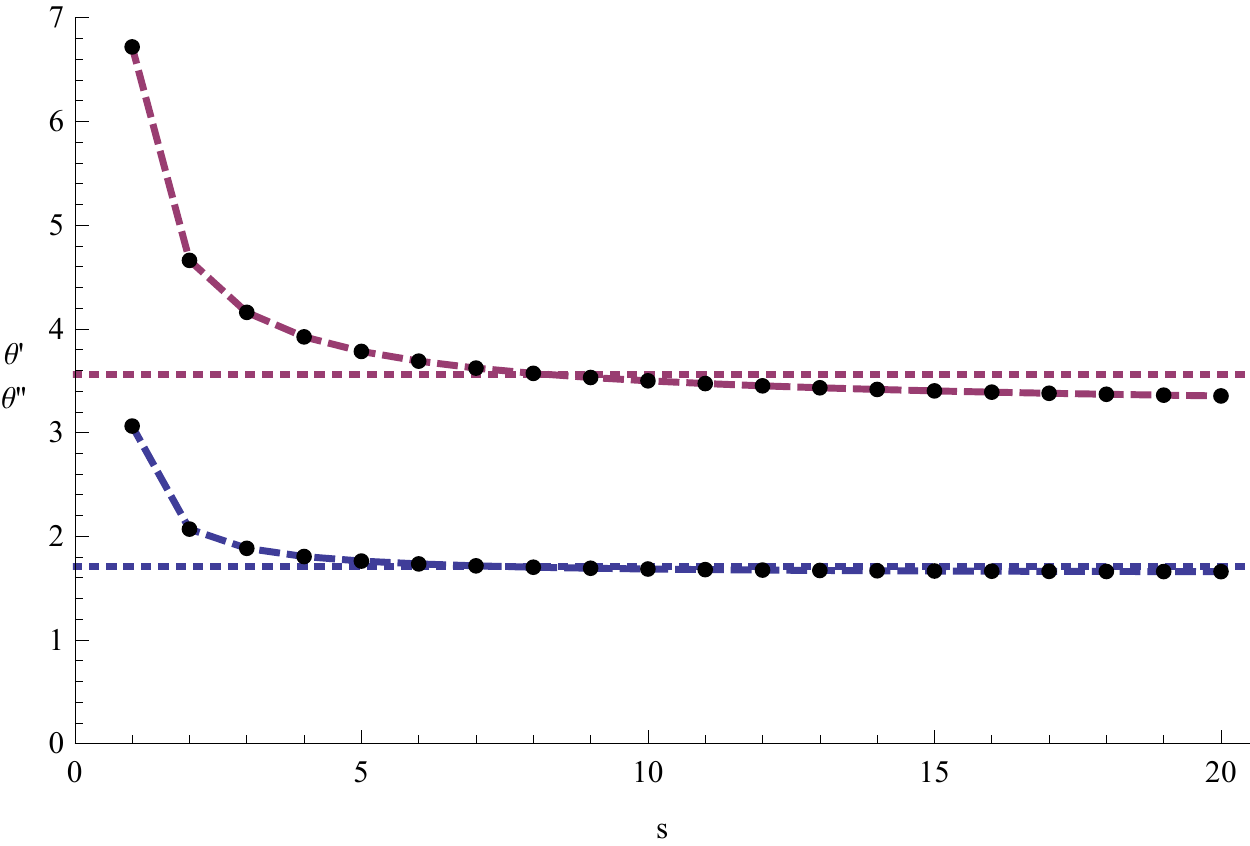}
		\label{fig:thetam5}}
	\subfigure[$\theta'$ (blue) and $\theta''$ (red) for $µ^2 = 10$.]{
		\centering
		\includegraphics[width=0.45\textwidth]{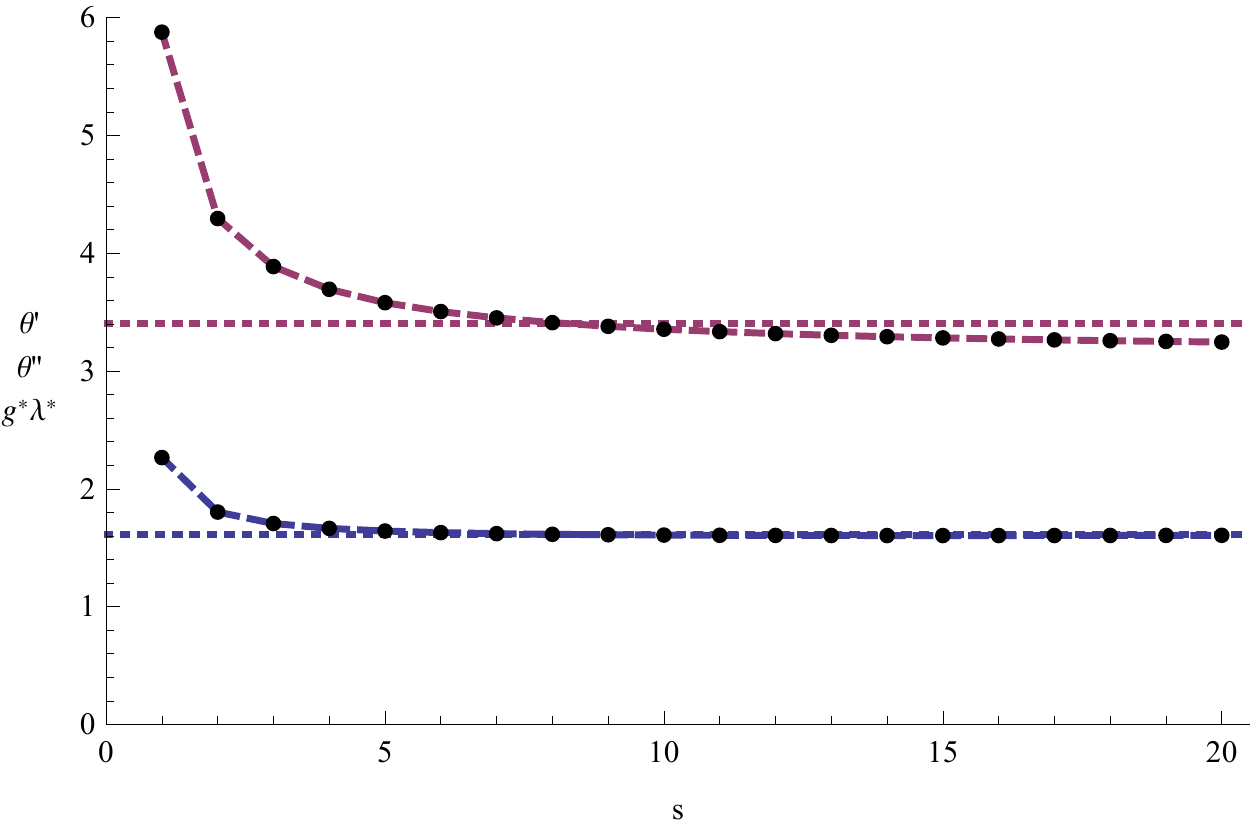}
		\label{fig:thetam10}}
	\caption{The plots show the $s$-dependence of the critical exponents for various fixed values of $µ^2$. At $µ = µ_{\mathrm{crit}}$ and $s = s_{\mathrm{crit}}$ the critical exponents $\theta = \theta' + i \theta''$ split into real and imaginary part (blue and red line, respectively). The case of two real real critical exponents $(µ < µ_{\mathrm{crit}}$ and $s < s_{\mathrm{crit}})$ is pictured for $µ^2 = \frac{1}{1.9}$ and $s=1$(green dots). The horizontal lines represent the values obtained for the optimized shape function.}
	\label{fig:critexpshape}
\end{figure}

\begin{figure}[H]
	\centering
	\subfigure[$\lambda^{*}g^{*}$ for $µ^2 = \frac{1}{1.9}$.]{
		\centering
		\includegraphics[width=0.45\textwidth]{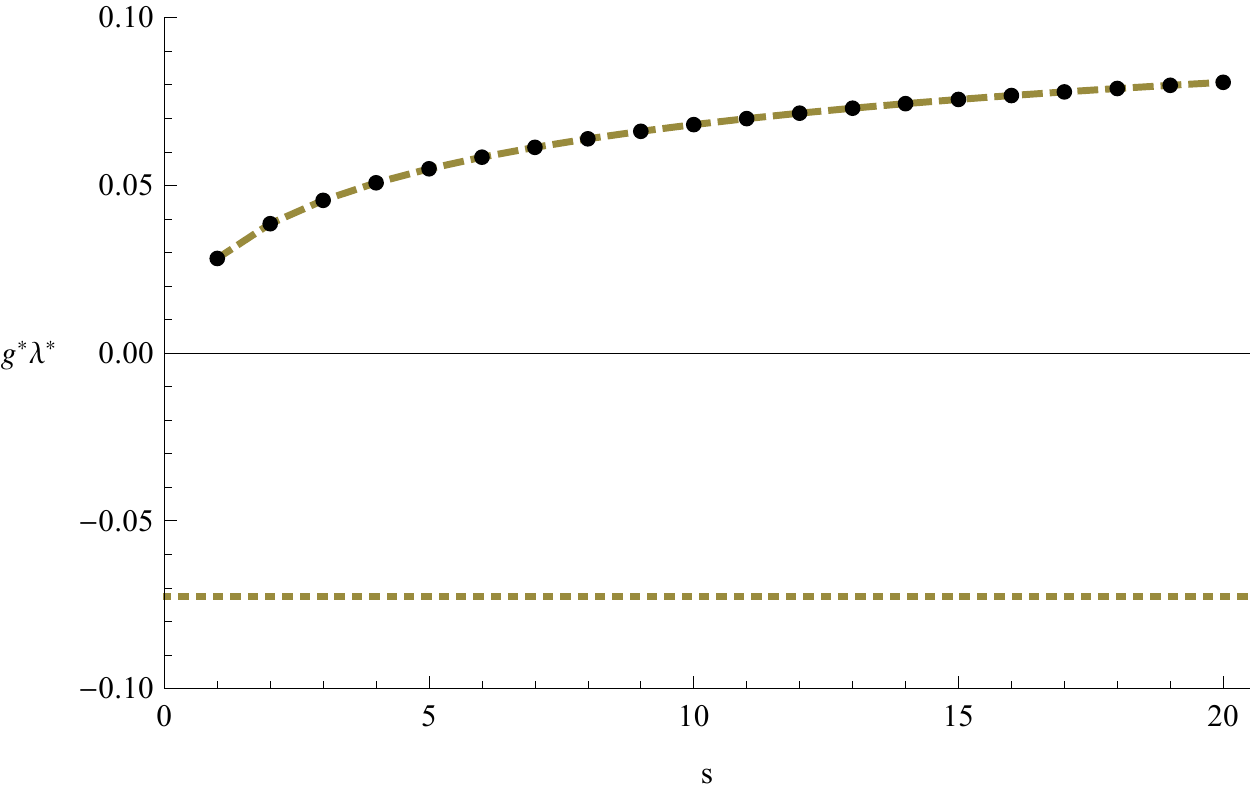}
		\label{fig:lgm119exp}}
	\subfigure[$\lambda^{*}g^{*}$ for $µ^2 = 1$.]{
		\centering
		\includegraphics[width=0.45\textwidth]{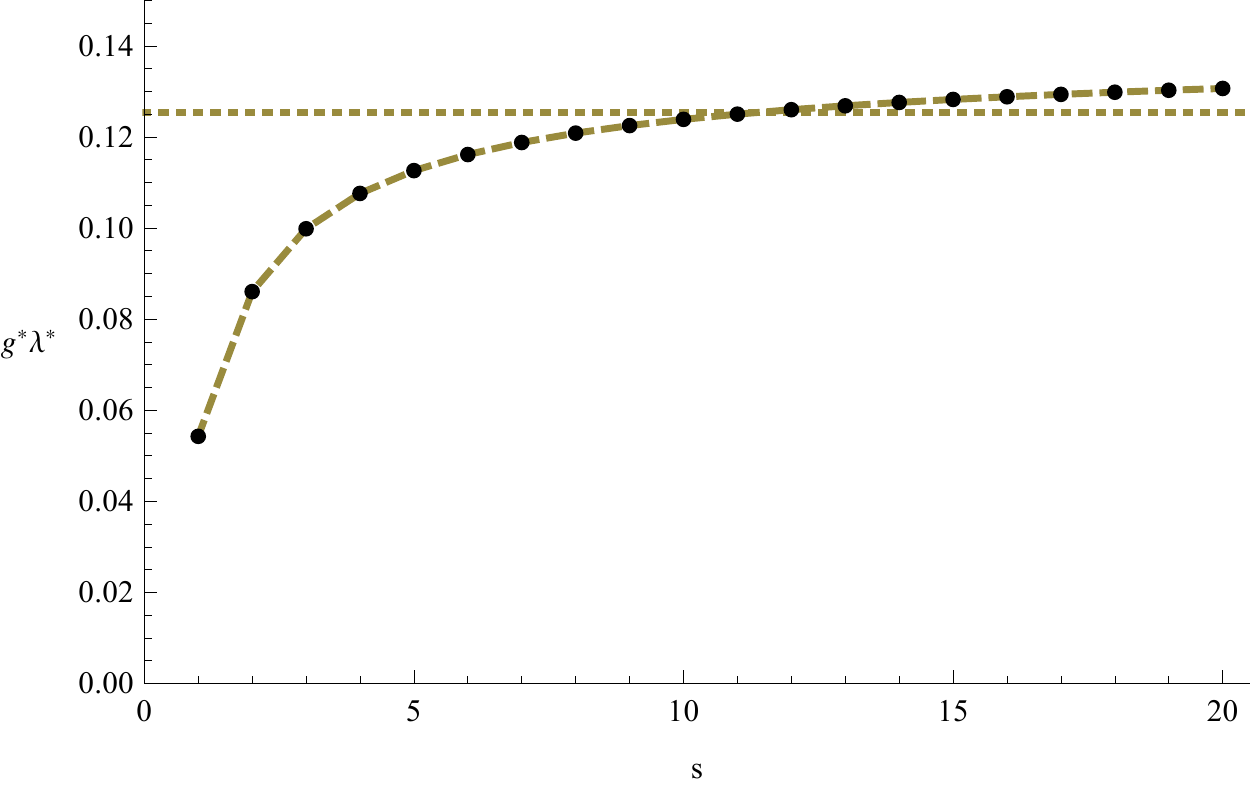}
		\label{fig:lgm1exp}}
	\subfigure[$\lambda^{*}g^{*}$ for $µ^2 = 2$.]{
		\centering
		\includegraphics[width=0.45\textwidth]{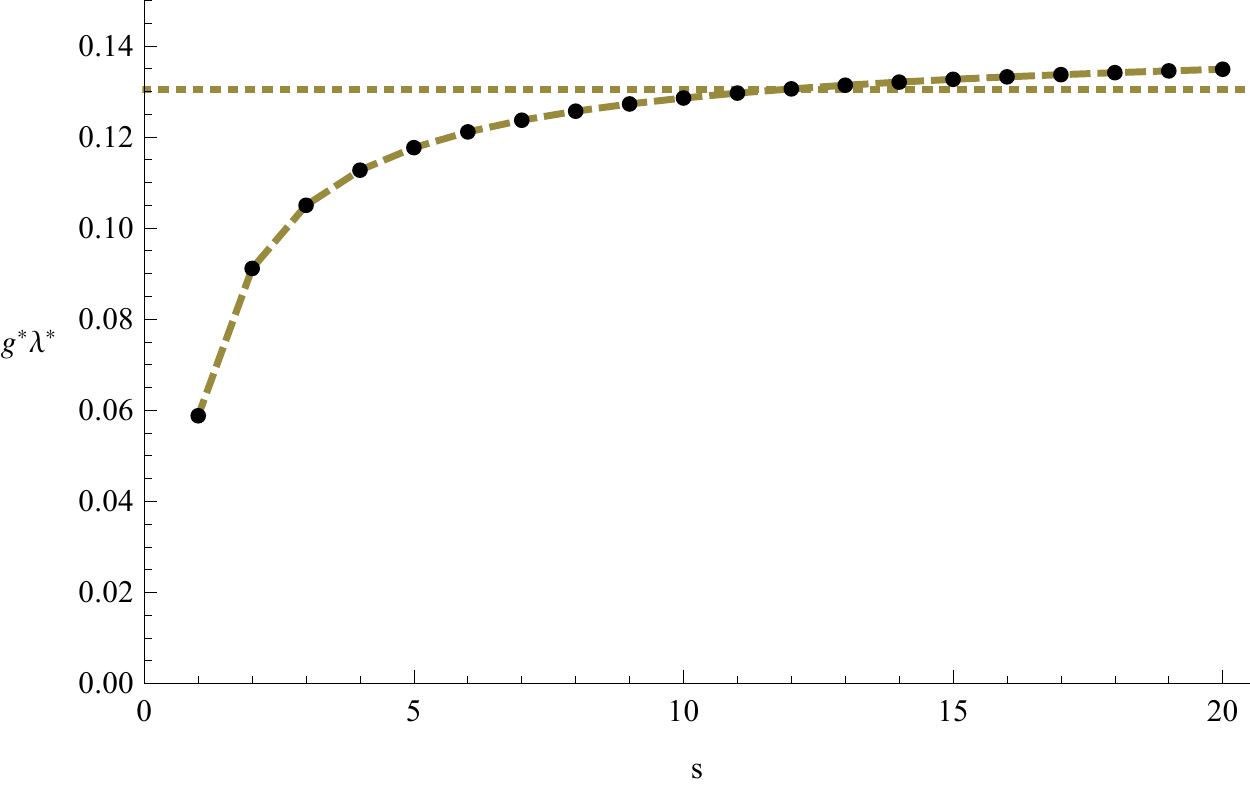}
		\label{fig:lgm2exp}}
	\subfigure[$\lambda^{*}g^{*}$ for $µ^2 = 5$.]{
		\centering
		\includegraphics[width=0.45\textwidth]{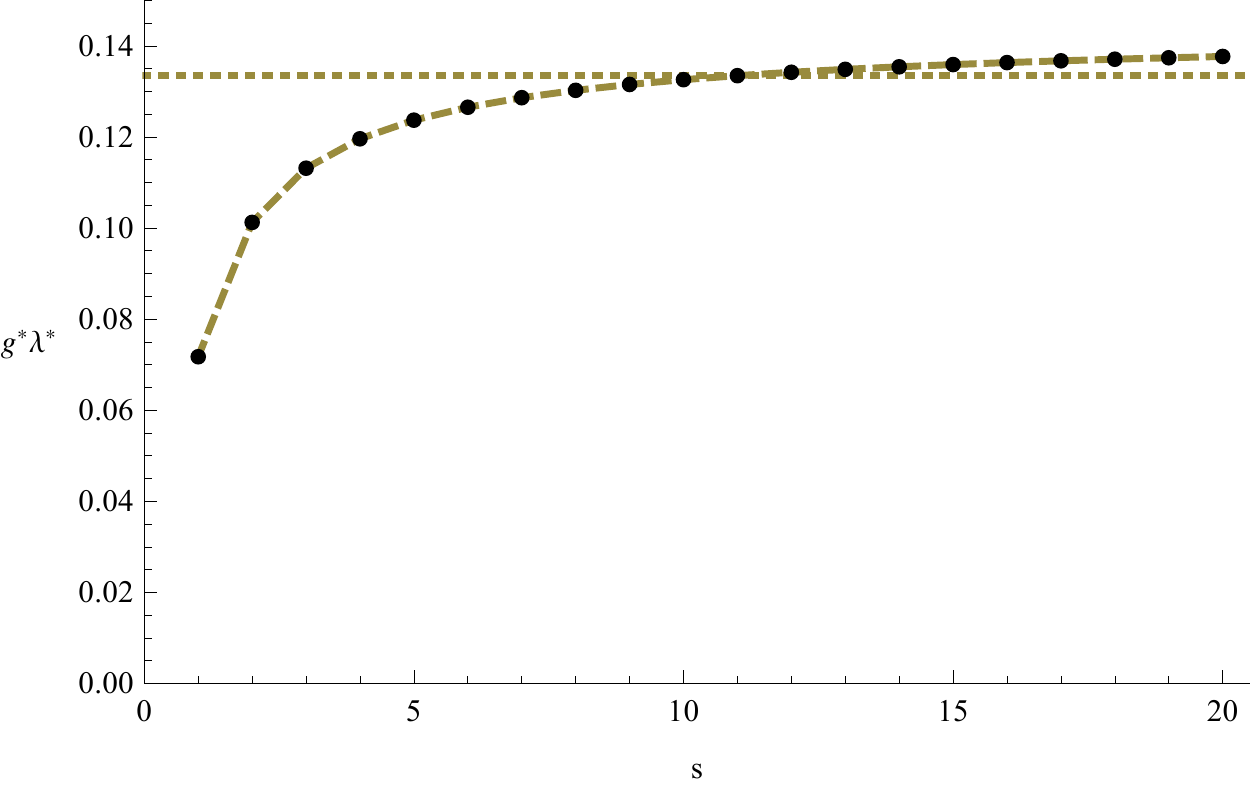}
		\label{fig:lgm5exp}}
	\subfigure[$\lambda^{*}g^{*}$ for $µ^2 = 10.$]{
		\centering
		\includegraphics[width=0.45\textwidth]{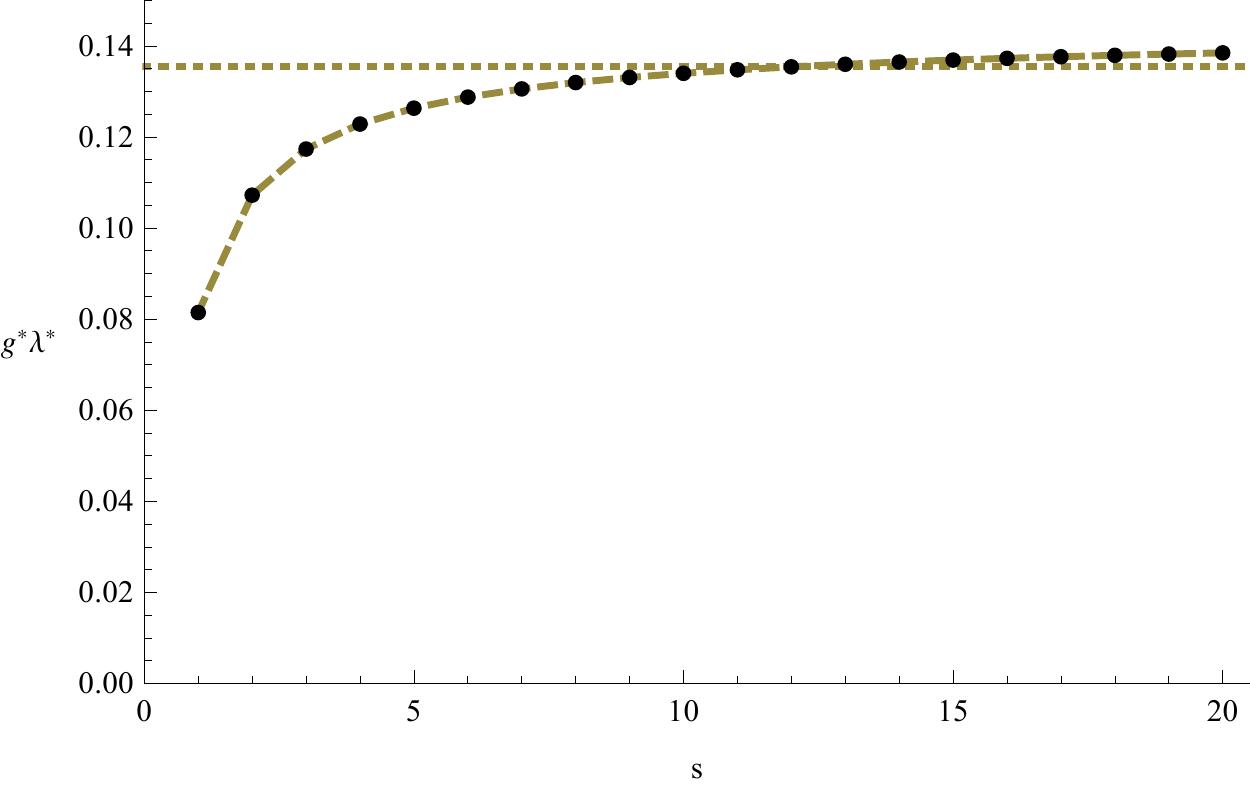}
		\label{fig:lgm10exp}}
	\caption{The coordinate product $\lambda^{*}g^{*}$  of the NGFP as a function of the shape parameter $s$, for selected values of $µ^2$. The horizontal lines represent the values obtained with the optimized shape function.}
	\label{fig:fixshape}
\end{figure}

We present two sets of diagrams showcasing our results. Figures \ref{fig:critexpshape} and \ref{fig:fixshape} display the $s$-dependence of the universal quantities for selected choices of $µ$. The set of Figures \ref{fig:NGFPm119} - \ref{fig:NGFPm10} depicts the corresponding phase portraits, ordered from small to large values of the shape parameter $s$.\\
\\
\textbf{(A)} Within a sufficiently good approximation to the exact flow, we expect $\lambda^{*}g^{*}$ and the critical exponents to be approximately universal quantities and to show only little $s$-dependence, as well as no qualitative changes of the phase portrait for different shape parameters. However, we observe that there is a pronounced dependence on $s$ for every given value of the mass parameter. While the existence of the \textbf{NGFP} still seems universal, not only its (non-universal) location but also its characteristic quantities change notably. 

\noindent \textbf{(i)} For $µ \gtrsim 0.9$ we again find the \textbf{NGFP} in the first quadrant with the critical exponents forming a complex conjugated pair. Compared to the previous results with $R^{(0)}_{\mathrm{opt}}$ the values of the real and imaginary part, $\theta'$ and $\theta''$ respectively, are notably larger for small $s$. With increasing shape parameter the critical exponents become smaller and converge to values very close to the ones found for $R^{(0)}_{\mathrm{opt}}$. 

The product $\lambda^{*}g^{*}$ exhibits the opposite behavior. It starts small and increases as a function of the shape parameter, eventually exceeding the previously obtained $R^{(0)}_{\mathrm{opt}}$-value. While the magnitude of the product $\lambda^{*}g^{*}$ still remains in a very narrow band, a closer look at the phase portraits reveals that the coordinates of the fixed point change strongly. The fixed point wanders ``up and to the left'', meaning smaller $\lambda$- and increasingly larger $g$-values. As a function of $s$, the coordinates $g^{*}$ and $\lambda^{*}$ change in opposing directions, but their product stays relatively similar. The entire region before running into the singularity gets ``squished'' with increasing shape parameter $s$. We additionally observe a rising numerical instability of the flow, especially notable close to the $\eta_N$-singularity.

\noindent \textbf{(ii)} For $µ \lesssim 0.9$ the picture changes even more severely. The optimized cutoff led to the \textbf{NGFP} in the second quadrant, and two real critical exponents. Now for every value of $s$, this fixed point switches to positive $\lambda^{*}$, i.e. it lies in the \textit{first} quadrant. While we retain the real critical exponents for $s \approx 1$, any larger value of the shape parameter leads to a complex pair. We again find a strong $s$-dependence of the fixed point's location; it increases with increasing shape parameter. \\
\\
\textbf{(B)} Utilizing the generalized exponential cutoff $R^{(0)}_{\mathrm{exp}}$, the resulting sequence of RG flows loses its resemblance with the cases of QECG and Tetrad Gravity. While it still mostly agrees with the results of QEG, it is less stable.\\
\\
\textbf{(C)} A comment on the case of shape parameter $s < 1$ is in order here. Technically $s<1$ is a valid choice. We find, however, that the calculations become exceedingly unsightly due to the involved numerics. In the presented case of $s = \frac{1}{2}$ we can still find a fixed point, but it lies extremely close to the $\eta_N$-singularity. This results in a highly unstable flow with no complete trajectories. One should not take this as a sign of the non existence of the \textbf{NGFP} though. The same phenomenon is well known to occur also in the Einstein-Hilbert \cite{Reuter:2001ag} and $R^2$-truncations \cite{Lauscher:2002sq} of QEG. \\
\\
\textbf{(D)} Taken together these observations show that the results are severely cutoff scheme dependent, making it difficult to draw any general conclusion besides the very existence of the \textbf{NGFP}. We must emphasize, however, \textit{the existence of the non-Gaussian fixed point is already a highly non-trivial result.} Nevertheless, the comparatively strong scheme dependence calls for an explanation, and this is the topic of the next section.

\subsection[The torsion invariants \texorpdfstring{$S^2$}{S2}, \texorpdfstring{$T^2$}{T2}, and \texorpdfstring{$q^2$}{q2} considered separately]{The torsion invariants \texorpdfstring{$\bm{S^2}$}{S2}, \texorpdfstring{$\bm{T^2}$}{T2}, and \texorpdfstring{$\bm{q^2}$}{q2} considered separately}\label{subsec:inditorinv}

As we discuss next, it is quite illuminating to understand how the individual torsion fields, $S_µ$, $T_µ$ and $\tensor{q}{^{\lambda}_{µ \nu}}$, each influence the RG flow. We analyze the three cases where only one of these fields is retained in the truncation at a time. The necessary $\beta$-functions are easily obtained by discarding the terms in (\ref{eq:B12}) and (\ref{eq:B34}) stemming from the two unwanted traces $\Delta \mathcal{B}_{SS}^{\mathrm{Tor}}$, or $\Delta \mathcal{B}_{TT}^{\mathrm{Tor}}$, or $\Delta \mathcal{B}_{qq}^{\mathrm{Tor}}$, respectively. The $\beta$-functions (\ref{eq:set}) thus retain their shape and only the $\check{B}$-functions get modified each time. We use both $R^{(0)}_{\mathrm{opt}}$ and the generalized exponential cutoff $R^{(0)}_{\mathrm{exp}}$ with the shape parameter $s$ ranging from $1$ to $20$.

We stress that we are not interested in a full analysis of a different theory space here that would depend on only one of the irreducible torsion tensors. Instead we seek a better comprehension how these fields separately impact the features of the entire RG flow of $\mathcal{T}_{\mathrm{dtor}}$ which we obtained in the previous sections, an obvious question being, for instance, which field(s) is(are) responsible for the migration of the fixed point into the second quadrant.

\subsubsection[The \texorpdfstring{$S^2$}{S2}-invariant]{The \texorpdfstring{$\bm{S^2}$}{S2}-invariant}\label{subsubsec:S2invariant}
First let us examine the influence of the axial-vector $S_µ$ corresponding to the totally antisymmetric part of the torsion tensor $\tensor{T}{^{\lambda}_{µ \nu}}$. 

Taking only $S_µ$ into account amounts to making the following ansatz for the action
\begin{equation}
	\Gamma_{k}^{S} \left[ g,S \right] = \frac{1}{16 \pi G_k} \int \mathrm{d}^{4}x\, \sqrt{g} \left[ - R_{\LC} + 2 \bar{\lambda}_k + \frac{1}{24} f_{k}^{S} \Bigl( g^{µ \nu} S_µ S_{\nu} \Bigr) \right] \ .
\end{equation}
It leads to the following modified $\check{B}$-functions:
\begin{subequations}
\begin{align}
	&\check{B}_{1}(µ^2) = \frac{2}{3 \pi} \check{\Phi}_{1}^{1}(\tfrac{1}{6} µ^2) , &{}&\check{B}_{2}(µ^2) = - \frac{1}{3 \pi} \check{\tilde{\Phi}}_{1}^{1}(\tfrac{1}{6} µ^2) , \\
	&\check{B}_{3}(µ^2) = 4 \check{\Phi}_{2}^{1}(\tfrac{1}{6} µ^2) , &{}&\check{B}_{4}(µ^2) = - 2 \check{\tilde{\Phi}}_{2}^{1}(\tfrac{1}{6} µ^2) .
\end{align}
\end{subequations} 
It is remarkable that now the argument of the threshold functions, due to always being $\frac{1}{6} µ^2$, is \textit{positive} throughout. This entails that we do not have to restrict our analysis to the range $µ^2 \geq \frac{1}{2}$, as in the general case.\\
\\
\textbf{(A)} In Figure \ref{fig:NGFPSS} we plot the resulting phase portrait for the optimized cutoff and various increasing values of $µ^2$. An UV attractive non-Gaussian fixed point is found in the first quadrant for every studied value of the mass parameter. The critical exponents form a complex pair, leading to spiraling RG trajectories. The coordinates of the non-Gaussian fixed point are visibly dependent on the mass parameter. The wandering of the fixed point's location for increasing $µ$ falls in between the ``converging'' case for $R^{(0)}_{\mathrm{opt}}$ in Section \ref{subsec:anal} and the ``diverging'' case for $R^{(0)}_{\mathrm{exp}}$ in Section \ref{subsec:shape}. We thus conclude that this fixed point of the ``$S^2$-only'' calculation corresponds to the \textbf{NGFP} of the truncation incorporating all three torsion fields.\\
\\
\textbf{(B)} For exceptionally small $µ < \frac{1}{2}$, however, two additional fixed points emerge, which we will denote by \textbf{NGFP}$\bm{^{\oplus}}$ and \textbf{NGFP}$\bm{^{\ominus}}$ in the following. The first fixed point, \textbf{NGFP}$\bm{^{\oplus}}$, is also located in the first quadrant, though at considerably larger $g^{*}$ and $\lambda^{*}$ values than the \textbf{NGFP}. The coordinates of \textbf{NGFP}$\bm{^{\oplus}}$ diverge rapidly for increasing mass parameter. As a consequence it is always screened from the \textbf{GFP} by the $\eta_N$-singularity and no connecting RG trajectory exists, as seen in Figure \ref{fig:PDSSm110v3}. Therefore \textbf{NGFP}$\bm{^{\oplus}}$ is not a suitable candidate for the construction of an asymptotically safe quantum field theory, and we discard it in the following.\\
\\
\textbf{(C)} The second new fixed point, \textbf{NGFP}$\bm{^{\ominus}}$, is located in the \textit{third} quadrant, meaning both $g^{*}$ and $\lambda^{*}$ are negative. It is UV attractive with a complex conjugated pair of critical exponents. Furthermore, \textbf{NGFP}$\bm{^{\ominus}}$ is indeed connected to the \textbf{GFP}: we find a second ``separatrix'', hitting the \textbf{GFP} at $k \rightarrow 0$, and separating trajectories with positive or negative IR values for the cosmological constant $\lambda$. 

Thus the fixed point \textit{\textbf{NGFP}$\bm{^{\ominus}}$ seems suitable for the Asymptotic Safety construction.} This is an important new feature of the ``$S^2$-only'' flow that was not observed in the full $S,T,q$-system. Instead of only one non-Gaussian fixed point, \textbf{NGFP}, we are now presented with a second possibility for the Asymptotic safety construction, namely \textbf{NGFP}$\bm{^{\ominus}}$. 

Taking a closer look at the phase portrait in Figure \ref{fig:PDSSm110v2}, we observe that it strongly resembles the RG flow of QECG in the $(\lambda,g)$-plane as found in \cite{Harst:2012phd,Harst:2014vca}. However, the fixed point \textbf{NGFP}$\bm{^{\ominus}}$ shows the same behavior as \textbf{NGFP}$\bm{^{\oplus}}$ and its location starts to run to infinity with an increase of the mass parameter. For $µ^2 > \frac{1}{10}$ the fixed point is already located ``behind'' the singularity, depicted in Figure \ref{fig:PDSSm310v2} for $µ^2 = \frac{3}{10}$. Hence the fixed point \textit{\textbf{NGFP}$\bm{^{\ominus}}$ is only accessible in a very narrow range of the mass parameter.}\\
\\
\textbf{(D)} Repeating the analysis with the generalized exponential cutoff leads to very similar results. We recover both fixed points, \textbf{NGFP} and \textbf{NGFP}$\bm{^{\ominus}}$, for every considered value of the shape parameter $s$. While the ``new'' fixed point \textbf{NGFP}$\bm{^{\ominus}}$ is now slightly more stable, and hence accessible for larger values of the mass parameter up to $µ^2 \approx \frac{7}{10}$, the fixed point \textbf{NGFP} is located notably closer to the singularity. Furthermore we again observe the strong wandering of the \textbf{NGFP} with increasing shape parameter $s$ due to the ``squishing'' of the region before the singularity similarly found in Section \ref{subsec:shape} for the flow incorporating all torsion fields. As in the full $S,T,q$-system, the RG flow is markedly more unstable when employing the generalized exponential cutoff $R^{(0)}_{\mathrm{exp}}$, though the cutoff scheme dependence is significantly weaker. 

To illustrate the results we have plotted the phase portrait for various values of $µ^2$ and the shape parameter $s=2$ in Figure \ref{fig:NGFPSSshape}.

\subsubsection[The \texorpdfstring{$T^2$}{T2}-invariant]{The \texorpdfstring{$\bm{T^2}$}{T2}-invariant}\label{subsubsec:T2invariant}

Taking only $T_µ$ into account amounts to the ansatz
\begin{equation}
	\Gamma_{k}^{T} \left[ g,T \right] = \frac{1}{16 \pi G_k} \int \mathrm{d}^{4}x\, \sqrt{g} \left[ - R_{\LC} + 2 \bar{\lambda}_k + \frac{2}{3} f_{k}^{T} \Bigl( g^{µ \nu} T_µ T_{\nu} \Bigr) \right] \ .
\end{equation}
It leads to the $\check{B}$-functions
\begin{subequations}
\begin{align}
	&\check{B}_{1}(µ^2) = \frac{2}{3 \pi} \check{\Phi}_{1}^{1}(\tfrac{8}{3} µ^2) , &{}&\check{B}_{2}(µ^2) = - \frac{1}{3 \pi} \check{\tilde{\Phi}}_{1}^{1}(\tfrac{8}{3} µ^2) , \\
	&\check{B}_{3}(µ^2) = 4 \check{\Phi}_{2}^{1}(\tfrac{8}{3} µ^2) , &{}&\check{B}_{4}(µ^2) = - 2 \check{\tilde{\Phi}}_{2}^{1}(\tfrac{8}{3} µ^2) .
\end{align}
\end{subequations}
Due to the positive argument $\frac{8}{3} µ^2$ in all threshold functions, there are no restrictions on the mass parameter, as was also the case for $S_µ$ alone. 

The $T^2$-invariant displays the same general behavior as the $S^2$-invariant, including an identical fixed point structure. As a consequence of the larger factor in the argument of the threshold functions, $\frac{8}{3} µ^2$ instead of $\frac{1}{6} µ^2$, the region for which a complete trajectory, connecting the fixed point \textbf{NGFP}$\bm{^{\ominus}}$ and the \textbf{GFP}, exists is even narrower. The larger argument also stabilizes the entire flow, which is especially notable when utilizing the exponential cutoff as seen in Figure \ref{fig:NGFPTTshape}.

We will not discuss the $T^2$-invariant in further detail, as its flow has essentially the same properties as $S^2$ considered in isolation.

\subsubsection[The \texorpdfstring{$q^2$}{q2}-invariant]{The \texorpdfstring{$\bm{q^2}$}{q2}-invariant}
Last, let us analyze the role played by the tensor $\tensor{q}{^{\lambda}_{µ \nu}}$, making the ansatz
\begin{equation}
	\Gamma_{k}^{q} \left[ g,q \right] = \frac{1}{16 \pi G_k} \int \mathrm{d}^{4}x\, \sqrt{g} \left[ - R_{\LC} + 2 \bar{\lambda}_k - \frac{1}{2} f_{k}^{q} \Bigl( g^{µ \nu} g^{\rho \sigma} g^{\alpha \beta} q_{µ \rho \alpha} q_{\nu \sigma \beta} \Bigr) \right] \ ,
\end{equation}
resulting in the following $\check{B}$-functions:
\begin{subequations}
\begin{align}
	&\check{B}_{1}(µ^2) = \frac{8}{3 \pi} \check{\Phi}_{1}^{1}(- 2 µ^2) , &{}&\check{B}_{2}(µ^2) = - \frac{4}{3 \pi} \check{\tilde{\Phi}}_{1}^{1}(-2 µ^2) , \\
	&\check{B}_{3}(µ^2) = 16 \check{\Phi}_{2}^{1}(-2 µ^2) , &{}&\check{B}_{4}(µ^2) = - 8 \check{\tilde{\Phi}}_{2}^{1}(-2 µ^2) .
\end{align}
\end{subequations}
Here we rediscover threshold functions $\check{\Phi}$ that actually constrain the range of the mass parameter to $µ^2 > \frac{1}{2}$, due to the minus sign in their argument $-2 µ^2$. As an immediate consequence, the additional fixed point \textbf{NGFP}$\bm{^{\ominus}}$, found in the ``$S^2$-only'' and ``$T^2$-only'' cases, is not present for this invariant. 

We again present two sets of phase portraits for increasing values of $µ^2$. Figure \ref{fig:NGFPqq} depicts the result for the optimized cutoff and Figure \ref{fig:NGFPqqshape} for the exponential one with shape parameter $s=2$. \\
\\
\textbf{(A)} As it turns out, the $q^2$-invariant has the most profound impact on the RG flow. We find an UV attractive non-Gaussian fixed point for any value of the mass parameter. For $µ \lesssim 1.5$ this fixed point lies in the second quadrant $(g^{*} >0 , \lambda^{*} < 0)$ when employing the optimized cutoff, and is located in the first quadrant when $µ \gtrsim 1.5$. The $q^2$-block, therefore, is the source for the migration of the non-Gaussian fixed point \textbf{NGFP} into the second quadrant that we had found in Section \ref{subsec:anal} for the full $S,T,q$-system. 

This behavior can be traced back to the negative sign of $q_{µ \nu \rho} q^{µ \nu \rho}$ relative to $S_µ S^µ$ and $T_µ T^µ$ in the decomposition of the action, see (\ref{eq:holsttruncation}). \\
\\
\textbf{(B)} Furthermore we observe that for $µ$-values slightly larger than the transition point $µ \approx 1.5$ the UV repulsive direction of the \textbf{GFP} still points into the negative $\lambda$-halfplane, even though the \textbf{NGFP} is located in the first quadrant. A similar effect occurs in the case of Tetrad Gravity \cite{Harst:2012ni,Harst:2012phd}; we already mentioned this fact in Section \ref{subsec:anal} as one of the properties which is different in the $\mathcal{T}_{\mathrm{dtor}}$ flow incorporating all three torsion fields on one side, and Tetrad Gravity on the other. \\
\\
\textbf{(C)} In Section \ref{subsec:shape} we extensively discussed the discrepancies between the optimized and the exponential cutoff. For the latter, the \textbf{NGFP} was located in the first quadrant for any value of the mass parameter. Interestingly enough, the $q^2$-invariant alone displays a substantially different behavior. Replacing $R^{(0)}_{\mathrm{opt}}$ by $R^{(0)}_{\mathrm{exp}}$ does not change the fixed point structure, and for either cutoff the \textbf{NGFP} is located in the second quadrant for $µ \lesssim 1.8$. Increasing $µ$ above the transition point $µ \approx 1.8$ the IR attractive eigen-direction of the \textbf{GFP} retains its orientation for slightly larger values of the mass parameter, again forcing the ``separatrix'' to curve around the $g$-axis. This feature is more pronounced for $R^{(0)}_{\mathrm{exp}}$ then $R^{(0)}_{\mathrm{opt}}$. Furthermore, the fixed point's location once again reveals a diverging nature similar to the one observed in Section \ref{subsec:shape}. It never crosses the $g$-axis a second time, however, and remains in the respective quadrant for all values of the shape parameter.

\subsubsection{Discussion}
The goal of this section was to gain a deeper insight into how the peculiar features of the RG flow on $\mathcal{T}_{\mathrm{dtor}}$ come about. The results can be summarized as follows.\\
\\
\textbf{(A)} The $S_µ S^µ$-invariant in isolation generally features a second non-Gaussian fixed point for small values of the mass parameter. The same is true for the $T_µ T^µ$-invariant, but it restricts the accessible region of the $g-\lambda$-plane to a range of $µ$ that has to be excluded when incorporating the $q_{µ \nu \rho} q^{µ \nu \rho}$-block. This explains why the fixed point \textbf{NGFP}$\bm{^{\ominus}}$ does not occur in the full $S,T,q$-truncation. \\
\\
\textbf{(B)} In Section \ref{subsec:anal} we found the \textbf{NGFP} of the full system at $\lambda^{*} < 0$ for small mass parameters $µ \lesssim 1$. The analysis of the individual fields revealed that we can ascribe the migration of the \textbf{NGFP} into the second quadrant to the \textit{negative (``tachyonic'') sign in front of the $q^2$-invariant.} Retaining only the $q^2$-block leads to a \textit{negative} fixed point value $\lambda^{*} < 0$ even for the most natural choice of mass parameter, $µ = 1$. Only by incorporating $S^2$ and $T^2$ as well, we recover a situation similar to the purely metric theory, QEG. \\
\\
\textbf{(C)} Interestingly, \textit{the individual invariants do not exhibit the significant cutoff dependence of the full $S,T,q$ flow.} Despite some (admittedly large) quantitative deviations, we observe no cutoff dependence at the qualitative level. Most importantly, the fixed point structure implied by the individual invariants is the same for the optimized and the generalized exponential cutoff. The flow incorporating all three torsion fields, however, looses the overturning of the \textbf{NGFP}. This suggests that \textit{the interplay between the three invariants holds the responsibility for the instability of the full flow,} in particular the diverging $g^{*}$ and $\lambda^{*}$ coordinates with increasing $s$.

\subsection[Identifying \texorpdfstring{$\bar{µ}$}{mu} with the running Planck mass]{Identifying \texorpdfstring{$\bm{\bar{µ}}$}{mu} with the running Planck mass}\label{subsec:planck}

The running of the dimensionfull Newton constant $G_k$ leads to a running Planck mass, $M_{\mathrm{Pl}}(k)$, which in natural units is given by 
\begin{equation}
	M_{\mathrm{Pl}}(k) \equiv \frac{1}{\sqrt{G_k}} = \frac{k}{\sqrt{g_k}} \ .
\end{equation}
In \cite{Daum:2010phd} it was proposed to identify $\bar{µ}$ with this mass scale. This amounts to the following choice for the dimensionless mass parameter
\begin{equation}
	µ \equiv µ(k) = \frac{1}{\sqrt{g_k}} \ .
\end{equation}
Inserting this identification into the $\beta$-functions (\ref{eq:set}) they read
\begin{subequations}
	\begin{equation}
		\beta_{g}\left(g,\lambda\right) = \Bigl[ 2 + \eta_{N} \Bigr] g \ ,
	\end{equation}
	\begin{equation}
		\beta_{\lambda}\left(g,\lambda\right) = - \Bigl( 2 - \eta_{N} \Bigr) \lambda + \frac{g}{2 \pi} \left[ \vphantom{\frac{1}{3}} \left( B_{3}(\lambda) + \check{B}_{3}(g^{-1}) \right) + \eta_{N} \left( B_{4}(\lambda) + \check{B}_{4}(g^{-1})  \right) \right] \ ,
	\end{equation}
	with the anomalous dimension
	\begin{equation}
		\eta_{N} = \frac{g \left( B_{1}(\lambda) + \check{B}_{1}(g^{-1}) \right)}{1 - g \left( B_{2}(\lambda) + \check{B}_{2}(g^{-1}) \right)} \ .
	\end{equation}
	\label{eq:betamg}
\end{subequations}
We are again confronted with the problem that the threshold functions $\check{\Phi}_{n}^{p}(w)$ are not well-defined for all values $µ \in \mathds{R}$. \\
\\
\textbf{(A)} Implementing the restrictions discussed in Section \ref{subsec:general} for the set of $\beta$-functions (\ref{eq:betamg}) and the optimized cutoff $R^{(0)}_{\mathrm{opt}}$, we observe that we have to exclude the regions $g \geq 2$ and $g \leq - \frac{1}{6}$ due to the factors $\ln\left(1- \frac{g}{2} \right)$ and $\ln\left(1+ 6 g \right)$, respectively\footnote{The factor $\ln\left(1+ \frac{3 g}{8} \right)$ introduces another restriction, $g \geq - \frac{8}{3}$, but this region has to be excluded due to the previously mentioned term already.}.
Despite these restrictions, we still find the Gaussian fixed point and an UV attractive non-Gaussian fixed point in the ``allowed'' region $- \frac{1}{6} < g < 2$. \\
\\
\textbf{(B)} Comparing the coordinates and universal quantities of this NGFP given in Tab. \ref{tab:propertiesmg} to those for $µ = \mathrm{const}$ summarized in Tab. \ref{tab:properties1}, we find they are \textit{very similar in the range $2 \lesssim µ^2 \lesssim 6$.}
\begin{table}[H]
	\centering
		\begin{tabular}{r|cc|ccc|l}
		$µ^2$ & $\lambda^{*}$ & $g^{*}$ & $\lambda^{*} g^{*}$ & $\theta'$ & $\theta''$ & phase portrait \\
		\hline \hline
		$\frac{1}{\sqrt{g}}$ & 0.226112 & 0.575602 & 0.130151 & 1.72306 & 3.52354 & Fig. \ref{fig:PDgtest3}
		\end{tabular}
	\caption{Properties of the NGFP for the identification $µ \equiv \frac{1}{\sqrt{g_k}}$.}
	\label{tab:propertiesmg}
\end{table} 
\noindent \textbf{(C)} The phase portrait for $µ \equiv \frac{1}{\sqrt{g_k}}$ is plotted in Figure \ref{fig:NGFPmg}. It closely resembles the flow of metric gravity, QEG. The NGFP is located in the first quadrant with the RG trajectories spiraling into it due to the imaginary part of the critical exponent. It is connected to the GFP via a ``separatrix''.\\
\\
\textbf{(D)} The identification of the mass parameter with the Planck mass was already considered before, namely in QECG. In \cite{Daum:2010phd} the identification was analysed in the $(\lambda,g)$-subsystem of the Holst truncation, where indeed a NGFP was recovered. However, those results differ considerably from ours. The NGFP found in \cite{Daum:2010phd} is located in the second quadrant; it has two real critical exponents, one positive and one negative. As this result also deviated from the case $µ = \mathrm{const.}$ considered in the same work, the authors concluded that in view of the universal quantities the case of constant mass parameters was more suitable for a comparison with QEG, as its scheme-dependence was weaker. 

Our findings are in much closer correspondence to the RG flow of QEG, making the identification a tempting choice, as it links the seemingly unphysical mass parameter to an actual physical observable. The universal existence of the NGFP when employing this identification definitely indicates that it is a viable choice and should be considered in any future studies involving a mass parameter of present kind.\\
\\
\textbf{(E)} Repeating the evaluation with $R^{(0)}_{\mathrm{exp}}$ we obtain similar results. The NGFP is again located in the ``allowed'' region of the first quadrant. However, the necessity to evaluate the involved integrals numerically, complicates a conclusive analysis, as we can not pin down the exact restrictions on the argument like we did before. In the previous sections we found no notable disparity of the constraints between the optimized and the generalized exponential cutoff. Rough numerical test for the current investigation agree with these results, such that we conclude to enforce the same restrictions, i.e. exclude the regions $g \geq 2$ and $g \leq - \frac{1}{6}$. 

We observe the - by now expected - strong $s$-dependence of the NGFP location with increasing shape parameter. It presents a bigger problem here than before with $µ = \mathrm{const}$. Choosing the shape parameter too large leads to a ``disappearance'' of the fixed point, as at around $s \gtrsim 19$ it wanders outside the well-defined region $g \in (- \frac{1}{6},2)$. Figure \ref{fig:PDmgs20} demonstrates this for $s=20$ and additionally shows the typical numerical problems that arise. One should not mistake the remnants above the $g=2$-line for actual RG trajectories or a true non-Gaussian fixed point, they rather correspond to a break down of the numerics and can not be taken at face value. Numerically solving the $\beta$-functions (\ref{eq:betamg}) for shape parameter $s=20$ yields a ``false'' NGFP at $(\lambda^{*},g^{*})=(0.0009,2.3819)$, which has to be discarded after taking a closer look at the entire phase portrait. 

Nonetheless, the existence of a suitable non-Gaussian fixed point for a considerable range of the shape parameter is a further encouragement and motivates the identification of $\bar{µ}$ with the running Planck mass in future studies. 

\subsection[Choosing a negative cutoff action for \texorpdfstring{$q^2$}{q2}]{Choosing a negative cutoff action for \texorpdfstring{$\bm{q^2}$}{q2}}\label{subsec:negacut}

In the previous sections we found that the most striking feature, the ``tipping'' of the \textbf{NGFP} from the second to the first quadrant, was on account of the differing sign in front of the $q_{\lambda µ \nu} q^{\lambda µ \nu}$ term. However, in Section \ref{sec:H0} we discussed the ambiguity in choosing the sign of the cutoff action for this field mode. Up to now all calculations and resulting findings were for an entirely positive cutoff action, i.e. $\xi = +1$ in (\ref{eq:zdefine}). In this section we change to the opposite sign of the $q$-cutoff and examine the flow, when one sets $\xi = -1$. 

As a result of a negative cutoff, the minus sign that previously appeared in the threshold functions $\check{\Phi}_{n}^{p}(w)$ and $\check{\tilde{\Phi}}_{n}^{p}(w)$ for the $\mathpzc{q}^2$-block disappears now. Therefore, the restriction on the range of the mass parameter does no longer apply and we can evaluate the $\beta$-functions (\ref{eq:set}) for all $µ \in \mathds{R}\setminus\{0\}$. 

For the full set of $µ^2$ dependent threshold functions appearing in the $\beta$-functions see paragraphs $\bm{(4)}$ and $\bm{(6)}$ in Appendix \ref{ch:threshold}. \\
\\
\textbf{(A)} We encountered the current situation before already, in Sections \ref{subsubsec:S2invariant} and \ref{subsubsec:T2invariant}, while analysing the $S^2$- and $T^2$-invariants separately. The unrestricted mass parameter $\bar{µ}$ led to the emergence of three non-Gaussian fixed points, denoted \textbf{NGFP}, \textbf{NGFP}$\bm{^{\oplus}}$ and \textbf{NGFP}$\bm{^{\ominus}}$, in addition to the standard \textbf{GFP}. Repeating the calculation for the full truncation with $\xi = -1$ reproduces this solution. Not only do we find all three NGFP's just mentioned, but also their respective behavior is identical to the one found in \ref{subsubsec:S2invariant} and \ref{subsubsec:T2invariant}, respectively. \\
\\
\textbf{(B)} We find the well known fixed point, \textbf{NGFP}, \textit{in the first quadrant for all values of the mass parameter.} It exhibits the same wandering of its location for increasing $µ$ observed in all previous calculations. The critical exponents always form a complex conjugated pair, leading to the characteristic, spiraling RG trajectories throughout. With increasing mass parameter the values of the critical exponents freeze out and converge to a fixed value.\\
\\
\textbf{(C)} The fixed point, \textbf{NGFP}$\bm{^{\oplus}}$, also located in the first quadrant, is always ``hidden'' behind the singularity caused by the anomalous dimension and its coordinates, $(\lambda^{*},g^{*})$, rapidly diverge with increasing mass parameter. As no RG trajectories connect it to the classical regime, it is not a suitable candidate for the Asymptotic Safety construction.\\
\\
\textbf{(D)} In contrast, the third NGFP, \textbf{NGFP}$\bm{^{\ominus}}$, located in the third quadrant, is considerably more stable. While it still displays a ``diverging'' nature, i.e. its $(\lambda^{*},g^{*})$-coordinates tend to infinity, for increasing mass parameter values, its range of accessibility is larger now. Most importantly it is accessible, and thus a viable choice for the Asymptotic Safety construction, for the ``natural'' choice of mass parameter $µ = 1$, and only lies ``behind'' the $\eta_N$-singularity for $µ \gtrsim 1.2$. 

It is UV attractive in both eigendirections with two real critical exponents. One of these quickly diverges, when the fixed point position approaches the singularity with increasing $µ$.  

As with \textbf{NGFP}$\bm{^{\oplus}}$, the physical relevance of this fixed point has to be called into question. Due to the structure of the $\beta$-functions (\ref{eq:system}) it is impossible for any RG trajectory to cross the $\lambda$-axis. This separation of theory space in two half planes is a well known feature of all known RG flows of quantum gravity \cite{Reuter:2001ag}. Consequently any trajectory, whose UV limit is defined at \textbf{NGFP}$\bm{^{\ominus}}$, cannot reach the classical regime, located in the \textit{first} quadrant, even if it connects to the \textbf{GFP}. As such \textbf{NGFP}$\bm{^{\ominus}}$ seems unlikely to be physically important. The fixed point might define a \textit{mathematically} sound field theory, but most probably not the \textit{physically} relevant one.\\
\\
\textbf{(E)} Repeating the analysis with the generalized exponential cutoff confirms these results. All three non-Gaussian fixed points are once again recovered with their respective, already discussed, features. 

Once more the (for the exponential cutoff) typical ``rising'' of the fixed point coordinates of \textbf{NGFP} with $µ$ to smaller $\lambda_{*}$- and increasingly higher $g_{*}$-values is observed. Its overall properties do not differ from the previously examined cases and therefore, again, is found suitable for an UV-completion of gravity, by the Asymptotic Safety construction.

We will not discuss the other two fixed points, \textbf{NGFP}$\bm{^{\oplus}}$ and \textbf{NGFP}$\bm{^{\ominus}}$, any further and only mention that they are also recovered when using the exponential cutoff.\\
\\
\textbf{(F)} Last let us return to the identification of the mass parameter with the running Planck mass, leading to $µ = \frac{1}{\sqrt{g_k}}$, this time with the negative cutoff action for the $q^2$-sector. The negative cutoff does not change the overall result and effectively only changes the restriction on Newton's coupling from $g < 2$ to $g > -2$, meaning that now the entire upper half-plane is accessible, resulting in the range $g \in (- \frac{1}{6},+\infty)$\footnote{The two previous constraints $g > - \frac{1}{6}$ and $g > - \frac{8}{3}$ obviously still apply, independent of the choice of $\xi$.}. Hence we again find an UV attractive non-Gaussian fixed point in the (changed) ``allowed'' region and all the comments made in the previous Section \ref{subsec:planck} still hold.\\
\\
\textbf{Discussion.} These results obviously raise the question, which type of cutoff action is to be preferred. Both choices yield a non-Gaussian fixed point suitable for the Asymptotic Safety construction, namely \textbf{NGFP}, but the properties of this fixed point are distinctly different in the two cases. 

\noindent \textbf{(i)} A completely positive cutoff action for the torsion fields, defined according to the \textbf{``$\bm{\mathcal{Z}_k = |\zeta_k|}$-rule''}, leads to the remarkable occurrence of a ``running'' fixed point, starting in the second quadrant for small $µ$, and then migrating over to the second quadrant with increasing mass parameter. This cutoff introduces a limitation on the range of said mass parameter, due to a singularity appearing in the threshold functions. Additionally, different shape functions $R^{(0)}$ can lead to qualitatively different flows, as the calculations with the exponential cutoff showed no sign of this wandering, instead showcasing the fixed point \textbf{NGFP} to be rooted firmly in the first quadrant.

\noindent \textbf{(ii)} Following the \textbf{``$\bm{\mathcal{Z}_k = \zeta_k}$-rule''} instead, i.e. choosing a negative cutoff action for $q_{\lambda µ \nu}$, avoids these difficulties and leads to a positive $\lambda_{*}$, as in the pure metric theory. Most of the fixed point properties are more stable and the discrepancy between different shape functions is resolved. This cutoff avoids the problem of ill-defined threshold functions for certain values of $µ$ and allows for a complete analysis in the entire range of the mass parameter $-\infty \leq µ \leq \infty$.

\noindent \textbf{(iii)} Comparing the two result, the latter choice, namely the ``$\mathcal{Z}_k = \zeta_k$-rule'', is clearly favored. While the possibility of a change in the sign of the cosmological constant is an intriguing phenomenon in its own right, it is harder to reconcile with other Asymptotic Safety studies and with cosmological observations. Therefore, the greater stability and conceptual uniformity of the second choice seem the more desirable option.

\begin{figure}[htbp]
	\centering
	\subfigure[$\theta'$ (blue solid), $\theta''$ (red dashed) and $\lambda^{*}g^{*}$ (yellow dot-dashed) as a function of $µ^2$.]{
		\centering
		\includegraphics[width=0.5\textwidth]{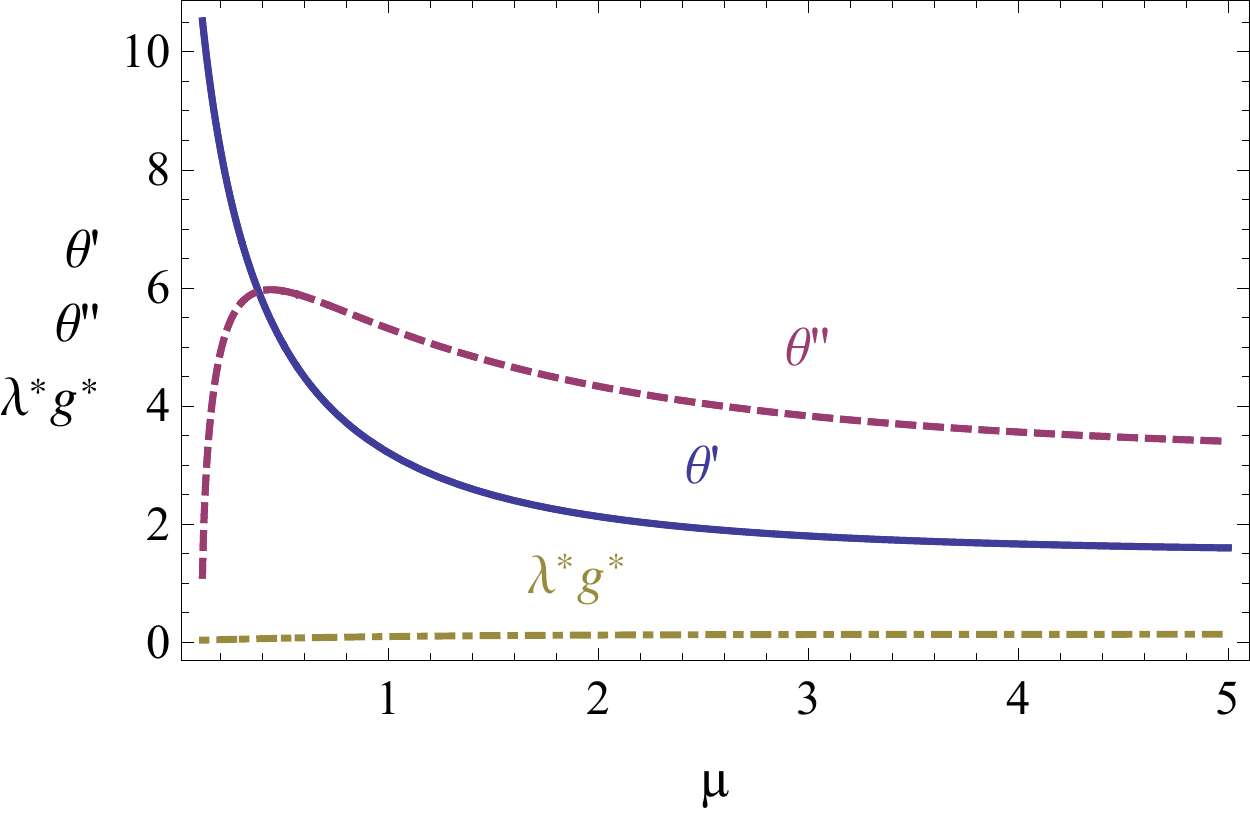}
		\label{fig:CriticalExponentsOptNegCut}}
	\subfigure[$\theta'$]{
		\centering
		\includegraphics[width=0.45\textwidth]{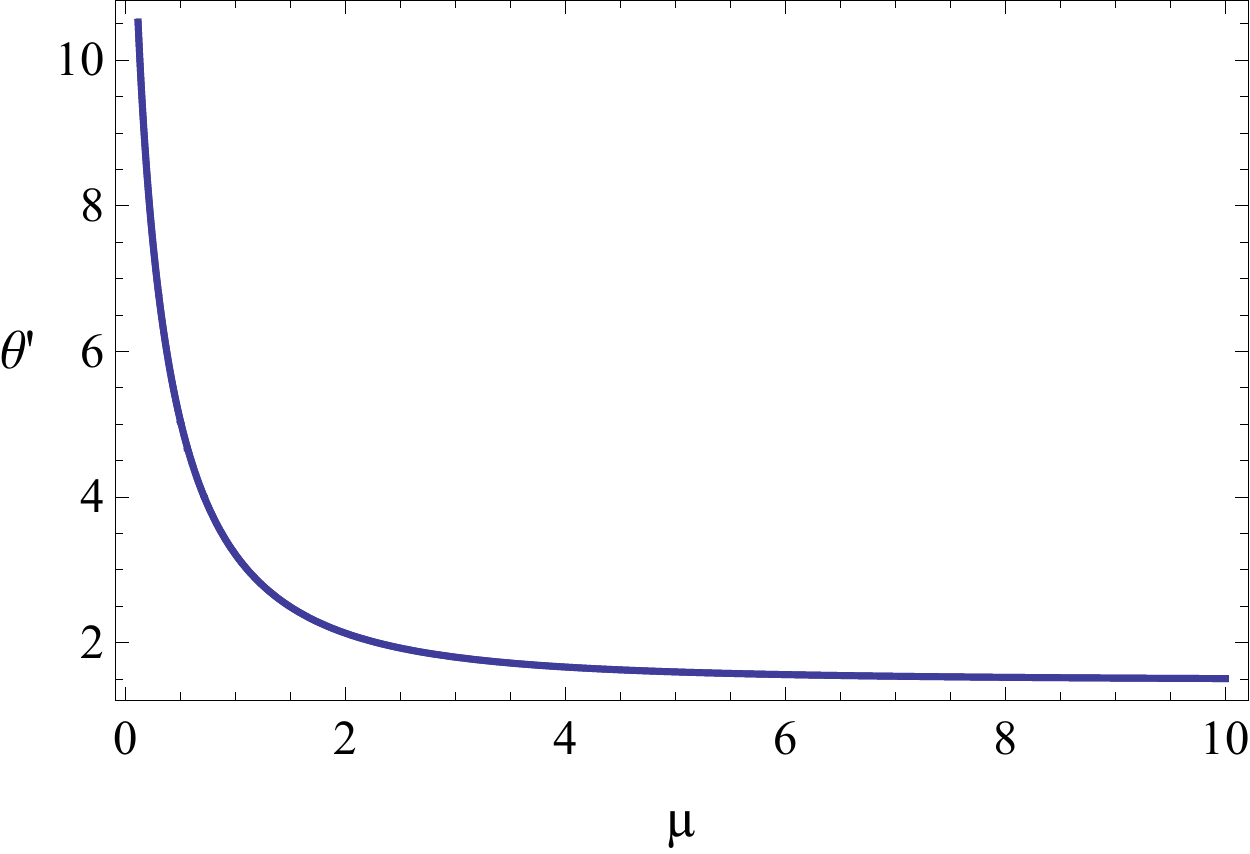}
		\label{fig:CriticalExponentsReOptNegCut}}
	\subfigure[$\theta''$]{
		\centering
		\includegraphics[width=0.45\textwidth]{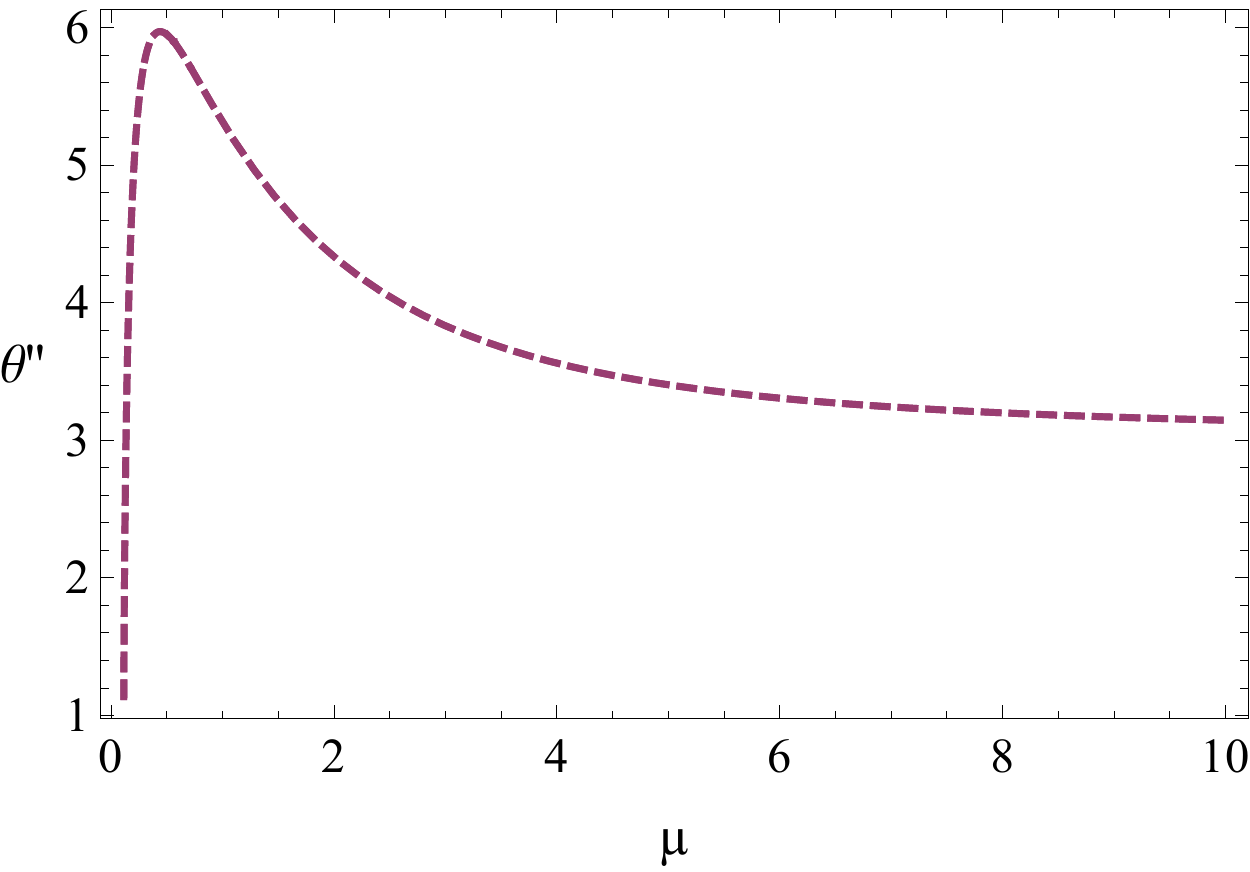}
		\label{fig:CriticalExponentsImOptNegCut}}
	\subfigure[$\lambda^{*}g^{*}$]{
		\centering
		\includegraphics[width=0.48\textwidth]{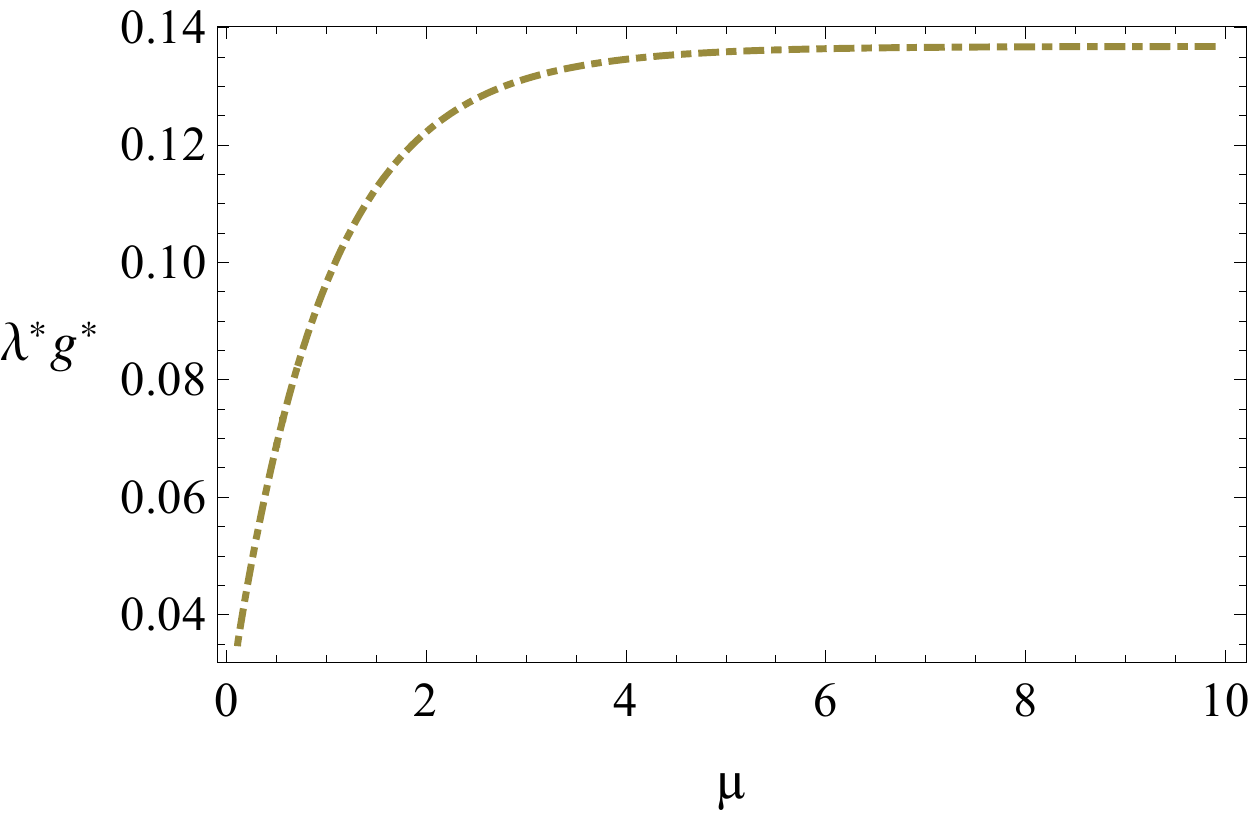}
		\label{fig:lgplotoptNegCut}}
	\caption{Critical exponents and $\lambda^{*}g^{*}$ of the \textbf{NGFP} as functions of $µ$. The critical exponents $\theta = \theta' + i \theta''$ (blue and red line, respectively), and the product $\lambda^{*}g^{*}$ (yellow line).}
	\label{fig:CritExpNegCut}
\end{figure}
\begin{figure}[htbp]
	\centering
	\subfigure[$\theta'$ (blue solid), $\theta''$ (red dashed) and $\lambda^{*}g^{*}$ (yellow dot-dashed) as a function of $µ^2$.]{
		\centering
		\includegraphics[width=0.5\textwidth]{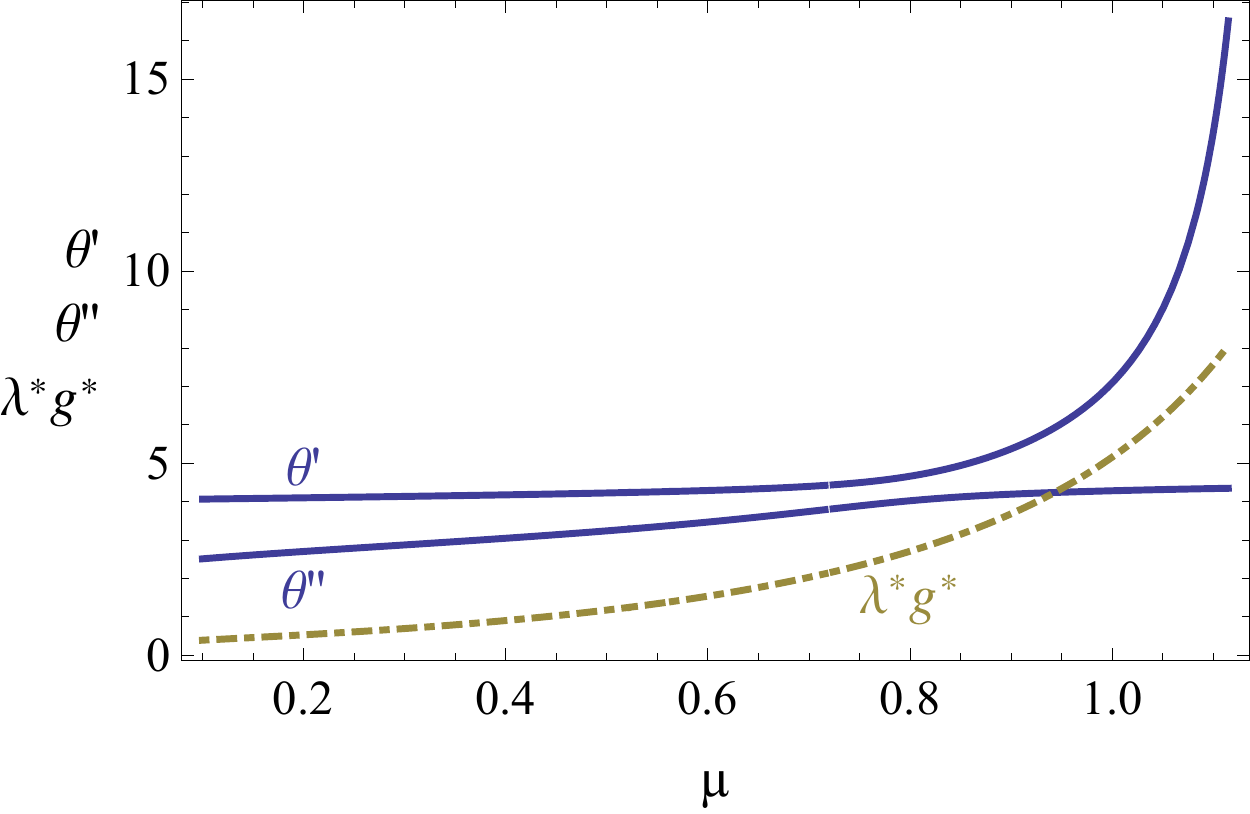}
		\label{fig:CriticalExponentsOptNegCutNGFPMinus}}
	\caption{Critical exponents and $\lambda^{*}g^{*}$ of the \textbf{NGFP}$\bm{^{\ominus}}$ as functions of $µ$. The critical exponents $\theta = \theta' + i \theta''$ (blue and red line, respectively), and the product $\lambda^{*}g^{*}$ (yellow line).}
	\label{fig:CritExpNegCutNGFPMinus}
\end{figure}

%% file: Summary.tex
In this paper we searched for asymptotically safe quantum gravity theories on a hitherto unexplored theory space. While earlier investigations tried the metric, or tetrads together with the spin-connection, or only tetrads as the carrier fields of the gravitational degrees of freedom, we used the metric in combination with the torsion tensor, the latter expressed in terms of its irreducible constituents. If there exist asymptotically safe theories on the new theory space they may, or may not be quantum mechanically equivalent to those found on the other spaces.

We found that within the (parity-even) Holst truncation the RG flow displays indeed the kind of non-Gaussian fixed point that we need for a non-perturbatively renormalizable theory. However, we also saw that the sign of the fixed point coordinate $\lambda^{*}$ depends on how the cutoff for one of the torsion fields is chosen. The Holst action implies non-derivative, mass like quadratic terms $(S_µ)^2$, $(T_µ)^2$, and $(q_{\lambda µ \nu})^2$ for the components of the torsion tensor, which are positive for $(S_µ)^2$ and $(T_µ)^2$, but \textit{the invariant $\int \intd^4 x \sqrt{\bar{g}} q_{\lambda µ \nu} q^{\lambda µ \nu}$ has a negative prefactor.} This leads to a similar difficulty as the ``wrong-sign'' kinetic term of the conformal factor. On top of it, a new and somewhat unusual type of threshold functions make their appearance since the torsion fields have no derivative term at all.

Negative quadratic terms in the action suggest two essentially different ways of adjusting the cutoff operator. \\
\noindent \textbf{(i)} One might apply the ``$\mathcal{Z}_k = \zeta_k$-rule'' nevertheless \cite{Reuter:1996cp}. For a mode with a term $\zeta_k (p^2 + m^2)$ in $\Gamma_k^{(2)}$ this rule proposes to use the operator $\mathcal{R}_k \equiv \mathcal{Z}_k k^2 R^{(0)}$ where
\begin{equation}
	\mathcal{Z}_k = \zeta_k \quad \text{both for} \quad \zeta_k > 0 \ \text{and} \ \zeta_k < 0 \ .
	\label{eq:zkchoice1}
\end{equation}
\noindent \textbf{(ii)} One may instead insist on a positive contribution of this mode to $\frac{1}{2} \widehat{\varphi} \mathcal{R}_k \widehat{\varphi}$ and apply the ``$\mathcal{Z}_k = |\zeta_k|$-rule'', setting
\begin{equation}
	\mathcal{Z}_k =
  \begin{cases}
    \zeta_k   & \quad \text{if } \ \zeta_k > 0 \\
    - \zeta_k	& \quad \text{if } \ \zeta_k < 0
  \end{cases}
  \label{eq:zkchoice2}
\end{equation}
We explored both options, the detailed results being described in Section \ref{sec:analysis}. We saw in particular that $\lambda^{*} < 0$ and $\lambda^{*} > 0$ if, respectively, the rules (\ref{eq:zkchoice1}) and (\ref{eq:zkchoice2}) are applied.

A negative value of the cosmological constant at the NGFP is particularly remarkable. In fact, the same unexpected result had also been found in Einstein-Cartan quantum gravity, while comparable calculations in metric gravity always yield $\lambda^{*} > 0$. Therefore, in \cite{Daum:2010phd,Daum:2013fu} and \cite{Harst:2012phd,Harst:2014vca} doubt had been cast on the possibility that the asymptotically safe theories in QEG and QECG could be equivalent. Even though our present results that were obtained with the structurally exact form of the FRGE cannot be compared directly to those from the proper-time flow equations employed in the earlier studies \cite{Daum:2010phd,Daum:2013fu,Harst:2012phd,Harst:2014vca}, we believe that the issue of $\lambda^{*} < 0$ vs. $\lambda^{*} > 0$ is still open, both on the side of $\mathcal{T}_{\mathrm{dtor}}$ and the Einstein-Cartan theory space\footnote{In the way the previous calculations are set up the indefinite nature of the torsion terms in $\Gamma_k^{(2)}$ and their implications are far from obvious.}. One should not dismiss the possibility of equivalent quantum theories too early, and first see how the results develop when the truncations are extended. 

Furthermore, in this paper we discussed in detail why the FRGE for the EAA always requires, as an explicit input, the specification of an isomorphism between the tangent and cotangent spaces of the field manifold, $\mathcal{F}$. Such an isomorphism is provided by a metric on $\mathcal{F}$, for instance. We showed that the form of the FRGE changes if one employs a scale dependent metric; or equivalently, we may write the flow equation in its traditional form but then, in the calculation of $\partial_k \mathcal{R}_k$, we must delete ``by hand'' all nonzero terms where $\partial_k$ hits the metric components. In the present paper, the freedom to choose the metric on $\mathcal{F}$ arbitrarily was utilized to uniformize the canonical dimensions of the multiplet of fields in question, but other applications are conceivable as well. In \cite{Percacci:2011uf} for instance a nontrivial metric was used to make the measure of the functional integral Weyl invariant.
\clearpage

%% file: AppendixComp.tex
\section*{Appendix}

\section{The Holst action in various field bases}\label{ch:holstaction}
The Holst action was originally defined in terms of the vielbein $\tensor{e}{^{a}_{µ}}$ and the spin connection $\tensor{\omega}{^{a b}_{µ}}$ of the Cartan formalism and takes the form
\begin{equation}
\label{eq:holstactionoriginal2}
	S_{\mathrm{Ho}} = - \frac{1}{16 \pi G} \int \mathrm{d}^{4}x \ e \left[ \tensor{e}{_{a}^{µ}} \tensor{e}{_{b}^{\nu}} \left( \tensor{F(e,\omega)}{^{ab}_{µ \nu}} - \frac{1}{2 \gamma} \tensor{\epsilon}{^{ab}_{cd}} \tensor{F(e,\omega)}{^{cd}_{µ \nu}} \right) - 2 \lambda \right].
\end{equation}
Using the identities 
\begin{equation*}
	\tensor{F}{^{ab}_{µ \nu}} = \tensor{e}{^{a}_{\rho}} \tensor{e}{^{b \sigma}} \tensor{R}{_{µ \nu}^{\rho}_{\sigma}} \ , \quad	\tensor{\epsilon}{^{µ \nu \rho \sigma}} = e\ \tensor{e}{_{a}^{µ}} \tensor{e}{_{b}^{\nu}} \tensor{e}{_{c}^{\rho}} \tensor{e}{_{d}^{\sigma}} \epsilon^{a b c d} \ ,
\end{equation*}
we can express the field strength $F(e,\omega)$ by the Riemann curvature tensor $R(\Gamma)$, with the connection coefficients $\Gamma$ including torsion, to obtain 
\begin{align}
	\begin{split}
	S_{\mathrm{Ho}} &= - \frac{1}{16 \pi G} \int \mathrm{d}^{4}x \left[ e\ \tensor{e}{_{a}^{µ}} \tensor{e}{_{b}^{\nu}} \tensor{e}{^{a}_{\rho}} \tensor{e}{^{b \sigma}} \tensor{R}{_{µ \nu}^{\rho}_{\sigma}} - \frac{1}{2 \gamma} e\ \tensor{e}{_{a}^{µ}} \tensor{e}{_{b}^{\nu}} \tensor{\epsilon}{^{ab}_{cd}} \tensor{e}{^{c}_{\rho}} \tensor{e}{^{d \sigma}} \tensor{R}{_{µ \nu}^{\rho}_{\sigma}} - e\ 2 \lambda \right] \\
	&= - \frac{1}{16 \pi G} \int \mathrm{d}^{4}x \left[ \sqrt{g}\ \delta^{µ}_{\rho} \delta^{\nu \sigma} \tensor{R}{_{µ \nu}^{\rho}_{\sigma}} - \frac{1}{2 \gamma} \tensor{\epsilon}{^{µ \nu}_{\rho}^{\sigma}} \tensor{R}{_{µ \nu}^{\rho}_{\sigma}} - \sqrt{g}\ 2 \lambda \right] \\
	&= - \frac{1}{16 \pi G} \int \mathrm{d}^{4}x\ \sqrt{g} \left[ R - \frac{1}{2 \gamma \sqrt{g}} \tensor{\epsilon}{^{µ \nu}_{\rho}^{\sigma}} \tensor{R}{_{µ \nu}^{\rho}_{\sigma}} - 2 \lambda \right].
	\end{split}
\end{align}
In the next step we employ the decomposition of the Riemann curvature tensor and the curvature scalar into its Levi-Civita counterparts and contorsion terms according to (\ref{eq:Riemdecomp}) and (\ref{eq:Rdecomb}) resulting in
\begin{multline}
 S_{\mathrm{Ho}} = - \frac{1}{16 \pi G} \int \mathrm{d}^{4}x\ \sqrt{g} \left[ \vphantom{\frac{\tensor{\epsilon}{^{µ \nu}_{\rho}^{\sigma}}}{\sqrt{g}}} R_{\LC} + 2 \tensor{(D_{\LC})}{_{ \kappa }} \tensor{K}{^{\kappa}_{\lambda}^{\lambda}} + 2 \tensor{K}{^{\kappa}_{ [ \kappa | \rho |}} \tensor{K}{^{\rho}_{\lambda ]}^{\lambda}} \right.
 \\ \left. - \frac{1}{2 \gamma} \frac{\tensor{\epsilon}{^{µ \nu}_{\rho}^{\sigma}}}{\sqrt{g}} \left( \tensor{(R_{\LC})}{_{µ \nu}^{\rho}_{\sigma}} + 2 \tensor{(D_{\LC})}{_{ [ µ }} \tensor{K}{^{\rho}_{\nu ] \sigma}} + 2 \tensor{K}{^{\rho}_{ [ µ | \tau |}} \tensor{K}{^{\tau}_{\nu ] \sigma}} \right) - 2 \lambda \right].
\end{multline}
First we note that the contraction of the Levi-Civita symbol and the Riemann curvature tensor vanishes due to the first Bianchi identity. The Levi-Civita symbol also ``cancels'' the antisymmetrization in the other two terms and we remain with
\begin{multline}
 S_{\mathrm{Ho}} = - \frac{1}{16 \pi G} \int \mathrm{d}^{4}x\ \sqrt{g} \left[ \vphantom{\frac{\tensor{\epsilon}{^{µ \nu}_{\rho}^{\sigma}}}{\sqrt{g}}} R_{\LC} + 2 \tensor{(D_{\LC})}{_{ \kappa }} \tensor{K}{^{\kappa}_{\lambda}^{\lambda}} + \tensor{K}{^{\kappa}_{ \kappa \rho}} \tensor{K}{^{\rho}_{\lambda}^{\lambda}} -  \tensor{K}{^{\kappa}_{ \lambda \rho}} \tensor{K}{^{\rho}_{\kappa}^{\lambda}} \right.
 \\ \left. - \frac{1}{\gamma} \frac{\tensor{\epsilon}{^{µ \nu}_{\rho}^{\sigma}}}{\sqrt{g}} \Bigl(\tensor{(D_{\LC})}{_{ µ }} \tensor{K}{^{\rho}_{\nu \sigma}} + \tensor{K}{^{\rho}_{µ \tau}} \tensor{K}{^{\tau}_{\nu \sigma}} \Bigr) - 2 \lambda \right]. \label{eq:SHOcont}
\end{multline}
Now we switch from the contorsion tensor $\tensor{K}{^{\lambda}_{µ \nu}}$ back to the torsion tensor $\tensor{T}{^{\lambda}_{µ \nu}}$ via (\ref{eq:contorsion}) and after some calculation the action takes the form
\begin{multline}
 S_{\mathrm{Ho}} = - \frac{1}{16 \pi G} \int \mathrm{d}^{4}x\ \sqrt{g} \left[ \vphantom{\frac{\tensor{\epsilon}{^{µ \nu}_{\rho}^{\sigma}}}{\sqrt{g}}} R_{\LC} + 2 \tensor{(D_{\LC})}{_{ \kappa }} \tensor{T}{^{\lambda \kappa}_{\lambda}} + \frac{1}{4} T_{\kappa \lambda \rho} T^{\kappa \lambda \rho} + \frac{1}{2} T_{\kappa \lambda \rho} T^{\lambda \kappa \rho} - \tensor{T}{^{\kappa \rho}_{\kappa}} \tensor{T}{^{\lambda}_{\rho \lambda}} \right.
 \\ \left. - \frac{1}{\gamma} \frac{\tensor{\epsilon}{^{µ \nu \rho \sigma}}}{\sqrt{g}} \Bigl( - \frac{1}{2} \tensor{(D_{\LC})}{_{ µ }} T_{\nu \rho \sigma} + \frac{1}{4} \tensor{T}{^{\tau}_{µ \nu}} \tensor{T}{_{\tau \rho \sigma}} \Bigr) - 2 \lambda \right]. \label{eq:SHOtor}
\end{multline}
Finally we decompose the torsion tensor into its irreducible parts according to (\ref{eq:Tdecomp}) and obtain after a somewhat tedious calculation and some index relabeling the desired result:
\begin{multline}
 S_{\mathrm{Ho}} = - \frac{1}{16 \pi G} \int \mathrm{d}^{4}x\ \sqrt{g} \left[ \vphantom{\frac{\tensor{\epsilon}{^{µ \nu}_{\rho}^{\sigma}}}{\sqrt{g}}} R_{\LC} + 2 \tensor{(D_{\LC})}{_{ µ }} T^{µ} - \frac{2}{3} T_µ T^µ - \frac{1}{24} S_µ S^µ + \frac{1}{2} q_{µ \nu \rho} q^{µ \nu \rho} \right.
 \\ \left. - \frac{1}{\gamma} \left( \frac{1}{2} \tensor{(D_{\LC})}{_{ µ }} S^µ - \frac{1}{3} T_µ S^µ + \frac{\tensor{\epsilon}{^{µ \nu \rho \sigma}}}{\sqrt{g}} \tensor{q}{_{µ \nu}^{\tau}} q_{\rho \sigma \tau} \right) - 2 \lambda \right].
 \label{eq:holstaction2}
\end{multline}
In this form the action is used in the main text.

\section{Threshold functions}\label{ch:threshold}
{\allowdisplaybreaks
In this Appendix we list the evaluated form of the threshold functions and the resulting $B$- and $\check{B}$-functions for both the optimized shape function, $R^{(0)}_{\mathrm{opt}}(z) = (1-z)\Theta(1-z)$, and generalized exponential shape function, $R^{(0)}_{\mathrm{exp}}(z;s) = \frac{s z}{e^{s z}-1}$.

\noindent \textbf{(1)} The threshold functions (\ref{eq:thresholdcheck}) for the optimized cutoff $R^{(0)}_{\mathrm{opt}}(z)$ and $w \in \mathds{R}\setminus[-1,0] $:
\begin{subequations}
	\begin{align*}
		\check{\Phi}_{1}^{1}(w) &= \int_{0}^{\infty} \mathrm{d}z\ \frac{(1-z)\Theta(1-z) - z\, \partial_{z}(1-z)\Theta(1-z)}{(1-z)\Theta(1-z) + w} \\
		&= \ln\left( 1 + \frac{1}{w} \right) \\
		\check{\tilde{\Phi}}_{1}^{1}(w) &= \int_{0}^{\infty} \mathrm{d}z\ \frac{(1-z)\Theta(1-z)}{(1-z)\Theta(1-z) + w} \\
		&= 1- w \ln\left(1+ \frac{1}{w} \right) \\
		\check{\Phi}_{2}^{1}(w) &= \int_{0}^{\infty} \mathrm{d}z\ z\ \frac{(1-z)\Theta(1-z) - z\, \partial_{z}(1-z)\Theta(1-z)}{(1-z)\Theta(1-z) + w} \\
		&= -1 +(1+w)\ln\left(1+ \frac{1}{w} \right) \\
		\check{\tilde{\Phi}}_{2}^{1}(w) &= \int_{0}^{\infty} \mathrm{d}z\ z\ \frac{(1-z)\Theta(1-z)}{(1-z)\Theta(1-z) + w} \\
		&= \frac{1}{2}+w-\left(w+w^2\right)\ln\left(1+ \frac{1}{w}\right)
	\end{align*}
\end{subequations}

\noindent \textbf{(2)} The $B$-functions (\ref{eq:B12}) and (\ref{eq:B34}) for the optimized cutoff $R^{(0)}_{\mathrm{opt}}(z)$:
\begin{subequations}
\begin{align*}
	B_{1}(\lambda_k) &= \frac{1}{3 \pi }\left(\frac{5}{1-2 \lambda_k } -\frac{9}{(1-2 \lambda_k)^2} -7\right) \\
	B_{2}(\lambda_k) &= -\frac{1}{6 \pi } \left(\frac{5}{2-4 \lambda_k}-\frac{3}{(1-2 \lambda_k )^2} \right) \\
	B_{3}(\lambda_k) &= \frac{10}{2-4 \lambda_k } -4\\
	B_{4}(\lambda_k) &= -\frac{5}{6-12 \lambda_k }
\end{align*}
\end{subequations}

\noindent \textbf{(3)} The $\check{B}$-functions (\ref{eq:B12}) and (\ref{eq:B34}) for the optimized cutoff $R^{(0)}_{\mathrm{opt}}(z)$ with $\xi = +1$ and $µ \in \mathds{R}\setminus[- \frac{1}{\sqrt{2}},\frac{1}{\sqrt{2}}]$:
\begin{subequations}
\begin{align*}
		\check{B}_{1}(µ^2) = &+ \frac{1}{3 \pi} \left( 2 \ln\left(1+ \tfrac{6}{µ^2}\right) + 2 \ln\left(1+ \tfrac{3}{8µ^2} \right) + 8 \ln\left(1- \tfrac{1}{2µ^2}\right) \vphantom{\frac{1}{2}}\right) \\
		\check{B}_{2}(µ^2) = &- \frac{1}{3 \pi} \left( 6 - \frac{µ^2}{6} \ln\left(1+ \tfrac{6}{µ^2} \right) - \frac{8µ^2}{3} \ln\left(1+ \tfrac{3}{8µ^2} \right) +8µ^2 \ln\left(1- \tfrac{1}{2µ^2} \right) \right) \\
		\check{B}_{3}(µ^2) = &-24 + \left( 4 + \frac{2µ^2}{3} \right)\ln\left(1+ \tfrac{6}{µ^2} \right) + \left(4+ \frac{32 µ^2}{3} \right) \ln\left(1+ \tfrac{3}{8µ^2} \right) \\
		&+ \left(16-32µ^2 \vphantom{\frac{1}{2}}\right)\ln\left(1- \tfrac{1}{2µ^2} \right) \\
		\check{B}_{4}(µ^2) = &-6 + \frac{31µ^2}{3} + \left( \frac{µ^2}{3}+ \frac{µ^4}{18} \right)\ln\left(1+ \tfrac{6}{µ^2} \right) + \left( \frac{16µ^2}{3} + \frac{128µ^4}{9} \right) \ln\left(1+ \tfrac{3}{8µ^2} \right) \\
		&- \left(16µ^2 -32µ^4 \vphantom{\frac{1}{2}}\right)\ln\left(1- \tfrac{1}{2µ^2} \right)
	\end{align*}
\end{subequations}

\noindent \textbf{(4)} The $\check{B}$-functions (\ref{eq:B12}) and (\ref{eq:B34}) for the optimized cutoff $R^{(0)}_{\mathrm{opt}}(z)$ with $\xi = -1$ and $µ \in \mathds{R}\setminus\{0\}$:
\begin{subequations}
\begin{align*}
		\check{B}_{1}(µ^2) = &+ \frac{1}{3 \pi} \left( 2 \ln\left(1+ \tfrac{6}{µ^2}\right) + 2 \ln\left(1+ \tfrac{3}{8µ^2} \right) + 8 \ln\left(1+ \tfrac{1}{2µ^2}\right) \vphantom{\frac{1}{2}}\right) \\
		\check{B}_{2}(µ^2) = &- \frac{1}{3 \pi} \left( 6 - \frac{µ^2}{6} \ln\left(1+ \tfrac{6}{µ^2} \right) - \frac{8µ^2}{3} \ln\left(1+ \tfrac{3}{8µ^2} \right) -8µ^2 \ln\left(1+ \tfrac{1}{2µ^2} \right) \right) \\
		\check{B}_{3}(µ^2) = &-24 + \left( 4 + \frac{2µ^2}{3} \right)\ln\left(1+ \tfrac{6}{µ^2} \right) + \left(4+ \frac{32 µ^2}{3} \right) \ln\left(1+ \tfrac{3}{8µ^2} \right) \\
		&+ \left(16+32µ^2 \vphantom{\frac{1}{2}}\right)\ln\left(1+ \tfrac{1}{2µ^2} \right) \\
		\check{B}_{4}(µ^2) = &-6 - \frac{65µ^2}{3} + \left( \frac{µ^2}{3}+ \frac{µ^4}{18} \right)\ln\left(1+ \tfrac{6}{µ^2} \right) + \left( \frac{16µ^2}{3} + \frac{128µ^4}{9} \right) \ln\left(1+ \tfrac{3}{8µ^2} \right) \\
		&+ \left(16µ^2 +32µ^4 \vphantom{\frac{1}{2}}\right)\ln\left(1+ \tfrac{1}{2µ^2} \right)
	\end{align*}
\end{subequations}

\noindent \textbf{(5)} The $B$-functions (\ref{eq:B12}) and (\ref{eq:B34}) for the generalized exponential cutoff $R^{(0)}_{\mathrm{exp}}(z;s)$:
\begin{subequations}
\begin{align*}
	B_{1}(\lambda_k ;s) = \frac{1}{3 \pi} \left( \vphantom{\int_{0}^{\infty}\mathrm{d}z\left[\frac{z \left(\frac{s z}{-1+e^{s z}}-z \left(\frac{s}{-1+e^{s z}}-\frac{e^{s z} s^2 z}{\left(-1+e^{s z}\right)^2}\right)\right)}{\left(z+\frac{s z}{-1+e^{s z}}-2 \lambda_k \right)^2}\right]} \right. 5 &\int_{0}^{\infty}\mathrm{d}z\left[\frac{\frac{s z}{-1+e^{s z}}-z \left(\frac{s}{-1+e^{s z}}-\frac{e^{s z} s^2 z}{\left(-1+e^{s z}\right)^2}\right)}{z+\frac{s z}{-1+e^{s z}}-2 \lambda_k }\right] \\
	- 18 &\int_{0}^{\infty}\mathrm{d}z\left[\frac{z \left(\frac{s z}{-1+e^{s z}}-z \left(\frac{s}{-1+e^{s z}}-\frac{e^{s z} s^2 z}{\left(-1+e^{s z}\right)^2}\right)\right)}{\left(z+\frac{s z}{-1+e^{s z}}-2 \lambda_k \right)^2}\right] \\
	- 4 &\int_{0}^{\infty}\mathrm{d}z\left[\frac{\frac{s z}{-1+e^{s z}}-z \left(\frac{s}{-1+e^{s z}}-\frac{e^{s z} s^2 z}{\left(-1+e^{s z}\right)^2}\right)}{z+\frac{s z}{-1+e^{s z}}}\right] \\
	- 6 &\int_{0}^{\infty}\mathrm{d}z\left[\frac{z \left(\frac{s z}{-1+e^{s z}}-z \left(\frac{s}{-1+e^{s z}}-\frac{e^{s z} s^2 z}{\left(-1+e^{s z}\right)^2}\right)\right)}{\left(z+\frac{s z}{-1+e^{s z}}\right)^2}\right] \left. \vphantom{\int_{0}^{\infty}\mathrm{d}z\left[\frac{z \left(\frac{s z}{-1+e^{s z}}-z \left(\frac{s}{-1+e^{s z}}-\frac{e^{s z} s^2 z}{\left(-1+e^{s z}\right)^2}\right)\right)}{\left(z+\frac{s z}{-1+e^{s z}}-2 \lambda_k \right)^2}\right]} \right) \\
	B_{2}(\lambda_k ;s) = -\frac{1}{6 \pi } \left( \vphantom{\int_{0}^{\infty}\mathrm{d}z\left[\frac{s z^2}{\left(-1+e^{s z}\right) \left(z+\frac{s z}{-1+e^{s z}}-2 \lambda_k \right)^2}\right]} \right. 5 &\int_{0}^{\infty}\mathrm{d}z\left[\frac{s z}{\left(-1+e^{s z}\right) \left(z+\frac{s z}{-1+e^{s z}}-2 \lambda_k \right)}\right] \\
	-18 &\int_{0}^{\infty}\mathrm{d}z\left[\frac{s z^2}{\left(-1+e^{s z}\right) \left(z+\frac{s z}{-1+e^{s z}}-2 \lambda_k \right)^2}\right] \left. \vphantom{\int_{0}^{\infty}\mathrm{d}z\left[\frac{s z^2}{\left(-1+e^{s z}\right) \left(z+\frac{s z}{-1+e^{s z}}-2 \lambda_k \right)^2}\right]} \right) \\
	B_{3}(\lambda_k ;s) = 10 &\int_{0}^{\infty}\mathrm{d}z\left[\frac{z \left(\frac{s z}{-1+e^{s z}}-z \left(\frac{s}{-1+e^{s z}}-\frac{e^{s z} s^2 z}{\left(-1+e^{s z}\right)^2}\right)\right)}{z+\frac{s z}{-1+e^{s z}}-2 \lambda_k }\right] \\
	- 8 &\int_{0}^{\infty}\mathrm{d}z\left[\frac{z \left(\frac{s z}{-1+e^{s z}}-z \left(\frac{s}{-1+e^{s z}}-\frac{e^{s z} s^2 z}{\left(-1+e^{s z}\right)^2}\right)\right)}{z+\frac{s z}{-1+e^{s z}}}\right] \\
	B_{4}(\lambda_k ;s) = -5 &\int_{0}^{\infty}\mathrm{d}z\left[\frac{s z^2}{\left(-1+e^{s z}\right) \left(z+\frac{s z}{-1+e^{s z}}-2 \lambda_k \right)}\right]
\end{align*}
\end{subequations}

\noindent \textbf{(6)} The $\check{B}$-functions (\ref{eq:B12}) and (\ref{eq:B34}) for the generalized exponential cutoff $R^{(0)}_{\mathrm{exp}}(z;s)$:
\begin{subequations}
\begin{align*}
	\check{B}_{1}(µ^2 ;s) = \frac{1}{3 \pi } \left(\vphantom{\int_{0}^{\infty}\left[\frac{\frac{s z}{-1+e^{s z}}-z \left(\frac{s}{-1+e^{s z}}-\frac{e^{s z} s^2 z}{\left(-1+e^{s z}\right)^2}\right)}{\frac{s z}{-1+e^{s z}}-2 \mu ^2}\right]}\right. 2 &\int_{0}^{\infty}\mathrm{d}z\left[\frac{\frac{s z}{-1+e^{s z}}-z \left(\frac{s}{-1+e^{s z}}-\frac{e^{s z} s^2 z}{\left(-1+e^{s z}\right)^2}\right)}{\frac{s z}{-1+e^{s z}}+\frac{\mu ^2}{6}}\right] \\
	+ 2 &\int_{0}^{\infty}\mathrm{d}z\left[\frac{\frac{s z}{-1+e^{s z}}-z \left(\frac{s}{-1+e^{s z}}-\frac{e^{s z} s^2 z}{\left(-1+e^{s z}\right)^2}\right)}{\frac{s z}{-1+e^{s z}}+\frac{8 \mu ^2}{3}}\right]\\
	+ 8 &\int_{0}^{\infty}\mathrm{d}z\left[\frac{\frac{s z}{-1+e^{s z}}-z \left(\frac{s}{-1+e^{s z}}-\frac{e^{s z} s^2 z}{\left(-1+e^{s z}\right)^2}\right)}{\frac{s z}{-1+e^{s z}}-2 \xi \mu ^2}\right] \left.\vphantom{\int_{0}^{\infty}\left[\frac{\frac{s z}{-1+e^{s z}}-z \left(\frac{s}{-1+e^{s z}}-\frac{e^{s z} s^2 z}{\left(-1+e^{s z}\right)^2}\right)}{\frac{s z}{-1+e^{s z}}-2 \xi \mu ^2}\right]}\right) \\
	\check{B}_{2}(µ^2 ;s) = -\frac{1}{3 \pi } \left(\vphantom{\int_{0}^{\infty}\mathrm{d}z\left[\frac{s z}{\left(-1+e^{s z}\right) \left(\frac{s z}{-1+e^{s z}}-2 \mu ^2\right)}\right]}\right. &\int_{0}^{\infty}\mathrm{d}z\left[\frac{s z}{\left(-1+e^{s z}\right) \left(\frac{s z}{-1+e^{s z}}+\frac{\mu ^2}{6}\right)}\right]\\
	+ &\int_{0}^{\infty}\mathrm{d}z\left[\frac{s z}{\left(-1+e^{s z}\right) \left(\frac{s z}{-1+e^{s z}}+\frac{8 \mu ^2}{3}\right)}\right]\\
	+ 4 &\int_{0}^{\infty}\mathrm{d}z\left[\frac{s z}{\left(-1+e^{s z}\right) \left(\frac{s z}{-1+e^{s z}}-2 \xi \mu ^2\right)}\right] \left.\vphantom{\int_{0}^{\infty}\mathrm{d}z\left[\frac{s z}{\left(-1+e^{s z}\right) \left(\frac{s z}{-1+e^{s z}}-2 \xi \mu ^2\right)}\right]}\right) \\
	\check{B}_{3}(µ^2 ;s) = 4 &\int_{0}^{\infty}\mathrm{d}z\left[\frac{z \left(\frac{s z}{-1+e^{s z}}-z \left(\frac{s}{-1+e^{s z}}-\frac{e^{s z} s^2 z}{\left(-1+e^{s z}\right)^2}\right)\right)}{\frac{s z}{-1+e^{s z}}+\frac{\mu ^2}{6}}\right]\\
	+4 &\int_{0}^{\infty}\mathrm{d}z\left[\frac{z \left(\frac{s z}{-1+e^{s z}}-z \left(\frac{s}{-1+e^{s z}}-\frac{e^{s z} s^2 z}{\left(-1+e^{s z}\right)^2}\right)\right)}{\frac{s z}{-1+e^{s z}}+\frac{8 \mu ^2}{3}}\right]\\
	+ 16 &\int_{0}^{\infty}\mathrm{d}z\left[\frac{z \left(\frac{s z}{-1+e^{s z}}-z \left(\frac{s}{-1+e^{s z}}-\frac{e^{s z} s^2 z}{\left(-1+e^{s z}\right)^2}\right)\right)}{\frac{s z}{-1+e^{s z}}-2 \xi \mu ^2}\right] \\
	\check{B}_{4}(µ^2 ;s) = -2 &\int_{0}^{\infty}\mathrm{d}z\left[\frac{s z^2}{\left(-1+e^{s z}\right) \left(\frac{s z}{-1+e^{s z}}+\frac{\mu ^2}{6}\right)}\right]\\
	-2 &\int_{0}^{\infty}\mathrm{d}z\left[\frac{s z^2}{\left(-1+e^{s z}\right) \left(\frac{s z}{-1+e^{s z}}+\frac{8 \mu ^2}{3}\right)}\right]\\
	-8 &\int_{0}^{\infty}\mathrm{d}z\left[\frac{s z^2}{\left(-1+e^{s z}\right) \left(\frac{s z}{-1+e^{s z}}-2 \xi \mu ^2\right)}\right]
\end{align*}
\end{subequations}
These integrals defining the various $\check{B}$-functions have been evaluated numerically in the course of (numerically) solving the RG equations in Section \ref{sec:analysis}. Note in particular the sign structure of the various denominators.}

\section{Phase portraits}\label{sec:PP}
In this appendix we collect the phase portraits discussed in the main text in a systematic way.
\begin{figure}[H]
	\centering
	\subfigure[\textbf{NGFP} for $µ^2 = \frac{1}{1.9}$.]{
		\centering
		\includegraphics[width=0.3\textwidth]{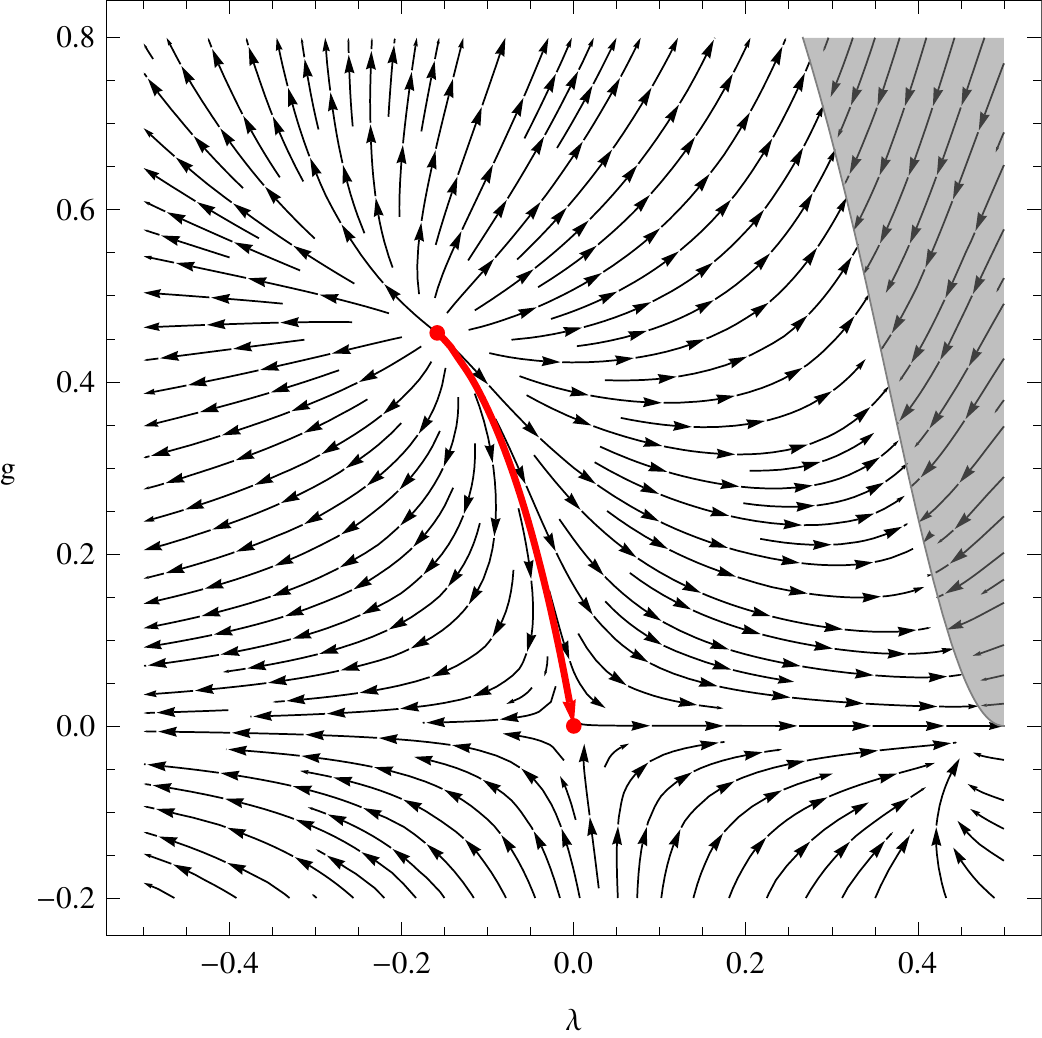}
		\label{fig:PDopt119}}
	\subfigure[\textbf{NGFP} for $µ^2 = \frac{1}{1.5}$.]{
		\centering
		\includegraphics[width=0.3\textwidth]{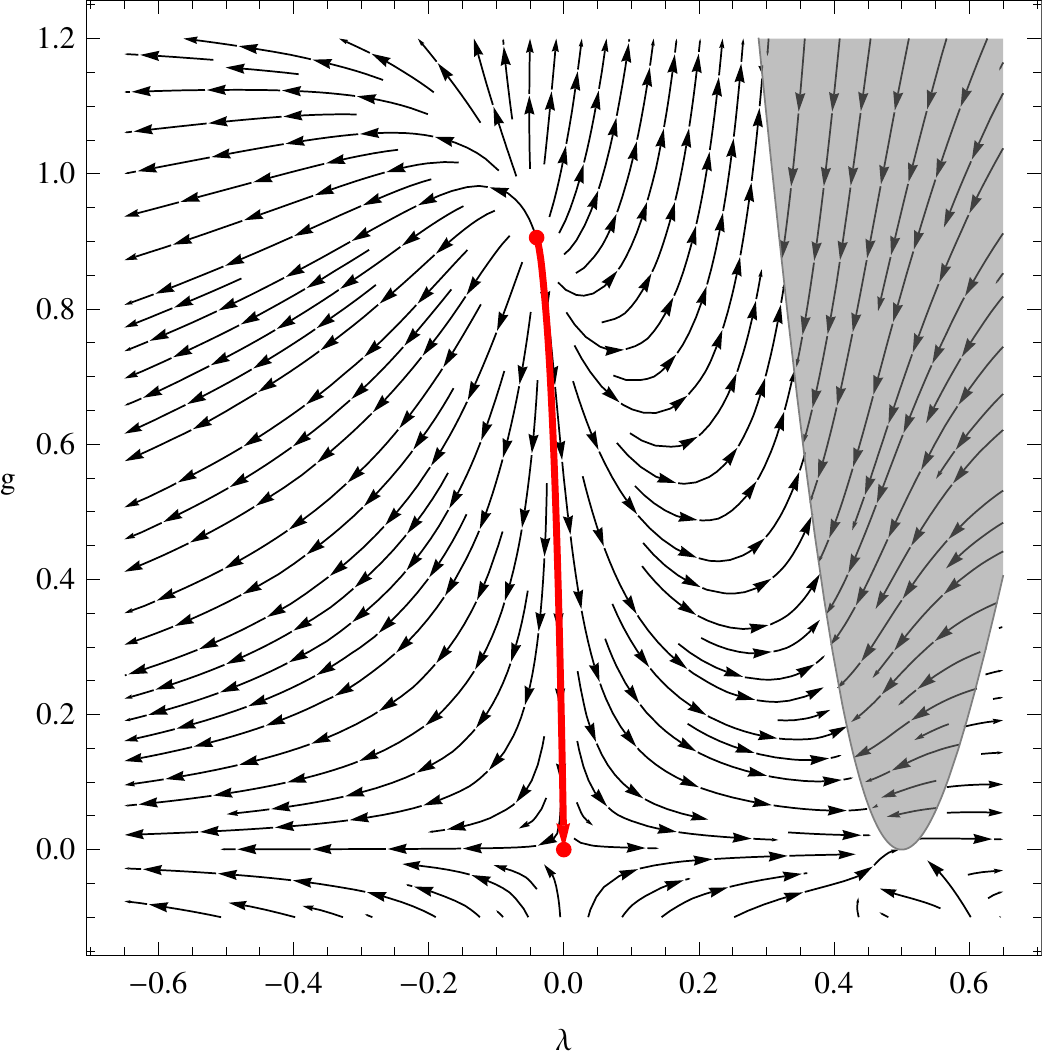}
		\label{fig:PDopt115}}
	\subfigure[\textbf{NGFP} for $µ^2 \approx 0.695309$.]{
		\centering
		\includegraphics[width=0.3\textwidth]{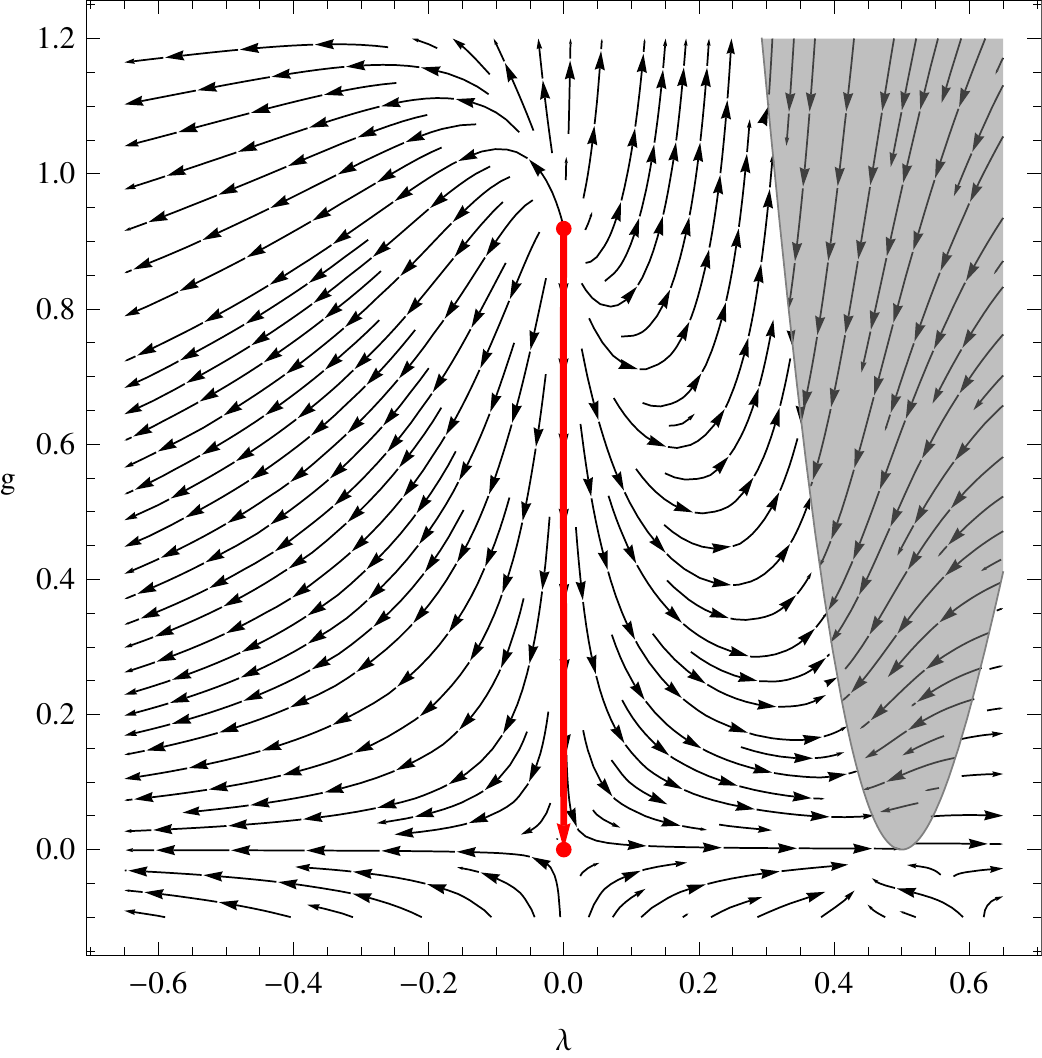}
		\label{fig:PDopttest}}
	\subfigure[\textbf{NGFP} for $µ^2 \approx 0.698591$.]{
		\centering
		\includegraphics[width=0.3\textwidth]{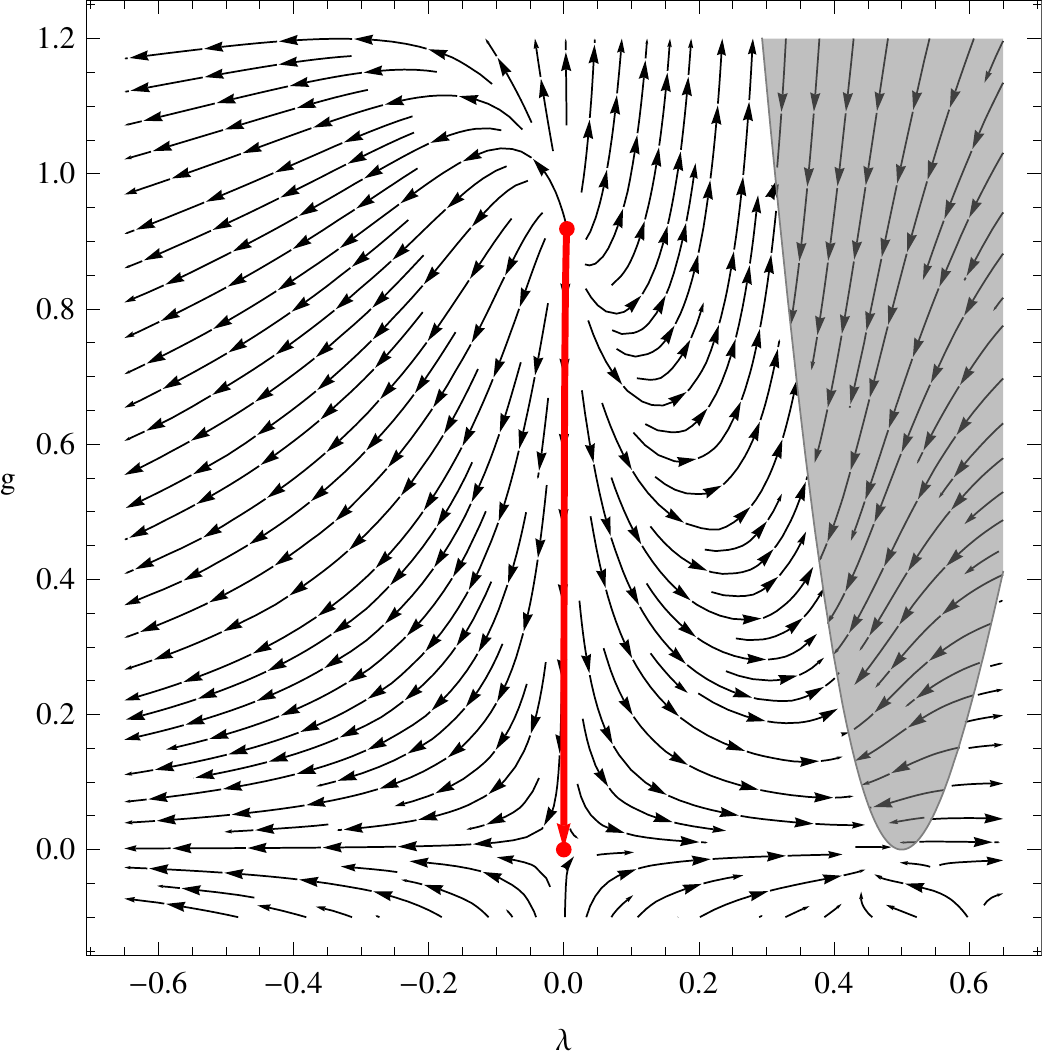}
		\label{fig:PDoptReal}}
	\subfigure[\textbf{NGFP} for $µ^2 = \frac{1}{1.1}$.]{
		\centering
		\includegraphics[width=0.3\textwidth]{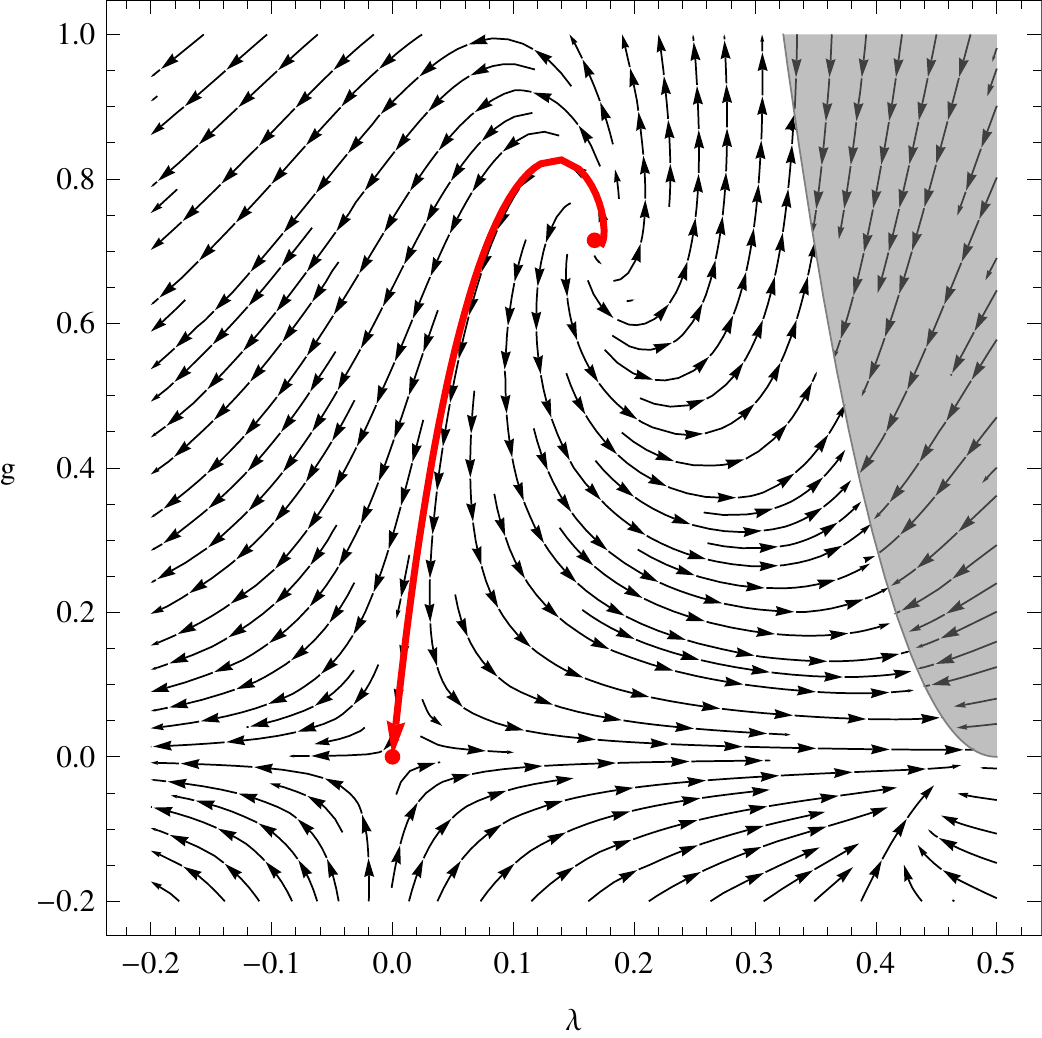}
		\label{fig:PDopt111v3}}
	\subfigure[\textbf{NGFP} for $µ^2 = 1$.]{
		\centering
		\includegraphics[width=0.3\textwidth]{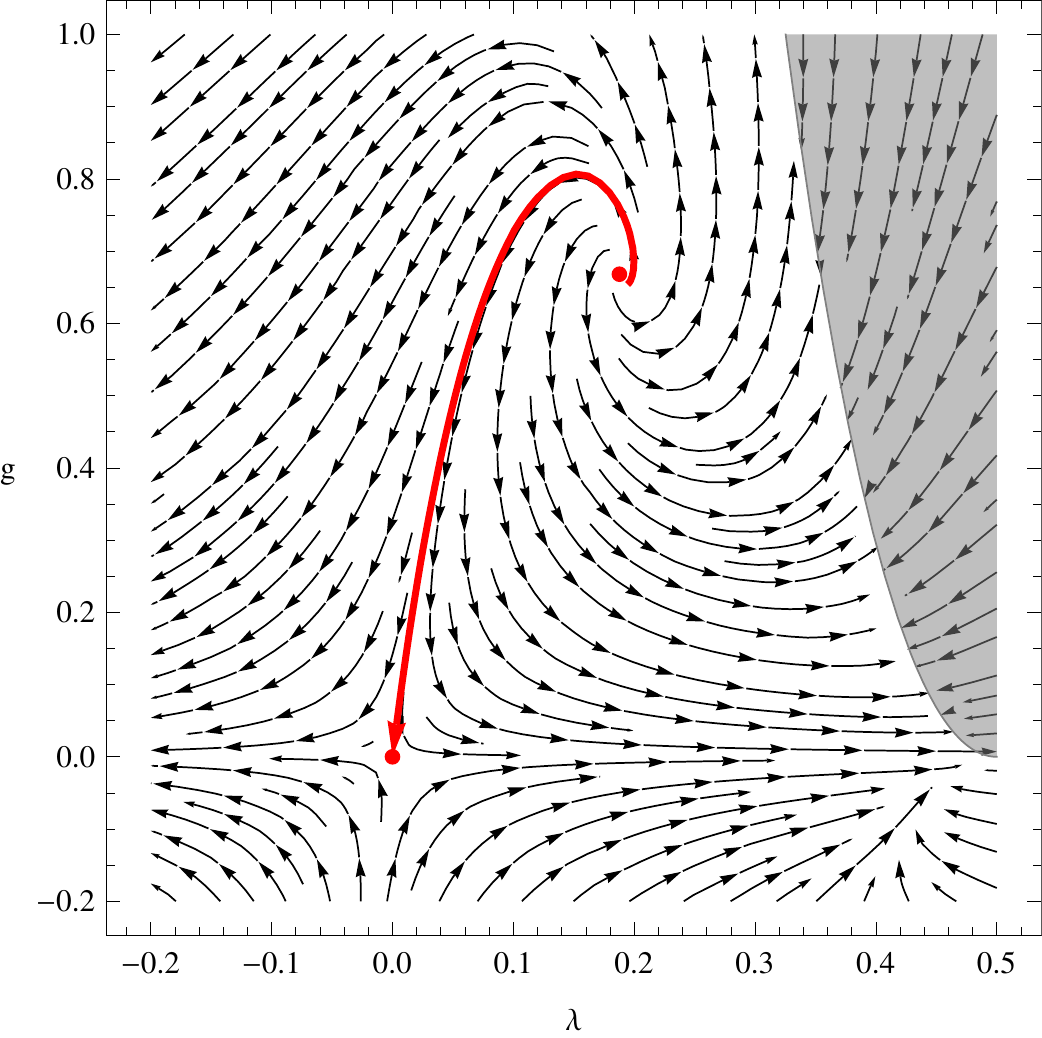}
		\label{fig:PDopt1v3}}
	\subfigure[\textbf{NGFP} for $µ^2 = 2$.]{
		\centering
		\includegraphics[width=0.3\textwidth]{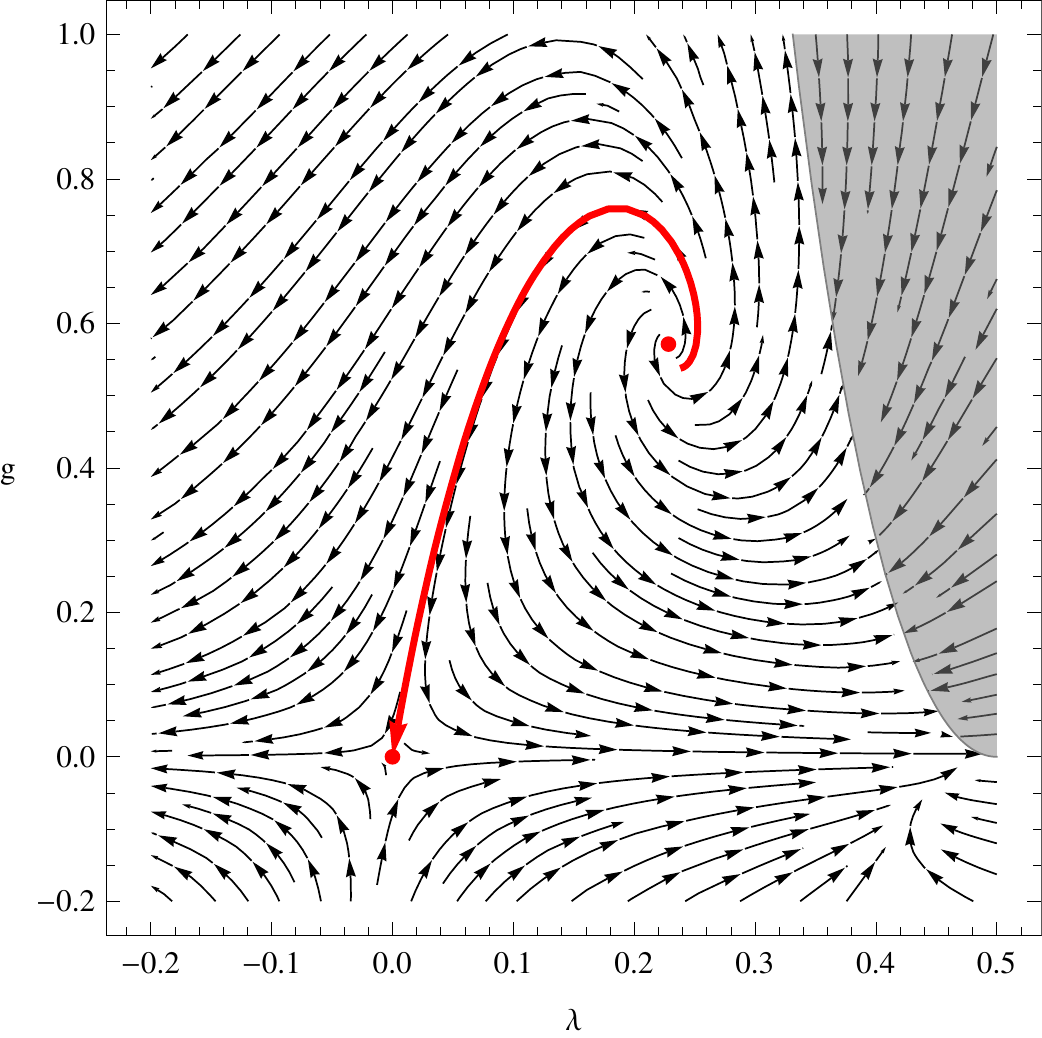}
		\label{fig:PDopt2v3}}
	\subfigure[\textbf{NGFP} for $µ^2 = 3$.]{
		\centering
		\includegraphics[width=0.3\textwidth]{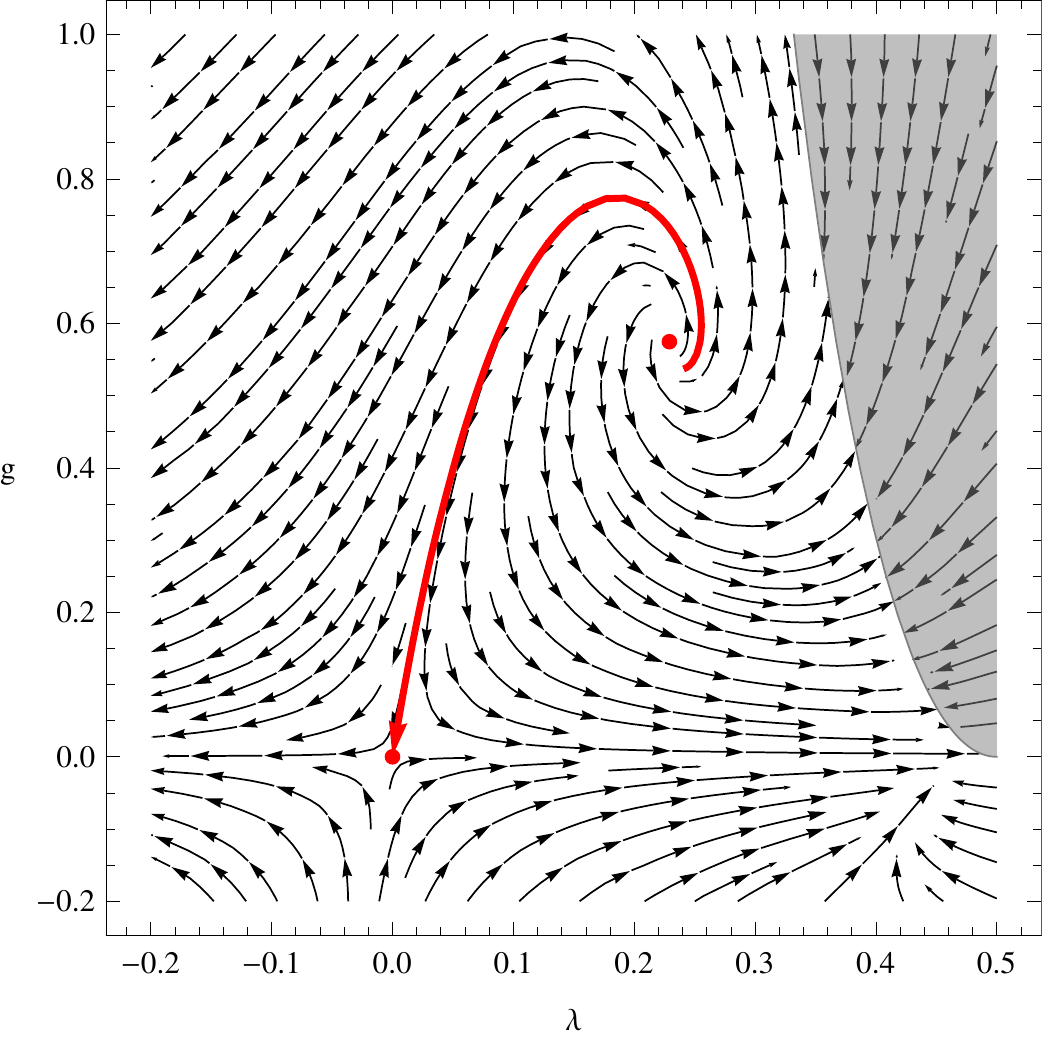}
		\label{fig:PDopt3v3}}
	\subfigure[\textbf{NGFP} for $µ^2 = 4$.]{
		\centering
		\includegraphics[width=0.3\textwidth]{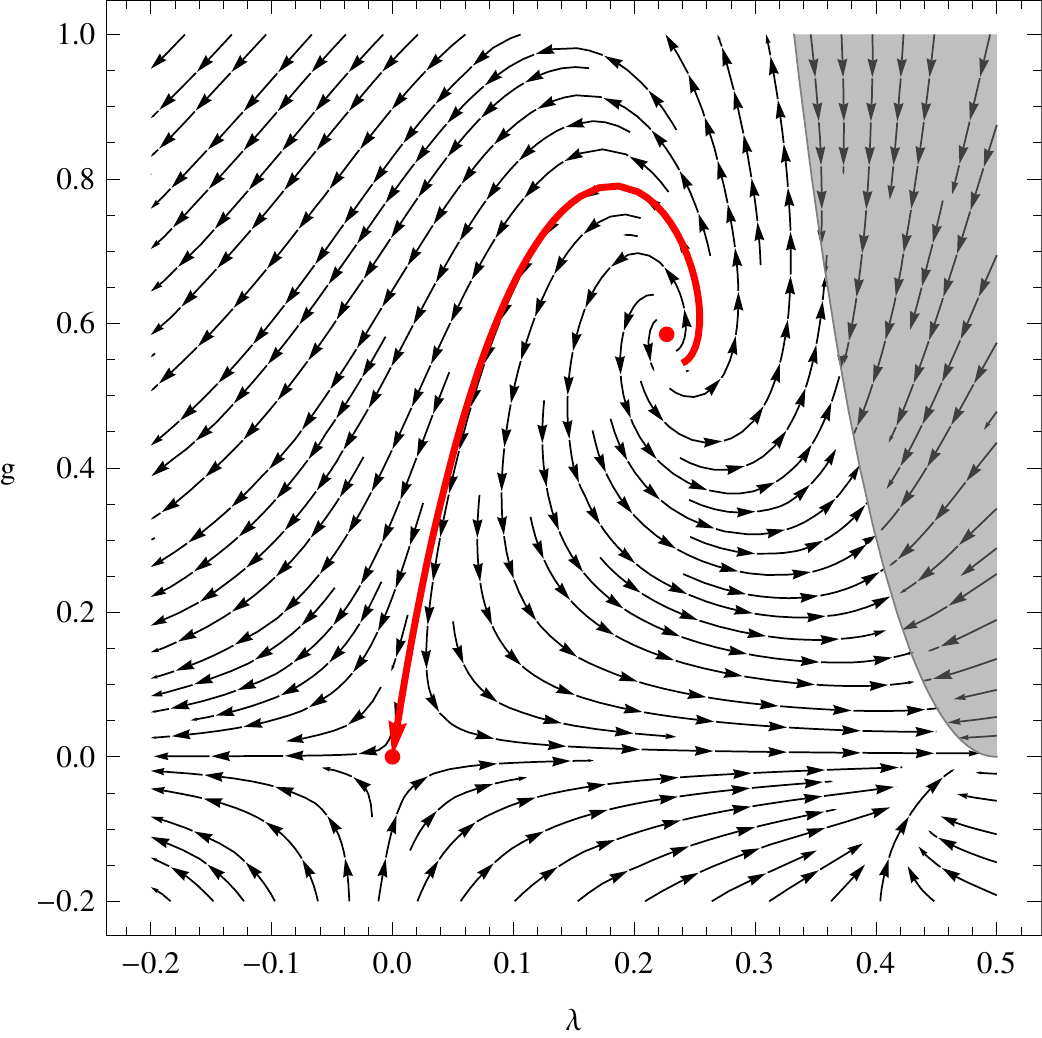}
		\label{fig:PDopt4v3}}
	\caption{RG phase portrait for different values of the squared mass parameter $µ^2$. The figures display the influence of the mass parameter: For small $µ$ we find phase diagrams similar to the ones from QECG, while for larger we obtain the situation known from QEG. The entire sequence strongly resembles the flow obtained in Tetrad Gravity.}
	\label{fig:NGFPfullTor}
\end{figure}
\clearpage

\begin{figure}[htbp]
	\centering
	\subfigure[\textbf{NGFP} for $µ^2 = 5$.]{
		\centering
		\includegraphics[width=0.3\textwidth]{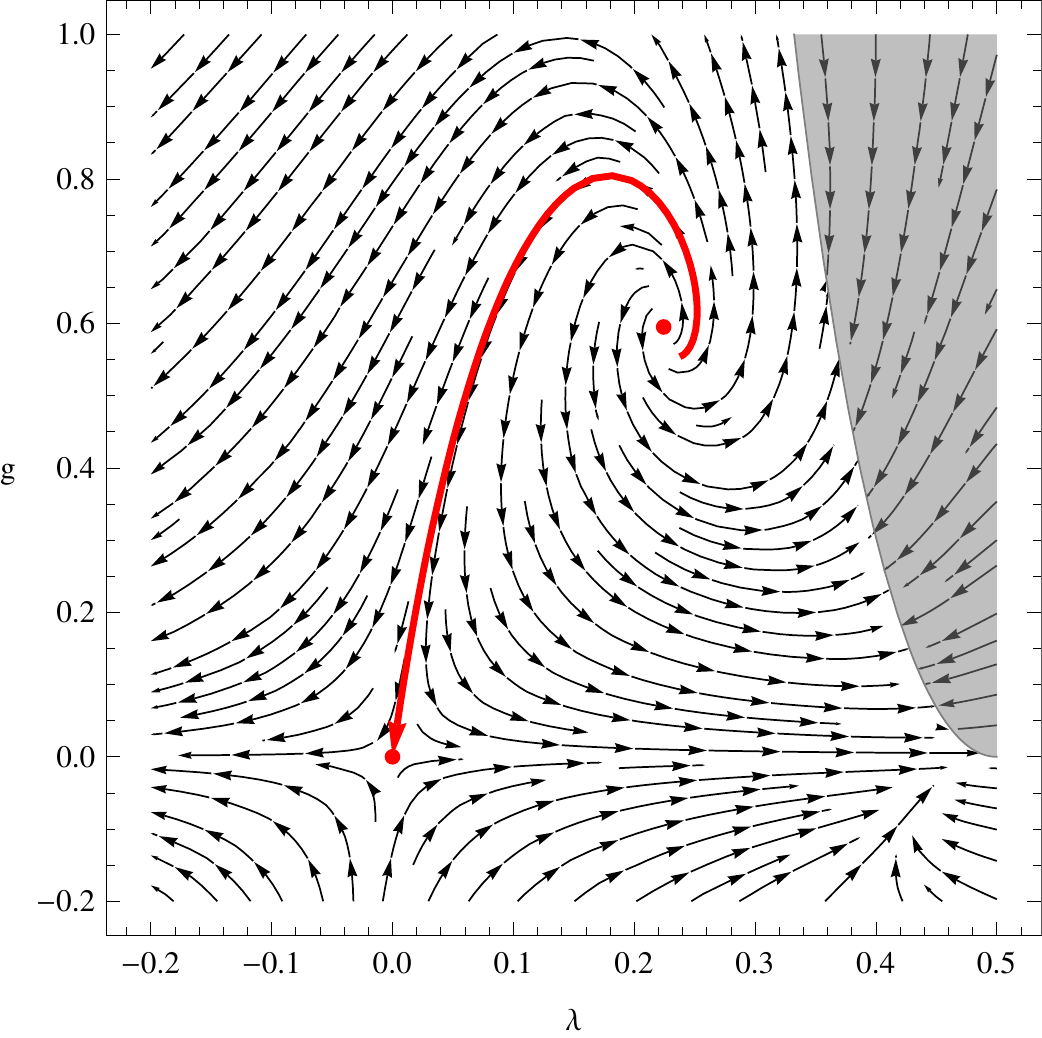}
		\label{fig:PDopt5v3}}
	\subfigure[\textbf{NGFP} for $µ^2 = 6$.]{
		\centering
		\includegraphics[width=0.3\textwidth]{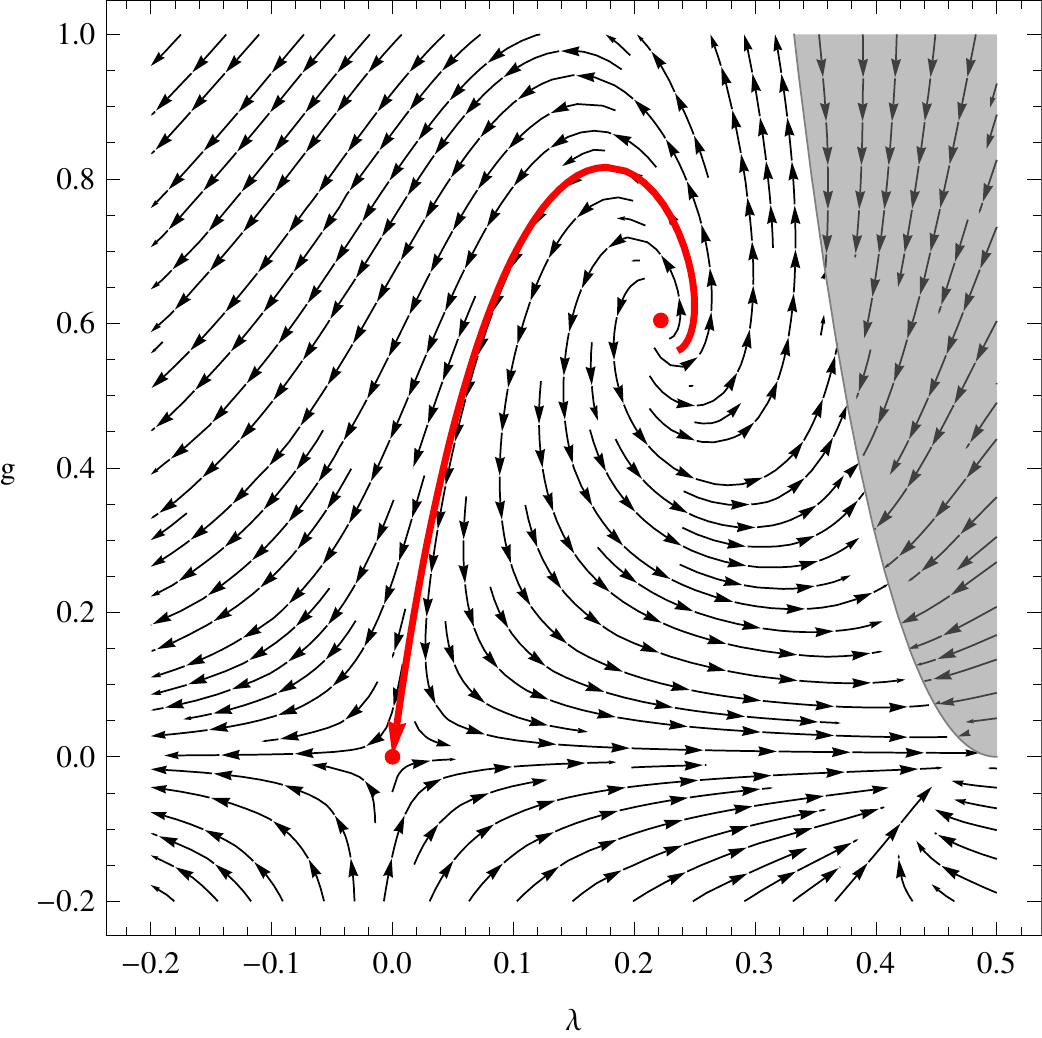}
		\label{fig:PDopt6v3}}
	\subfigure[\textbf{NGFP} for $µ^2 = 7$.]{
		\centering
		\includegraphics[width=0.3\textwidth]{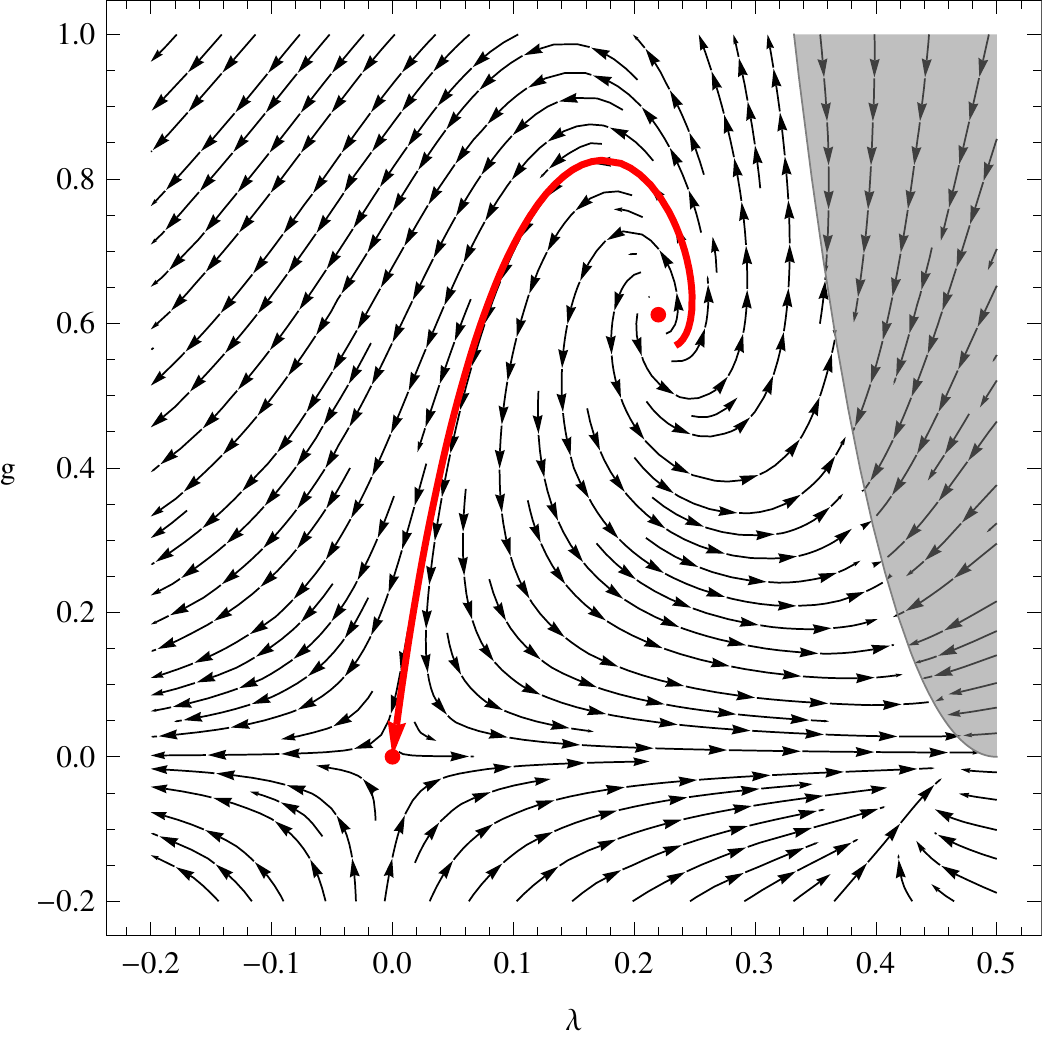}
		\label{fig:PDopt7v3}}
	\subfigure[\textbf{NGFP} for $µ^2 = 8$.]{
		\centering
		\includegraphics[width=0.3\textwidth]{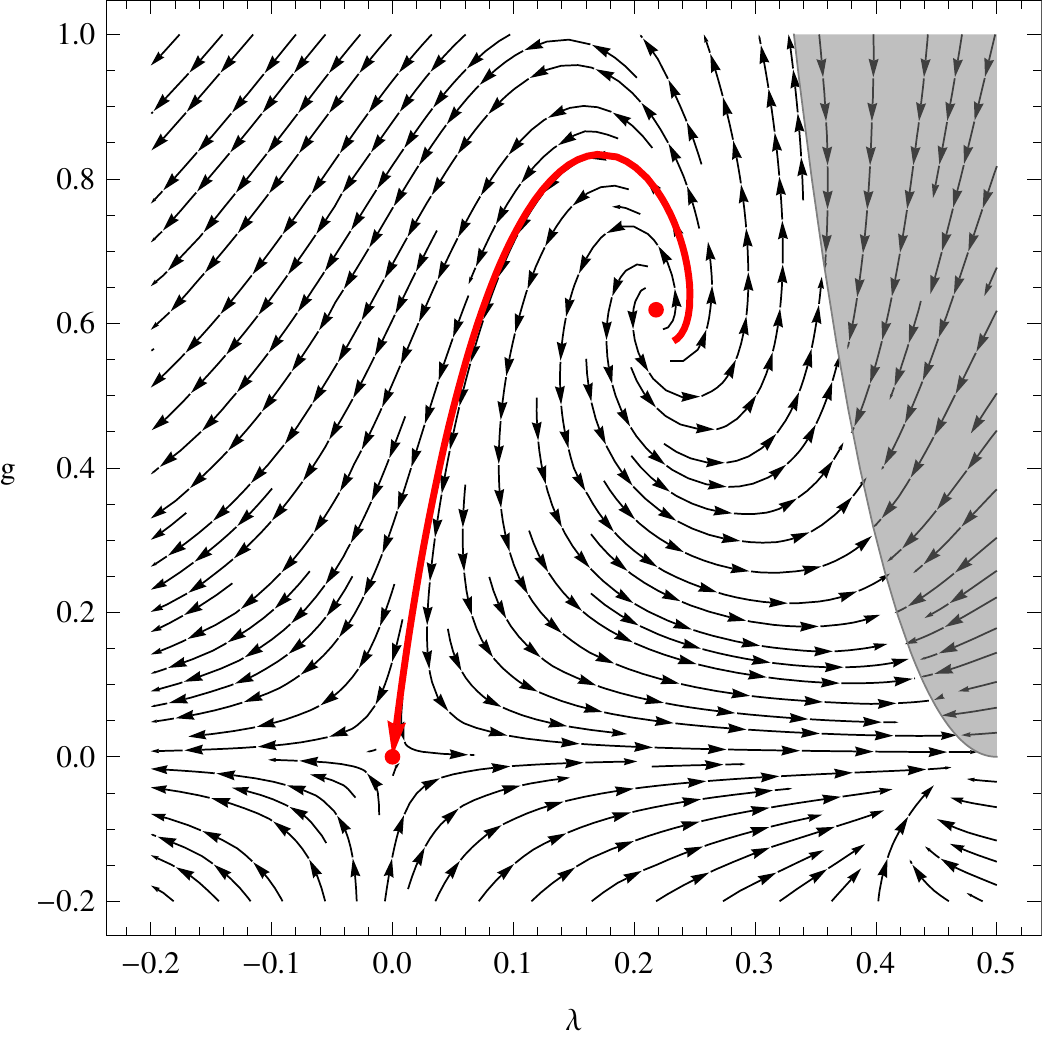}
		\label{fig:PDopt8v3}}
	\subfigure[\textbf{NGFP} for $µ^2 = 9$.]{
		\centering
		\includegraphics[width=0.3\textwidth]{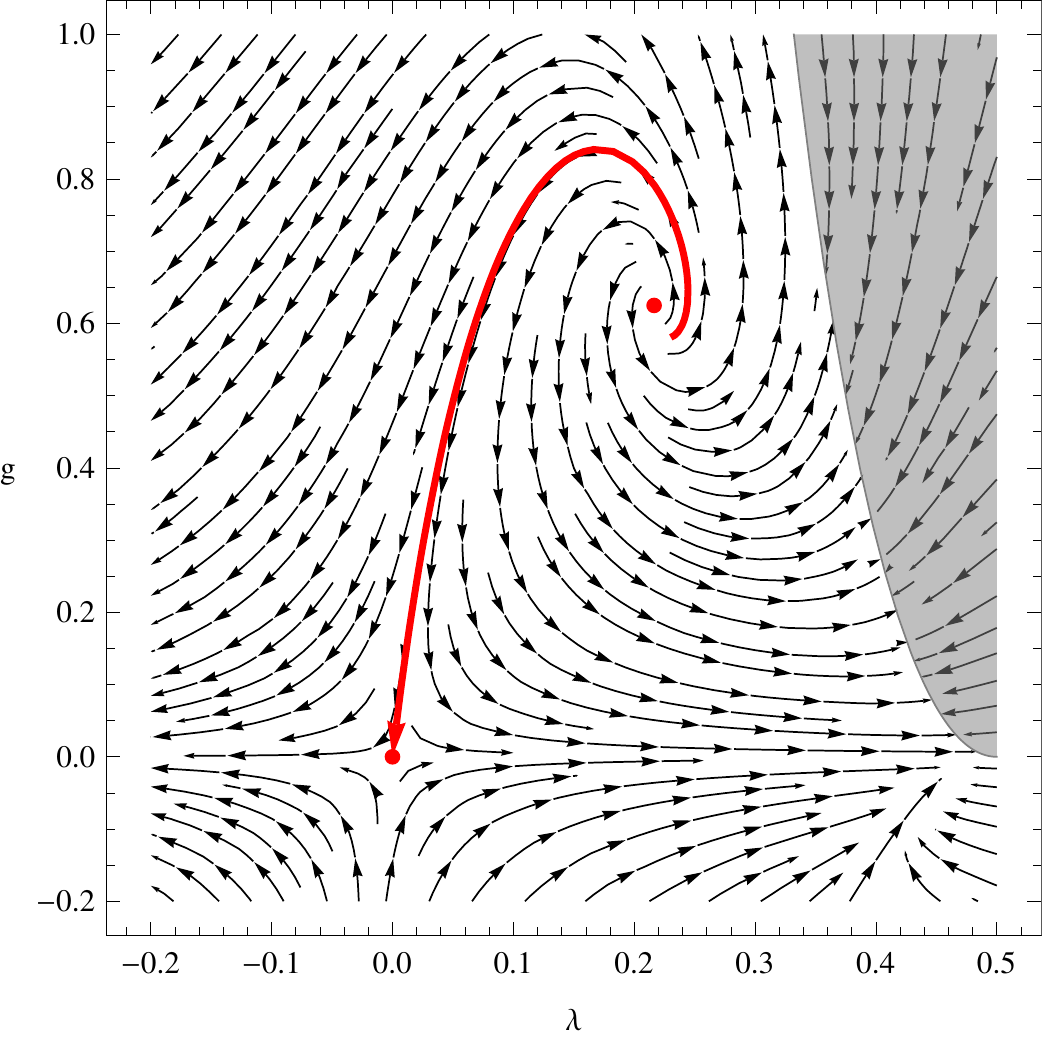}
		\label{fig:PDopt9v3}}
	\subfigure[\textbf{NGFP} for $µ^2 = 10$.]{
		\centering
		\includegraphics[width=0.3\textwidth]{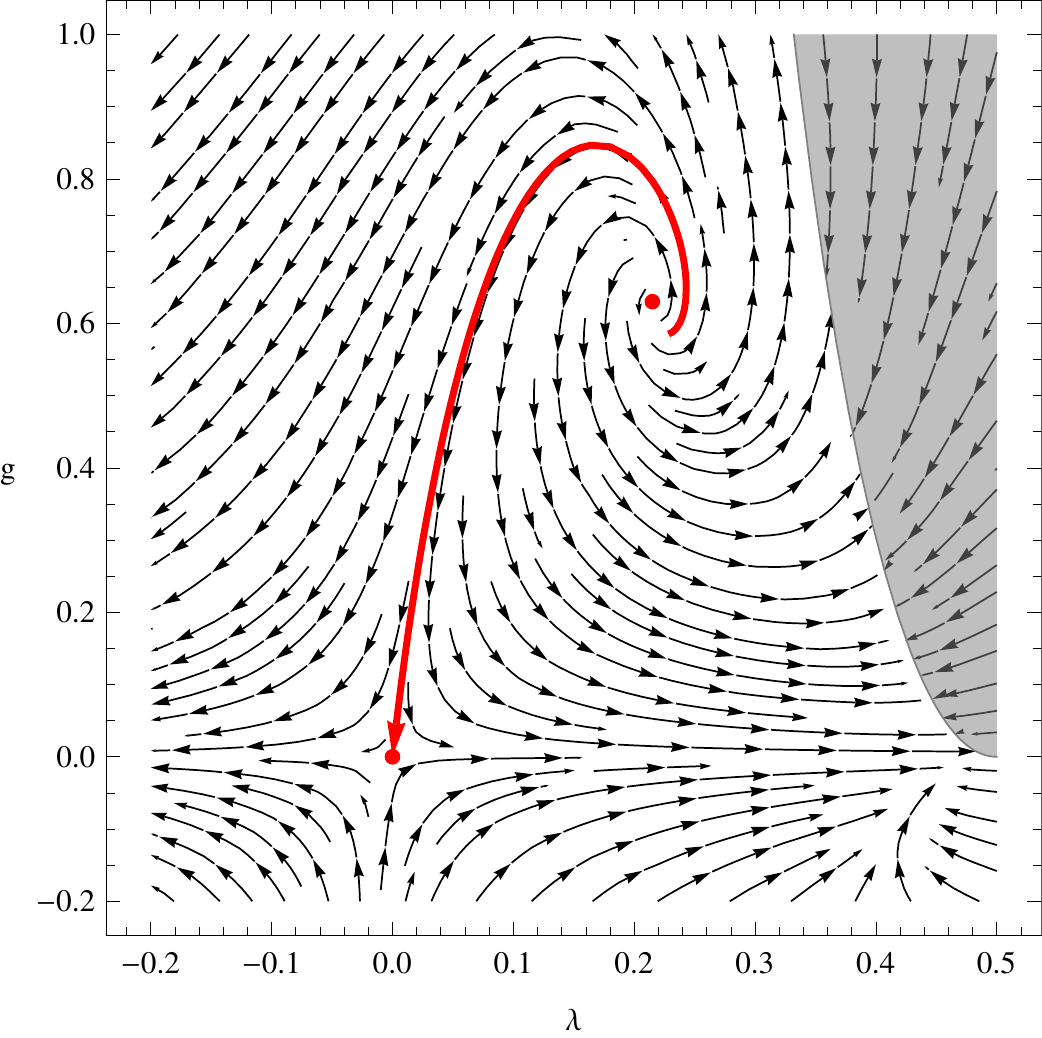}
		\label{fig:PDopt10v3}}
	\subfigure[\textbf{NGFP} for $µ^2 = 25$.]{
		\centering
		\includegraphics[width=0.3\textwidth]{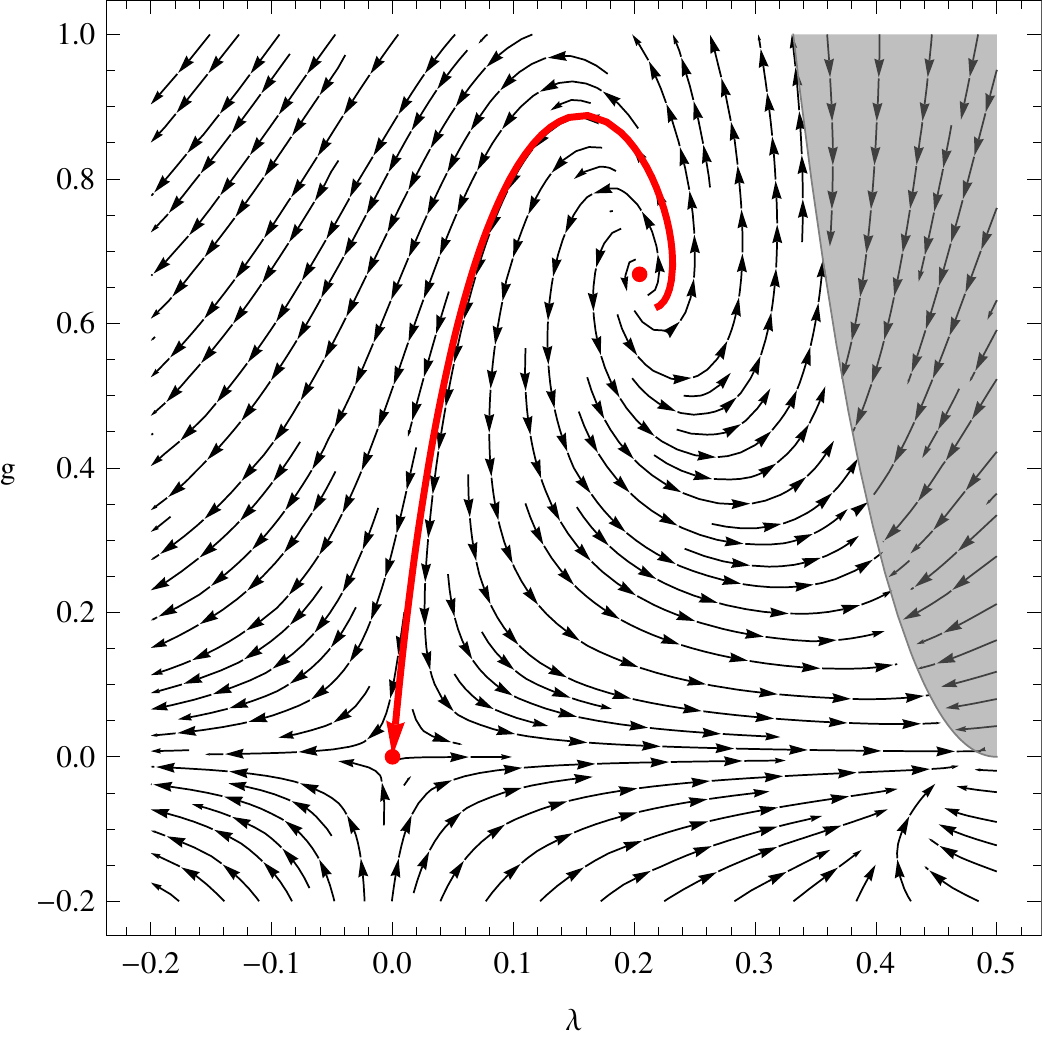}
		\label{fig:PDopt25v3}}
	\subfigure[\textbf{NGFP} for $µ^2 = 50$.]{
		\centering
		\includegraphics[width=0.3\textwidth]{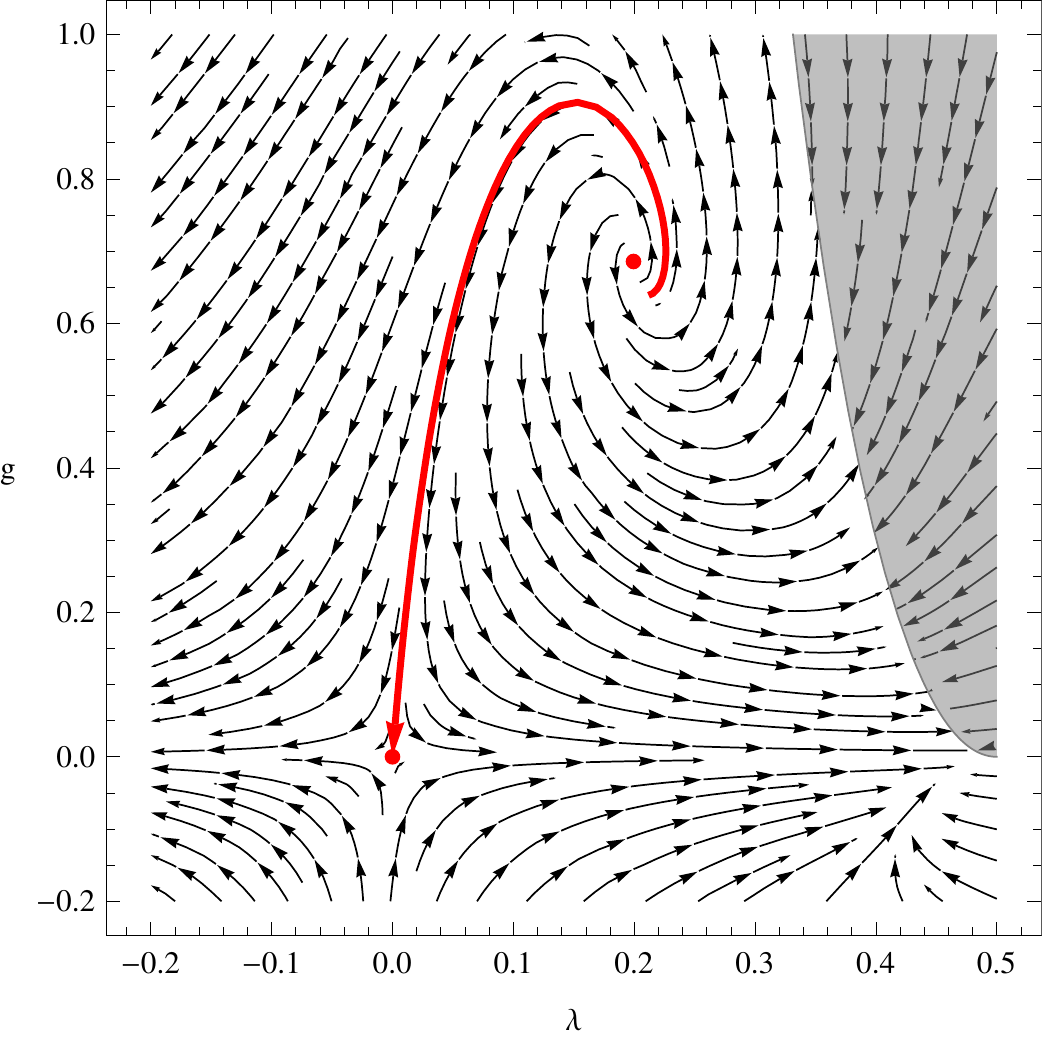}
		\label{fig:PDopt50v3}}
	\subfigure[\textbf{NGFP} for $µ^2 = 100$.]{
		\centering
		\includegraphics[width=0.3\textwidth]{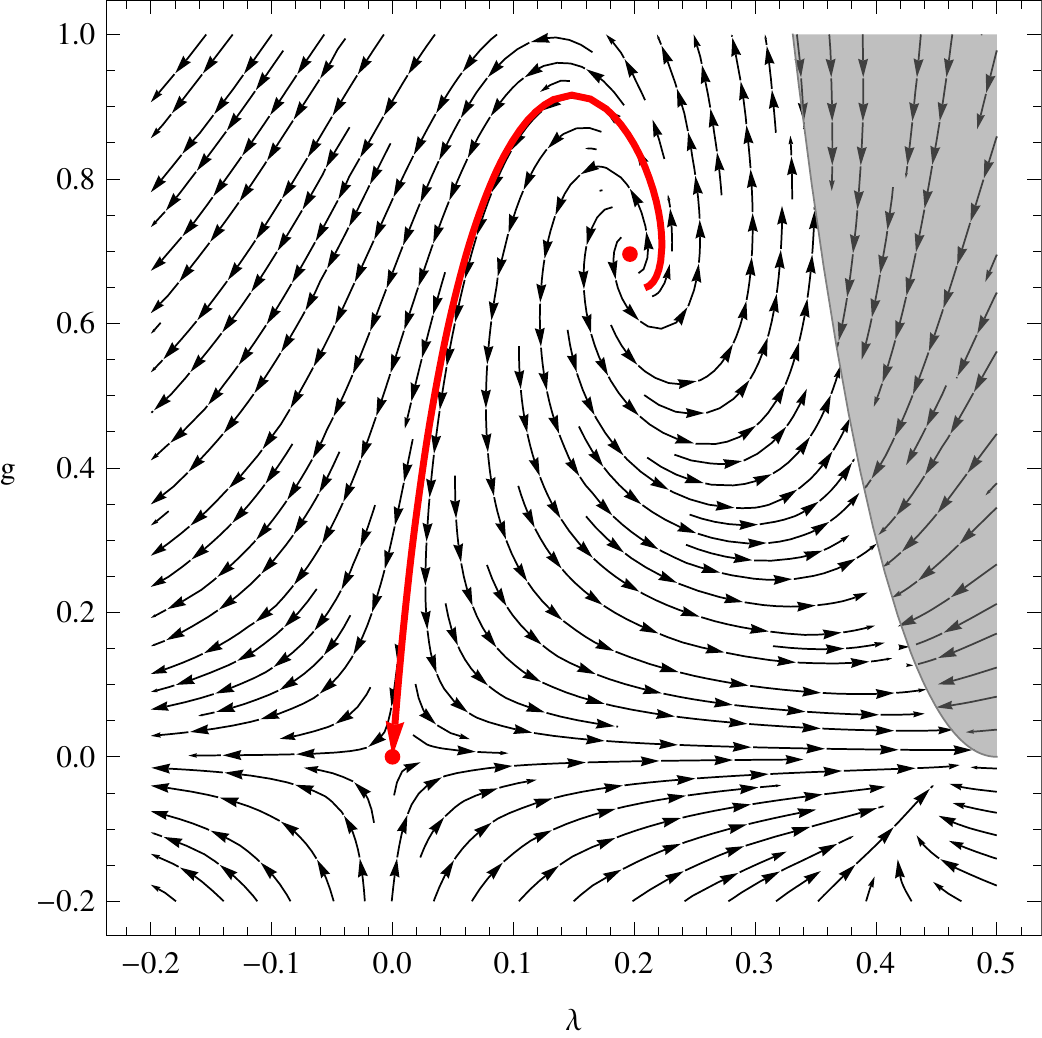}
		\label{fig:PDopt100v3}}
	\caption{RG phase portrait for different values of the squared mass parameter $µ^2$. For larger values of $µ$ the dependence on the mass parameter weakens considerably.}
	\label{fig:NGFPfullTor2}
\end{figure}
\clearpage

\begin{figure}[htbp]
	\centering
	\subfigure[\textbf{NGFP} for $s = \frac{1}{2}$.]{
		\centering
		\includegraphics[width=0.3\textwidth]{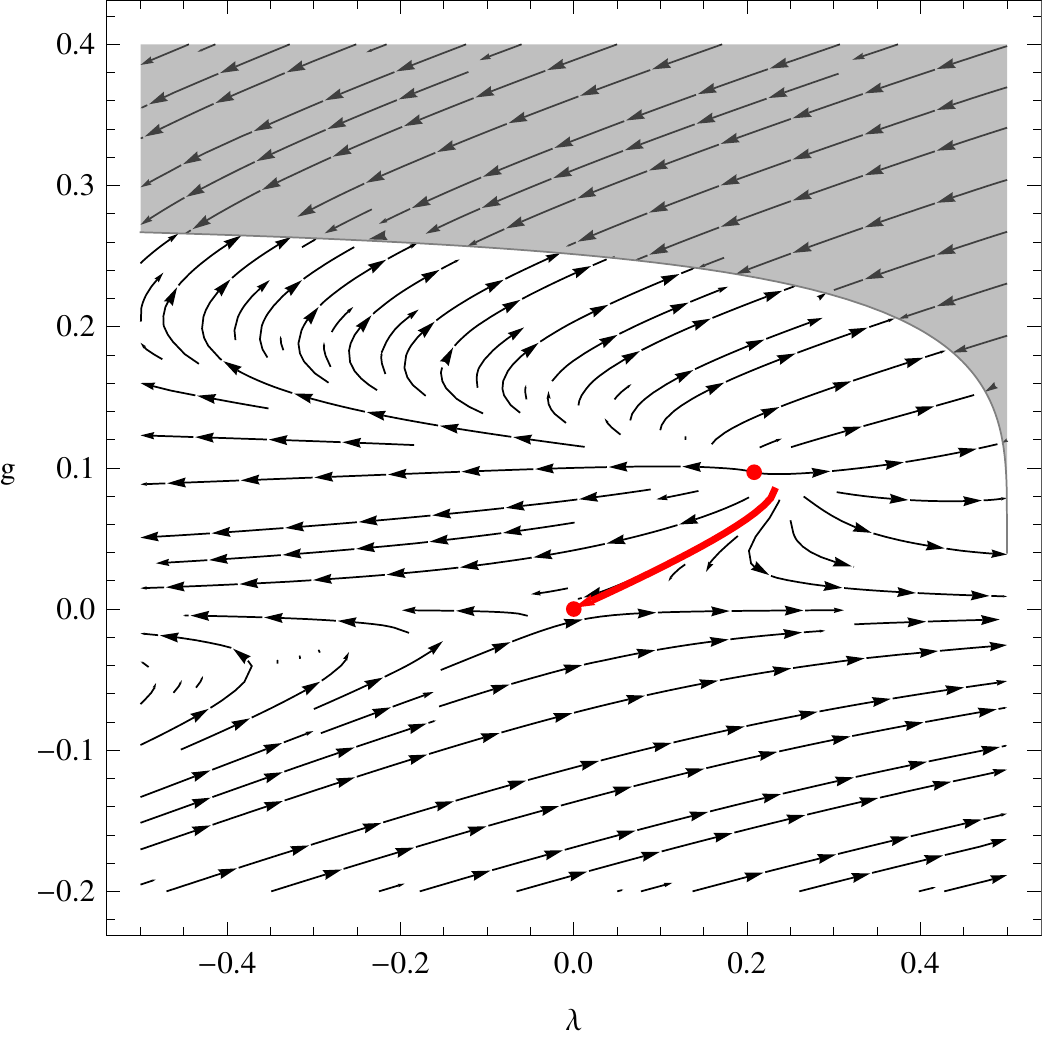}
		\label{fig:PDm119s12v2}}
	\subfigure[\textbf{NGFP} for $s = 1$.]{
		\centering
		\includegraphics[width=0.3\textwidth]{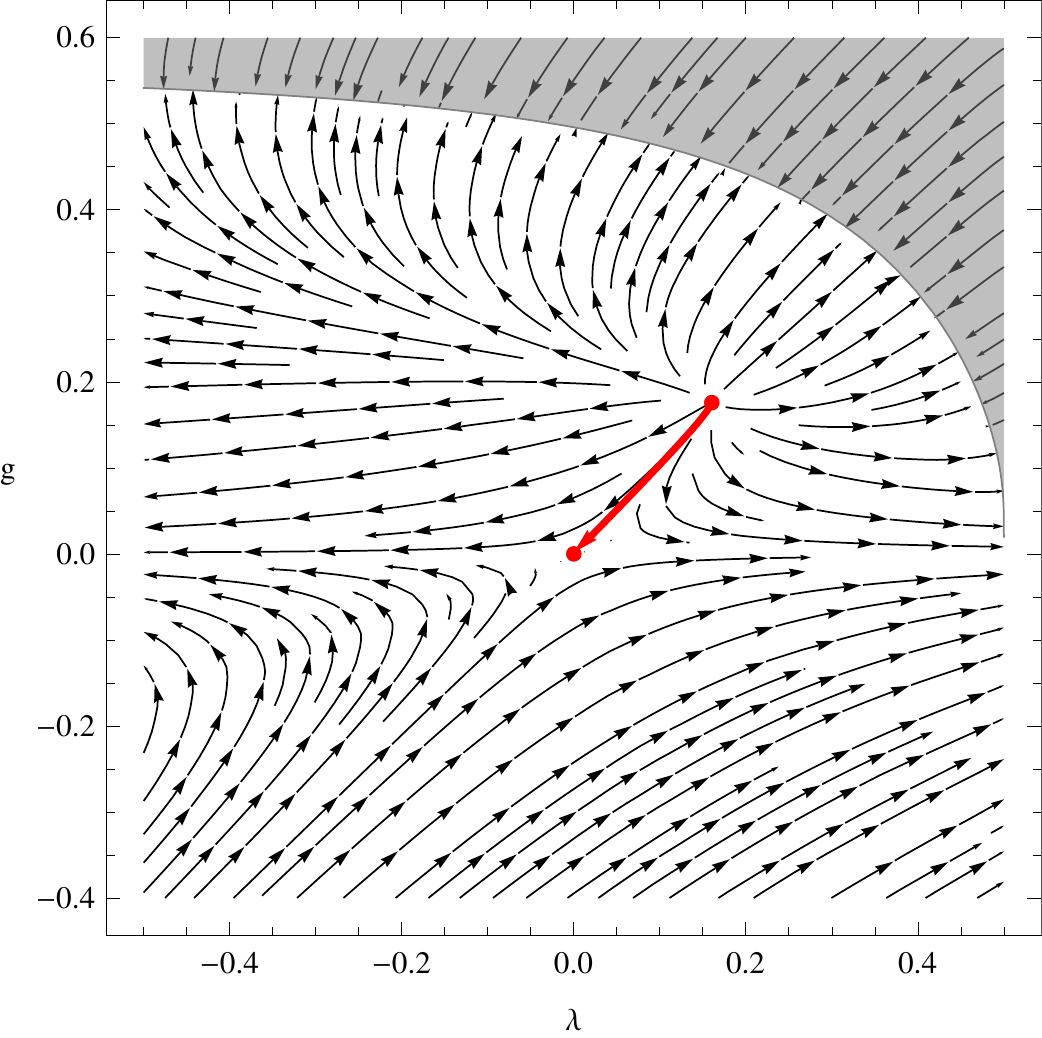}
		\label{fig:PDm119s1v2}}
	\subfigure[\textbf{NGFP} for $s = 2$.]{
		\centering
		\includegraphics[width=0.3\textwidth]{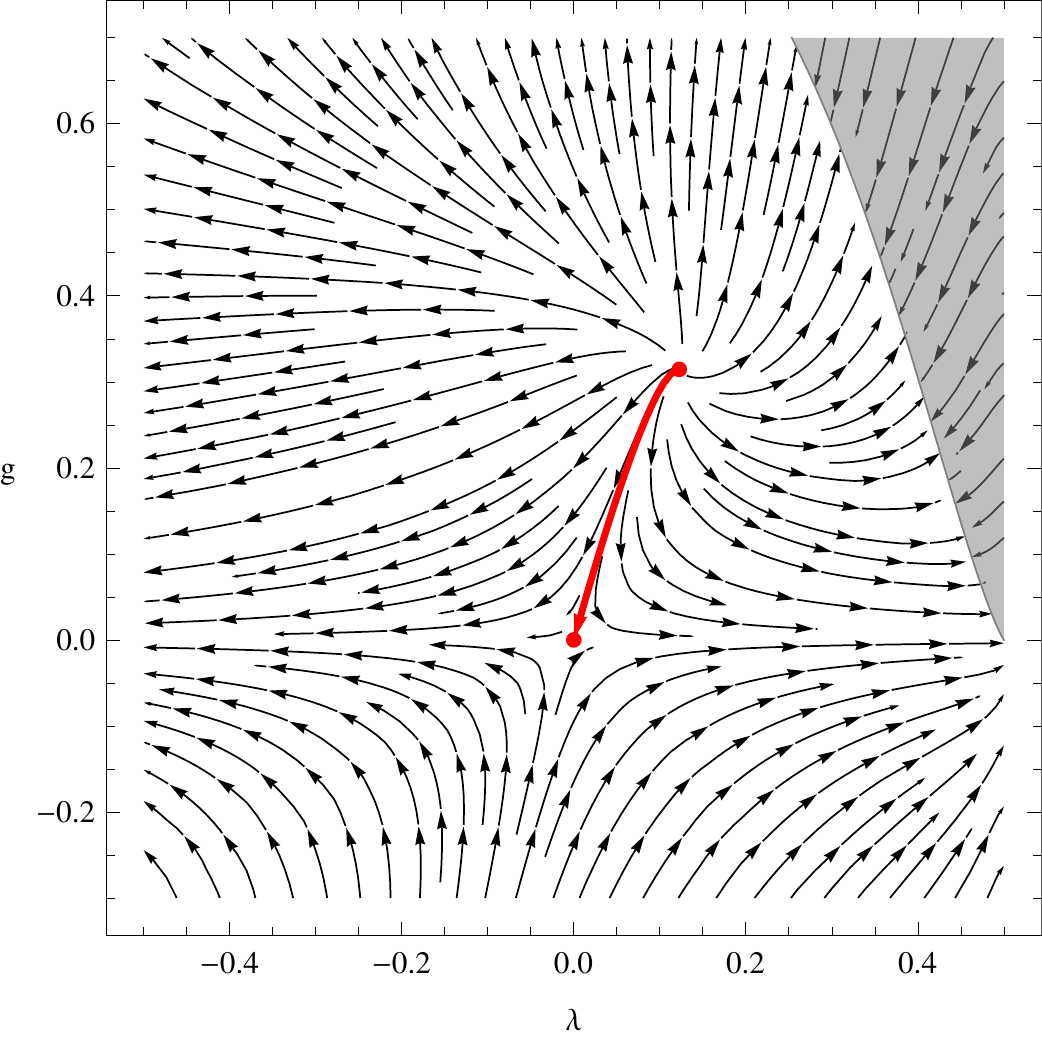}
		\label{fig:PDm119s2v2}}
	\subfigure[\textbf{NGFP} for $s = 3$.]{
		\centering
		\includegraphics[width=0.3\textwidth]{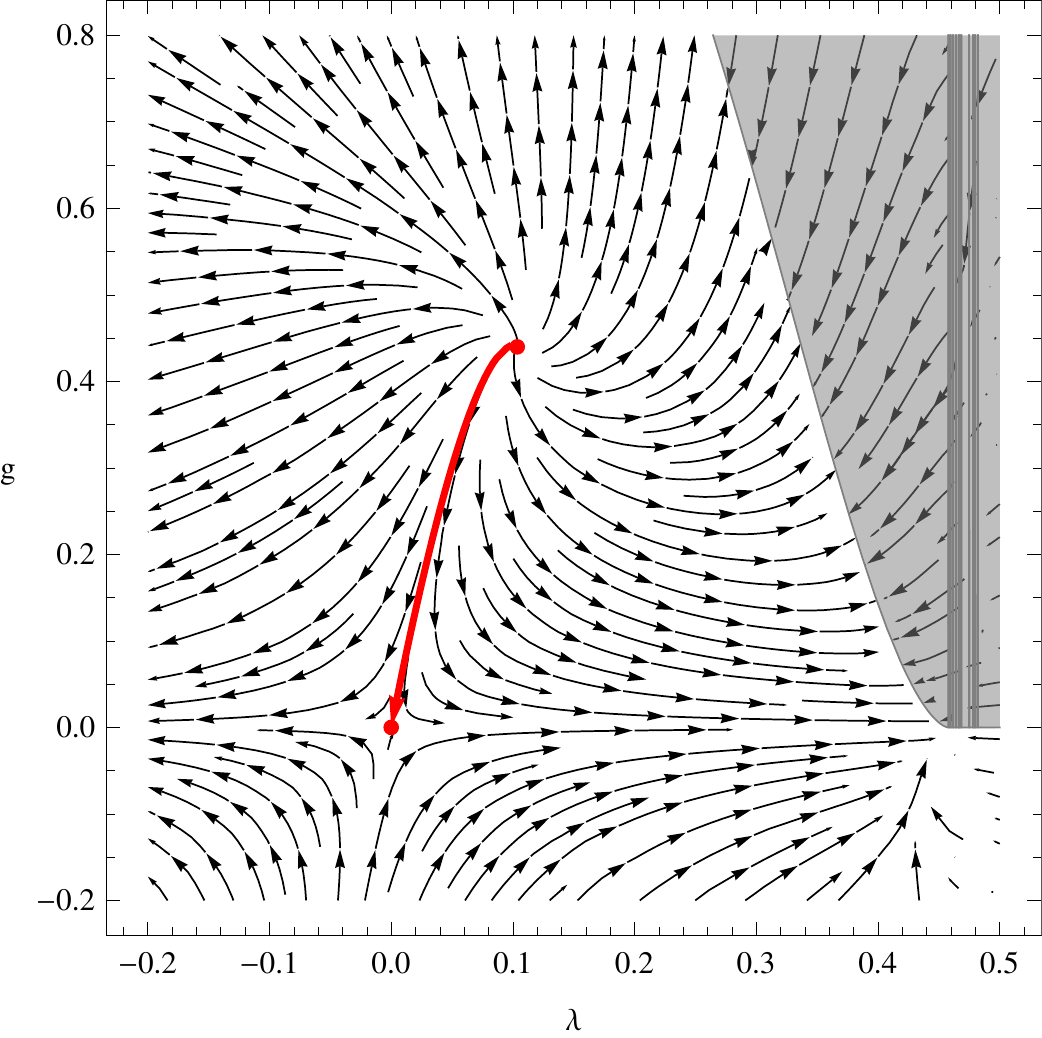}
		\label{fig:PDm119s3v2}}
	\subfigure[\textbf{NGFP} for $s = 4$.]{
		\centering
		\includegraphics[width=0.3\textwidth]{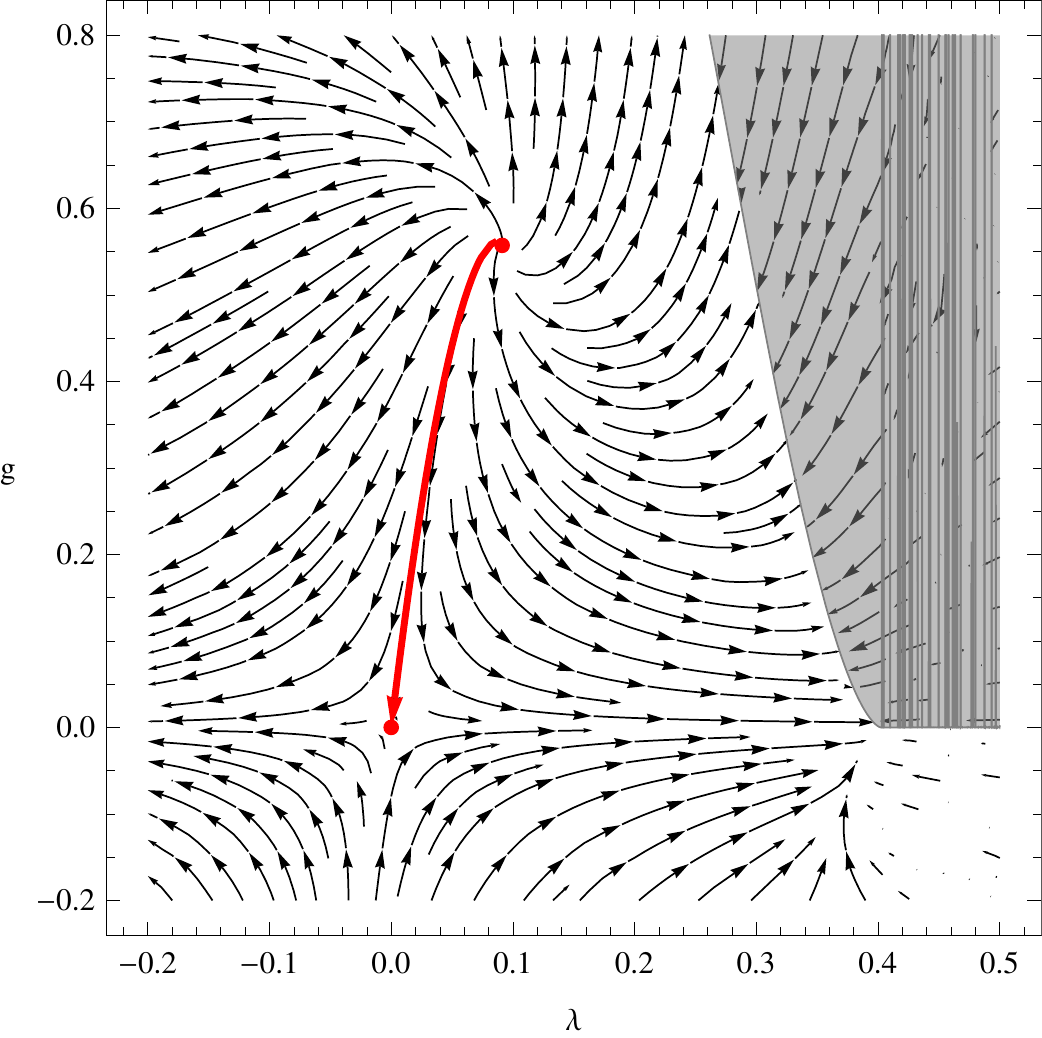}
		\label{fig:PDm119s4v2}}
	\subfigure[\textbf{NGFP} for $s = 5$.]{
		\centering
		\includegraphics[width=0.3\textwidth]{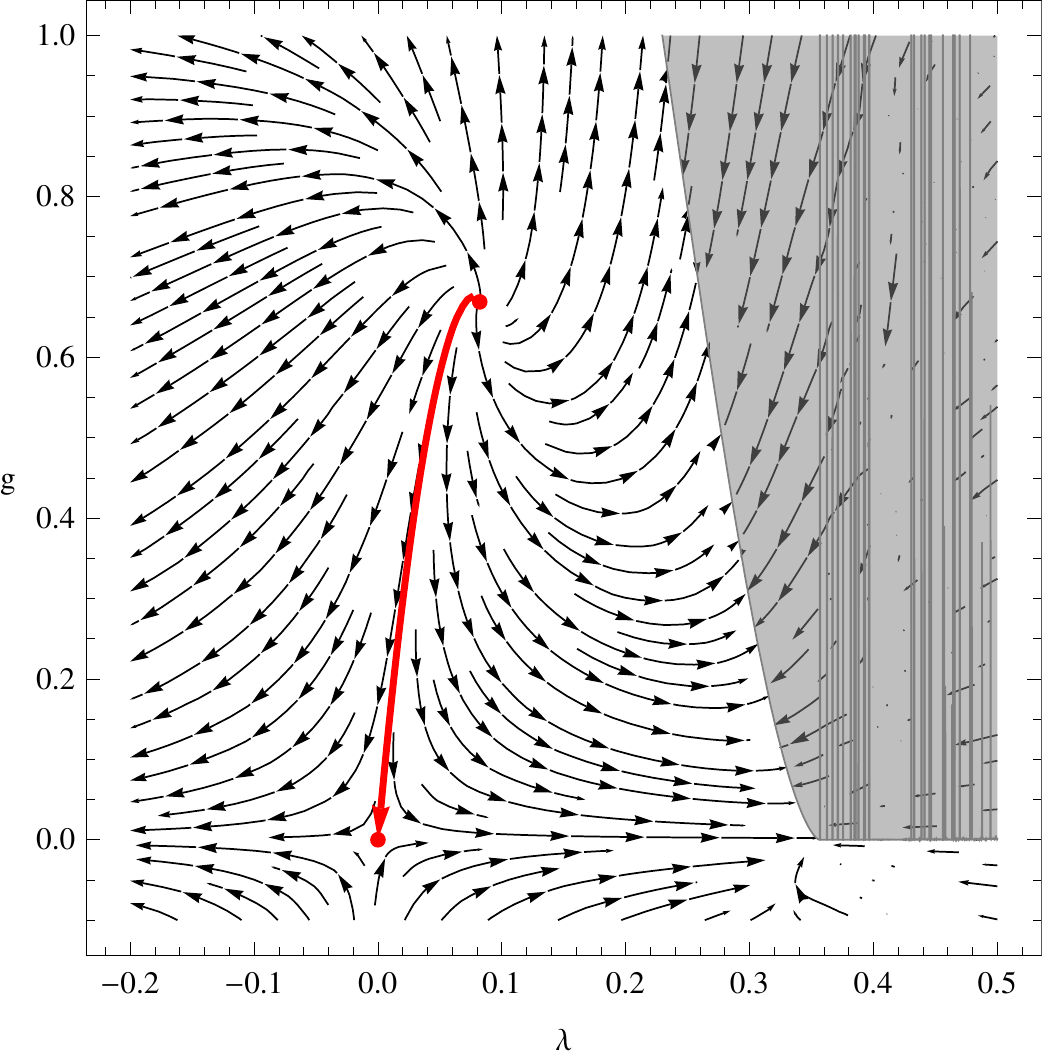}
		\label{fig:PDm119s5v2}}
	\subfigure[\textbf{NGFP} for $s = 6$.]{
		\centering
		\includegraphics[width=0.3\textwidth]{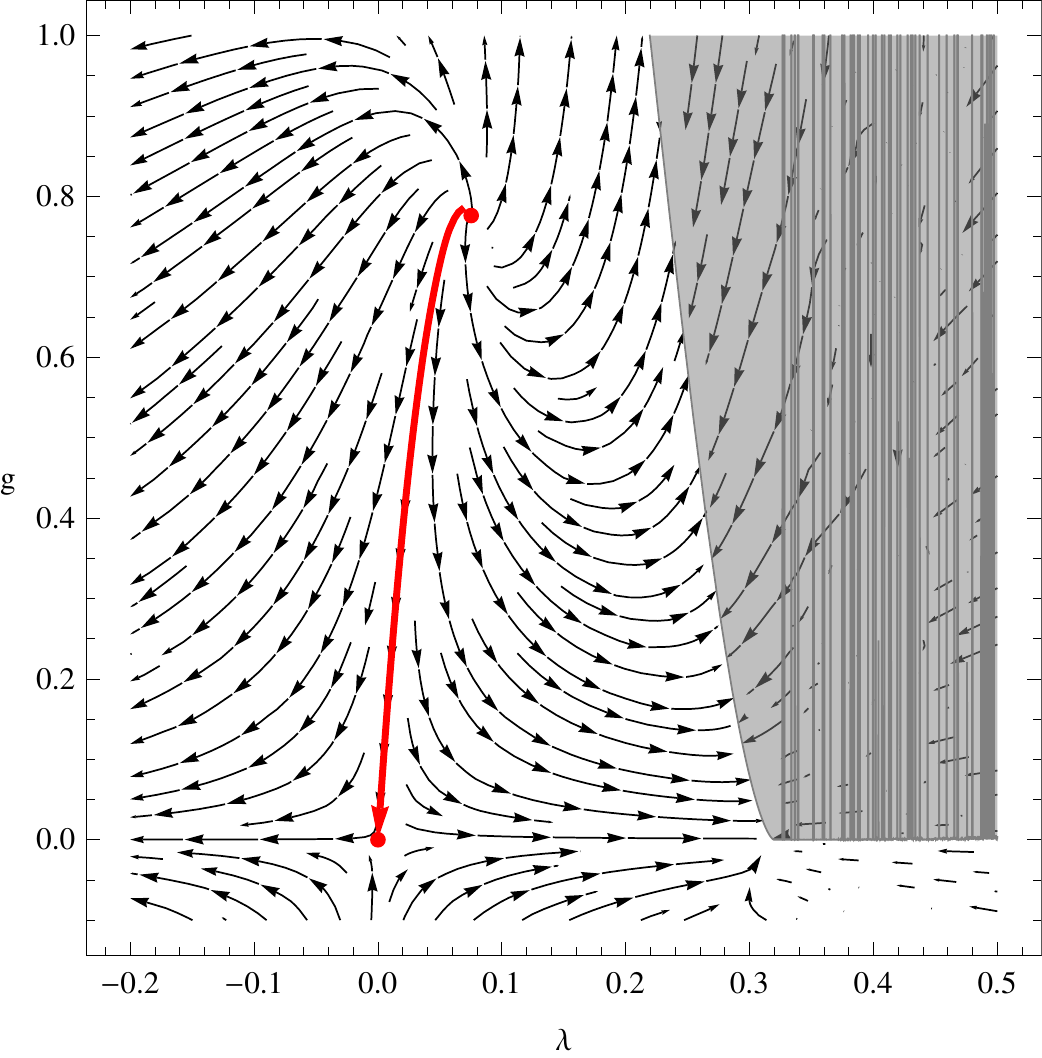}
		\label{fig:PDm119s6v2}}
	\subfigure[\textbf{NGFP} for $s = 7$.]{
		\centering
		\includegraphics[width=0.3\textwidth]{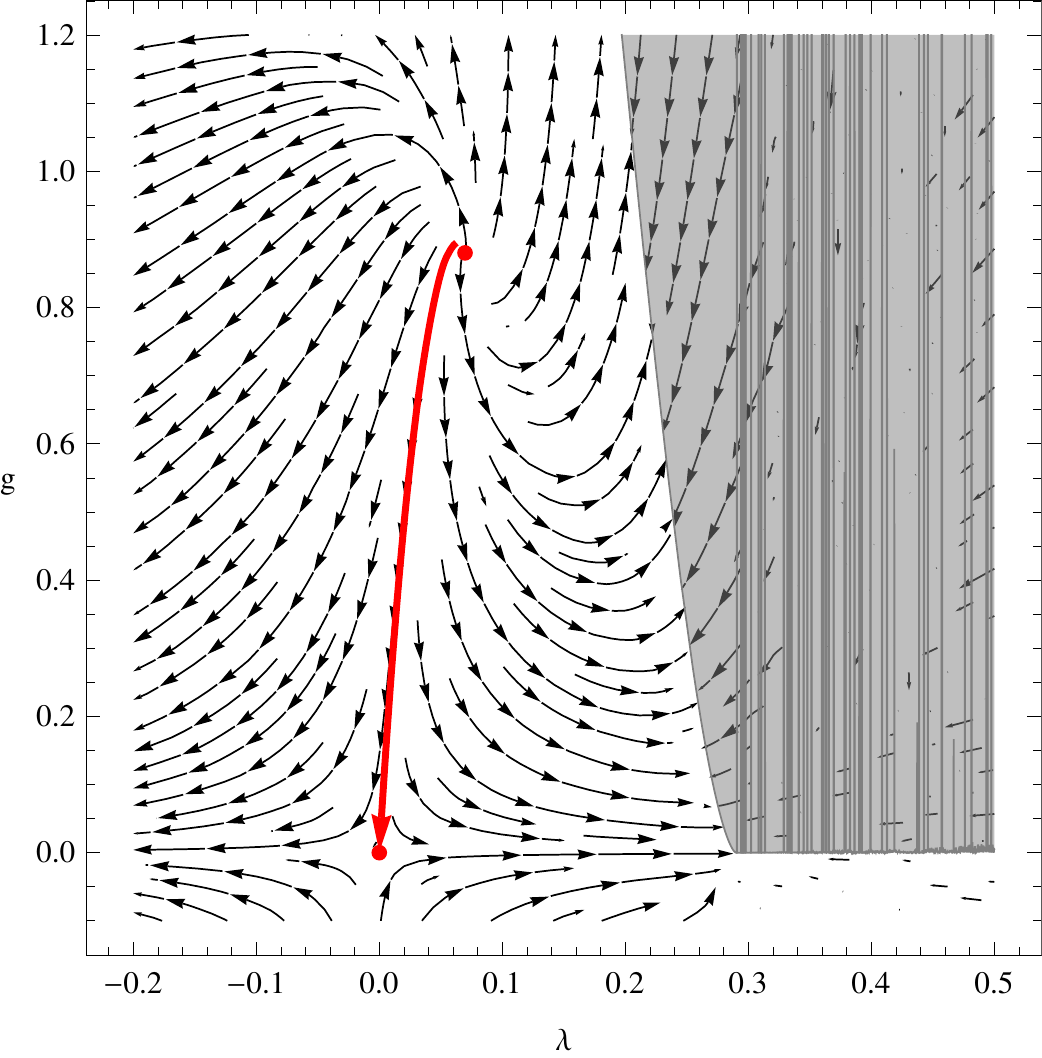}
		\label{fig:PDm119s7v2}}
	\subfigure[\textbf{NGFP} for $s = 8$.]{
		\centering
		\includegraphics[width=0.3\textwidth]{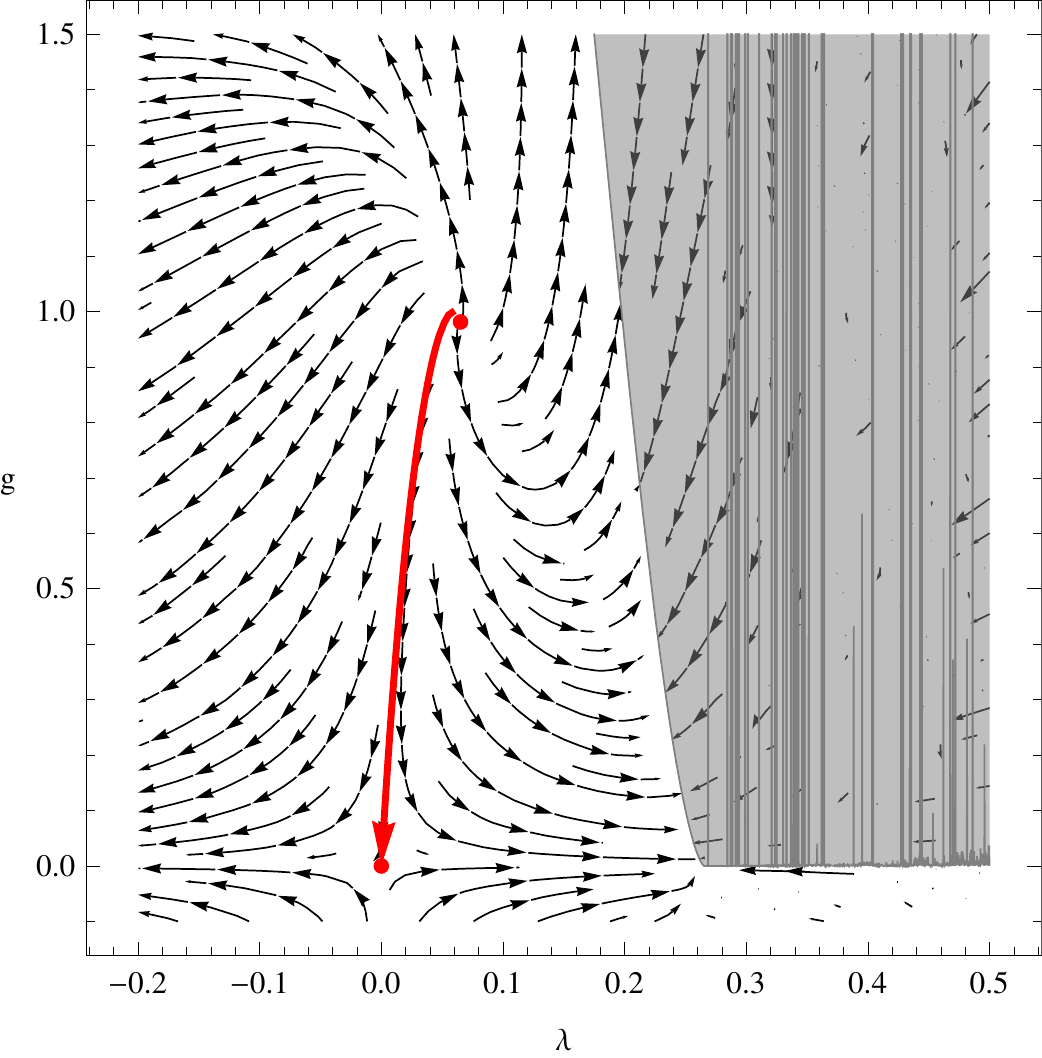}
		\label{fig:PDm119s8v2}}
	\subfigure[\textbf{NGFP} for $s = 9$.]{
		\centering
		\includegraphics[width=0.3\textwidth]{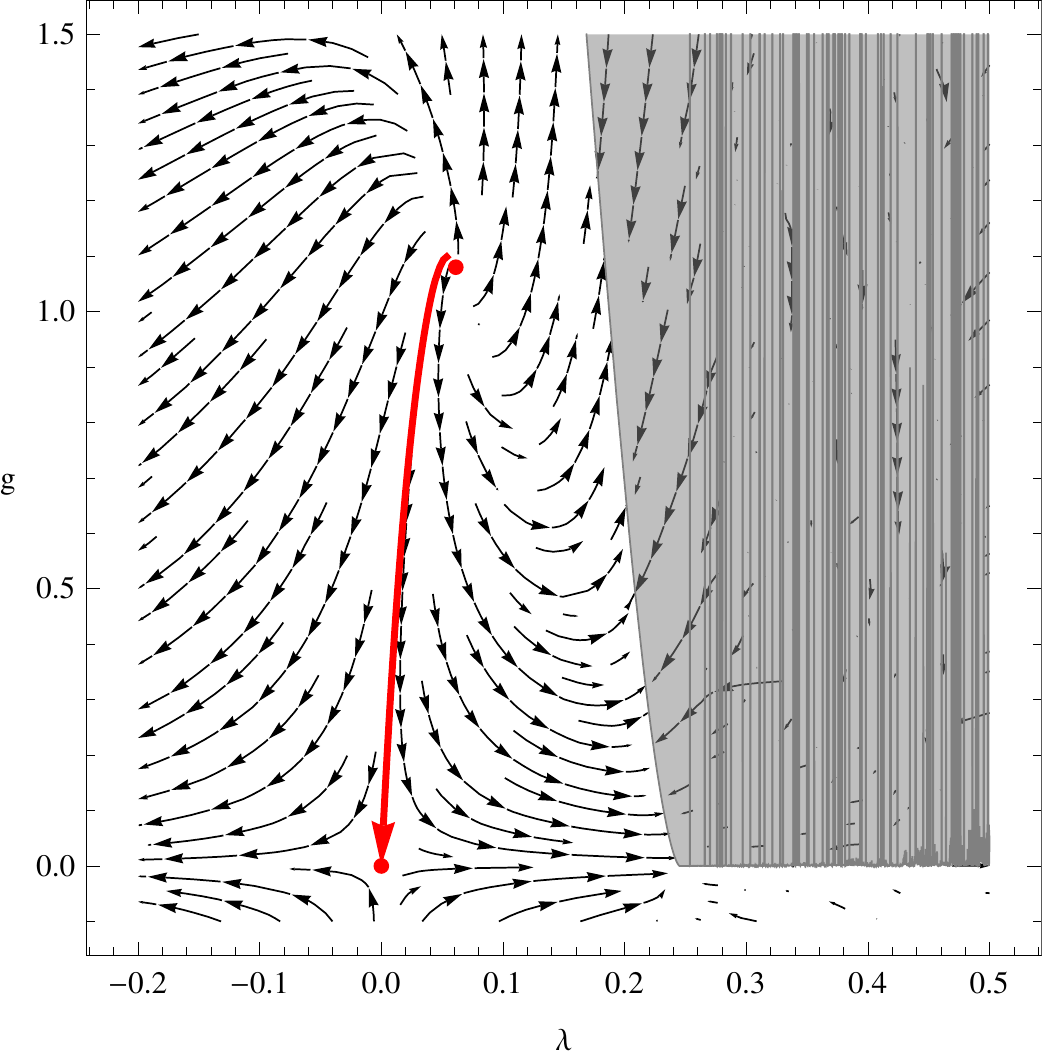}
		\label{fig:PDm119s9v2}}
	\subfigure[\textbf{NGFP} for $s = 10$.]{
		\centering
		\includegraphics[width=0.3\textwidth]{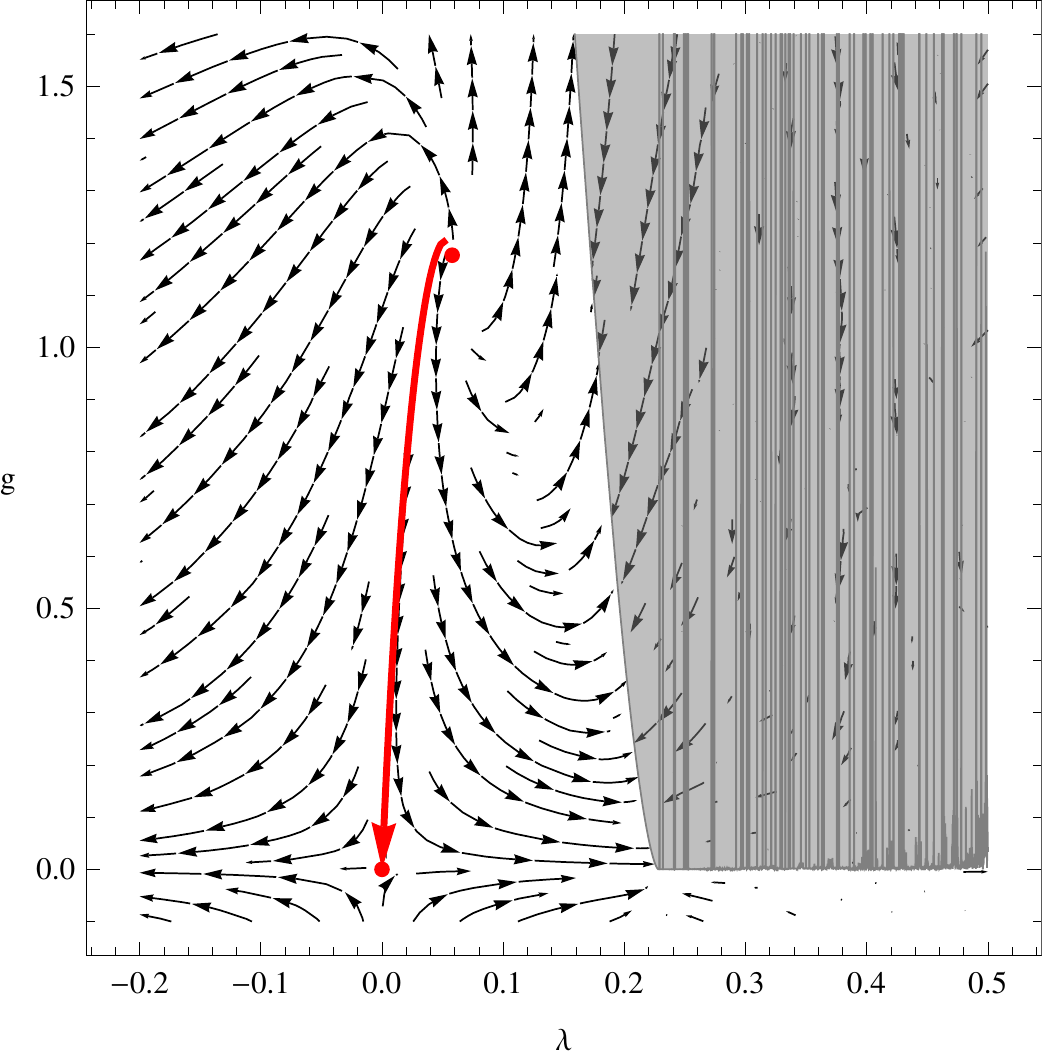}
		\label{fig:PDm119s10v2}}
	\subfigure[\textbf{NGFP} for $s = 20$.]{
		\centering
		\includegraphics[width=0.3\textwidth]{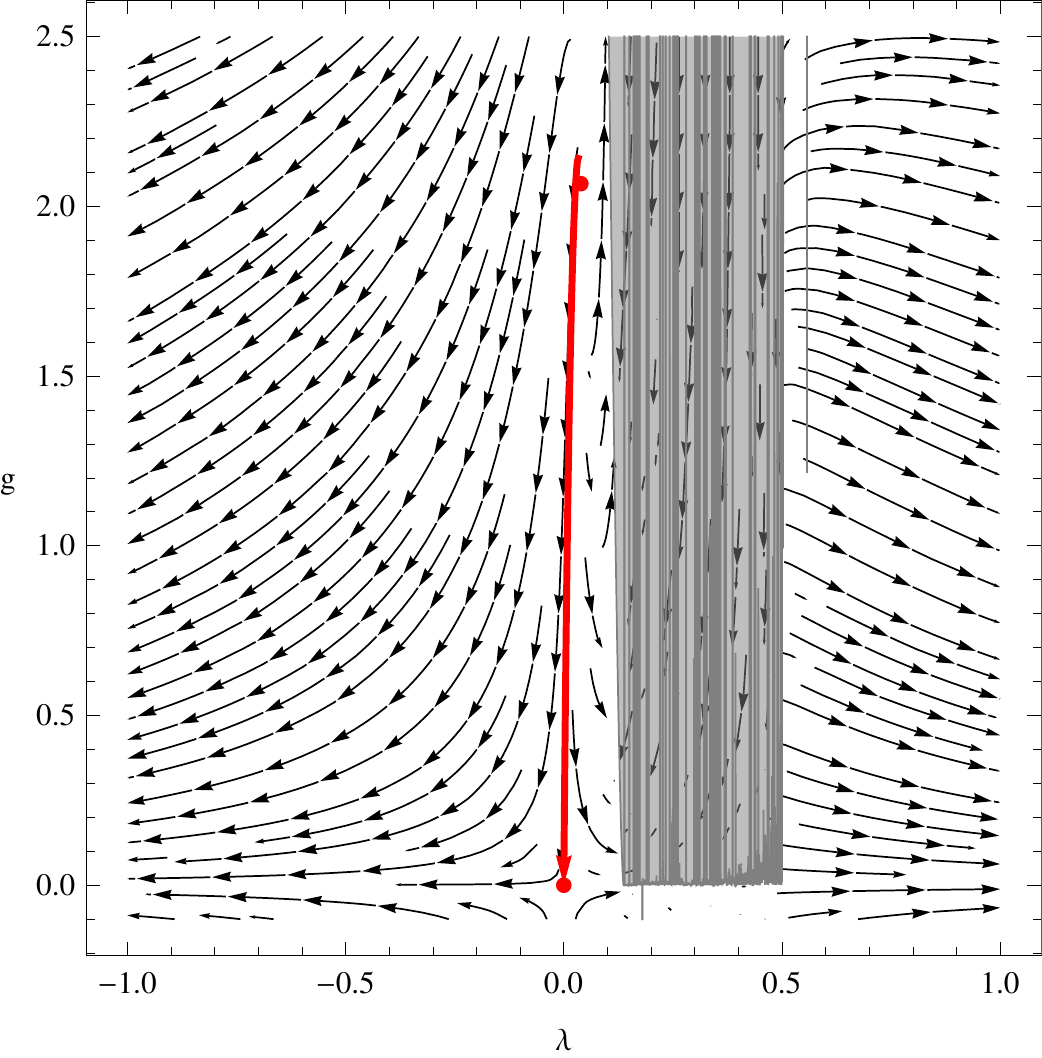}
		\label{fig:PDm119s20v3}}
	\caption{RG phase portrait for the squared mass parameter $µ^2 = \frac{1}{1.9}$ and different values of the shape parameter $s$, ranging from $s=\frac{1}{2}$ to $s=20$.}
	\label{fig:NGFPm119}
\end{figure}
\clearpage

\begin{figure}[htbp]
	\centering
	\subfigure[\textbf{NGFP} for $s = \frac{1}{2}$.]{
		\centering
		\includegraphics[width=0.3\textwidth]{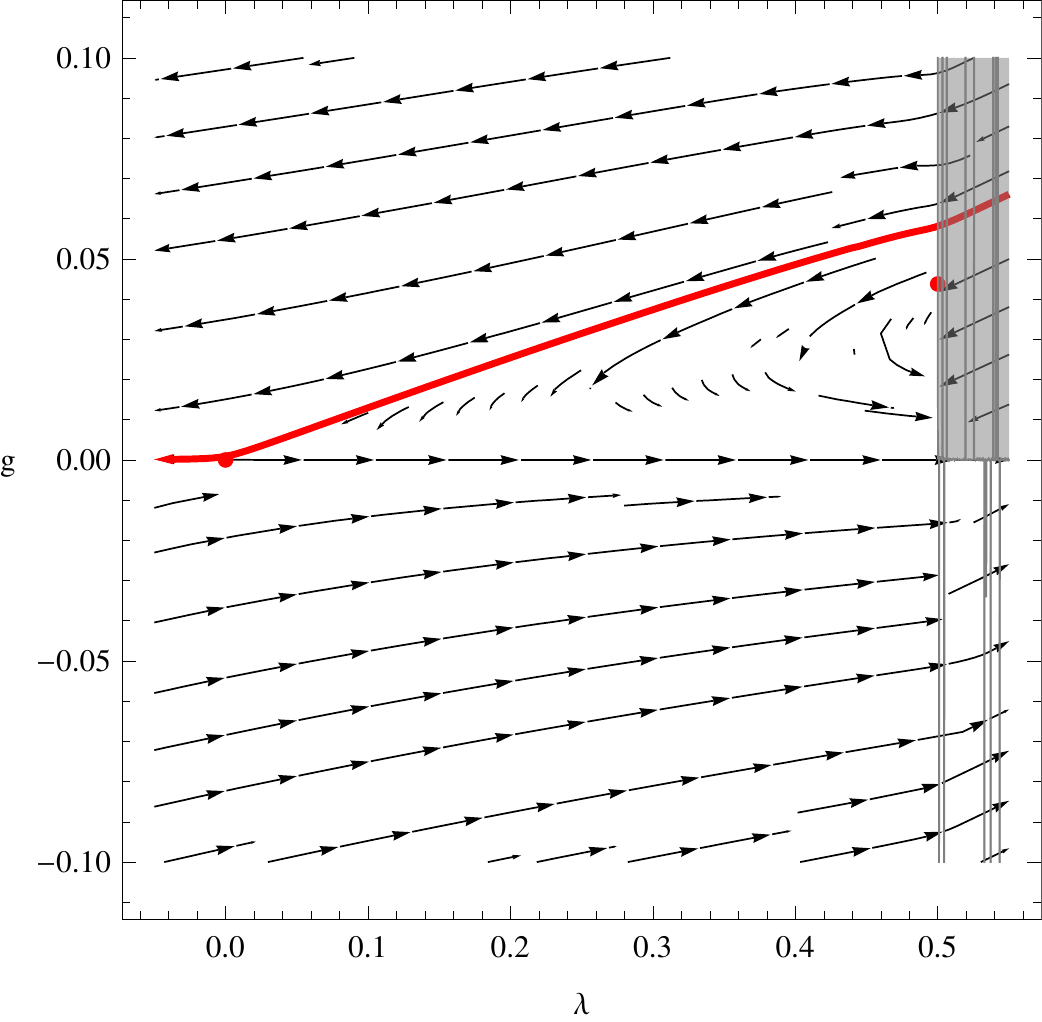}
		\label{fig:PDm1s12}}
	\subfigure[\textbf{NGFP} for $s = 1$.]{
		\centering
		\includegraphics[width=0.3\textwidth]{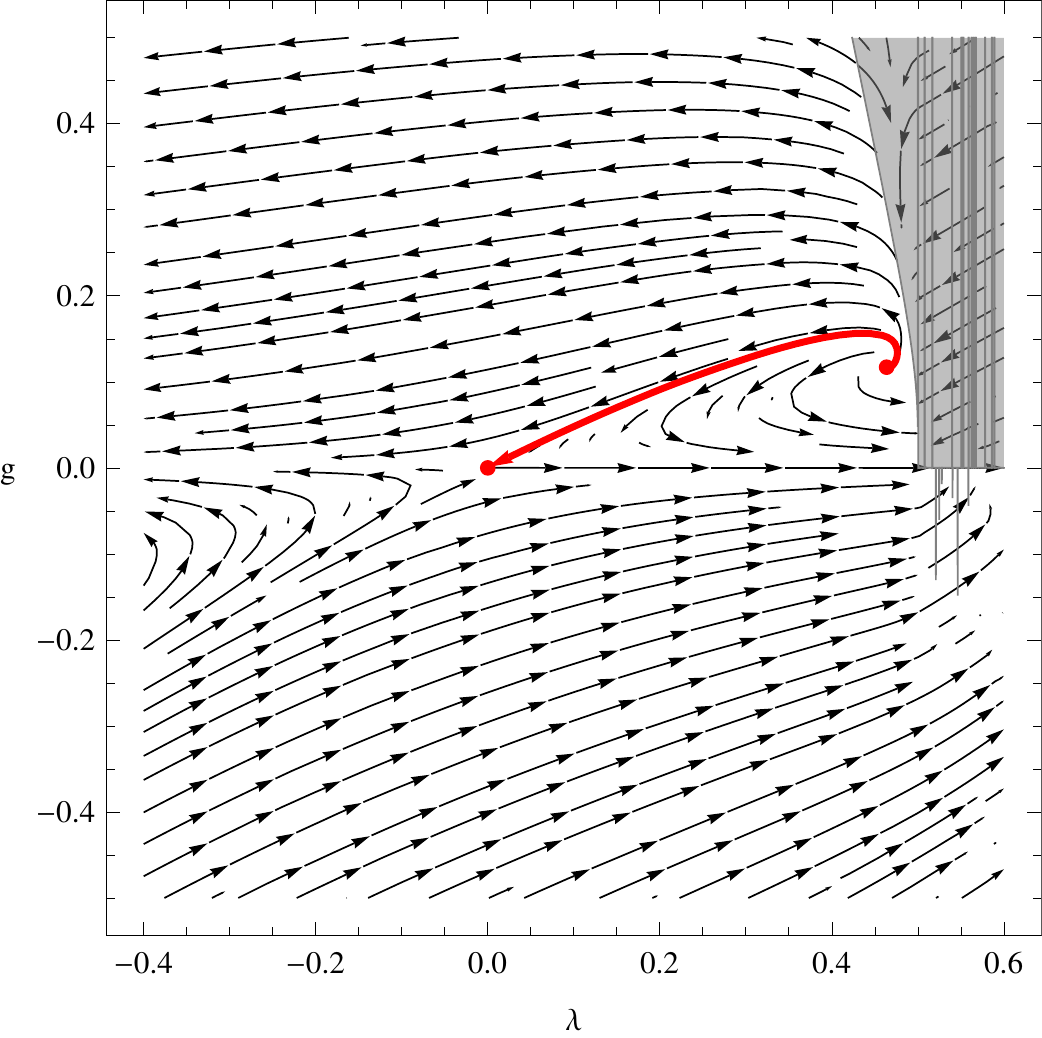}
		\label{fig:PDm1s1}}
	\subfigure[\textbf{NGFP} for $s = 2$.]{
		\centering
		\includegraphics[width=0.3\textwidth]{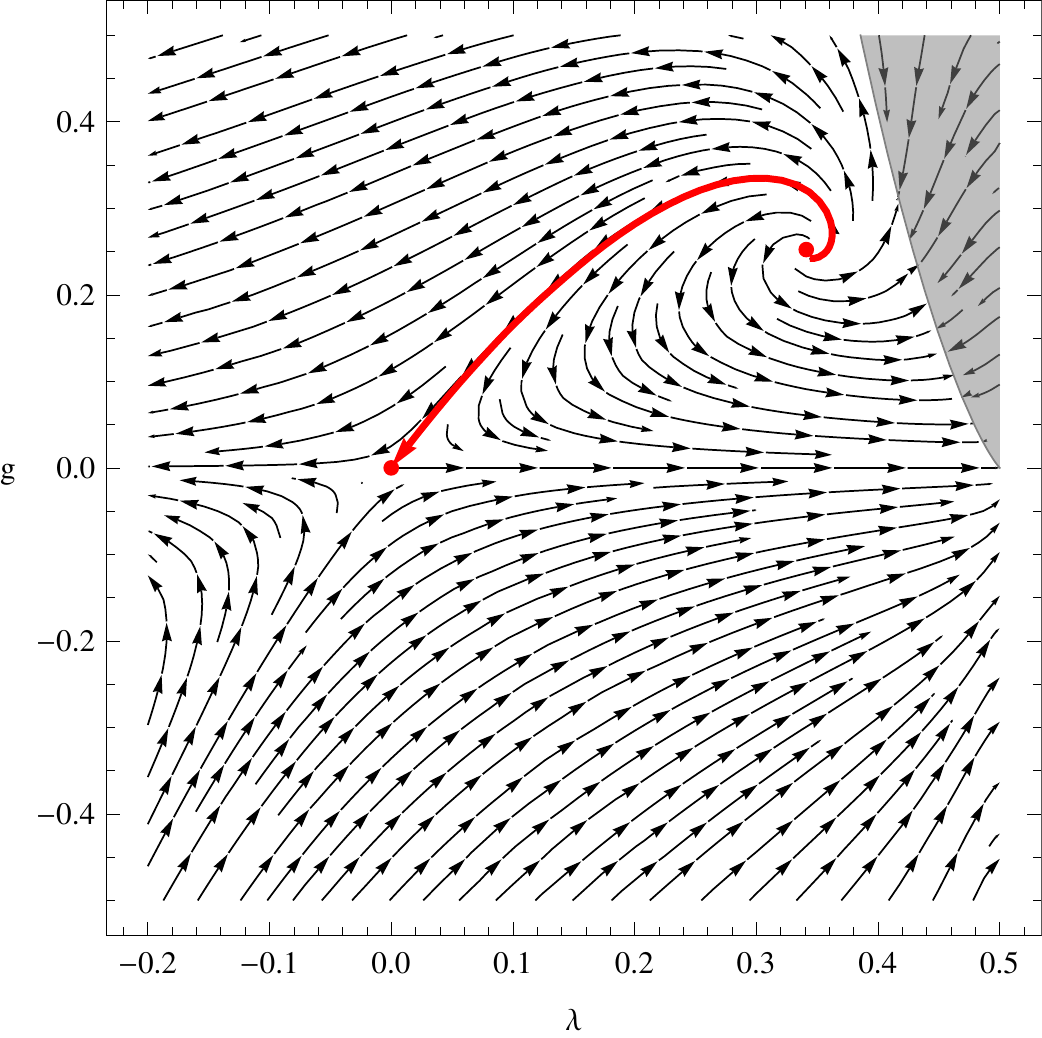}
		\label{fig:PDm1s2v2}}
	\subfigure[\textbf{NGFP} for $s = 3$.]{
		\centering
		\includegraphics[width=0.3\textwidth]{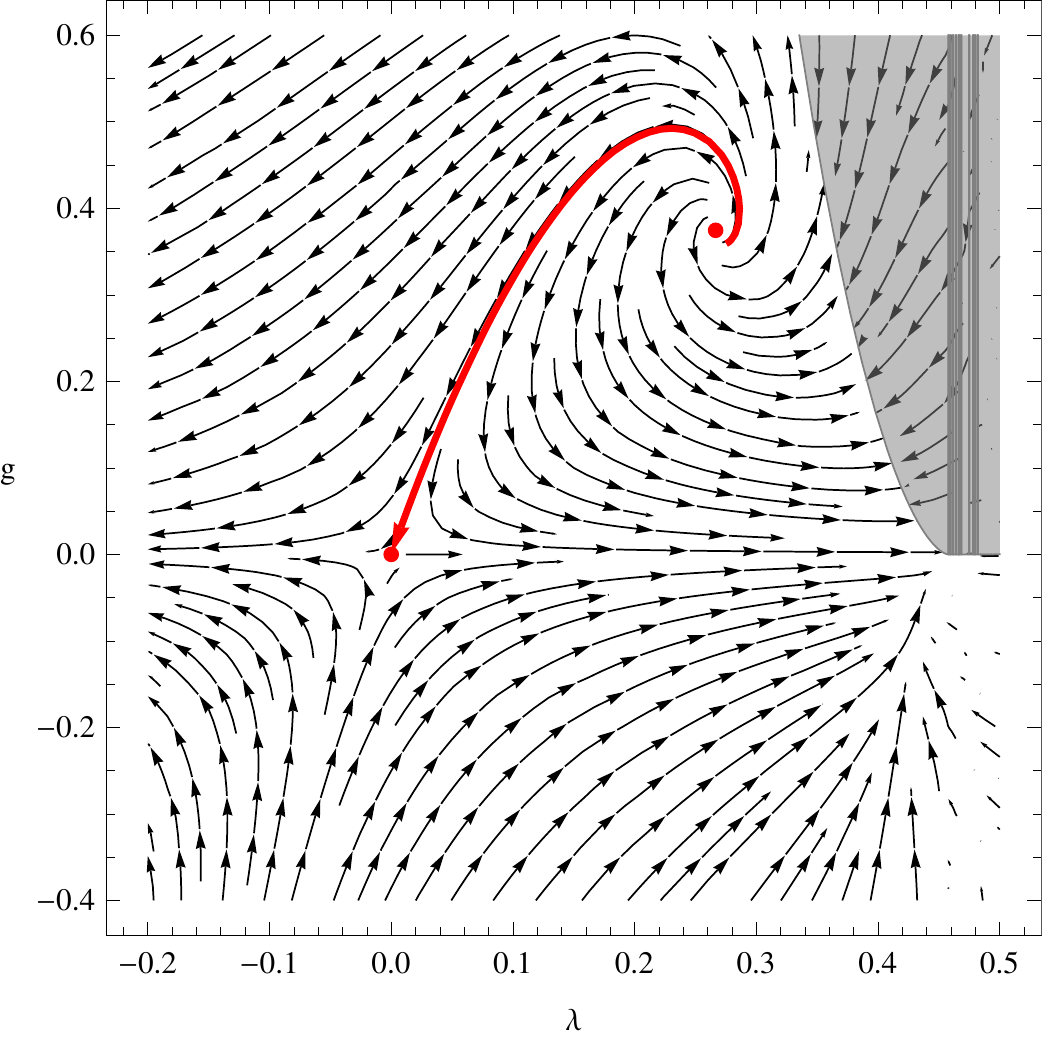}
		\label{fig:PDm1s3v2}}
	\subfigure[\textbf{NGFP} for $s = 4$.]{
		\centering
		\includegraphics[width=0.3\textwidth]{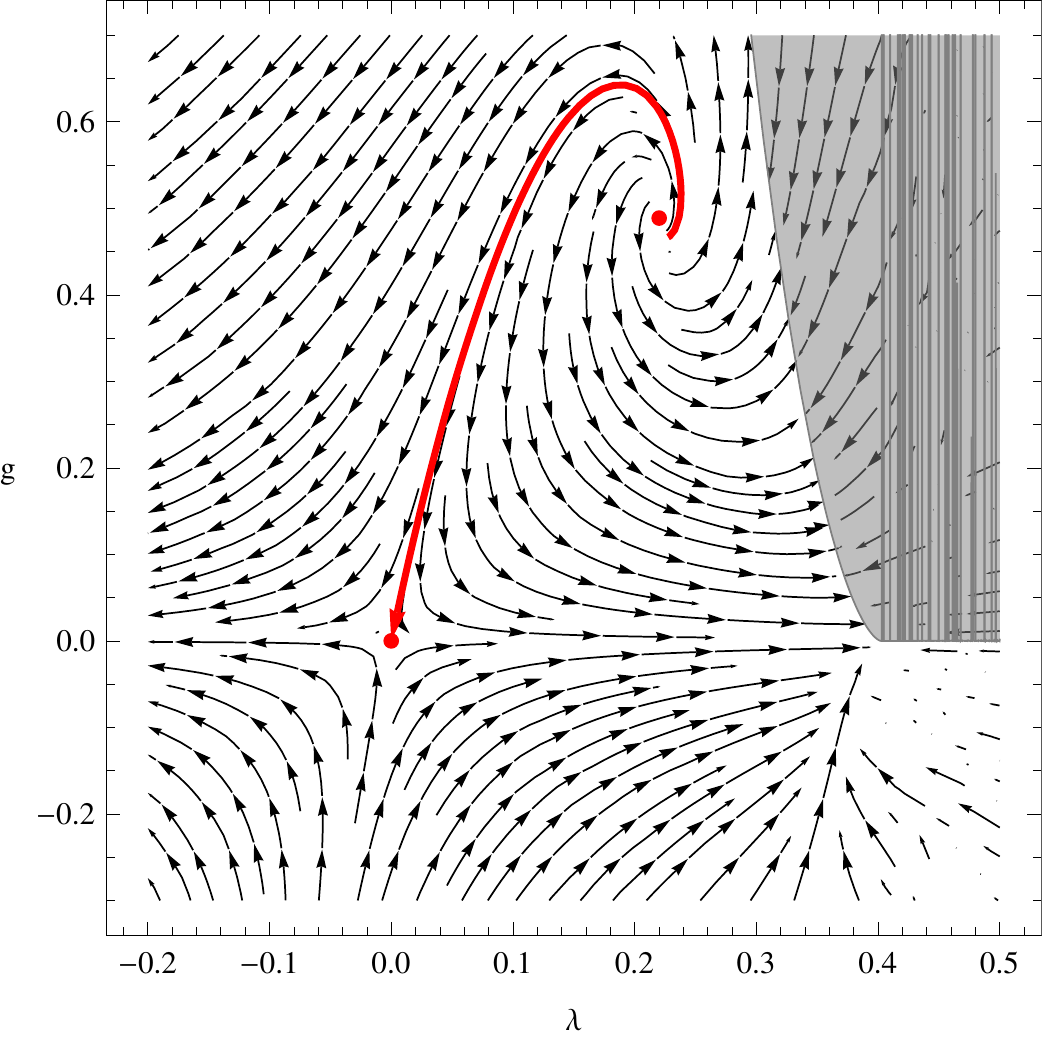}
		\label{fig:PDm1s4v2}}
	\subfigure[\textbf{NGFP} for $s = 5$.]{
		\centering
		\includegraphics[width=0.3\textwidth]{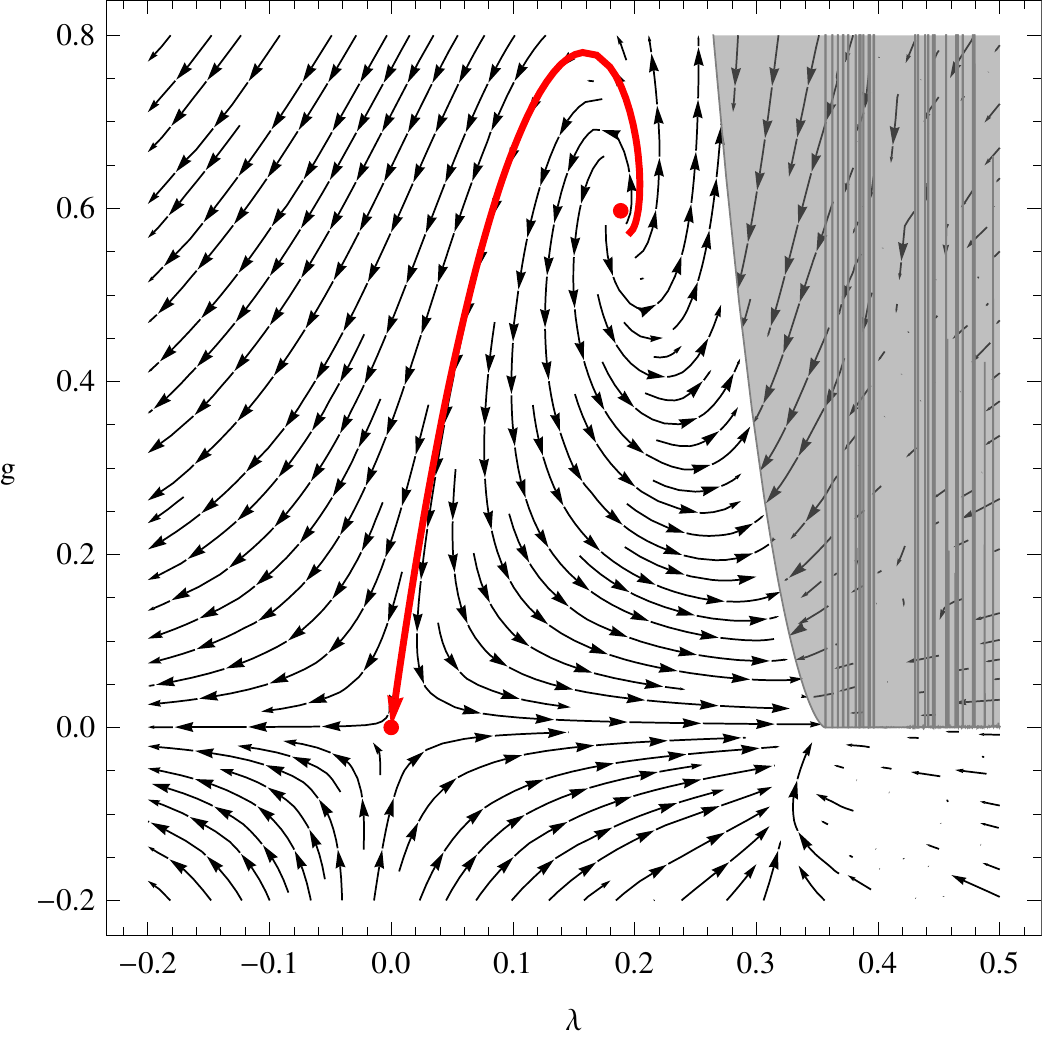}
		\label{fig:PDm1s5v2}}
	\subfigure[\textbf{NGFP} for $s = 6$.]{
		\centering
		\includegraphics[width=0.3\textwidth]{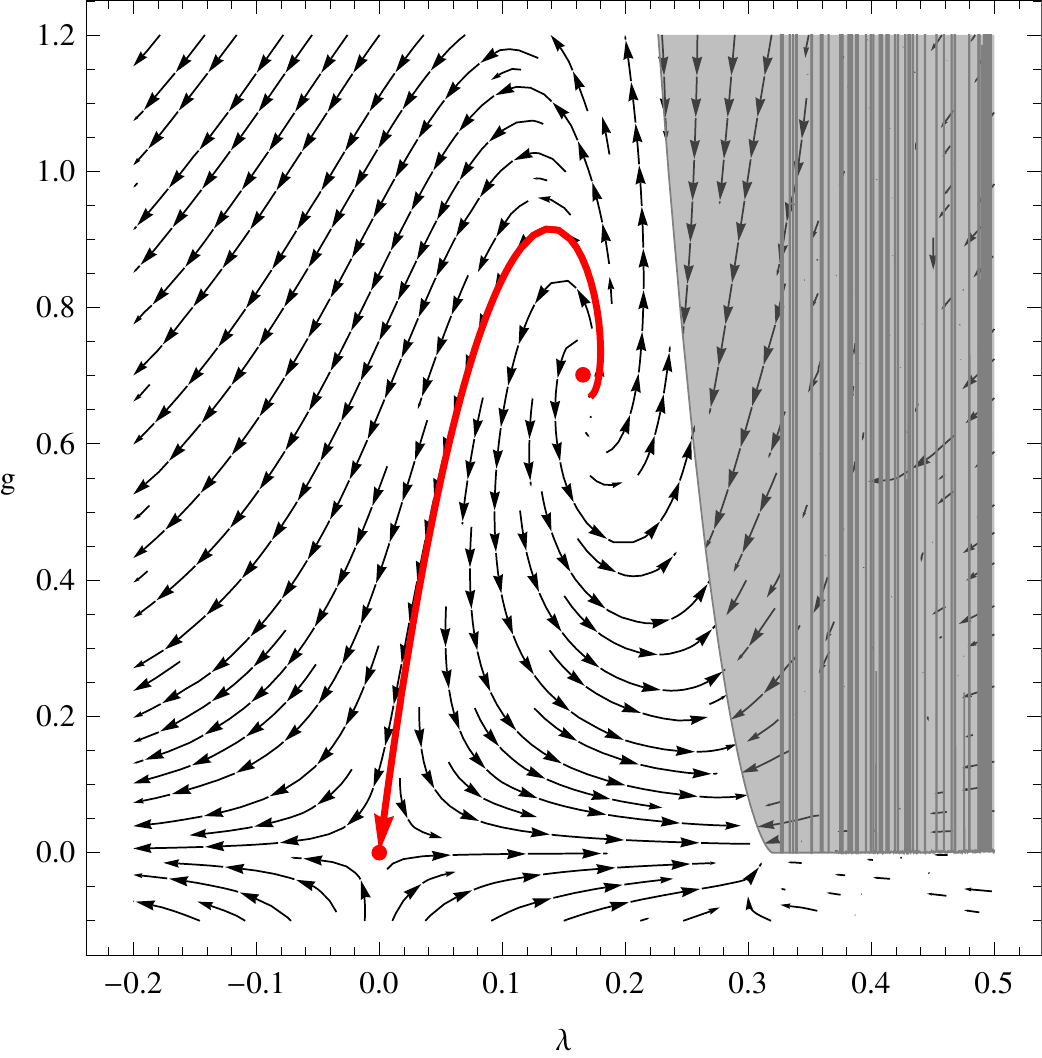}
		\label{fig:PDm1s6v2}}
	\subfigure[\textbf{NGFP} for $s = 7$.]{
		\centering
		\includegraphics[width=0.3\textwidth]{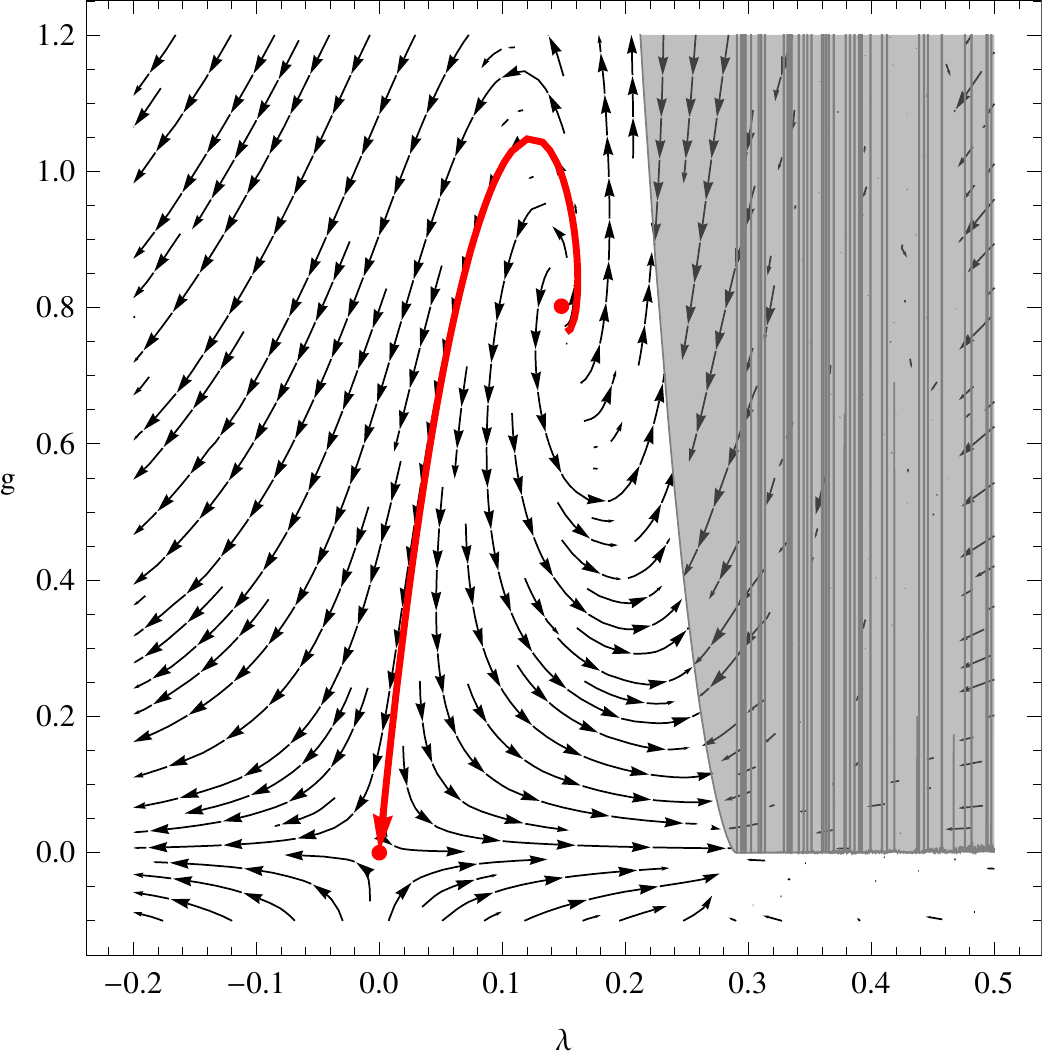}
		\label{fig:PDm1s7v2}}
	\subfigure[\textbf{NGFP} for $s = 8$.]{
		\centering
		\includegraphics[width=0.3\textwidth]{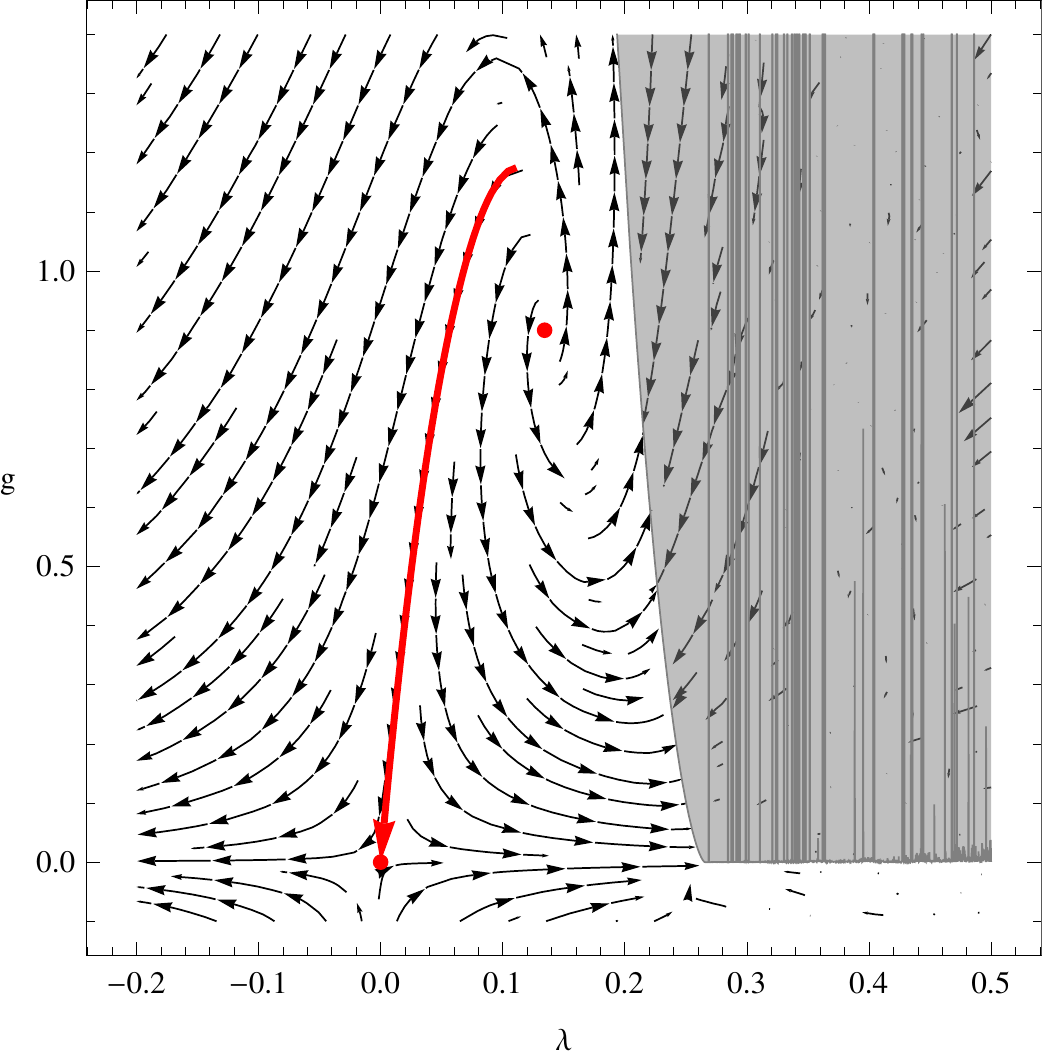}
		\label{fig:PDm1s8v2}}
	\subfigure[\textbf{NGFP} for $s = 9$.]{
		\centering
		\includegraphics[width=0.3\textwidth]{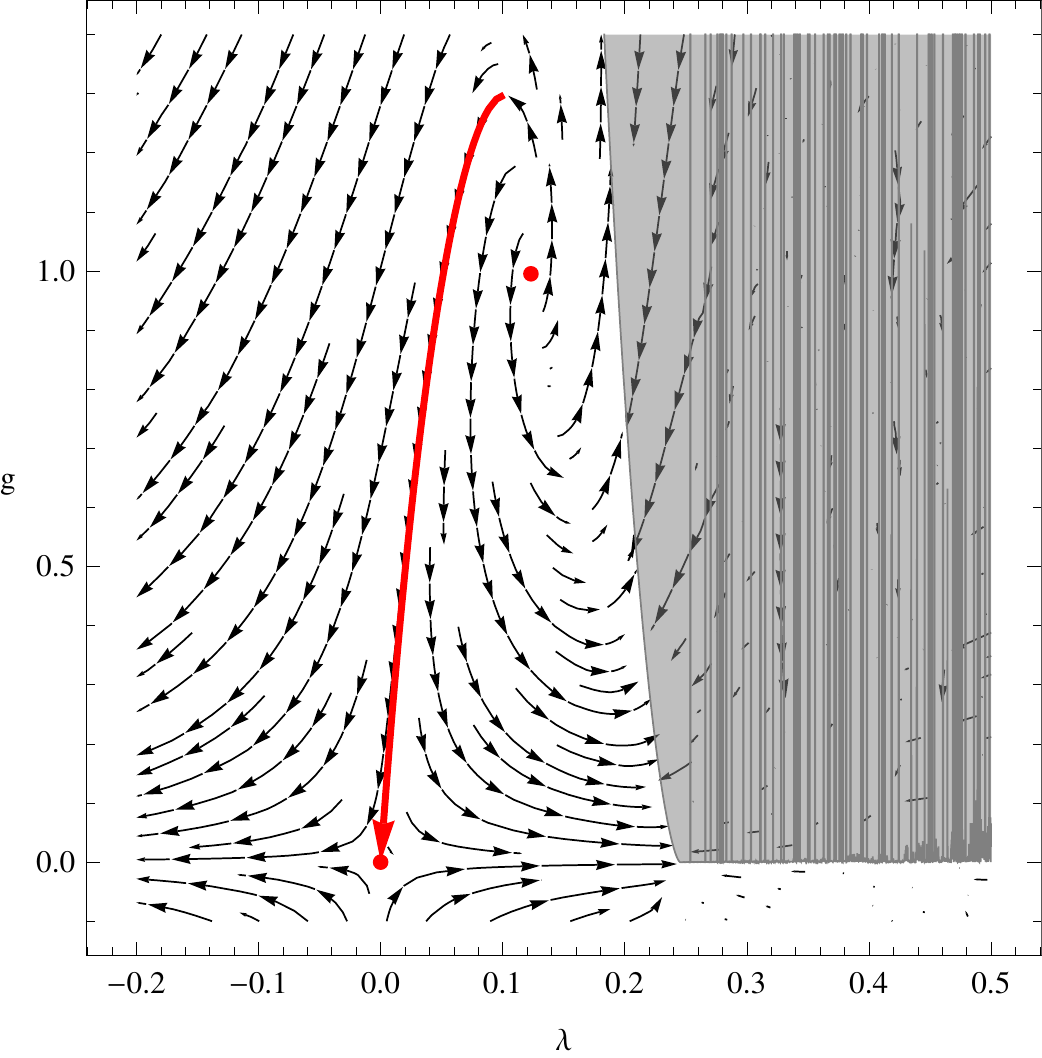}
		\label{fig:PDm1s9v2}}
	\subfigure[\textbf{NGFP} for $s = 10$.]{
		\centering
		\includegraphics[width=0.3\textwidth]{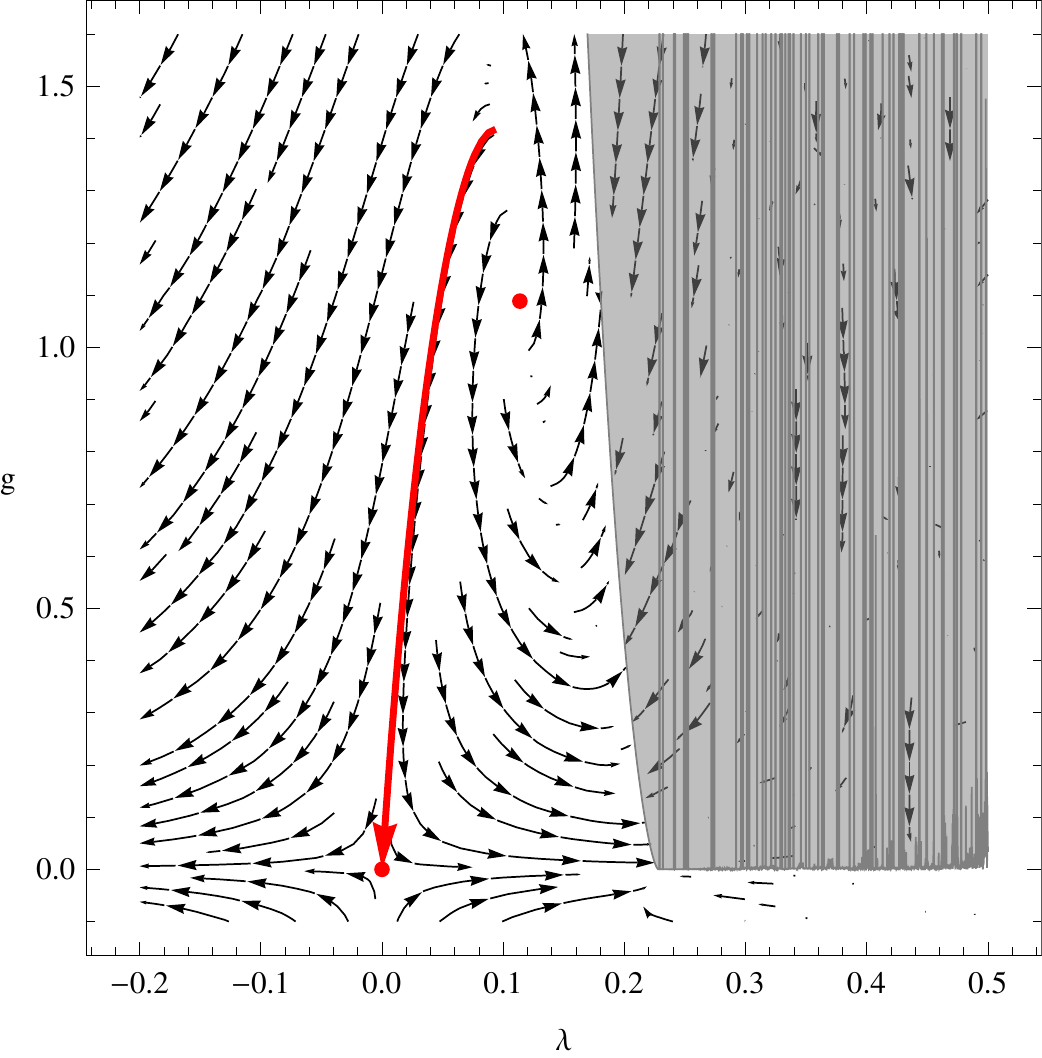}
		\label{fig:PDm1s10v2}}
	\subfigure[\textbf{NGFP} for $s = 20$.]{
		\centering
		\includegraphics[width=0.3\textwidth]{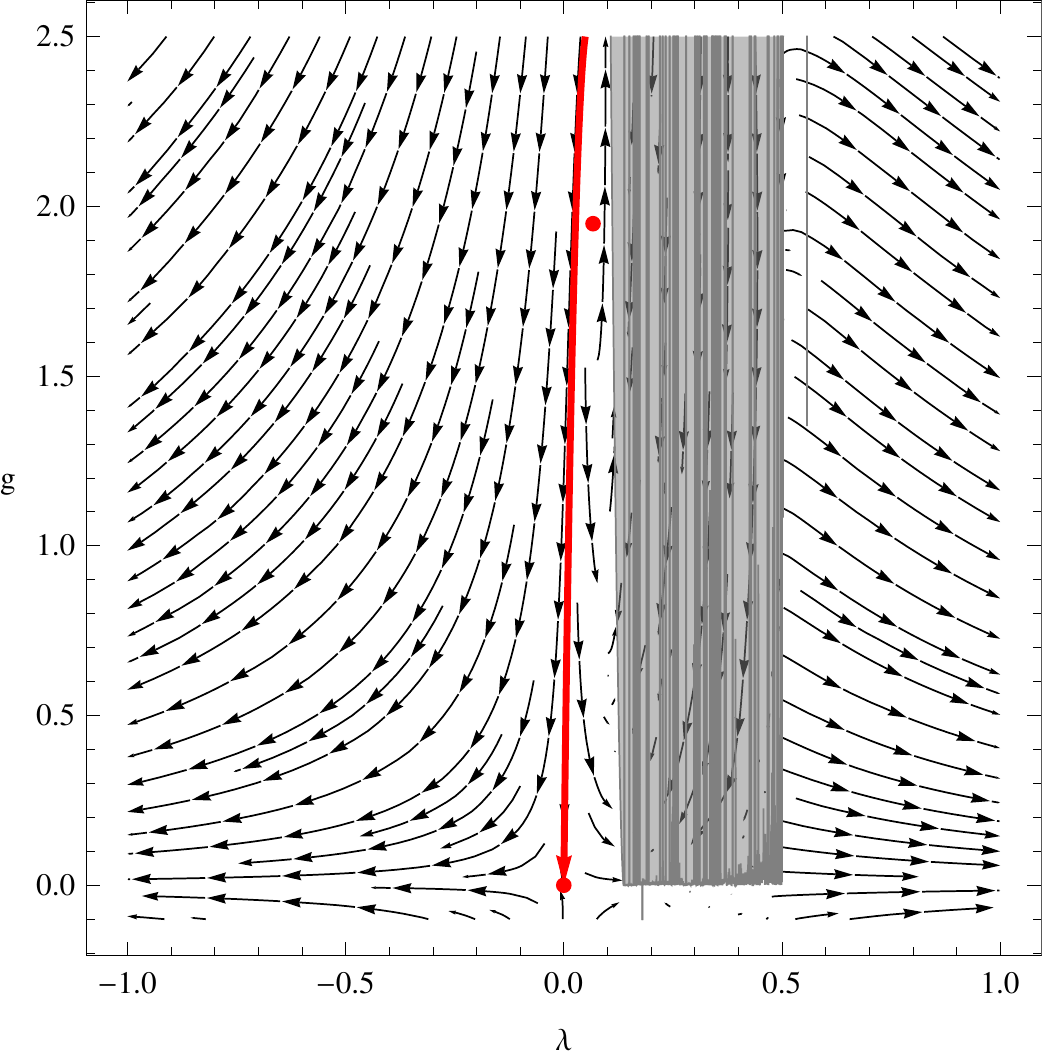}
		\label{fig:PDm1s20v3}}
	\caption{RG phase portrait for the squared mass parameter $µ^2 = 1$ and different values of the shape parameter $s$, ranging from $s=\frac{1}{2}$ to $s=20$.}
	\label{fig:NGFPm1}
\end{figure}
\clearpage

\begin{figure}[htbp]
	\centering
	\subfigure[\textbf{NGFP} for $s = \frac{1}{2}$.]{
		\centering
		\includegraphics[width=0.3\textwidth]{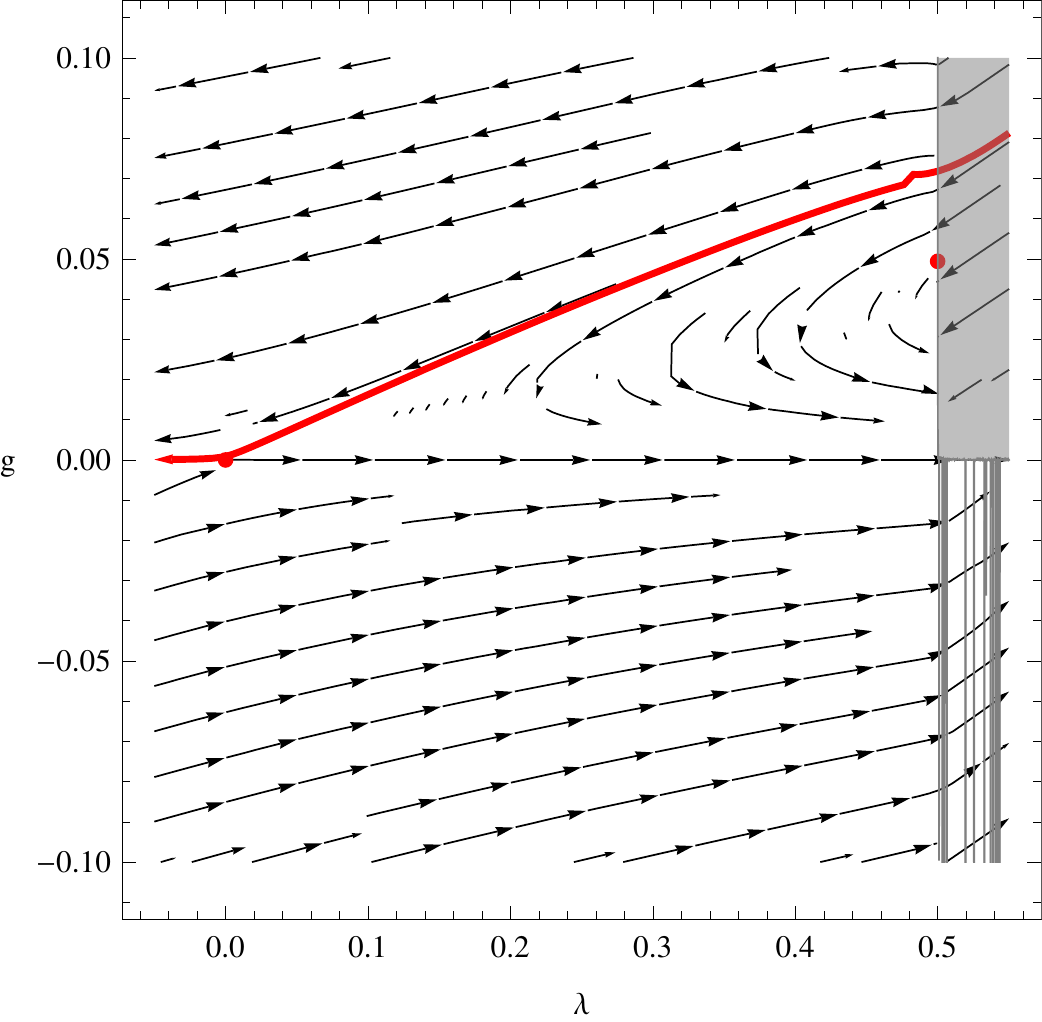}
		\label{fig:PDm2s12}}
	\subfigure[\textbf{NGFP} for $s = 1$.]{
		\centering
		\includegraphics[width=0.3\textwidth]{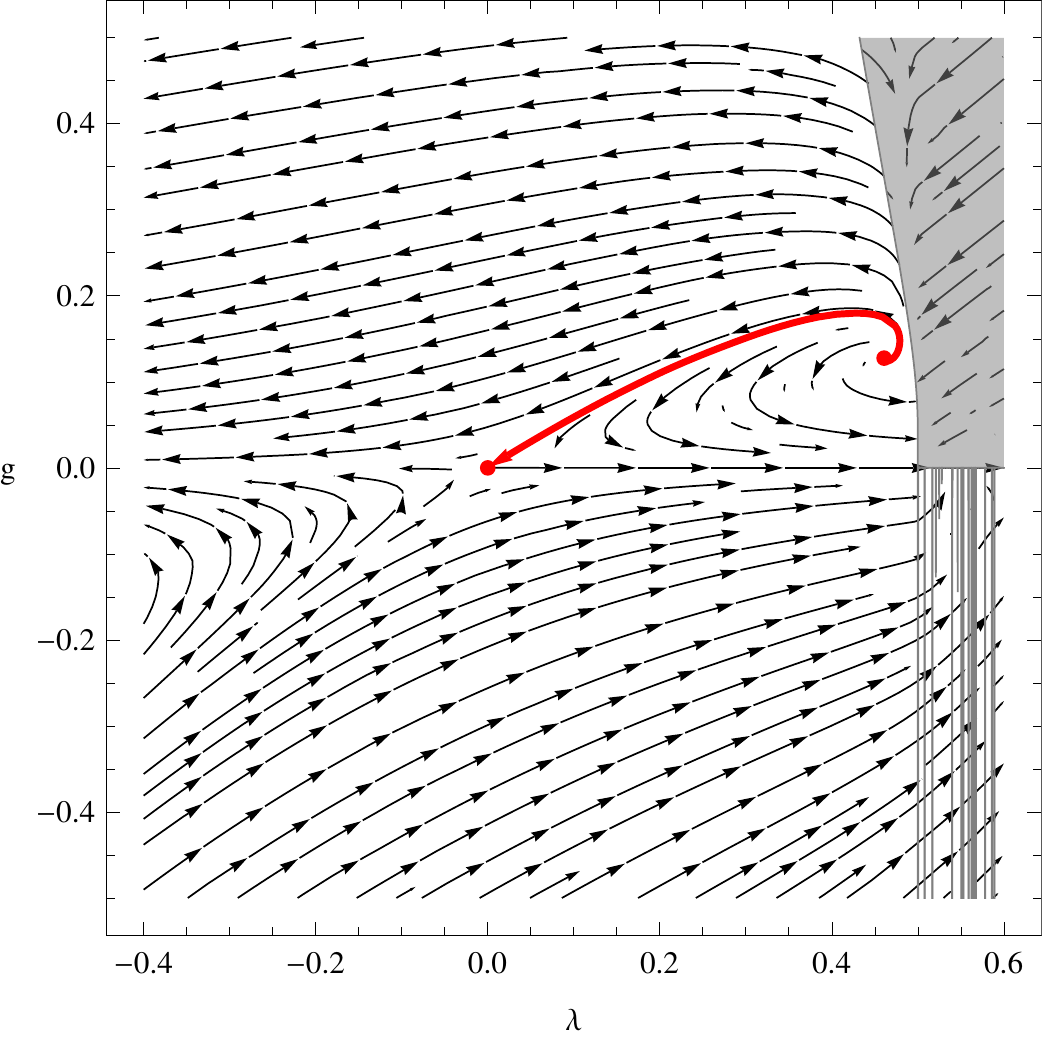}
		\label{fig:PDm2s1}}
	\subfigure[\textbf{NGFP} for $s = 2$.]{
		\centering
		\includegraphics[width=0.3\textwidth]{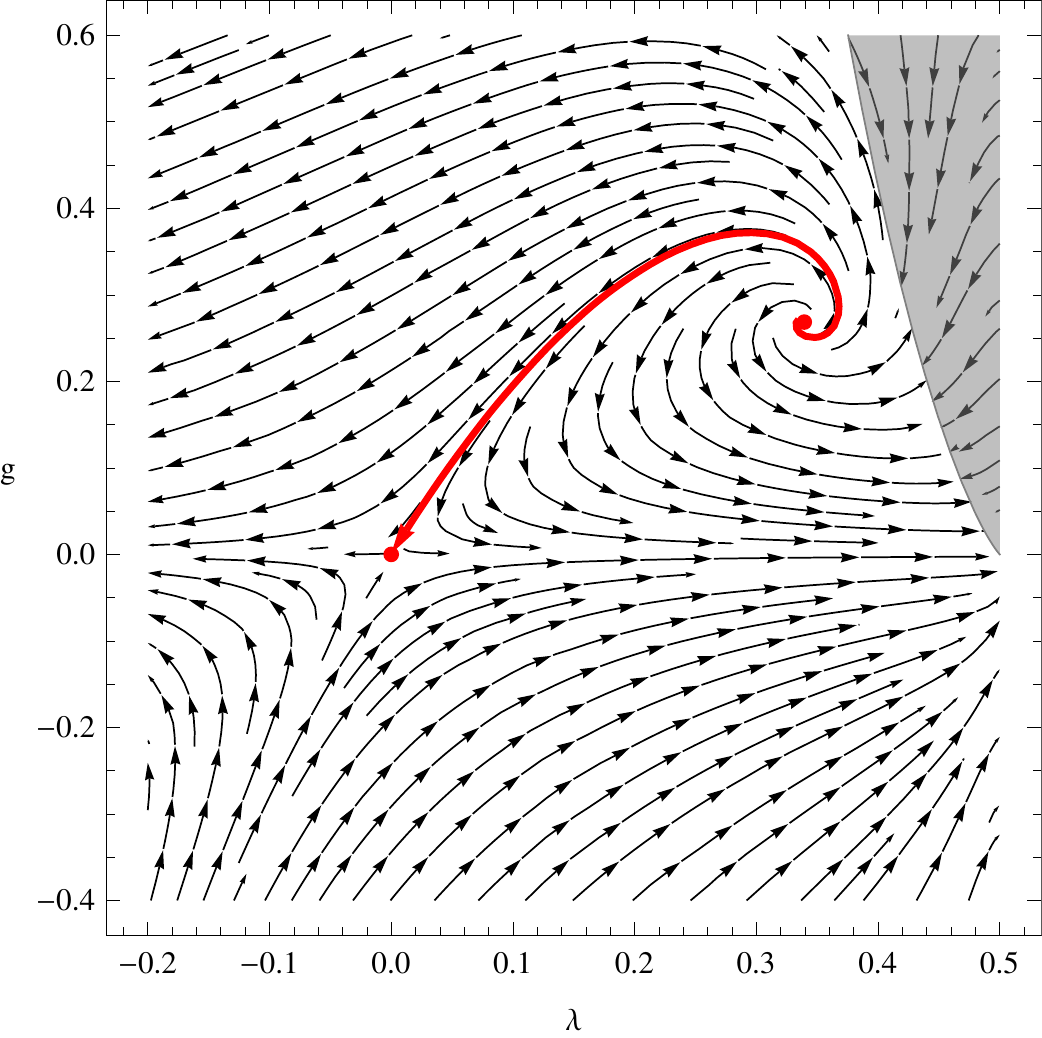}
		\label{fig:PDm2s2}}
	\subfigure[\textbf{NGFP} for $s = 3$.]{
		\centering
		\includegraphics[width=0.3\textwidth]{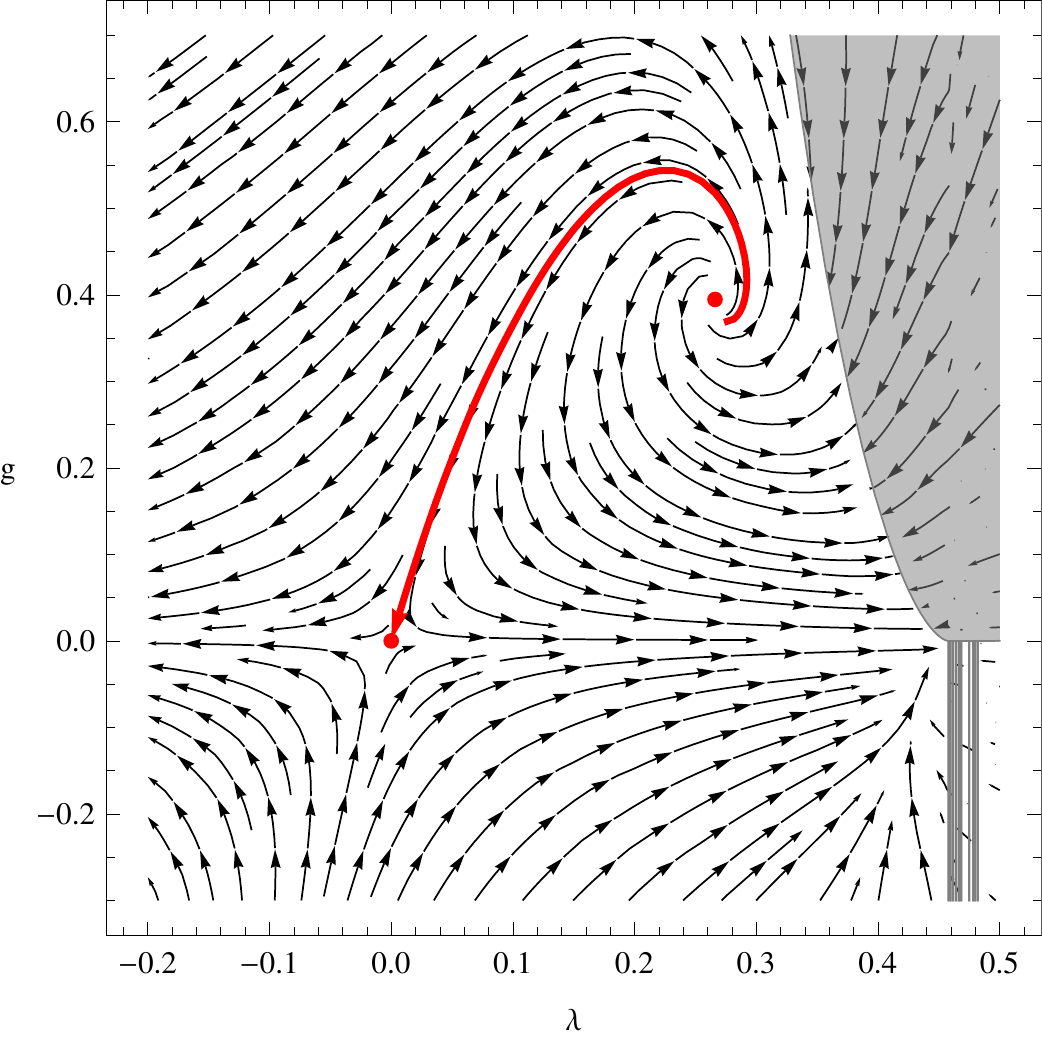}
		\label{fig:PDm2s3}}
	\subfigure[\textbf{NGFP} for $s = 4$.]{
		\centering
		\includegraphics[width=0.3\textwidth]{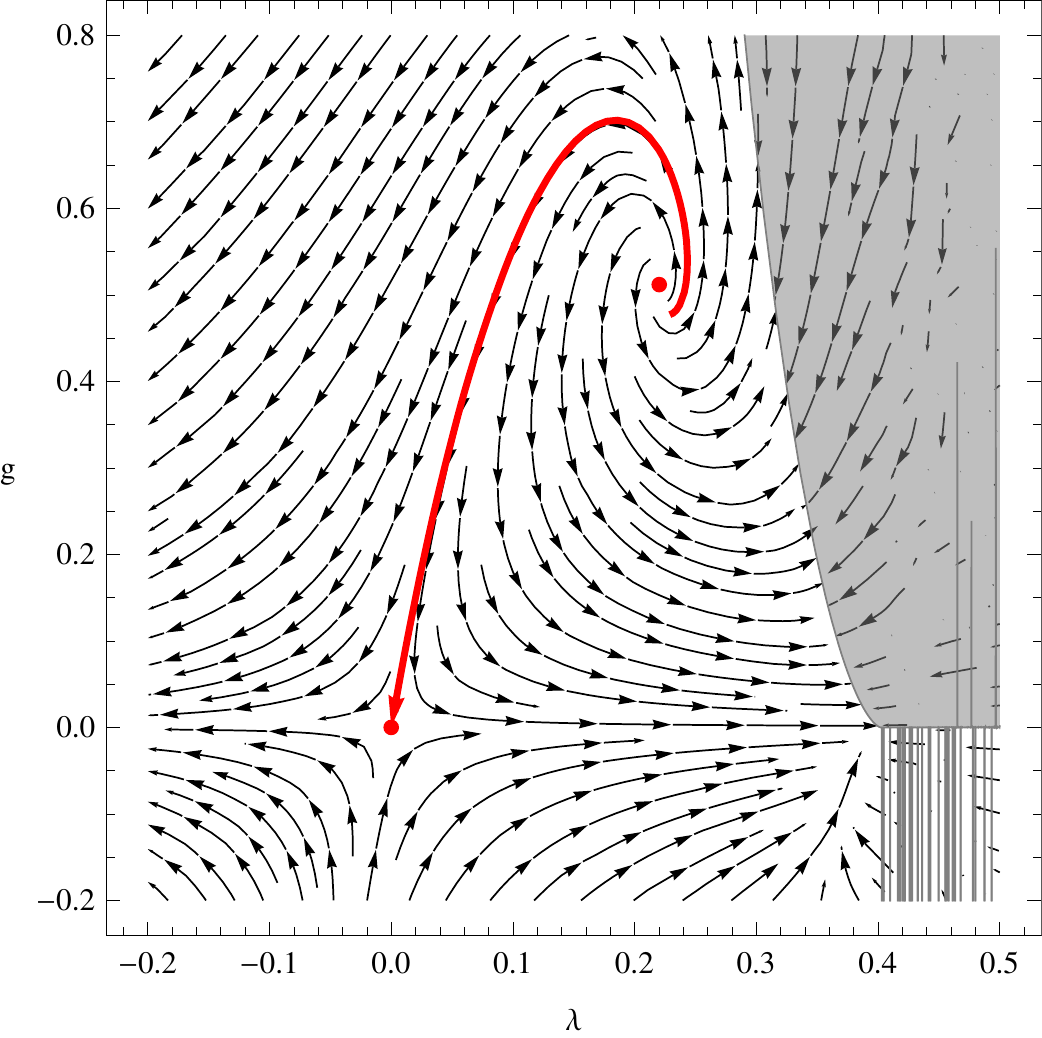}
		\label{fig:PDm2s4}}
	\subfigure[\textbf{NGFP} for $s = 5$.]{
		\centering
		\includegraphics[width=0.3\textwidth]{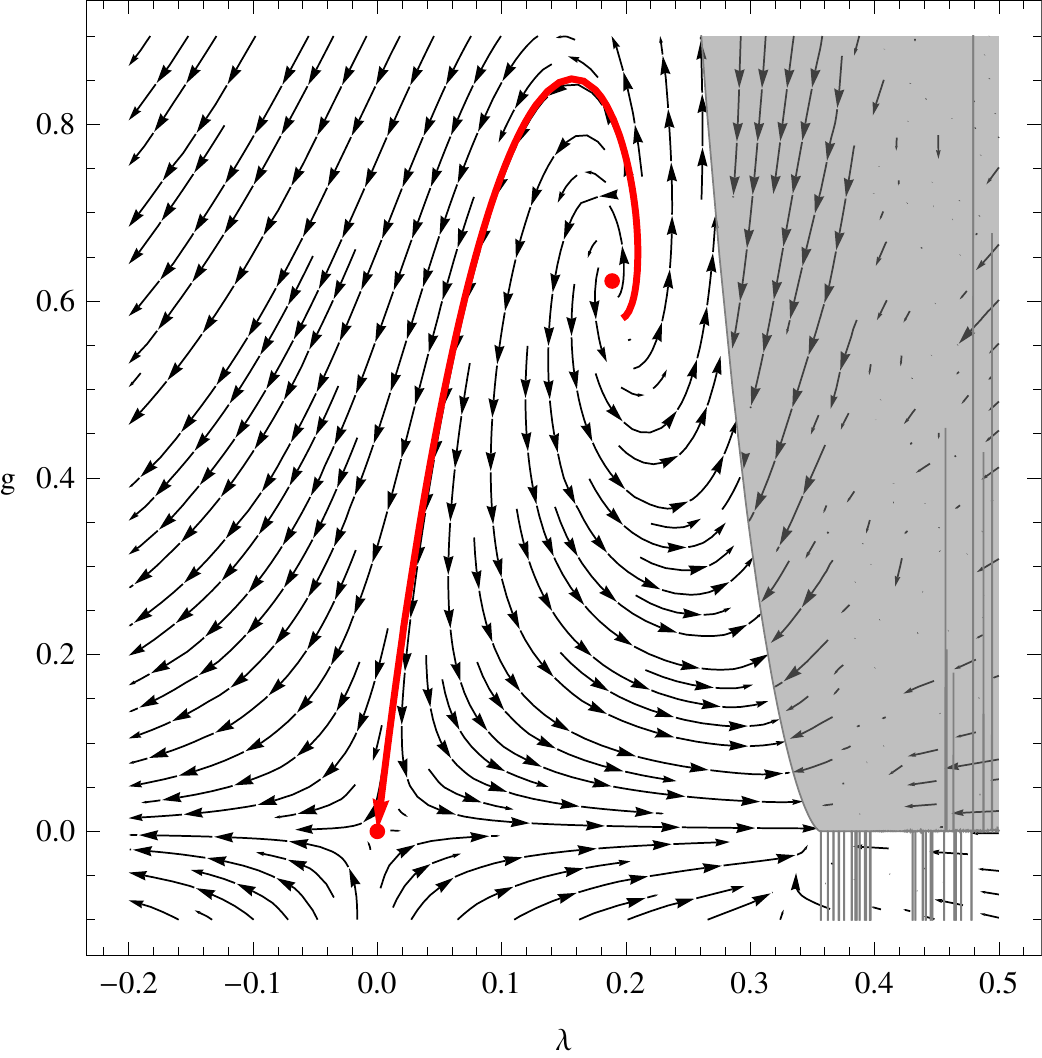}
		\label{fig:PDm2s5}}
	\subfigure[\textbf{NGFP} for $s = 6$.]{
		\centering
		\includegraphics[width=0.3\textwidth]{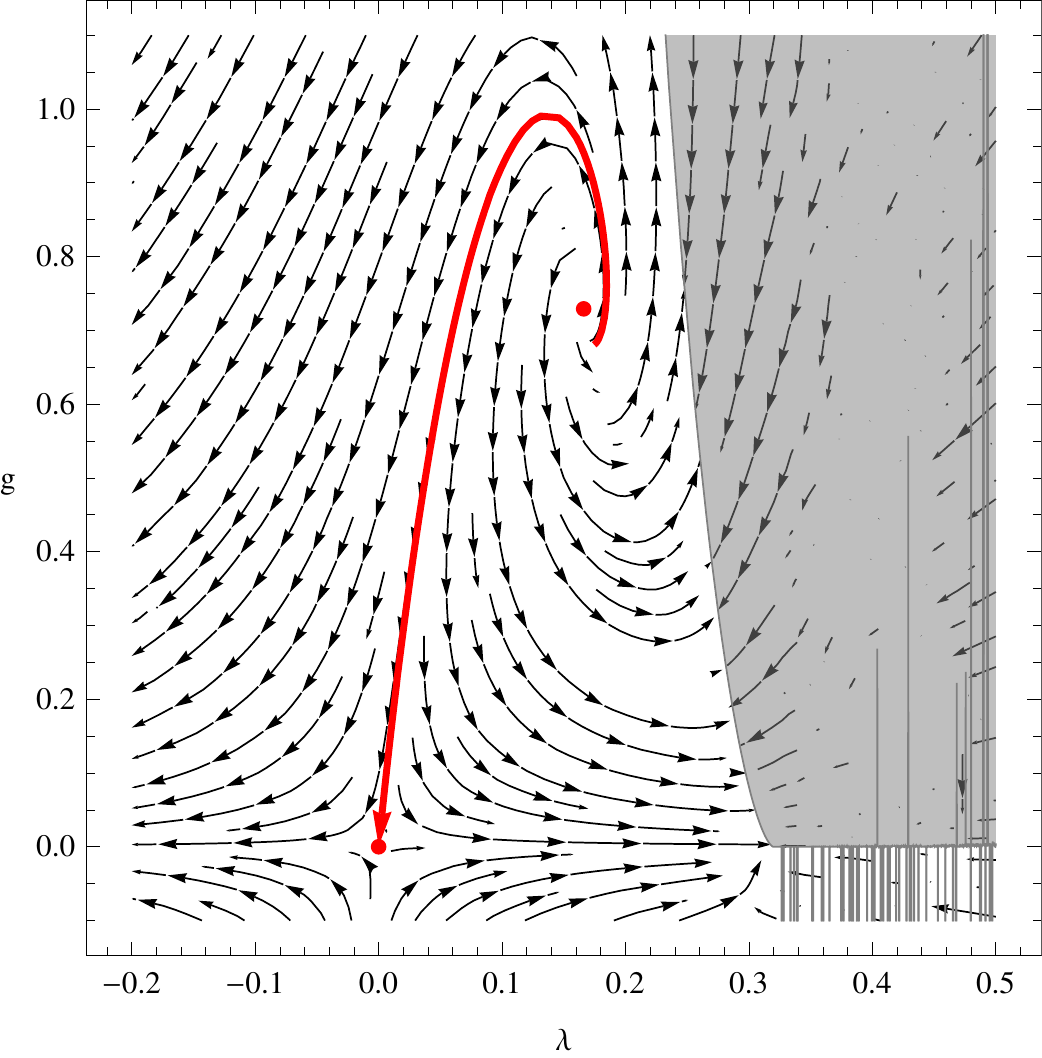}
		\label{fig:PDm2s6}}
	\subfigure[\textbf{NGFP} for $s = 7$.]{
		\centering
		\includegraphics[width=0.3\textwidth]{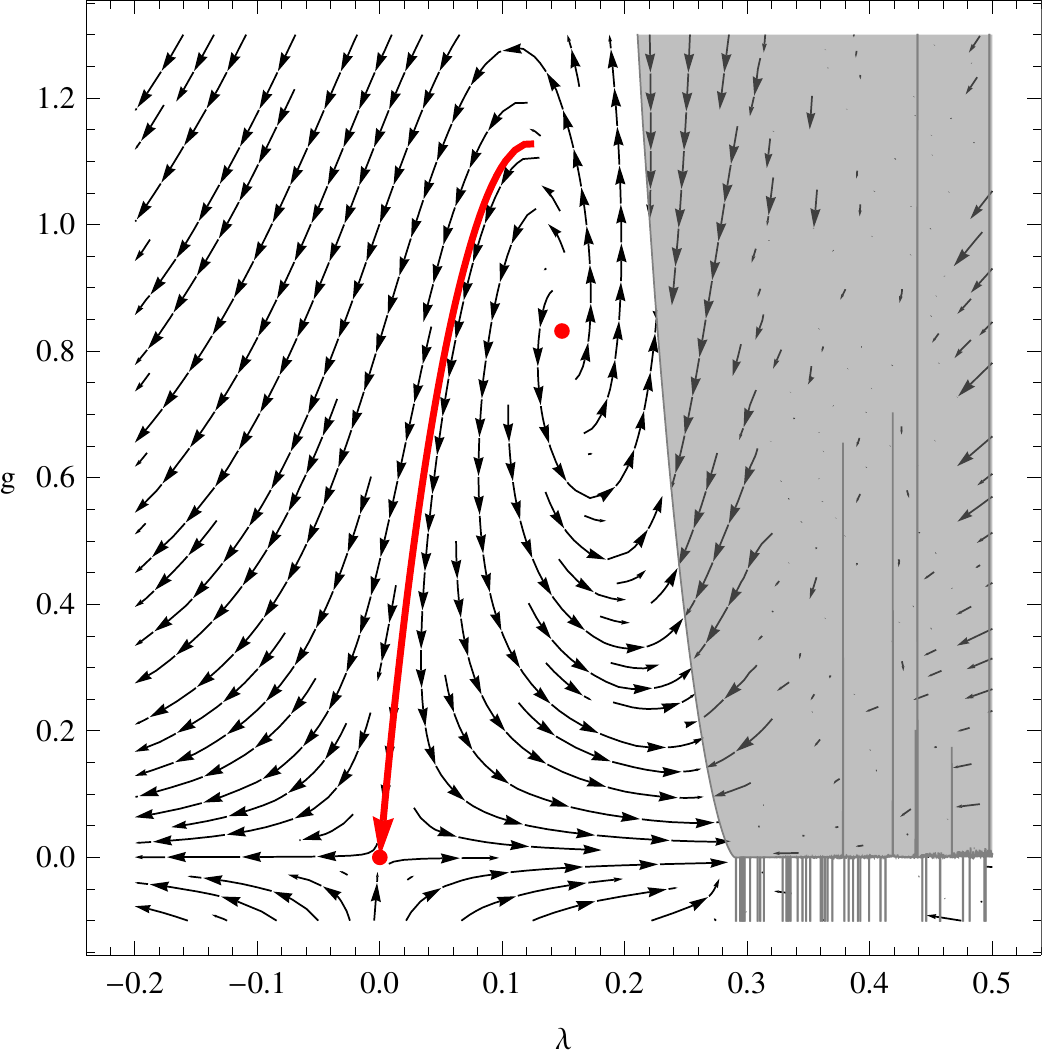}
		\label{fig:PDm2s7}}
	\subfigure[\textbf{NGFP} for $s = 8$.]{
		\centering
		\includegraphics[width=0.3\textwidth]{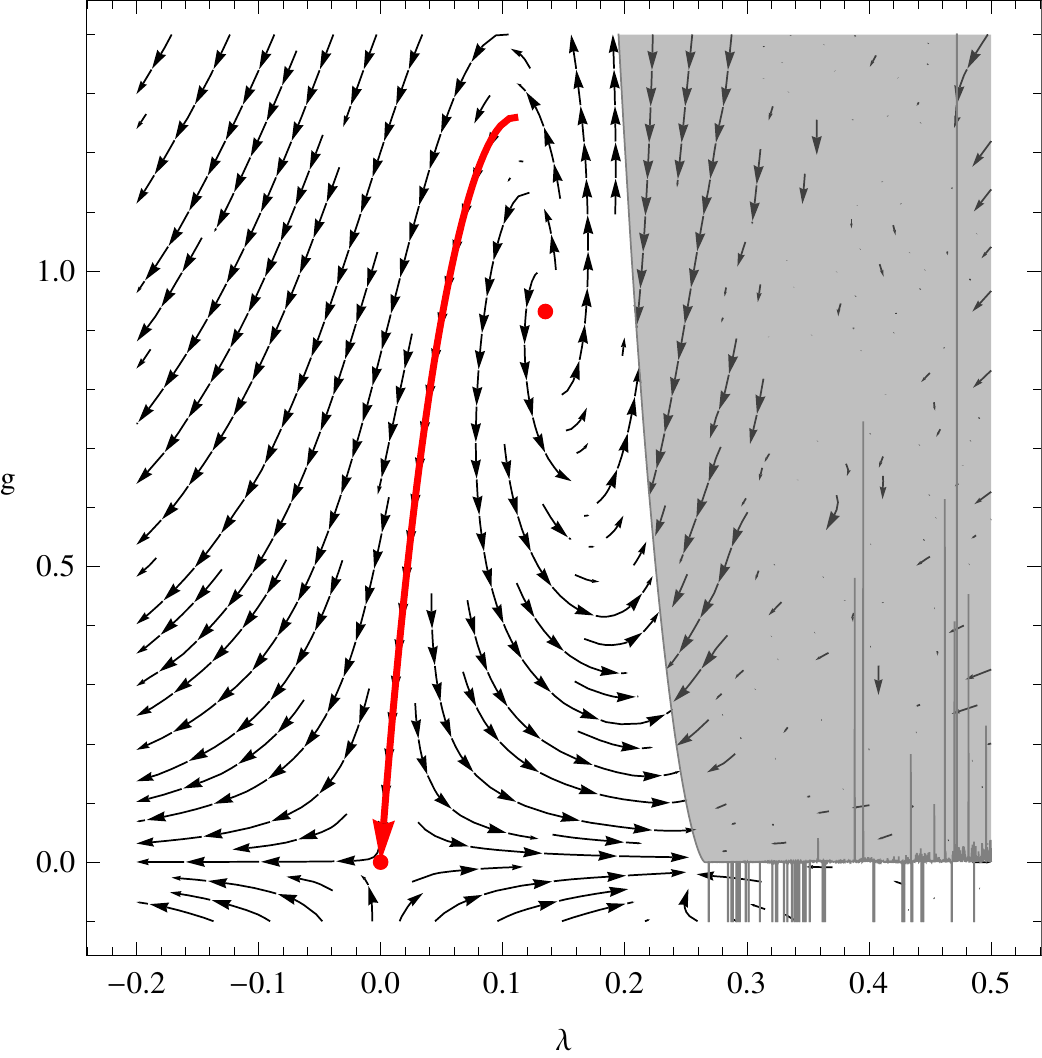}
		\label{fig:PDm2s8}}
	\subfigure[\textbf{NGFP} for $s = 9$.]{
		\centering
		\includegraphics[width=0.3\textwidth]{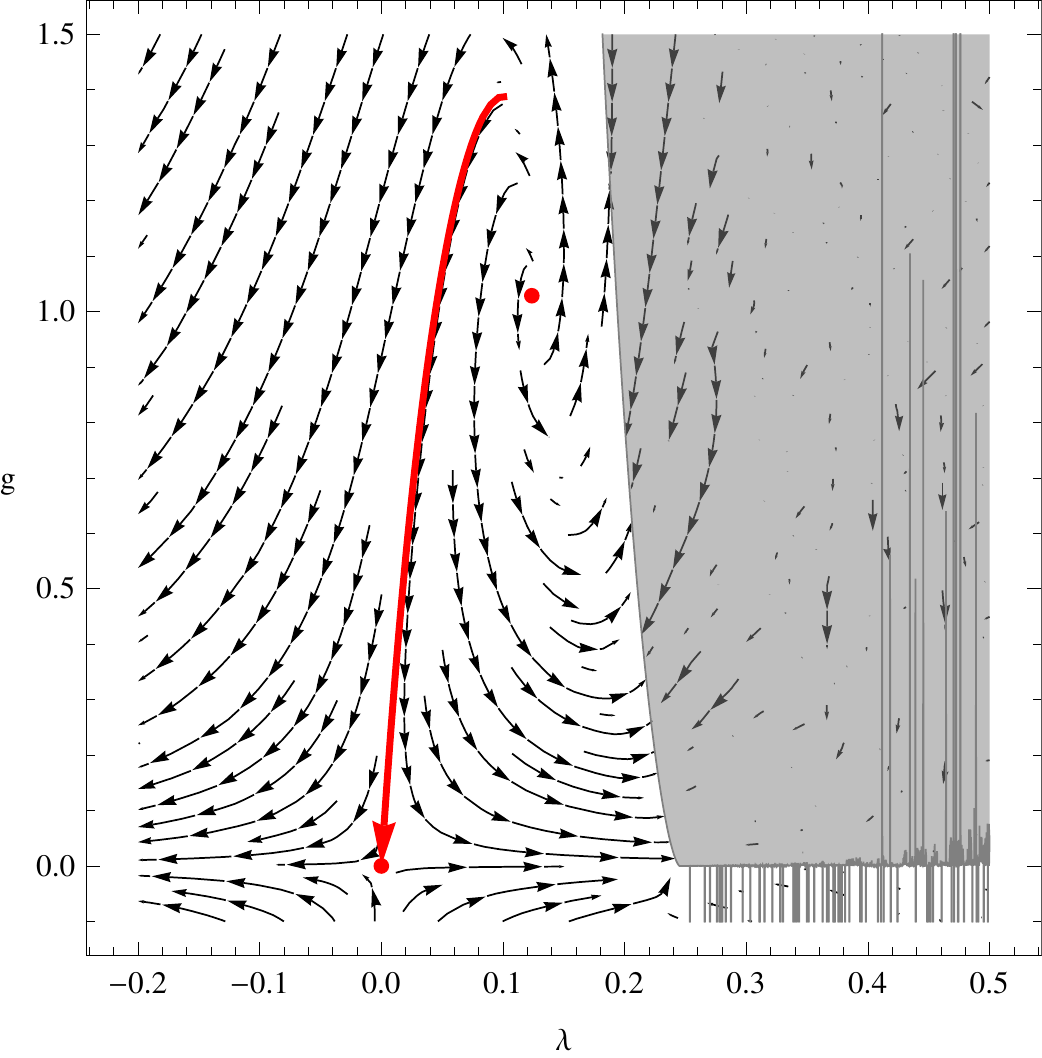}
		\label{fig:PDm2s9}}
	\subfigure[\textbf{NGFP} for $s = 10$.]{
		\centering
		\includegraphics[width=0.3\textwidth]{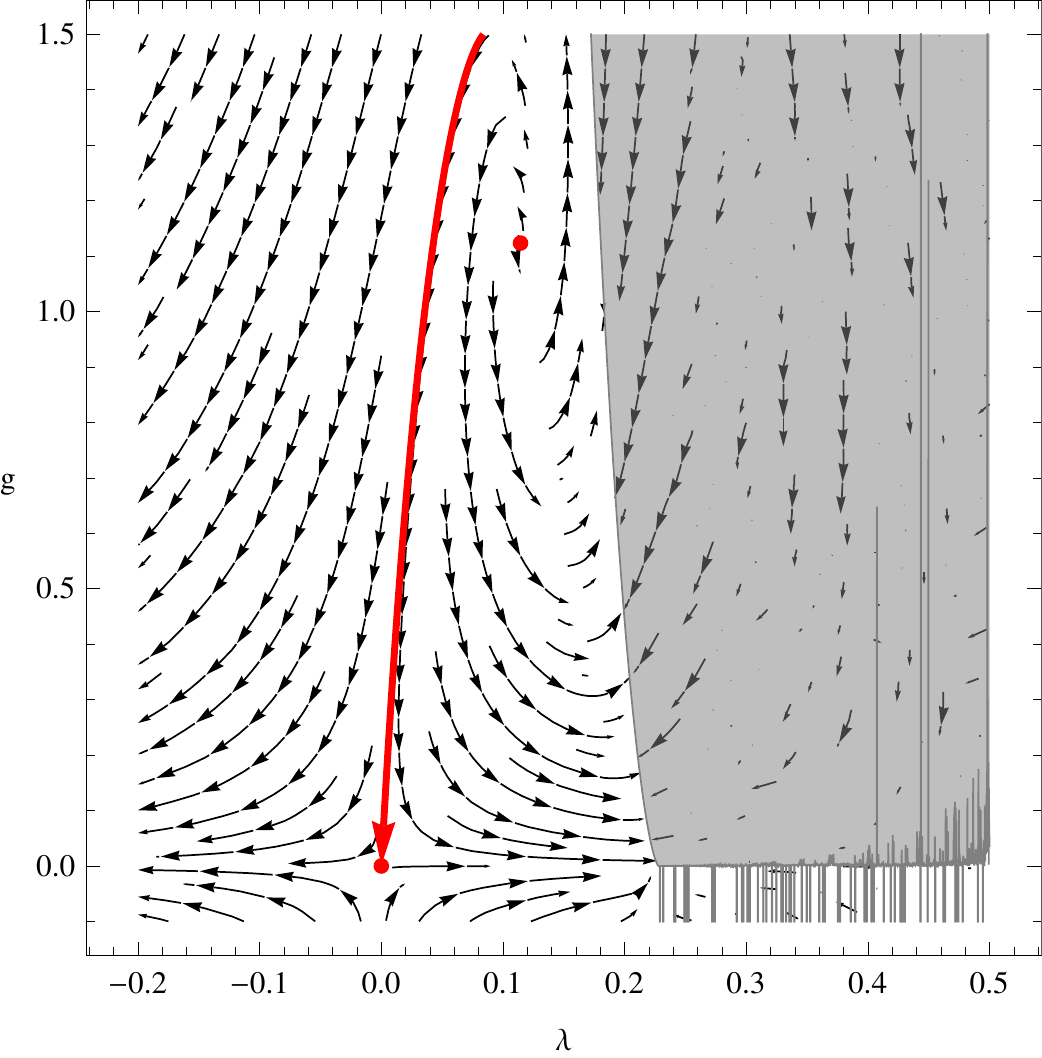}
		\label{fig:PDm2s10}}
	\subfigure[\textbf{NGFP} for $s = 20$.]{
		\centering
		\includegraphics[width=0.3\textwidth]{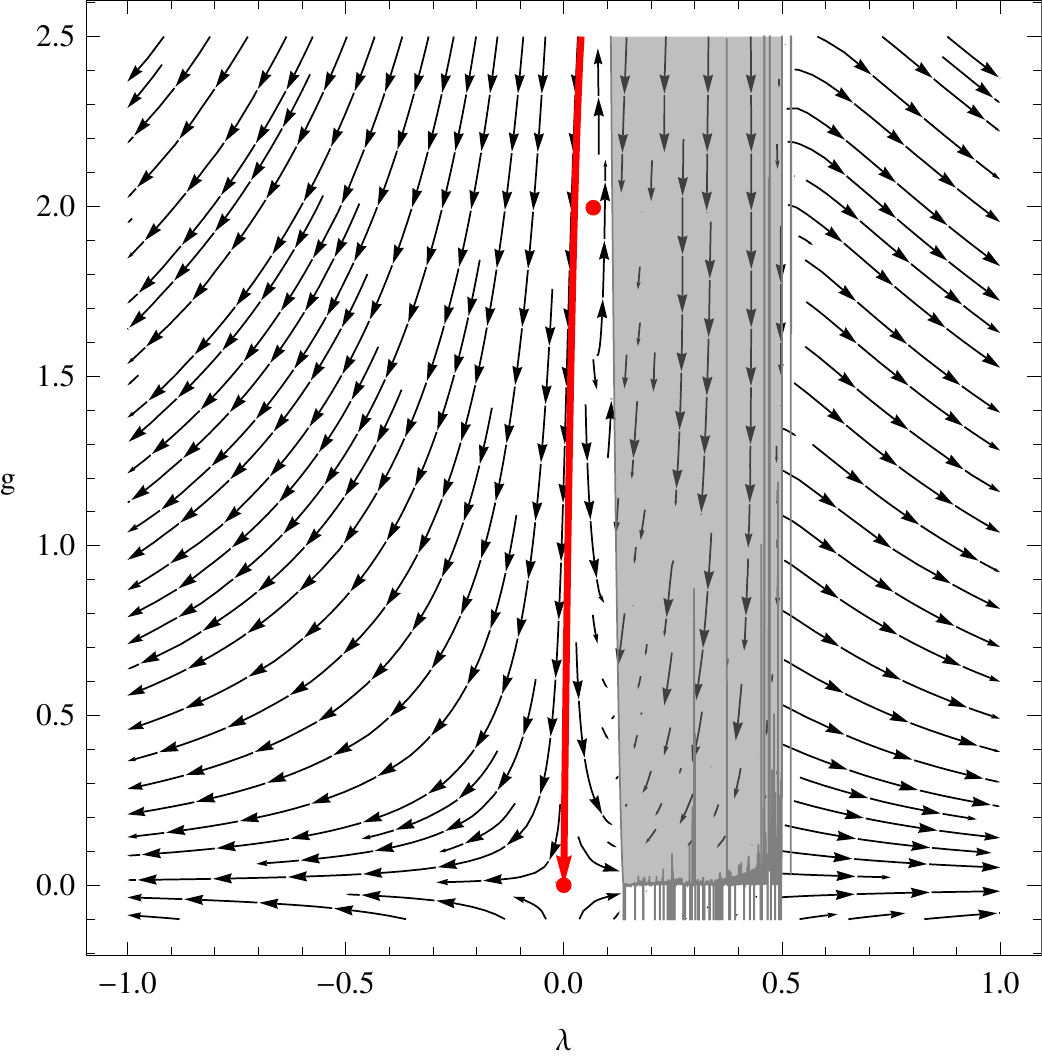}
		\label{fig:PDm2s20}}
	\caption{RG phase portrait for the squared mass parameter $µ^2 = 2$ and different values of the shape parameter $s$, ranging from $s=\frac{1}{2}$ to $s=20$.}
	\label{fig:NGFPm2}
\end{figure}
\clearpage

\begin{figure}[htbp]
	\centering
	\subfigure[\textbf{NGFP} for $s = \frac{1}{2}$.]{
		\centering
		\includegraphics[width=0.3\textwidth]{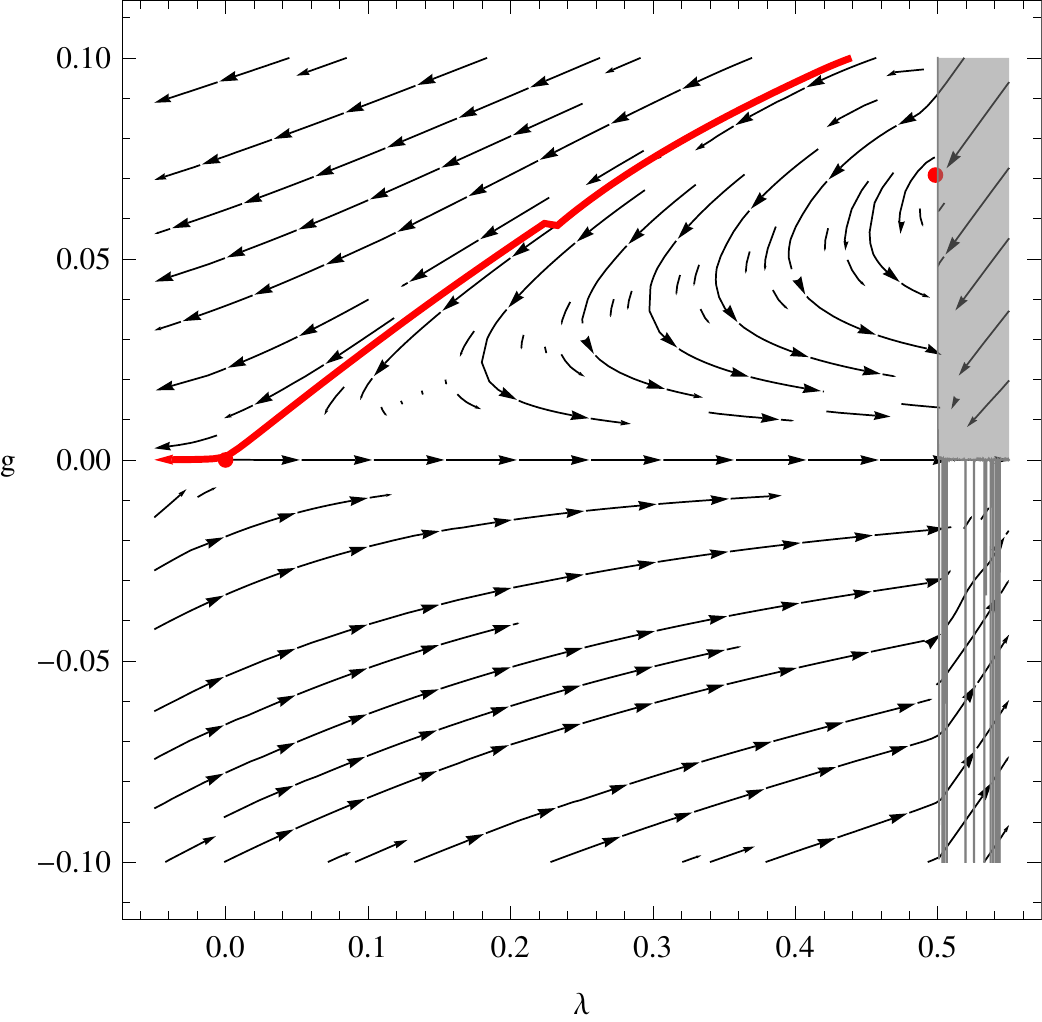}
		\label{fig:PDm5s12}}
	\subfigure[\textbf{NGFP} for $s = 1$.]{
		\centering
		\includegraphics[width=0.3\textwidth]{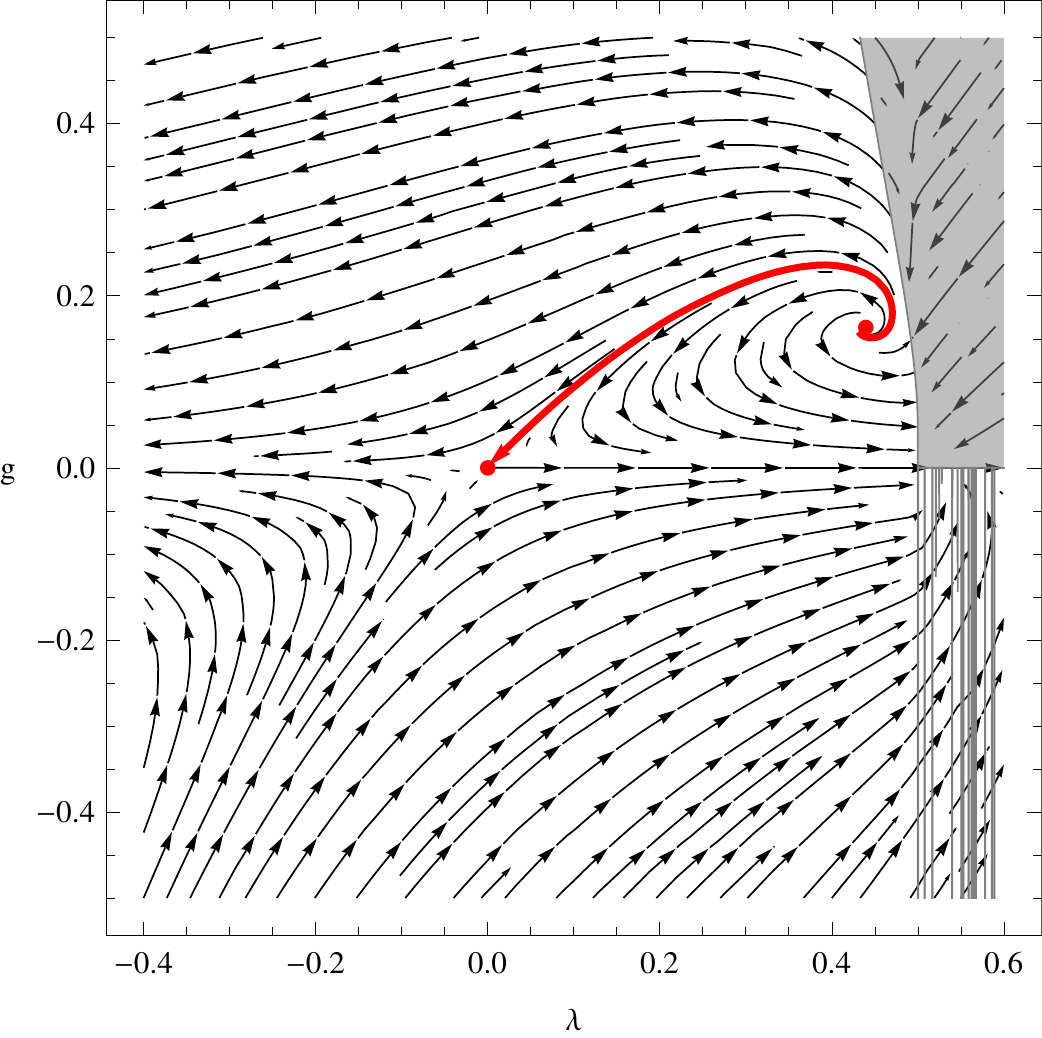}
		\label{fig:PDm5s1v2}}
	\subfigure[\textbf{NGFP} for $s = 2$.]{
		\centering
		\includegraphics[width=0.3\textwidth]{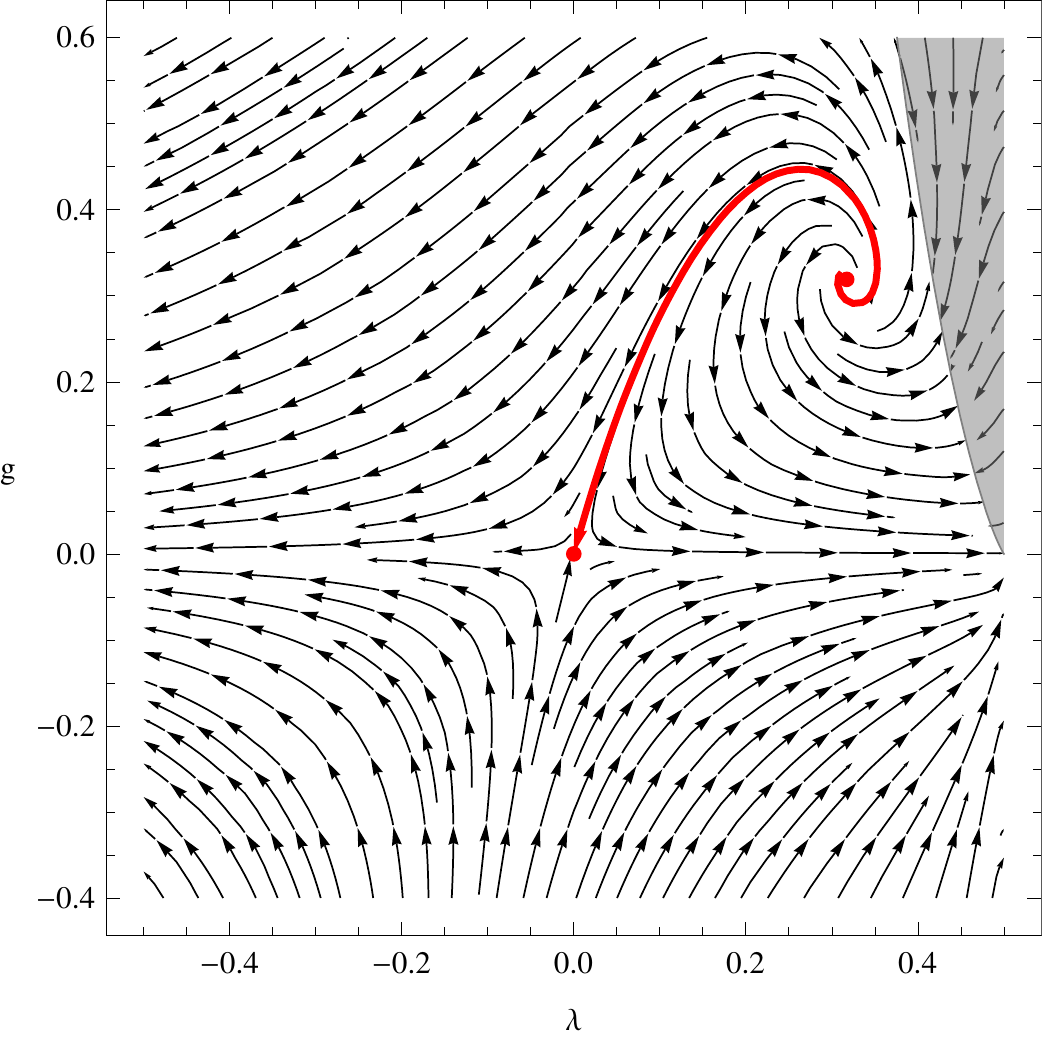}
		\label{fig:PDm5s2v2}}
	\subfigure[\textbf{NGFP} for $s = 3$.]{
		\centering
		\includegraphics[width=0.3\textwidth]{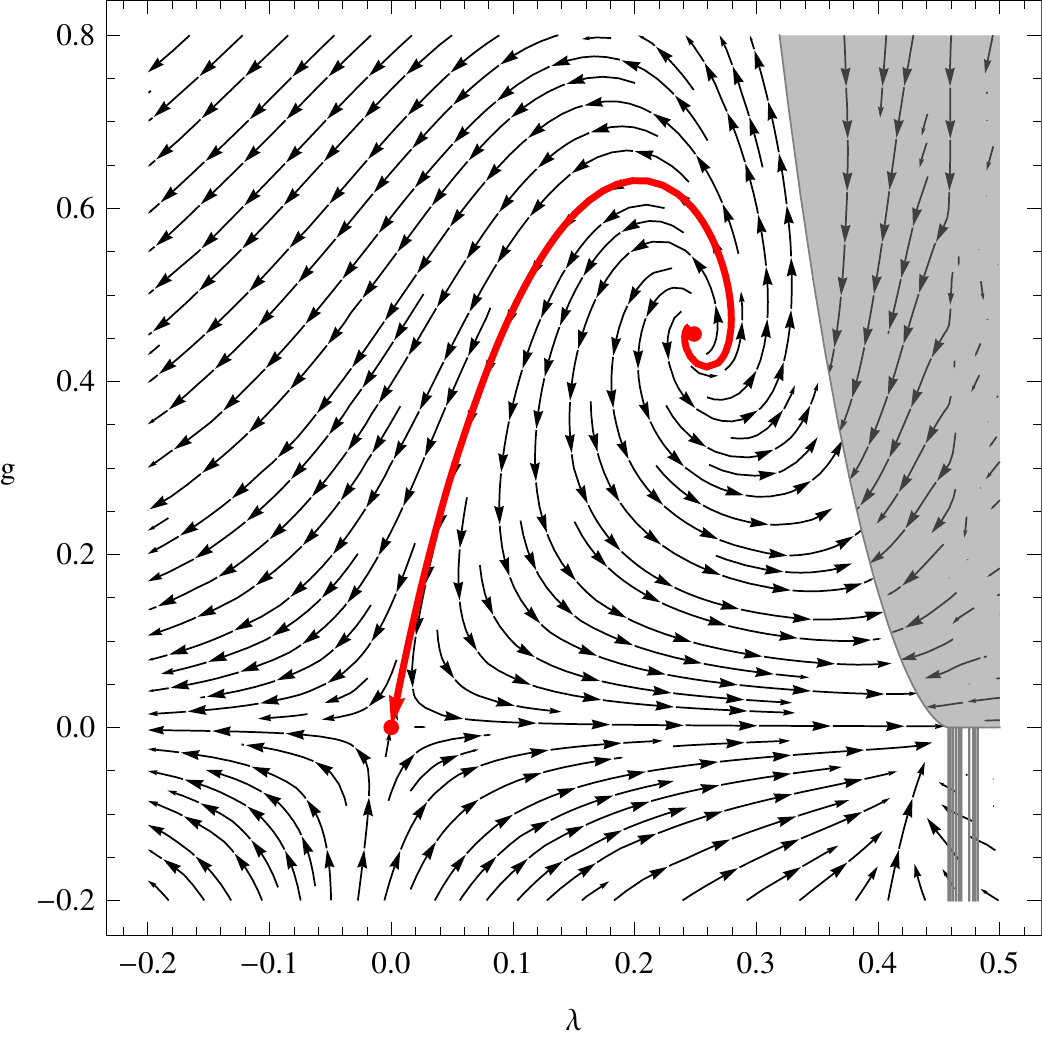}
		\label{fig:PDm5s3v2}}
	\subfigure[\textbf{NGFP} for $s = 4$.]{
		\centering
		\includegraphics[width=0.3\textwidth]{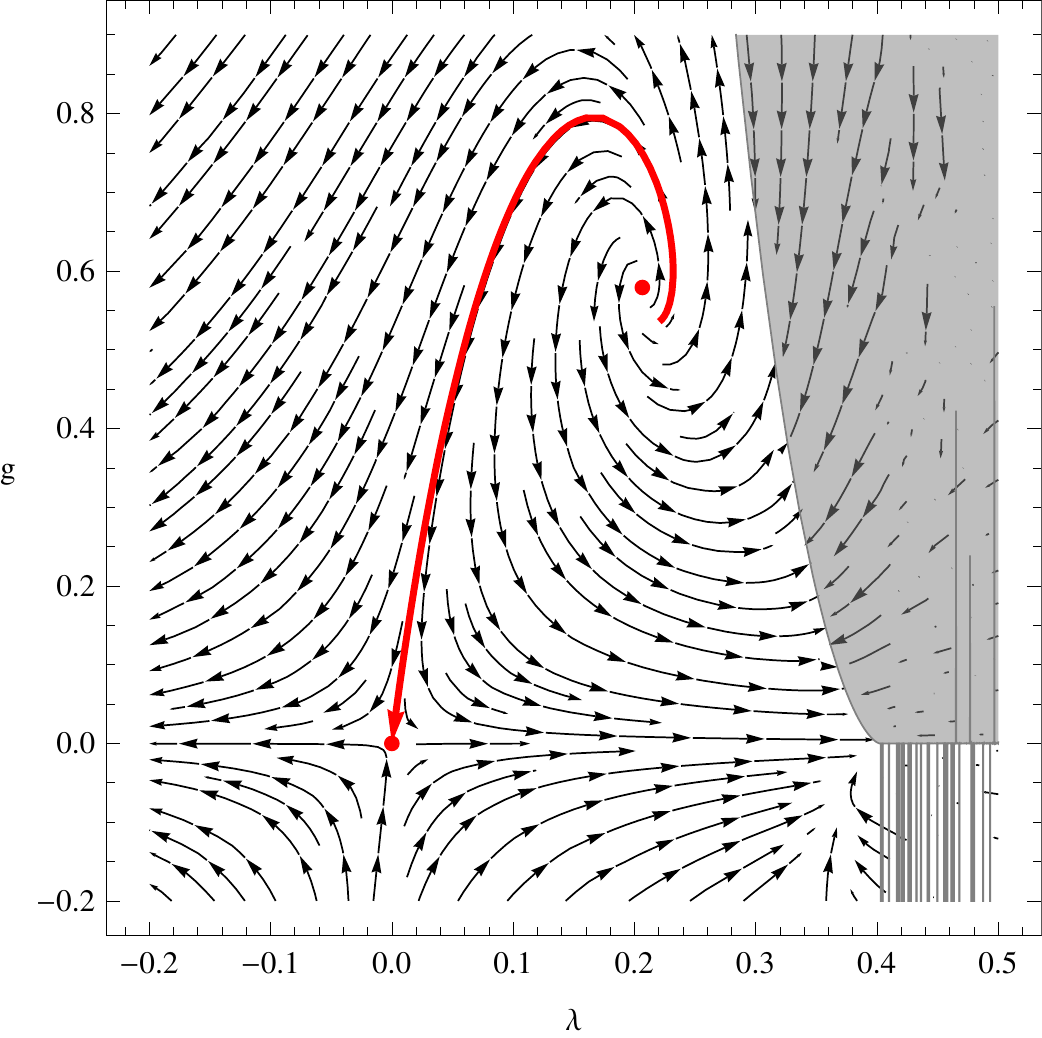}
		\label{fig:PDm5s4v2}}
	\subfigure[\textbf{NGFP} for $s = 5$.]{
		\centering
		\includegraphics[width=0.3\textwidth]{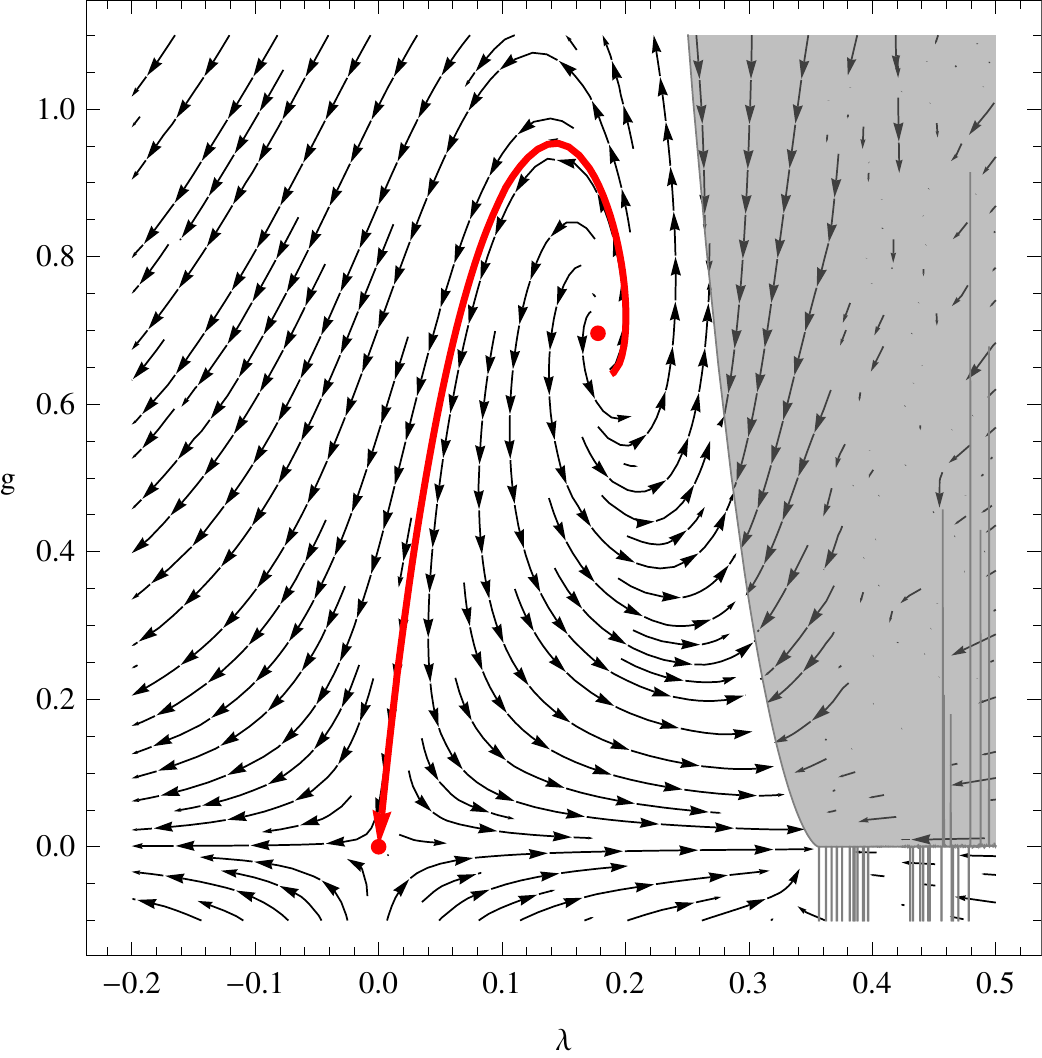}
		\label{fig:PDm5s5v2}}
	\subfigure[\textbf{NGFP} for $s = 6$.]{
		\centering
		\includegraphics[width=0.3\textwidth]{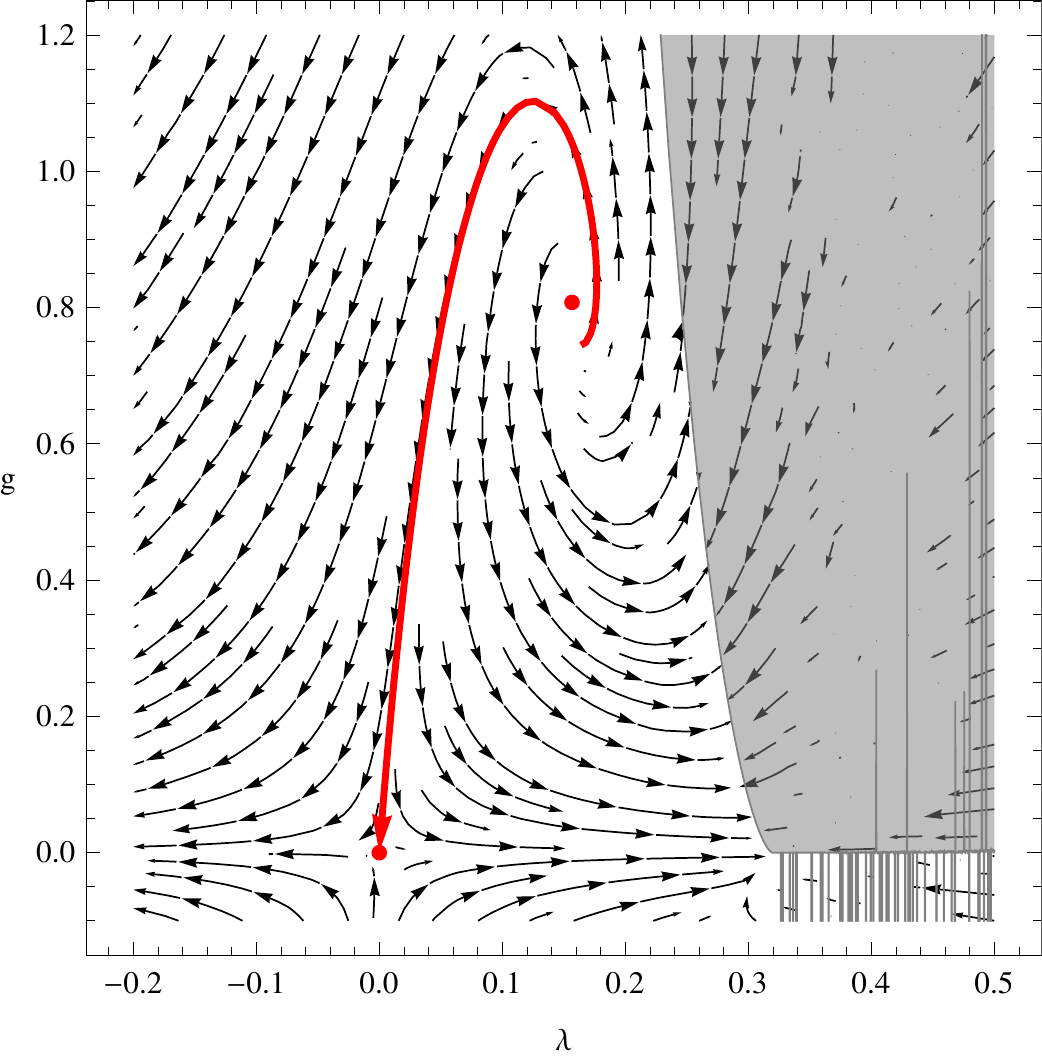}
		\label{fig:PDm5s6v2}}
	\subfigure[\textbf{NGFP} for $s = 7$.]{
		\centering
		\includegraphics[width=0.3\textwidth]{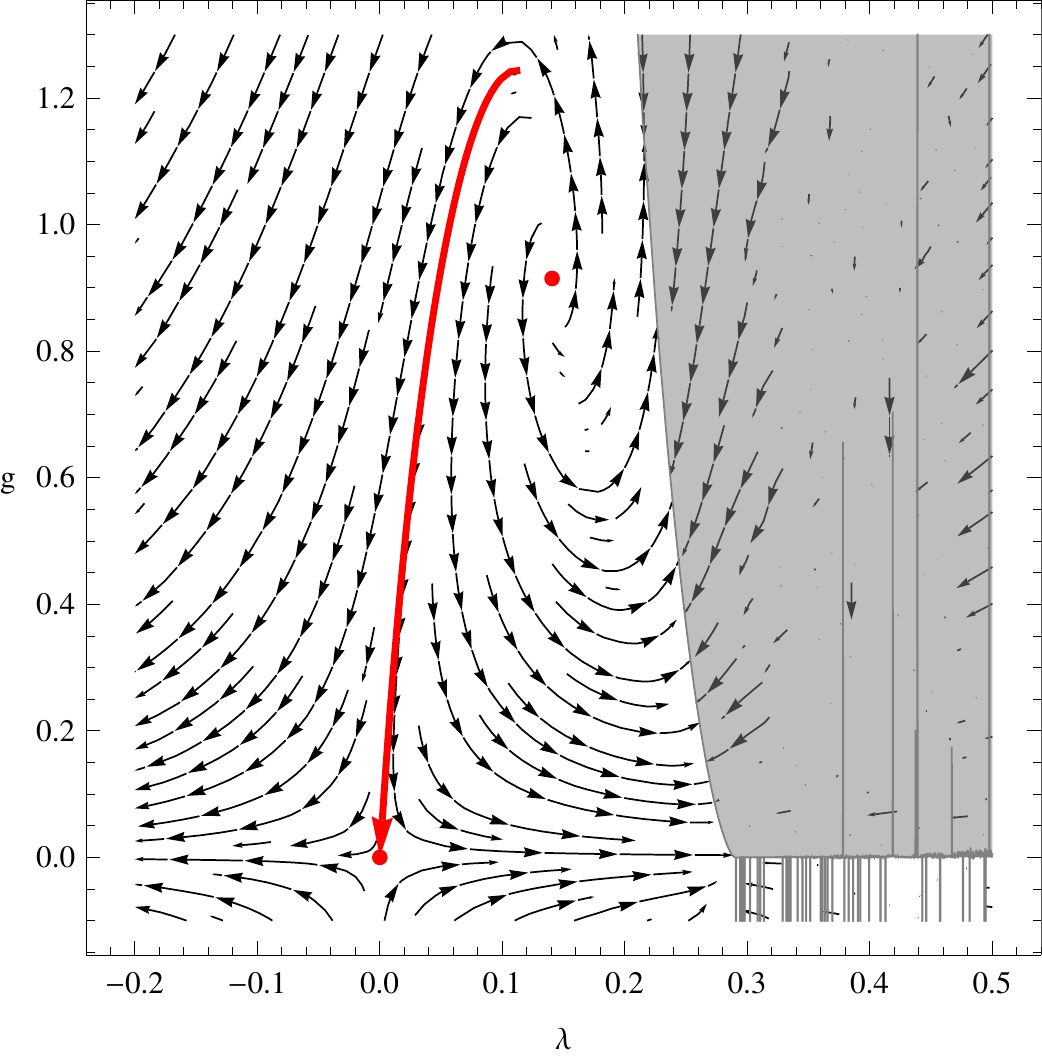}
		\label{fig:PDm5s7v2}}
	\subfigure[\textbf{NGFP} for $s = 8$.]{
		\centering
		\includegraphics[width=0.3\textwidth]{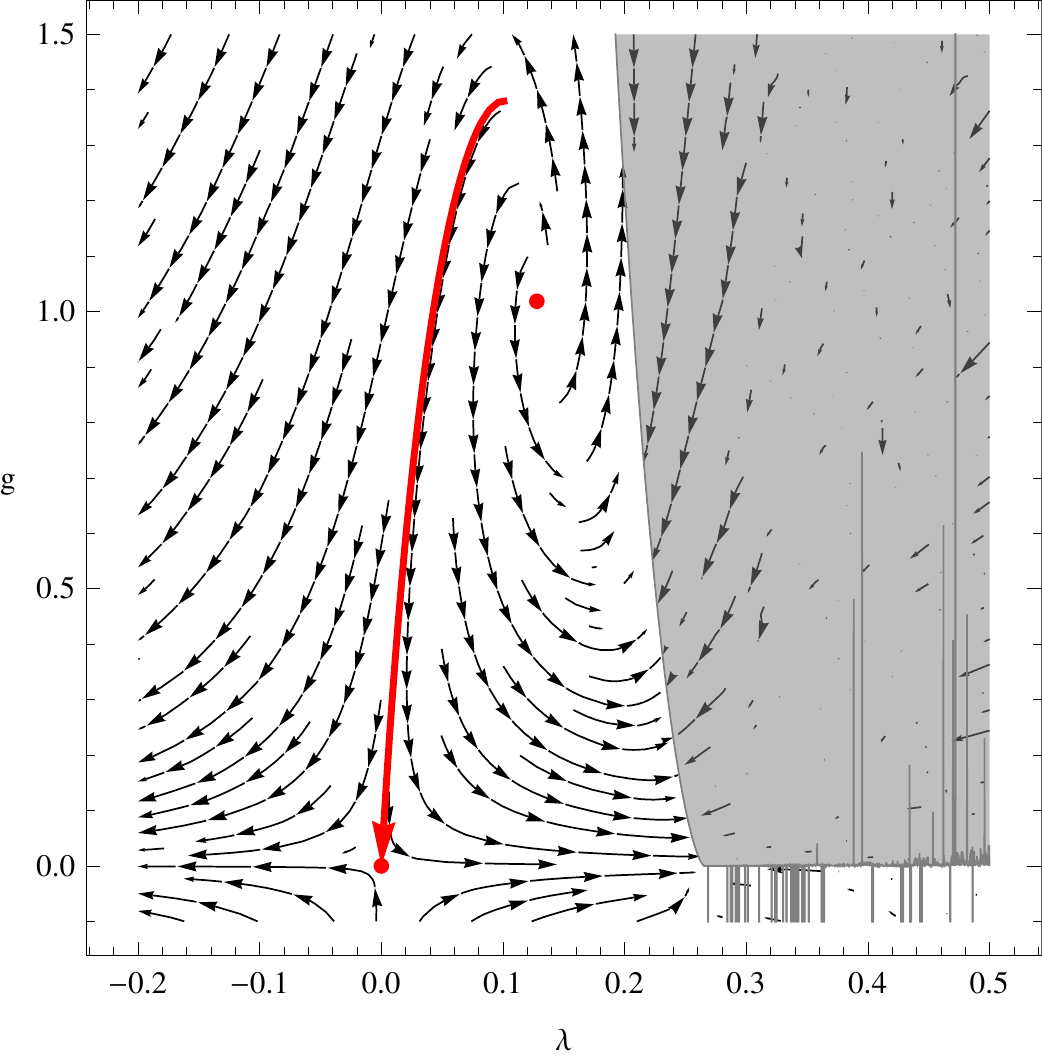}
		\label{fig:PDm5s8v2}}
	\subfigure[\textbf{NGFP} for $s = 9$.]{
		\centering
		\includegraphics[width=0.3\textwidth]{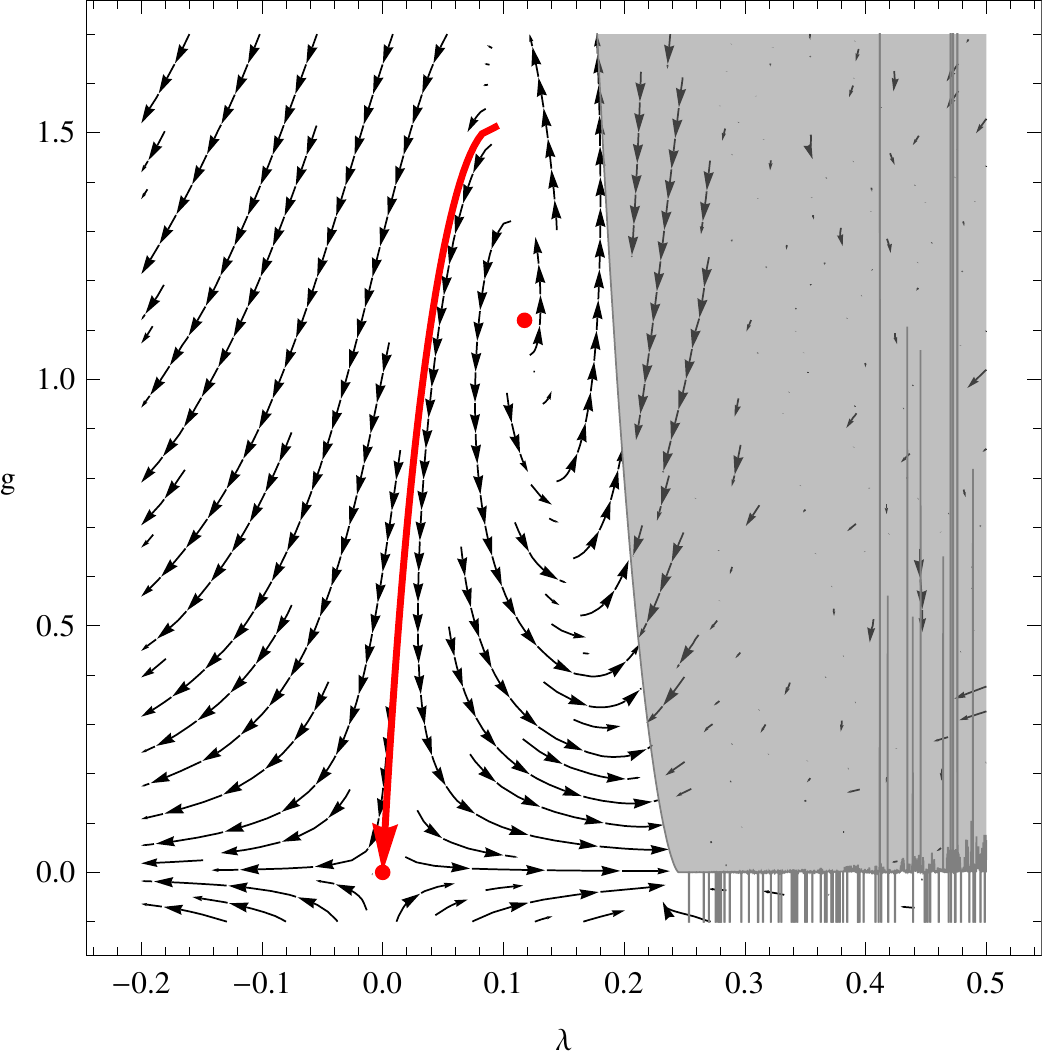}
		\label{fig:PDm5s9v2}}
	\subfigure[\textbf{NGFP} for $s = 10$.]{
		\centering
		\includegraphics[width=0.3\textwidth]{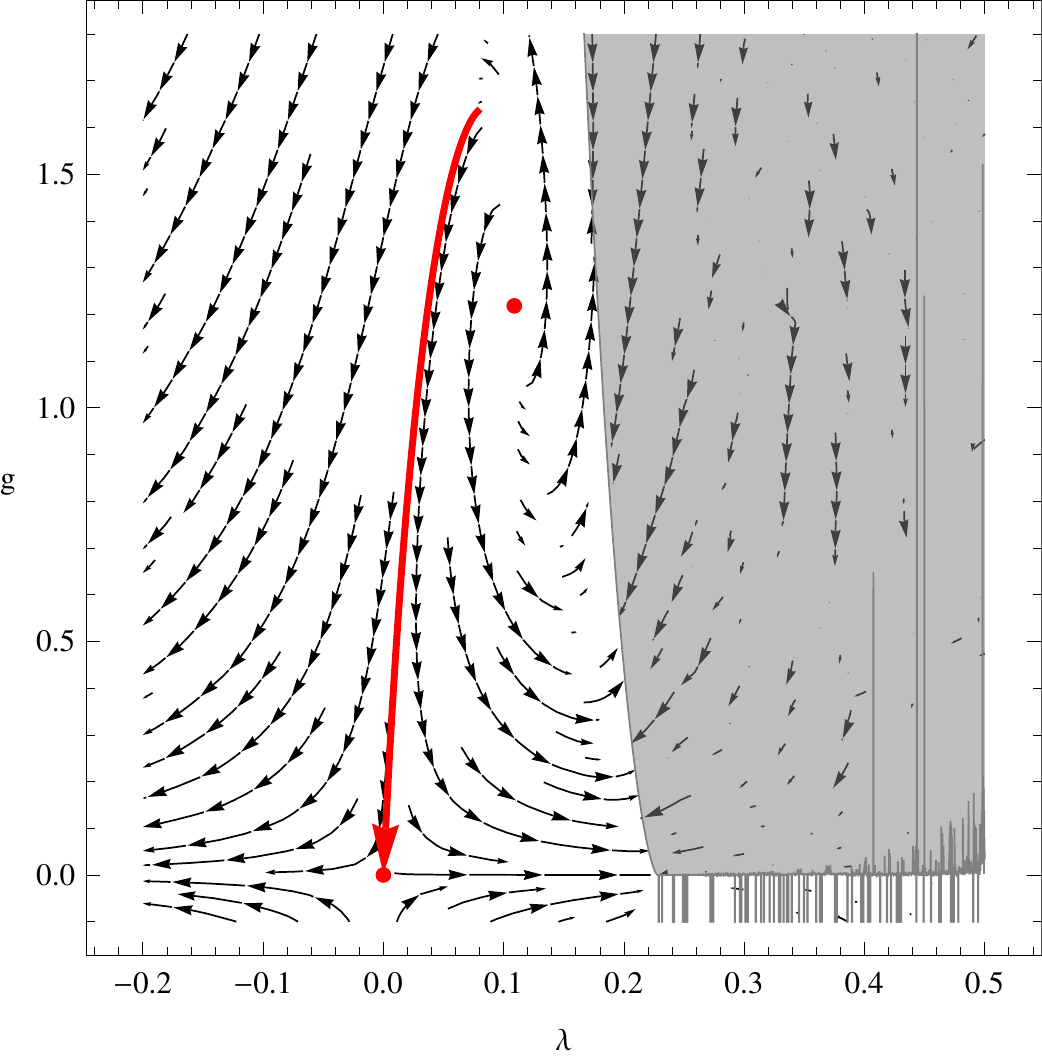}
		\label{fig:PDm5s10v2}}
	\subfigure[\textbf{NGFP} for $s = 20$.]{
		\centering
		\includegraphics[width=0.3\textwidth]{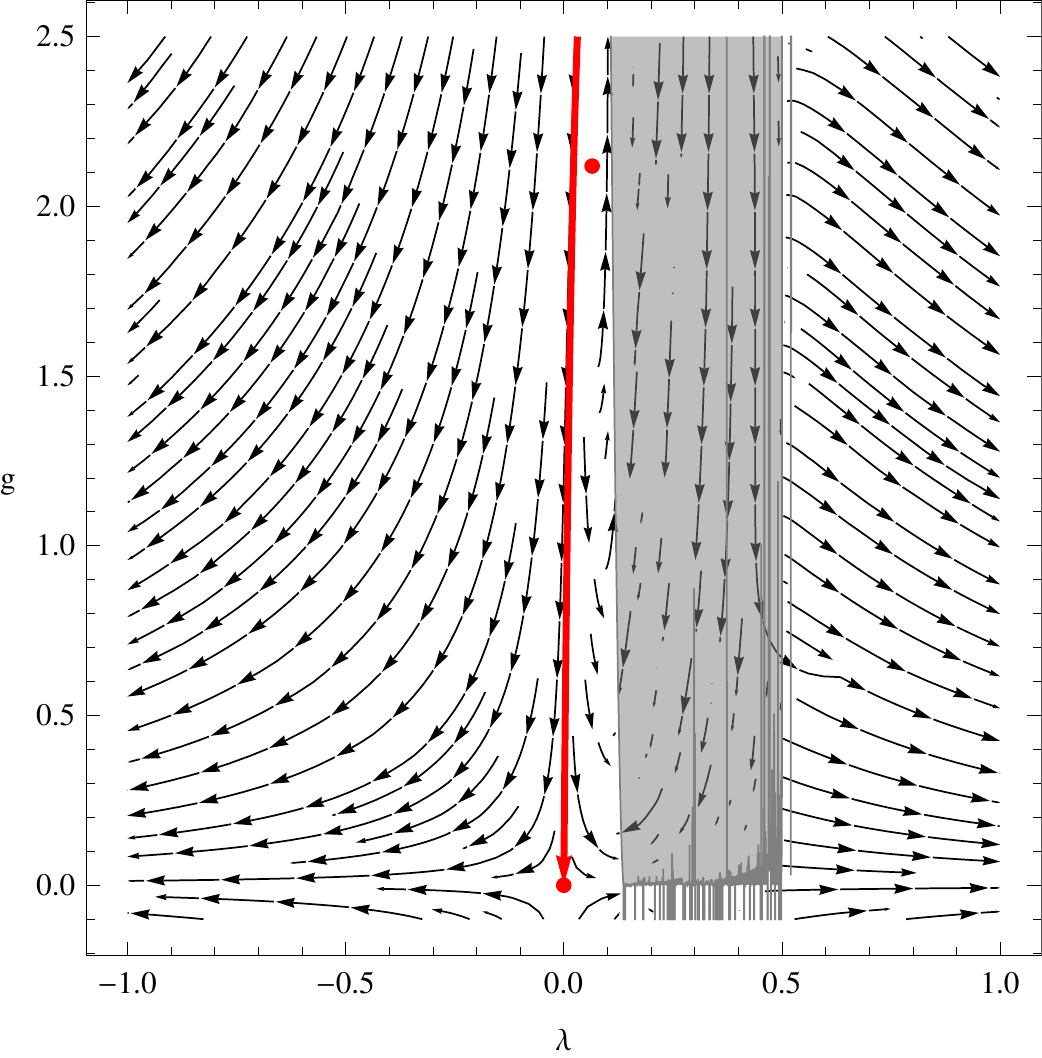}
		\label{fig:PDm5s20}}
	\caption{RG phase portrait for the squared mass parameter $µ^2 = 5$ and different values of the shape parameter $s$, ranging from $s=\frac{1}{2}$ to $s=20$.}
	\label{fig:NGFPm5}
\end{figure}
\clearpage

\begin{figure}[htbp]
	\centering
	\subfigure[\textbf{NGFP} for $s = \frac{1}{2}$.]{
		\centering
		\includegraphics[width=0.3\textwidth]{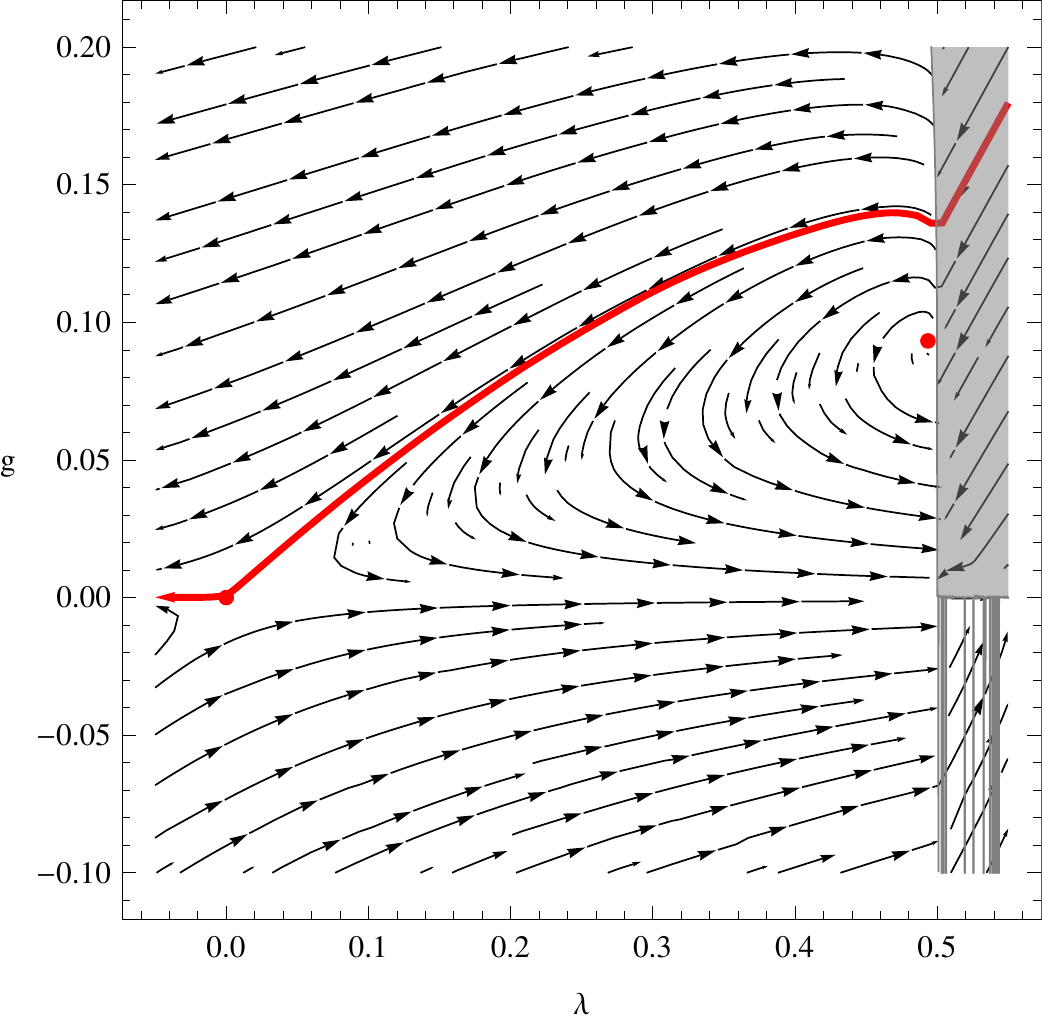}
		\label{fig:PDm10s12}}
	\subfigure[\textbf{NGFP} for $s = 1$.]{
		\centering
		\includegraphics[width=0.3\textwidth]{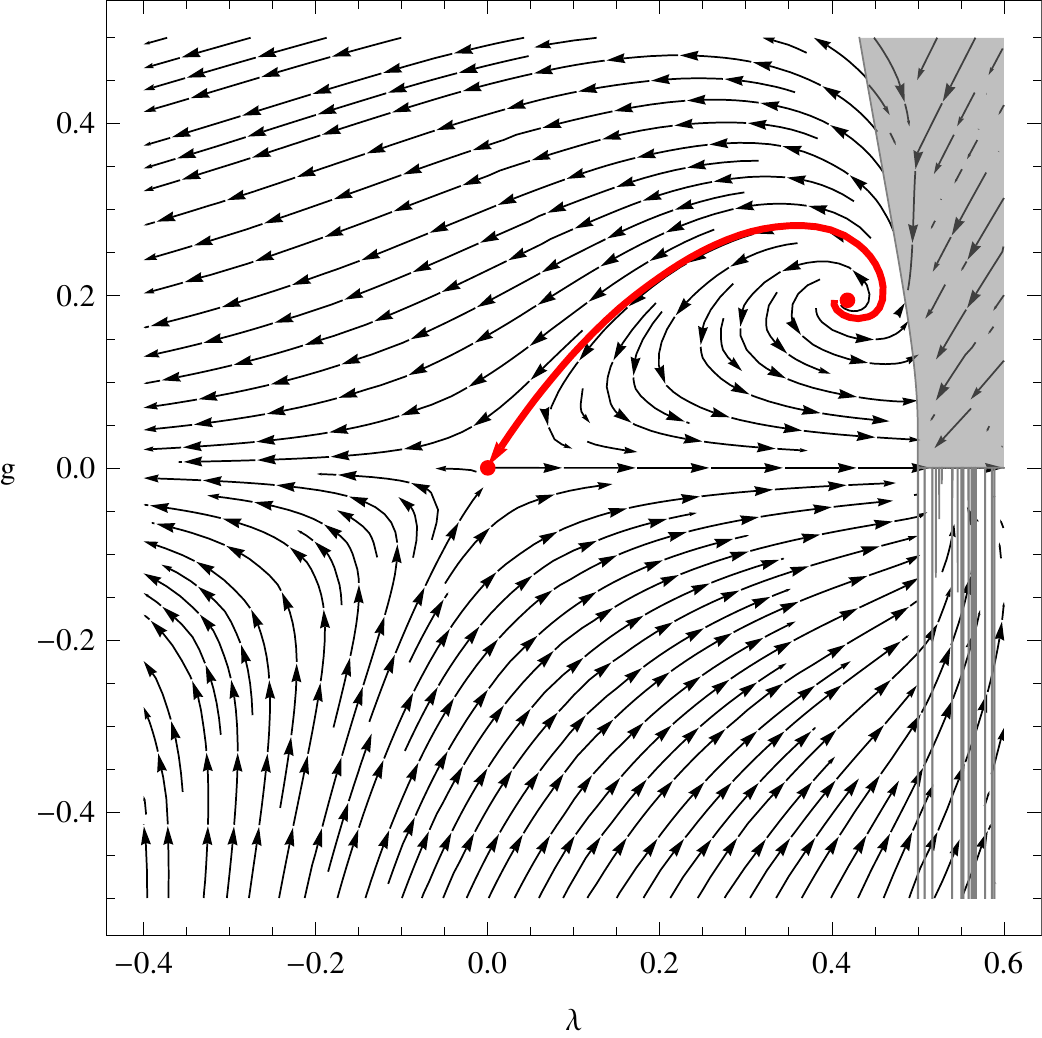}
		\label{fig:PDm10s1}}
	\subfigure[\textbf{NGFP} for $s = 2$.]{
		\centering
		\includegraphics[width=0.3\textwidth]{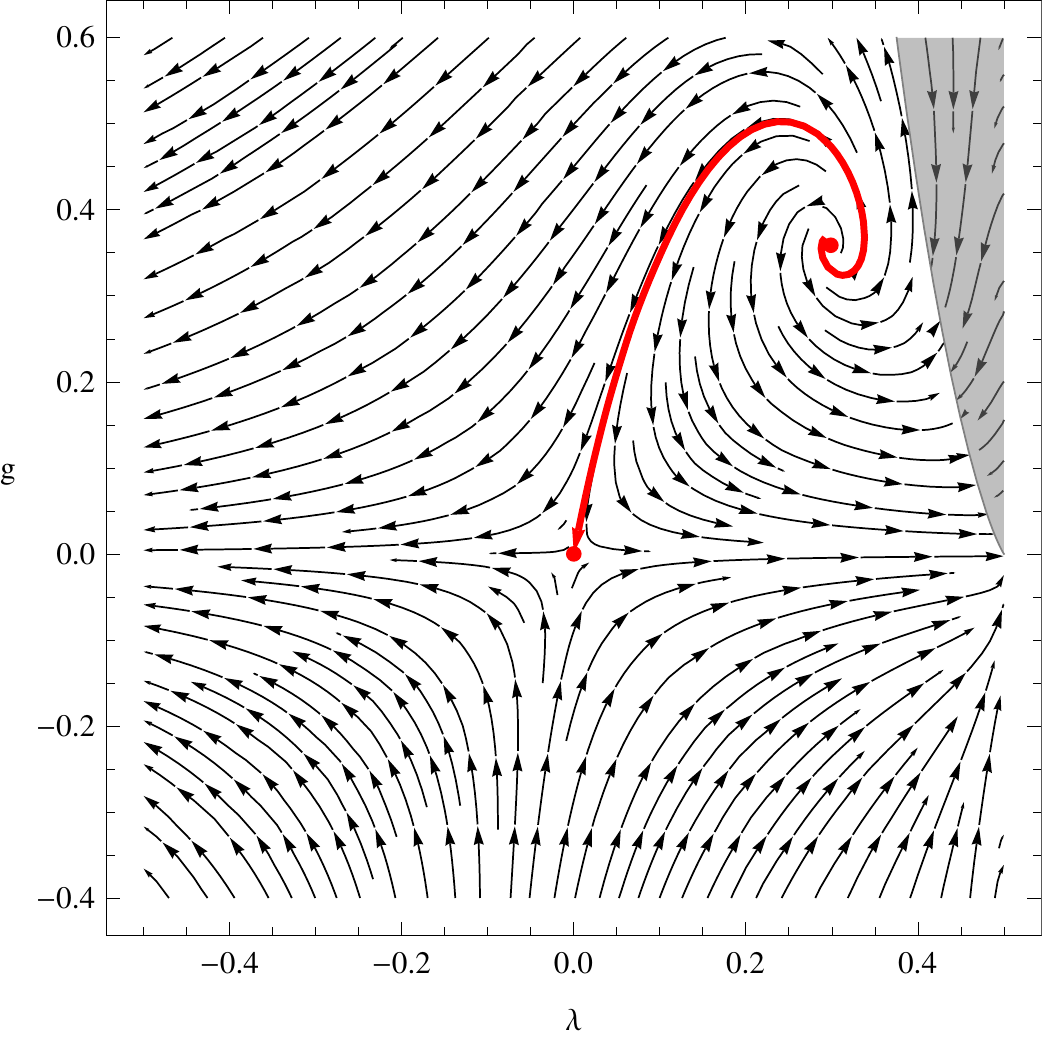}
		\label{fig:PDm10s2}}
	\subfigure[\textbf{NGFP} for $s = 3$.]{
		\centering
		\includegraphics[width=0.3\textwidth]{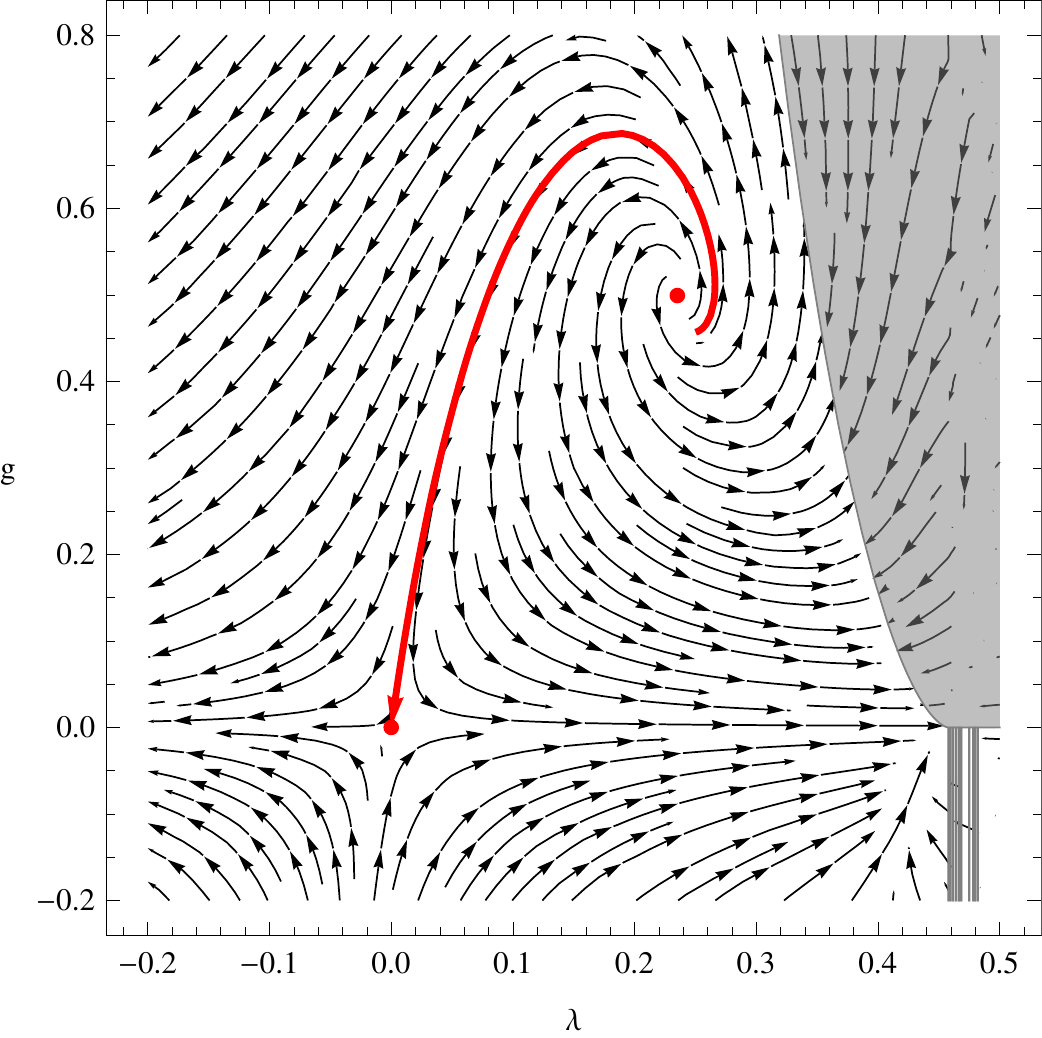}
		\label{fig:PDm10s3}}
	\subfigure[\textbf{NGFP} for $s = 4$.]{
		\centering
		\includegraphics[width=0.3\textwidth]{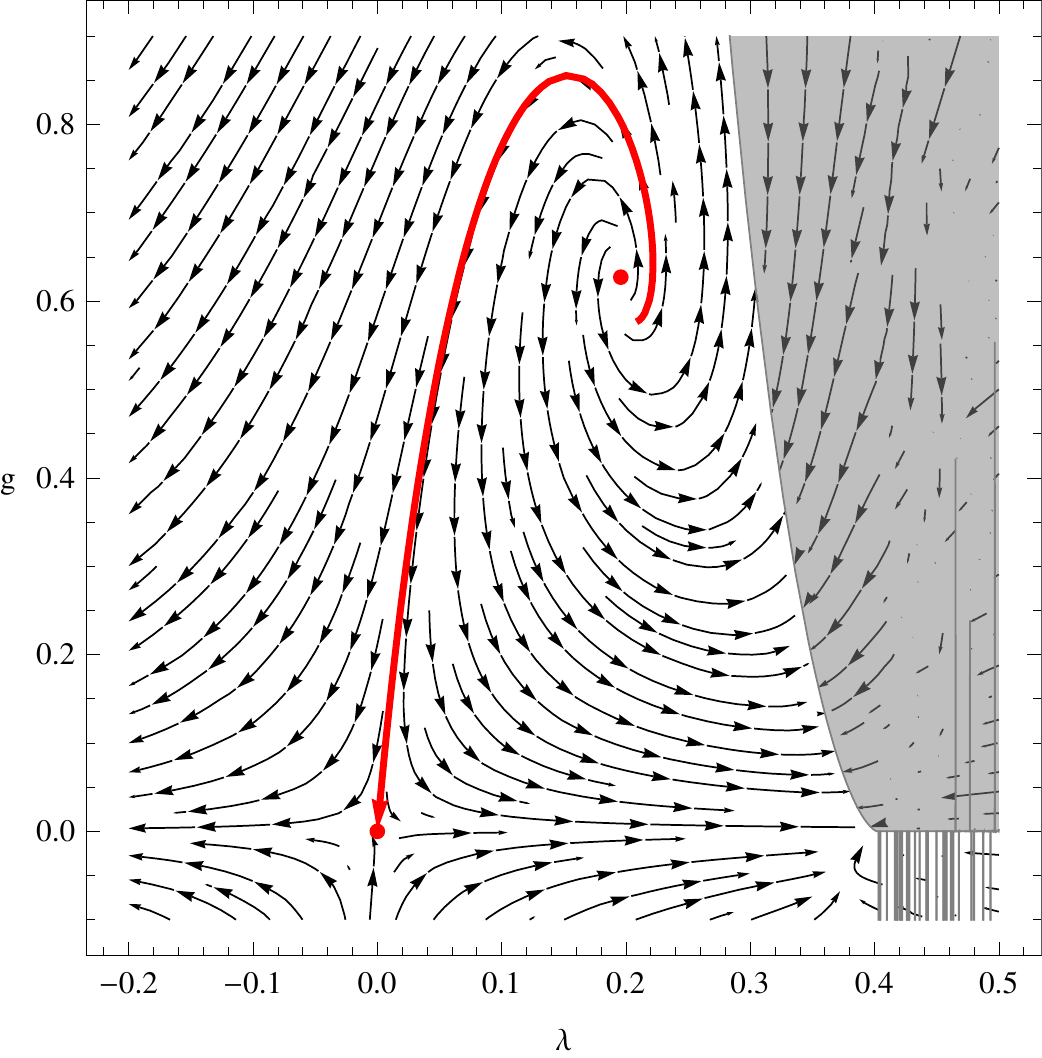}
		\label{fig:PDm10s4}}
	\subfigure[\textbf{NGFP} for $s = 5$.]{
		\centering
		\includegraphics[width=0.3\textwidth]{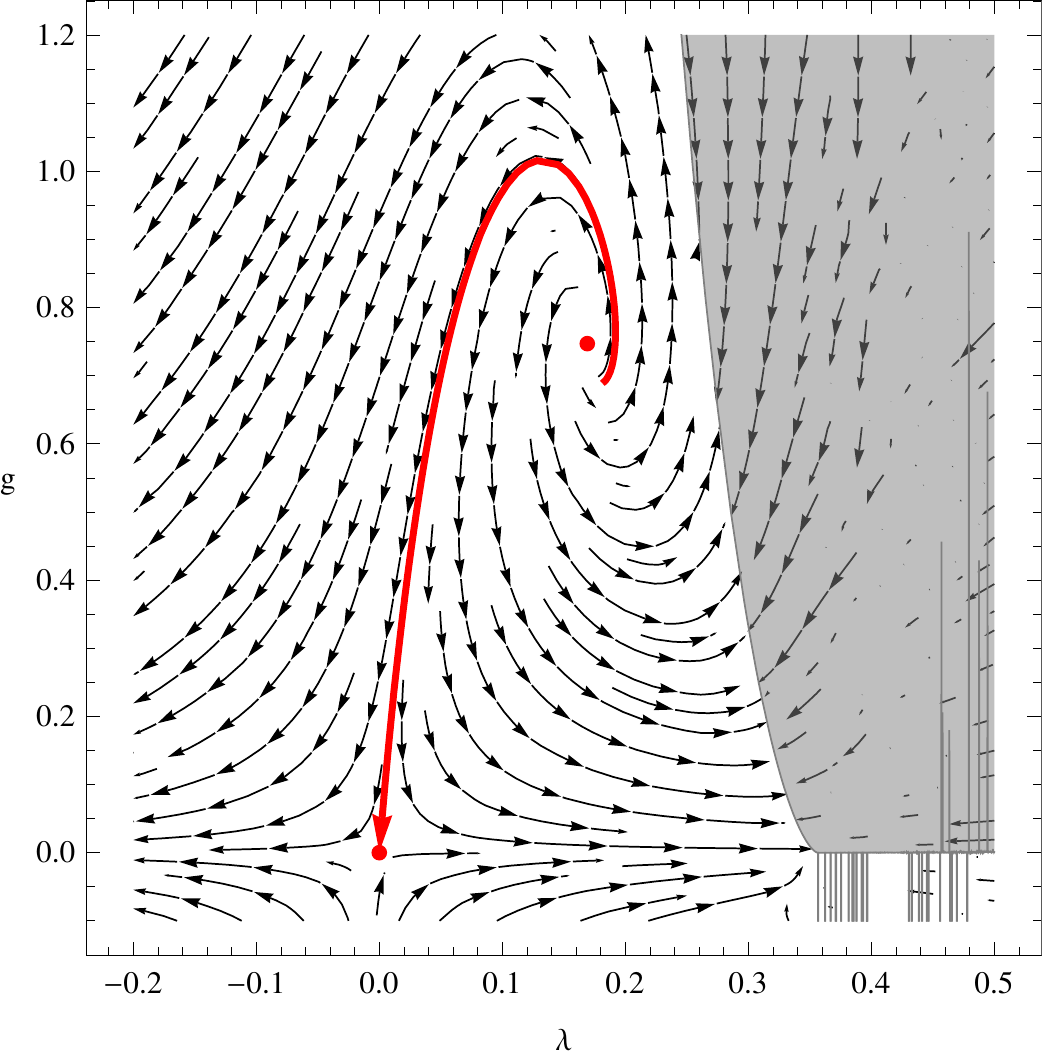}
		\label{fig:PDm10s5}}
	\subfigure[\textbf{NGFP} for $s = 6$.]{
		\centering
		\includegraphics[width=0.3\textwidth]{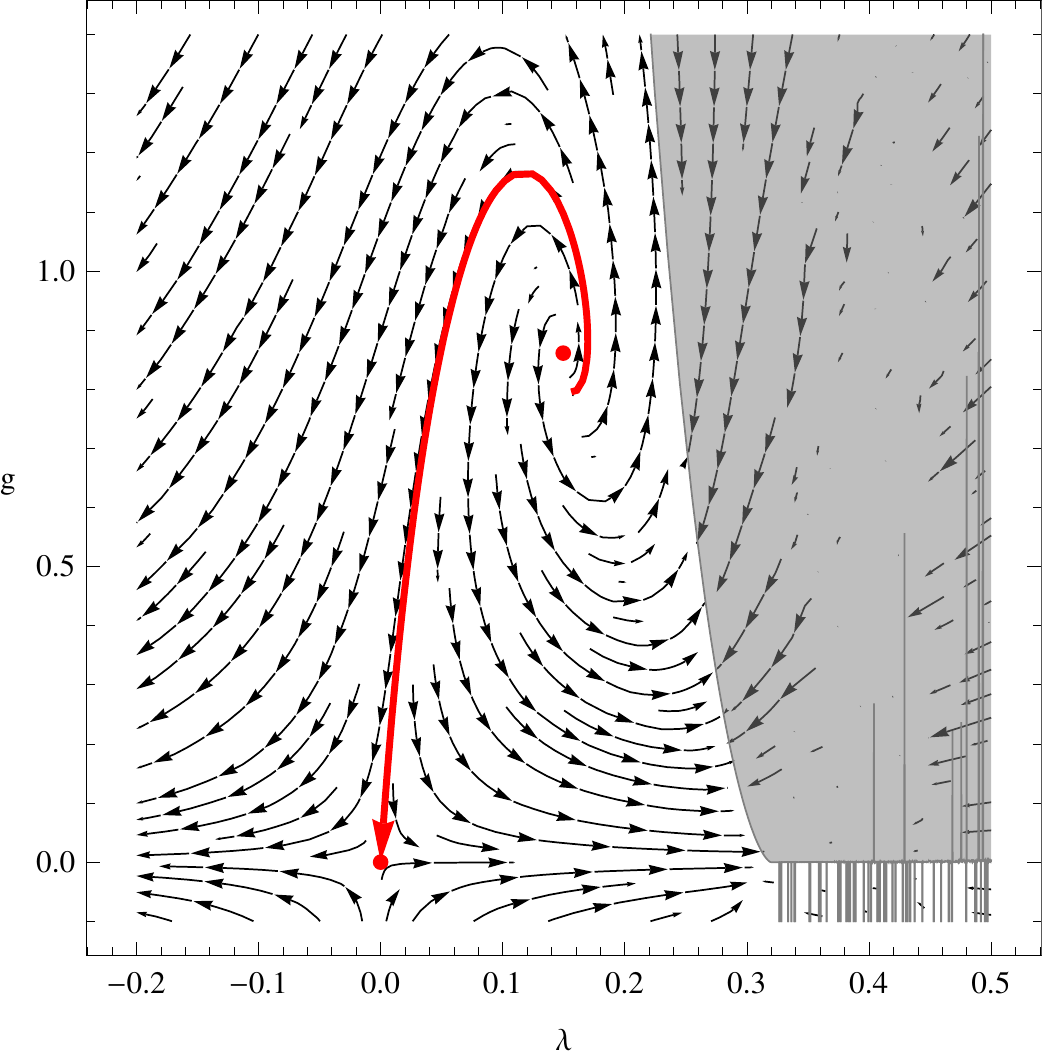}
		\label{fig:PDm10s6}}
	\subfigure[\textbf{NGFP} for $s = 7$.]{
		\centering
		\includegraphics[width=0.3\textwidth]{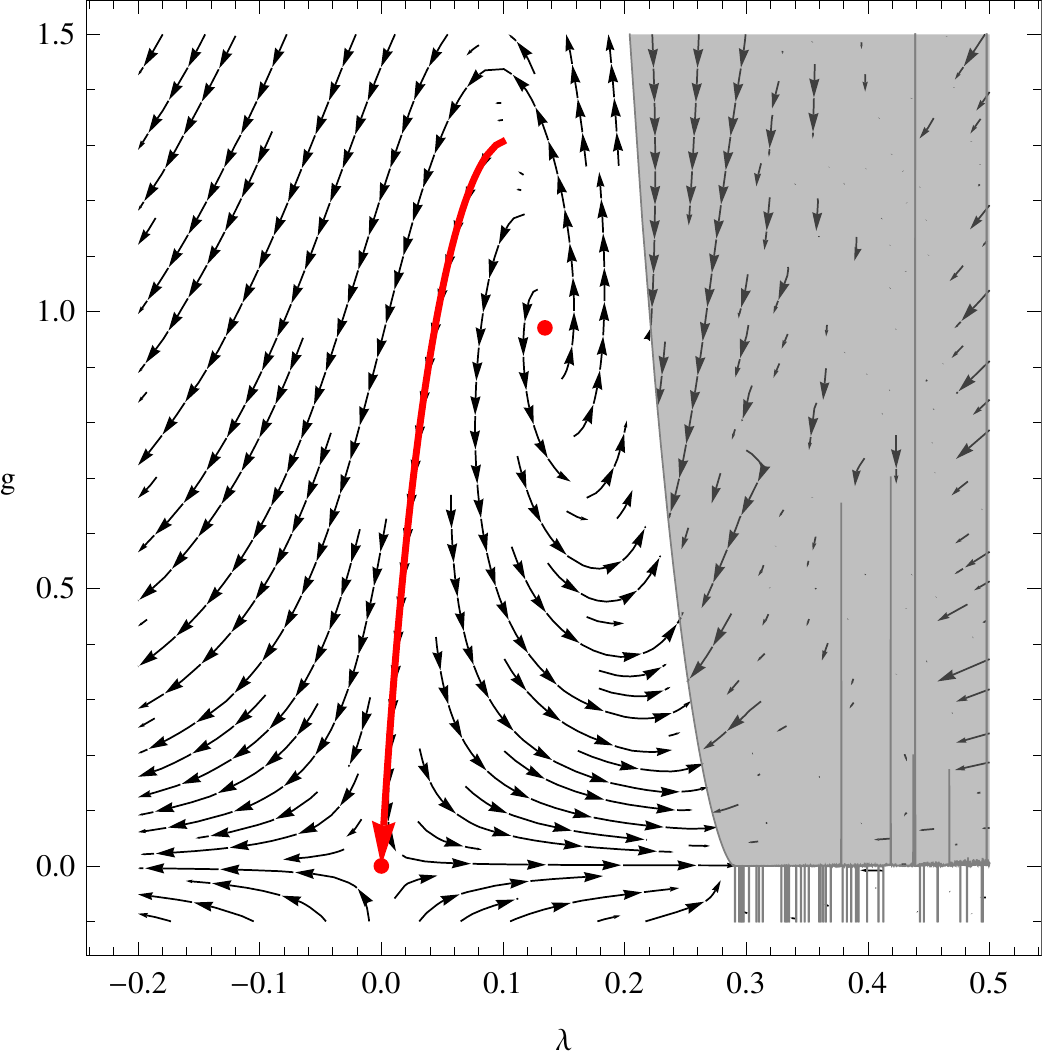}
		\label{fig:PDm10s7}}
	\subfigure[\textbf{NGFP} for $s = 8$.]{
		\centering
		\includegraphics[width=0.3\textwidth]{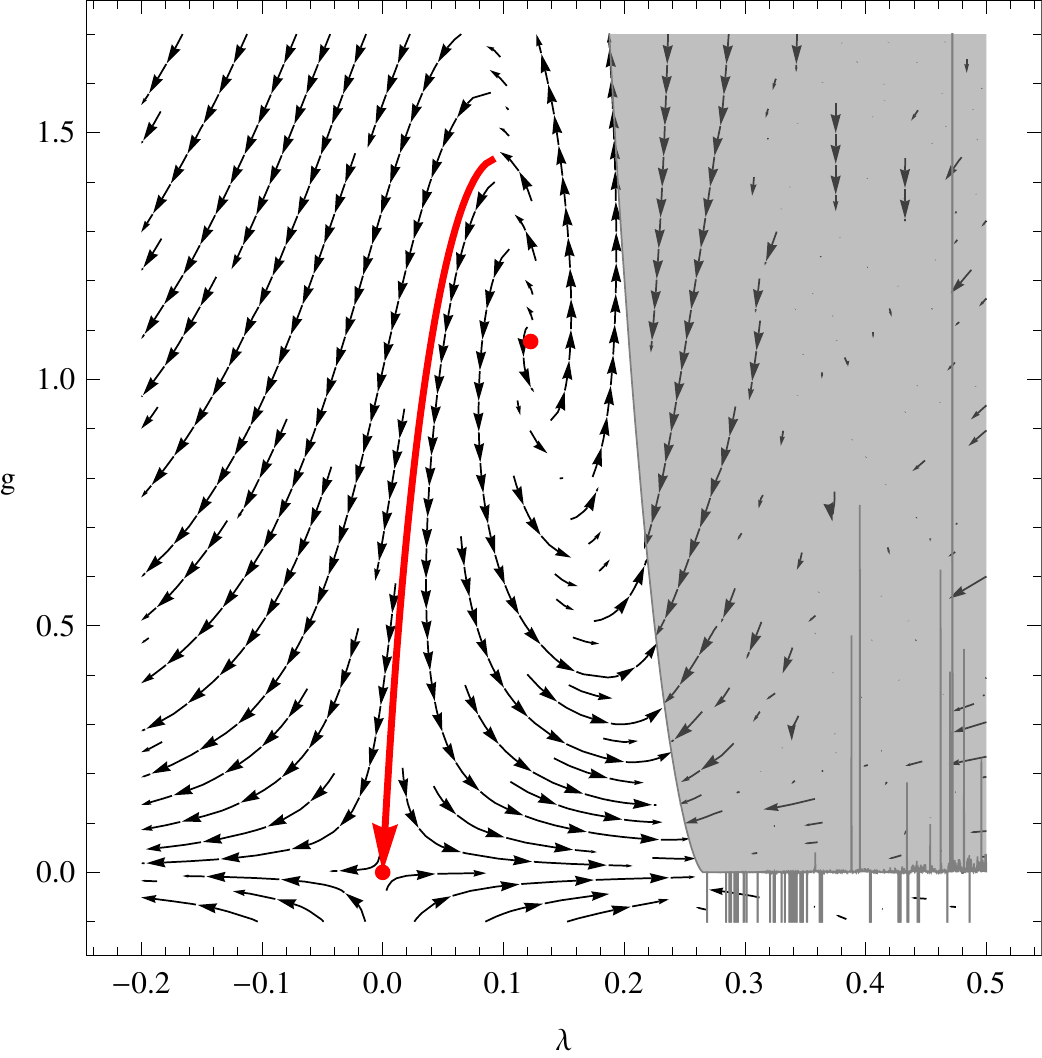}
		\label{fig:PDm10s8}}
	\subfigure[\textbf{NGFP} for $s = 9$.]{
		\centering
		\includegraphics[width=0.3\textwidth]{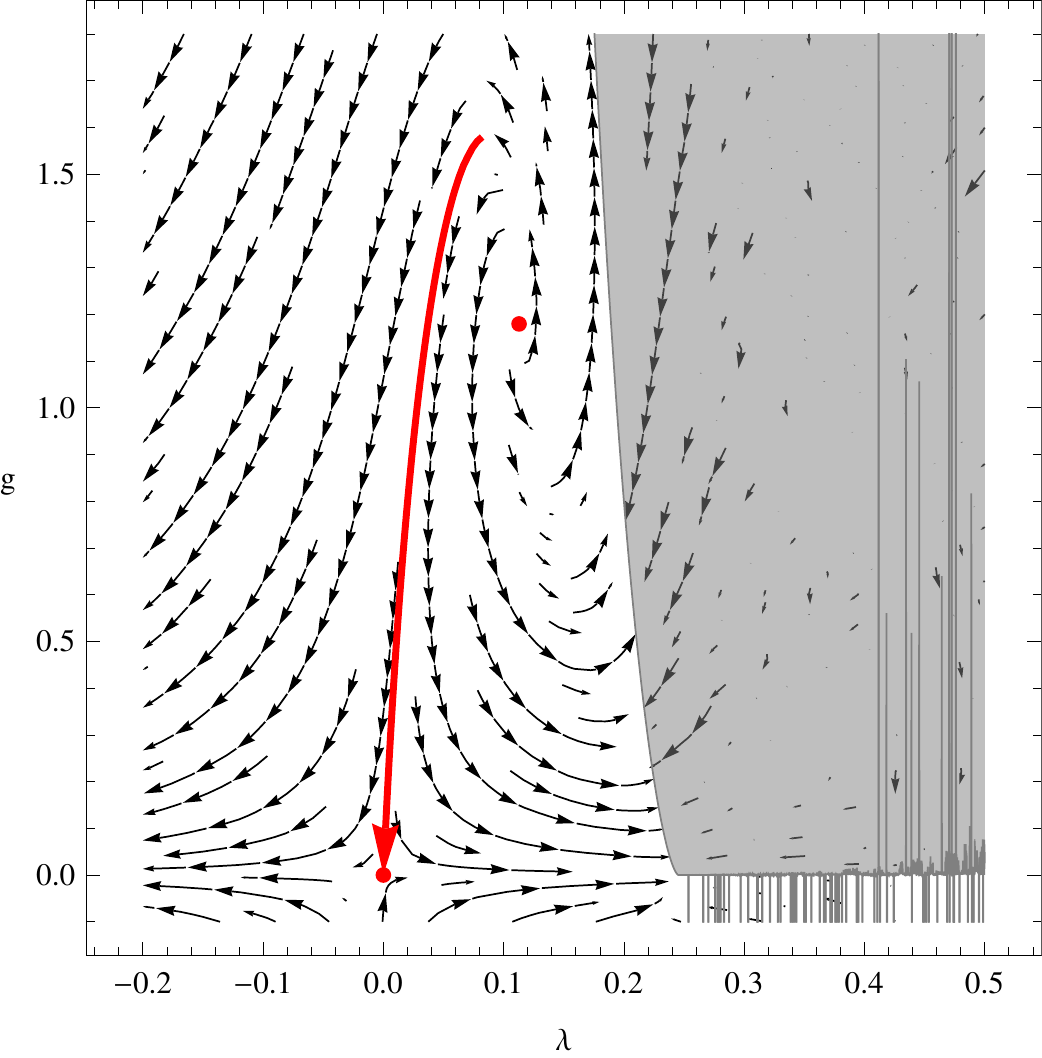}
		\label{fig:PDm10s9}}
	\subfigure[\textbf{NGFP} for $s = 10$.]{
		\centering
		\includegraphics[width=0.3\textwidth]{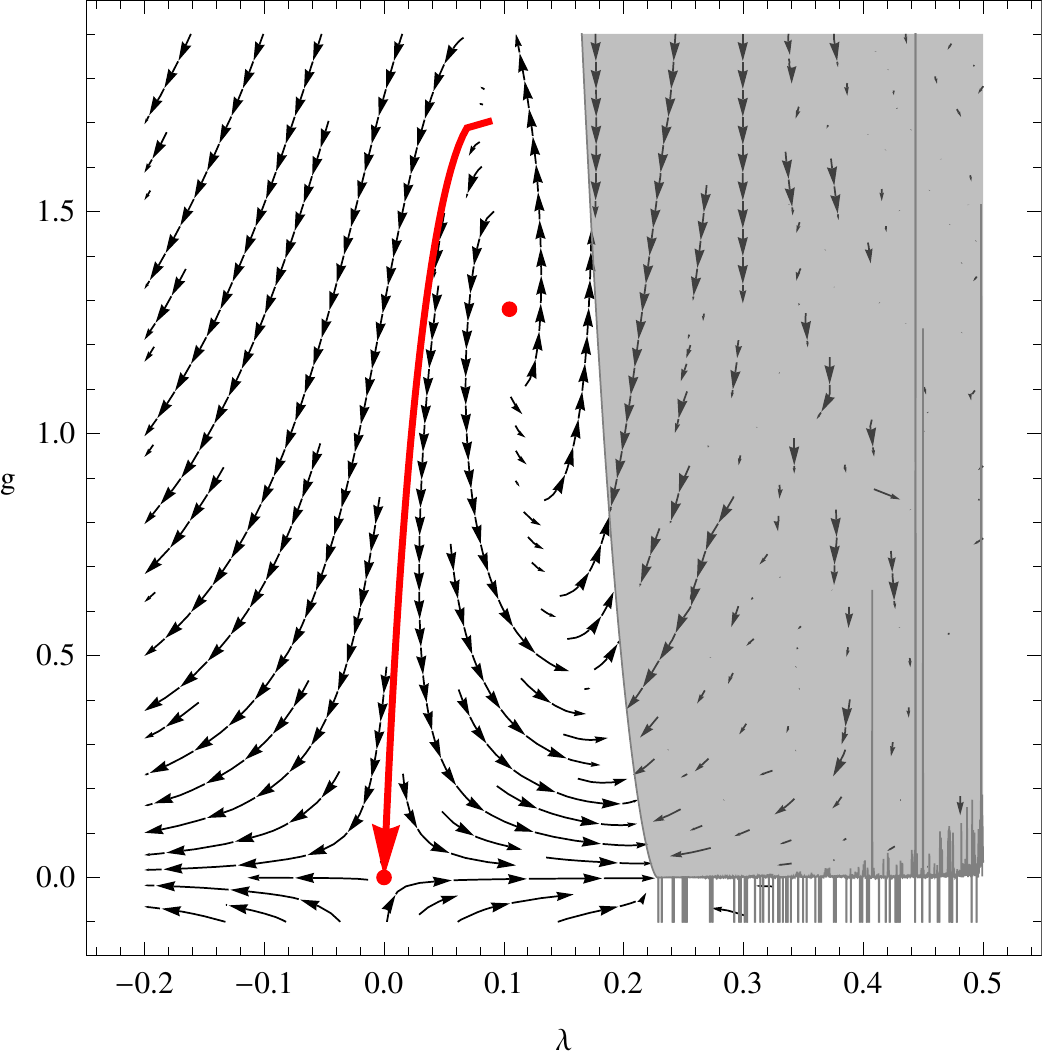}
		\label{fig:PDm10s10}}
	\subfigure[\textbf{NGFP} for $s = 20$.]{
		\centering
		\includegraphics[width=0.3\textwidth]{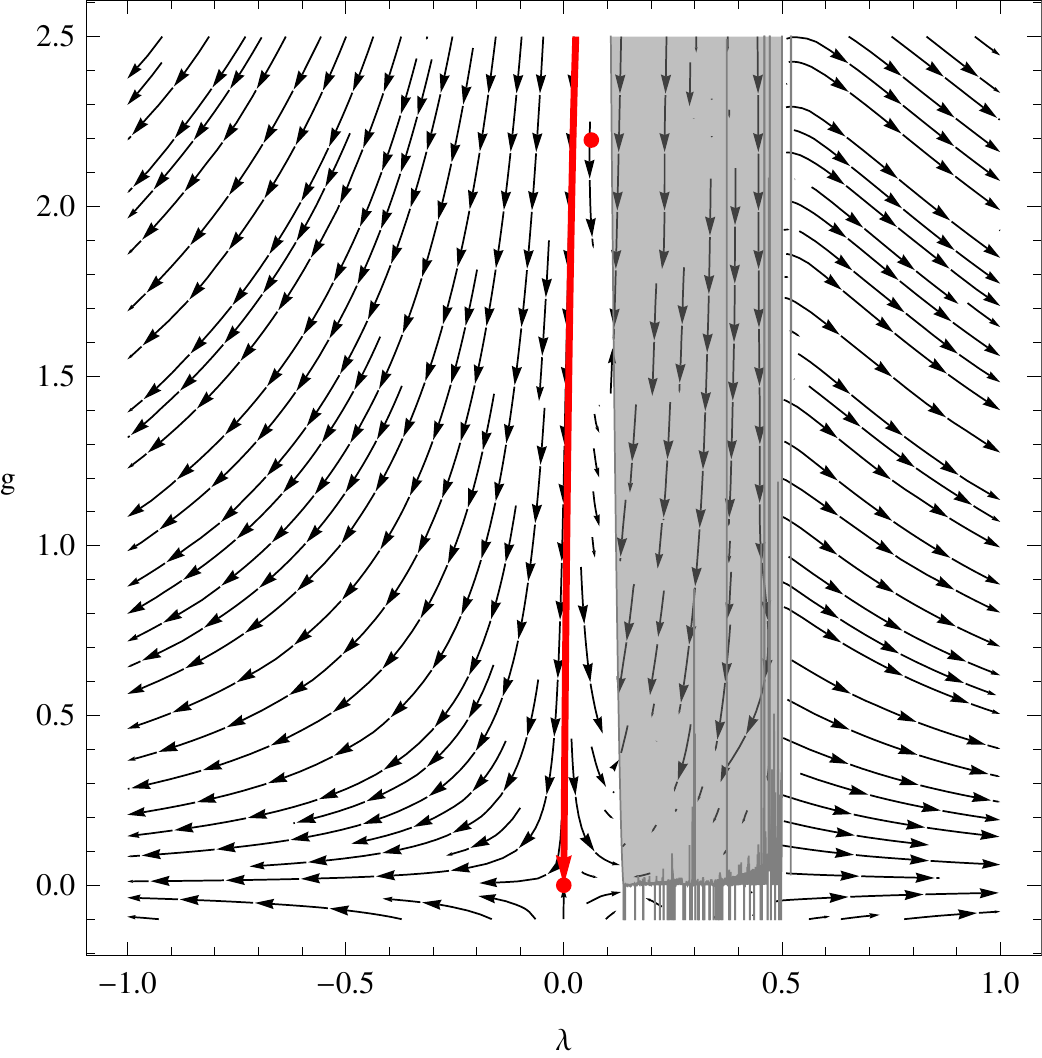}
		\label{fig:PDm10s20}}
	\caption{RG phase portrait for the squared mass parameter $µ^2 = 10$ and different values of the shape parameter $s$, ranging from $s=\frac{1}{2}$ to $s=20$.}
	\label{fig:NGFPm10}
\end{figure}
\clearpage

\begin{figure}[htbp]
	\centering
	\subfigure[\textbf{NGFP} for $µ^2 = \frac{1}{10}$.]{
		\centering
		\includegraphics[width=0.3\textwidth]{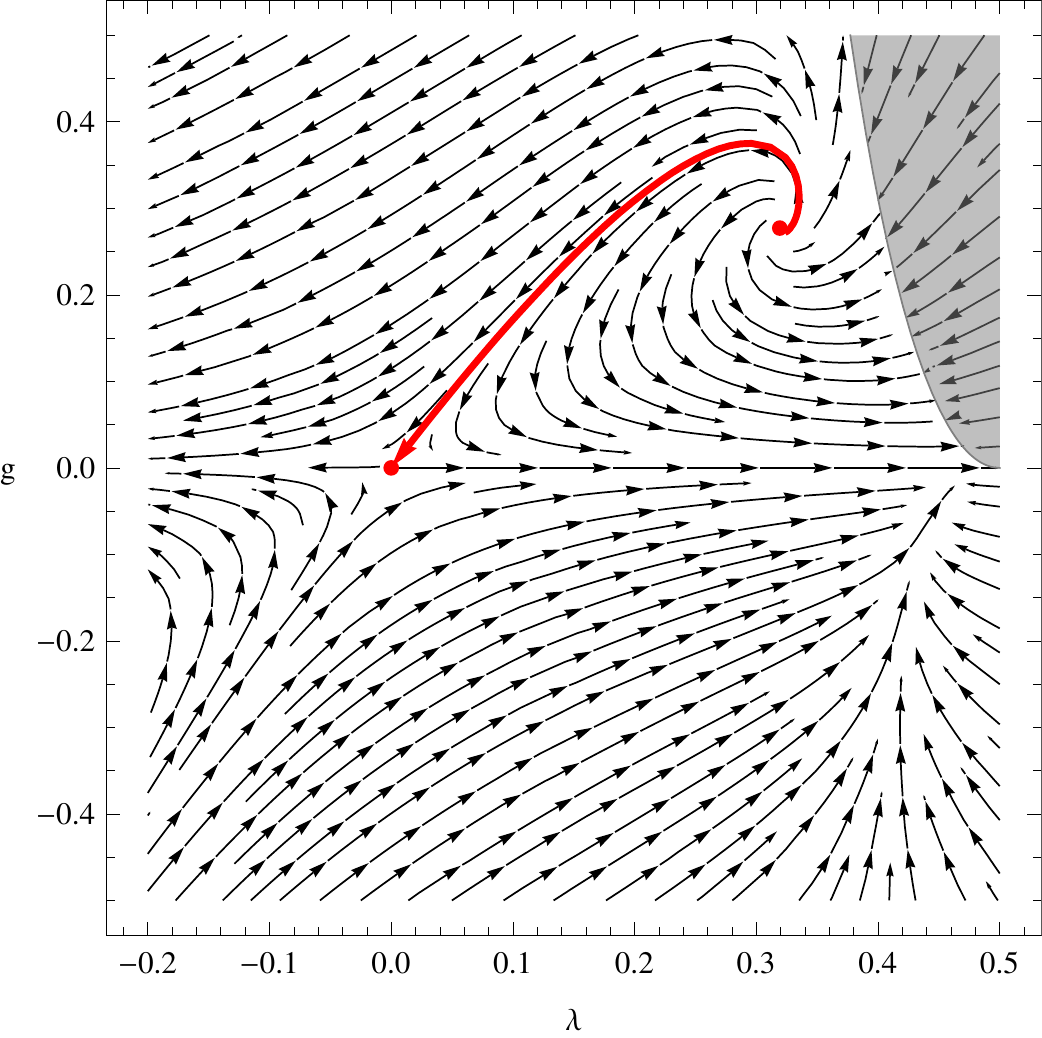}
		\label{fig:PDSSm110}}
	\subfigure[\textbf{NGFP}$\bm{^{\ominus}}$ for $µ^2 = \frac{1}{10}$.]{
		\centering
		\includegraphics[width=0.3\textwidth]{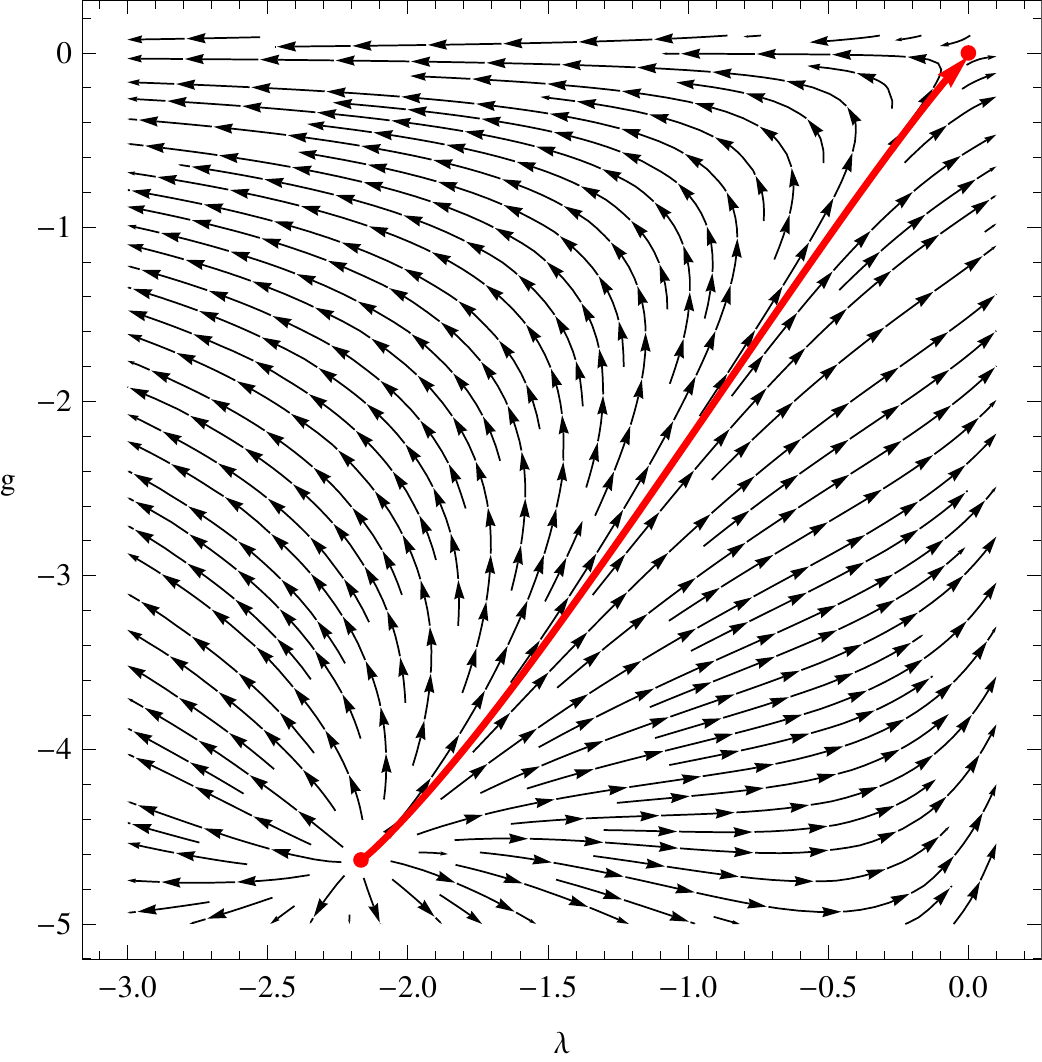}
		\label{fig:PDSSm110v2}}
	\subfigure[\textbf{NGFP}$\bm{^{\oplus}}$ for $µ^2 = \frac{1}{10}$.]{
		\centering
		\includegraphics[width=0.3\textwidth]{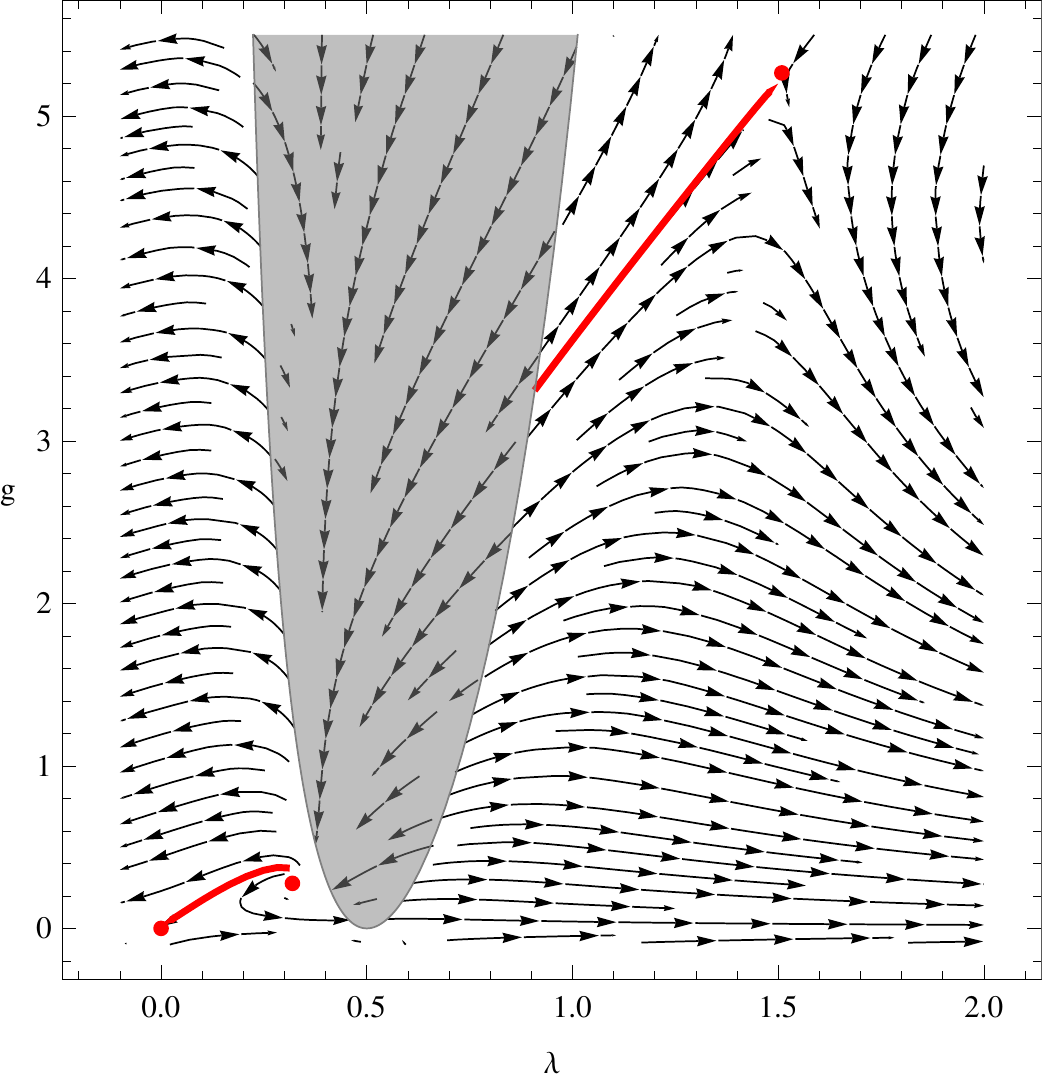}
		\label{fig:PDSSm110v3}}
	\subfigure[\textbf{NGFP} for $µ^2 = \frac{3}{10}$.]{
		\centering
		\includegraphics[width=0.3\textwidth]{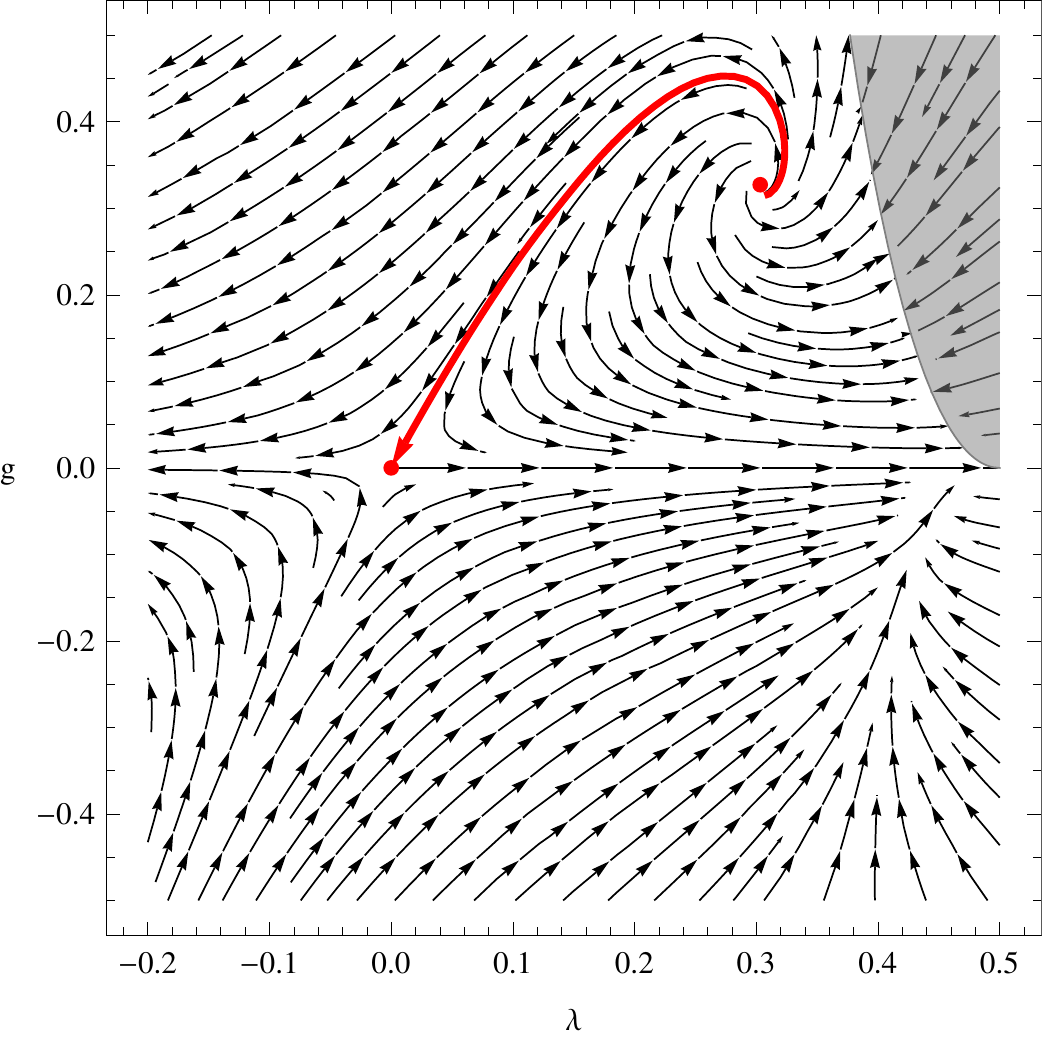}
		\label{fig:PDSSm310}}
	\subfigure[\textbf{NGFP}$\bm{^{\ominus}}$ for $µ^2 = \frac{3}{10}$.]{
		\centering
		\includegraphics[width=0.3\textwidth]{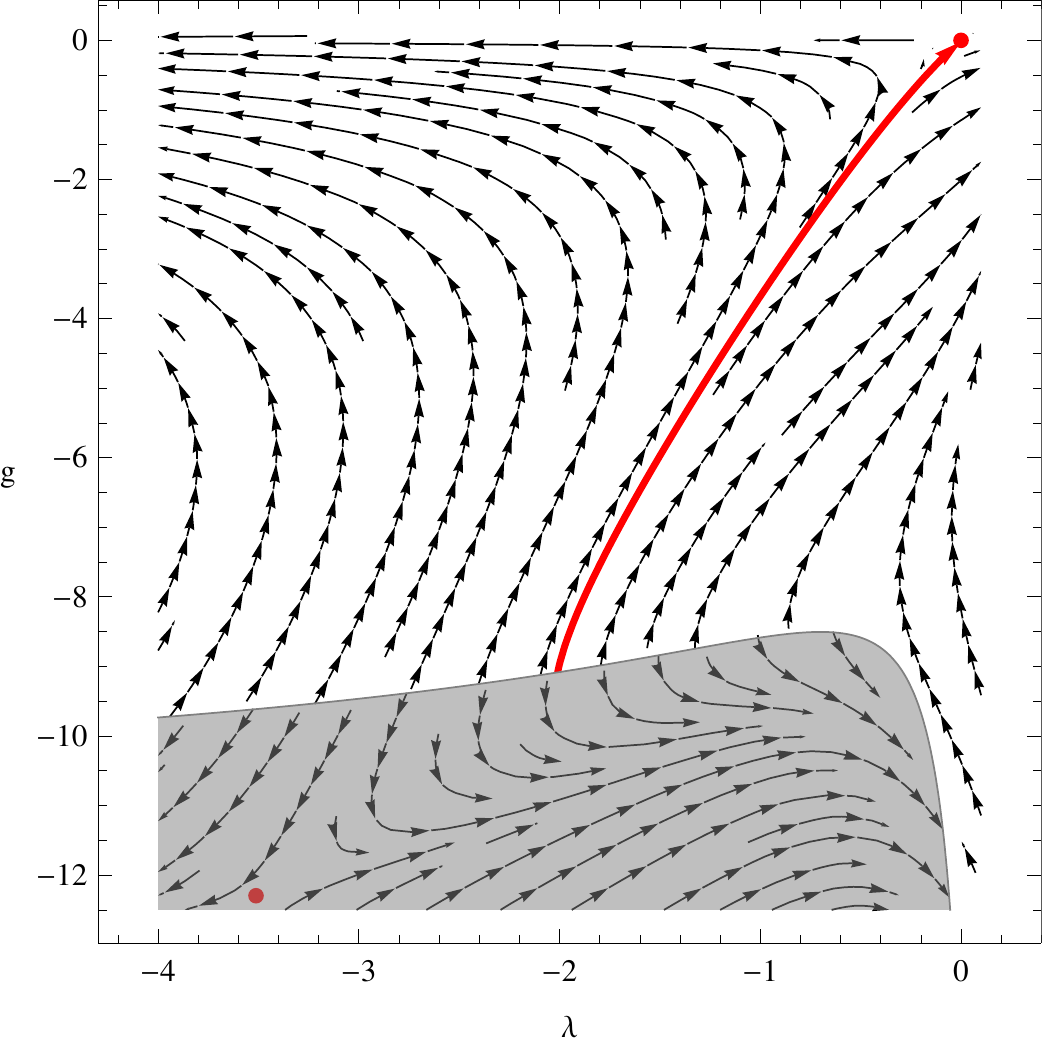}
		\label{fig:PDSSm310v2}}
	\subfigure[\textbf{NGFP} for $µ^2 = \frac{1}{2}$.]{
		\centering
		\includegraphics[width=0.3\textwidth]{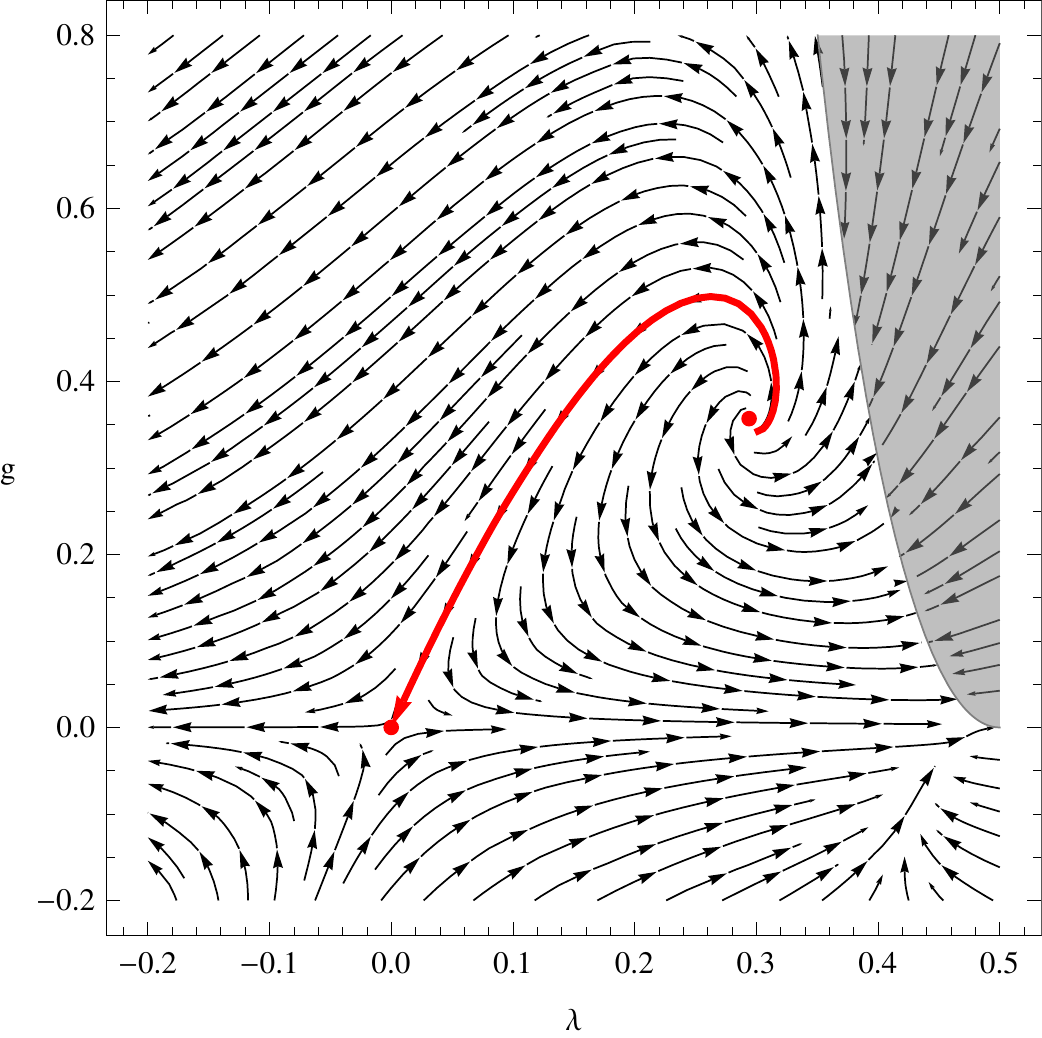}
		\label{fig:PDSSm12}}
	\subfigure[\textbf{NGFP} for $µ^2 = 1$.]{
		\centering
		\includegraphics[width=0.3\textwidth]{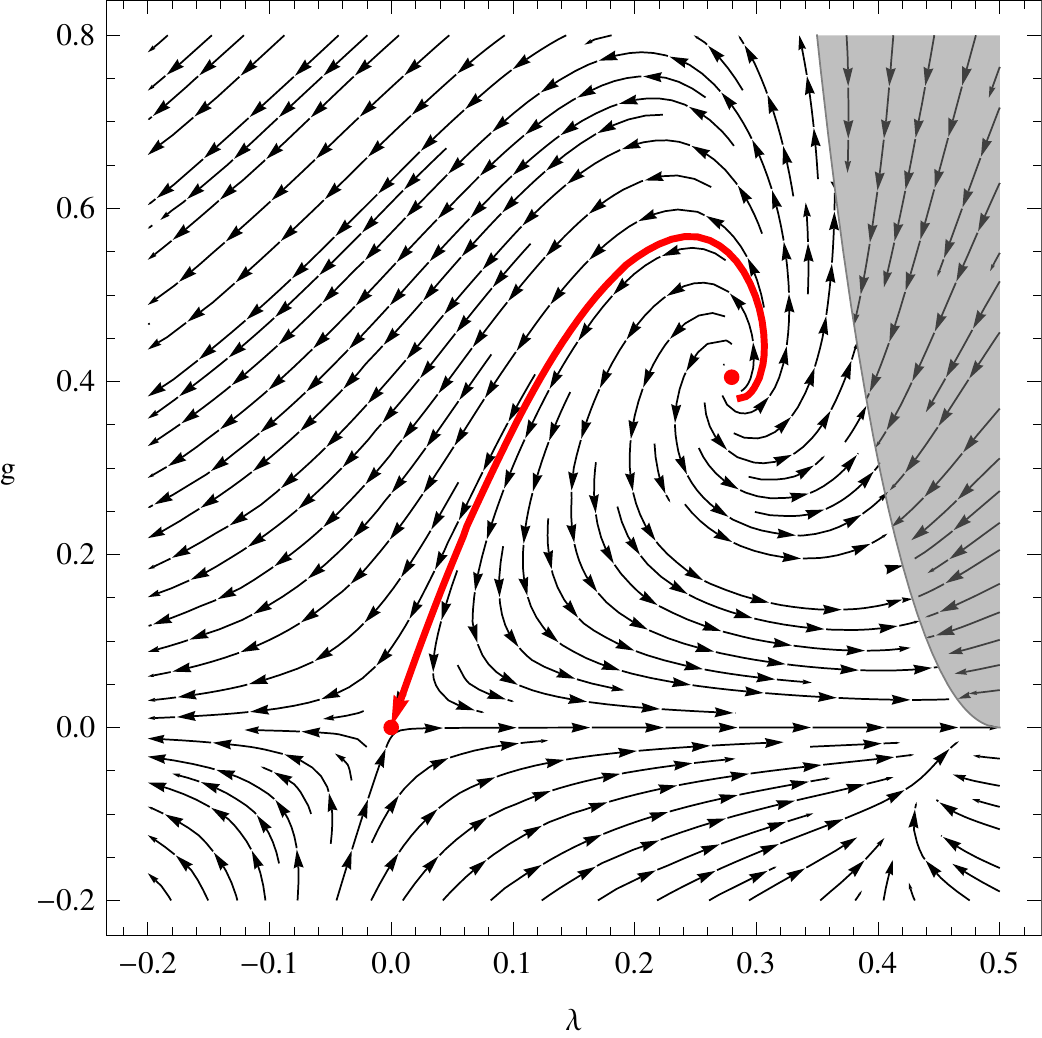}
		\label{fig:PDSSm1}}
	\subfigure[\textbf{NGFP} for $µ^2 = 2$.]{
		\centering
		\includegraphics[width=0.3\textwidth]{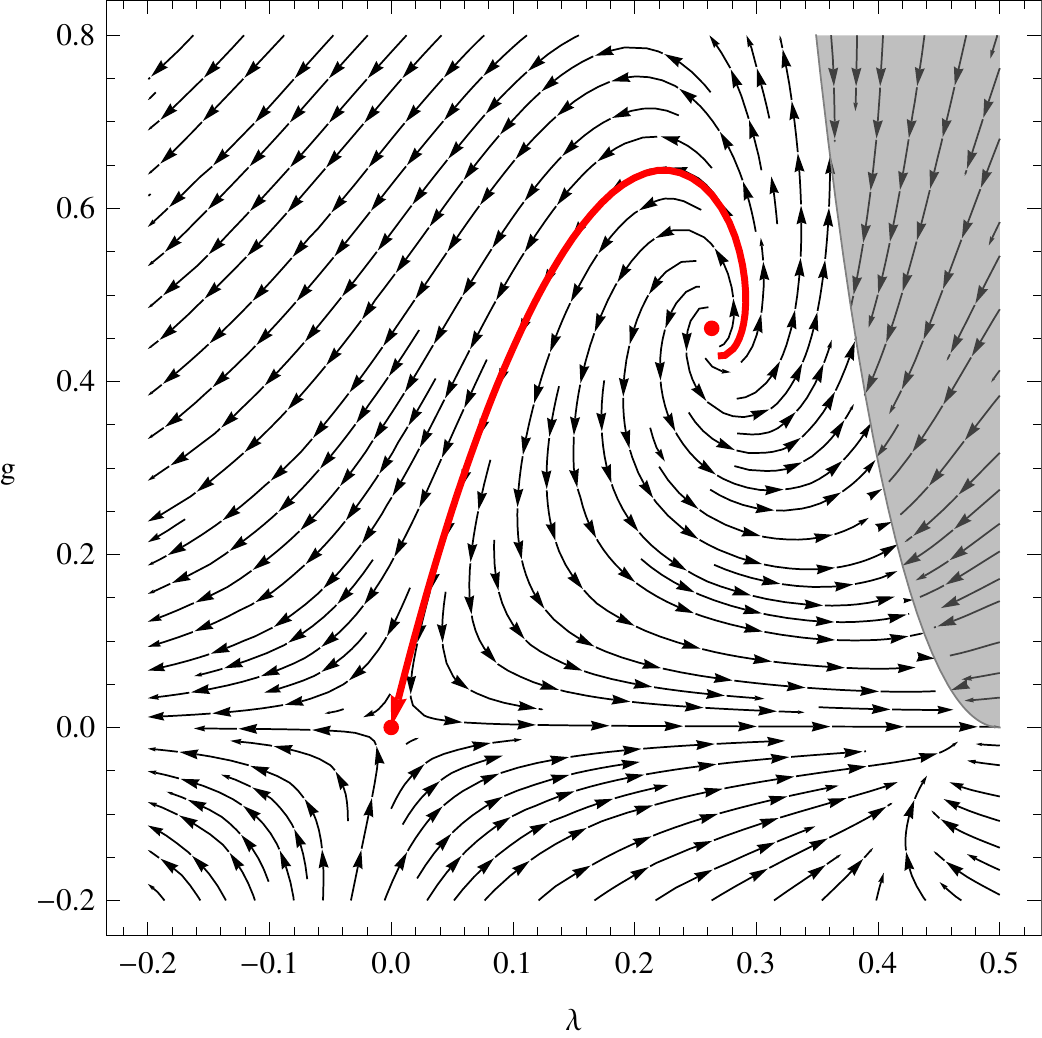}
		\label{fig:PDSSm2}}
	\subfigure[\textbf{NGFP} for $µ^2 = 10$.]{
		\centering
		\includegraphics[width=0.3\textwidth]{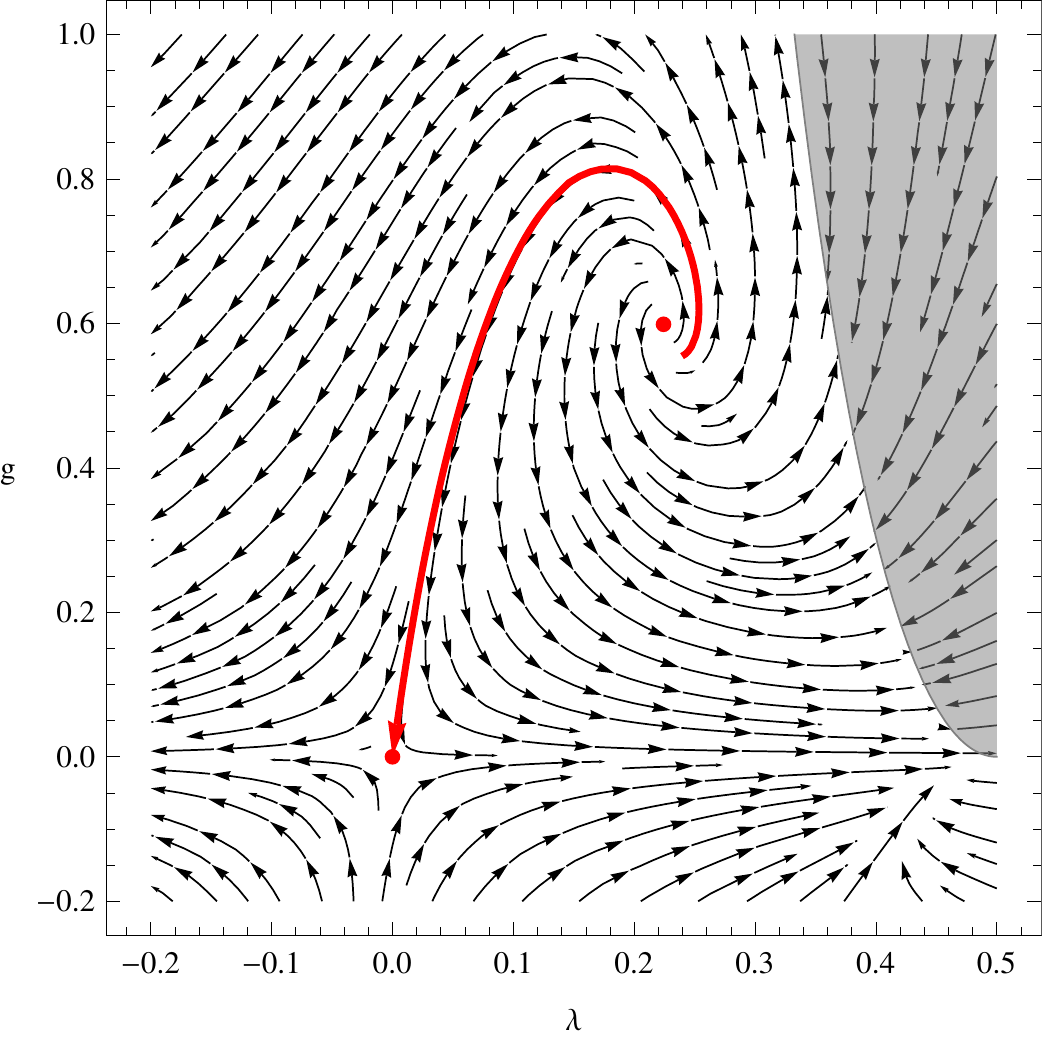}
		\label{fig:PDSSm10}}
	\subfigure[\textbf{NGFP} for $µ^2 = 100$.]{
		\centering
		\includegraphics[width=0.3\textwidth]{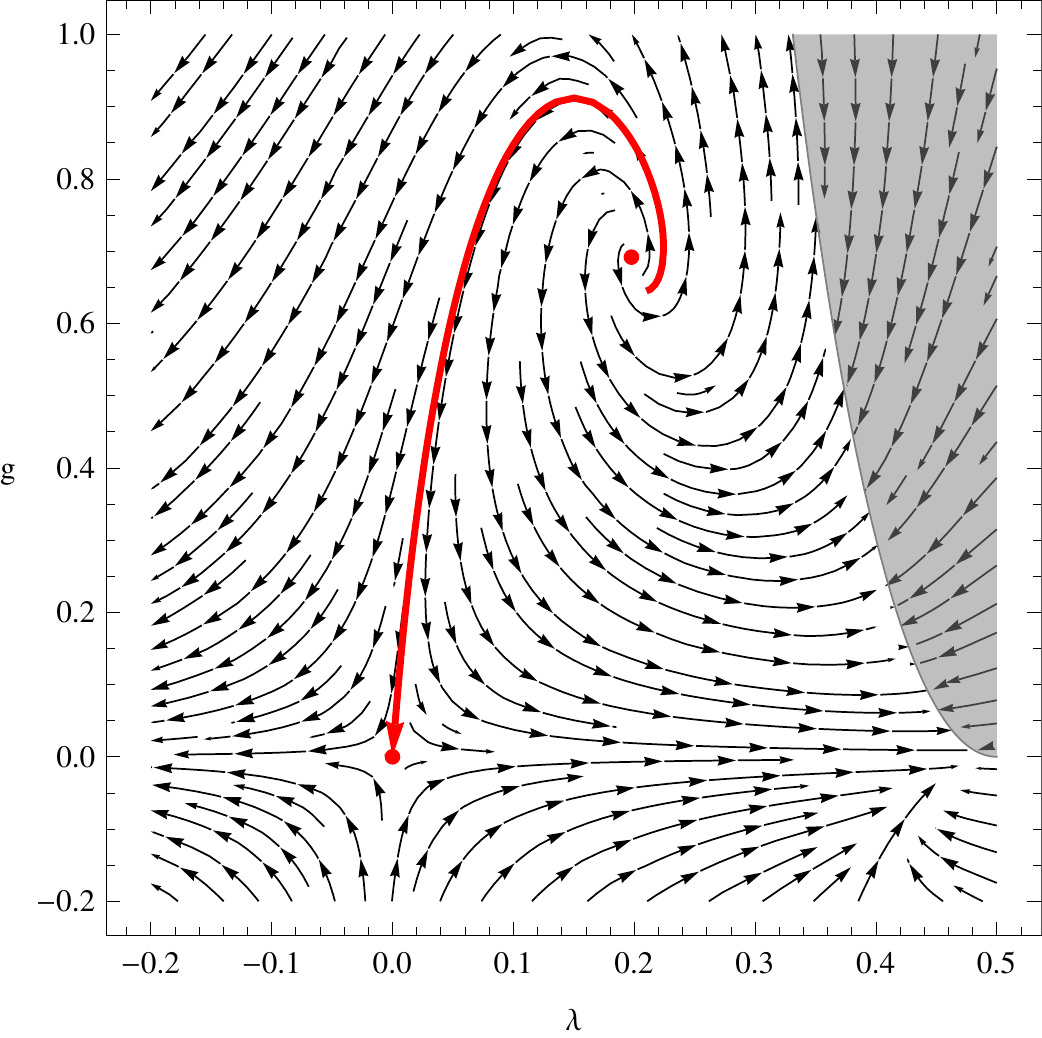}
		\label{fig:PDSSm100}}
	\caption{Phase portrait of the RG flow only incorporating $S_µ$ for different values of the squared mass parameter $µ^2$.}
	\label{fig:NGFPSS}
\end{figure}
\clearpage

\begin{figure}[htbp]
	\centering
	\subfigure[\textbf{NGFP} for $µ^2 = \frac{1}{10}$.]{
		\centering
		\includegraphics[width=0.3\textwidth]{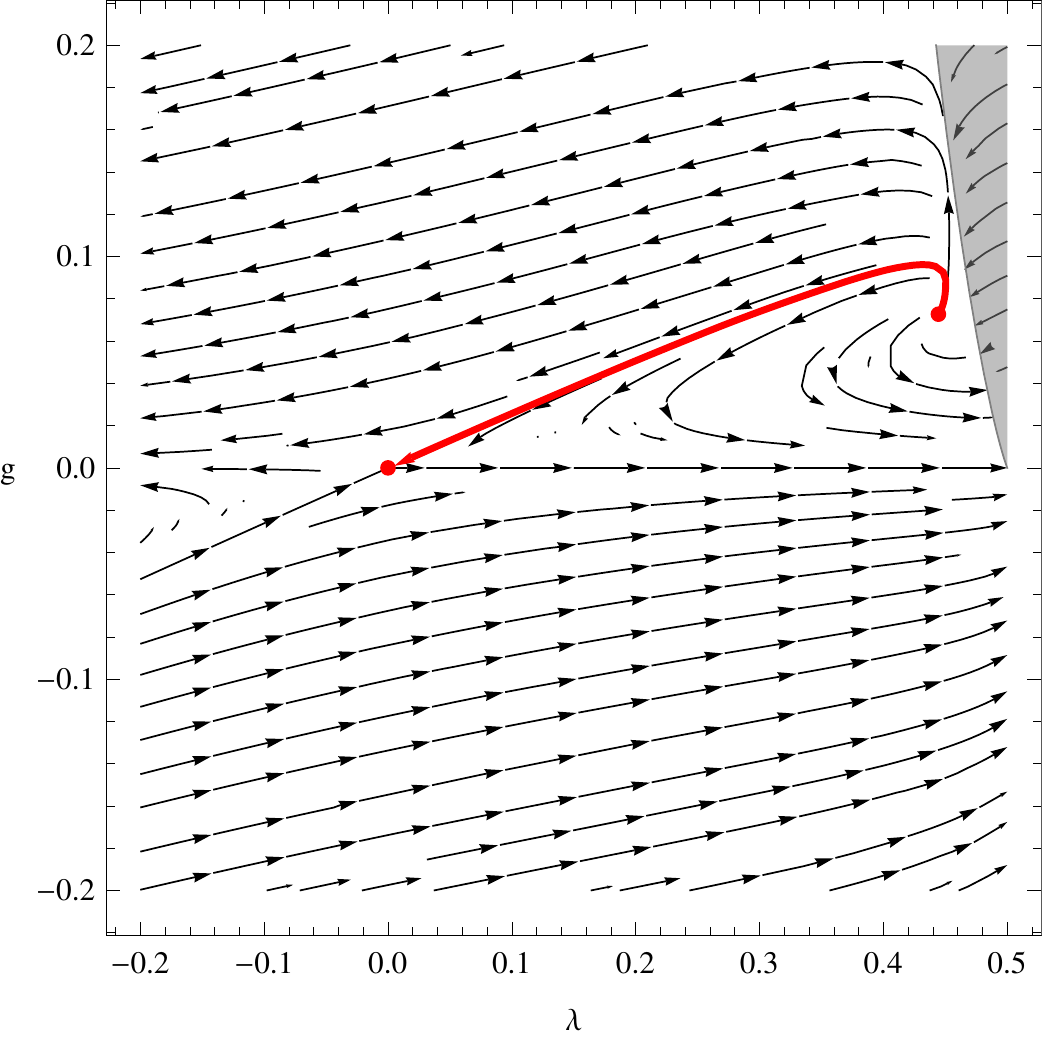}
		\label{fig:PDm110s2SS}}
	\subfigure[\textbf{NGFP} for $µ^2 = \frac{3}{10}$.]{
		\centering
		\includegraphics[width=0.3\textwidth]{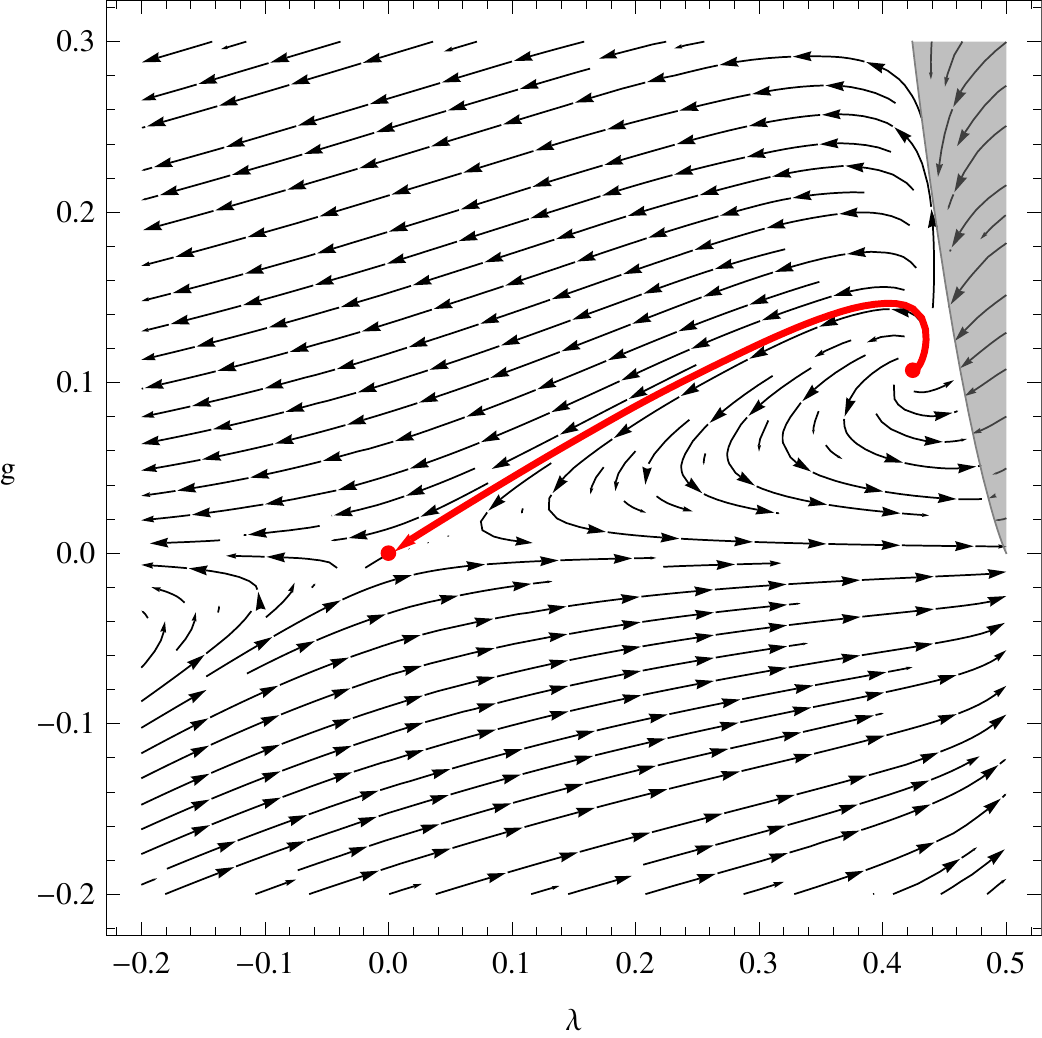}
		\label{fig:PDm310s2SS}}
	\subfigure[\textbf{NGFP}$\bm{^{\ominus}}$ for $µ^2 = \frac{3}{10}$.]{
		\centering
		\includegraphics[width=0.3\textwidth]{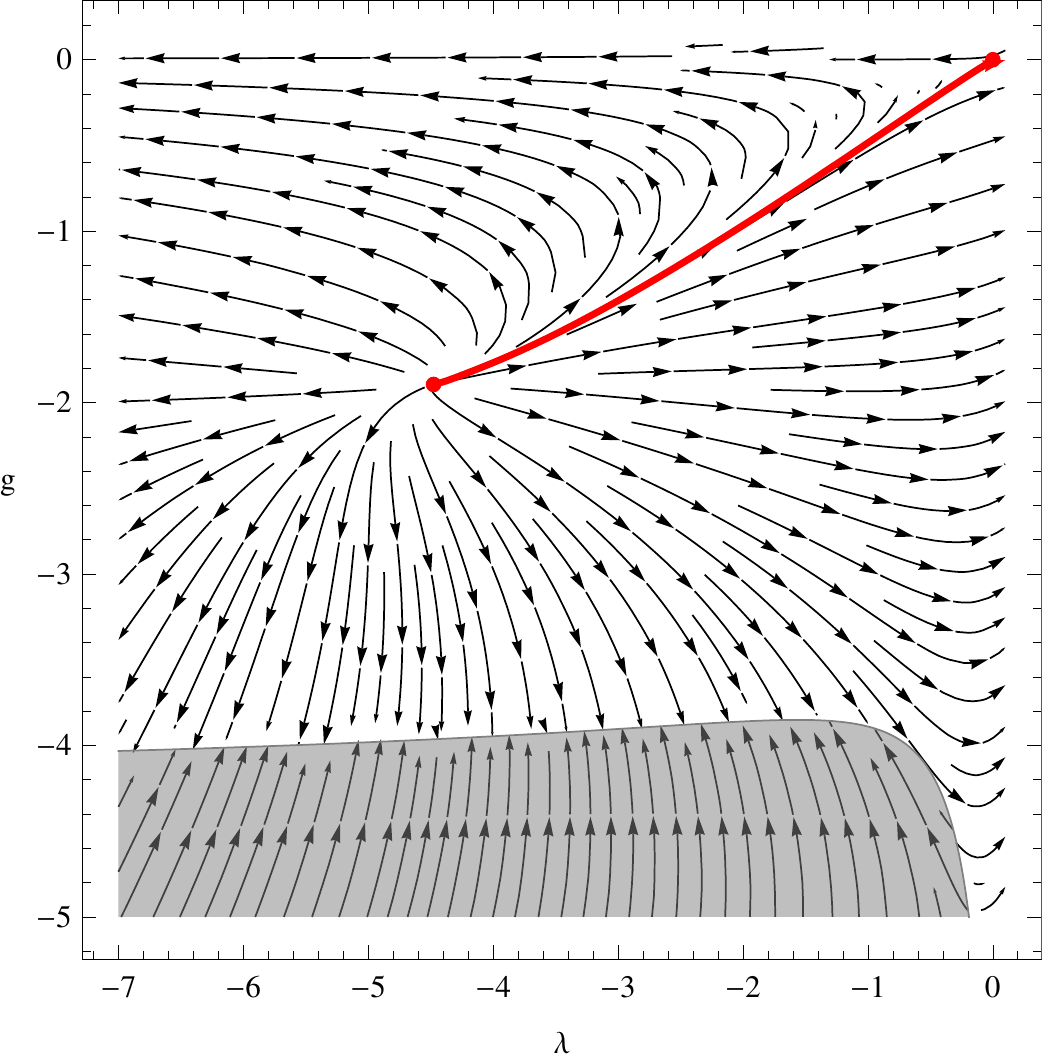}
		\label{fig:PDm310s2SSv3}}
	\subfigure[\textbf{NGFP} for $µ^2 = \frac{1}{2}$.]{
		\centering
		\includegraphics[width=0.3\textwidth]{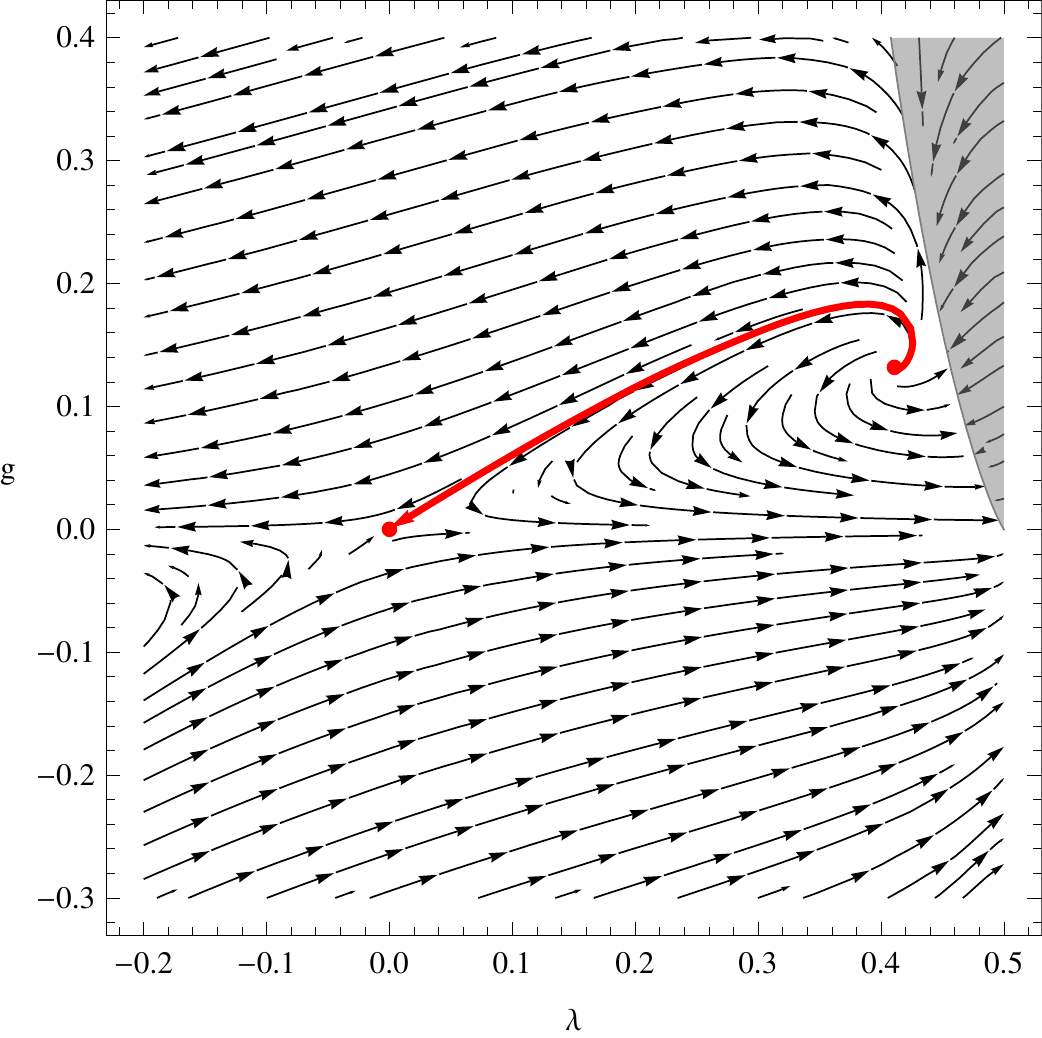}
		\label{fig:PDm119s2SS}}
	\subfigure[\textbf{NGFP} for $µ^2 = 1$.]{
		\centering
		\includegraphics[width=0.3\textwidth]{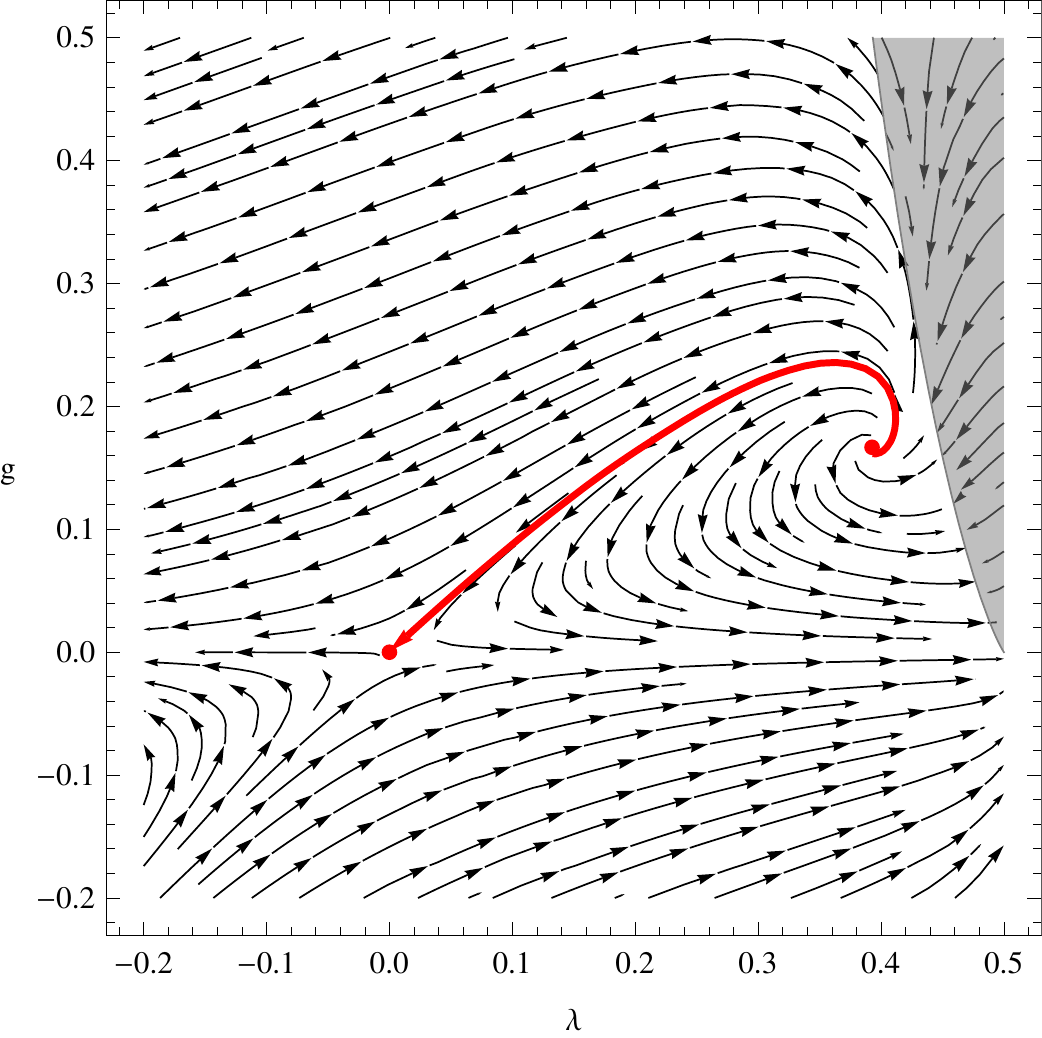}
		\label{fig:PDm1s2SS}}
	\subfigure[\textbf{NGFP} for $µ^2 = 2$.]{
		\centering
		\includegraphics[width=0.3\textwidth]{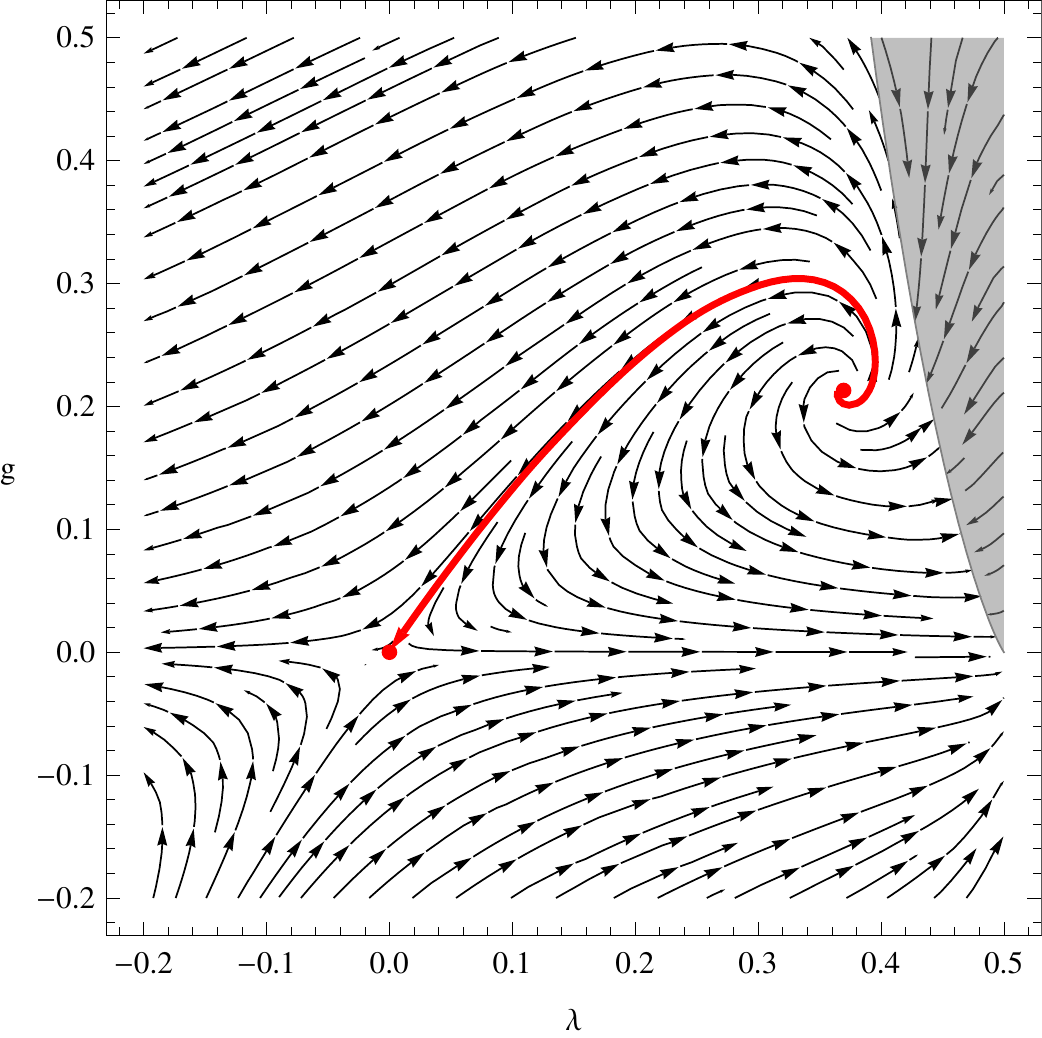}
		\label{fig:PDm2s2SS}}
	\subfigure[\textbf{NGFP} for $µ^2 = 10$.]{
		\centering
		\includegraphics[width=0.3\textwidth]{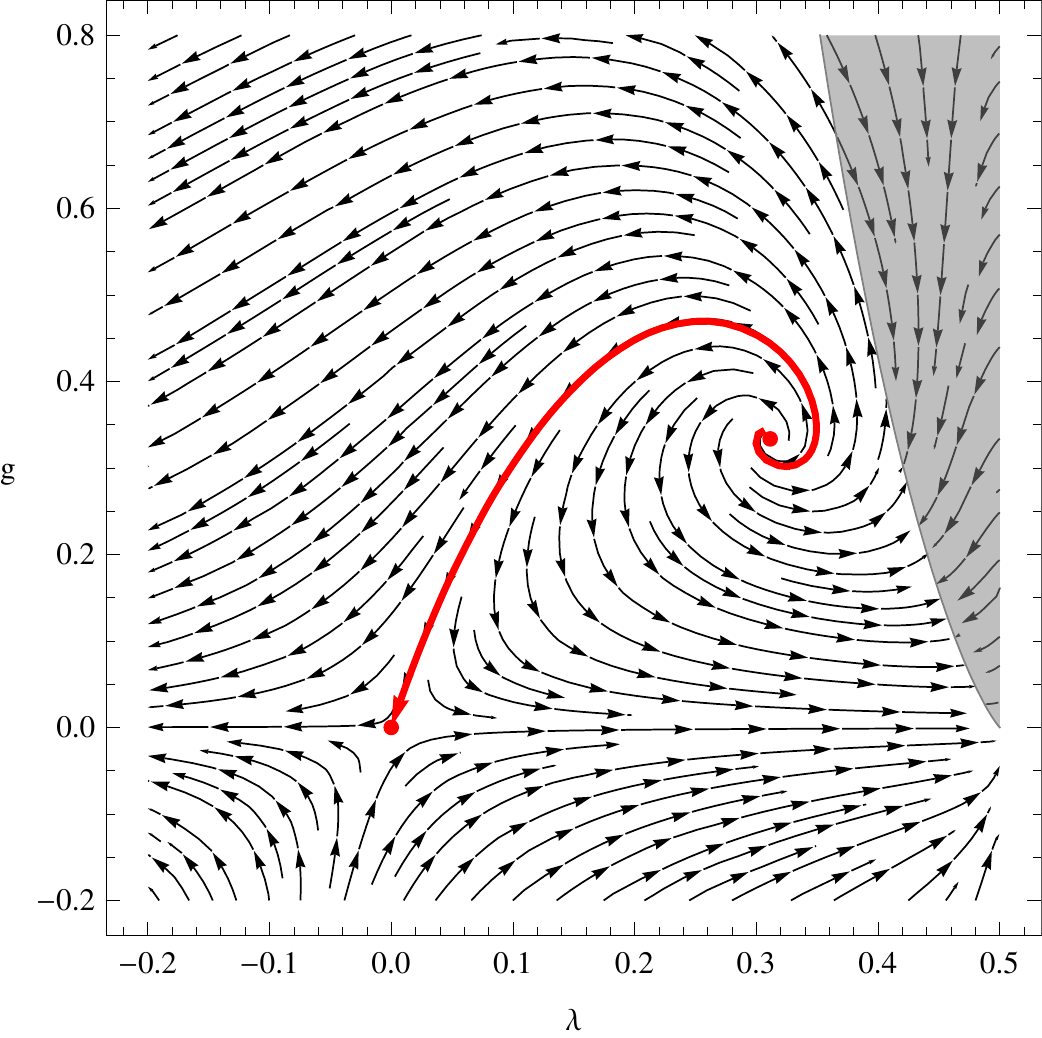}
		\label{fig:PDm10s2SS}}
	\subfigure[\textbf{NGFP} for $µ^2 = 100$.]{
		\centering
		\includegraphics[width=0.3\textwidth]{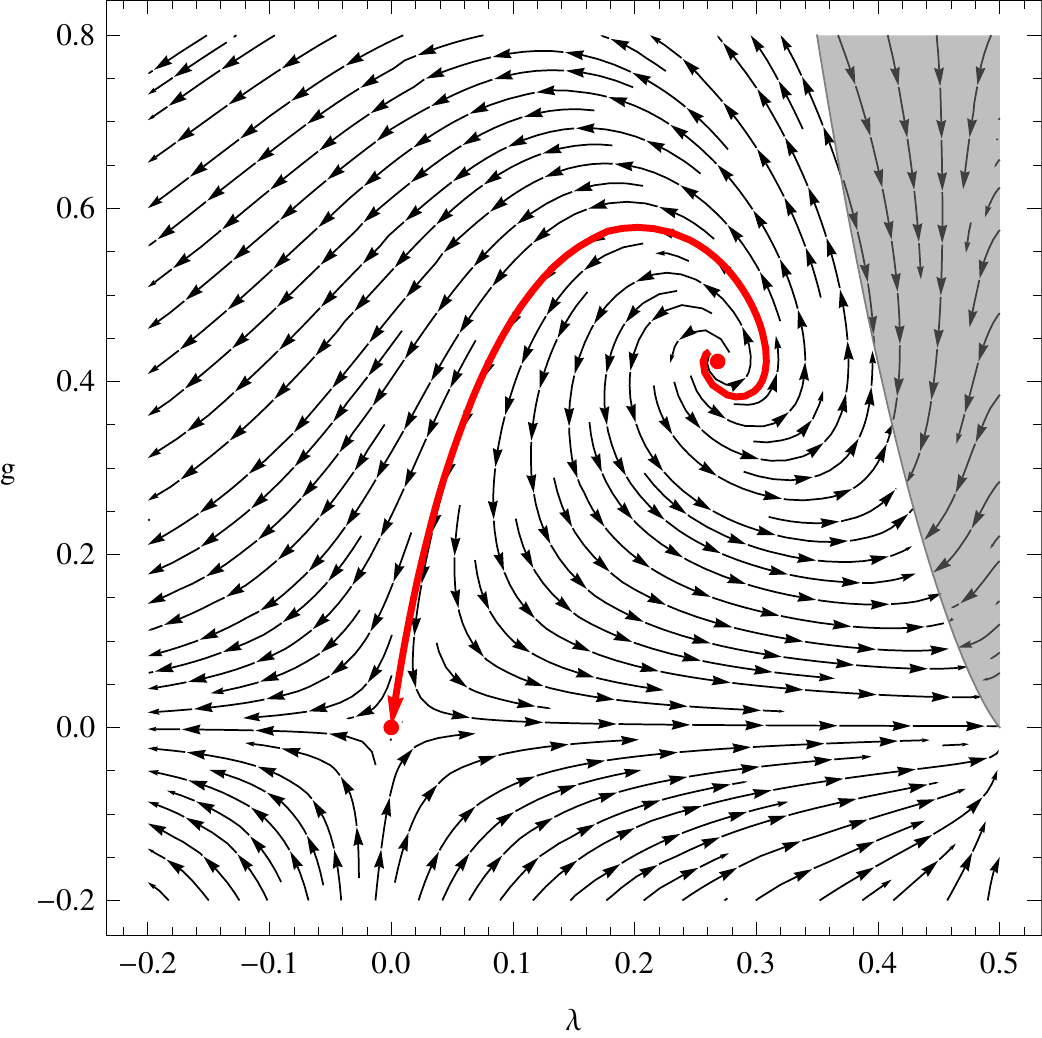}
		\label{fig:PDm100s2SS}}
	\caption{Phase portrait of the RG flow only incorporating $S_µ$ for different values of the squared mass parameter $µ^2$ and employing the generalized exponential cutoff with shape parameter $s=2$.}
	\label{fig:NGFPSSshape}
\end{figure}
\clearpage

\begin{figure}[htbp]
	\centering
	\subfigure[\textbf{NGFP} for $µ^2 = \frac{1}{100}$.]{
		\centering
		\includegraphics[width=0.3\textwidth]{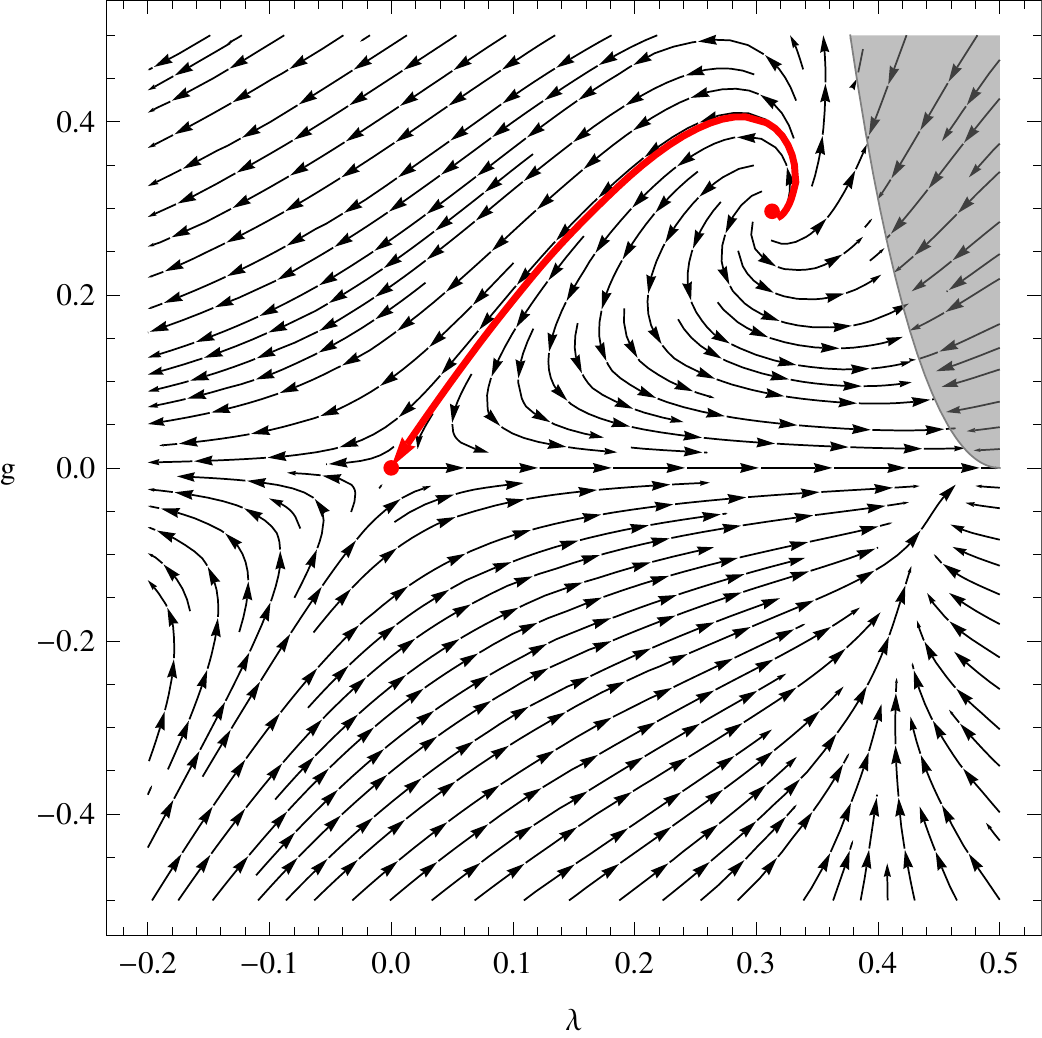}
		\label{fig:PDTTm1100}}
	\subfigure[\textbf{NGFP}$\bm{^{\ominus}}$ for $µ^2 = \frac{1}{100}$.]{
		\centering
		\includegraphics[width=0.3\textwidth]{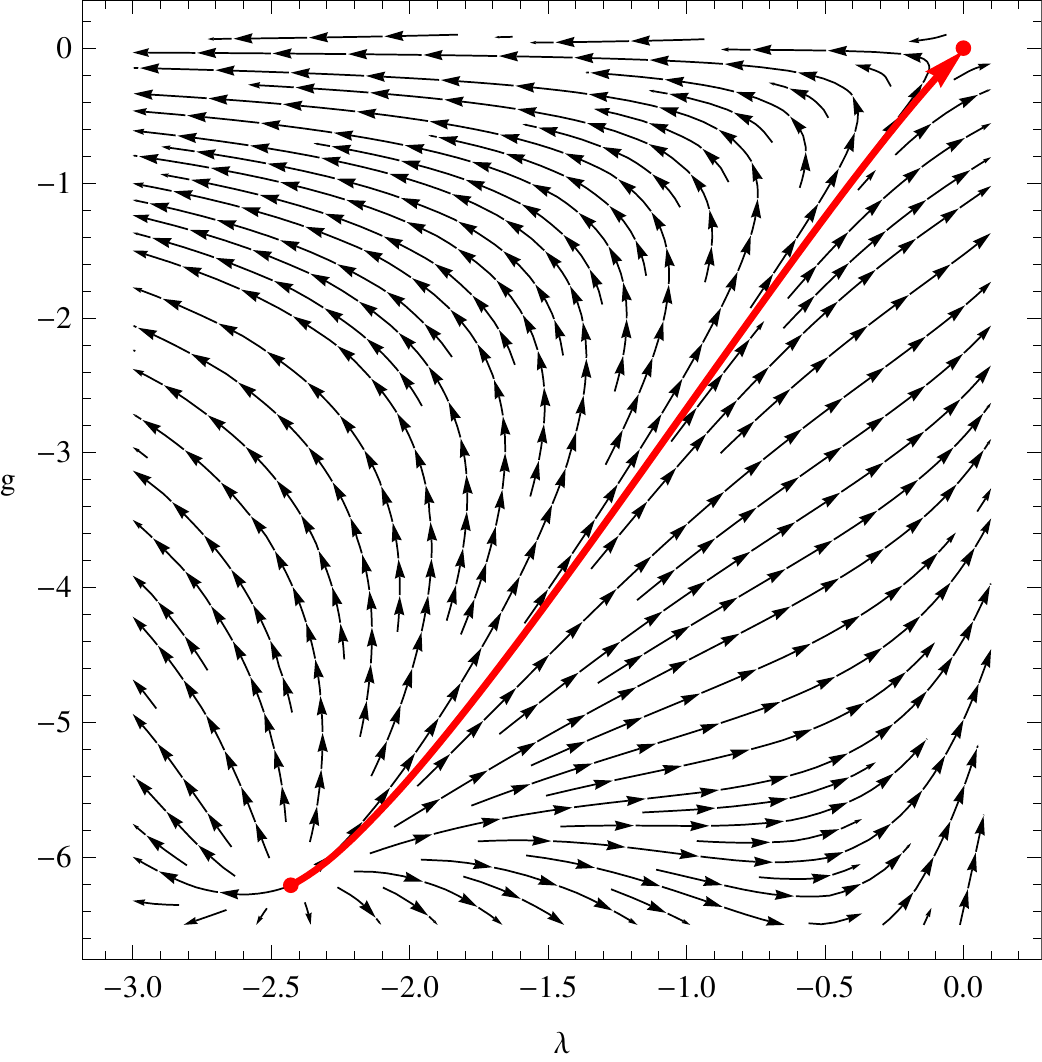}
		\label{fig:PDTTm1100v2}}
	\subfigure[\textbf{NGFP} for $µ^2 = \frac{1}{10}$.]{
		\centering
		\includegraphics[width=0.3\textwidth]{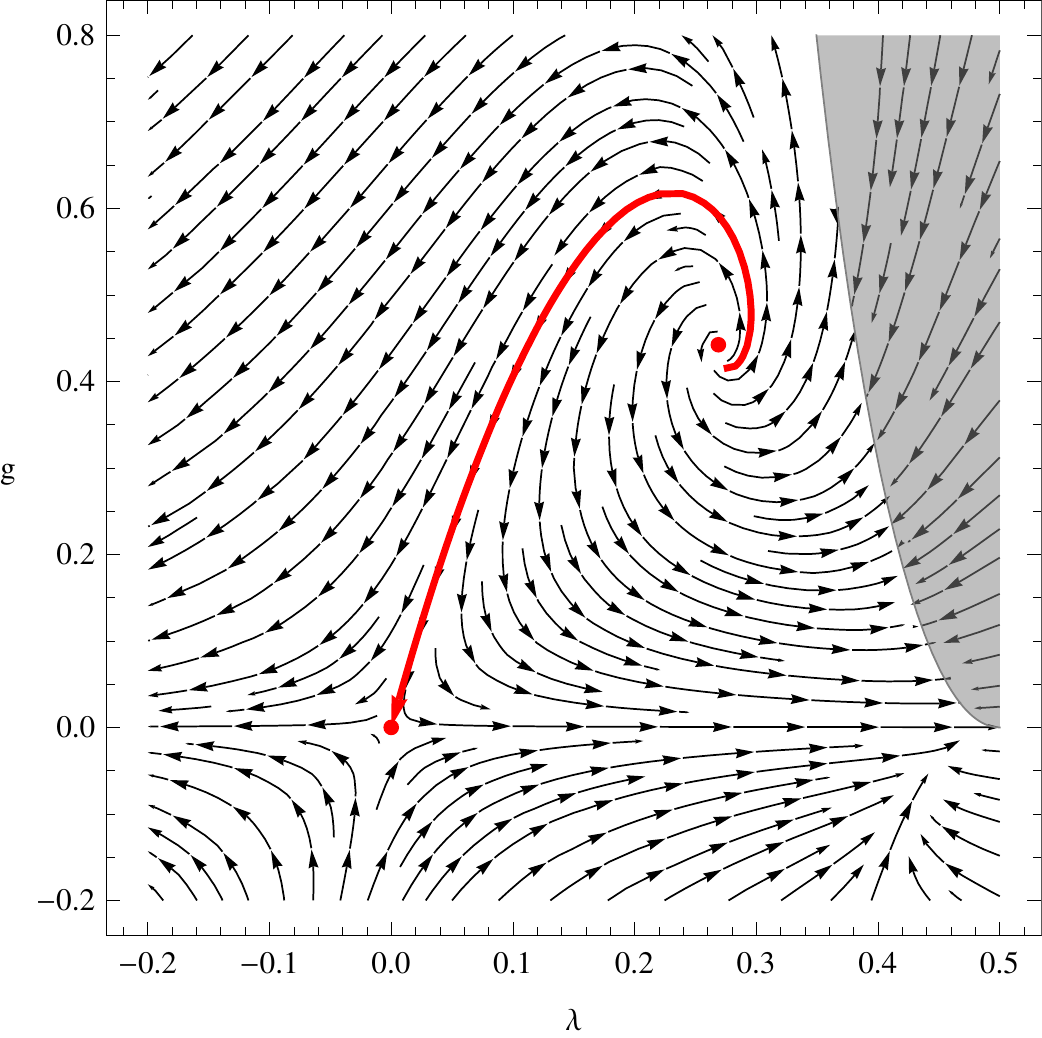}
		\label{fig:PDTTm110}}
	\subfigure[\textbf{NGFP} for $µ^2 = \frac{3}{10}$.]{
		\centering
		\includegraphics[width=0.3\textwidth]{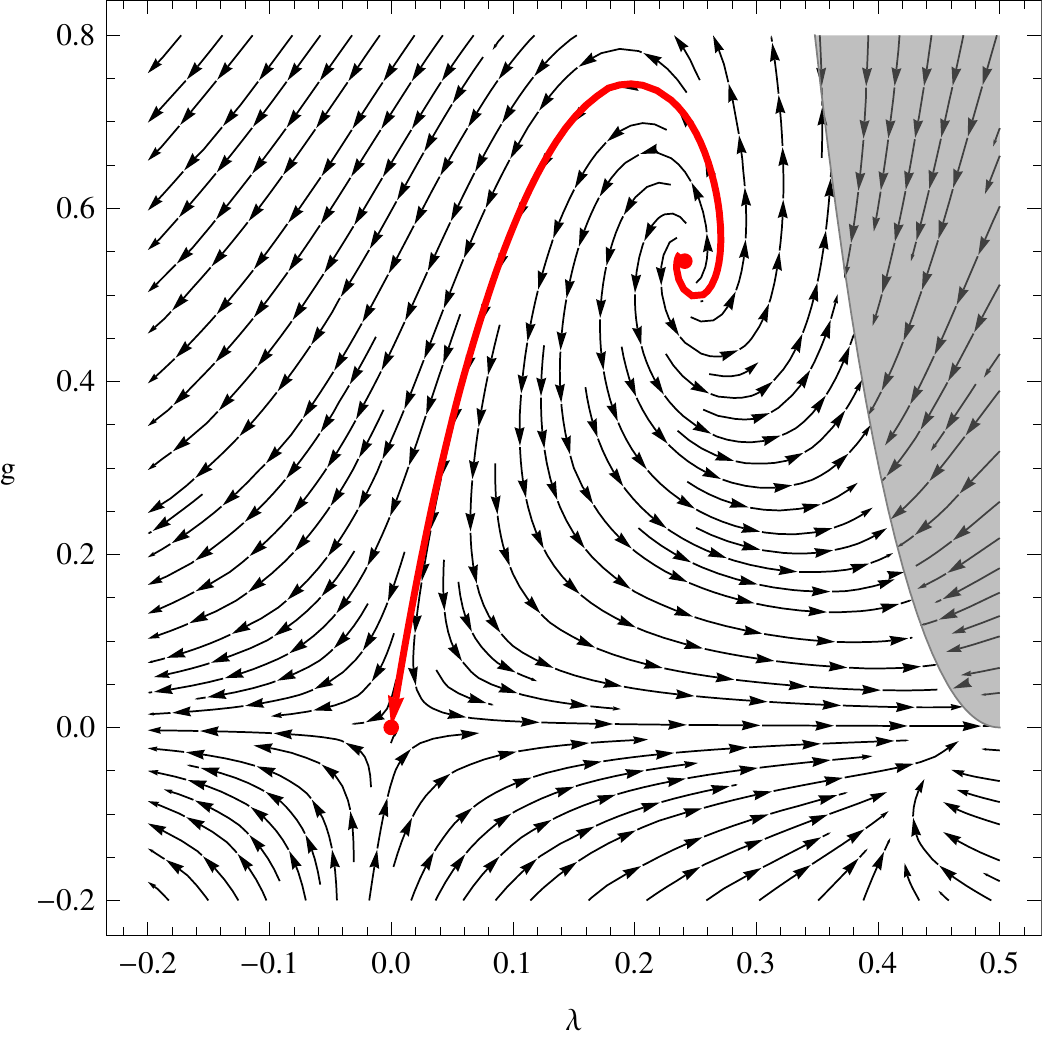}
		\label{fig:PDTTm310}}
	\subfigure[\textbf{NGFP} for $µ^2 = \frac{1}{2}$.]{
		\centering
		\includegraphics[width=0.3\textwidth]{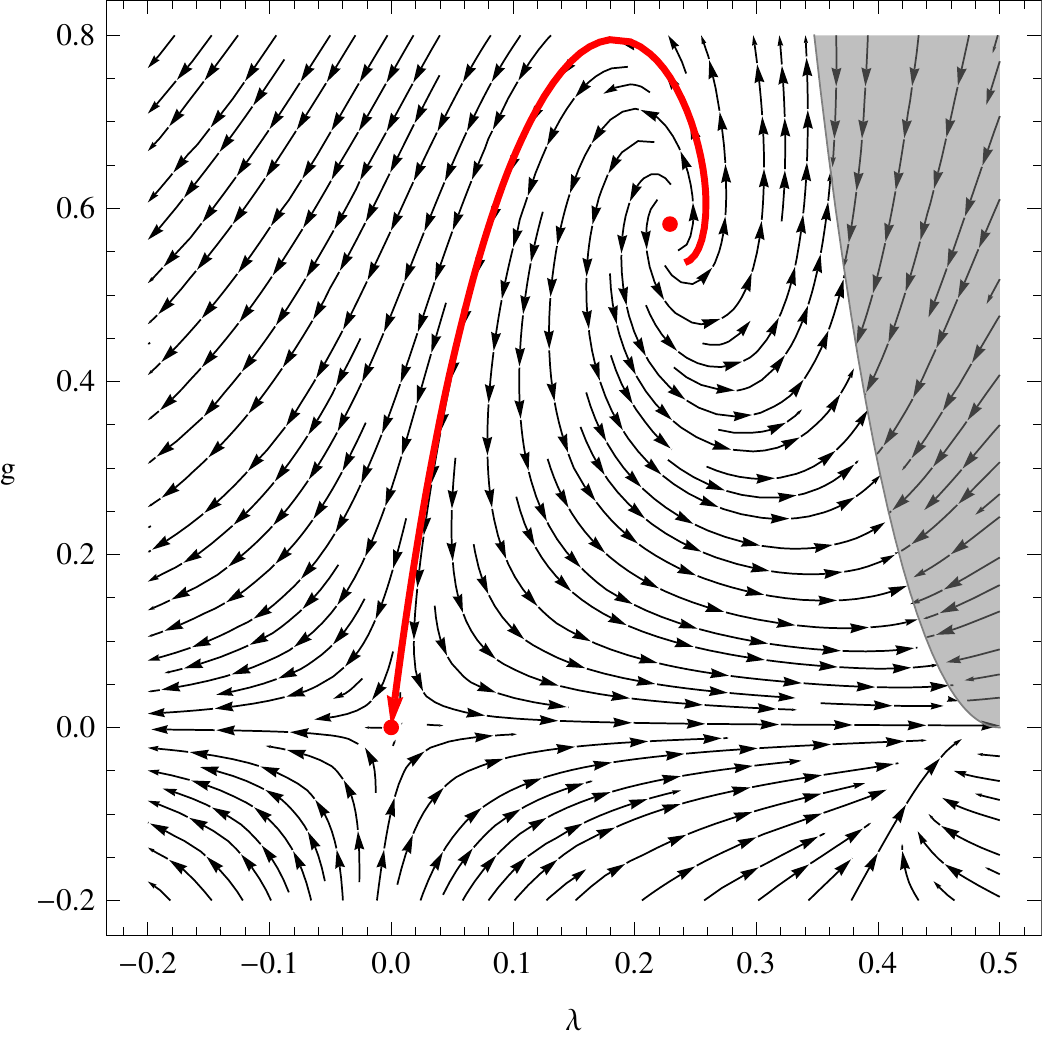}
		\label{fig:PDTTm12}}
	\subfigure[\textbf{NGFP} for $µ^2 = 1$.]{
		\centering
		\includegraphics[width=0.3\textwidth]{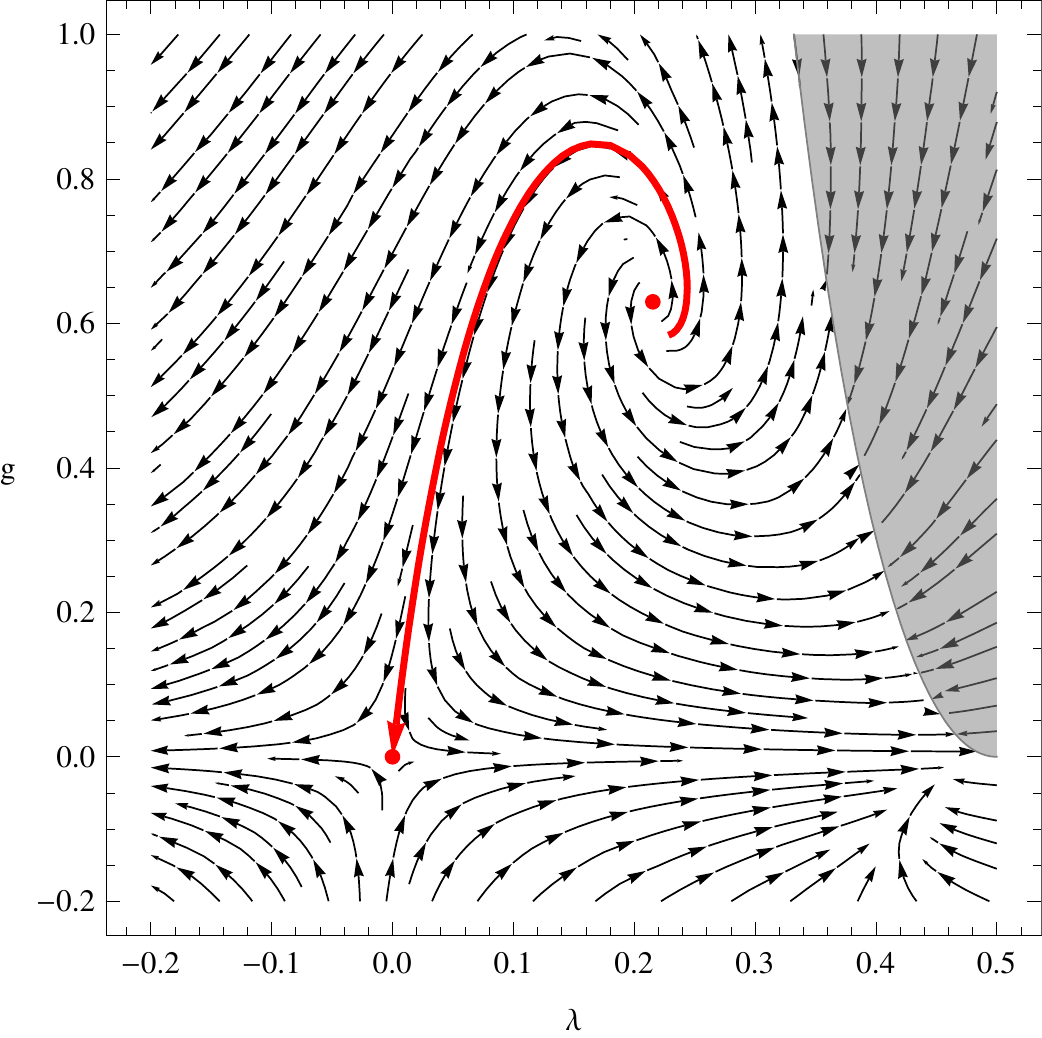}
		\label{fig:PDTTm1}}
	\subfigure[\textbf{NGFP} for $µ^2 = 2$.]{
		\centering
		\includegraphics[width=0.3\textwidth]{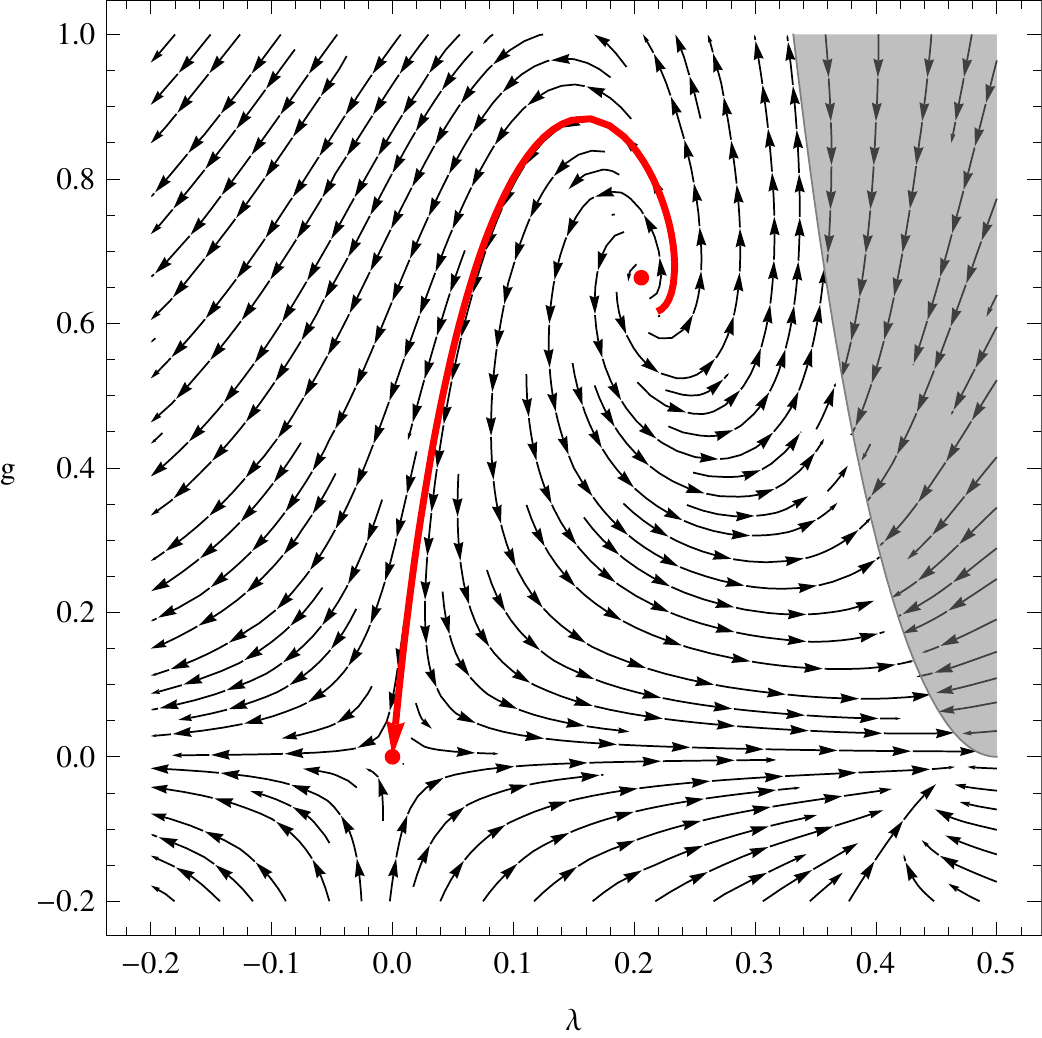}
		\label{fig:PDTTm2}}
	\subfigure[\textbf{NGFP} for $µ^2 = 10$.]{
		\centering
		\includegraphics[width=0.3\textwidth]{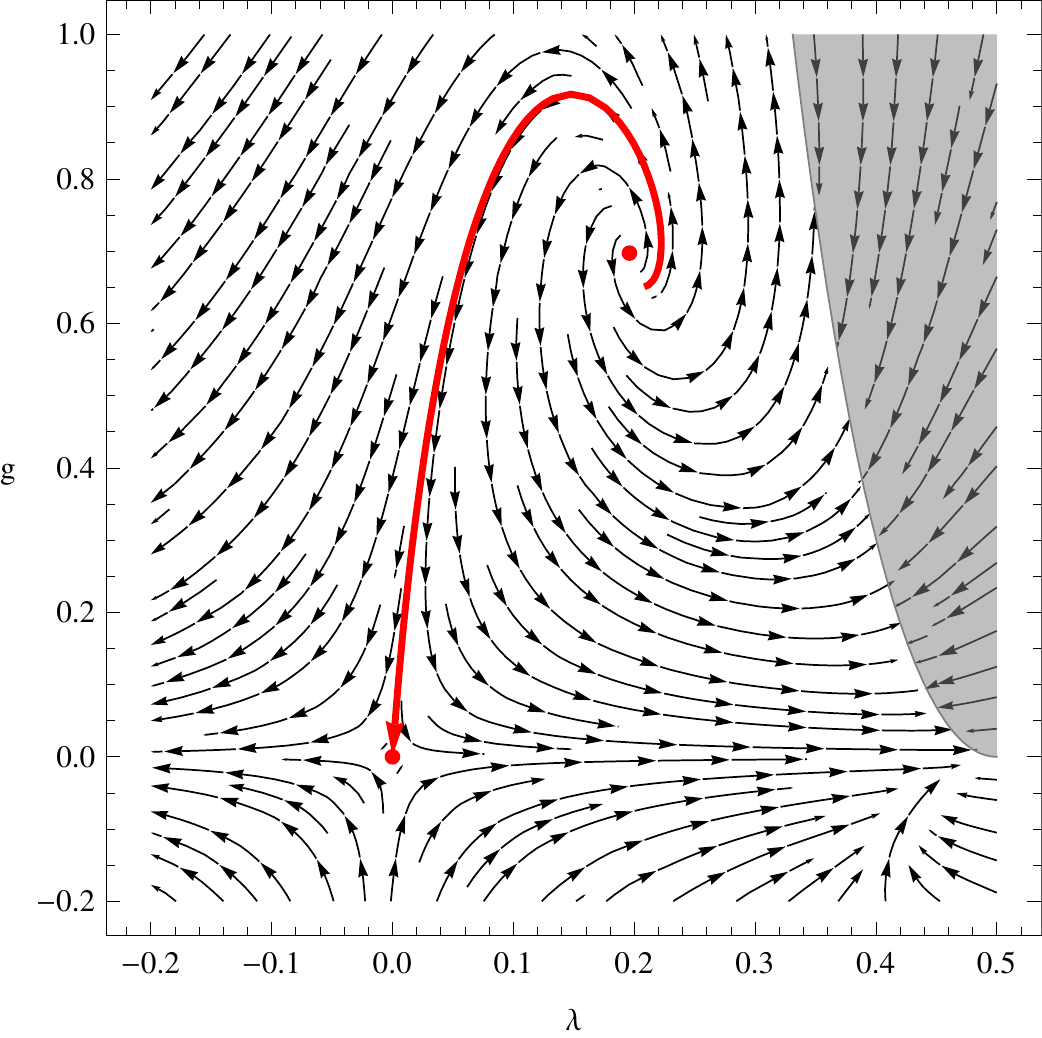}
		\label{fig:PDTTm10}}
	\subfigure[\textbf{NGFP} for $µ^2 = 100$.]{
		\centering
		\includegraphics[width=0.3\textwidth]{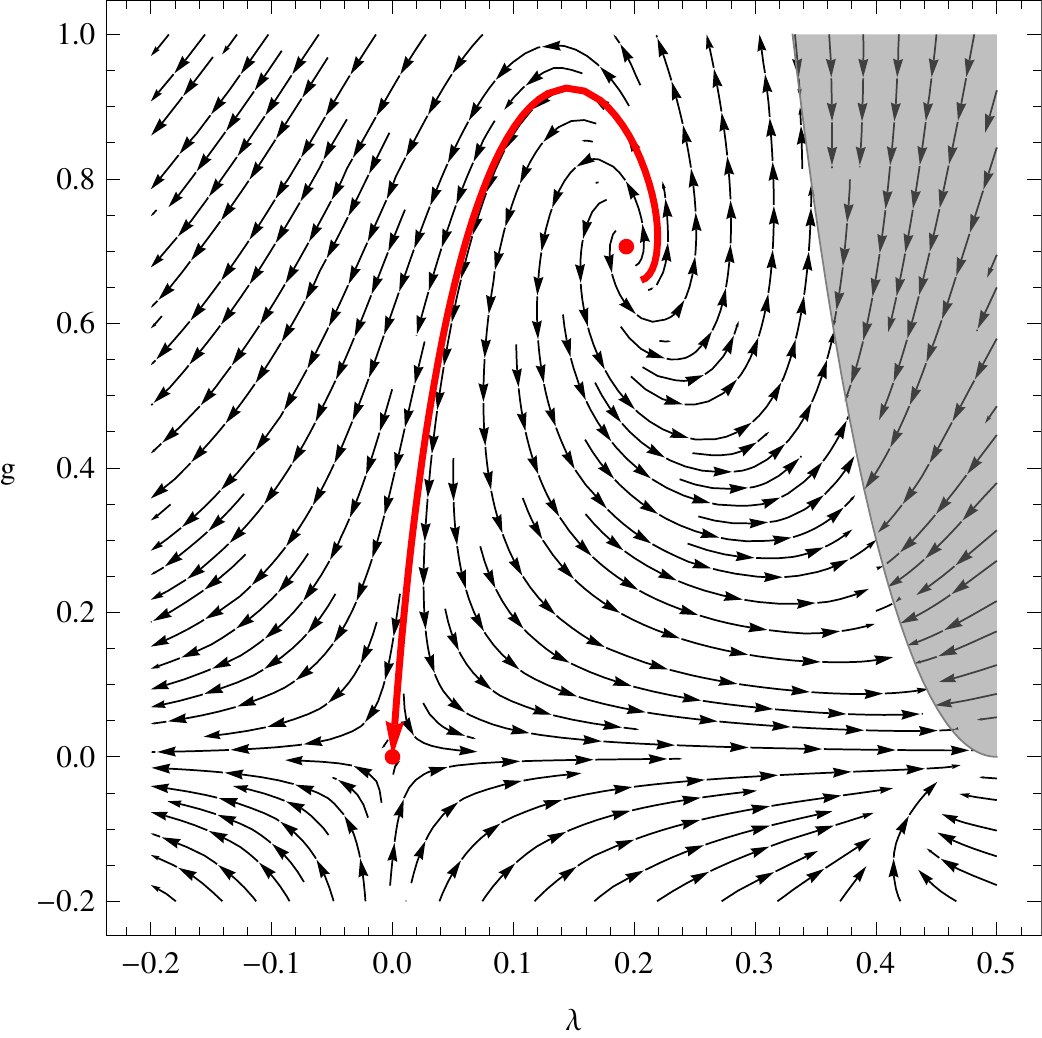}
		\label{fig:PDTTm100}}
	\caption{Phase portrait of the RG flow only incorporating $T_µ$ for different values of the squared mass parameter $µ^2$.}
	\label{fig:NGFPTT}
\end{figure}
\clearpage

\begin{figure}[htbp]
	\centering
	\subfigure[\textbf{NGFP}$\bm{^{\ominus}}$ for $µ^2 = \frac{1}{100}$.]{
		\centering
		\includegraphics[width=0.3\textwidth]{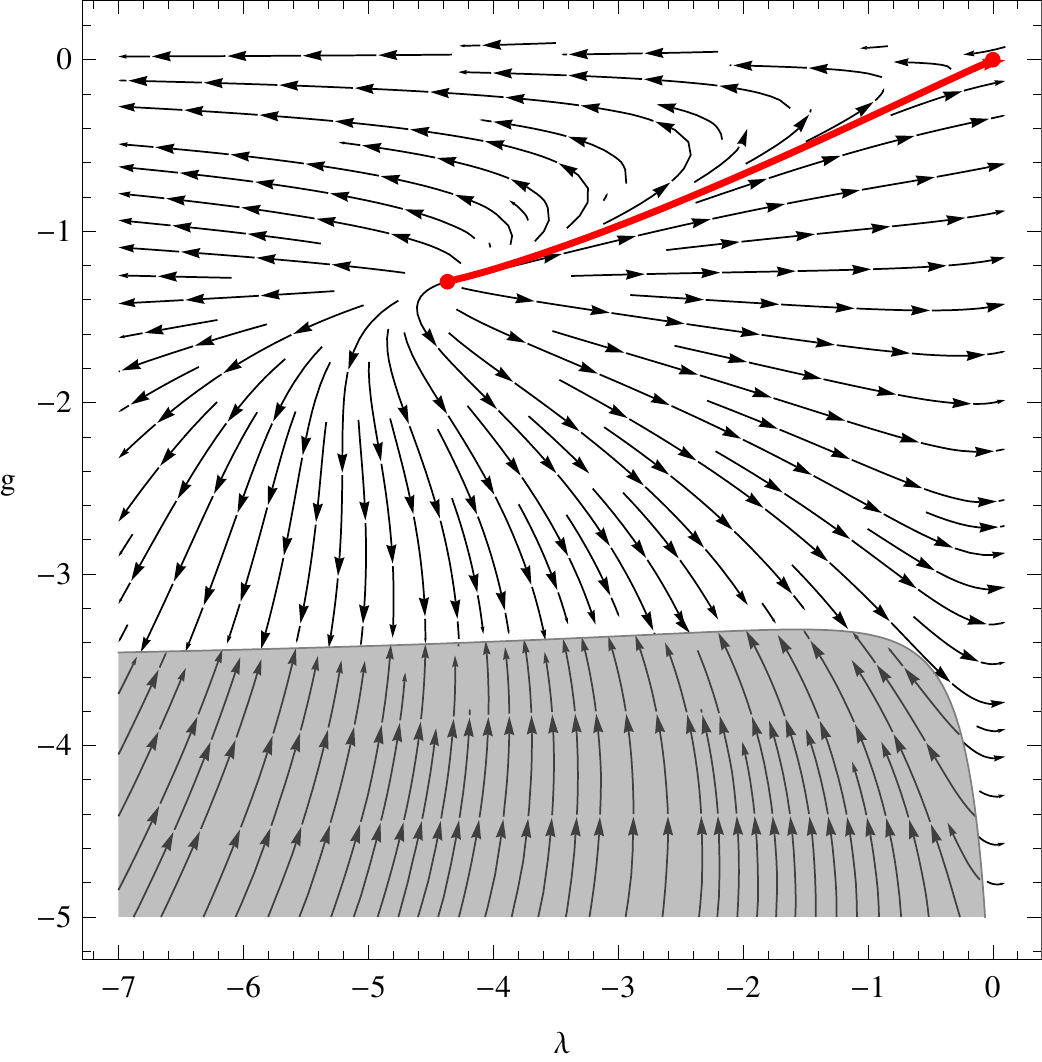}
		\label{fig:PDm1100s2TT}}
	\subfigure[\textbf{NGFP} for $µ^2 = \frac{3}{10}$.]{
		\centering
		\includegraphics[width=0.3\textwidth]{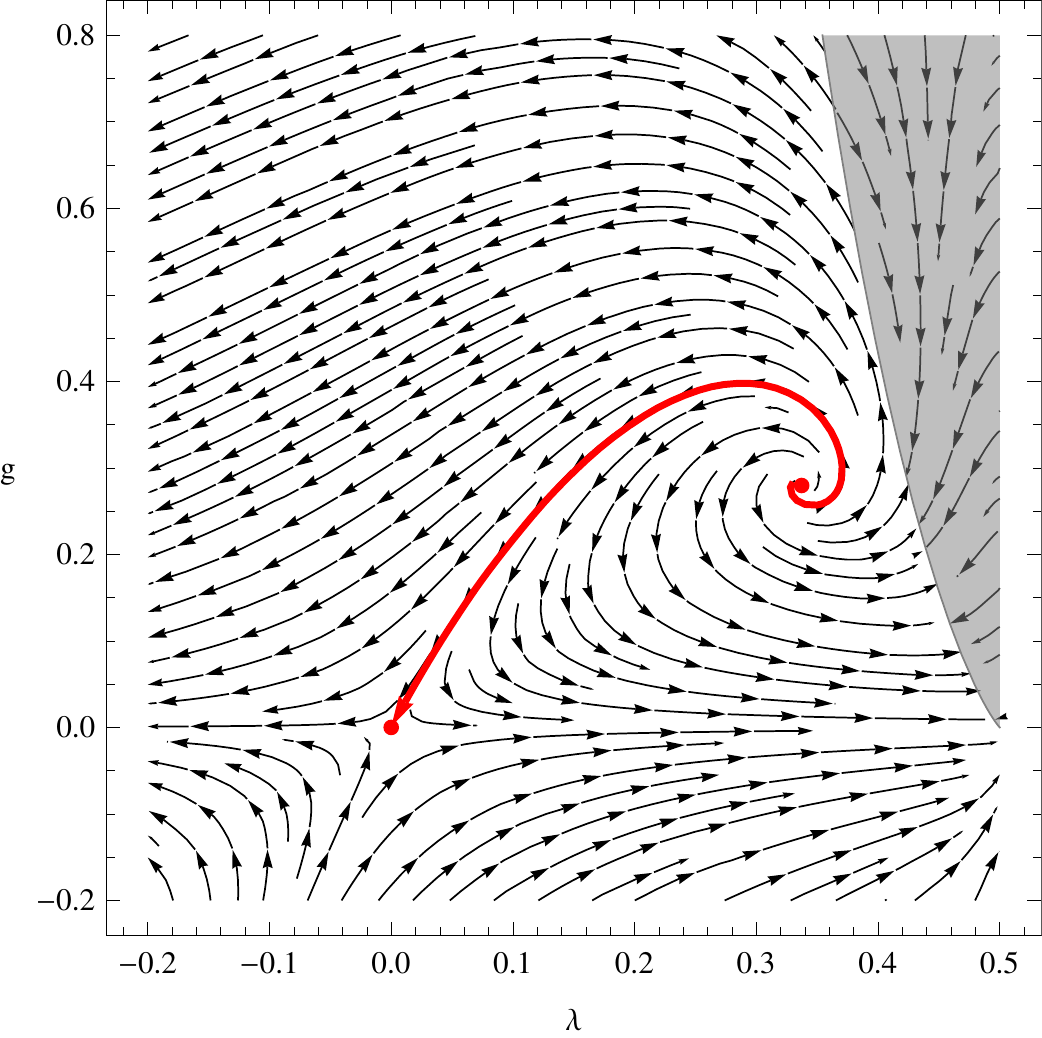}
		\label{fig:PDm310s2TT}}
	\subfigure[\textbf{NGFP} for $µ^2 = \frac{1}{2}$.]{
		\centering
		\includegraphics[width=0.3\textwidth]{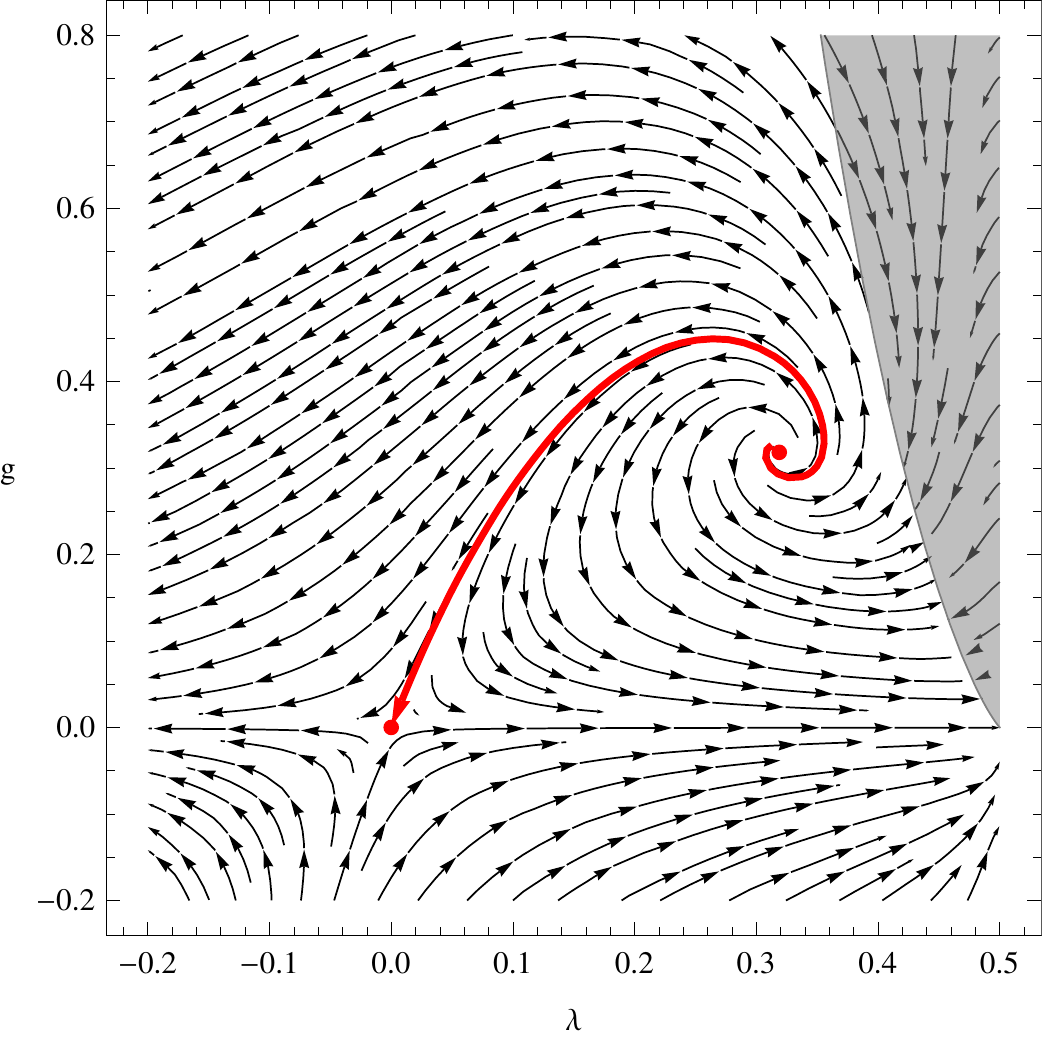}
		\label{fig:PDm12s2TT}}
	\subfigure[\textbf{NGFP} for $µ^2 = 1$.]{
		\centering
		\includegraphics[width=0.3\textwidth]{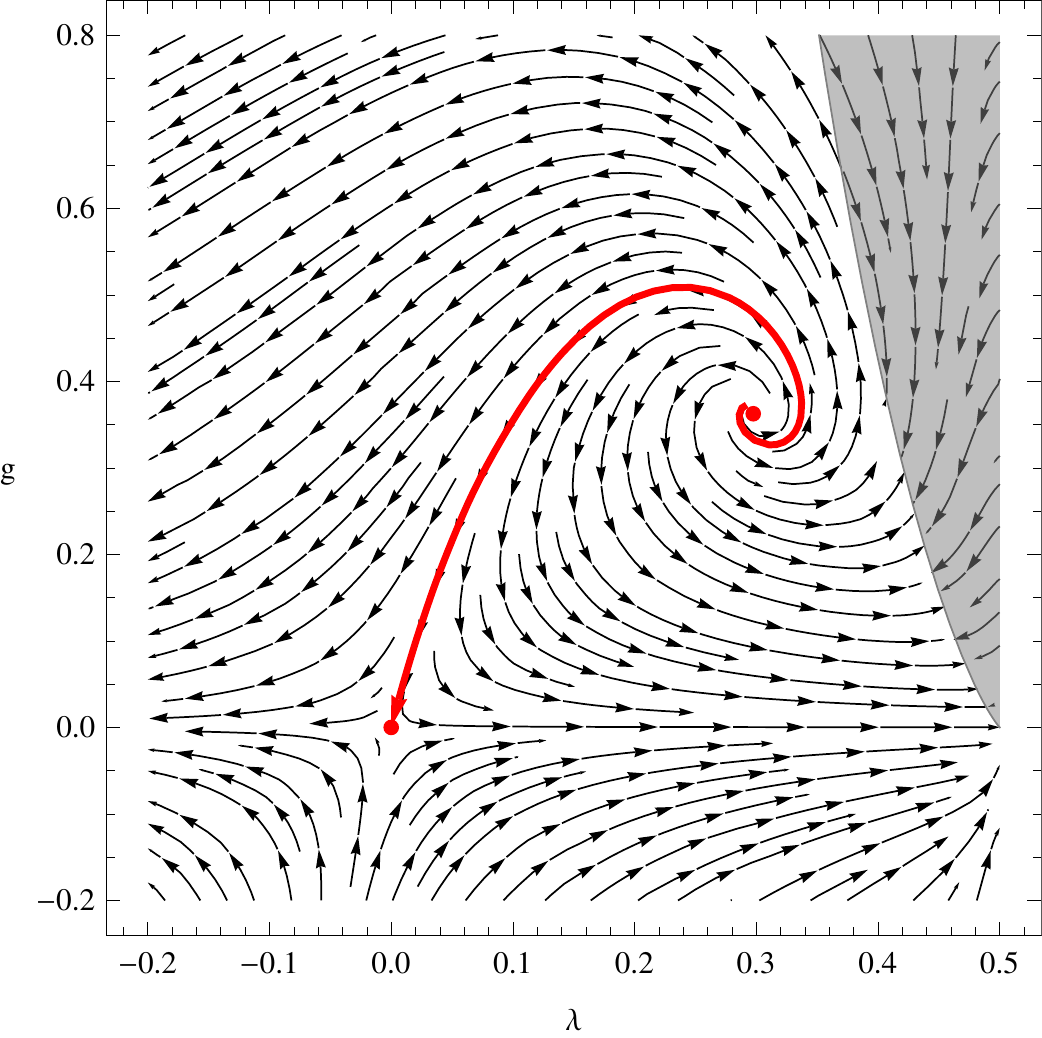}
		\label{fig:PDm1s2TT}}
	\subfigure[\textbf{NGFP} for $µ^2 = 2$.]{
		\centering
		\includegraphics[width=0.3\textwidth]{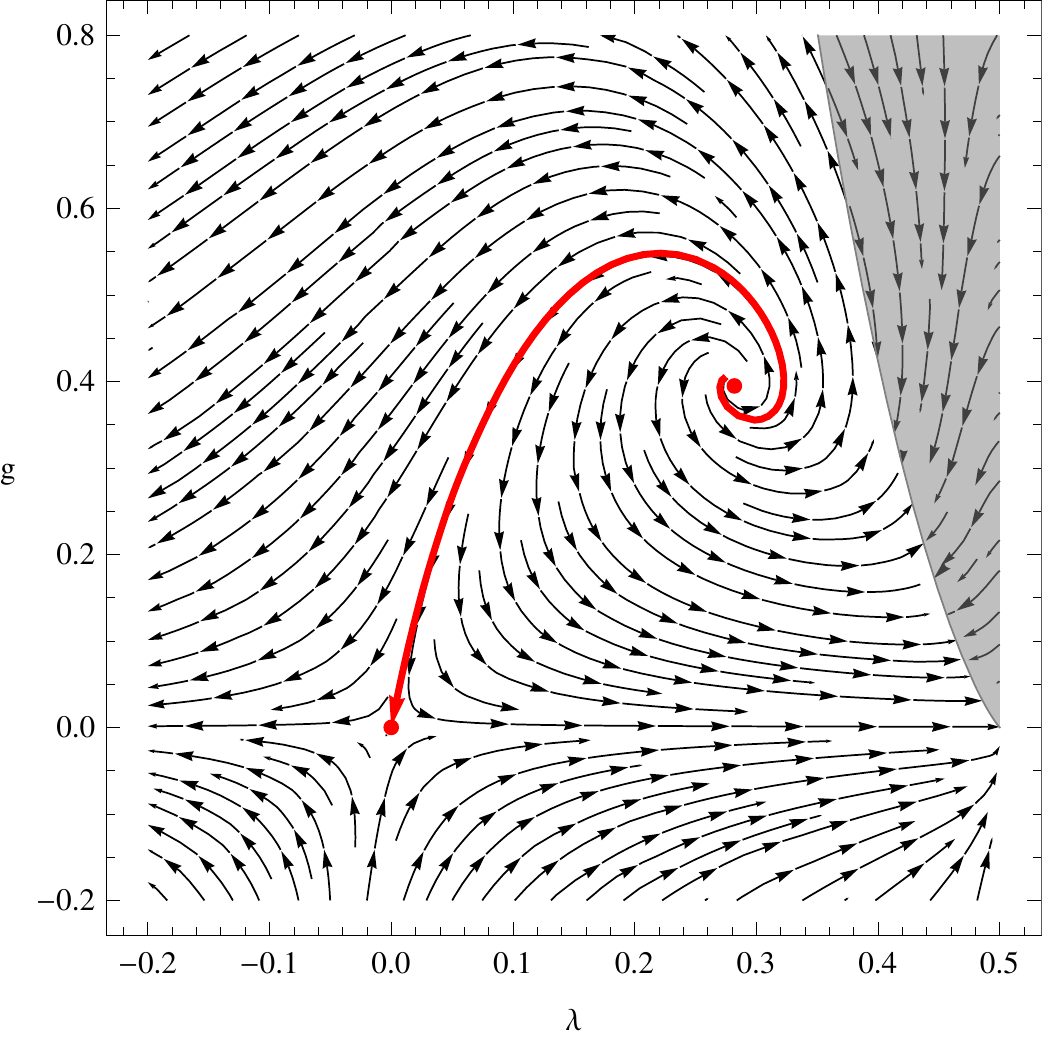}
		\label{fig:PDm2s2TT}}
	\subfigure[\textbf{NGFP} for $µ^2 = 10$.]{
		\centering
		\includegraphics[width=0.3\textwidth]{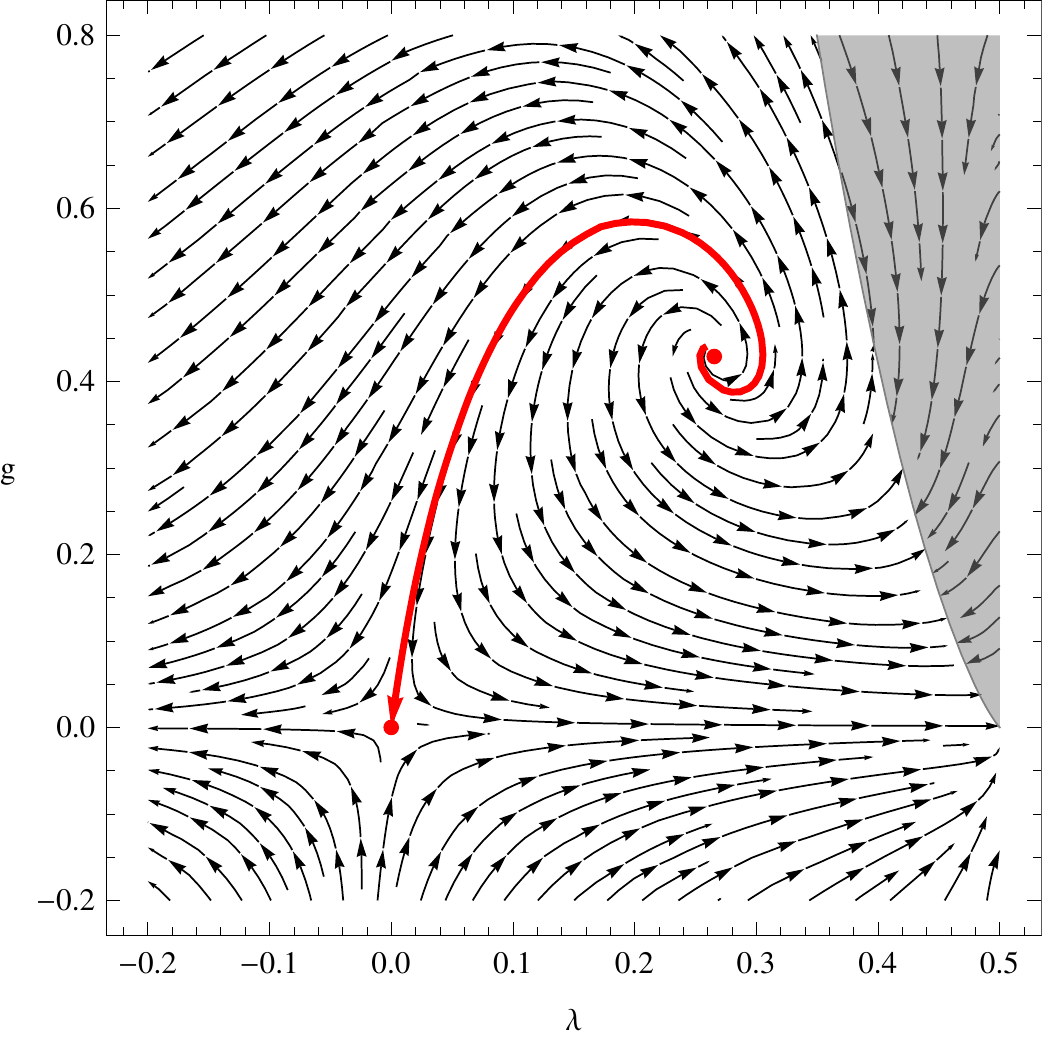}
		\label{fig:PDm10s2TT}}
	\subfigure[\textbf{NGFP} for $µ^2 = 100$.]{
		\centering
		\includegraphics[width=0.3\textwidth]{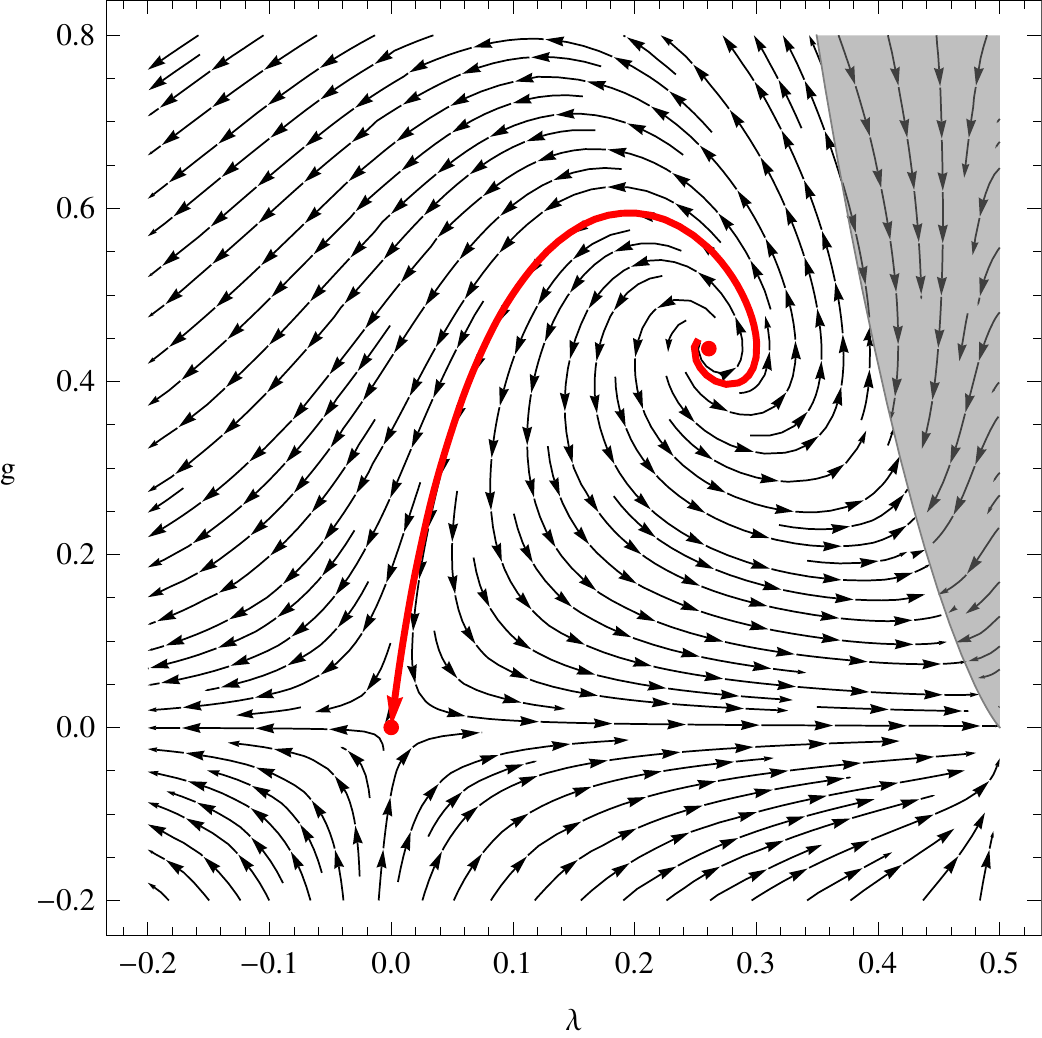}
		\label{fig:PDm100s2TT}}
	\caption{Phase portrait of the RG flow only incorporating $T_µ$ for different values of the squared mass parameter $µ^2$ and employing the generalized exponential cutoff with shape parameter $s=2$.}
	\label{fig:NGFPTTshape}
\end{figure}
\clearpage

\begin{figure}[htbp]
	\centering
	\subfigure[\textbf{NGFP} for $µ^2 = \frac{1}{1.9}$.]{
		\centering
		\includegraphics[width=0.3\textwidth]{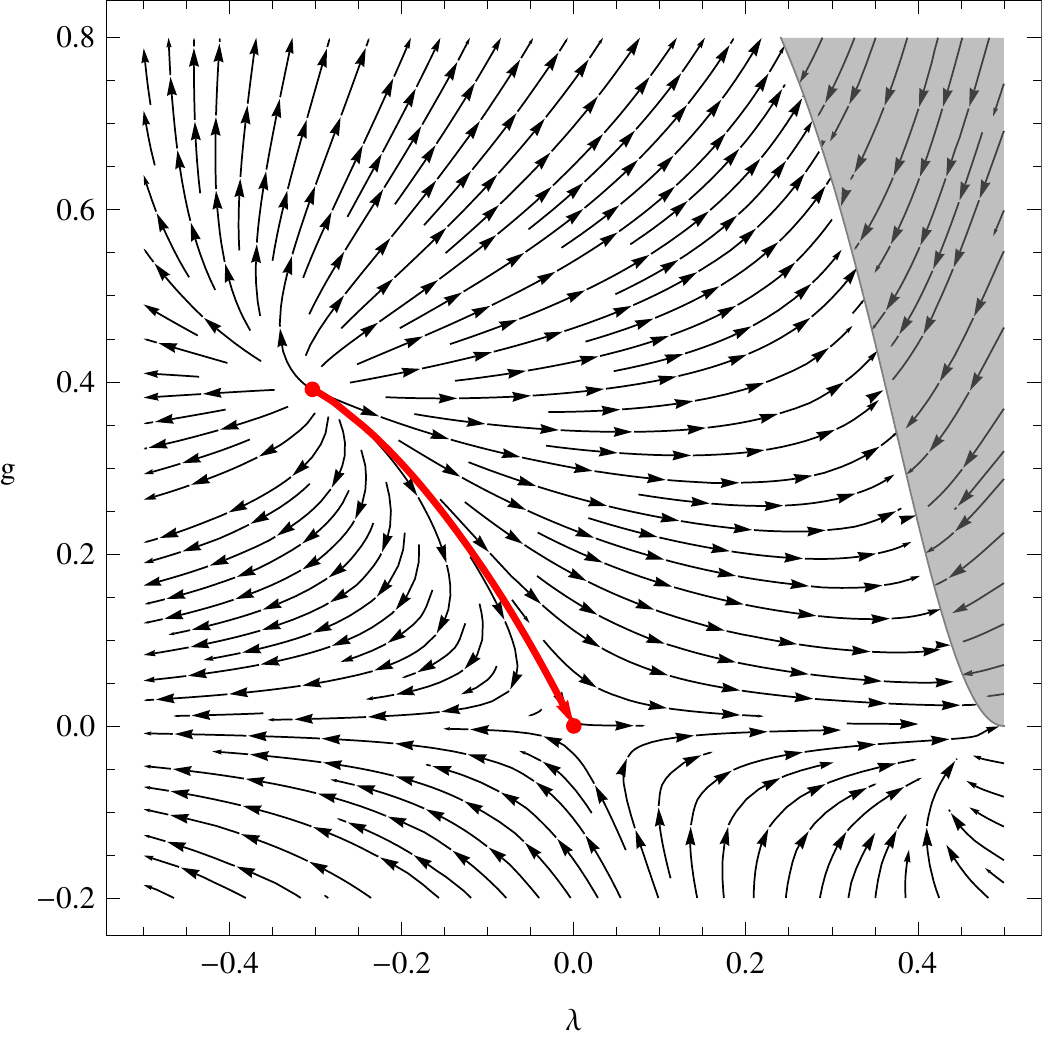}
		\label{fig:PDqqm119}}
	\subfigure[\textbf{NGFP} for $µ^2 = \frac{1}{1.5}$.]{
		\centering
		\includegraphics[width=0.3\textwidth]{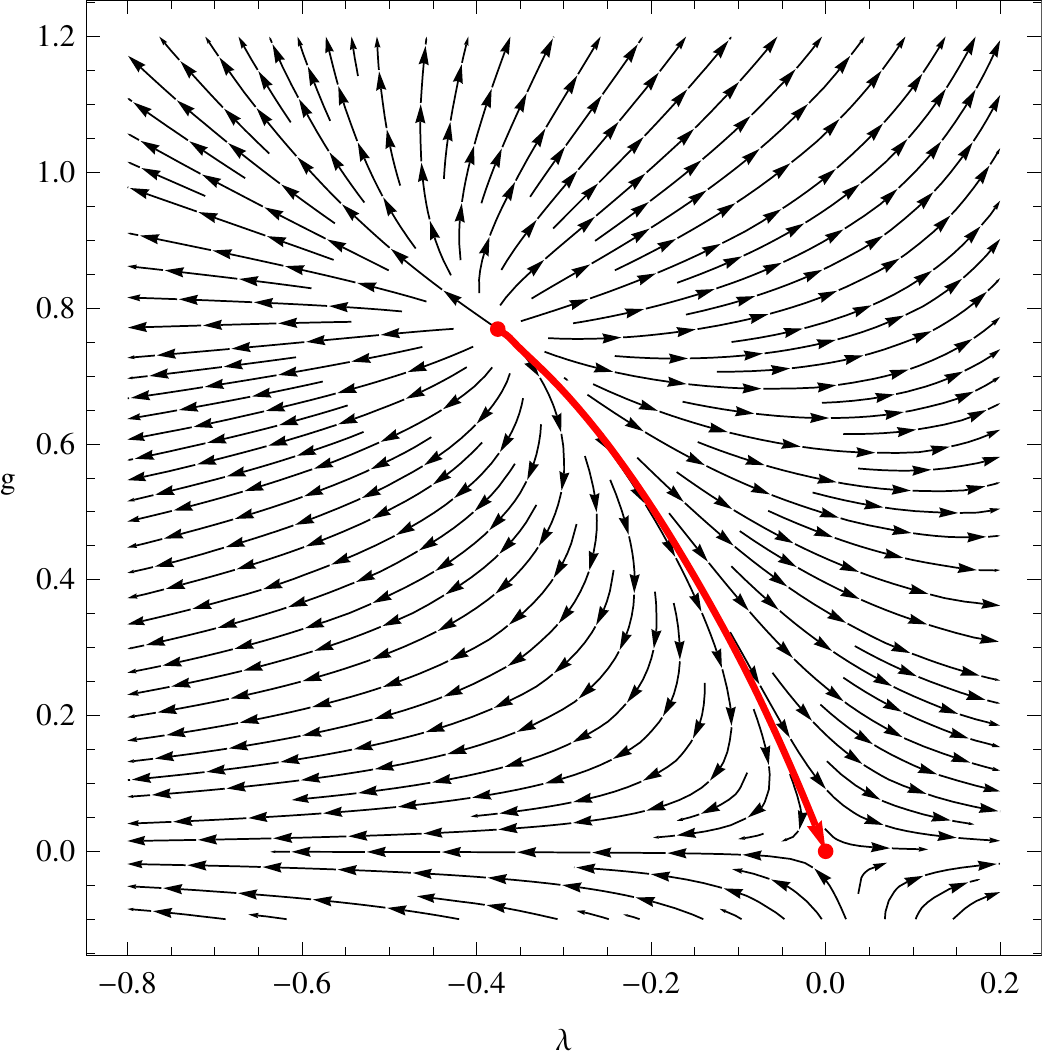}
		\label{fig:PDqqm115}}
	\subfigure[\textbf{NGFP} for $µ^2 = \frac{1}{1.1}$.]{
		\centering
		\includegraphics[width=0.3\textwidth]{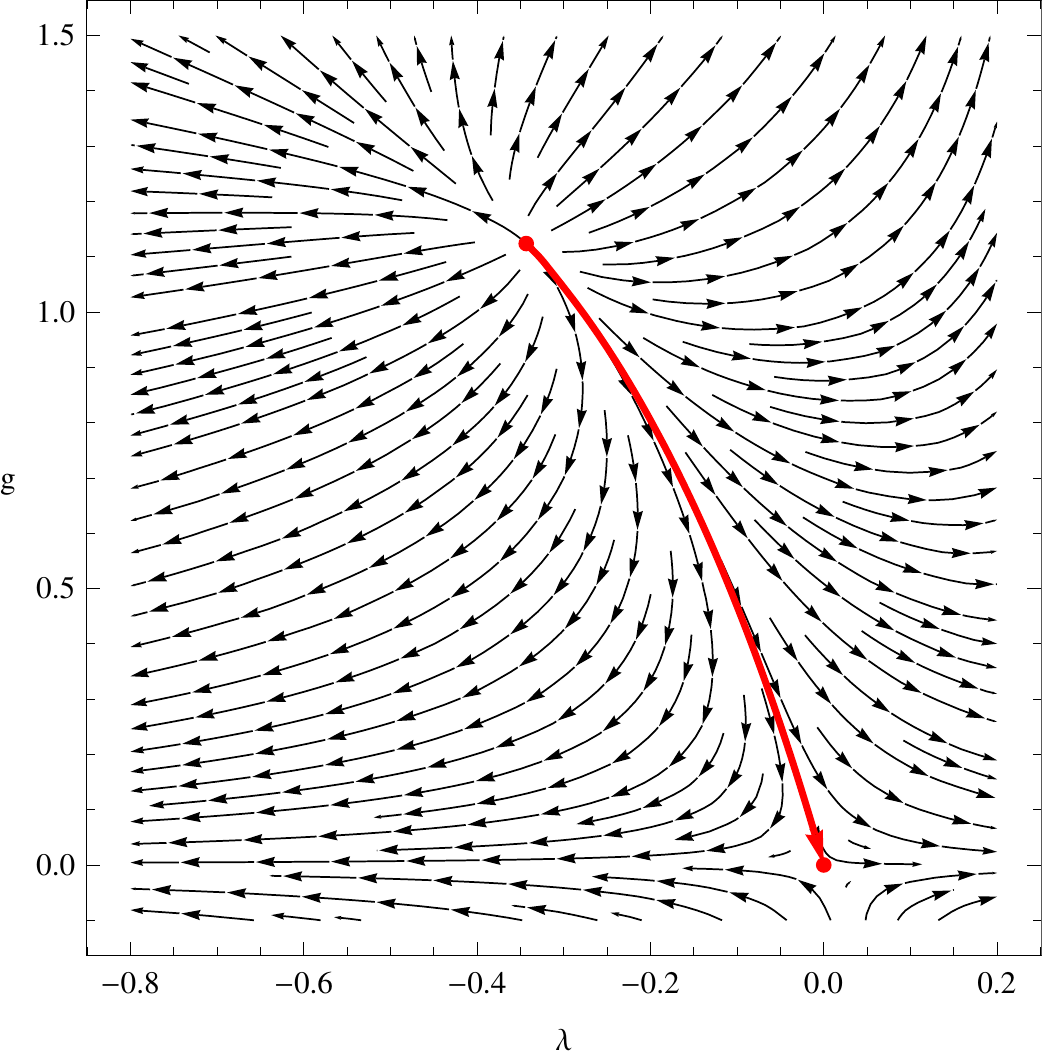}
		\label{fig:PDqqm111}}
	\subfigure[\textbf{NGFP} for $µ^2 = 1$.]{
		\centering
		\includegraphics[width=0.3\textwidth]{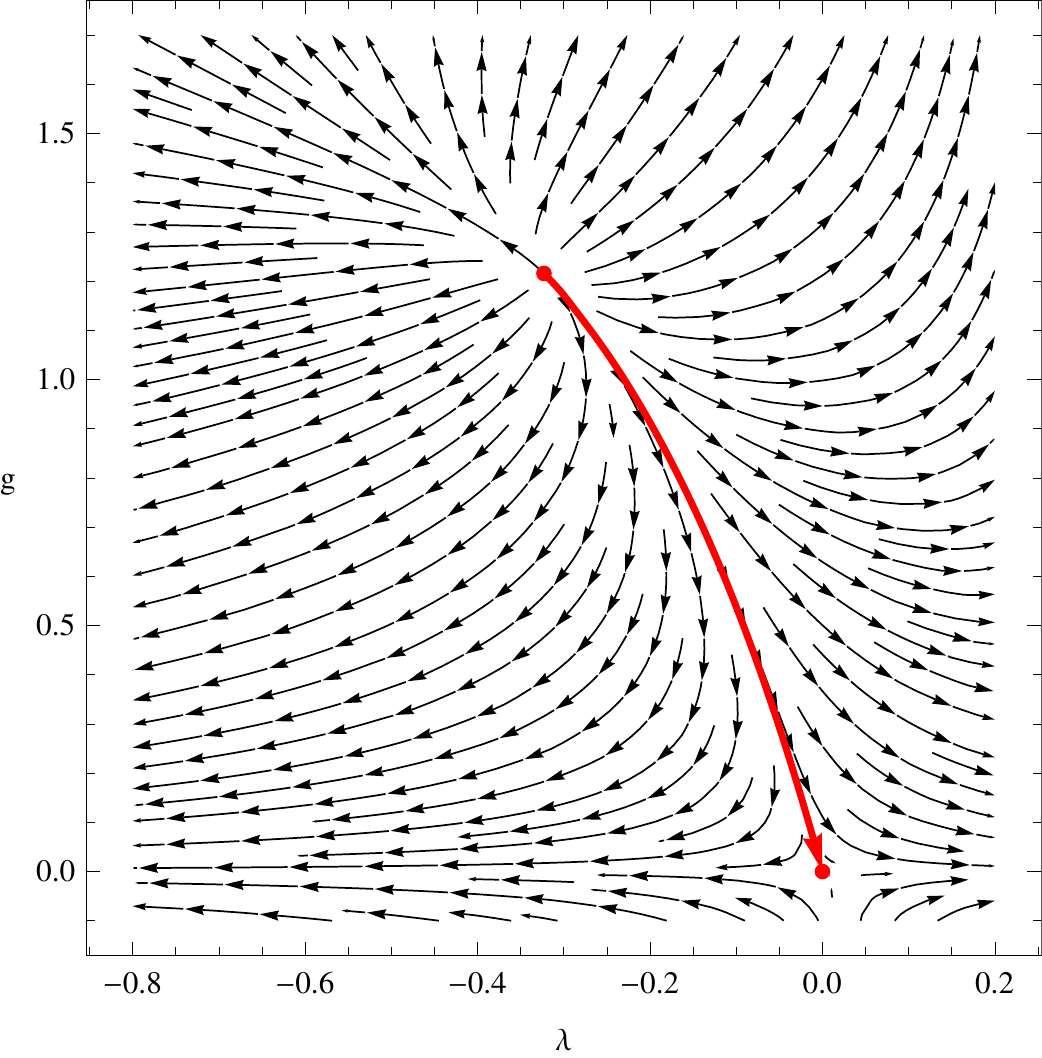}
		\label{fig:PDqqm1}}
	\subfigure[\textbf{NGFP} for $µ^2 = 2$.]{
		\centering
		\includegraphics[width=0.3\textwidth]{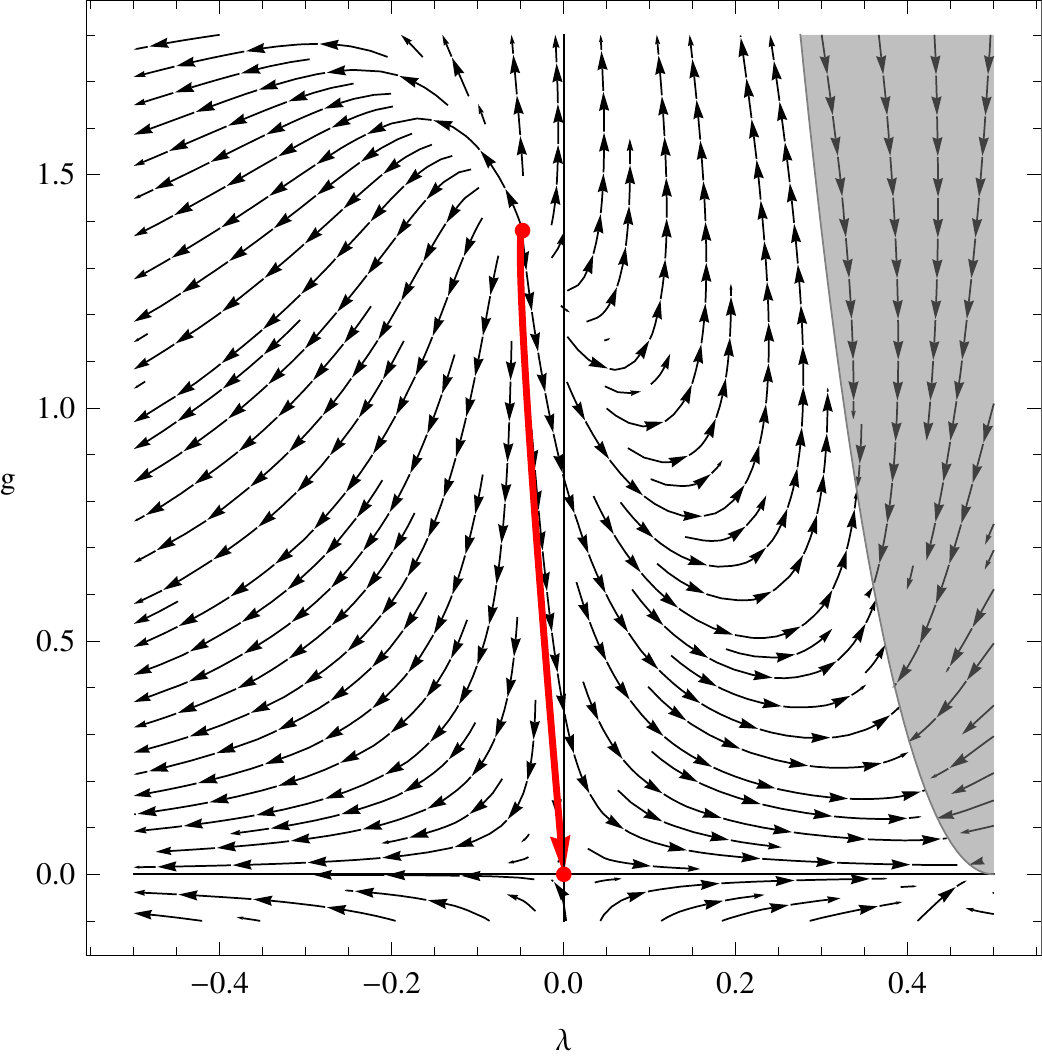}
		\label{fig:PDqqm2line}}
	\subfigure[\textbf{NGFP} for $µ^2 = 2.5$.]{
		\centering
		\includegraphics[width=0.3\textwidth]{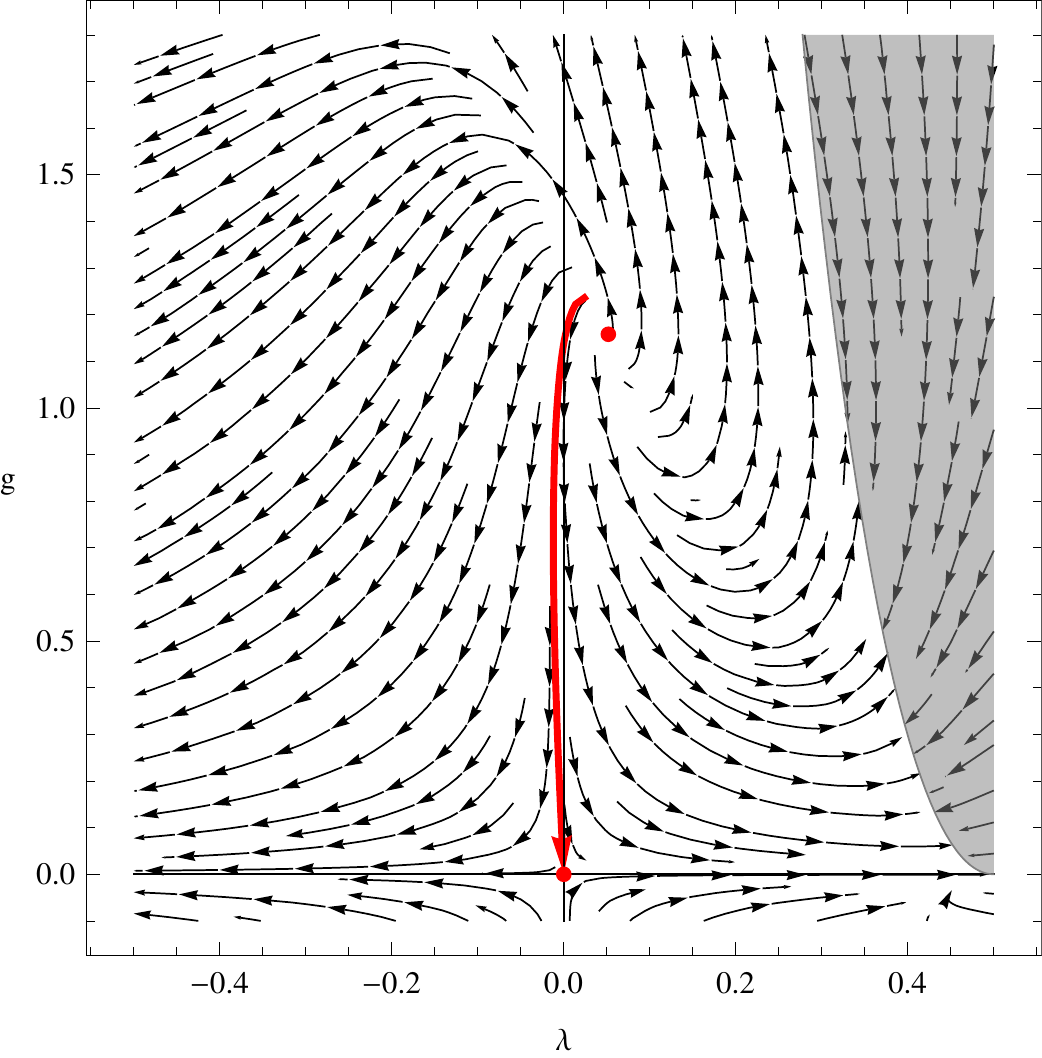}
		\label{fig:PDqqm25line}}
	\subfigure[\textbf{NGFP} for $µ^2 = 3$.]{
		\centering
		\includegraphics[width=0.3\textwidth]{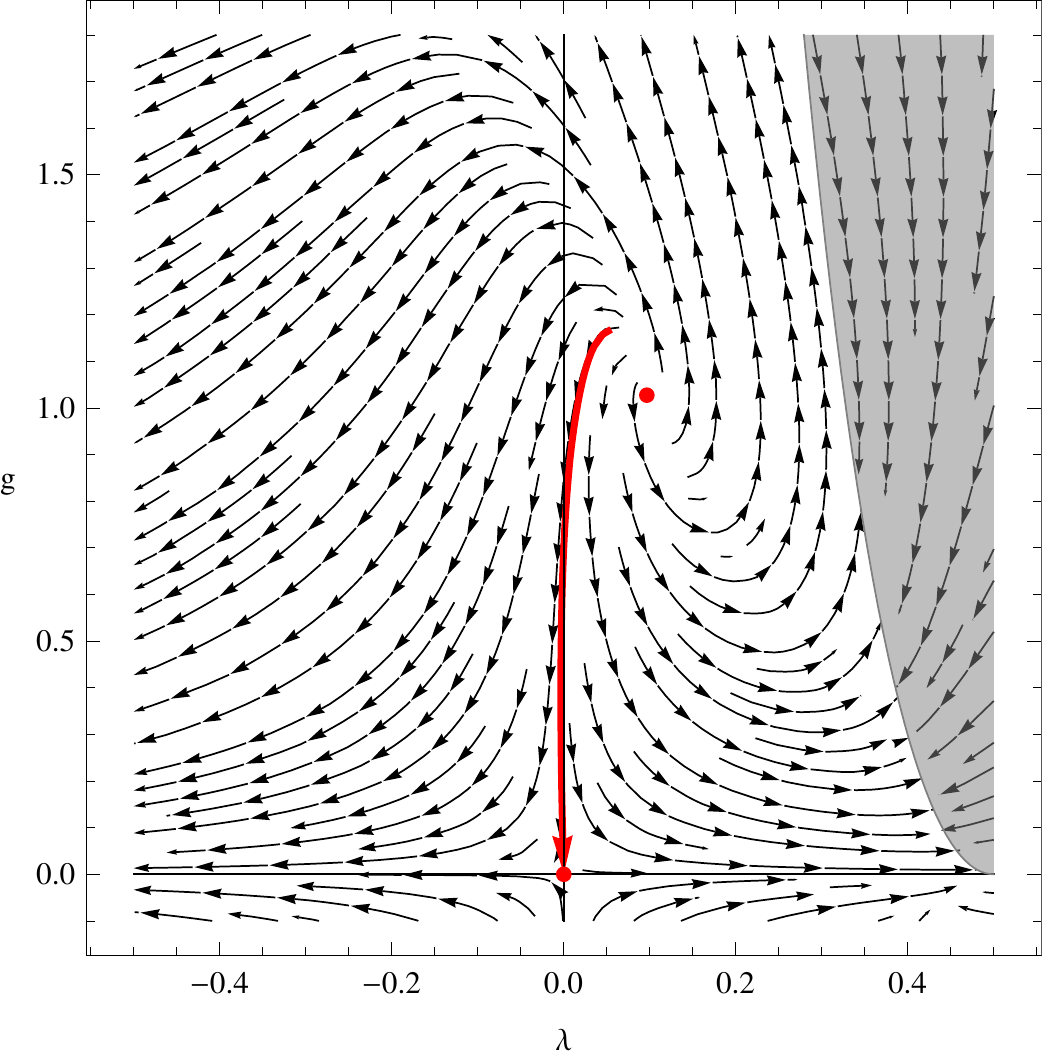}
		\label{fig:PDqqm3line}}
	\subfigure[\textbf{NGFP} for $µ^2 = 10$.]{
		\centering
		\includegraphics[width=0.3\textwidth]{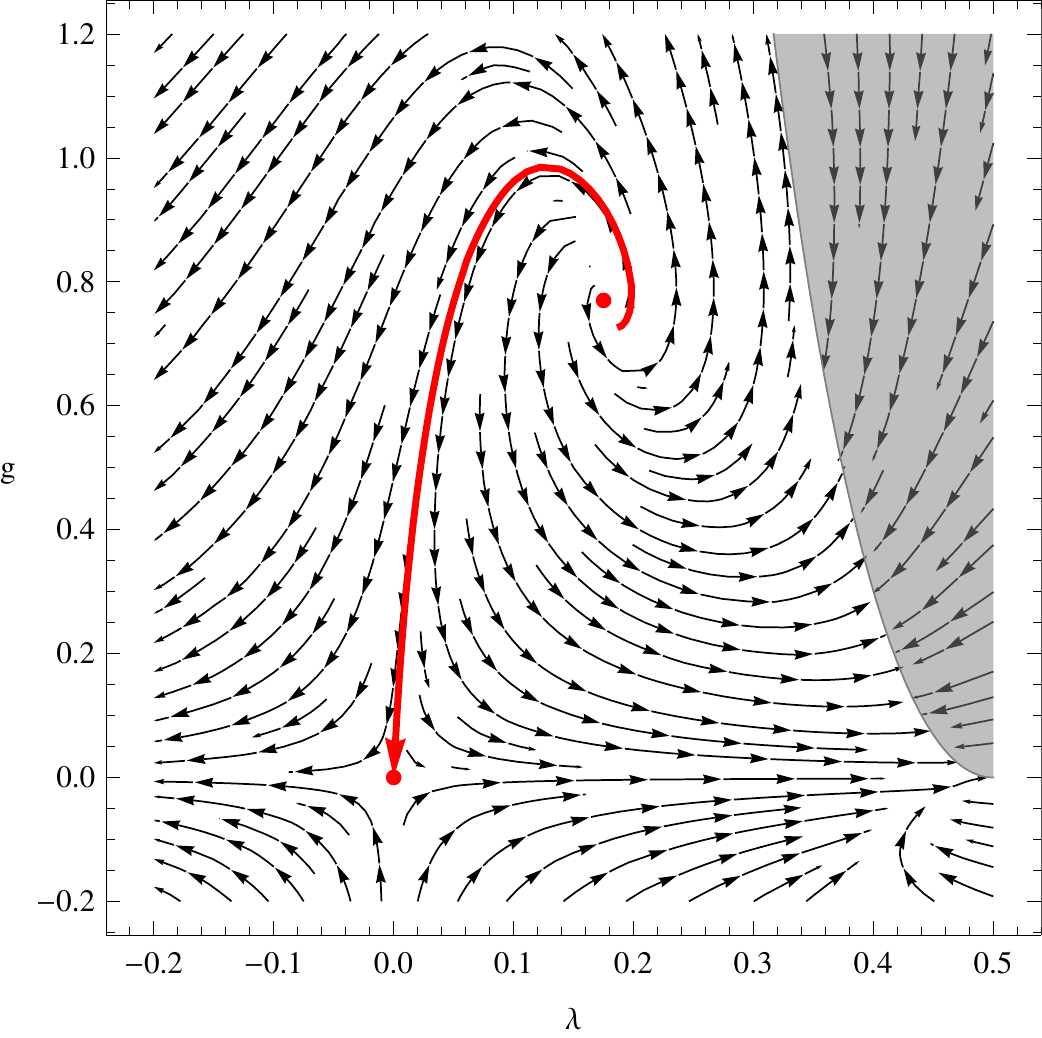}
		\label{fig:PDqqm10}}
	\subfigure[\textbf{NGFP} for $µ^2 = 100$.]{
		\centering
		\includegraphics[width=0.3\textwidth]{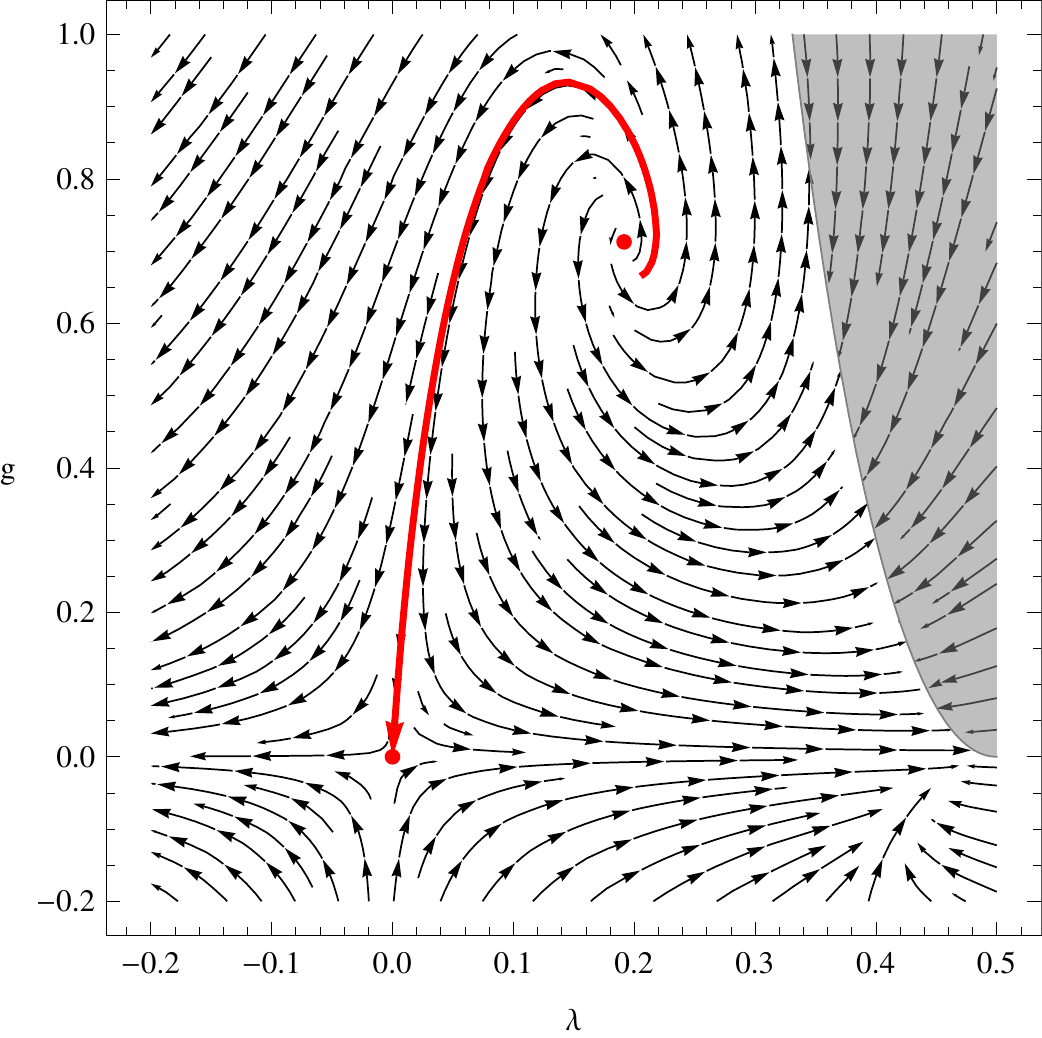}
		\label{fig:PDqqm100}}
	\caption{Phase portrait of the RG flow only incorporating $\tensor{q}{^{\lambda}_{µ \nu}}$ for different values of the squared mass parameter $µ^2$.}
	\label{fig:NGFPqq}
\end{figure}
\clearpage

\begin{figure}[htbp]
	\centering
	\subfigure[\textbf{NGFP} for $µ^2 = \frac{1}{1.9}$.]{
		\centering
		\includegraphics[width=0.3\textwidth]{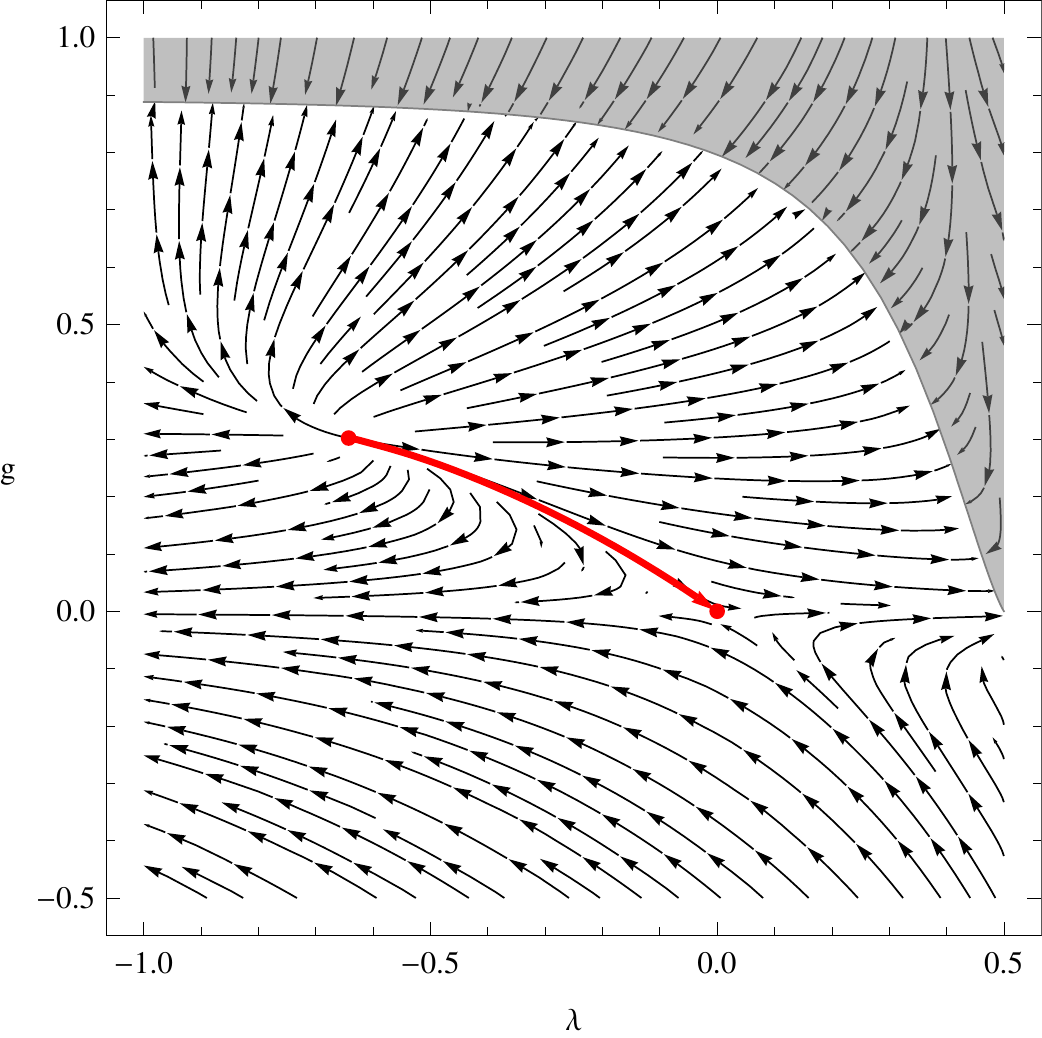}
		\label{fig:PDm119s2qq}}
	\subfigure[\textbf{NGFP} for $µ^2 = \frac{1}{1.5}$.]{
		\centering
		\includegraphics[width=0.3\textwidth]{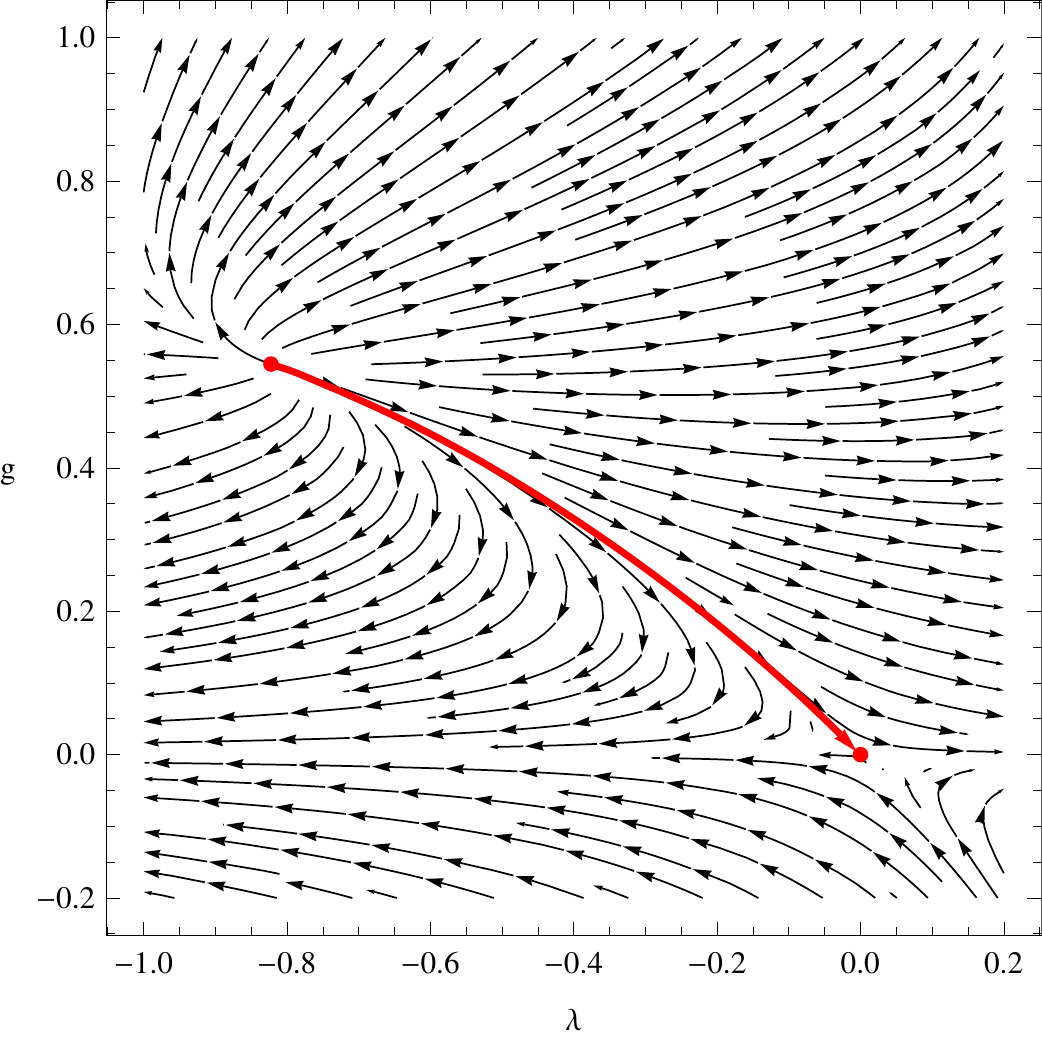}
		\label{fig:PDm115s2qq}}
	\subfigure[\textbf{NGFP} for $µ^2 = \frac{1}{1.1}$.]{
		\centering
		\includegraphics[width=0.3\textwidth]{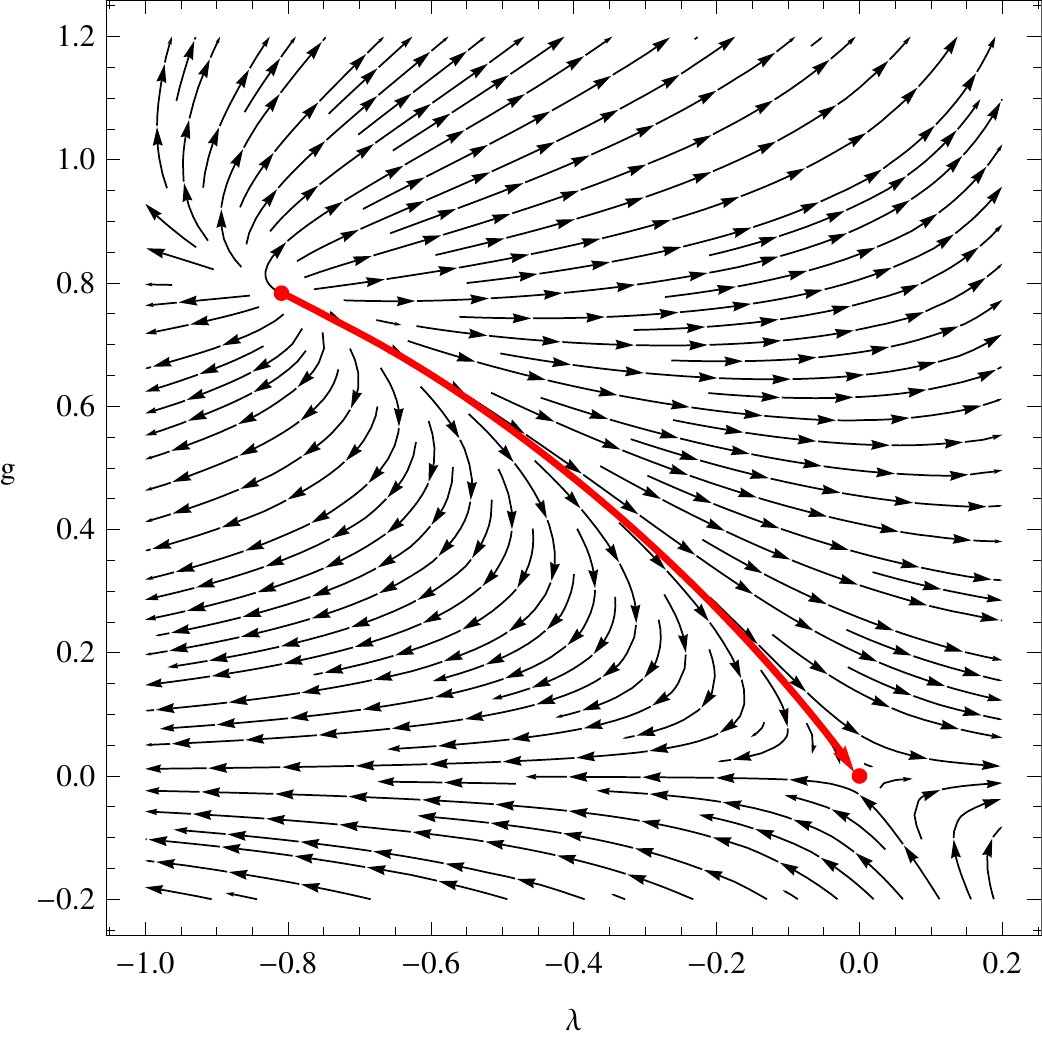}
		\label{fig:PDm111s2qq}}
	\subfigure[\textbf{NGFP} for $µ^2 = 1$.]{
		\centering
		\includegraphics[width=0.3\textwidth]{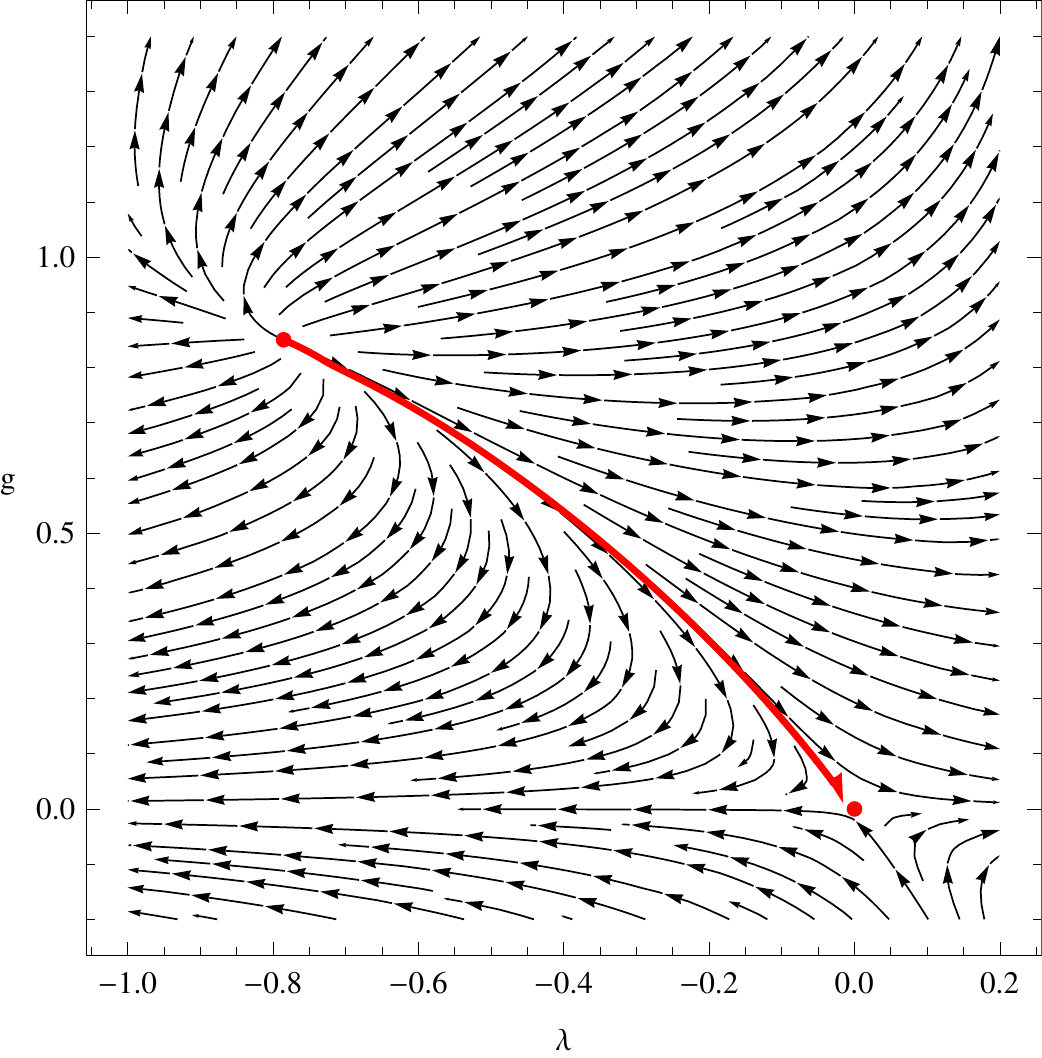}
		\label{fig:PDm1s2qq}}
	\subfigure[\textbf{NGFP} for $µ^2 = 2$.]{
		\centering
		\includegraphics[width=0.3\textwidth]{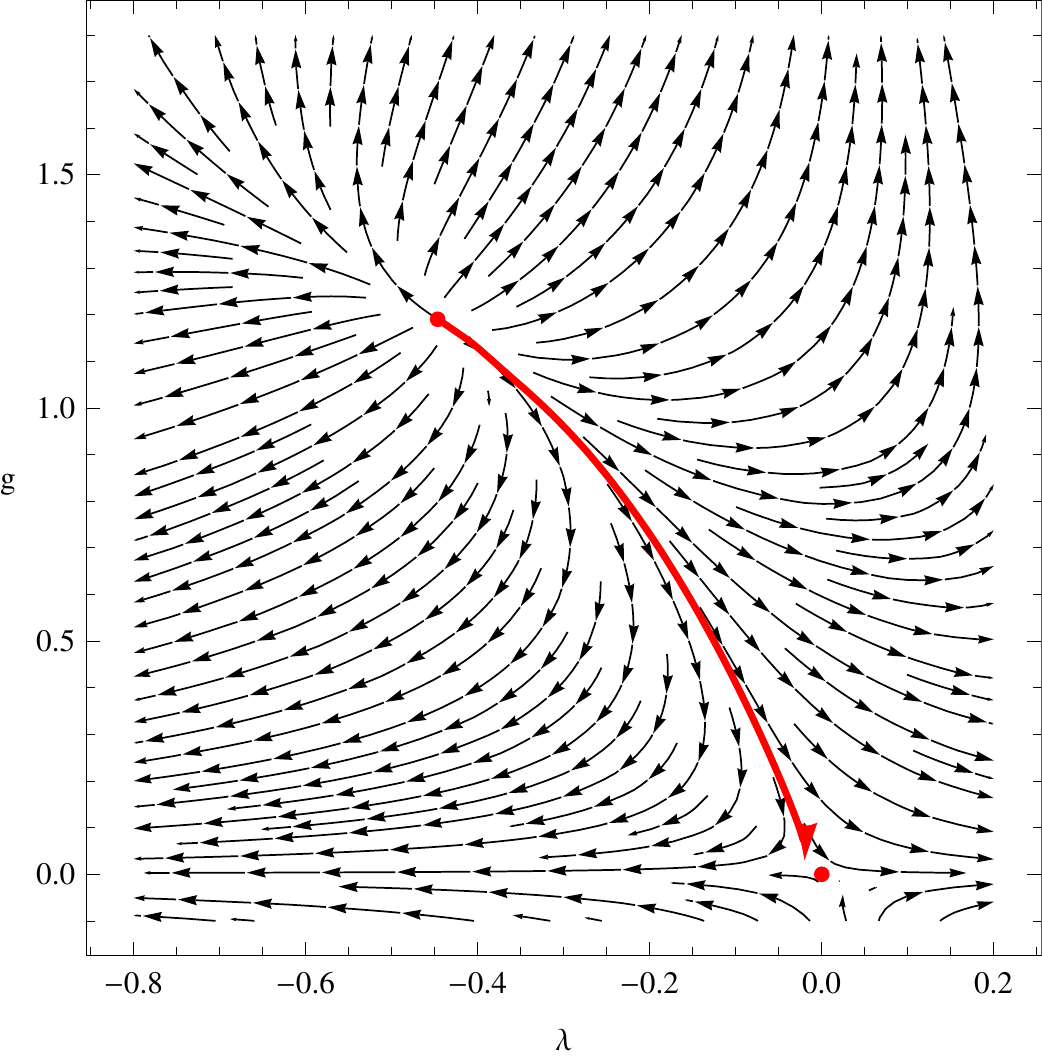}
		\label{fig:PDm2s2qq}}
	\subfigure[\textbf{NGFP} for $µ^2 = 2.5$.]{
		\centering
		\includegraphics[width=0.3\textwidth]{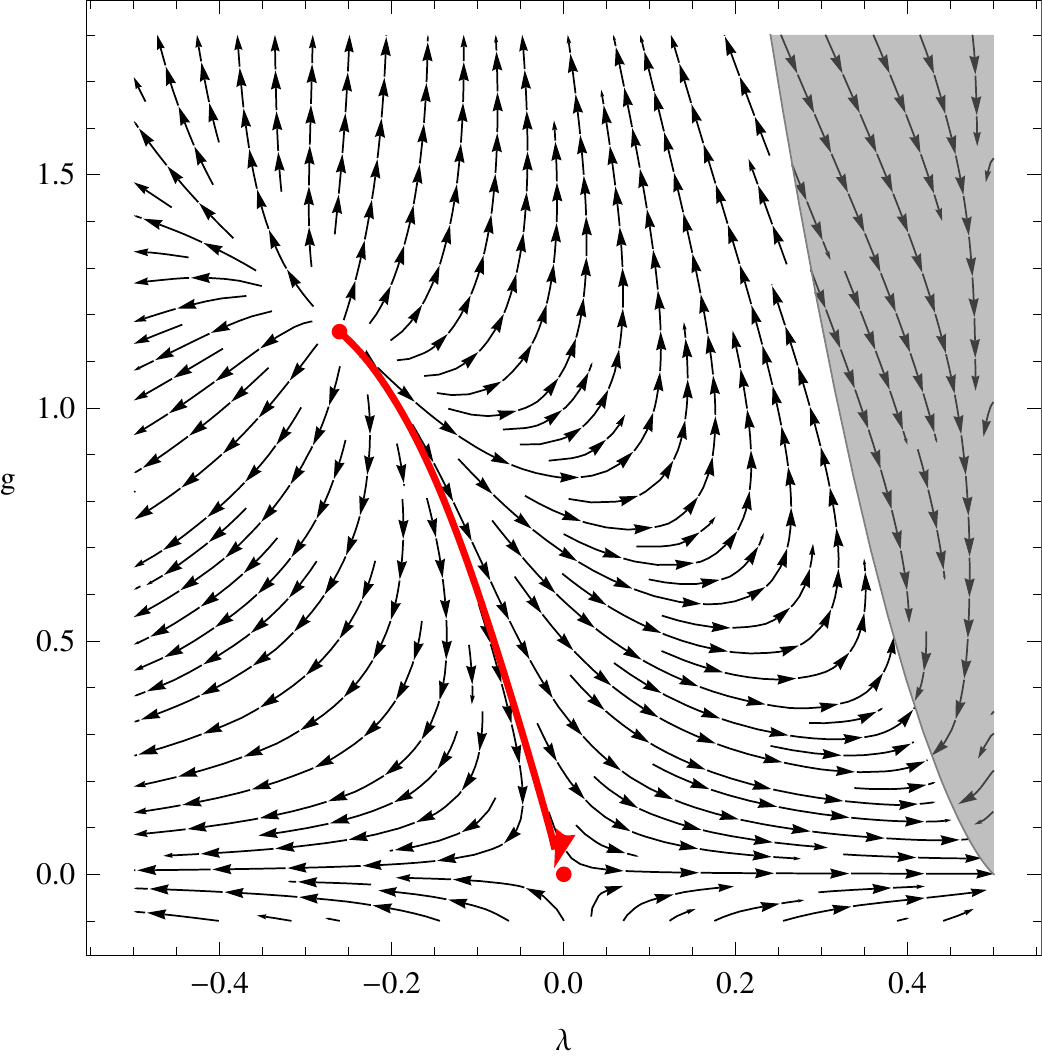}
		\label{fig:PDm25s2qq}}
	\subfigure[\textbf{NGFP} for $µ^2 = 3$.]{
		\centering
		\includegraphics[width=0.3\textwidth]{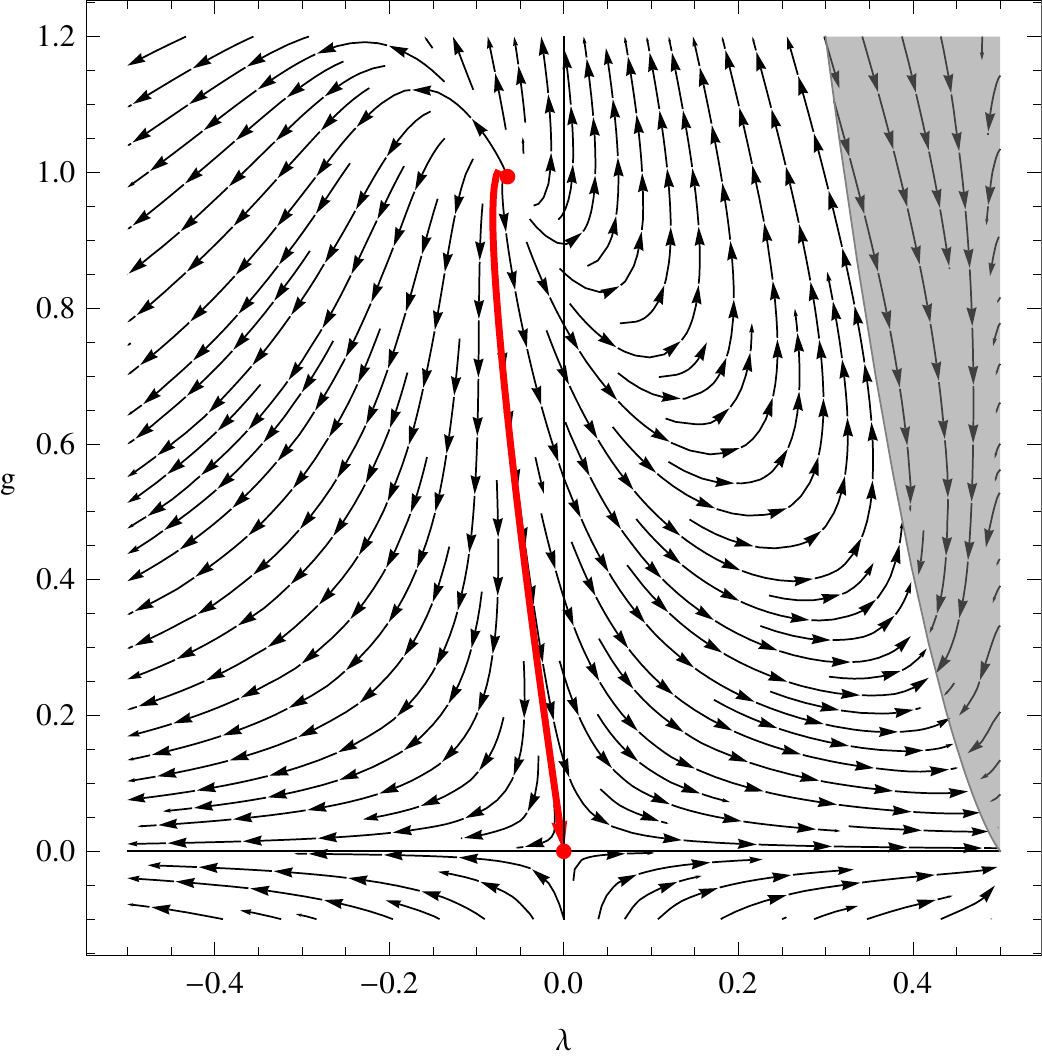}
		\label{fig:PDm3s2qq}}
	\subfigure[\textbf{NGFP} for $µ^2 = 3.5$.]{
		\centering
		\includegraphics[width=0.3\textwidth]{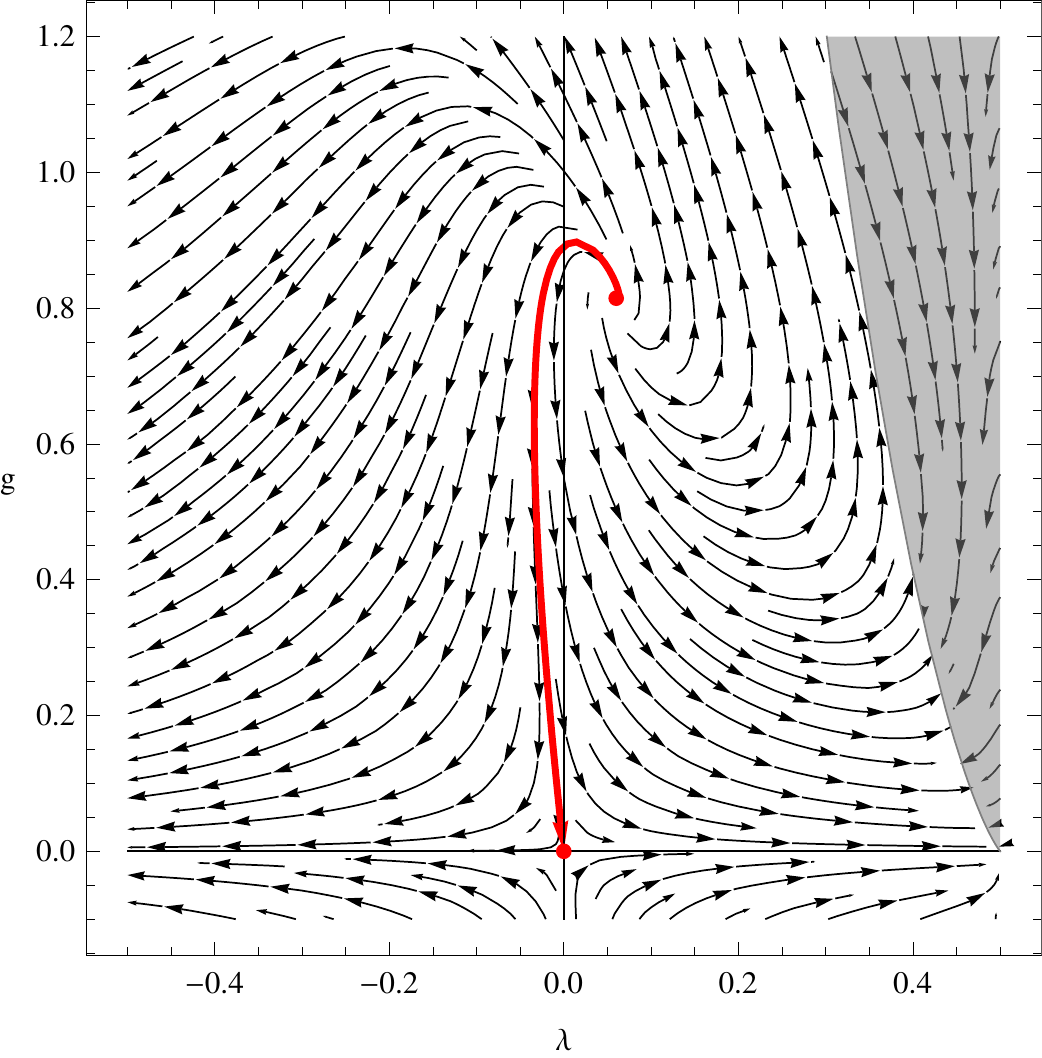}
		\label{fig:PDm35s2qq}}
	\subfigure[\textbf{NGFP} for $µ^2 = 4$.]{
		\centering
		\includegraphics[width=0.3\textwidth]{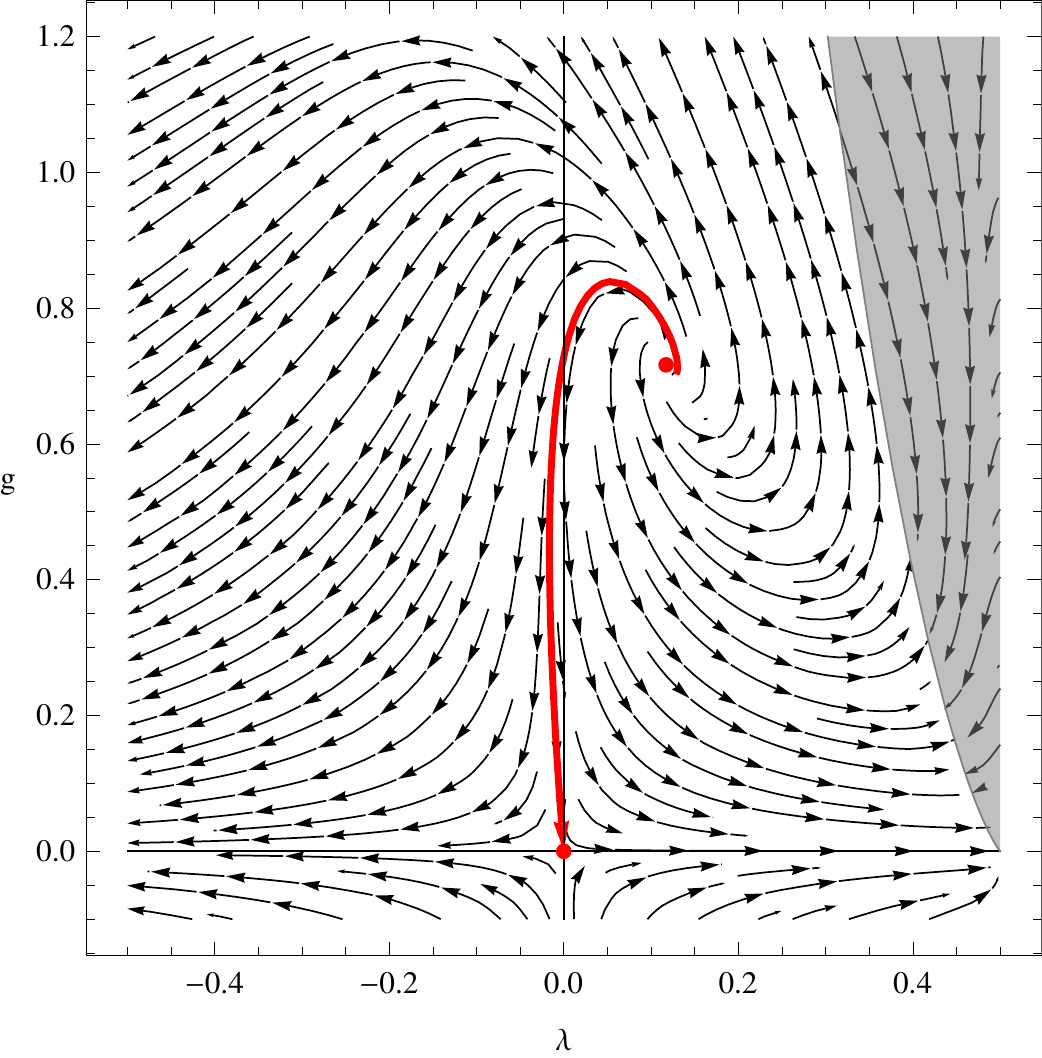}
		\label{fig:PDm4s2qq}}
	\subfigure[\textbf{NGFP} for $µ^2 = 5$.]{
		\centering
		\includegraphics[width=0.3\textwidth]{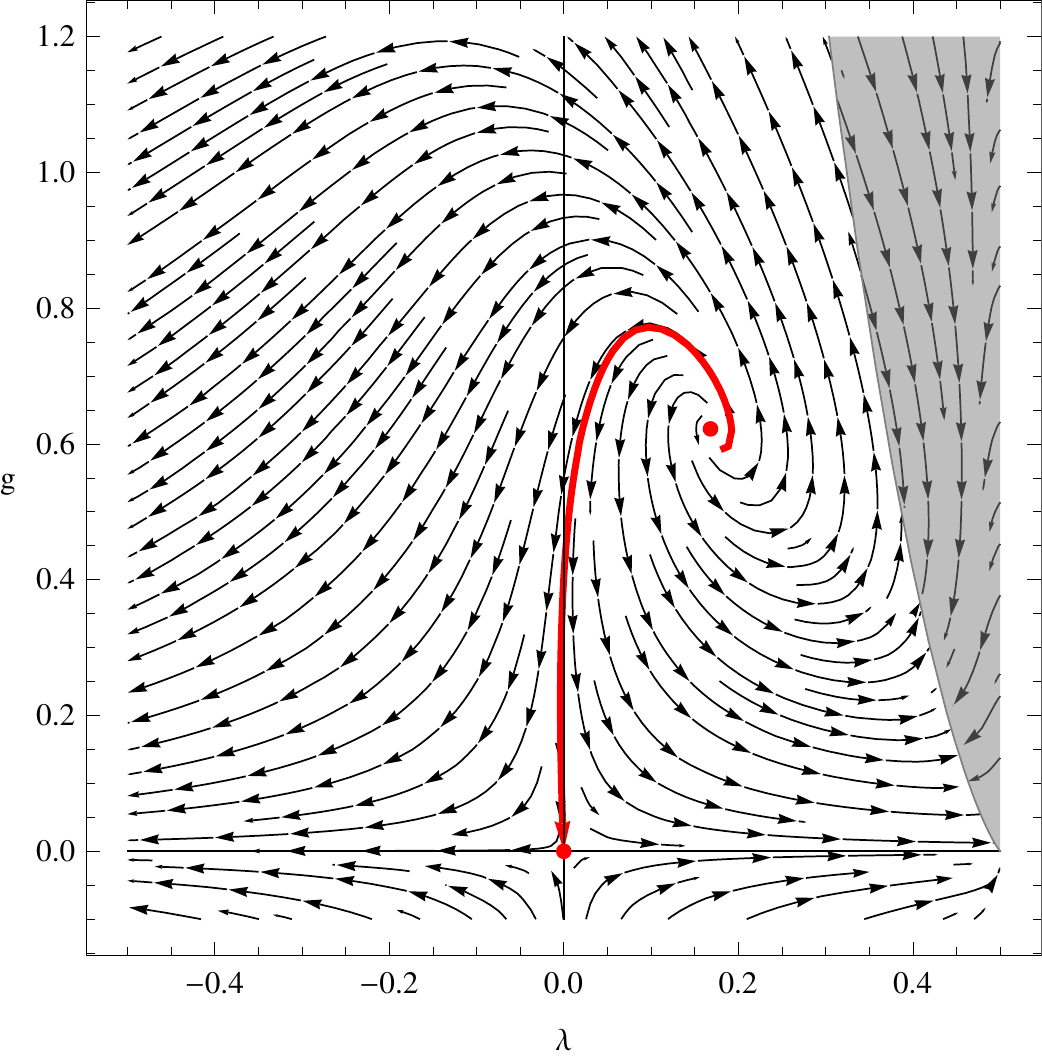}
		\label{fig:PDm5s2qq}}
	\subfigure[\textbf{NGFP} for $µ^2 = 10$.]{
		\centering
		\includegraphics[width=0.3\textwidth]{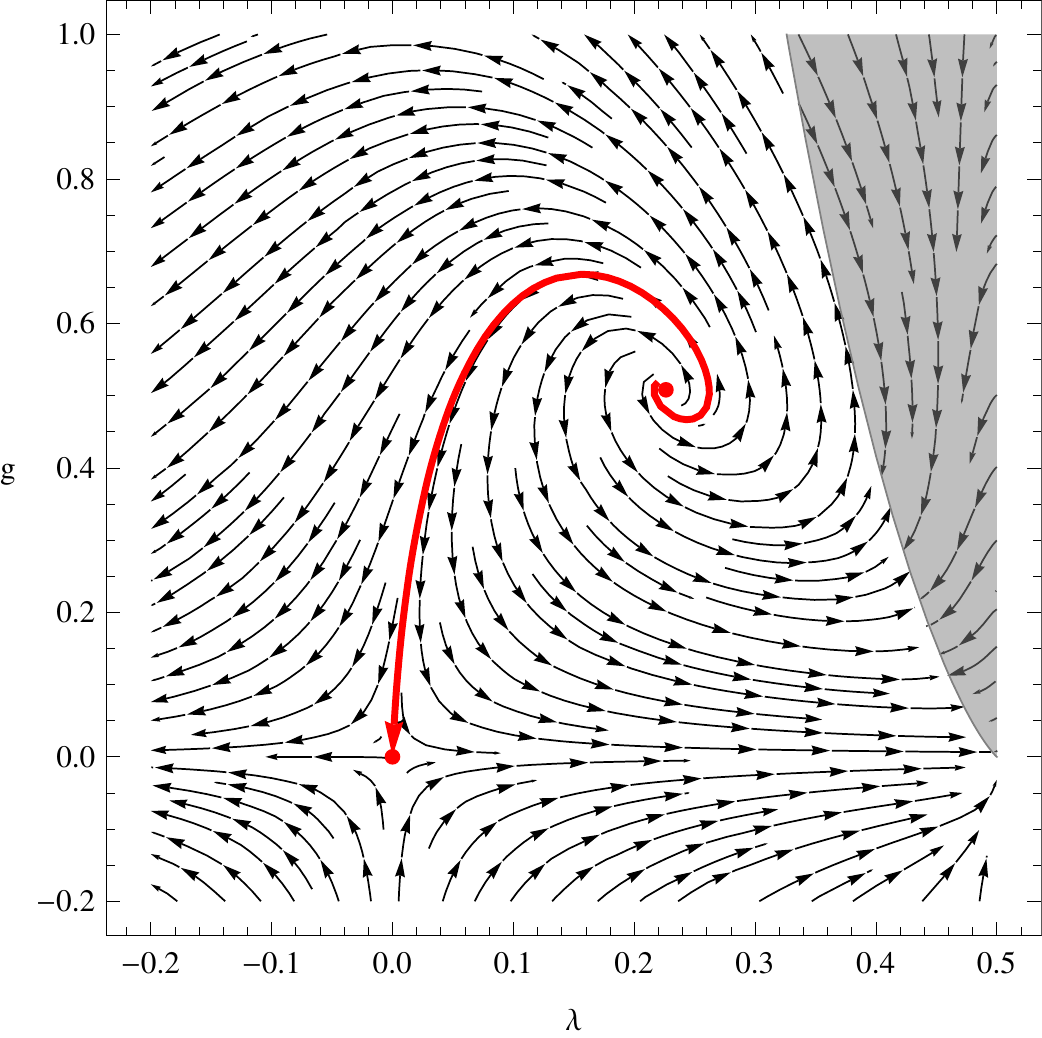}
		\label{fig:PDm10s2qq}}
	\subfigure[\textbf{NGFP} for $µ^2 = 100$.]{
		\centering
		\includegraphics[width=0.3\textwidth]{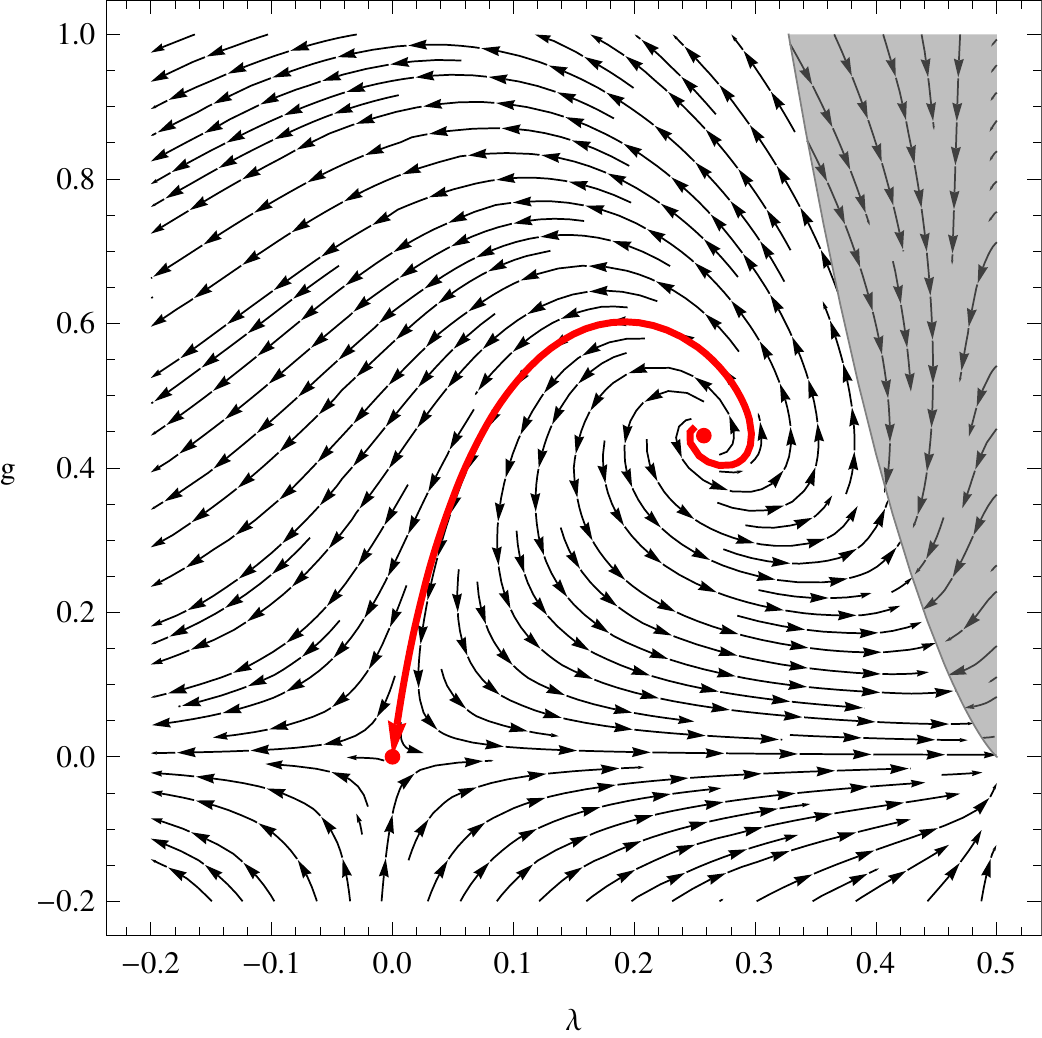}
		\label{fig:PDm100s2qq}}
	\caption{Phase portrait of the RG flow only incorporating $\tensor{q}{^{\lambda}_{µ \nu}}$ for different values of the squared mass parameter $µ^2$ and employing the generalized exponential cutoff with shape parameter $s=2$.}
	\label{fig:NGFPqqshape}
\end{figure}
\clearpage
\begin{figure}[htbp]
	\centering
	\subfigure[Positive $g$-halfplane ending at $g= +2$.]{
		\centering
		\includegraphics[width=0.3\textwidth]{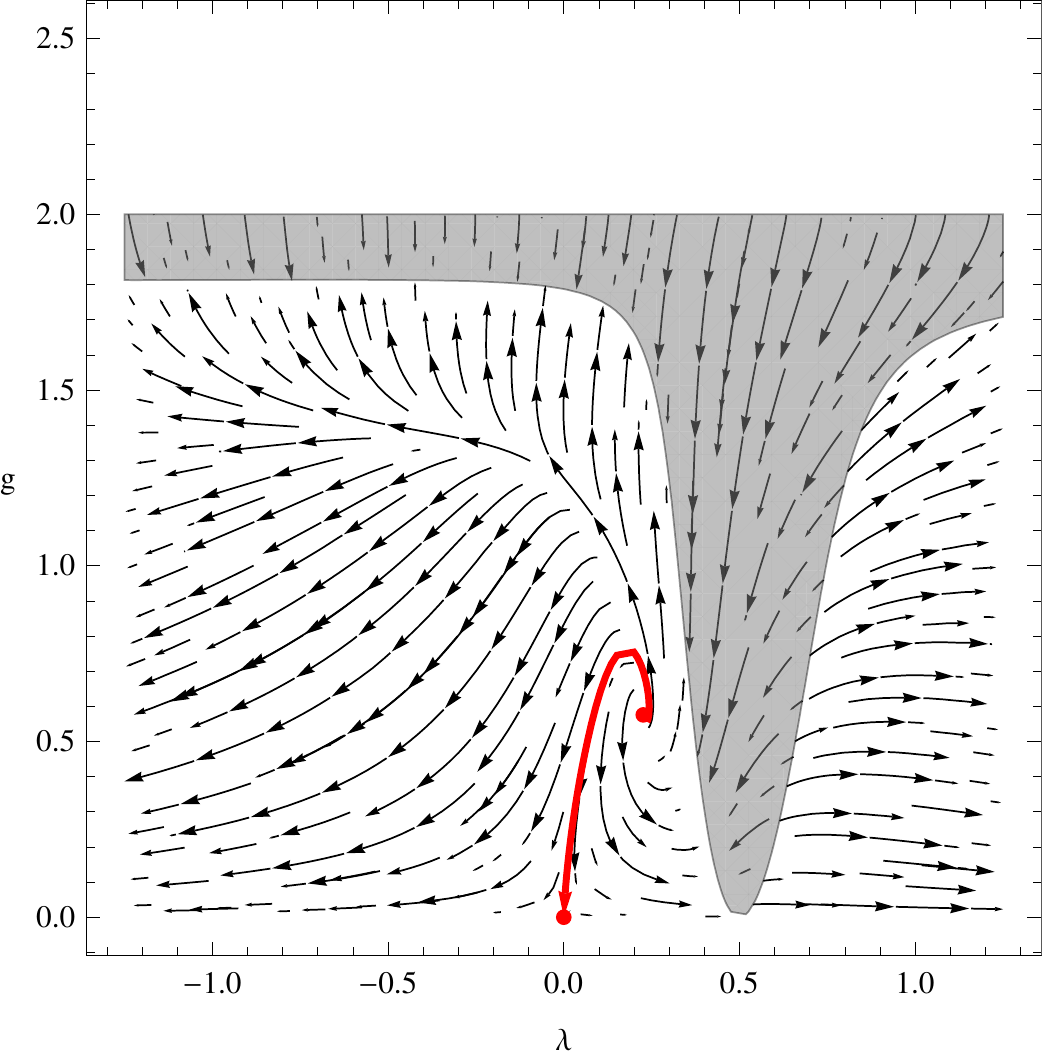}
		\label{fig:PDgtestv2}}
	\subfigure[Negative $g$-halfplane ending at $g = - \frac{1}{6}$.]{
		\centering
		\includegraphics[width=0.3\textwidth]{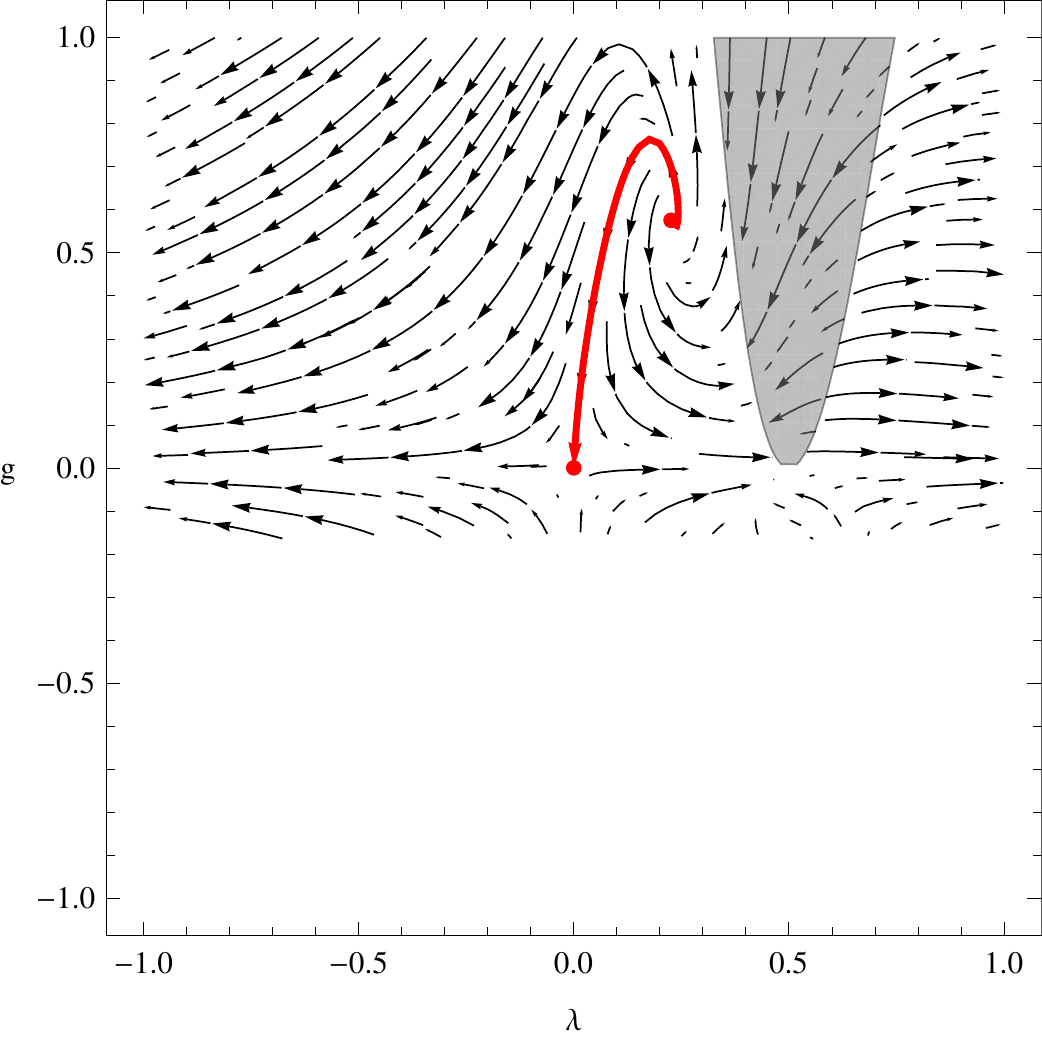}
		\label{fig:PDgtest2v2}}
	\subfigure[The \textbf{NGFP} in the ``allowed'' region.]{
		\centering
		\includegraphics[width=0.3\textwidth]{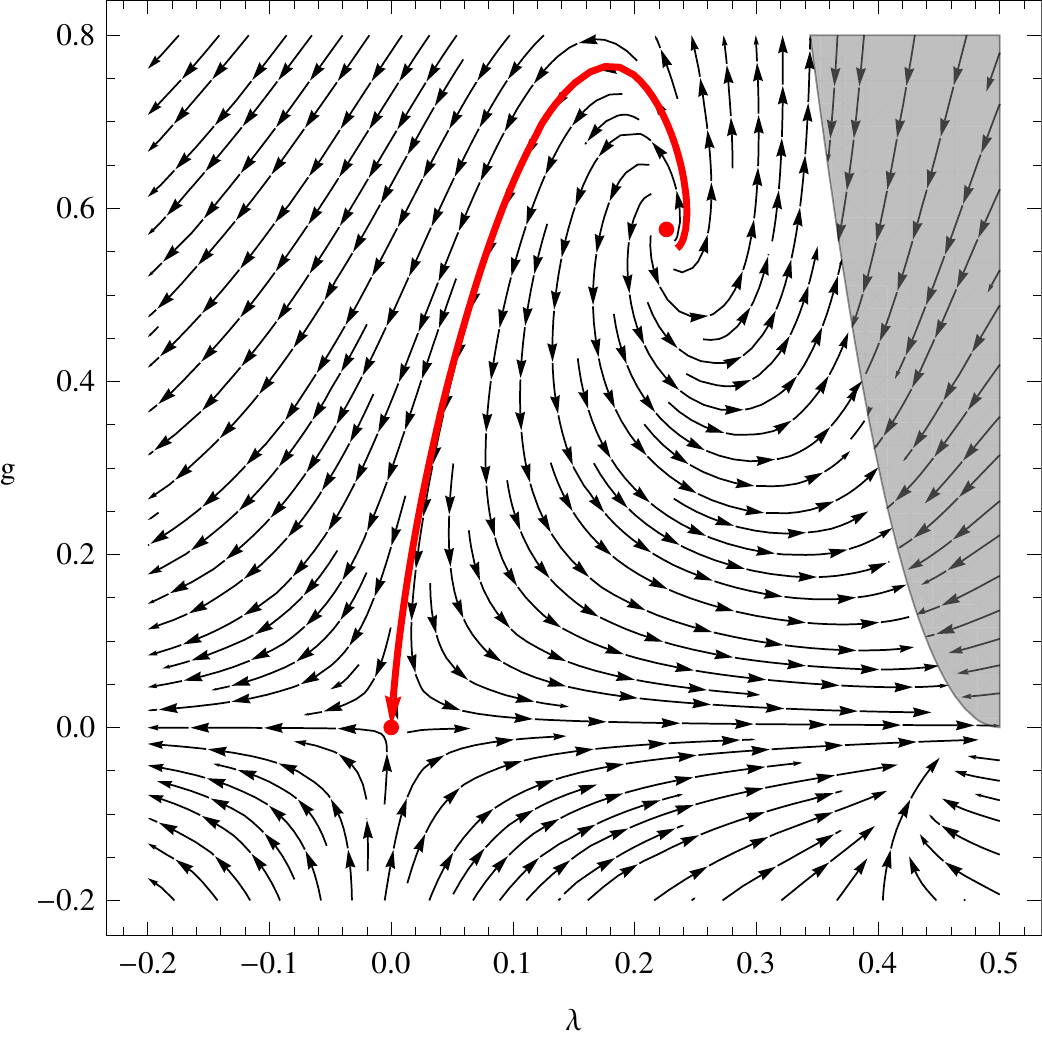}
		\label{fig:PDgtest3}}
	\caption{Phase portrait for the identification $µ \equiv \frac{1}{\sqrt{g_k}}$.}
	\label{fig:NGFPmg}
\end{figure}
\begin{figure}[H]
	\centering
	\subfigure[\textbf{NGFP} for $s = 2$.]{
		\centering
		\includegraphics[width=0.3\textwidth]{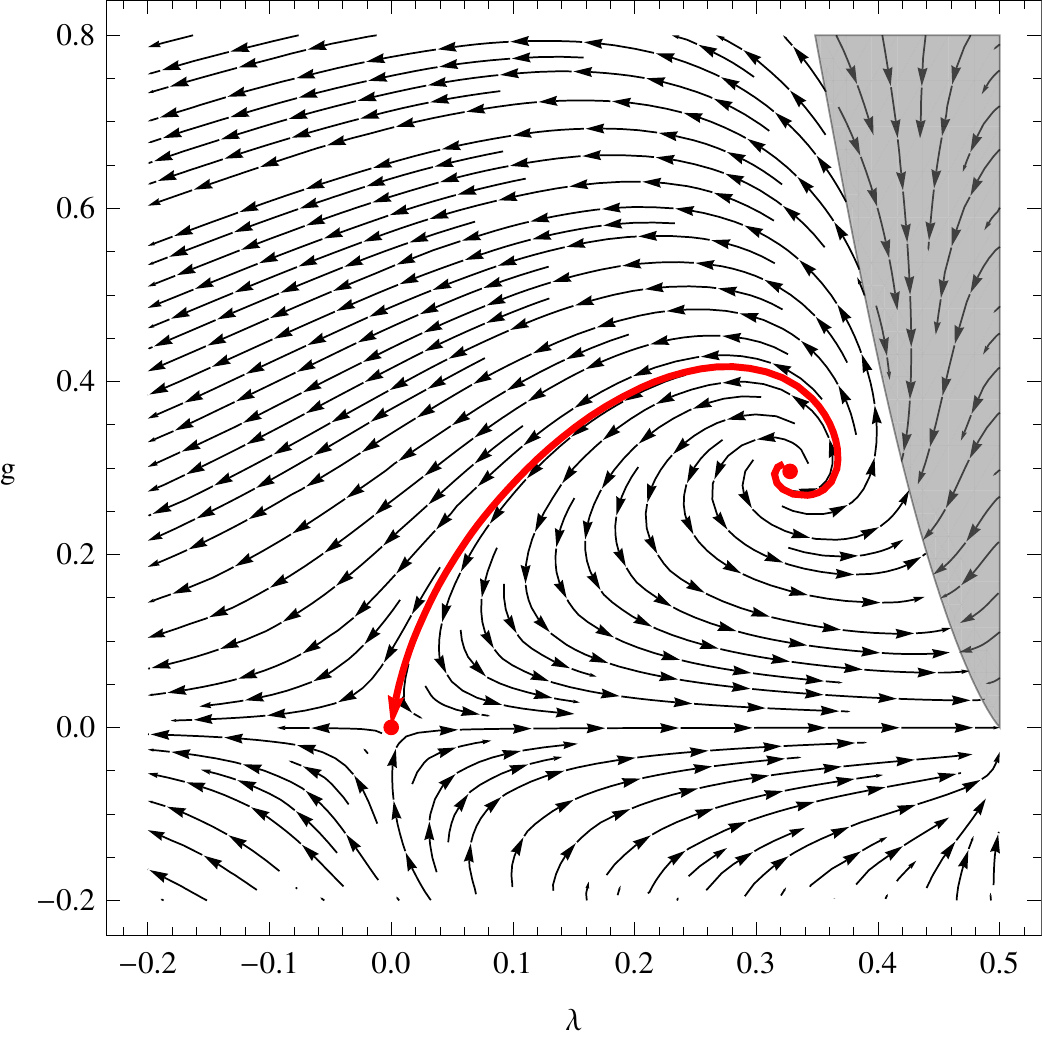}
		\label{fig:PDmgs2}}
	\subfigure[\textbf{NGFP} for $s = 5$.]{
		\centering
		\includegraphics[width=0.3\textwidth]{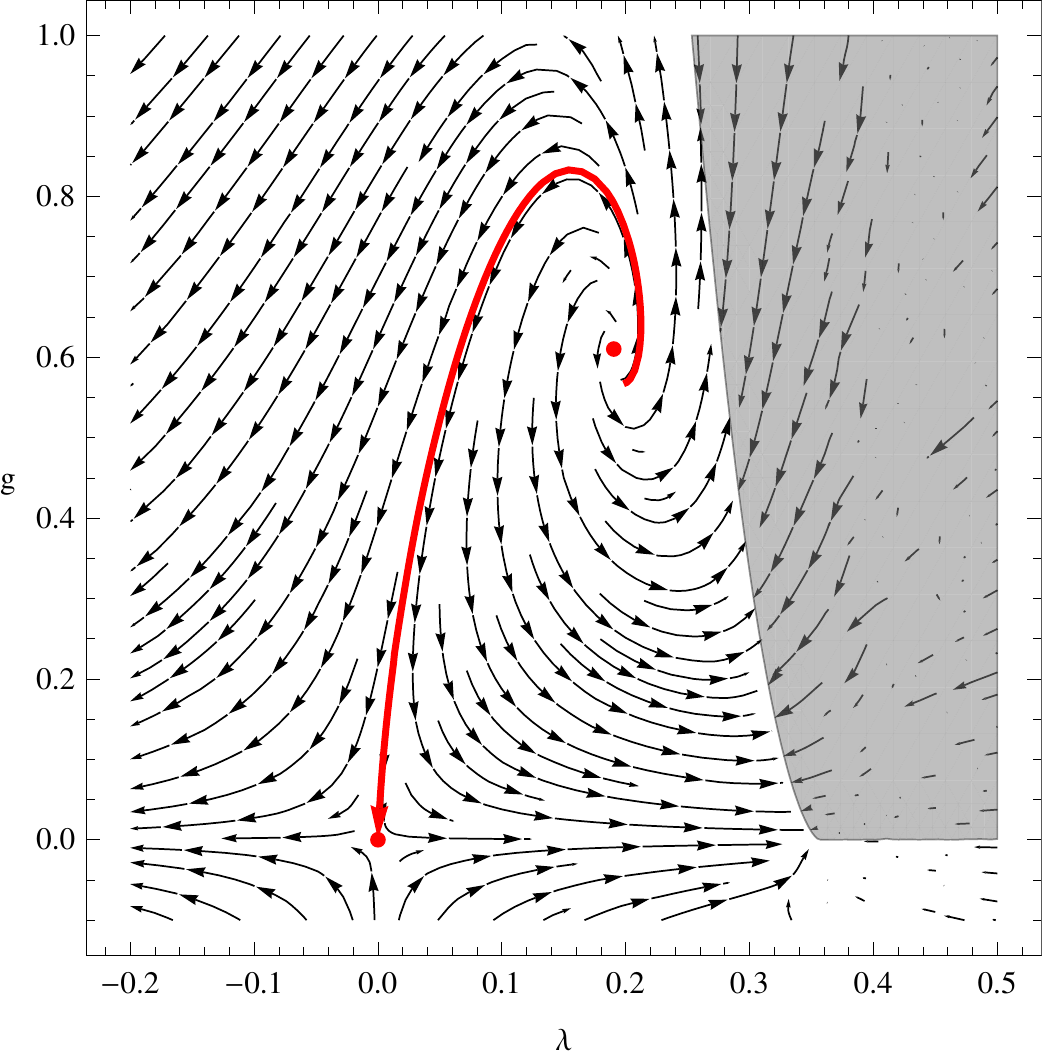}
		\label{fig:PDmgs5}}
	\subfigure[\textbf{NGFP} for $s = 7$.]{
		\centering
		\includegraphics[width=0.3\textwidth]{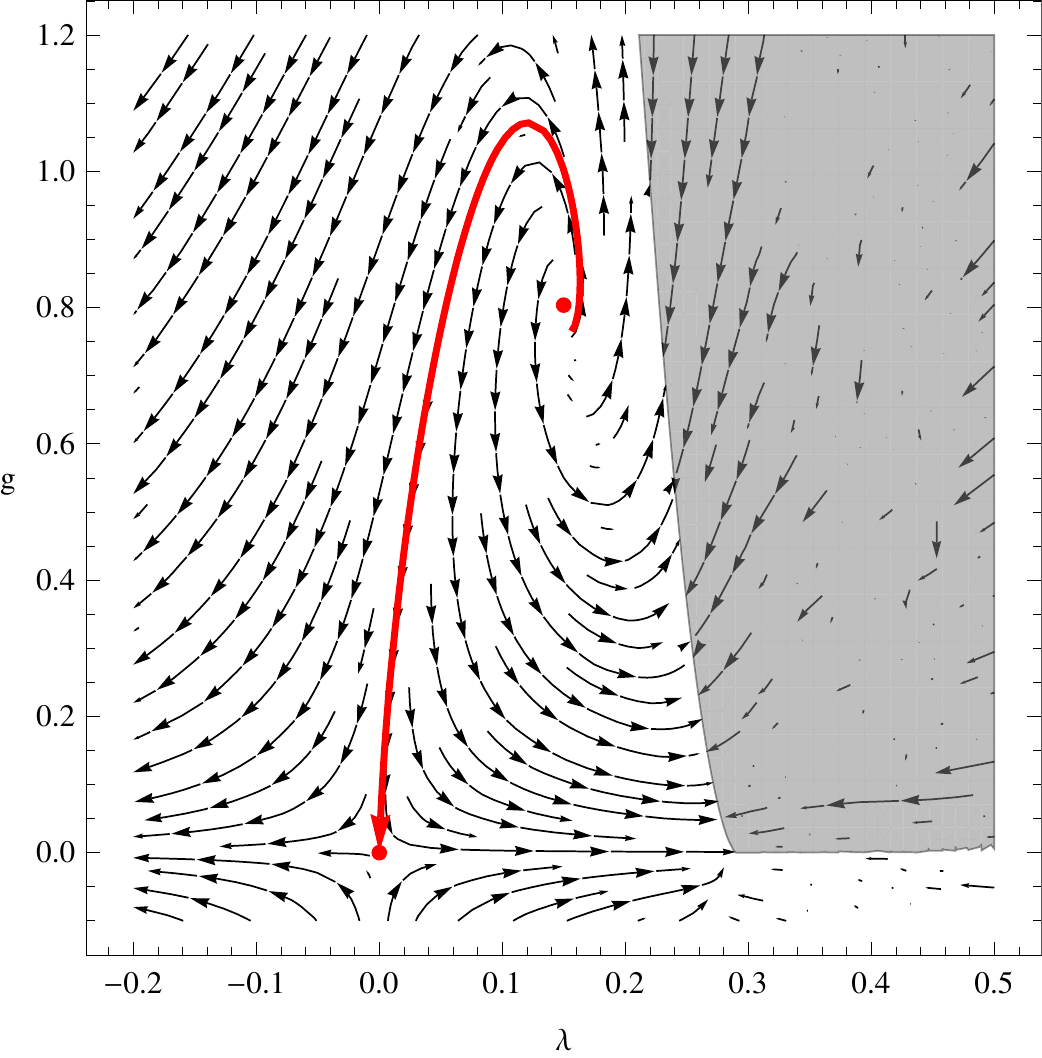}
		\label{fig:PDmgs7}}
	\subfigure[\textbf{NGFP} for $s = 10$.]{
		\centering
		\includegraphics[width=0.3\textwidth]{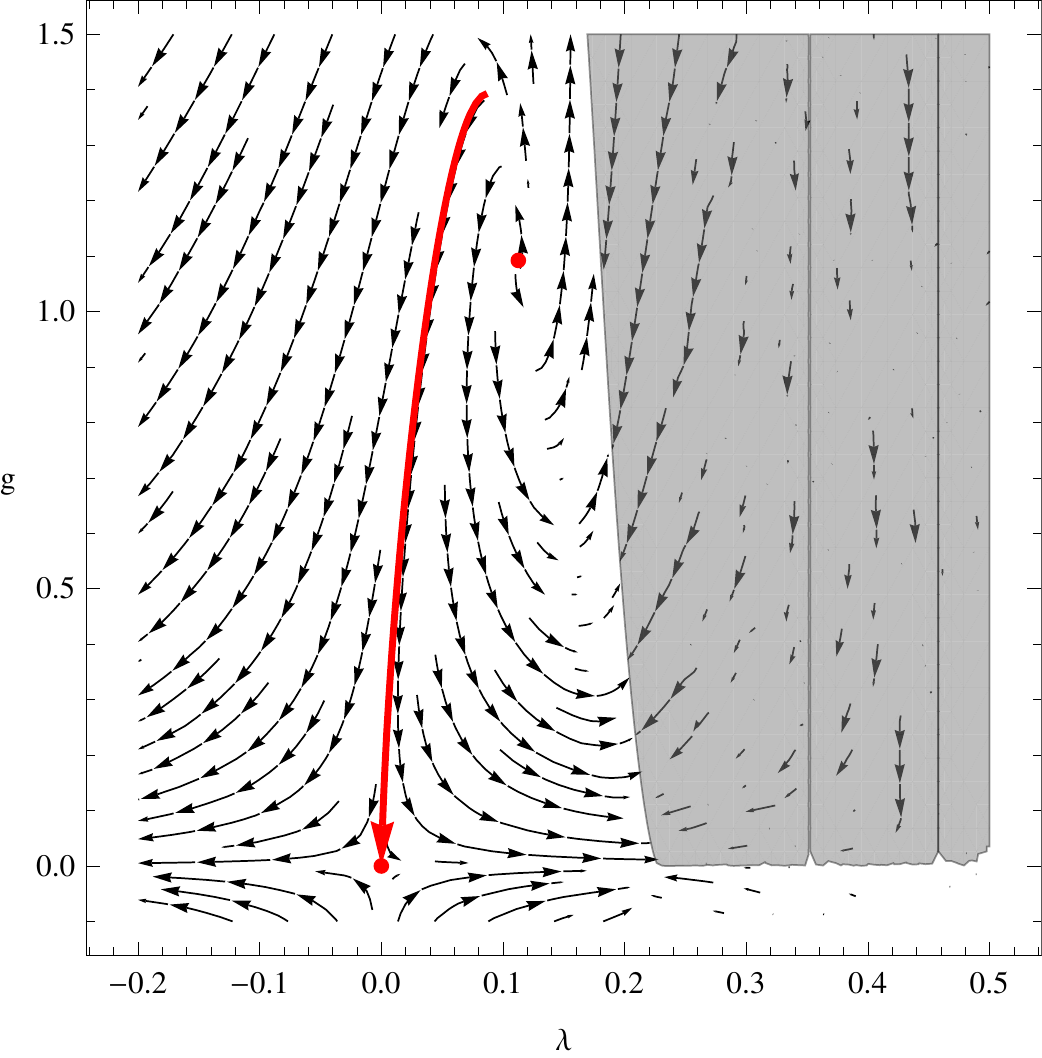}
		\label{fig:PDmgs10}}
	\subfigure[\textbf{NGFP} for $s = 15$.]{
		\centering
		\includegraphics[width=0.3\textwidth]{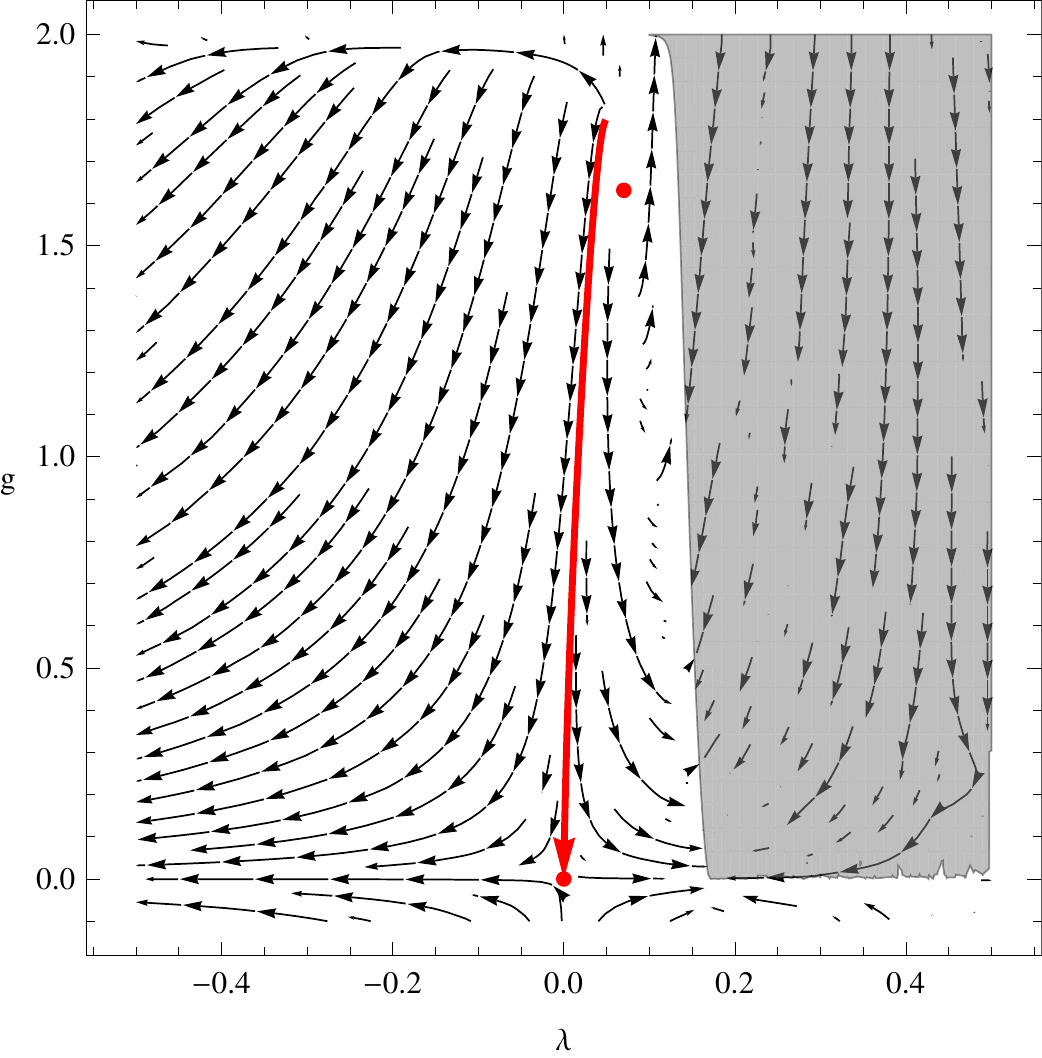}
		\label{fig:PDmgs15}}
	\subfigure[``False'' NGFP for $s = 20$.]{
		\centering
		\includegraphics[width=0.3\textwidth]{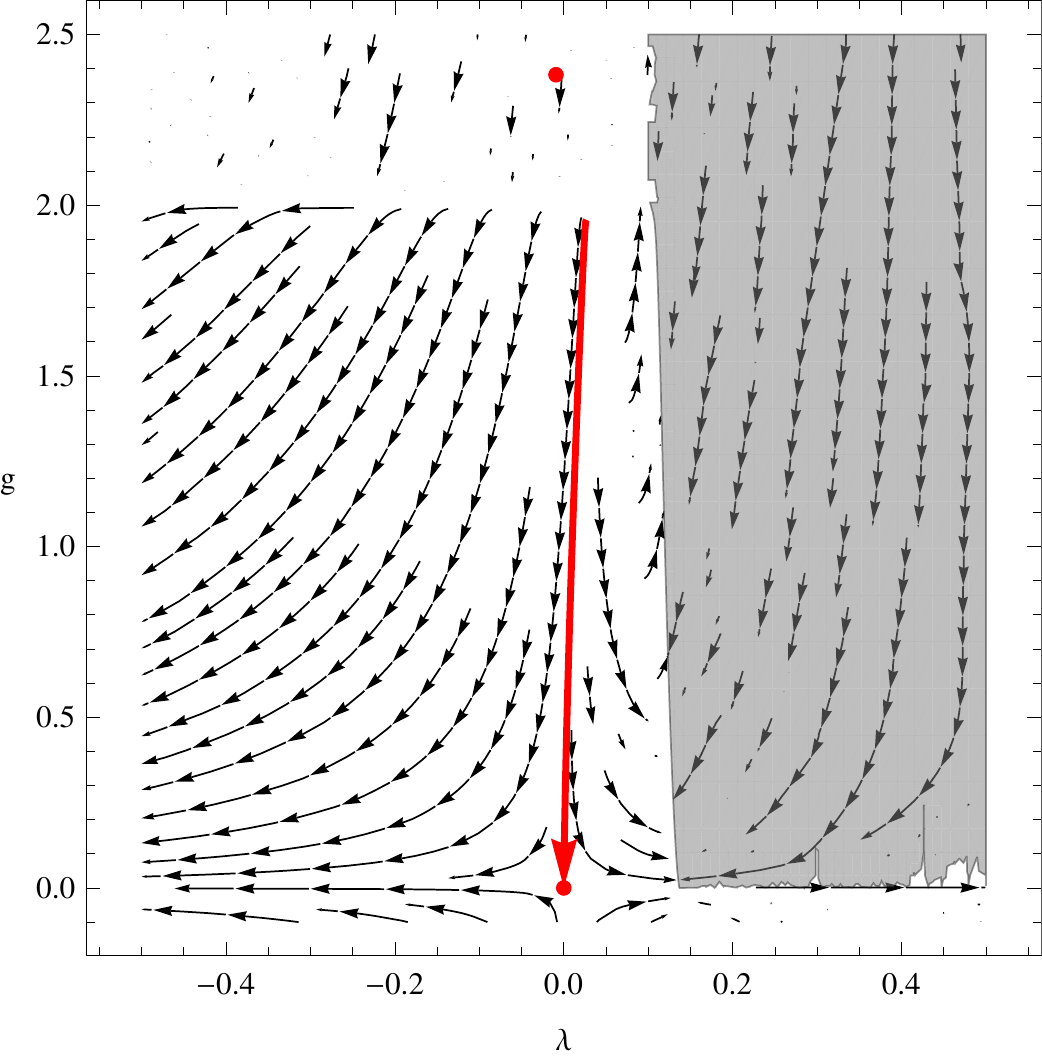}
		\label{fig:PDmgs20}}
	\caption{Phase portrait for the identification $µ \equiv \frac{1}{\sqrt{g_k}}$ and different shape parameters $s$.}
	\label{fig:NGFPmgshape}
\end{figure}
\clearpage

\begin{figure}[H]
	\centering
	\subfigure[\textbf{NGFP}$\bm{^{\oplus}}$ for $µ^2 = \frac{1}{100}$.]{
		\centering
		\includegraphics[width=0.3\textwidth]{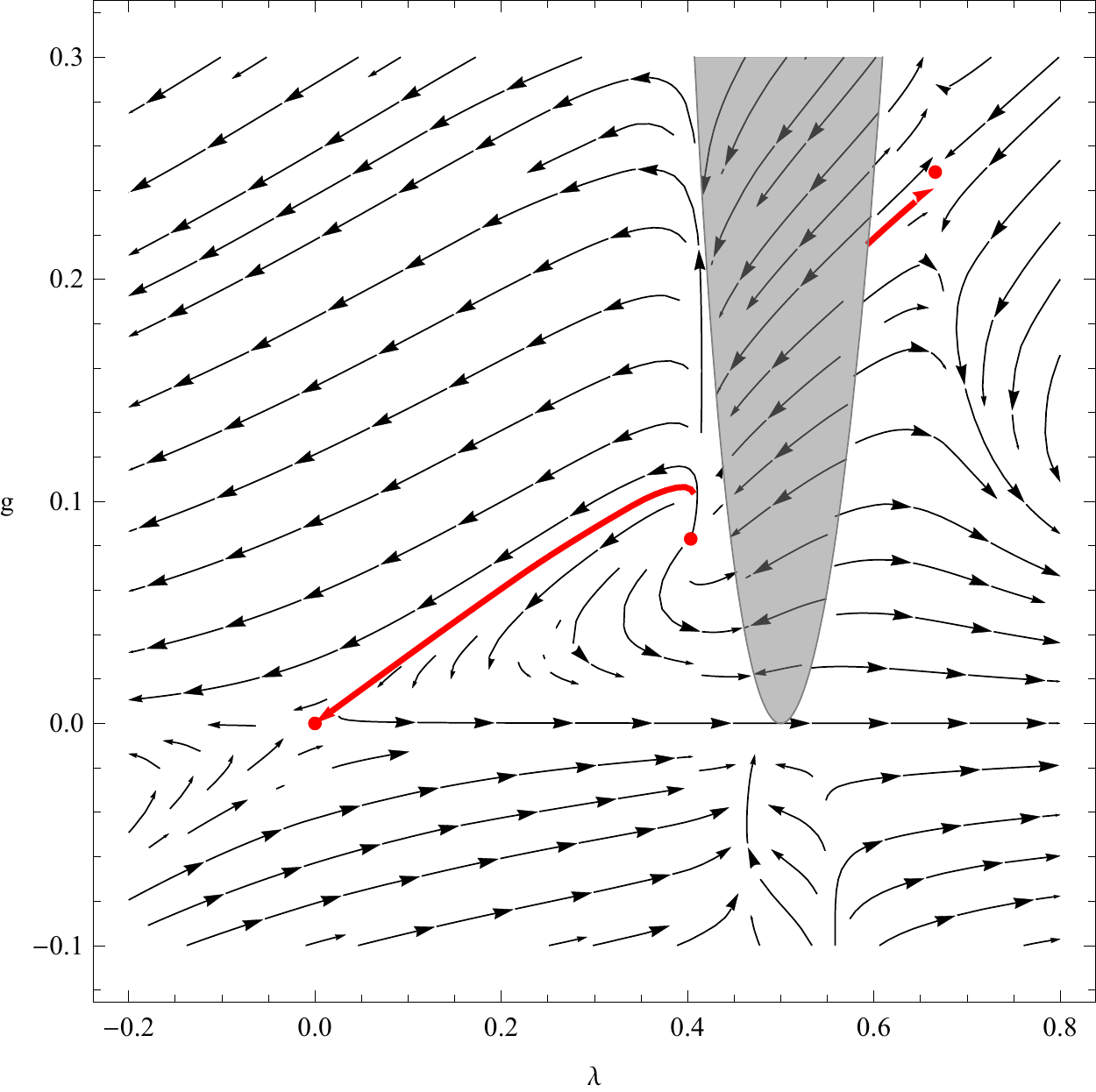}
		\label{fig:PDneg1100v2}}
	\subfigure[\textbf{NGFP} for $µ^2 = \frac{1}{100}$.]{
		\centering
		\includegraphics[width=0.3\textwidth]{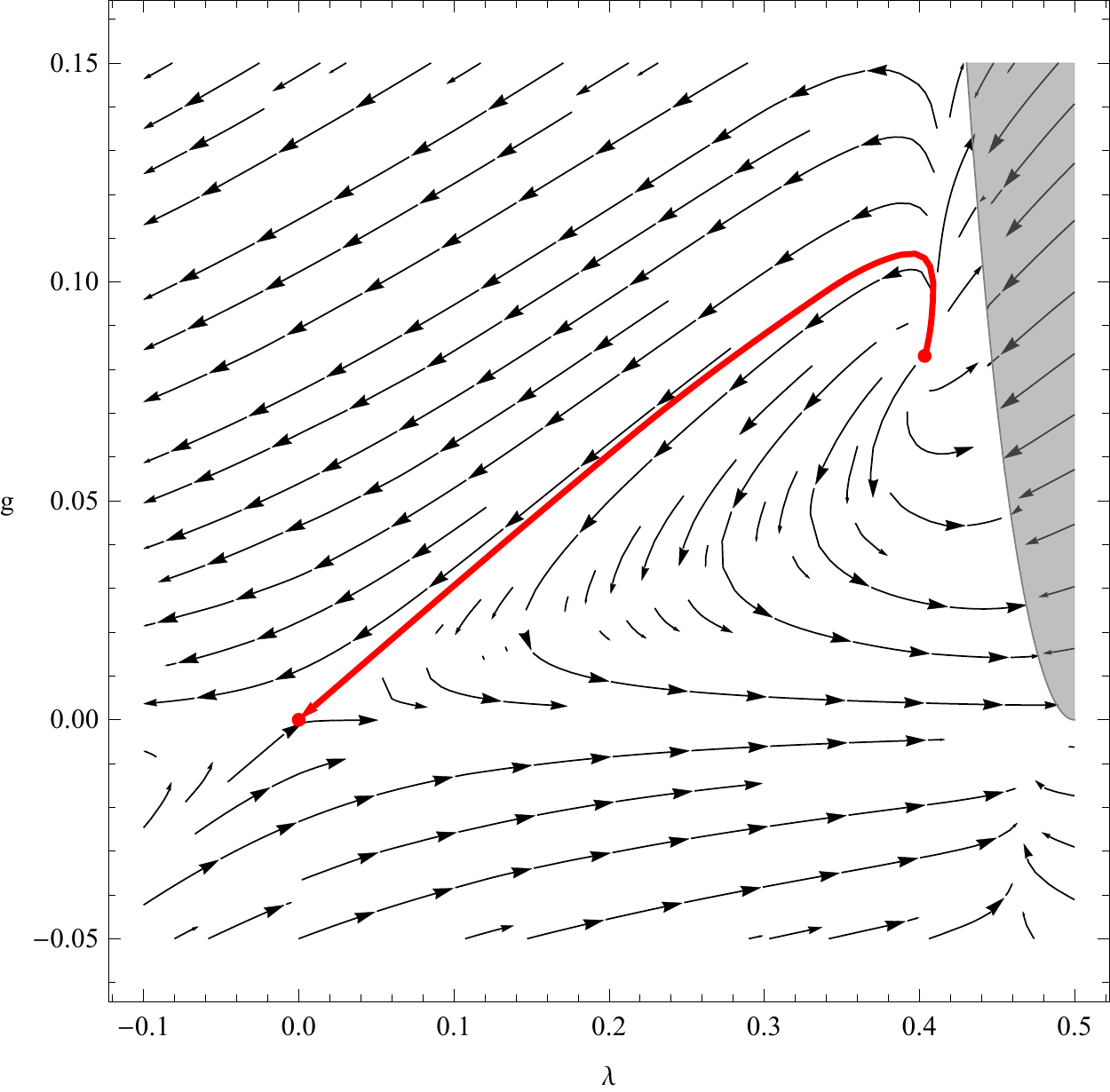}
		\label{fig:PDneg1100}}
	\subfigure[\textbf{NGFP}$\bm{^{\oplus}}$ for $µ^2 = \frac{1}{10}$.]{
		\centering
		\includegraphics[width=0.3\textwidth]{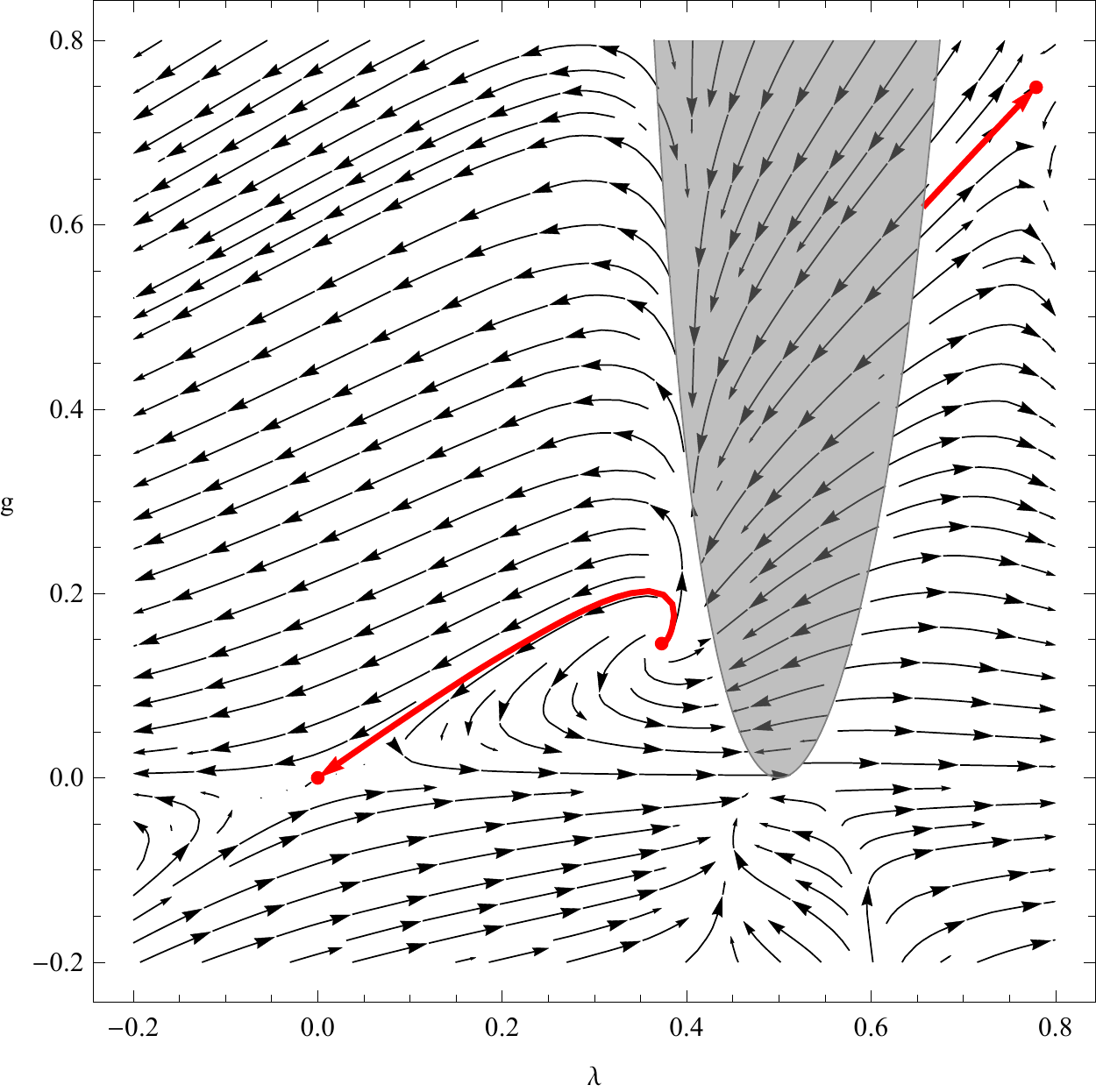}
		\label{fig:PDneg110v2}}
	\subfigure[\textbf{NGFP} for $µ^2 = \frac{1}{10}$.]{
		\centering
		\includegraphics[width=0.3\textwidth]{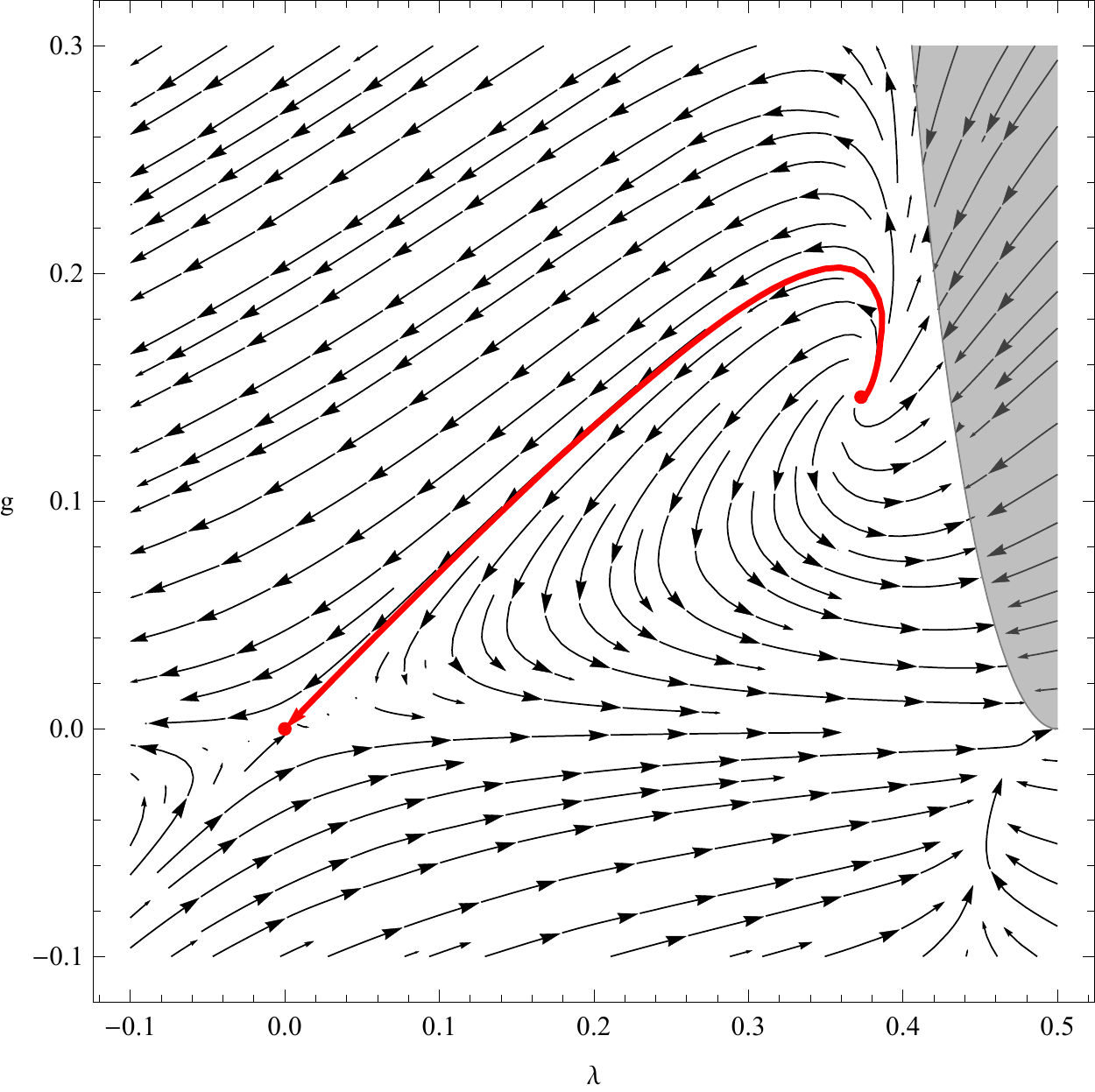}
		\label{fig:PDneg110}}
	\subfigure[\textbf{NGFP}$\bm{^{\oplus}}$ for $µ^2 = \frac{1}{2}$.]{
		\centering
		\includegraphics[width=0.3\textwidth]{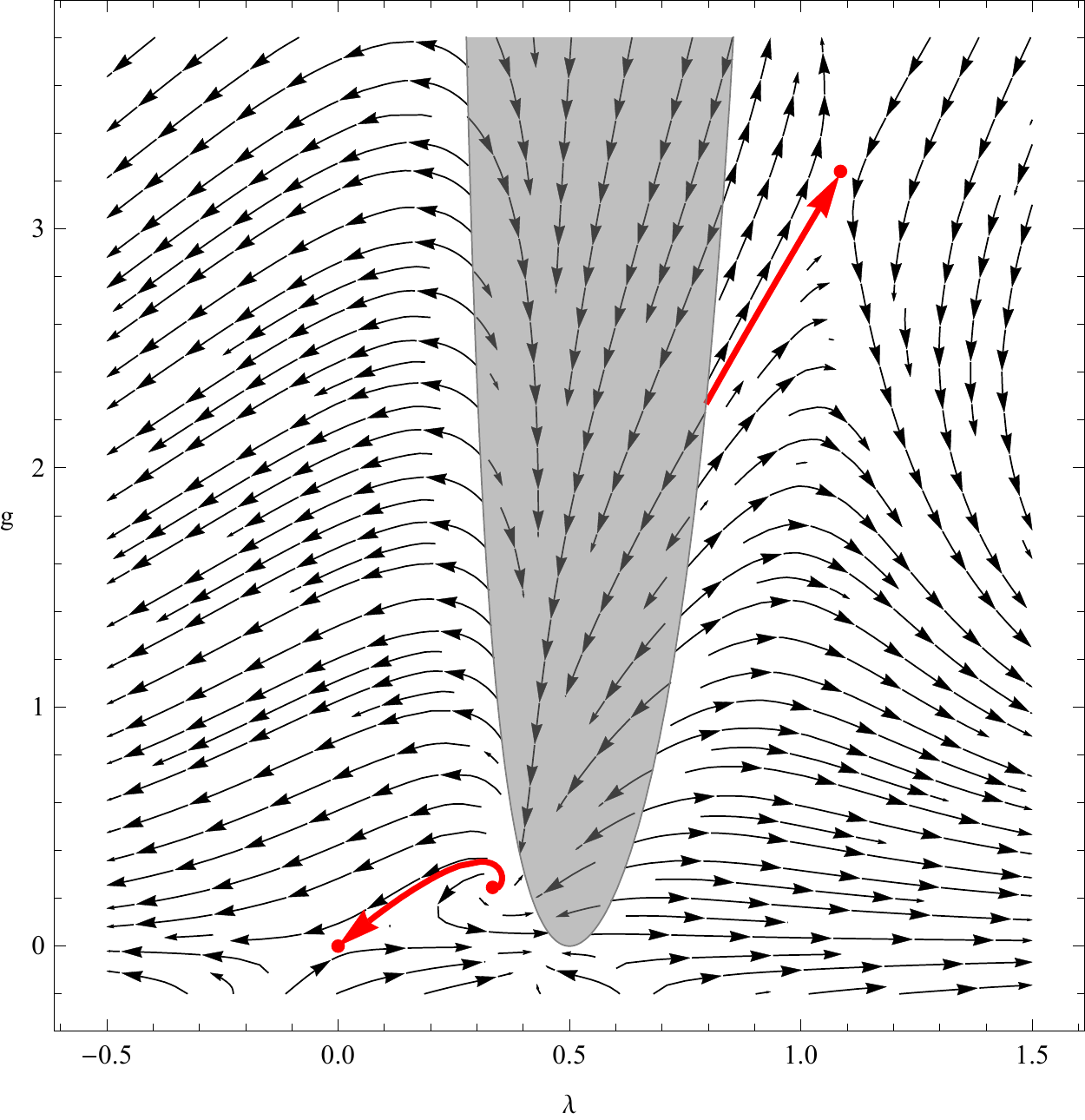}
		\label{fig:PDneg12v2}}
	\subfigure[\textbf{NGFP} for $µ^2 = \frac{1}{2}$.]{
		\centering
		\includegraphics[width=0.3\textwidth]{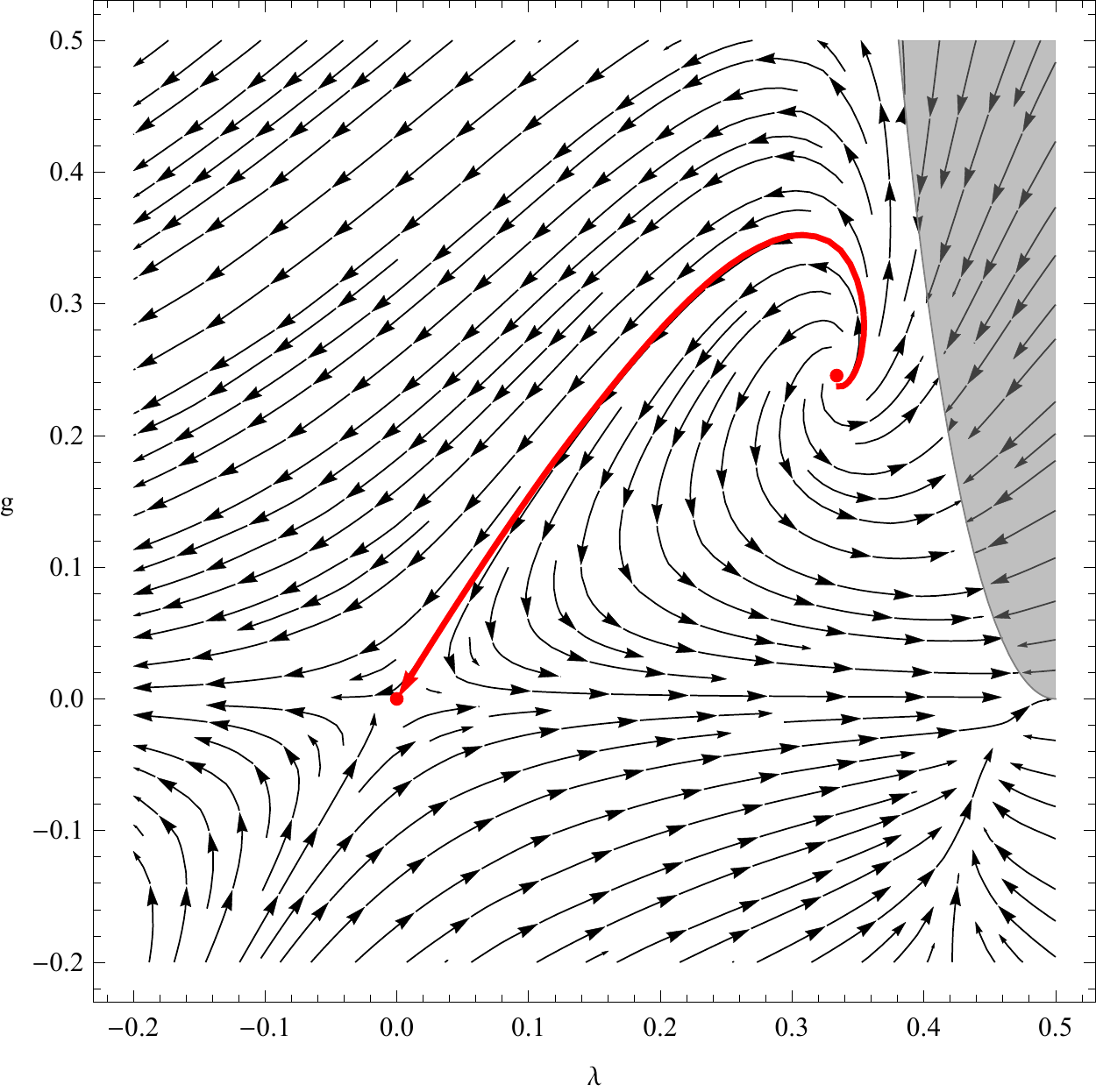}
		\label{fig:PDneg12}}
	\subfigure[\textbf{NGFP}$\bm{^{\oplus}}$ for $µ^2 = 1$.]{
		\centering
		\includegraphics[width=0.3\textwidth]{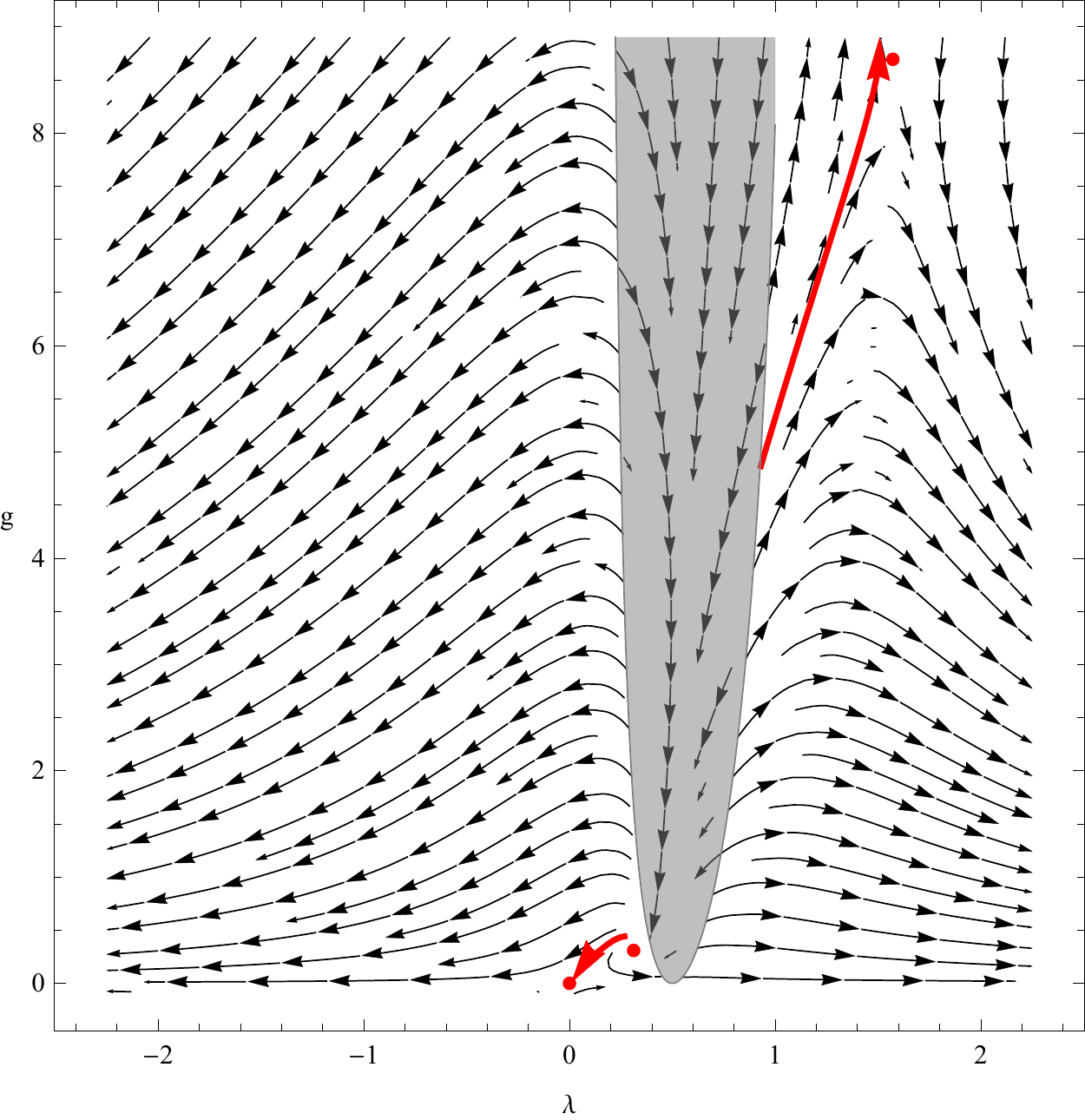}
		\label{fig:PDneg1v2}}
	\subfigure[\textbf{NGFP} for $µ^2 = 1$.]{
		\centering
		\includegraphics[width=0.3\textwidth]{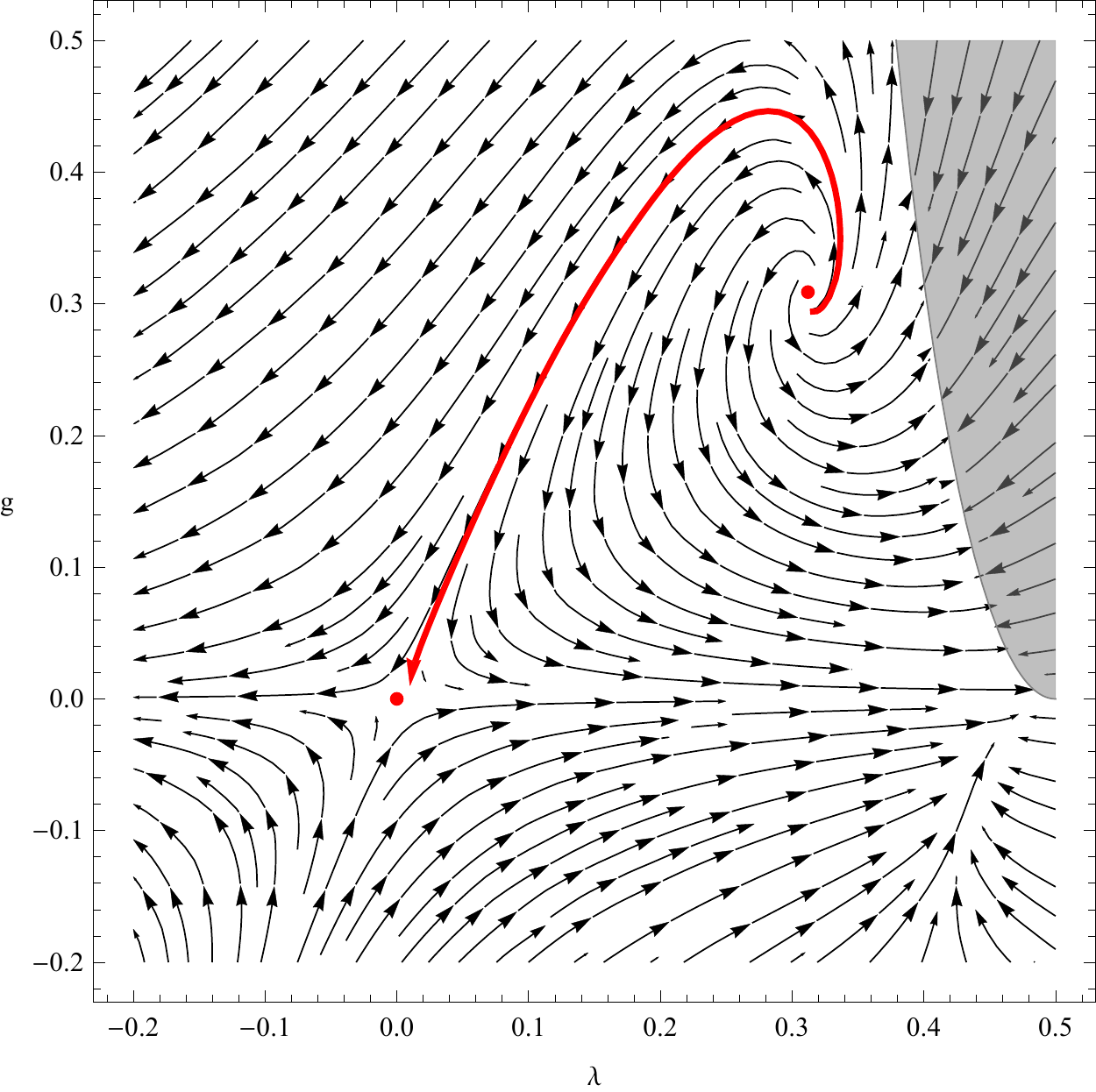}
		\label{fig:PDneg1}}
	\subfigure[\textbf{NGFP} for $µ^2 = 2$.]{
		\centering
		\includegraphics[width=0.3\textwidth]{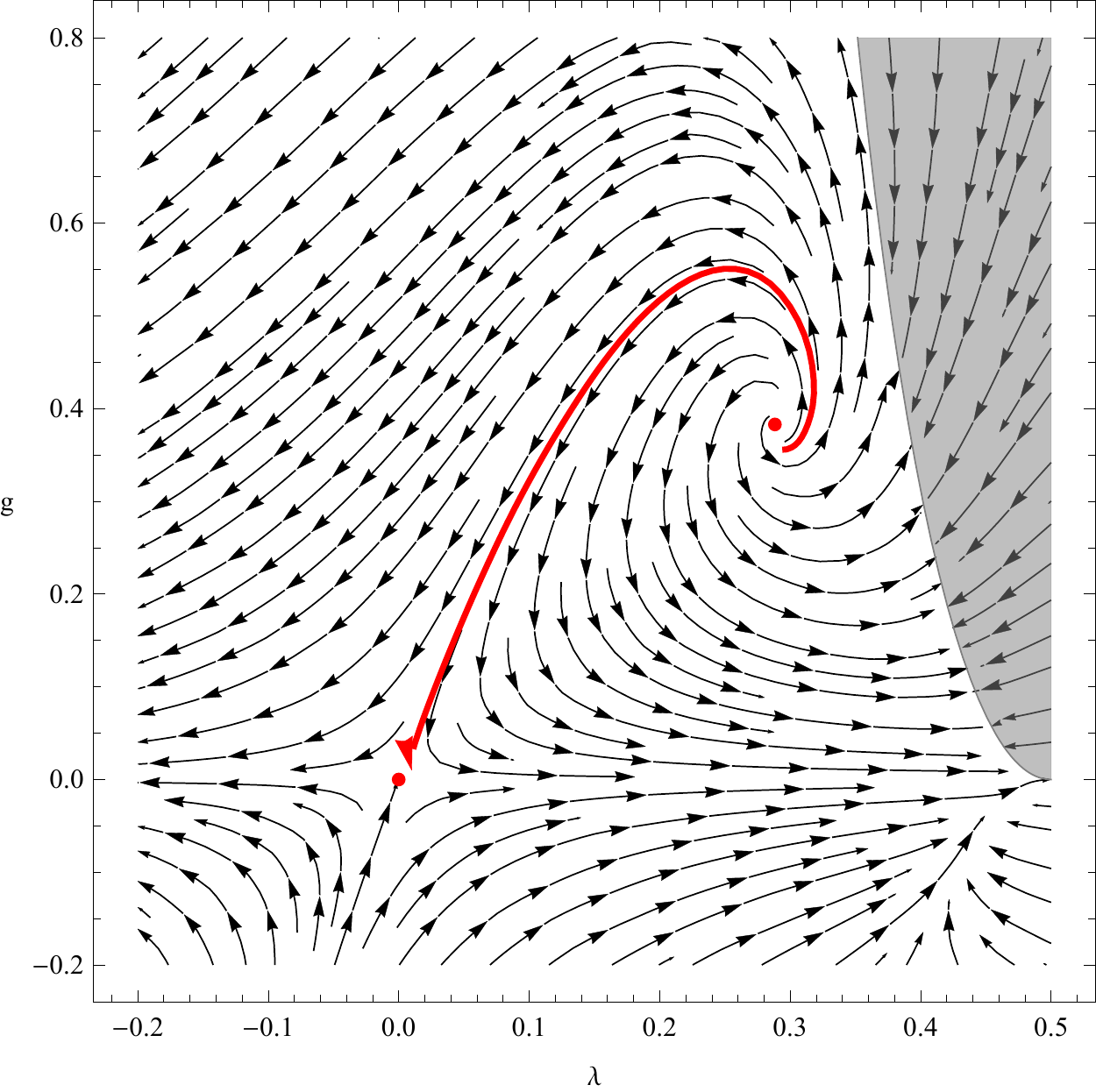}
		\label{fig:PDneg2}}
	\caption{RG phase portrait for different values of the squared mass parameter $µ^2$. Choosing the normalization factor according to the ``$\mathcal{Z}_k = \zeta_k$-rule'' leads to the emergence of two fixed points, \textbf{NGFP} and \textbf{NGFP}$\bm{^{\oplus}}$, the latter of which is always located ``behind'' the singularity. The previously found running of the fixed point coordinates, namely the change of sign for the cosmological constant, no longer occurs and the \textbf{NGFP} is always located in the first quadrant.}
	\label{fig:NGFPneg}
\end{figure}
\clearpage

\begin{figure}[htbp]
	\centering
	\subfigure[\textbf{NGFP} for $µ^2 = 3$.]{
		\centering
		\includegraphics[width=0.3\textwidth]{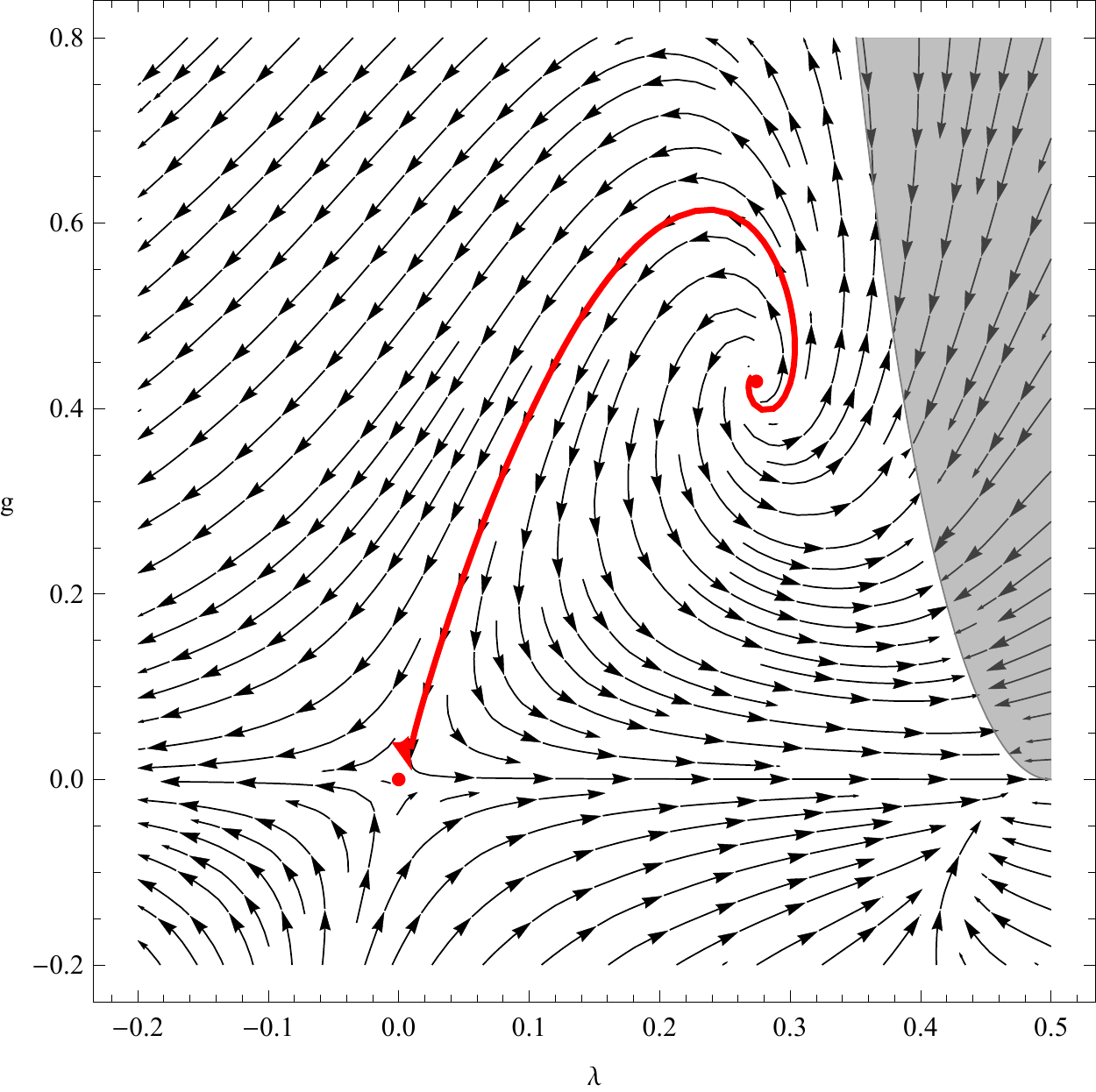}
		\label{fig:PDneg3}}
	\subfigure[\textbf{NGFP} for $µ^2 = 4$.]{
		\centering
		\includegraphics[width=0.3\textwidth]{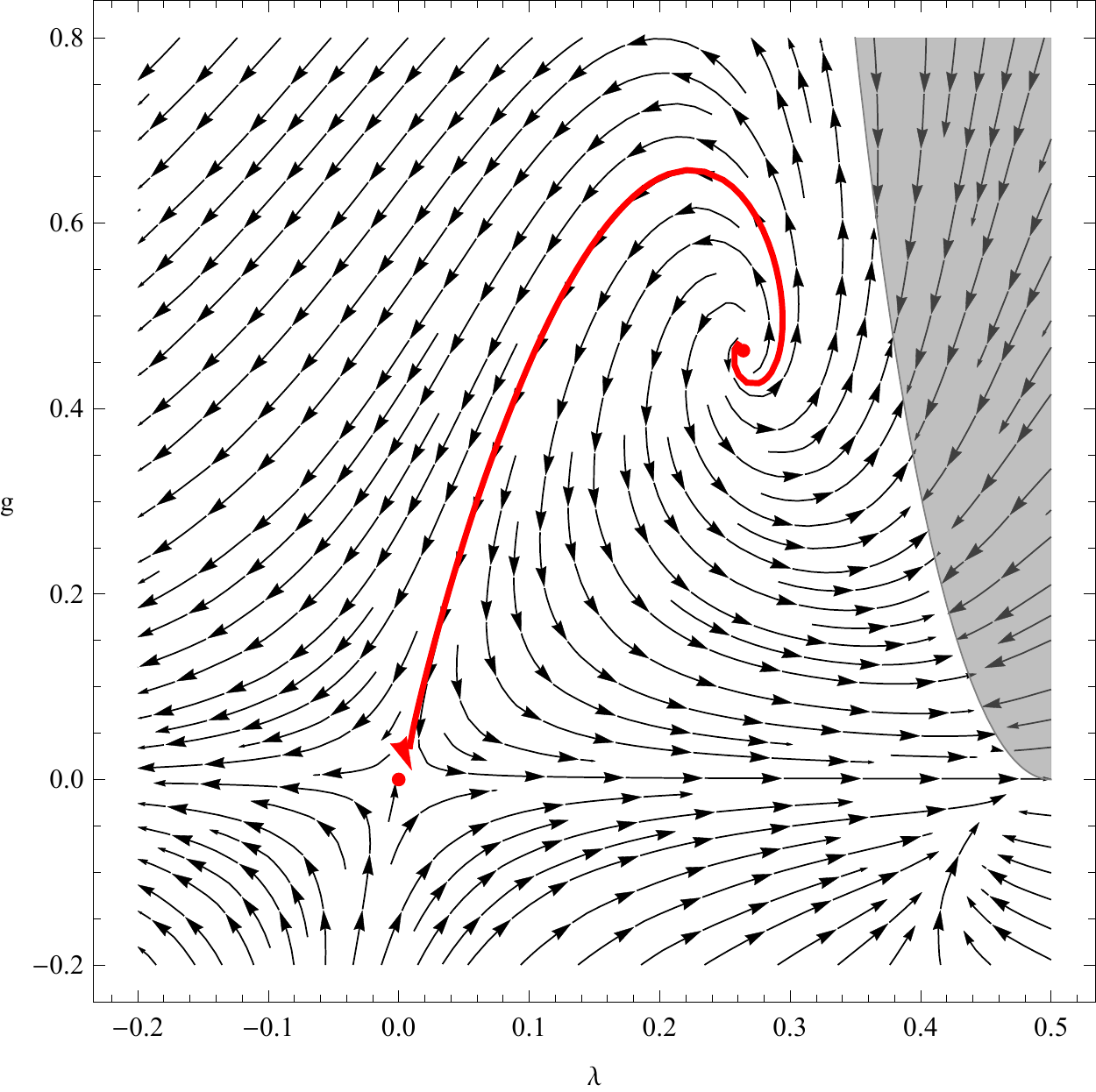}
		\label{fig:PDneg4}}
	\subfigure[\textbf{NGFP} for $µ^2 = 5$.]{
		\centering
		\includegraphics[width=0.3\textwidth]{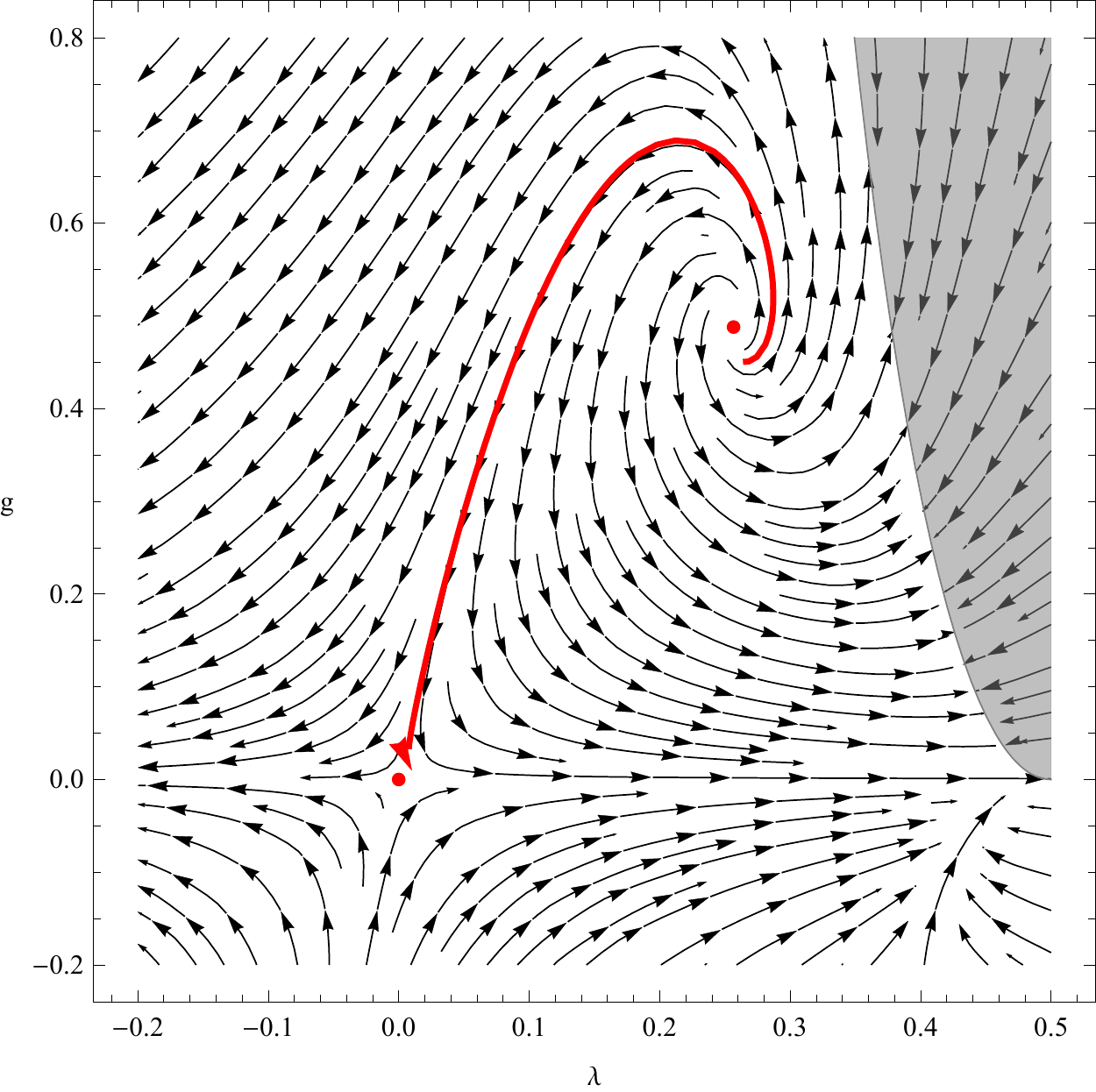}
		\label{fig:PDneg5}}
	\subfigure[\textbf{NGFP} for $µ^2 = 6$.]{
		\centering
		\includegraphics[width=0.3\textwidth]{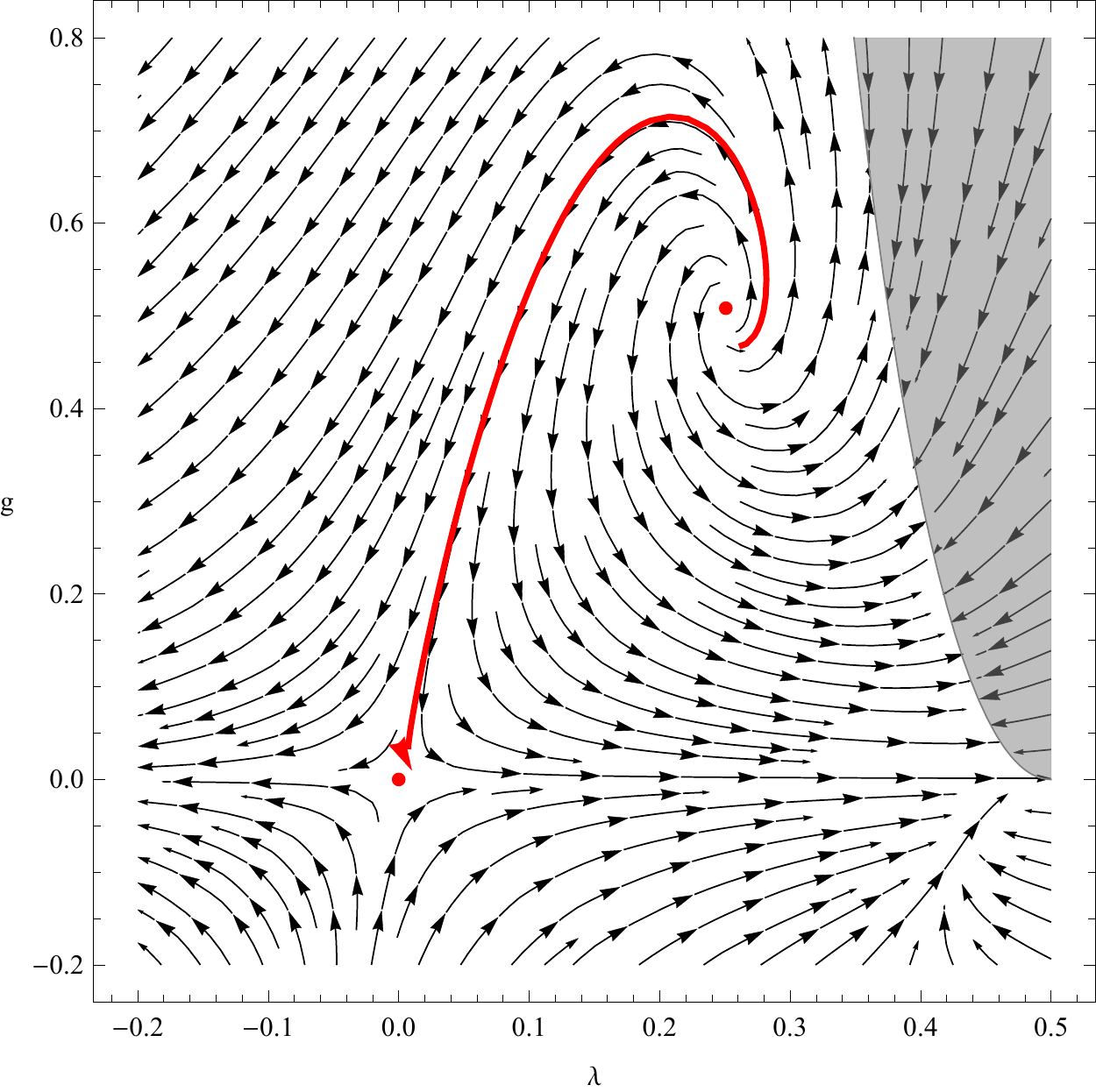}
		\label{fig:PDneg6}}
	\subfigure[\textbf{NGFP} for $µ^2 = 7$.]{
		\centering
		\includegraphics[width=0.3\textwidth]{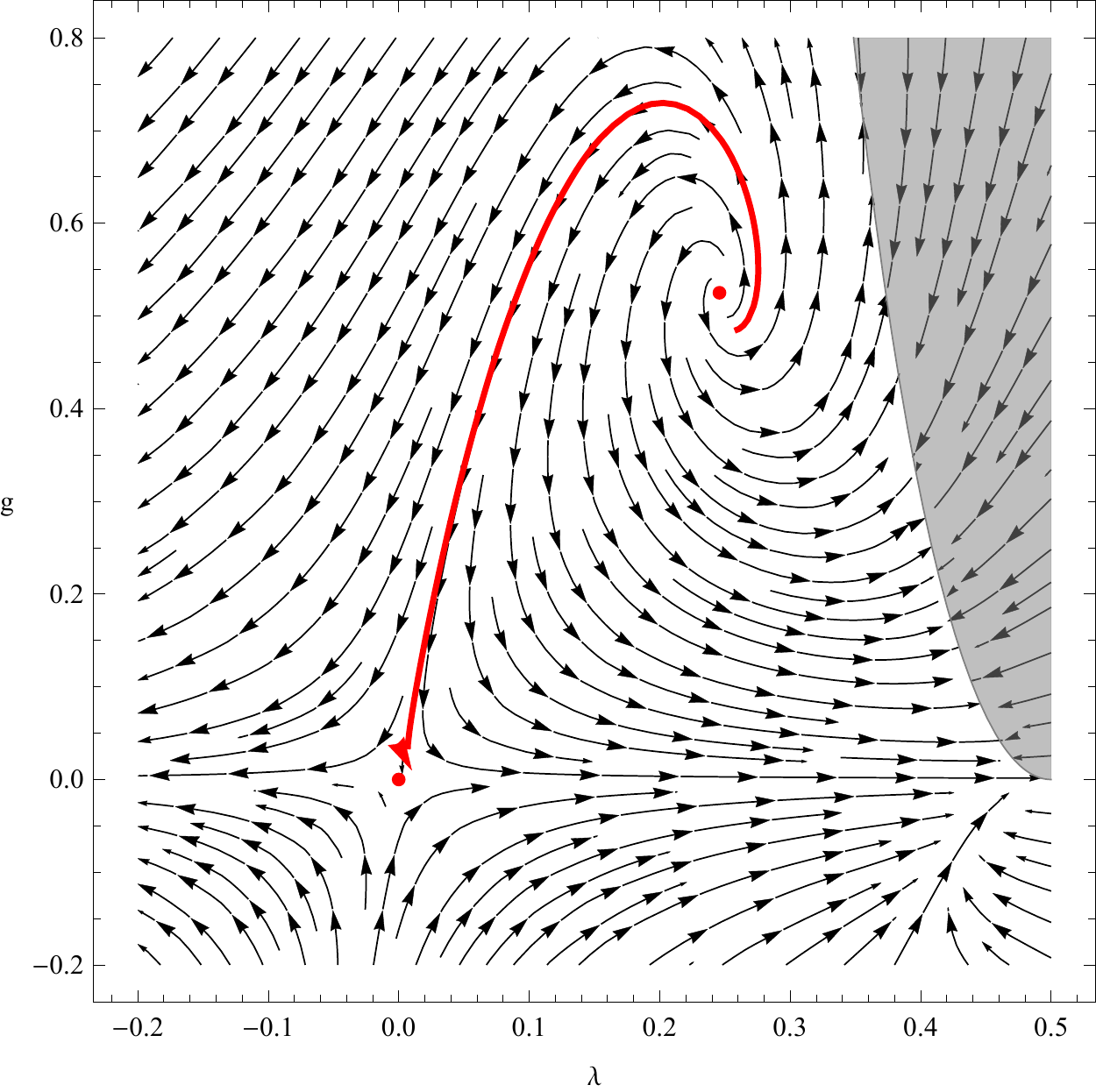}
		\label{fig:PDneg7}}
	\subfigure[\textbf{NGFP} for $µ^2 = 8$.]{
		\centering
		\includegraphics[width=0.3\textwidth]{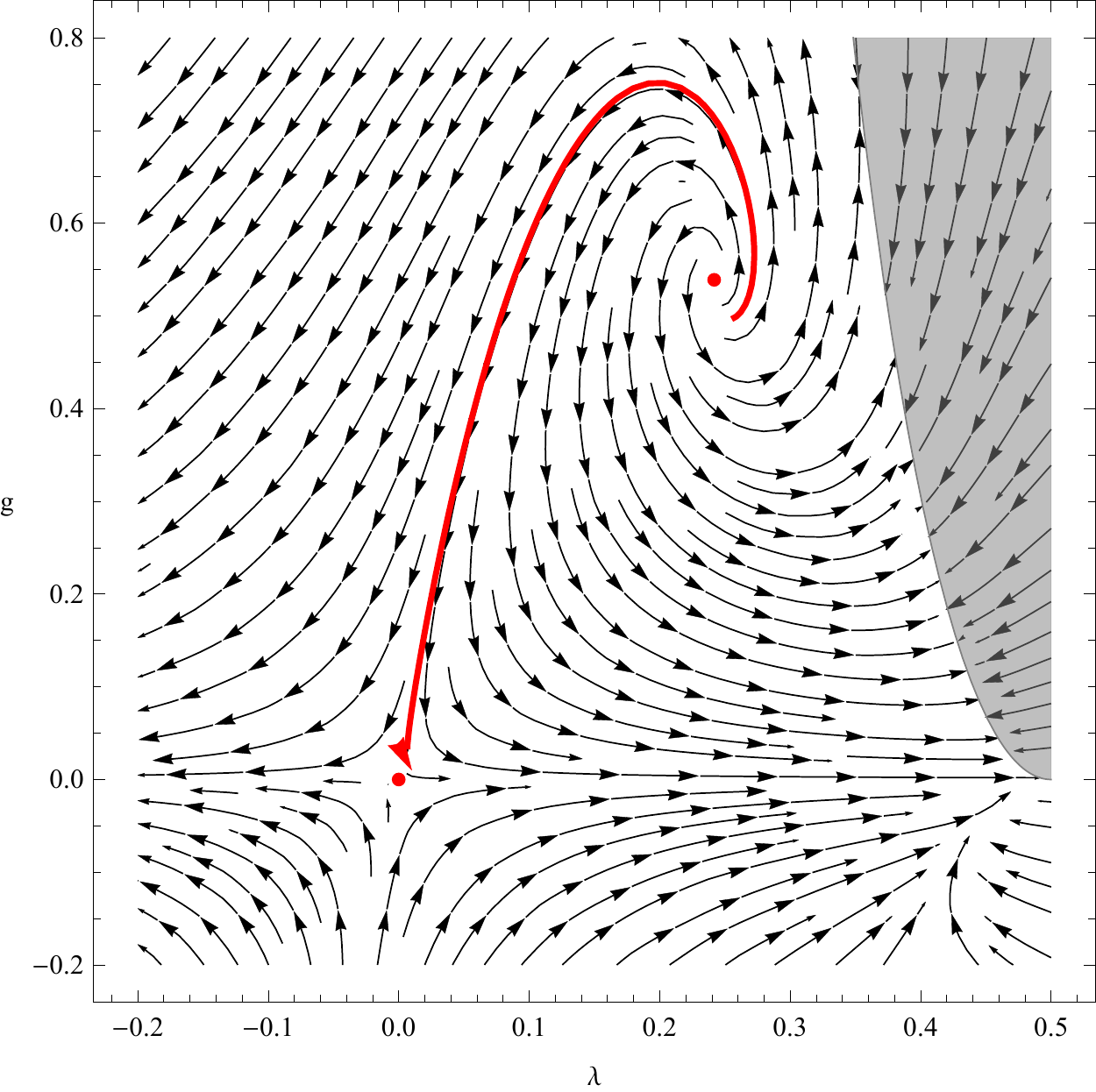}
		\label{fig:PDneg8}}
	\subfigure[\textbf{NGFP} for $µ^2 = 9$.]{
		\centering
		\includegraphics[width=0.3\textwidth]{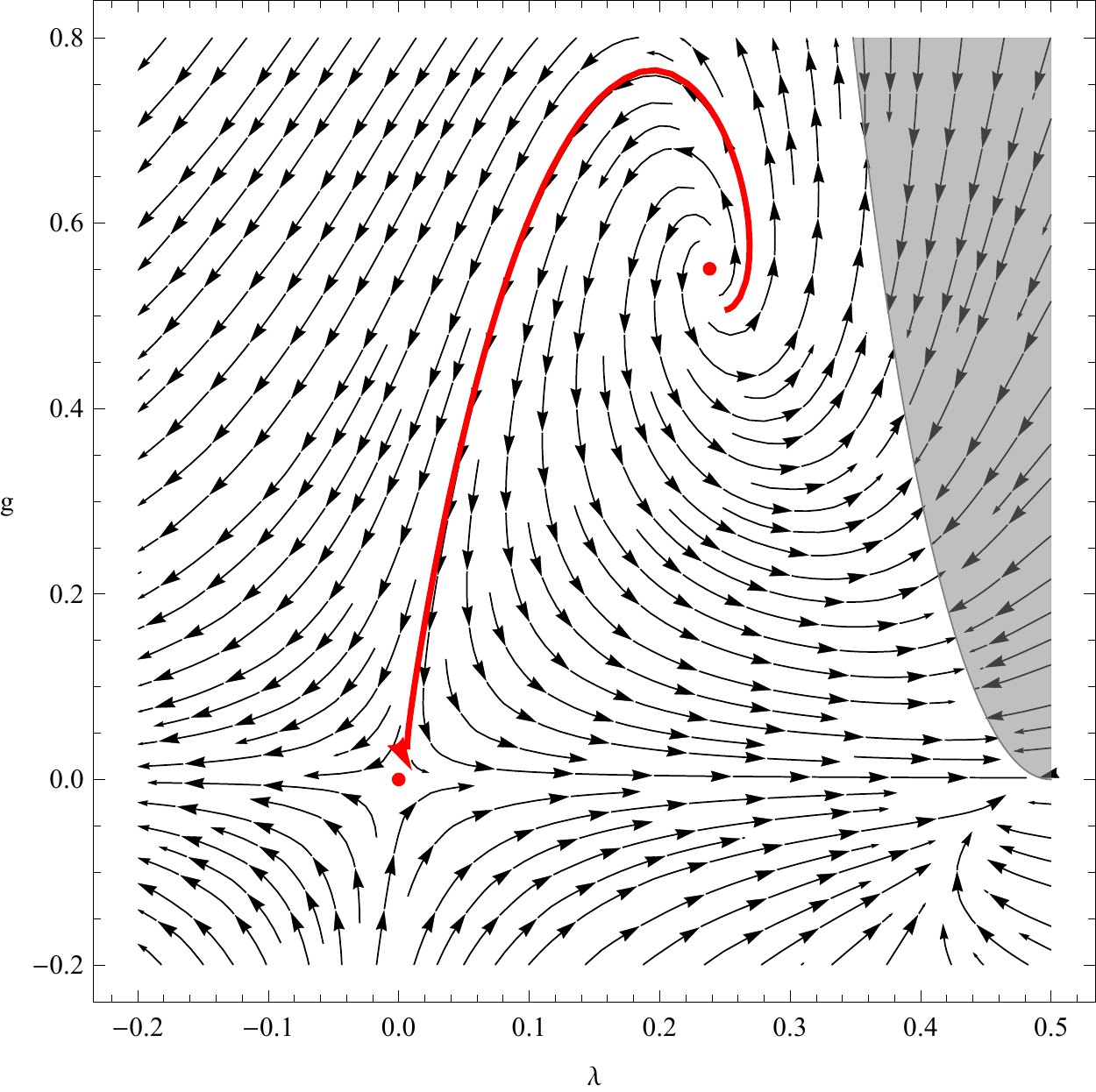}
		\label{fig:PDneg9}}
	\subfigure[\textbf{NGFP} for $µ^2 = 10$.]{
		\centering
		\includegraphics[width=0.3\textwidth]{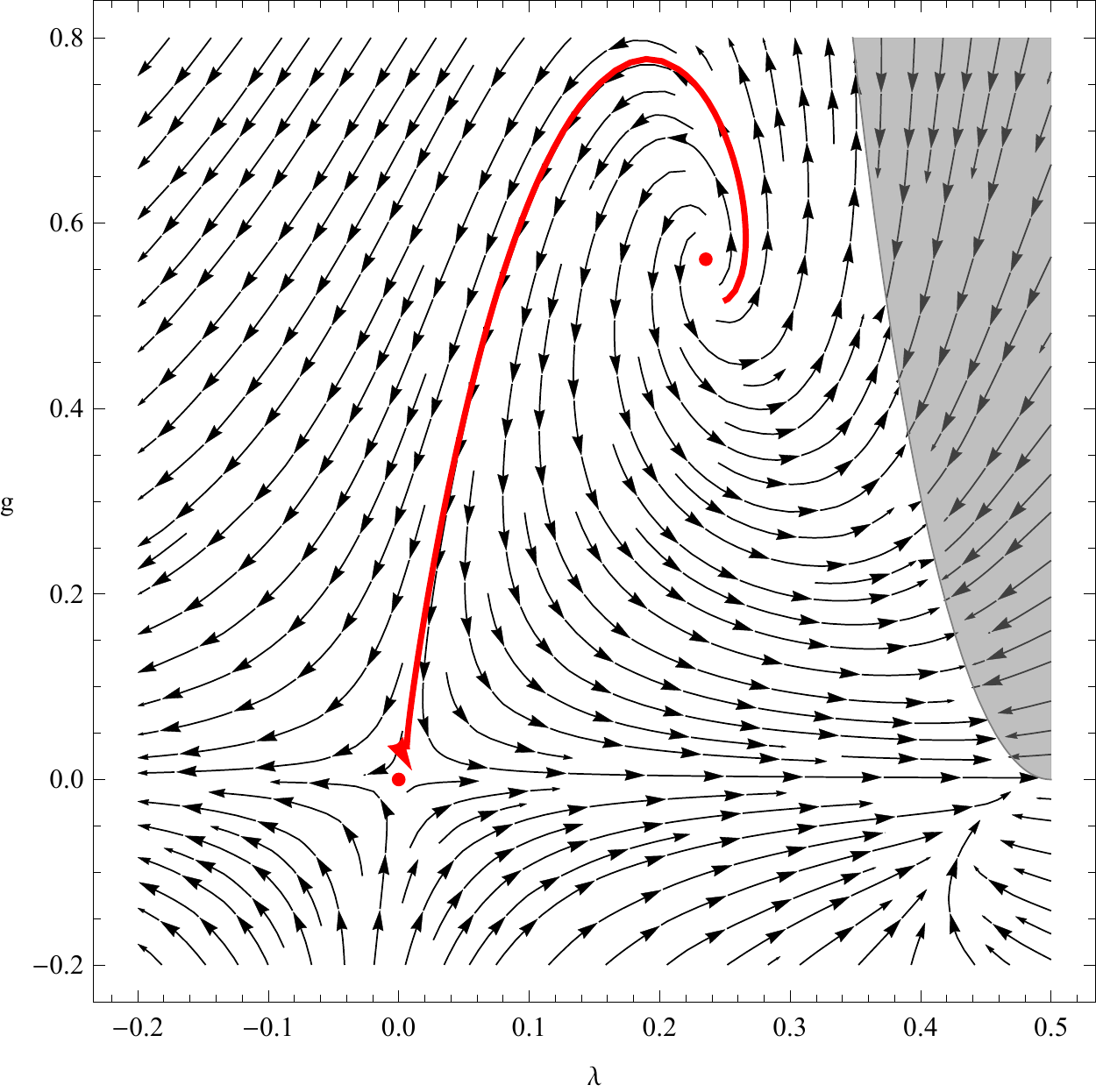}
		\label{fig:PDneg10}}
	\subfigure[\textbf{NGFP} for $µ^2 = 100$.]{
		\centering
		\includegraphics[width=0.3\textwidth]{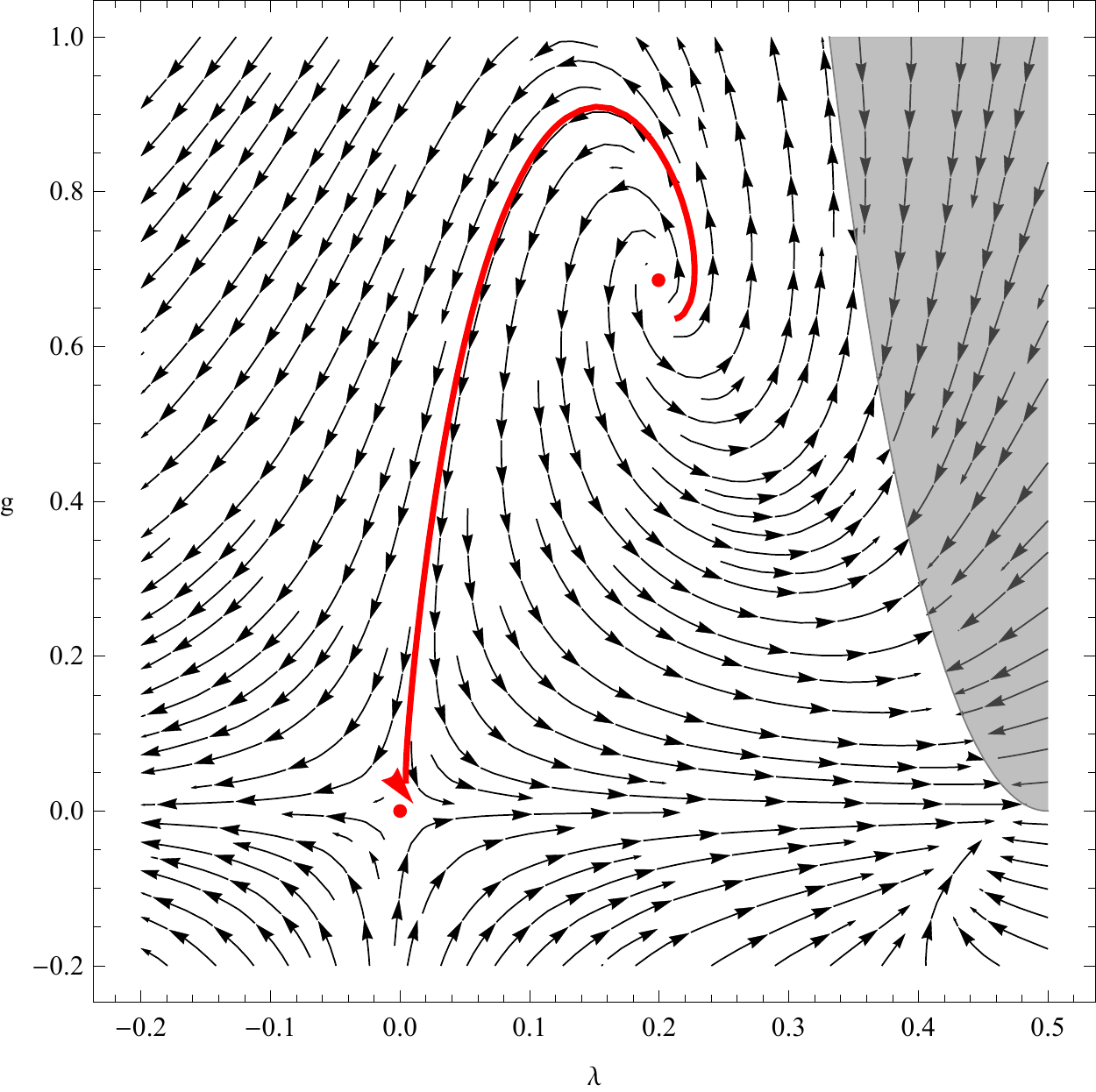}
		\label{fig:PDneg100}}
	\caption{RG phase portrait for different values of the squared mass parameter $µ^2$. In the same manner as before the dependence on the mass parameter weakens considerably for larger values of $µ$.}
	\label{fig:NGFPneg2}
\end{figure}
\clearpage

\begin{figure}[H]
	\centering
	\subfigure[\textbf{NGFP}$\bm{^{\ominus}}$ for $µ^2 = \frac{1}{100}$.]{
		\centering
		\includegraphics[width=0.3\textwidth]{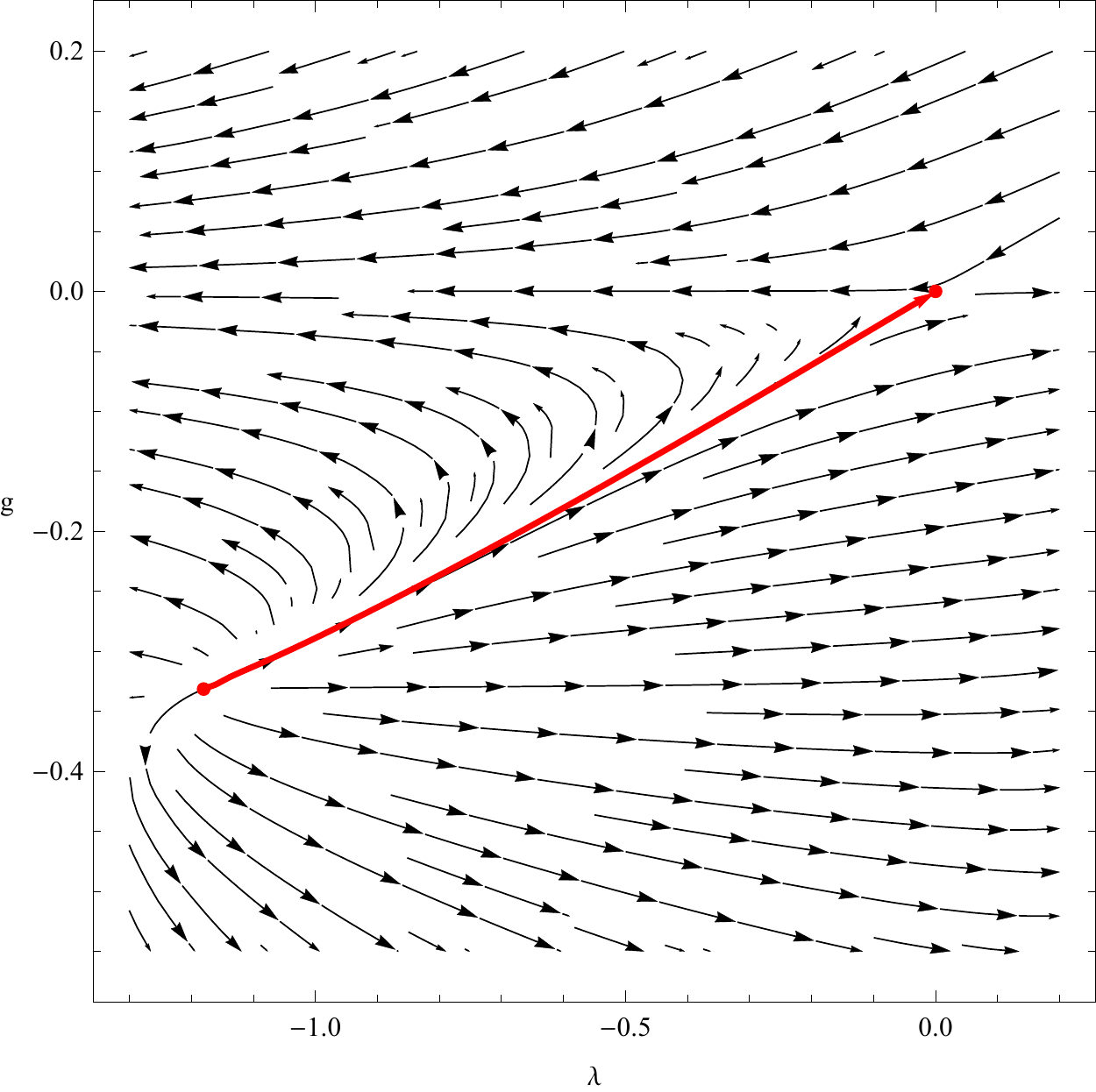}
		\label{fig:PDneg1100v3}}
	\subfigure[\textbf{NGFP}$\bm{^{\ominus}}$ for $µ^2 = \frac{1}{2}$.]{
		\centering
		\includegraphics[width=0.3\textwidth]{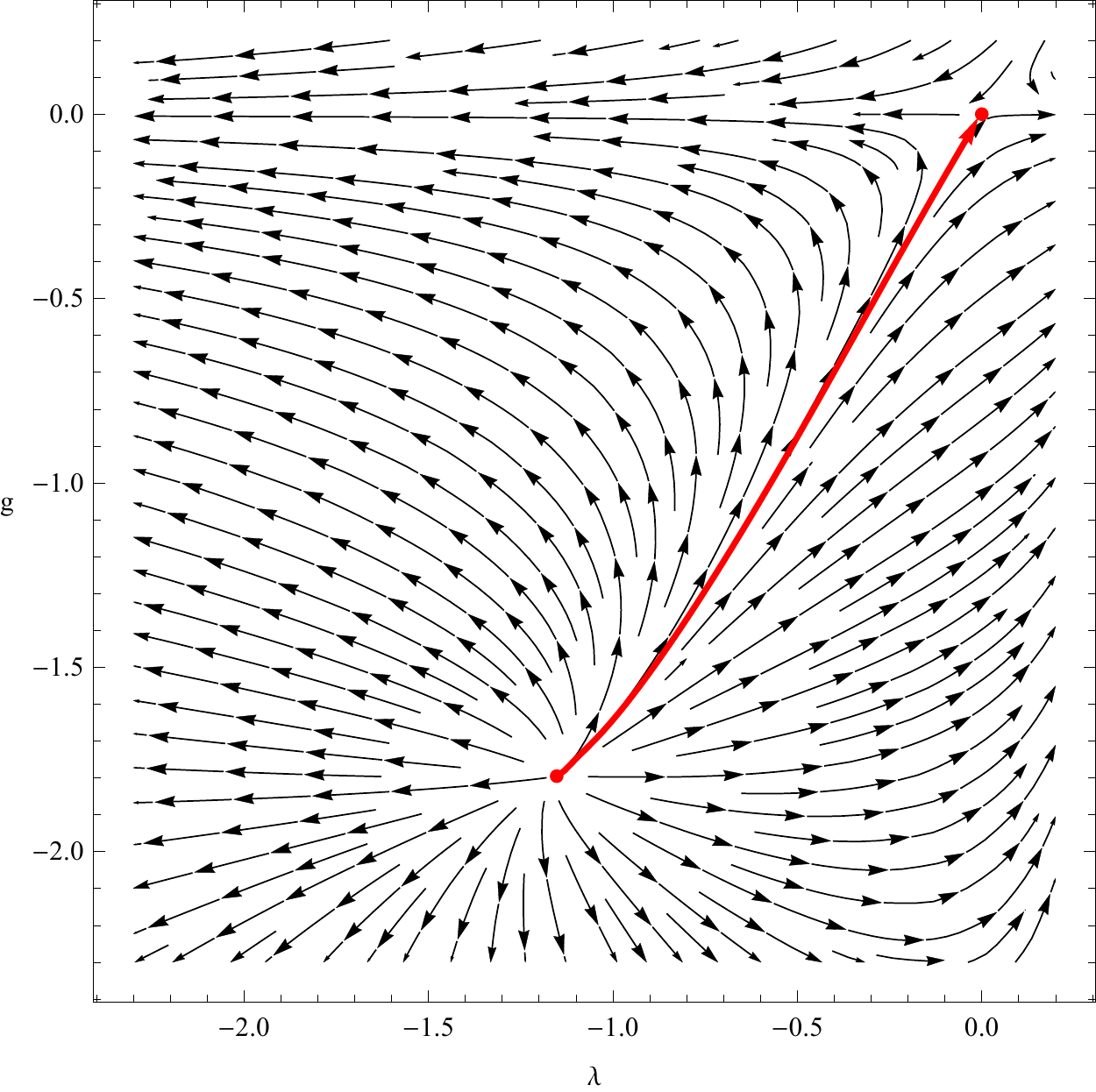}
		\label{fig:PDneg12v3}}
	\subfigure[\textbf{NGFP}$\bm{^{\ominus}}$ for $µ^2 = 1$.]{
		\centering
		\includegraphics[width=0.3\textwidth]{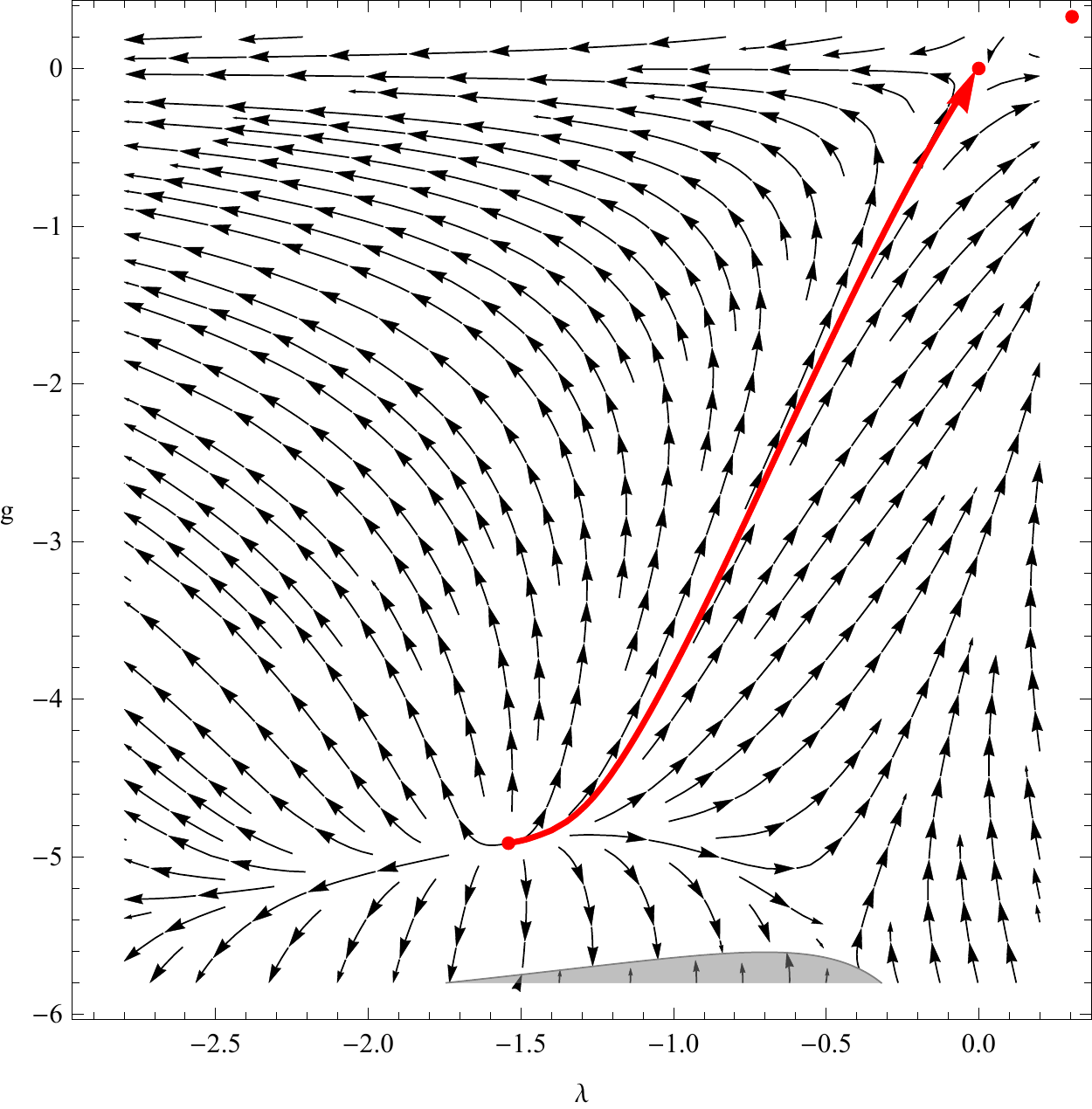}
		\label{fig:PDneg1v4}}
	\subfigure[\textbf{NGFP}$\bm{^{\ominus}}$ for $µ^2 = 1.44$.]{
		\centering
		\includegraphics[width=0.3\textwidth]{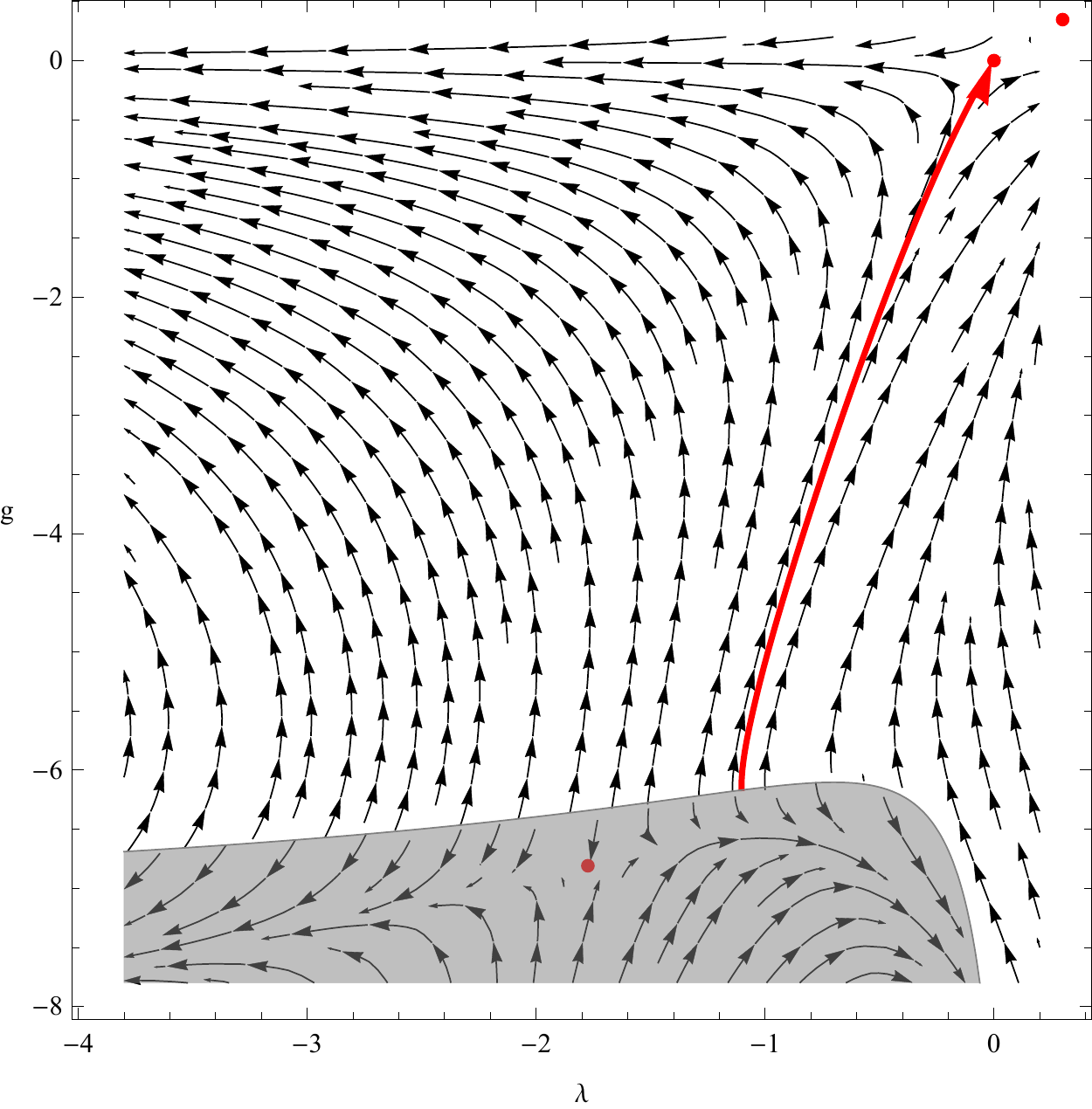}
		\label{fig:PDneg1v5}}
	\caption{Phase portrait depicting the third non-Gaussian fixed point \textbf{NGFP}$\bm{^{\ominus}}$ for different values of $µ^2$. For $µ^2 \lesssim 1.44$ this fixed point might define a mathematically complete field theory, but most likely not a physically relevant one.}
	\label{fig:NGFPnegMinus}
\end{figure}

\begin{figure}[htbp]
	\centering
	\subfigure[The \textbf{NGFP} in the ``allowed'' region $g > - \frac{1}{6}$.]{
		\centering
		\includegraphics[width=0.3\textwidth]{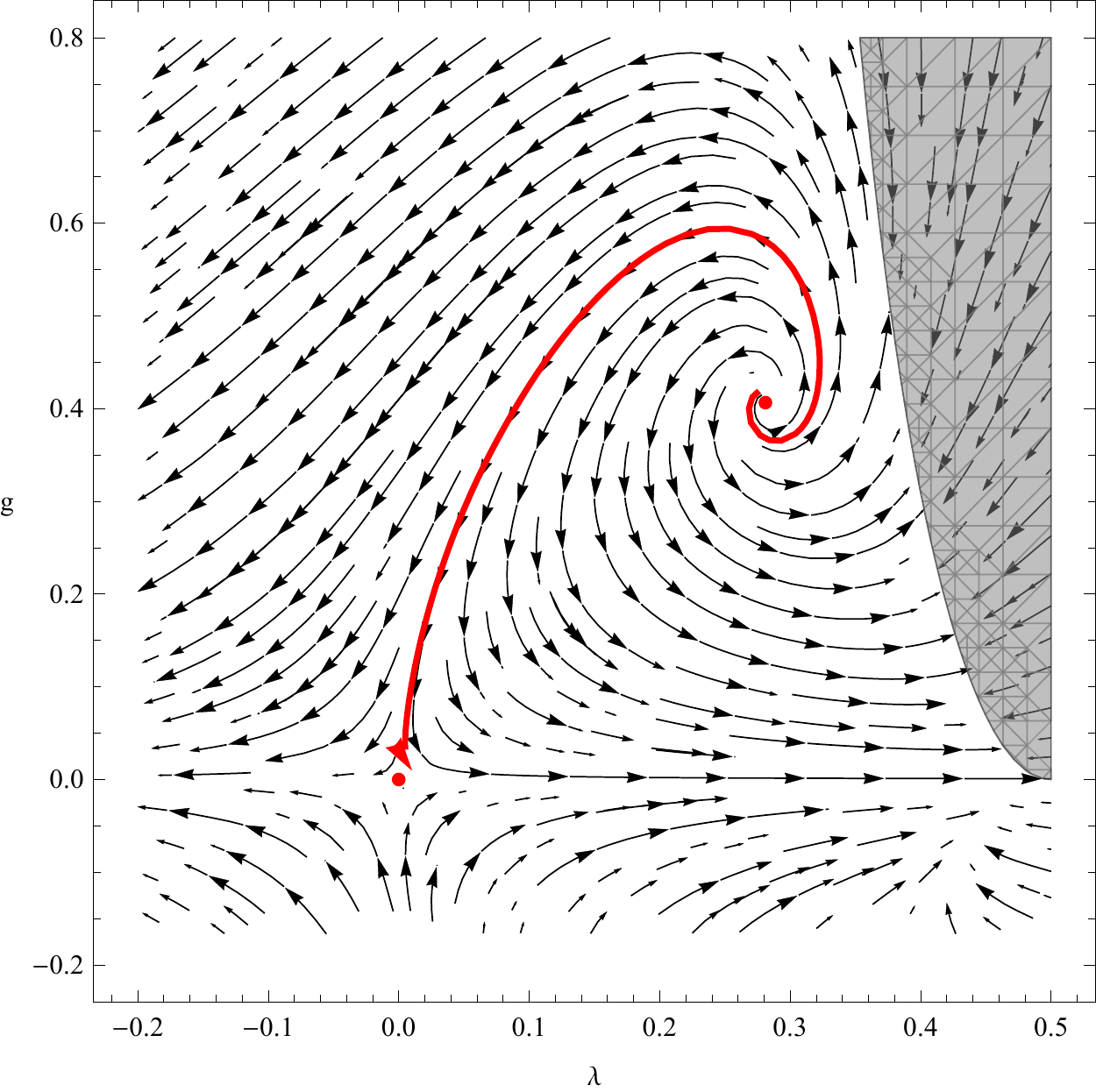}
		\label{fig:PDnegMpl}}
	\caption{Phase portrait for the identification $µ \equiv \frac{1}{\sqrt{g_k}}$ in the case of the ``$\mathcal{Z}_k = \zeta_k$-rule'', i.e. $\xi = -1$.}
	\label{fig:NGFPnegMPl}
\end{figure}
\clearpage

%% file: Paper2014.bbl
\begin{thebibliography}{99}

\bibitem{Ashtekar:2004eh} 
  A.~Ashtekar and J.~Lewandowski,
  Background independent Quantum Gravity: A Status Report,
  Class.\ Quant.\ Grav.\  {\bf 21}, R53 (2004),
  [\href{http://arxiv.org/abs/gr-qc/0404018}{{\tt arXiv:gr-qc/0404018}}].

\bibitem{Ashtekar:2014kba} 
  A.~Ashtekar, M.~Reuter and C.~Rovelli,
  From General Relativity to Quantum Gravity,
  in: A. Ashtekar, B. Berger, J. Isenberg and M. MacCallum (Eds.), {\it General Relativity and Gravitation: A Centennial Perspective},
  Cambridge University Press, Cambridge (2015),
  [\href{http://arxiv.org/abs/1408.4336}{{\tt arXiv:1408.4336}}].

\bibitem{Baekler:2011jt} 
  P.~Baekler and F.~W.~Hehl,
  Beyond Einstein-Cartan gravity: Quadratic torsion and curvature invariants with even and odd parity including all boundary terms,
  Class.\ Quant.\ Grav.\  {\bf 28}, 215017 (2011),
  [\href{http://arxiv.org/abs/1105.3504}{{\tt arXiv:1105.3504}}].

\bibitem{Becker:2012js} 
  D.~Becker and M.~Reuter,
  Running boundary actions, Asymptotic Safety, and black hole thermodynamics,
  JHEP {\bf 1207}, 172 (2012),
  [\href{http://arxiv.org/abs/1205.3583}{{\tt arXiv:1205.3583}}].

\bibitem{Becker:2014qya} 
  D.~Becker and M.~Reuter,
  En route to Background Independence: Broken split-symmetry, and how to restore it with bi-metric average actions,
  Annals Phys.\  {\bf 350}, 225 (2014),
  [\href{http://arxiv.org/abs/1404.4537}{{\tt arXiv:1404.4537}}].

\bibitem{Benedetti:2011nd} 
  D.~Benedetti and S.~Speziale,
  Perturbative quantum gravity with the Immirzi parameter,
  JHEP {\bf 1106}, 107 (2011),
  [\href{http://arxiv.org/abs/1104.4028}{{\tt arXiv:1104.4028}}].

\bibitem{Benedetti:2011yb} 
  D.~Benedetti and S.~Speziale,
  Perturbative running of the Immirzi parameter,
  J.\ Phys.\ Conf.\ Ser.\  {\bf 360}, 012011 (2012),
  [\href{http://arxiv.org/abs/1111.0884}{{\tt arXiv:1111.0884}}].

\bibitem{Capozziello:2001mq} 
  S.~Capozziello, G.~Lambiase and C.~Stornaiolo,
  Geometric classification of the torsion tensor in space-time,
  Annalen Phys.\  {\bf 10}, 713 (2001),
  [\href{http://arxiv.org/abs/gr-qc/0101038}{{\tt arXiv:gr-qc/0101038}}].

\bibitem{Daum:2010phd}
	J.~E.~Daum,
	Konstruktion und Analyse einer funktionalen Renormierungsgruppengleichung für Gravitation im Einstein-Cartan-Zugang,
	PhD thesis, Johannes Gutenberg-Universität in Mainz (2010),
	[online: \href{http://ubm.opus.hbz-nrw.de/volltexte/2011/2773/}{2011/2773/}].

\bibitem{Daum:2011bs} 
  J.~E.~Daum and M.~Reuter,
  Running Immirzi Parameter and Asymptotic Safety,
  PoS CNCFG {\bf 2010}, 003 (2010),
  [\href{http://arxiv.org/abs/1111.1000}{{\tt arXiv:1111.1000}}].

\bibitem{Daum:2010qt} 
  J.-E.~Daum and M.~Reuter,
  Renormalization Group Flow of the Holst Action,
  Phys.\ Lett.\ B {\bf 710}, 215 (2012),
  [\href{http://arxiv.org/abs/1012.4280}{{\tt arXiv:1012.4280}}].

\bibitem{Daum:2013fu} 
  J.~E.~Daum and M.~Reuter,
  Einstein-Cartan gravity, Asymptotic Safety, and the running Immirzi parameter,
  Annals Phys.\  {\bf 334}, 351 (2013),
  [\href{http://arxiv.org/abs/1301.5135}{{\tt arXiv:1301.5135}}].

\bibitem{DeWitt:1965jb} 
  B.~S.~DeWitt,
  {\it Dynamical Theory of Groups and Fields},
  Gordon and Breach, New York (1965).

\bibitem{DeWitt:1967yk} 
  B.~S.~DeWitt,
  Quantum Theory of Gravity. 1. The Canonical Theory,
  Phys.\ Rev.\  {\bf 160}, 1113 (1967).

\bibitem{DeWitt:1967ub} 
  B.~S.~DeWitt,
  Quantum Theory of Gravity. 2. The Manifestly Covariant Theory,
  Phys.\ Rev.\  {\bf 162}, 1195 (1967).

\bibitem{DeWitt:1967uc} 
  B.~S.~DeWitt,
  Quantum Theory of Gravity. 3. Applications of the Covariant Theory,
  Phys.\ Rev.\  {\bf 162}, 1239 (1967).

\bibitem{DeWitt:1985bc} 
  B.~S.~DeWitt,
  The spacetime approach to quantum field theory,
  in: DeWitt, B.S. and Stora, R.  {\it Relativity, Groups and Topology II} (Les Houches, Session XL, 1983), p. 381-738, North Holland (1984).

\bibitem{Dona:2013qba} 
  P.~Donà, A.~Eichhorn and R.~Percacci,
  Matter matters in asymptotically safe quantum gravity,
  Phys.\ Rev.\ D {\bf 89}, no. 8, 084035 (2014),
  [\href{http://arxiv.org/abs/1311.2898}{{\tt arXiv:1311.2898}}].

\bibitem{Dou:1997fg} 
  D.~Dou and R.~Percacci,
  The running gravitational couplings,
  Class.\ Quant.\ Grav.\  {\bf 15}, 3449 (1998),
  [\href{http://arxiv.org/abs/hep-th/9707239}{{\tt arXiv:hep-th/9707239}}].

\bibitem{Harst:2012phd}
	U.~Harst,
	Investigations on Asymptotic Safety of Metric, Tetrad and Einstein-Cartan Gravity,
	PhD thesis, Johannes Gutenberg-Universität in Mainz (2012),
	[online: \href{http://ubm.opus.hbz-nrw.de/volltexte/2013/3358/}{2013/3358/}].

\bibitem{Harst:2012ni} 
  U.~Harst and M.~Reuter,
  The ``tetrad only'' theory space: Nonperturbative renormalization flow and Asymptotic Safety,
  JHEP {\bf 1205}, 005 (2012),
  [\href{http://arxiv.org/abs/1203.2158}{{\tt arXiv:1203.2158}}].

\bibitem{Harst:2014vca} 
  U.~Harst and M.~Reuter,
  A new functional flow equation for Einstein–Cartan quantum gravity,
  Annals Phys.\  {\bf 354}, 637 (2015),
  [\href{http://arxiv.org/abs/1410.7003}{{\tt arXiv:1410.7003}}].

\bibitem{Harst:TBA} 
  U.~Harst and M.~Reuter,
  in preparation.

\bibitem{Holst:1995pc} 
  S.~Holst,
  Barbero's Hamiltonian derived from a generalized Hilbert-Palatini action,
  Phys.\ Rev.\ D {\bf 53}, 5966 (1996),
  [\href{http://arxiv.org/abs/gr-qc/9511026}{{\tt arXiv:gr-qc/9511026}}].

\bibitem{Lauscher:2002sq} 
  O.~Lauscher and M.~Reuter,
  Flow equation of quantum Einstein gravity in a higher derivative truncation,
  Phys.\ Rev.\ D {\bf 66}, 025026 (2002),
  [\href{http://arxiv.org/abs/hep-th/0205062}{{\tt arXiv:hep-th/0205062}}].

\bibitem{Litim:2000ci} 
  D.~F.~Litim,
  Optimization of the exact renormalization group,
  Phys.\ Lett.\ B {\bf 486}, 92 (2000),
  [\href{http://arxiv.org/abs/hep-th/0005245}{{\tt arXiv:hep-th/0005245}}].

\bibitem{Manrique:2009uh} 
  E.~Manrique and M.~Reuter,
  Bimetric Truncations for Quantum Einstein Gravity and Asymptotic Safety,
  Annals Phys.\  {\bf 325}, 785 (2010),
  [\href{http://arxiv.org/abs/0907.2617}{{\tt arXiv:0907.2617}}].

\bibitem{Manrique:2010mq} 
  E.~Manrique, M.~Reuter and F.~Saueressig,
  Matter Induced Bimetric Actions for Gravity,
  Annals Phys.\  {\bf 326}, 440 (2011),
  [\href{http://arxiv.org/abs/1003.5129}{{\tt arXiv:1003.5129}}].

\bibitem{Manrique:2010am} 
  E.~Manrique, M.~Reuter and F.~Saueressig,
  Bimetric Renormalization Group Flows in Quantum Einstein Gravity,
  Annals Phys.\  {\bf 326}, 463 (2011),
  [\href{http://arxiv.org/abs/1006.0099}{{\tt arXiv:1006.0099}}].

\bibitem{Niedermaier:2006wt} 
  M.~Niedermaier and M.~Reuter,
  The Asymptotic Safety Scenario in Quantum Gravity,
  Living Rev.\ Rel.\  {\bf 9}, 5 (2006).

\bibitem{ONeill:1983} 
  B.~O'Neill,
  {\it Semi-Riemannian Geometry},
  Academic Press, San Diego (1983).

\bibitem{Pagani:2015ema} 
  C.~Pagani and R.~Percacci,
  Quantum gravity with torsion and non-metricity,
  preprint,
  [\href{http://arxiv.org/abs/1506.02882}{{\tt arXiv:1506.02882}}].

\bibitem{Parker:2009uva} 
  L.~E.~Parker and D.~J.~Toms,
  {\it Quantum Field Theory in Curved Spacetime: Quantized Fields and Gravity},
  Cambridge University Press, Cambridge (2009).

\bibitem{Percacci:2011uf} 
  R.~Percacci,
  Renormalization group flow of Weyl invariant dilaton gravity,
  New J.\ Phys.\  {\bf 13}, 125013 (2011),
  [\href{http://arxiv.org/abs/1110.6758}{{\tt arXiv:1110.6758}}].

\bibitem{Percacci:2002ie} 
  R.~Percacci and D.~Perini,
  Constraints on matter from asymptotic safety,
  Phys.\ Rev.\ D {\bf 67}, 081503 (2003),
  [\href{http://arxiv.org/abs/hep-th/0207033}{{\tt arXiv:hep-th/0207033}}].

\bibitem{Percacci:2003jz} 
  R.~Percacci and D.~Perini,
  Asymptotic safety of gravity coupled to matter,
  Phys.\ Rev.\ D {\bf 68}, 044018 (2003),
  [\href{http://arxiv.org/abs/hep-th/0304222}{{\tt arXiv:hep-th/0304222}}].

\bibitem{Perez:2003vx} 
  A.~Perez,
  Spin foam models for quantum gravity,
  Class.\ Quant.\ Grav.\  {\bf 20}, R43 (2003),
  [\href{http://arxiv.org/abs/gr-qc/0301113}{{\tt arXiv:gr-qc/0301113}}].

\bibitem{Reuter:1996cp} 
  M.~Reuter,
  Nonperturbative Evolution Equation for Quantum Gravity,
  Phys.\ Rev.\ D {\bf 57}, 971 (1998),
  [\href{http://arxiv.org/abs/hep-th/9605030}{{\tt arXiv:hep-th/9605030}}].

\bibitem{Reuter:2001ag} 
  M.~Reuter and F.~Saueressig,
  Renormalization group flow of quantum gravity in the Einstein-Hilbert truncation,
  Phys.\ Rev.\ D {\bf 65}, 065016 (2002),
  [\href{http://arxiv.org/abs/hep-th/0110054}{{\tt arXiv:hep-th/0110054}}].

\bibitem{Reuter:2012id} 
  M.~Reuter and F.~Saueressig,
  Quantum Einstein Gravity,
  New J.\ Phys.\  {\bf 14}, 055022 (2012),
  [\href{http://arxiv.org/abs/1202.2274}{{\tt arXiv:1202.2274}}].

\bibitem{Rogers:1979vp} 
  A.~Rogers,
  A Global Theory of Supermanifolds,
  J.\ Math.\ Phys.\  {\bf 21}, 1352 (1980).

\bibitem{Rovelli:2004tv} 
  C.~Rovelli,
  {\it Quantum Gravity},
  Cambridge University Press, Cambridge (2004).

\bibitem{Shapiro:2001rz} 
  I.~L.~Shapiro,
  Physical aspects of the space-time torsion,
  Phys.\ Rept.\  {\bf 357}, 113 (2002),
  [\href{http://arxiv.org/abs/hep-th/0103093}{{\tt arXiv:hep-th/0103093}}].

\bibitem{Shapiro:2014kma} 
  I.~L.~Shapiro and P.~M.~Teixeira,
  Quantum Einstein-Cartan theory with the Holst term,
  Class.\ Quant.\ Grav.\  {\bf 31}, 185002 (2014),
  [\href{http://arxiv.org/abs/1402.4854}{{\tt arXiv:1402.4854}}].

\bibitem{Thiemann:2008} 
  T.~Thiemann,
  {\it Modern Canonical Quantum General Relativity},
  Cambridge University Press, Cambridge (2008).

\bibitem{Vacca:2010mj} 
  G.~P.~Vacca and O.~Zanusso,
  Asymptotic Safety in Einstein Gravity and Scalar-Fermion Matter,
  Phys.\ Rev.\ Lett.\  {\bf 105}, 231601 (2010),
  [\href{http://arxiv.org/abs/1009.1735}{{\tt arXiv:1009.1735}}].

\bibitem{Vassilevich:2003xt} 
  D.~V.~Vassilevich,
  Heat kernel expansion: User's manual,
  Phys.\ Rept.\  {\bf 388}, 279 (2003),
  [\href{http://arxiv.org/abs/hep-th/0306138}{{\tt arXiv:hep-th/0306138}}].

\bibitem{Vilkovisky:1984st} 
  G.~A.~Vilkovisky,
  The Unique Effective Action in Quantum Field Theory,
  Nucl.\ Phys.\ B {\bf 234}, 125 (1984).

\bibitem{Vilkoviskii:1984un} 
  G.~A.~Vilkoviskii,
  The Gospel According To DeWitt,
  in: S.M. Christensen ( Ed.), {\it Quantum Theory of Gravity. Essays in honor of the 60th birthday of Bryce S. DeWitt}, p. 169-209,
  Adam Hilger Ltd, Bristol (1985).

\bibitem{Wilson:1974mb} 
  K.~G.~Wilson,
  The Renormalization Group: Critical Phenomena and the Kondo Problem,
  Rev.\ Mod.\ Phys.\  {\bf 47}, 773 (1975).

\bibitem{Wilson:1973jj} 
  K.~G.~Wilson and J.~B.~Kogut,
  The Renormalization group and the $\epsilon$ expansion,
  Phys.\ Rept.\  {\bf 12}, 75 (1974).

\bibitem{Wipf:2013vp} 
  A.~Wipf,
  {\it Statistical Approach to Quantum Field Theory},
  Lect.\ Notes Phys.\  {\bf 864}, Springer (2013).


\end{thebibliography}
